\pdfoutput=1
\documentclass[11pt,letterpaper]{article}

\usepackage[utf8]{inputenc}
\usepackage[T1]{fontenc}
\usepackage{lmodern}
\usepackage[letterpaper,margin=1in]{geometry}
\usepackage{microtype}
\usepackage{amsmath,amssymb,amsthm,mathtools}
\usepackage{thmtools}
\usepackage{needspace}
\usepackage{booktabs,array,longtable}
\usepackage{enumitem}
\usepackage{tikz}
\usepackage{pgfplots}
\pgfplotsset{compat=1.18}
\usepackage[table]{xcolor}
\usepackage[disable]{todonotes}
\usepackage[hidelinks,bookmarksdepth=3,hypertexnames=false]{hyperref}
\usepackage[capitalize,nameinlink]{cleveref}

\theoremstyle{plain}
\newtheorem{theorem}{Theorem}[section]
\newtheorem{lemma}[theorem]{Lemma}
\newtheorem{proposition}[theorem]{Proposition}
\newtheorem{corollary}[theorem]{Corollary}
\newtheorem{claim}[theorem]{Claim}
\newtheorem{conjecture}[theorem]{Conjecture}
\theoremstyle{definition}
\newtheorem{definition}[theorem]{Definition}

\theoremstyle{remark}
\newtheorem{remark}[theorem]{Remark}
\newtheorem*{remark*}{Remark}

\crefname{conjecture}{Conjecture}{Conjectures}
\crefname{claim}{Claim}{Claims}
\makeatletter
\@for\thm@env:=lemma,proposition,corollary,claim,conjecture,definition,example,remark\do{%
\edef\thm@tmp{\noexpand\AddToHook{env/\thm@env/begin}{\noexpand\crefalias{theorem}{\thm@env}}}\thm@tmp}
\makeatother

\numberwithin{equation}{section}

\DeclareMathOperator{\pw}{pw}
\DeclareMathOperator{\tw}{tw}
\DeclareMathOperator{\Corr}{Corr}
\DeclareMathOperator{\Cov}{Cov}
\DeclareMathOperator{\Var}{Var}
\DeclareMathOperator{\dist}{dist}
\DeclareMathOperator{\sech}{sech}
\DeclareMathOperator{\clip}{clip}
\DeclareMathOperator{\poly}{poly}
\newcommand{\E}{\mathbb{E}}
\newcommand{\R}{\mathbb{R}}
\newcommand{\N}{\mathbb{N}}
\renewcommand{\Pr}{\mathbb{P}}
\newcommand{\Ostar}{O^{*}}
\newcommand{\eps}{\varepsilon}
\newcommand{\ind}{\mathbf{1}}
\newcommand{\outfr}{\partial^{+}}
\newcommand{\infr}{\partial^{-}}
\newcommand{\seqC}{\mathsf{C}}
\newcommand{\seqM}{\mathsf{M}}
\newcommand{\seqF}{\mathsf{F}}
\newcommand{\tripP}{\mathsf{P}}
\newcommand{\tripR}{\mathsf{R}}
\newcommand{\tripS}{\mathsf{S}}
\newcommand{\cthree}{c_{3}}
\definecolor{loud}{HTML}{FFD400}
\definecolor{muted}{HTML}{8A94A6}
\newcommand{\lo}[1]{\cellcolor{loud}{\boldmath$#1$}}
\newcommand{\bs}[1]{\textcolor{muted}{$#1$}}
\newcommand{\lobs}[1]{\cellcolor{loud}\textcolor{muted}{\boldmath$#1$}}

\title{Path Decompositions of Sparse Graphs via Gaussian Sweeps\\
and Faster Exponential Algorithms}
\author{Serge Gaspers\\[2pt]
  \small UNSW Sydney, Australia\\
\small\texttt{serge.gaspers@unsw.edu.au}}
\date{}

\begin{document}
\pagenumbering{roman}
\maketitle
\thispagestyle{empty}

\begin{abstract}
  \normalsize
  Using the bisection bound of Monien and Preis, Fomin and H{\o}ie proved in 2006 that
every graph of maximum degree~$3$ on $n$ vertices has pathwidth at most $n/6+o(n)$.
We prove that the pathwidth, bisection width and cutwidth of these graphs are at most
$\cthree\,n+o(n)$, where $\cthree=\frac{3}{4\pi}\arccos\frac56\approx0.1398$. For
bisection width this improves the bound $\frac n6+o(n)$ of Monien and Preis (2006) and
extends to all such graphs the bound that Lichev and Mitsche (2023) obtained for random
cubic graphs. For maximum degree~$4$, we prove that bisection width and cutwidth are at
most $\frac n3+o(n)$, improving the bound $\frac{2n}5+o(n)$ of Monien and Preis for
$4$-regular graphs, and that pathwidth is at most $0.2458n$, improving the bound
$\frac n3$ of Fomin, Gaspers, Saurabh and Stepanov (2009). We show that every graph with
$m$ edges has pathwidth at most $\frac{7}{48}m+o(n)$, improving the bound $\frac{13}{75}m$ of
Scott and Sorkin (2007), and we give several further bounds, none best for all graphs,
of the form $\sum_{d\ge3}a_d\,n_d+o(n)$ with coefficients $a_d<1$, where $n_d$ is the number of vertices of degree~$d$. The path
decompositions, and linear orderings of the stated cutwidth, can be computed in
polynomial time.

Instead of deriving pathwidth bounds from bisection width, we give each vertex a
Gaussian label, a weighted sum of independent standard normal variables, one for each
vertex of a ball around it, so that adjacent vertices have strongly correlated labels. Sweeping a threshold through
the labels gives a path decomposition whose bags are frontiers: the vertices
above the threshold that have a neighbor below it. We bound the probability that a vertex lies in a frontier from above by
comparison with an explicit Gaussian model of a vertex and its neighbors; a
refinement of this sweep gives the width $\cthree\,n+o(n)$ for maximum degree~$3$.

Faster exact algorithms follow. For graphs of bounded maximum or average degree, they
include algorithms for feedback vertex set and the traveling salesperson problem. For
general graphs, a branching algorithm that finishes sparse instances by dynamic
programming becomes faster for minimum weight dominating set, and the bound in $m$
gives faster algorithms for Max 2-CSP and for counting subgraphs with $k$ edges.
{}
\end{abstract}

\vfill
\noindent{\small Code and formal proofs: \cite{GaspersCompanion2026},
\url{https://serggasp.github.io/pathwidth-site/}.}

\clearpage
\setcounter{tocdepth}{1}
\begingroup
\let\tocaddvspace\addvspace
\renewcommand*{\addvspace}[1]{\tocaddvspace{0.6em}}
\tableofcontents
\endgroup

\clearpage
\pagenumbering{arabic}

\section{Introduction}\label{sec:intro}

How large can the pathwidth of a sparse graph be? For graphs of maximum degree~$3$ the best upper
bound has stood for twenty years: Fomin and H{\o}ie~\cite{FominHoie2006} proved that
they have pathwidth at most $\frac16n+o(n)$, where $n$ is the number of vertices. We prove that graphs of maximum degree~$3$
have pathwidth at most
\[
  \cthree\, n+o(n),\qquad \cthree=\frac{3}{4\pi}\arccos\frac56\approx 0.13982,
\]
with corresponding improvements at every degree, per edge, and in the running times
of the exact algorithms built on these bounds.

The proof of the old bound cuts the graph into two halves along a bisection with
about $\frac n6$ edges, which exists by the theorem of Monien and
Preis~\cite[Theorem~1]{MonienPreis2006}, and then decomposes each half with a prescribed
boundary. Each of the two steps needs width about $\frac n6$. The per-degree bounds
of Fomin, Gaspers, Saurabh and Stepanov~\cite{FominGaspersSaurabhStepanov2009} and
of Edwards and McDermid~\cite{EdwardsM15} extend the bound to all graphs by deleting
vertices of large degree, and the per-edge bound $\frac{13}{75}m$ of Scott and
Sorkin~\cite{ScottSorkin2007}, as well as the bound $\frac m{5.769}$ of Kneis, M\"olle,
Richter and Rossmanith~\cite{KneisMRR2009}, also follows from these per-degree
bounds. Many exact
algorithms for sparse instances run a dynamic program over a decomposition of such a
width, and their running times are stated through these constants. We replace
bisection by a different construction: a threshold sweep through a correlated
Gaussian labeling of the vertices. Counting only the edges that cross a threshold of
this sweep already gives pathwidth at most
$\left(\frac{3}{2\pi}\arcsin\frac13\right)n+o(n)\approx0.1623n$, below $\frac n6$,
with a short complete proof whose probabilistic tools are the random-hyperplane lemma
of Goemans and Williamson~\cite{GoemansWilliamson1995} and Chebyshev's inequality
(\cref{thm:cut-sweep}).

\subsection{Width bounds}\label{sec:intro-bounds}

Throughout, graphs are finite and simple, $n=|V(G)|$, $m=|E(G)|$, $n_d=n_d(G)$ is
the number of vertices of degree $d$, and $\Delta(G)$ is the maximum degree. Every
remainder $o(n)$ (or $o(m)$) stands for a function of $n$ (or $m$) alone, the same
for all graphs in the stated class. Since every path decomposition is a tree
decomposition, all bounds also hold for treewidth.

\begin{theorem}[bounded degree]\label{thm:intro-bounded}
  Let $G$ be a graph with $n$ vertices.
  \begin{enumerate}[label=\textup{(\alph*)},align=left,leftmargin=2.6em,labelwidth=1.6em,labelsep=1em,itemsep=2pt]
    \item\label{thm:intro-bounded:a} If $\Delta(G)\le3$, then
      $\pw(G)\le\cthree\,n+o(n)$, where
      $\cthree=\frac{3}{4\pi}\arccos\frac56\approx0.13982$.
    \item\label{thm:intro-bounded:b} If $\Delta(G)\le4$, then
      $\pw(G)\le F_4\,n+o(n)$, where $F_4\approx0.24577$ is the maximum over the
      threshold of an explicit Gaussian integral, the ideal $4$-star function
      \eqref{eq:overview-f}.
  \end{enumerate}
\end{theorem}

Part~\ref{thm:intro-bounded:a} is \cref{thm:cubic}, proved in \cref{sec:cubic}, and
part~\ref{thm:intro-bounded:b} is \cref{thm:deg4}\ref{thm:deg4:a}, proved in
\cref{sec:plain-sweep}.
The constant $\cthree$ also bounds two other width parameters. The \emph{bisection
width} $\operatorname{bw}(G)$ is the minimum number of edges between the two sides of
a partition of $V(G)$ into parts whose sizes differ by at most one, and the
\emph{cutwidth} $\operatorname{cw}(G)$ is the minimum, over all linear orderings of
$V(G)$, of the maximum number of edges joining a vertex before a gap of the ordering
to a vertex after it.

\begin{corollary}[bisection width and cutwidth]\label{cor:bisection}
  Every graph $G$ with $\Delta(G)\le3$ satisfies
  $\operatorname{bw}(G)\le\operatorname{cw}(G)\le\cthree\,n+o(n)$.
\end{corollary}

The corollary follows from \cref{thm:intro-bounded}\ref{thm:intro-bounded:a}
through the chain $\operatorname{bw}(G)\le\operatorname{cw}(G)\le\pw(G)+2$ for
maximum degree~$3$: bisection width is at most
cutwidth~\cite[Lemma~2.4]{DiazPetitSerna2002}, cutwidth equals the search number for
maximum degree~$3$~\cite[Corollary~3.3]{MakedonSudborough1989}, the search number is
at most the vertex separation number plus~$2$~\cite[Theorem~2.1]{EllisSudboroughTurner1994}
(see also \cite{Bodlaender1998}), and vertex separation equals
pathwidth~\cite[Theorem~3.1]{Kinnersley1992}.

For bisection width, \cref{cor:bisection} improves the bound $\frac n6+o(n)$ that
Monien and Preis~\cite[Theorem~1]{MonienPreis2006} proved for cubic graphs.
For cutwidth, the previous bound $\frac n6+o(n)$ followed from the theorem of Fomin
and H{\o}ie by the same relations. \Cref{cor:alg-cutwidth} computes, in deterministic
time $n^{2+o(1)}$, an ordering of cutwidth at most $\cthree n+o(n)$ and a bisection
with at most that many edges between its sides.

The constant $\cthree$ is not new as a bisection constant: Lichev and
Mitsche~\cite{LichevMitsche2023} proved that \emph{random} cubic graphs have bisection
width at most $\cthree n+o(n)$ asymptotically almost surely (stated there as
$0.139822n$), and at least $0.103295n$. What is new is that the same constant holds
for every graph of maximum degree~$3$, and for pathwidth. The lower bound for random
cubic graphs gives pathwidth at least $0.103295n-2$.
We conjecture (\cref{conj:ideal-wave}) the upper bound
$\pw(G)\le(\cthree-3p_{\mathrm{ch}})n+o(n)\approx0.1322n+o(n)$ for cubic graphs.

The simplest form of our method, which counts the edges that cross a threshold,
also bounds the cutwidth for larger maximum degree. \Cref{thm:cut-sweep-p} in
\cref{sec:warmup} shows that every graph $G$ with $\Delta(G)\le p$ has cutwidth at most
$\gamma_pm+o(n)$, where $\gamma_p=\frac1\pi\arccos\frac{2\sqrt{p-1}}p$. For
$\Delta(G)\le4$ this is $\frac m6+o(n)\le\frac n3+o(n)$, which improves the second
bound of Monien and Preis~\cite[Theorem~2]{MonienPreis2006}, $\frac{2n}5+o(n)$ for
the bisection width of $4$-regular graphs; for $\Delta(G)\le5$ it is at most
$0.5121n+o(n)$. For $p\in\{4,5\}$ the orderings can be computed in deterministic
polynomial time (\cref{thm:alg-cut-sweep}). The bound also gives the edge isoperimetric number
(\cref{sec:warmup-remarks}).

For graphs with vertices of all degrees we prove one bound per vertex, with a
closed-form coefficient for every degree, and one bound per edge.

\begin{theorem}[all graphs]\label{thm:main}
  Every graph $G$ satisfies $\pw(G)\le\sum_{d\ge 3}a_d\,n_d+o(n)$ for the sequence
  \[
    a_d=
    \begin{cases}
      \dfrac{3d-4}{32} & \text{for } 3\le d\le 6,\\[8pt]
      1-\dfrac{111}{28\,(d+1)} & \text{for } 7\le d\le 10,\\[8pt]
      1-\dfrac{4}{d+1} & \text{for } d\ge 11.
    \end{cases}
  \]
\end{theorem}
The first coefficients are $(a_3,a_4,a_5,a_6)=\bigl(\tfrac{5}{32},\,\tfrac14,\,\tfrac{11}{32},\,\tfrac{7}{16}\bigr)$;
rounded up to four decimals, $(a_3,\dots,a_{11})$ is
$(0.1563,\,0.2500,\,0.3438,\,0.4375,\,0.5045,\,0.5596,\,0.6036,\,0.6397,\,0.6667)$.

\begin{corollary}[edges]\label{cor:edges}
  Every graph $G$ satisfies $\pw(G)\le \frac{7}{48}\, m+o(n)$.
\end{corollary}

Both are proved in \cref{sec:bootstrap}. \Cref{tab:width-bounds} compares the
coefficients with the previous ones. The relative improvement is largest at
degree~$4$: from $\frac13$ to $\frac14$ per vertex of degree~$4$, and to
$F_4\approx0.2458$ for the whole class $\Delta\le4$. The per-edge bound comes from
the degree $d$ that maximizes $\frac{2a_d}{d}$. For the previous coefficients this
is $d=5$, with $\frac{13}{75}\approx0.1733$; for ours it is $d=6$, with
$\frac{2a_6}{6}=\frac{7}{48}\approx0.1458$. For large degrees the coefficients of
\cref{thm:main} approach~$1$ as $1-\frac{4}{d+1}$, against
$1-\frac{3.5}{d+1}+O(d^{-3})$ for the coefficients $g_{1/6}(d)$ of Edwards and
McDermid~\cite{EdwardsM15}.

\begin{table}[tb]
  \centering
  \caption{Upper bounds on pathwidth, without the lower-order terms: per vertex for
    the classes $\Delta\le3$ and $\Delta\le4$ (\cref{thm:intro-bounded}), per vertex of
    degree $d$ (\cref{thm:main}) and per edge (\cref{cor:edges}). Previous bounds:
    \cite{FominHoie2006} for $\Delta\le3$; \cite{FominGaspersSaurabhStepanov2009} for
    $\Delta\le4$ and for degrees $3$ to $6$; the coefficients $g_{1/6}(d)$
    of~\cite{EdwardsM15} for $d\ge7$; and $\frac{13}{75}m$
    from~\cite{ScottSorkin2007} (\cite{KneisMRR2009} state $\frac m{5.769}$). Decimals are rounded to four places. In
    \cref{thm:main} the coefficient of degree~$3$ exceeds $\cthree$: the triple behind it
    trades a larger coefficient at degree~$3$ for smaller ones at degrees $4$ and $5$, and
  for $\Delta\le3$ \cref{thm:intro-bounded}\ref{thm:intro-bounded:a} applies.}
  \label{tab:width-bounds}
  \small
  \setlength{\tabcolsep}{4.5pt}
  \renewcommand{\arraystretch}{1.25}
  \begin{tabular}{@{}l cc ccccc c c@{}}
    \toprule
    & $\Delta\le3$ & $\Delta\le4$ & $d=3$ & $d=4$ & $d=5$ & $d=6$ & $d=7$ & $d\ge11$ & per edge\\
    \midrule
    previous & $\frac16$ & $\frac13$ & $\frac16$ & $\frac13$ & $\frac{13}{30}$ & $\frac{23}{45}$
    & $0.5698$ & $1-\frac{3.5}{d+1}+O(d^{-3})$ & $\frac{13}{75}$\\
    this paper & $0.1398$ & $0.2458$ & $\frac{5}{32}$ & $\frac14$ & $\frac{11}{32}$ & $\frac{7}{16}$
    & $0.5045$ & $1-\frac{4}{d+1}$ & $\frac{7}{48}$\\
    \bottomrule
  \end{tabular}
\end{table}

\Cref{thm:main} is one of several bounds of this form. \Cref{sec:sequences} states all
coefficient vectors of the paper as \emph{pathwidth sequences}, among them the pair
$(\cthree,0.2627)$ for maximum degree~$4$ (\cref{thm:switched4}), and extends each of
them to all degrees by a bootstrap lemma. It also shows that the bootstrap of
$(\frac5{32},\frac14,\frac{11}{32})$, which \cref{thm:main} rounds up to closed forms, has the
smallest coefficient at every degree $d\ge5$.

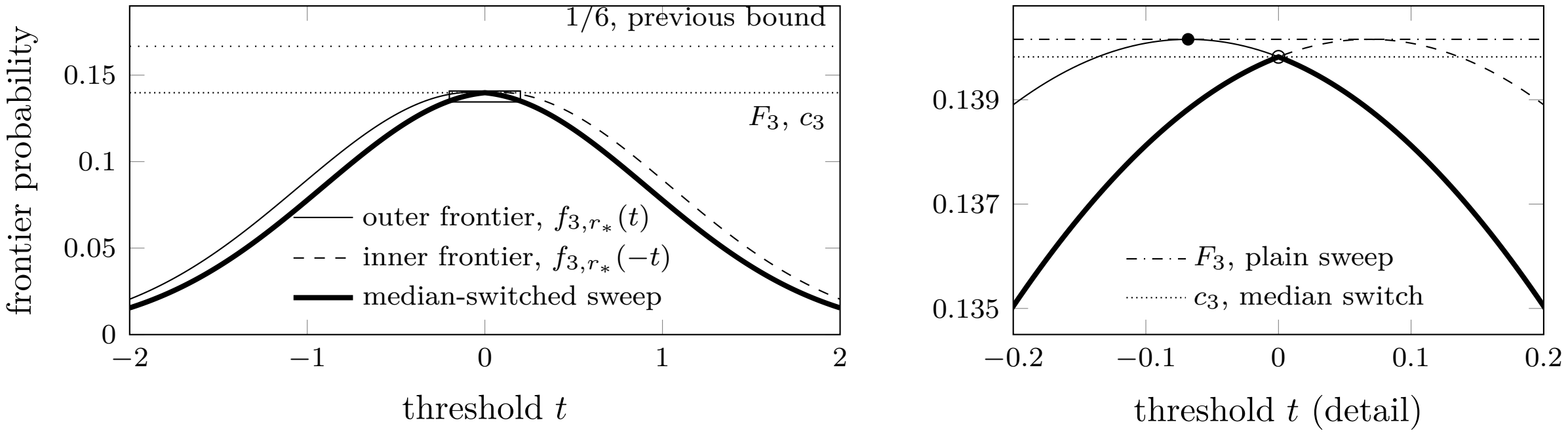
\begin{figure}[tb]
  \centering
  \begin{tikzpicture}
    \begin{axis}[name=full,
        width=0.53\textwidth, height=4.9cm,
        xmin=-2, xmax=2, ymin=0, ymax=0.19,
        xlabel={threshold $t$}, ylabel={frontier probability},
        xlabel style={font=\small}, ylabel style={font=\small},
        tick label style={font=\scriptsize},
        ytick={0,0.05,0.10,0.15},
        yticklabel style={/pgf/number format/fixed,/pgf/number format/precision=2},
        legend style={font=\scriptsize, at={(0.5,0.03)}, anchor=south, draw=none, fill=none},
      legend cell align=left]
      \addplot[black, thin] table[x=t,y=f]{figures/ideal-star-curve.dat};
      \addlegendentry{outer frontier, $f_{3,r_*}(t)$}
      \addplot[black, thin, dashed] table[x=t,y=fneg]{figures/ideal-star-curve.dat};
      \addlegendentry{inner frontier, $f_{3,r_*}(-t)$}
      \addplot[black, line width=1.5pt] table[x=t,y=env]{figures/ideal-star-curve.dat};
      \addlegendentry{median-switched sweep}
      \addplot[black, dotted, domain=-2:2, samples=2] {1/6};
      \addplot[black, densely dotted, domain=-2:2, samples=2] {0.1398221};
      \node[font=\scriptsize, anchor=south east] at (axis cs:2,0.1667) {$1/6$, previous bound};
      \node[font=\scriptsize, anchor=north east] at (axis cs:2,0.1398) {$F_3$, $\cthree$};
      \draw[thin] (axis cs:-0.2,0.1345) rectangle (axis cs:0.2,0.1408);
    \end{axis}
    \begin{axis}[name=zoom, at={(full.right of south east)}, anchor=left of south west,
        xshift=0.6cm,
        width=0.42\textwidth, height=4.9cm,
        xmin=-0.2, xmax=0.2, ymin=0.1345, ymax=0.1408,
        xlabel={threshold $t$ (detail)},
        xlabel style={font=\small},
        tick label style={font=\scriptsize},
        xtick={-0.2,-0.1,0,0.1,0.2},
        ytick={0.135,0.137,0.139,0.141},
        yticklabel style={/pgf/number format/fixed,/pgf/number format/precision=3},
        scaled y ticks=false,
        legend style={font=\scriptsize, at={(0.5,0.03)}, anchor=south, draw=none, fill=none},
      legend cell align=left]
      \addplot[black, thin, forget plot] table[x=t,y=f]{figures/ideal-star-curve.dat};
      \addplot[black, thin, dashed, forget plot] table[x=t,y=fneg]{figures/ideal-star-curve.dat};
      \addplot[black, line width=1.5pt, forget plot] table[x=t,y=env]{figures/ideal-star-curve.dat};
      \addplot[black, dashdotted, domain=-0.2:0.2, samples=2] {0.1401583};
      \addlegendentry{$F_3$, plain sweep}
      \addplot[black, densely dotted, domain=-0.2:0.2, samples=2] {0.1398221};
      \addlegendentry{$\cthree$, median switch}
      \addplot[black, only marks, mark=*, mark size=1.6pt, forget plot] coordinates {(-0.06814,0.1401583)};
      \addplot[black, only marks, mark=o, mark size=1.8pt, forget plot] coordinates {(0,0.1398221)};
    \end{axis}
  \end{tikzpicture}
  \caption{The ideal $3$-star function $f_{3,r_*}$ at the cubic correlation
    $r_*=\frac{2\sqrt2}3$, computed from the ideal star function \eqref{eq:overview-f} of
    \cref{sec:overview}, and its reflection. The right panel enlarges the box drawn in the
    left panel, and the dotted line at $1/6$ is the previous bound for maximum
    degree~$3$~\cite{FominHoie2006}. The plain sweep has width at most
    $F_3n+o(n)\approx0.1402n$, where $F_3$ is the maximum of the thin solid curve,
    attained at $t\approx-0.068$ (filled dot). The median-switched sweep keeps inner frontiers for $t<0$ and outer
    frontiers for $t>0$. It has width at most $\cthree n+o(n)\approx0.1398n$, where
    $\cthree$ is the maximum of the thick envelope, attained at $t=0$ (open dot): by the switch lemma
    (\cref{sec:overview-switch}), the passage from inner to outer frontiers at $t=0$ adds
  nothing at leading order.}
  \label{fig:ideal-star}
\end{figure}

\paragraph{The method in one paragraph.} We give every vertex a Gaussian label, a
weighted sum of independent standard normal variables over a ball around the vertex
with critically decaying weights, and sweep a threshold through the labels. The
vertex separation layout turns the sweep into a path decomposition whose width is the
size of the largest \emph{frontier}, the set of vertices on one side of the threshold
with a neighbor on the other side. Whether a vertex lies in a frontier depends only on
the labels of its closed neighborhood, a \emph{star}. We bound that probability from above by
the one for an explicit \emph{ideal} star, using that the regular simplex minimizes
Gaussian orthant probabilities. A threshold has two frontiers: its \emph{outer}
frontier is the set of vertices above the threshold with a neighbor at or below it,
and its \emph{inner} frontier is the set of vertices at or below the threshold with a
neighbor above it. For maximum degree~$3$, outer frontiers at every threshold give
width about $0.1402n$; inner frontiers for thresholds below the median and outer
frontiers above it lower the width to $\cthree n+o(n)$.
\Cref{fig:ideal-star} shows the two frontier curves of the ideal star and the envelope
that bounds the median-switched sweep; \cref{sec:overview} develops these ideas.

{}

\subsection{Algorithms and consequences}\label{sec:intro-apps}

The proofs of the width bounds are probabilistic. \Cref{sec:algorithms} makes them
algorithmic: it approximates the Gaussian variables by a finite rational probability distribution and
derandomizes by the method of conditional expectations. \Cref{thm:algo-precise}
there is the precise form of the following theorem and states the computational
model.

\begin{theorem}[computing the decompositions]\label{thm:algo}
  Let $\beta(G)$ be any of the bounds of
  \cref{thm:intro-bounded,thm:main,cor:edges,thm:deg4,thm:deg5,thm:switched4,thm:pw-seq,cor:table}
  (for \cref{thm:deg4}\ref{thm:deg4:a}, $F_4n$), without its remainder term. For every
  graph $G$ in the class of the bound, a path
  decomposition of $G$ of width at most $\beta(G)+o(n)$ can be computed in deterministic time $n^{2+o(1)}+O(n(n+m))$, and
  by a Las Vegas algorithm in expected time $n^{1+o(1)}+O(n(n+m))$.
\end{theorem}

\Cref{part:apps} follows the new constants through the literature of exact
algorithms. The five most striking consequences are the following, and
\cref{tab:apps} lists forty-five rows of bounds that, as far as we know, two recent
preprints would not supersede (see the paragraph on the Exponential Time Hypothesis
below). To our knowledge, each bound of \cref{thm:apps-summary} and \cref{tab:apps}
is the fastest known for its problem and graph class (for a deterministic bound,
among deterministic algorithms), subject to the table notes.

\begin{theorem}[faster exact algorithms]\label{thm:apps-summary}
  Polynomial factors are omitted; $r$ is the domain size, $n$ the number of variables
  or vertices, $m$ the number of constraints or edges and $k$ the pattern size. The
  following problems can be solved in the stated time.
  \begin{enumerate}[label=\textup{(\roman*)},align=left,leftmargin=2.8em,labelwidth=1.8em,labelsep=1em,itemsep=2pt]
    \item Minimum Weight Dominating Set and \#Minimum Weight Dominating Set, for vertex
      weights with polynomially many distinct subset sums, in time $O(1.4815^n)$.
    \item Max $(r,2)$-CSP in time $r^{7m/48+o(m)}$.
    \item The number of subgraphs of a graph that are isomorphic to a given pattern
      with $k$ edges, in time $k^{O(k)}n^{7k/48+o(k)}$.
    \item Permutation Pattern Matching, in time $2^{O(k)}n^{253k/1215+o(k)}$, and $k$-opt
      Optimization for the traveling salesperson problem, in time $O(n^{(253/1215+\eps_k)k})$
      with $\eps_k\to0$ as $k\to\infty$.
    \item On graphs of maximum degree $4$, Feedback Vertex Set by a Monte Carlo
      algorithm in time $O(1.2600^n)$ and deterministically in time $O(1.4853^n)$, and the
      traveling salesperson problem deterministically in time $O(1.4214^n)$; on graphs of
      maximum degree $3$, Hamiltonian Cycle and Connected Dominating Set by Monte Carlo
      algorithms in time $O(1.1312^n)$ and $O(1.1661^n)$.
  \end{enumerate}
\end{theorem}

Items (i) to (v) are \cref{thm:ds}, \cref{thm:max2csp}\ref{it:max2csp-edges},
\cref{thm:subgraphs}, \cref{thm:perm} and \cref{thm:connectivity}.
\Cref{tab:apps} gives, for each bound, the previous bound, the work that proves it,
and the theorem that proves the new one. Polynomial factors are omitted, and every
exponent carries an additive $o(\cdot)$ term. The notation is that of
\cref{thm:apps-summary}; in addition, $\bar d$ is the average degree, and for Co-Path
Packing and $k$-Co-Path Set, $k$ is the number of deleted vertices or edges. The
algorithms of \cref{thm:connectivity} for connectivity and path problems are Monte
Carlo algorithms unless marked det.

\paragraph{The Exponential Time Hypothesis.} Two recent preprints, by
OpenAI~\cite{OpenAI2026ETH} and Anthropic~\cite{Claude2026ETH}, claim to refute the
Exponential Time Hypothesis (ETH)~\cite{ImpagliazzoP2001} by giving a
subexponential-time algorithm for \#3-SAT. As far as we know, they do not affect the
bounds of \cref{tab:apps}. Part~II contains seventeen further bounds that they would
supersede; its opening lists them (twelve in \cref{tab:apps-affected} and five in
polynomial space) and explains how the preprints reach them.

\subsection{Evidence and verification}\label{sec:intro-evidence}

\Cref{thm:cubic}, which is \cref{thm:intro-bounded}\ref{thm:intro-bounded:a}, and
\cref{cor:bisection} use no computation, and \cref{thm:cubic} has a complete formal
proof in the Lean proof assistant, without hypotheses. Their written proofs use the
simplex theorem for three Gaussian variables, which is a
case of Mulgund's theorem~\cite{Mulgund2026}, a preprint whose Lean formalization we
checked (\cref{app:formal}); for three variables it also has an elementary
proof (\cref{rem:stars-mulgund}), which the formal proof of \cref{thm:cubic} uses. The other width bounds also
use the cases of four and five variables, which rest on the same preprint; for four
variables the theorem also follows from the four-cap theorem of Fejes
T\'oth~\cite{FejesToth1950,Boroczky2004}. They rest in addition on computer-assisted certificates in ball
arithmetic and exact rational arithmetic (\cref{prop:F3F4,prop:certificate}, and two
certificates for \cref{thm:switched4}).
\Cref{thm:intro-bounded}\ref{thm:intro-bounded:b},
\cref{thm:deg4,thm:deg5,thm:switched4,thm:main}, \cref{cor:edges} and \cref{thm:pw-seq}
for its seeds other than $\seqC$ also have formal proofs in Lean, conditional on these
inputs; for the seed $\seqC$ the formal proof is unconditional. \Cref{app:formal} contains further details on the formal verification and states which result needs which input.

Throughout, ``the accompanying
code'' refers to the programs that perform the computer-assisted steps; they and the
Lean proofs are published on the companion website of this paper~\cite{GaspersCompanion2026}.

\subsection{Related work}\label{sec:intro-related}

\emph{Width bounds.} Fomin, Grandoni and Kratsch~\cite[\S4.3.1]{FominGK2005} asked
for the best possible upper bound on the pathwidth and treewidth of graphs of maximum
degree~$3$, and Fomin and H{\o}ie~\cite[\S5]{FominHoie2006} and Fomin and
Kratsch~\cite[notes to Chapter~5, p.~100]{FominKratsch2010} asked to reduce the gap
between the upper bound $\frac n6$ of Fomin and H{\o}ie, the best previous one for
maximum degree~$3$, and a lower bound $0.082n$. The question of the largest pathwidth
of an $n$-vertex cubic graph is listed among the unsolved problems of
Wikipedia~\cite{WikipediaUnsolved2026,WikipediaCubicGraph2026,WikipediaPathwidth2026}.
Van Rooij asked at a Dagstuhl seminar for stronger bounds on the treewidth of graphs
of bounded degree, with associated tree decompositions computable in polynomial
time~\cite{Dagstuhl08431OpenProblems}. He stated the cubic case as an open problem:
to prove that cubic graphs have treewidth at most $c_tn+o(n)$ for some constant
$c_t<\frac16$, with tree decompositions of this width computable in polynomial
time~\cite[Open Problem~14.2]{vanRooij2011}. \Cref{thm:intro-bounded,thm:algo} solve
Open Problem~14.2 with $c_t=\cthree$, even for path decompositions.

\emph{Gaussian methods on sparse graphs.} Thresholding Gaussian waves is a classical
local algorithm on random regular graphs and graphs of large
girth~\cite{CsokaGHV2015,Lyons2017CPC,HarangiVirag2015}. The cut density
$\frac{3}{2\pi}\arccos\frac{2\sqrt2}{3}$ of a single threshold appears in
Lyons~\cite{Lyons2017CPC}, and the correlation sequence of our labels on the
$3$-regular tree is the extremal one of Backhausz, Szegedy and
Vir\'ag~\cite{BackhauszSzegedyVirag2015} for factor of i.i.d.\ processes. Lichev and
Mitsche~\cite{LichevMitsche2023} used such waves, with a selection of helpful
vertices, for the bisection width of random cubic graphs. Their analysis uses that
random cubic graphs converge locally to the $3$-regular tree, where the correlations
of the wave are explicit. In an arbitrary graph, short cycles leave the correlation
of two neighbors of a vertex unknown; the star comparison of
\cref{sec:overview-stars} takes the place of the tree structure. Orderings by a Gaussian
projection with bounds on prefix boundaries appear in the approximation algorithms of
Feige~\cite{Feige2000} and of Louis, Raghavendra and
Vempala~\cite{LouisRaghavendraVempala2013}, with non-constant-factor losses. We are not aware of any previous work that uses a threshold-by-threshold Gaussian layout to bound a width parameter of all graphs in a class, nor of the star comparison.

{}
\section{Technical overview}\label{sec:overview}

\theoremstyle{plain}
\newtheorem*{ovlemma}{Lemma}
\newtheorem*{ovtheorem}{Theorem}

Our decompositions are of a different kind from the bisection-based ones of
\cref{sec:intro}. Every vertex receives a real label, a threshold sweeps through the
labels, and each bag is an outer frontier, the set of vertices above the threshold
with a neighbor at or below it, together with the vertex being introduced. The
analysis of such a sweep splits into a local question, how likely a given vertex is
to lie on the frontier of the current threshold, and a global step: the frontier
sizes concentrate around their means. This section explains the ideas without
proofs, in the order of Part~I, and points to the exact statements.
\Cref{sec:overview-sweeps,sec:overview-labels} describe the sweep and the labels,
\cref{sec:overview-stars} bounds the probability that a vertex lies on a frontier,
\cref{sec:overview-switch} improves the layout for maximum degree~$3$,
\cref{sec:overview-bootstrap} extends the bounds to mixed and large degrees, and
\cref{sec:overview-algorithms} makes them algorithmic and explains how the widths
become running times.

\subsection{Sweeps and the layout lemma}\label{sec:overview-sweeps}

Let $y\colon V(G)\to\R$ be an injective labeling. For a threshold $t\in\R$, the
sublevel set $S_t=\{v:y_v\le t\}$ has an \emph{outer frontier} $\outfr S_t$ and an
\emph{inner frontier} $\infr S_t$,
\[
  \outfr S_t=\{v:\ y_v>t,\ \exists u\in N(v):\ y_u\le t\},\qquad
  \infr S_t=\{u:\ y_u\le t,\ \exists v\in N(u):\ y_v>t\}.
\]
Every path decomposition in Part~I is built from the layouts of the following lemma.

\begin{ovlemma}[layout; \cref{lem:layout}]
  For every injective labeling $y$ of a graph $G$,
  $\pw(G)\le\max_t|\outfr S_t|$ and $\pw(G)\le\max_t|\infr S_t|$.
\end{ovlemma}

\begin{figure}[tb]
  \centering
  \begin{minipage}[c]{0.6\textwidth}
    \centering
    \begin{tikzpicture}[x=2.1cm,y=0.95cm,
        sv/.style={circle,draw,fill=black,inner sep=0pt,minimum size=5.5pt},
        ov/.style={circle,draw,fill=white,inner sep=0pt,minimum size=5.5pt},
        fv/.style={circle,draw,very thick,fill=white,inner sep=0pt,minimum size=5.5pt},
      lab/.style={font=\scriptsize}]
      \node[sv] (v1) at (-1.6,0) {}; \node[lab,above left=0pt] at (v1) {$v_1$};
      \node[sv] (v2) at (-1.3,1) {}; \node[lab,below right=0pt] at (v2) {$v_2$};
      \node[sv] (v3) at (-1.0,0) {}; \node[lab,above=1pt] at (v3) {$v_3$};
      \node[sv] (v4) at (-0.7,1) {}; \node[lab,below left=0pt] at (v4) {$v_4$};
      \node[fv] (v5) at (-0.4,0) {}; \node[lab,above right=0pt] at (v5) {$v_5$};
      \node[fv] (v6) at (-0.15,1) {}; \node[lab,below=1pt] at (v6) {$v_6$};
      \node[fv] (v7) at (0.1,0) {}; \node[lab,above=1pt] at (v7) {$v_7$};
      \node[fv] (v8) at (0.35,1) {}; \node[lab,below=1pt] at (v8) {$v_8$};
      \node[ov] (v9) at (0.6,0) {}; \node[lab,above=1pt] at (v9) {$v_9$};
      \node[ov] (v10) at (0.9,1) {}; \node[lab,below=1pt] at (v10) {$v_{10}$};
      \node[ov] (v11) at (1.2,0) {}; \node[lab,above left=0pt] at (v11) {$v_{11}$};
      \node[ov] (v12) at (1.5,1) {}; \node[lab,below right=0pt] at (v12) {$v_{12}$};
      \draw (v1)--(v2) (v1)--(v3) (v2)--(v4) (v6)--(v8) (v7)--(v9) (v9)--(v11)
      (v10)--(v12) (v11)--(v12);
      \draw (v6) to[bend left=60] (v10);
      \draw (v8) to[bend left=60] (v12);
      \draw (v5) to[bend right=60] (v9);
      \draw (v7) to[bend right=60] (v11);
      \draw[very thick] (v3)--(v5) (v4)--(v5);
      \draw[very thick] (v2) to[bend left=60] (v6);
      \draw[very thick] (v4) to[bend left=60] (v8);
      \draw[very thick] (v3) to[bend right=60] (v7);
      \draw[dashed] (-0.55,-1.0) -- (-0.55,1.75);
      \node[lab,anchor=north] at (-0.55,-1.05) {$t$};
      \draw[->] (-1.8,-1.0) -- (1.75,-1.0);
      \node[lab,anchor=north west] at (1.1,-1.05) {label $y_v$};
    \end{tikzpicture}
  \end{minipage}\hfill
  \begin{minipage}[c]{0.37\textwidth}
    \centering
    \begin{tikzpicture}[scale=0.9,
        v/.style={circle,draw,fill=white,inner sep=0pt,minimum size=5.5pt},
      lab/.style={font=\scriptsize}]
      \node[v] (c) at (0,0) {};
      \node[v] (a) at (90:1.1) {};
      \node[v] (b) at (210:1.1) {};
      \node[v] (d) at (330:1.1) {};
      \draw (c)--(a) (c)--(b) (c)--(d);
      \node[lab,right=2pt] at (c) {$Y^*_0=Z$};
      \node[lab,above=1pt] at (a) {$Y^*_1=rZ+s_rU^*_1$};
      \node[lab,below left=0pt] at (b) {$Y^*_2$};
      \node[lab,below right=0pt] at (d) {$Y^*_3$};
      \begin{scope}[xshift=3.3cm]
        \draw[thin] (0,0) circle (0.75);
        \draw[->] (0,0) -- (90:0.75); \draw[->] (0,0) -- (210:0.75); \draw[->] (0,0) -- (330:0.75);
        \node[lab] at (90:1.0) {$w_1$};
        \node[lab] at (210:1.02) {$w_2$};
        \node[lab] at (330:1.02) {$w_3$};
        \node[lab] at (0,-1.3) {$U^*_i=\langle w_i,G\rangle$};
      \end{scope}
    \end{tikzpicture}
  \end{minipage}
  \caption{Left: a sweep. The vertices are placed at their labels; the dashed line is
    the threshold $t$, the filled vertices form $S_t$, the thick edges cross the
    threshold, and the vertices with a thick border form $\outfr S_t=\{v_5,v_6,v_7,v_8\}$,
    while $\infr S_t=\{v_2,v_3,v_4\}$. Right: the ideal $3$-star of
    \cref{sec:overview-stars}. The center $Z$ and the planar standard Gaussian vector $G$
    are independent, $w_1,w_2,w_3$ are unit vectors at mutual angles $\frac{2\pi}3$, and
    $s_r=\sqrt{1-r^2}$; the neighbors have correlation $r$ with the center and sum to
  $3rZ$.}
  \label{fig:sweep}
\end{figure}
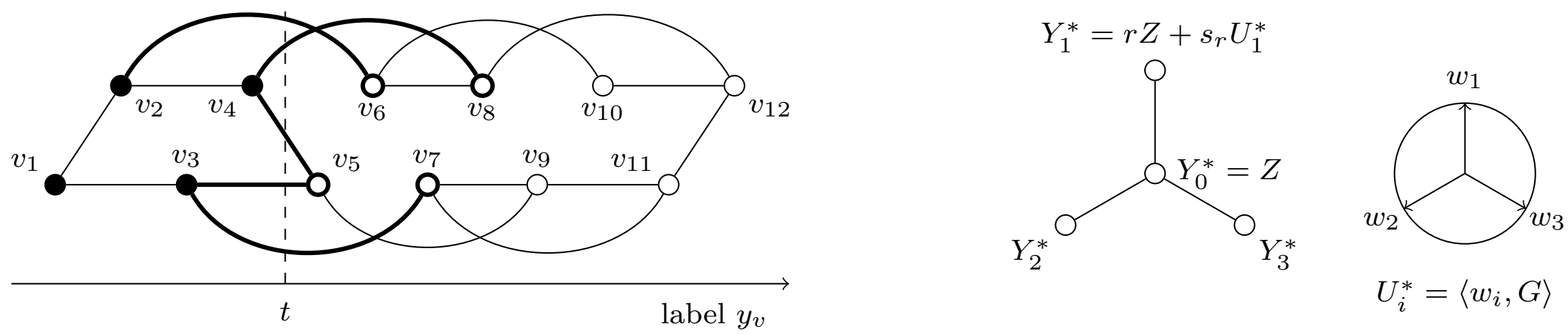

The decomposition behind the lemma is the vertex separation layout of
Kinnersley~\cite{Kinnersley1992}, shown in \cref{fig:sweep}: with $v_1,\dots,v_n$
the vertices in increasing order of their labels and $T_i=\{v_1,\dots,v_i\}$, its
bags are $W_i=\{v_i\}\cup\outfr T_i$. So we need labels for which every frontier is
small at every threshold. For random labels,
$\E|\outfr S_t|=\sum_v\Pr(v\in\outfr S_t)$; if each label depends only on a bounded
neighborhood, the count concentrates (\cref{lem:local-variance}), and a finite grid of
thresholds covers all thresholds (\cref{lem:grid}). Everything then rests on the
probability that one vertex lies on the frontier.

\subsection{Gaussian labels with critical decay}\label{sec:overview-labels}

A vertex lies on a frontier only if one of its edges crosses the threshold, that
is, only if the two ends of that edge fall on different sides of it. If two
neighbors had equal labels, their edge would never cross. For two standard normal labels with correlation $\rho$, the probability
that they fall on different sides of zero is $\arccos(\rho)/\pi$ (Sheppard's formula,
\cref{lem:sheppard}). So the labels should be as correlated as possible across every
edge, while each label keeps unit variance and depends only on a bounded
neighborhood, so that labels of far-apart vertices are independent. We take
independent standard normal variables $(Z_w)_{w\in V(G)}$ and, for a radius $R$, with $B_R(v)=\{w:\dist(v,w)\le R\}$ the ball of radius $R$ around~$v$,
\begin{equation}\label{eq:overview-labels}
  X_v=\sum_{w\in B_R(v)}2^{-\dist(v,w)/2}\,Z_w,\qquad Y_v=X_v/\sqrt{\Var X_v}
\end{equation}
(\cref{def:labels}). For maximum degree $p$ the base $2$ is replaced by $p-1$.

\paragraph{Why this choice of weights?} A label is a weighted sum of the variables around the
vertex. Normalizing does not change correlations, so it suffices to look at the sums
$X_u,X_v$ of~\eqref{eq:overview-labels} for two neighbors $u,v$. They share the
variables $Z_w$ of every vertex
$w$ in both balls, and the two weights of such a $Z_w$ differ by a factor at most
$\sqrt2$, because $\dist(u,w)$ and $\dist(v,w)$ differ by at most $1$. Weights with
such a bounded ratio give the shared part a correlation at least
$\sech(\frac12\ln2)=\frac{2\sqrt2}3$ (\cref{lem:sech-ratio}); with base $p-1$ the
factor is $\sqrt{p-1}$ and the bound $\frac{2\sqrt{p-1}}p$. The only loss comes from
the outer shell, the vertices in the ball of $u$ but not in that of $v$, and the decay
rate decides how much squared mass lies there. If the coefficients decay faster than
$2^{-\dist/2}$, the squared mass sits near the center, each label is close to $Z_v$
alone, and neighboring labels are nearly independent. If they decay more slowly, the
outermost sphere of the ball dominates, and that is where the balls of $u$ and $v$
differ. At the \emph{critical} rate $2^{-\dist/2}$, every sphere of a tree carries the
same squared mass, so the squared mass is spread evenly over the $R$ layers of the
ball, and the outer shell carries only a fraction $O(1/R)$ of the variance
(\cref{eq:critical}). In a graph that is not a tree the spheres are smaller, which
only helps; nothing is assumed about girth, regularity or connectivity.

\begin{ovlemma}[edge correlations; \cref{lem:correlation-padded}]
  Let $G$ be a graph with $\Delta(G)\le3$ and let $(Y_v)$ be the labels
  \eqref{eq:overview-labels} at radius $R$. Then every edge $uv$ satisfies
  $\Corr(Y_u,Y_v)\ge r_*(1-\frac1R)$, where $r_*=\frac{2\sqrt2}3\approx0.9428$. For
  maximum degree $p$ and base $p-1$ the bound is $\frac{2\sqrt{p-1}}p(1-\frac1R)$.
\end{ovlemma}

The constant is best possible for labels whose weights depend only on the distance:
on the infinite $p$-regular tree, the local limit of $p$-regular graphs of large
girth, no such \emph{radial} Gaussian labeling has a
neighbor correlation above $\frac{2\sqrt{p-1}}p$, and the labels
approximate the \emph{Gaussian wave}, which attains this value in the limit
(\cref{rem:labels-optimal}).

Vertices of degree at most $2$ can be removed by series--parallel reductions at a cost
of $O(\log n)$ in the pathwidth (\cref{lem:reduction}), so it suffices to bound graphs
with $\delta(G)\ge3$. For such graphs with vertices of several degrees we use \emph{mixed
labels}: a step away from $v$ through a vertex $x$ multiplies the coefficient by
$(d(x)-1)^{-1/2}$, so that the decay is critical at every vertex at once. An edge
whose ends have degrees $i$ and $j$ then has correlation at least
$r_{ij}=\sech\bigl(\tfrac14\ln((i-1)(j-1))\bigr)$ up to a factor $1-o(1)$
(\cref{lem:correlation}); $r_{33}=r_*$, $r_{44}=\frac{\sqrt3}2$ and $r_{55}=\frac45$.
For both kinds of labels, $Y_v$ depends only on the variables in a ball of bounded
radius around $v$, the labels are distinct with probability $1$, and frontier counts have
variance $O(n)$ (\cref{lem:locality,lem:local-variance}). The radius $R$ grows slowly
with $n$, so that the loss $\frac1R$ vanishes while the dependence stays local.

\subsection{The star comparison and the simplex theorem}\label{sec:overview-stars}

This subsection bounds the probability that a single vertex lies on an outer frontier,
using only the correlation bounds of \cref{sec:overview-labels}. Summed over the
vertices, these bounds give the plain sweep bounds at the end of the subsection, and
the median switch and the charging certificate of the next two subsections build on
them.

Fix a vertex $v$ of degree $d$ with neighbors $u_1,\dots,u_d$. The vector
$(Y_v,Y_{u_1},\dots,Y_{u_d})$ is a centered Gaussian vector with unit variances whose
\emph{incident} correlations $\Corr(Y_v,Y_{u_i})$, between the center and a neighbor,
are at least known lower bounds $r_i$; we call it a \emph{$d$-star with lower bounds
$r$} (\cref{def:star}), with center $Y_0=Y_v$ and neighbors $Y_i=Y_{u_i}$. These lower
bounds are all that the labels guarantee. The correlations $\Corr(Y_i,Y_j)$ between
two neighbors are not controlled: short cycles through $v$ may make them large or
small. The vertex lies in $\outfr S_\tau$ exactly when the \emph{frontier event}
$\{Y_0>\tau,\ \min_iY_i\le\tau\}$ occurs, and we bound its probability from above in
terms of $r$ alone, in three steps.

\emph{Reduction to the center} (\cref{lem:center}). The first step removes the
correlations between neighbors from the problem. The event $\{Y_0>\tau\}$ is the
disjoint union of the frontier event and the event that the whole star lies above
$\tau$. Let $\Phi$ be the standard normal distribution function. Since $Y_0$ is
standard normal and its distribution is symmetric about $0$,
$\Pr(Y_0>\tau)=1-\Phi(\tau)=\Phi(-\tau)$ for every $\tau\in\R$, positive or negative,
and so
\[
  \Pr\bigl(Y_0>\tau,\ \min_iY_i\le\tau\bigr)
  =\Phi(-\tau)-\Pr\bigl(Y_k>\tau\ \text{for}\ 0\le k\le d\bigr).
\]
It therefore suffices to bound the probability that the whole star lies above $\tau$
from below. Positively correlated variables tend to exceed a threshold together, so
this probability should be smallest when the correlations are smallest. Slepian's
inequality~\cite{Slepian1962} (\cref{lem:slepian}) makes this precise: if two centered
Gaussian vectors have the same variances and every covariance of the first is at
least the corresponding covariance of the second, then all coordinates of the first
exceed $\tau$ with at least the probability that all coordinates of the second do. So we
lower each incident correlation $\rho_i=\Corr(Y_0,Y_i)$ to exactly $r_i$, without
raising any other correlation. With fresh independent standard normal variables
$\xi_1,\dots,\xi_d$, put
\[
  Y'_0=Y_0,\qquad
  Y'_i=Y_i-(\rho_i-r_i)\,Y_0+\sqrt{\rho_i^2-r_i^2}\;\xi_i\qquad(1\le i\le d).
\]
Subtracting a multiple of the center removes the excess correlation with the center,
and the noise $\xi_i$ restores unit variance. A direct computation gives
$\Var Y'_i=1$, $\Corr(Y'_0,Y'_i)=r_i$ and
$\Cov(Y'_i,Y'_j)=\Cov(Y_i,Y_j)-\rho_i\rho_j+r_ir_j\le\Cov(Y_i,Y_j)$ for $i\ne j$. So the
correlations between neighbors go down as well, whatever they were, and by Slepian's
inequality the new star has at least the frontier probability of the old one.

The new star has a simple structure. Each neighbor is a multiple of the center plus
an independent remainder: $Y'_i=r_iY_0+s_iU_i$, with $s_i=\sqrt{1-r_i^2}$ and a
centered Gaussian vector $U=(U_1,\dots,U_d)$ with unit variances, independent of
$Y_0$. The incident correlations are now known exactly, and everything unknown about
the star sits in the correlation matrix of $U$. Given $Y_0=z$, the neighbor $Y'_i$
lies above $\tau$ exactly when $U_i>a_i(z)=(\tau-r_iz)/s_i$. These thresholds
decrease as $z$ grows: a high center pulls its neighbors up. Integrating over the
values $z>\tau$ of the center against its density $\varphi$, the standard normal
density, and taking for each $z$ the least favorable correlation matrix of $U$, gives
\[
  \Pr\bigl(Y_0>\tau,\ \min_iY_i\le\tau\bigr)\ \le\
  \Phi(-\tau)-\int_\tau^\infty\varphi(z)\,P_{\min}\bigl(a(z)\bigr)\,dz,\qquad
  a_i(z)=\frac{\tau-r_iz}{s_i},
\]
where $P_{\min}(a)$ is the infimum of $\Pr(U_i>a_i\ \forall i)$ over all correlation
matrices of $U$. The right-hand side depends only on $\tau$ and $r$, and the next
step evaluates $P_{\min}$.

\emph{The simplex theorem} (\cref{lem:simplex}). For equal thresholds
$a_1=\dots=a_d=a$ and $d\le5$ the infimum is attained by the most spread-out
configuration, the
\emph{regular simplex vector} $U^*$: $d$ standard normal variables with all
pairwise correlations $-\frac1{d-1}$, the most negative common value possible, at which
their sum is $0$. This is a case of a recent theorem of
Mulgund~\cite{Mulgund2026}, which proves a conjecture of Balitskiy, Karasev and
Tsigler~\cite{BalitskiyKarasevTsigler2017}; for three coordinates it also has an
elementary proof, and for four it also follows from the four-cap theorem of L.~Fejes
T\'oth~\cite{FejesToth1950,Boroczky2004}.
Write $S_d(a)=\Pr(U^*_i>a\ \forall i)$.

\emph{Ideal stars} (\cref{def:ideal-star}). The first two steps combine into one
explicit worst case. Suppose that all lower bounds are equal to some $r$, as on a
$d$-regular graph. Then all thresholds $a_i(z)$ are equal to $(\tau-rz)/s_r$, the
simplex theorem gives $P_{\min}(a(z))=S_d\bigl((\tau-rz)/s_r\bigr)$, and the bound
of the first step becomes an explicit integral. This integral is not only a bound:
it is the frontier probability of an actual star. Let $Z$ be standard normal and
$U^*$ the regular simplex vector in $d$ variables, independent of $Z$. The
\emph{ideal $d$-star} is $Y^*_0=Z$, $Y^*_i=rZ+s_rU^*_i$ (\cref{fig:sweep}, right).
Its incident correlations are exactly $r$, and its neighbors are spread out as far
as these correlations allow: the remainders $U^*_i$ sum to $0$, so the neighbors sum
to $drZ$, and any two neighbors have correlation $r^2-s_r^2/(d-1)=(dr^2-1)/(d-1)$,
the smallest common value possible. Given $Z=z$, all neighbors lie above $\tau$ with probability exactly
$S_d\bigl((\tau-rz)/s_r\bigr)$, so the computation of the first step holds with
equality for this star. Its frontier probability, the \emph{ideal star function}, is
therefore
\begin{equation}\label{eq:overview-f}
  f_{d,r}(\tau)=\Pr\bigl(Y^*_0>\tau,\ \min_iY^*_i\le\tau\bigr)
  =\Phi(-\tau)-\int_\tau^\infty\varphi(z)\,S_d\Bigl(\frac{\tau-rz}{s_r}\Bigr)dz .
\end{equation}
For $3\le d\le5$, every $d$-star whose lower bounds are all at least $r$ has frontier
probability at most $f_{d,r}(\tau)$ at every threshold $\tau$, and so does every
star with fewer than $d$ neighbors and such lower bounds, because adding neighbors
can only enlarge the frontier event (\cref{cor:stars-ideal}). As the ideal star is
itself a $d$-star with lower bounds $r$, no better bound follows from the lower
bounds alone.

The layout lemma takes the largest frontier over all thresholds, so what counts is
the worst threshold. The suprema
\[
  F_3=\sup_\tau f_{3,r_{33}}(\tau)\approx0.1401583,\qquad
  F_4=\sup_\tau f_{4,r_{44}}(\tau)\approx0.2457720
\]
are enclosed rigorously in ball arithmetic (\cref{prop:F3F4}); the first is attained
near $\tau\approx-0.068$, the second near $\tau\approx-0.164$. For a graph of maximum
degree $4$, the labels with base $3$ give every edge correlation at least
$r_{44}(1-\frac1R)$, so every vertex lies in $\outfr S_\tau$ with probability at most
$f_{4,r_{44}}(\tau)+o(1)\le F_4+o(1)$. By linearity of expectation,
$\E|\outfr S_\tau|\le F_4n+o(n)$ at every threshold, and concentration, the finite
grid of thresholds and the layout lemma turn this into a bound on the pathwidth. For
maximum degree~$3$ the same argument with the labels~\eqref{eq:overview-labels} gives $F_3$. This is the
\emph{plain sweep}, the layout lemma applied to the Gaussian labels.

\begin{ovtheorem}[plain sweep; \cref{thm:deg4}\ref{thm:deg4:a} and the case
  $n_4=0$ of \cref{thm:deg4}\ref{thm:deg4:b}]
  Every graph with $\Delta(G)\le4$ satisfies $\pw(G)\le F_4n+o(n)$, and every graph
  with $\Delta(G)\le3$ satisfies $\pw(G)\le F_3n+o(n)$.
\end{ovtheorem}

This is where the plain sweep stops. For maximum degree $3$, the coefficient $F_3$ is
the best that any analysis of the outer frontiers can give, because it is essentially
attained: on cubic graphs of large girth the stars of the labels are nearly ideal,
so the outer frontier at the threshold $\tau\approx-0.068$ has about $F_3n$ vertices.
Improving the bound requires changing the layout, not the analysis.

\subsection{The median switch}\label{sec:overview-switch}

The change is suggested by the shape of $f_{3,r}$ (\cref{fig:ideal-star}). At the
median threshold $0$ its value has a closed form. Since $Y^*_1+Y^*_2+Y^*_3=3rY^*_0$,
all three neighbors can be positive only when the center is. So $f_{3,r}(0)$ is one
half minus the probability that three standard normal variables with pairwise
correlation $b(r)=\frac{3r^2-1}2$ are all positive, and Sheppard's formula gives
\begin{equation}\label{eq:overview-kappa}
  f_{3,r}(0)=\kappa(r):=\frac3{4\pi}\arccos\frac{3r^2-1}2,\qquad
  \kappa(r_*)=\frac3{4\pi}\arccos\frac56=\cthree
\end{equation}
(\cref{lem:kappa}). Moreover $f_{3,r}(t)\le f_{3,r}(0)$ for every $t\ge0$
(\cref{lem:ideal-monotone}). So $f_{3,r}$ exceeds its value at the median only below
the median, and there we use the inner frontier instead.
Since $-Y$ has the same distribution as $Y$ (\emph{sign symmetry}), the event
$v\in\infr S_t$ for $Y$ is the event $v\in\outfr S_{-t}$ for $-Y$, whose probability
is at most $f_{3,r}(-t)\le\kappa(r)$ when $t\le0$ (\cref{prop:star-cubic}). A layout
with inner frontiers below zero and outer frontiers above zero therefore has expected
frontier size at most $\kappa(r)n$ at every nonzero threshold.

The difficulty lies at threshold zero, where the layout must pass from the inner
frontier $U=\infr S_0$ to the outer frontier $W=\outfr S_0$ while every edge between
$S_0$ and its complement stays covered. Putting $U\cup W$ into one bag would give a
bag of about $2\kappa(r)n$ vertices. The \emph{connector} does better: it adds the
vertices of $W$ one at a time, in an order given by a purely combinatorial lemma
(\cref{lem:connector}), and drops a vertex of $U$ as soon as all its neighbors above
zero are present. Only the vertices whose three neighbors all lie on the other side
of zero obstruct this exchange; let $D$ be their set.

\begin{ovlemma}[median-switched layout; \cref{lem:median-layout}]
  Let $\Delta(G)\le3$ and let $y$ be an injective labeling. If $|\infr S_t|\le w$ for
  all $t<0$, $|\outfr S_t|\le w$ for all $t>0$, and
  $\max\bigl(|U|+|D\cap W|,\ |W|+|D\cap U|\bigr)+1\le w$, then $\pw(G)\le w$.
\end{ovlemma}

For a vertex $v$ of degree $d\le3$, the events $v\in U$ and $v\in D\cap W$ are
$\{Y_v\le0<\max_iY_{u_i}\}$ and $\{d=3,\ Y_v>0\ge\max_iY_{u_i}\}$. The connector
adds nothing at leading order because of the following lemma, whose right side is
the same bound $\kappa(r)$ as at nonzero thresholds.

\begin{ovlemma}[switch lemma; \cref{lem:switch}]
  Let $(Y_0,Y_1,\dots,Y_d)$ be a $d$-star with all lower bounds equal to $r\in(0,1)$
  and $d\le3$. Then
  \[
    \Pr\bigl(Y_0\le0,\ \exists i:\ Y_i>0\bigr)
    +\ind_{d=3}\Pr\bigl(Y_0>0,\ Y_i\le0\text{ for all }i\bigr)\ \le\ \kappa(r).
  \]
\end{ovlemma}

At the ideal star the second event is impossible, since the neighbor labels sum to
$3r$ times the center label, and the first has probability exactly $\kappa(r)$; a star
that makes the second event possible loses at least as much in the first. After
conditioning on the center, the proof reduces to showing that the probability that at
least two of three standard normal residuals exceed a level $h>0$ is smallest for the
equilateral matrix of $U^*$ (\cref{lem:cubic-exceed}). This is not an orthant event,
so Slepian's inequality does not apply; a deformation monitored by Plackett's identity
(\cref{lem:plackett}) reduces it to the planar case. Concentration and $R\to\infty$
then give the bound with constant $\kappa(r_*)=\cthree$.

\begin{ovtheorem}[\cref{thm:cubic}]
  Every graph $G$ with $\Delta(G)\le3$ satisfies $\pw(G)\le\cthree\,n+o(n)$.
\end{ovtheorem}

The gain over the plain sweep is small, $F_3-\cthree\approx3.4\cdot10^{-4}$. But the
median switch replaces a numerically enclosed maximum by a closed form, and its proof
uses no computation.
The value $\cthree$ is exactly what the median-switched sweep gives on cubic graphs of
large girth, so any further improvement must change the rule at threshold zero
(\cref{sec:conclusion}). With mixed labels, the same layout gives
\cref{thm:switched4}.

\subsection{Mixed degrees, the charging certificate and the bootstrap}
\label{sec:overview-bootstrap}

\paragraph{Degrees 3 to 5.} With mixed labels, the lower bounds $r_{d(v)d(u)}$ of the
star of $v$ depend on the degrees of its neighbors. So the bound of the reduction to
the center depends on the \emph{type} $T=(d;k_3,k_4,k_5)$ of $v$, where $d$ is its
degree and $k_j$ counts its neighbors of degree $j$. We write this bound $C_T(\tau)$;
there are $46$ types for degrees $3$ to $5$ (\cref{def:admissible}). A vertex of
degree $3$ next to vertices of higher degree has weaker correlations than one in a
cubic region, and its bound exceeds the coefficient $a_3$ that we aim for. We move
the excess along the edges: a vertex of degree $i$ sends an amount $t_{ij}$ along each
edge to a neighbor of degree $j>i$. The transfers sum to zero over the graph
(\cref{lem:charging}). Hence $\E|\outfr S_\tau|\le\sum_da_dn_d$ whenever
$(a_3,a_4,a_5)$ is \emph{admissible}, that is, whenever for every $\tau\in[-3,3]$ some
transfer vector $(t_{34},t_{35},t_{45})$ brings the transferred bound of every type
below the coefficient of its degree. (Beyond $|\tau|=3$ the frontier probabilities
are below $5\Phi(-3)<0.007$.)

We verify admissibility by a computer-assisted proof (\cref{prop:certificate}). The
thresholds are partitioned into cells. Each cell has one transfer vector and, for
every type, a rational lower bound on the integral in $C_T(\tau)$, obtained in ball
arithmetic. The final inequalities are checked in exact rational arithmetic with
positive slack. This certifies the pathwidth sequences
\[
  \tripP=(F_3,\,0.2575,\,0.3667),\qquad
  \tripR=(0.1455,\,0.2530,\,0.3600),\qquad
  \tripS=\bigl(\tfrac5{32},\,\tfrac14,\,\tfrac{11}{32}\bigr)
\]
of \cref{thm:deg5}; the bounds $(F_3,0.2575)$ and $(\frac5{32},\frac14)$ for maximum
degree $4$ (\cref{thm:deg4}\ref{thm:deg4:b},\ref{thm:deg4:c}) are their restrictions to
graphs without vertices of degree $5$. The triple $\tripP$ is best when few vertices
have degree $4$ or $5$, $\tripS$ when many do. All coefficient vectors of the paper
are collected in \cref{thm:pw-seq}.

\paragraph{All graphs.} Beyond degree $5$ no probability is involved. A combinatorial
induction in the spirit of Fomin, Gaspers, Saurabh and
Stepanov~\cite{FominGaspersSaurabhStepanov2009} and of Edwards and
McDermid~\cite{EdwardsM15} deletes vertices of large degree one at a time
(\cref{prop:deletion}). Putting a deleted vertex back into every bag adds at most one to
the width, and we compare this increase with the decrease of the potential $\sum_vb_{d(v)}$. Deleting a
vertex $w$ of maximum degree $d$ removes the term $b_d$ of $w$ itself. It also lowers
the degree of each neighbor by one, which lowers the potential by an increment
$h_{d'}=b_{d'}-b_{d'-1}$, where $d'$ is the degree of the neighbor; this increment is
at least the running minimum $H_d=\min_{3\le i\le d}h_i$. With $d-1$ neighbors of
degree $d$ and one of smaller degree, the decrease balances the increase exactly when
$1-b_d=(d-1)H_d+H_{d-1}$, and the two worst cases give the recurrence
\begin{equation}\label{eq:overview-boot}
  b_d=\max\Bigl\{\frac{1+(d-1)\,b_{d-1}-H_{d-1}}{d},\;1-d\,H_{d-1}\Bigr\}\qquad(d>k)
\end{equation}
(\cref{eq:bootstrap-rec}). Vertices of degree at most $2$ are removed by the
classical series--parallel reductions after every deletion, and all reductions
together add $O(\log n)$ to the pathwidth (\cref{lem:reduction}).

The bootstrap lemma (\cref{lem:bootstrap}) gives the coefficients $b_d$ for $d>k$
from a base of maximum degree~$k$ by~\eqref{eq:overview-boot}.
When the increments are nonincreasing, the first term of the maximum is the larger
one, and \eqref{eq:overview-boot} is the recurrence
$d\,b_d=(d-2)\,b_{d-1}+b_{d-2}+1$ of Edwards and McDermid (\cref{rem:bootstrap-em}).
The recurrence also explains why a better cubic constant alone improves nothing at
higher degrees: with $k=3$ and $b_3=\cthree<\frac16$ the second term is the larger
one, and $b_4=1-4\cthree\approx0.4407$, worse than the known $\frac13$. \Cref{thm:main} is therefore bootstrapped from the degree-$5$ base
$\tripS$: its coefficients are $\frac{3d-4}{32}$ up to degree $6$, then
$1-\frac{111}{28(d+1)}$ up to degree $10$ and $1-\frac4{d+1}$ from degree $11$ on. The
edge bound follows because $2a_d/d\le\frac7{48}$ for every $d$, with equality only
at $d=6$, where $a_6=\frac7{16}$ (\cref{cor:edges}); applying this to every
component separately also gives a remainder $o(m)$ (\cref{rem:edges-m}).

\subsection{Algorithms and running times}\label{sec:overview-algorithms}\label{sec:overview-apps}

\Cref{sec:algorithms} turns the existence proofs into algorithms. It replaces the
Gaussian variables by a finite rational distribution coupled to the normal distribution, which changes
every per-vertex probability of Part~I by $o(1)$ uniformly in the threshold
(\cref{lem:alg-coupling,lem:alg-perturb}), and then fixes the variables one at a time
by the method of conditional expectations~\cite{ErdosSelfridge1973,Raghavan1988},
applied to finitely many local counts (\cref{lem:alg-derand}). The radius grows only
like $\frac18\log_2\log_2n$, so each conditional expectation is computable by local
enumeration in $n^{o(1)}$ time.

Part~II turns the widths into running times in three ways. \emph{Substitution}: a
dynamic program with $c^{w+1}\poly(n)$ steps over a path decomposition of width
$W(G)+o(n)$ runs in time $c^{W(G)+o(n)}\poly(n)$ (\cref{lem:apps-dp}); this gives most
bounds of \cref{sec:apps-dp} and, through Cut\&Count, the rank-based approach and
algorithms over orderings of small cutwidth, those of \cref{sec:apps-connectivity}. \emph{Rerunning a measure-and-conquer analysis}: in
polynomial space the widths enter Separate, Measure and
Conquer~\cite{GaspersSorkin2017} through balanced separators of size $cn$, and the
separator constant enters the mathematical program of the analysis in one or two places
(\cref{lem:lp}, \cref{thm:smc}). \emph{Moving a switch point}: with coefficient
$\frac14$ in place of $\frac13$ at degree~$4$, the algorithm of Nederlof, van Rooij
and van Dijk~\cite{NederlofRD2014} for counting dominating sets can hand the sets of
size~$4$ to its dynamic program, which lowers its running time from $1.5002^n$ to
$1.4815^n$ (\cref{thm:ds}).

\begin{table}[p]
  \centering
  \caption{Running-time bounds of this paper that, as far as we know, the
    preprints~\cite{OpenAI2026ETH,Claude2026ETH} that claim to refute ETH do not affect
  (\cref{sec:intro-apps}). Letters refer to the notes that follow the table.}
  \label{tab:apps}
  \newcommand*{\tabthm}[1]{\hyperref[#1]{Thm.~\ref*{#1}}}
\footnotesize
\setlength{\tabcolsep}{4pt}
\setlength{\aboverulesep}{0.2ex}\setlength{\belowrulesep}{0.3ex}
\renewcommand{\arraystretch}{0.905}
\begin{tabular}{@{}llll@{}}
  \toprule
  Problem & Previous & New & References\\
  \midrule
  \multicolumn{4}{@{}l}{\emph{Feedback vertex sets}}\\
  Feedback Vertex Set, $\Delta\le4$ & $1.443^n$ & $2^{n/3}<1.2600^n$ & \cite{CyganNPPRW2011arXiv}\textsuperscript{f}, \tabthm{thm:connectivity}\\
  Feedback Vertex Set, $\Delta\le4$, det. & $1.7117^n$ & $5^{F_4n}<1.4853^n$ & \cite{GaspersLee2017arXiv}, \tabthm{thm:connectivity}\\
  Feedback Vertex Set, $\Delta\le5$ & $1.61^n$ & $2^{0.5121n}<1.4262^n$ & \cite{CyganNPPRW2011arXiv}\textsuperscript{f}, \tabthm{thm:connectivity}\\
  Feedback Vertex Set, $\Delta\le6$ & $1.6297^n$ & $3^{7n/16}<1.6172^n$ & \cite{LiNederlof2019arXiv}\textsuperscript{f}, \tabthm{thm:connectivity}\\
  Connected FVS and others\textsuperscript{e}, $\Delta\le3$ & $1.26^n$ & $4^{\cthree n}<1.2139^n$ & \cite{CyganNPPRW2011arXiv}\textsuperscript{f}, \tabthm{thm:connectivity}\\
  Connected FVS and others\textsuperscript{e}, $\Delta\le4$ & $1.588^n$ & $4^{F_4n}<1.4060^n$ & \cite{CyganNPPRW2011arXiv}\textsuperscript{f}, \tabthm{thm:connectivity}\\
  Connected FVS and others\textsuperscript{e}, $\Delta\le5$ & $1.824^n$ & $4^{11n/32}<1.6105^n$ & \cite{CyganNPPRW2011arXiv}\textsuperscript{f}, \tabthm{thm:connectivity}\\
  \midrule
  \multicolumn{4}{@{}l}{\emph{Domination}}\\
  (\#)Min.\ Weight Dominating Set\textsuperscript{c} & $1.5002^n$ & $1.4815^n$ & \cite{NederlofRD2014}\textsuperscript{a}, \tabthm{thm:ds}\\
  (\#)Min.\ Weight Dominating Set, $\Delta\le3$ & $3^{n/6}<1.2010^n$ & $3^{\cthree n}<1.1661^n$ & \cite{FominHoie2006}, \tabthm{thm:bdd}\\
  Connected Dominating Set, $\Delta\le3$ & $1.26^n$ & $3^{\cthree n}<1.1661^n$ & \cite{CyganNPPRW2011arXiv}\textsuperscript{f}, \tabthm{thm:connectivity}\\
  Connected Dominating Set, $\Delta\le4$ & $1.588^n$ & $4^{F_4n}<1.4060^n$ & \cite{CyganNPPRW2011arXiv}\textsuperscript{f}, \tabthm{thm:connectivity}\\
  Connected Dominating Set, $\Delta\le5$ & $1.8139^n$ & $4^{11n/32}<1.6105^n$ & \cite{BinkeleRaibleF2012}, \tabthm{thm:connectivity}\\
  \midrule
  \multicolumn{4}{@{}l}{\emph{Independence and vertex covers}}\\
  (\#)Max.\ Weight Indep.\ Set, $\Delta\le3$ & $2^{n/6}<1.1225^n$ & $2^{\cthree n}<1.1018^n$ & \cite{FominHoie2006}, \tabthm{thm:bdd}\\
  (\#)Max.\ Weight Indep.\ Set, $\Delta\le4$ & $2^{n/3}<1.2600^n$ & $2^{F_4n}<1.1858^n$ & \cite{FominGaspersSaurabhStepanov2009}, \tabthm{thm:bdd}\\
  Connected Vertex Cover, $\Delta\le4$ & $1.443^n$ & $2^{n/3}<1.2600^n$ & \cite{CyganNPPRW2011arXiv}\textsuperscript{f}, \tabthm{thm:connectivity}\\
  Connected Vertex Cover, $\Delta\le5$ & $1.61^n$ & $2^{0.5121n}<1.4262^n$ & \cite{CyganNPPRW2011arXiv}\textsuperscript{f}, \tabthm{thm:connectivity}\\
  \midrule
  \multicolumn{4}{@{}l}{\emph{Max 2-CSP}}\\
  Max $(r,2)$-CSP, all instances & $r^{13m/75}<r^{m/5.769}$ & $r^{7m/48}<r^{m/6.857}$ & \cite{ScottSorkin2007}, \tabthm{thm:max2csp}\\
  Max $(r,2)$-CSP, $\Delta\le3$ & $r^{n/6}$ & $r^{\cthree n}<r^{0.13983n}$ & \cite{FominHoie2006}, \tabthm{thm:max2csp}\\
  Max $(r,2)$-CSP, $\Delta\le4$ & $r^{n/3}$, $r^{m/6}$ & $r^{F_4n}$, $r^{\frac{3F_4}{2+16F_4}m}$ & \cite{FominGaspersSaurabhStepanov2009}, \cite{ScottSorkin2007}, \tabthm{thm:max2csp}\\
  Max $(r,2)$-CSP\textsuperscript{b}, avg.\ degree $\bar d\ge7$ & $r^{n(1-3.5/(\bar d+1)+O(\bar d^{-3}))}$ & $r^{n(1-3.964/(\bar d+1))}$ & \cite{EdwardsM15}, \tabthm{thm:max2csp}\\
  Max $(r,2)$-CSP\textsuperscript{b}, avg.\ degree $\bar d\ge11$ & $r^{n(1-3.5/(\bar d+1)+O(\bar d^{-3}))}$ & $r^{n(1-4/(\bar d+1))}$ & \cite{EdwardsM15}, \tabthm{thm:max2csp}\\
  \midrule
  \multicolumn{4}{@{}l}{\emph{Patterns and packings}}\\
  Counting $k$-edge subgraphs\textsuperscript{d} & $n^{0.174k}$ & $n^{(7/48)k}<n^{0.1459k}$ & \cite{CurticapeanDM2017}, \tabthm{thm:subgraphs}\\
  Counting $k$-paths, $k$-cycles\textsuperscript{d} & $n^{0.1704k}$ & $n^{0.1459k}$ & \cite{BonnetIJK2019arXiv}, \tabthm{thm:subgraphs}\\
  Permutation Pattern Matching & $n^{k/4}$ & $2^{O(k)}n^{(253/1215)k}$ & \cite{BerendsohnKM2021}, \tabthm{thm:perm}\\
  $k$-opt for TSP, general graphs & $n^{(1/4+\eps_k)k}$ & $n^{(253/1215+\eps_k)k}$ & \cite{CyganKS2019}, \tabthm{thm:perm}\\
  $k$-opt for TSP, bounded degree & $n^{(23/135+\eps_k)k}$ & $n^{(7/48+\eps_k)k}$ & \cite{BonnetIJK2019}, \tabthm{thm:perm}\\
  Co-Path Packing & $2.9241^k$ & $2.8192^k$ & \cite{LiuXiao2026}, \tabthm{thm:connectivity}\\
  $k$-Co-Path Set & $1.588^k$ & $1.5423^k$ & \cite{SullivanvdP2016}, \tabthm{thm:connectivity}\\
  \midrule
  \multicolumn{4}{@{}l}{\emph{Cycles and paths}}\\
  Hamiltonian Cycle, $\Delta\le3$ & $1.1583^n$ & $1.1312^n$ & \cite{CyganKN2012arXiv}\textsuperscript{f}, \tabthm{thm:connectivity}\\
  Hamiltonian Cycle, $\Delta\le4$ & $1.5058^n$ & $1.3416^n$ & \cite{CyganKN2012arXiv}\textsuperscript{f,g}, \tabthm{thm:connectivity}\\
  Hamiltonian Cycle, $\Delta\le5$ & $1.6246^n$ & $1.5252^n$ & \cite{BjorklundKKZ2015arXiv}\textsuperscript{f}, \tabthm{thm:connectivity}\\
  Hamiltonian Cycle, avg.\ degree $d$ & $1.12^{dn}$ & $1.0937^{dn}$ & \cite{KowalikM2020arXiv}, \tabthm{thm:connectivity}\\
  TSP, $\Delta\le3$, det. & $1.2128^n$ & $1.1757^n$ & \cite{BodlaenderCKN2015}, \tabthm{thm:connectivity}\\
  TSP, $\Delta\le4$, det. & $1.6110^n$ & $1.4214^n$ & \cite{BodlaenderCKN2015}\textsuperscript{g}, \tabthm{thm:connectivity}\\
  TSP, $\Delta\le5$, det. & $1.8588^n$ & $1.6352^n$ & \cite{BodlaenderCKN2015}\textsuperscript{g}, \tabthm{thm:connectivity}\\
  TSP, $\Delta\le6$, det. & $1.9672^n$ & $1.8699^n$ & \cite{BjorklundHKK2012}, \tabthm{thm:connectivity}\\
  TSP, avg.\ degree $d$, det. & $1.1321^{dn}$ & $1.1100^{dn}$ & \cite{KowalikM2020arXiv}, \tabthm{thm:connectivity}\\
  Longest Path, $\Delta\le3$ & $1.26^n$ & $1.1874^n$ & \cite{CyganNPPRW2011arXiv}\textsuperscript{f}, \tabthm{thm:connectivity}\\
  Longest Path, $\Delta\le4$ & $1.5875^n$ & $1.3523^n$ & \cite{BjorklundKKZ2015arXiv}\textsuperscript{f}, \tabthm{thm:connectivity}\\
  Longest Path, $\Delta\le5$ & $1.6246^n$ & $1.5252^n$ & \cite{BjorklundKKZ2015arXiv}\textsuperscript{f}, \tabthm{thm:connectivity}\\
  \midrule
  \multicolumn{4}{@{}l}{\emph{Other connectivity problems}}\\
  Steiner Tree, $\Delta\le3$ & $1.201^n$ & $3^{\cthree n}<1.1661^n$ & \cite{CyganNPPRW2011arXiv}\textsuperscript{f}, \tabthm{thm:connectivity}\\
  Steiner Tree, $\Delta\le4$ & $1.3533^n$ & $3^{F_4n}<1.3100^n$ & \cite{FominGKLS2013}, \tabthm{thm:connectivity}\\
  Max Internal Spanning Tree, $\Delta\le6$ & $1.8423^n$ & $4^{7n/16}<1.8341^n$ & \cite{BjorklundKKZ2015arXiv}\textsuperscript{f}, \tabthm{thm:connectivity}\\
  \midrule
  \multicolumn{4}{@{}l}{\emph{Directed problems}}\\
  Directed problems\textsuperscript{h}, $\Delta\le3$ & $1.349^n$ & $6^{\cthree n}<1.2848^n$ & \cite{CyganNPPRW2011arXiv}\textsuperscript{f}, \tabthm{thm:connectivity}\\
  Directed problems\textsuperscript{h}, $\Delta\le4$ & $1.818^n$ & $6^{F_4n}<1.5533^n$ & \cite{CyganNPPRW2011arXiv}\textsuperscript{f}, \tabthm{thm:connectivity}\\
  \bottomrule
\end{tabular}
{}
\end{table}

\begin{table}[tb]
  \begin{minipage}{\textwidth}
    \textbf{Notes to \cref{tab:apps}.}
    \footnotesize
A letter after a citation concerns the previous bound, and a letter after a problem
concerns the whole row.
\textsuperscript{a}\,A gap in the analysis of this algorithm allowed us only to reconstruct a running time of $O(1.5004^n)$ (\cref{sec:nrd-gap}).
\textsuperscript{b}\,A record only for moderate $\bar d$; for large $\bar d$, the
algorithms of Williams~\cite{Williams2005}, for integer scores of polynomial size, and
of Golovnev and Kutzkov~\cite{GolovnevK14}, for Boolean domains, are faster
(\cref{thm:max2csp}).
\textsuperscript{c}\,When the set of possible weight sums has polynomial size, as in
\cite[Cor.~32]{NederlofRD2014} (minimum weight) and \cite[Cor.~8.14]{vanRooij2011}
(counting).
\textsuperscript{d}\,Both bounds carry a further factor $k^{O(k)}$.
\textsuperscript{e}\,The problems of \cref{thm:connectivity}\ref{it:conn-cds} other than
Connected Dominating Set: Connected Feedback Vertex Set, Connected Odd Cycle
Transversal, Exact $k$-Leaf Spanning Tree, Maximum Full Degree Spanning Tree, Graph
Metric Traveling Salesperson, and undirected Min Cycle Cover and Longest Cycle; Maximum
Leaf Spanning Tree also has the bounds of Connected Dominating Set.
\textsuperscript{f}\,By a Monte Carlo algorithm.
\textsuperscript{g}\,The cited algorithm, run with the pathwidth bounds
of~\cite{FominGaspersSaurabhStepanov2009}, and for TSP with $\omega\le\frac94$.
\textsuperscript{h}\,The directed versions of Min Cycle Cover, Longest Path and Longest
Cycle, and Exact $k$-Leaf Outbranching; $\Delta$ is the maximum degree of the
underlying undirected graph.

The bounds for TSP, previous and new, are evaluated with the matrix multiplication
exponent $\omega\le\frac94$~\cite{OpenAI2026omega,AmanovBZ2026} (\cref{sec:apps-connectivity}).
{}
  \end{minipage}
\end{table}
{}

\Needspace*{8\baselineskip}
\part{Width bounds}\label{part:width}
\section{Short proof of a pathwidth bound below \texorpdfstring{$\frac n6$}{n/6} for cubic graphs}\label{sec:warmup}

This section proves, without using the rest of the paper, that a short argument
improves on the bound $\frac n6$ for the pathwidth, the cutwidth and the bisection
width of graphs of maximum degree $3$. The algorithm behind it is simple: give every
vertex a random real label, sort the vertices by label, and use the sorted order as a
layout. The analysis counts, for every threshold $t$, the edges with one end labelled
at most $t$ and the other above $t$. The main results (\cref{sec:intro}) are stronger;
the point here is a short proof. The reader is assumed to know graphs, not
probability: the facts about random variables that we use are collected in
\cref{sec:warmup-toolkit}, and the only estimate on the graph is a count of the
vertices that the balls around the two ends of an edge share. Up to the end of
\cref{sec:warmup-proof} the section is self-contained;
\cref{sec:warmup-degree-p,sec:warmup-remarks} connect the argument to the rest of the
paper. The later sections of \cref{part:width} refine each step: the layout lemma,
Sheppard's formula and the concentration bound of \cref{sec:prelim} are the
general forms of \cref{lem:warmup-layout}, \cref{lem:crossing}\ref{lem:crossing:a}
and \cref{lem:warmup-variance}, and
\cref{sec:labels} generalizes the labels and \cref{lem:warmup-edges}.

Graphs are finite and simple. For a graph $G$, let $n=|V(G)|$ and $m=|E(G)|$, let
$\Delta(G)$ be its maximum degree, $N(v)$ the set of neighbors of a vertex $v$ and
$N(X)=\bigcup_{v\in X}N(v)$, let $\dist(v,w)$ be the distance in $G$, let
$B_R(v)=\{w:\dist(v,w)\le R\}$ be the ball of radius $R$ around $v$ and
$\{w:\dist(v,w)=k\}$ its \emph{sphere} of radius $k$, and for $X\subseteq V(G)$ let $e(X)$ be the number
of edges with exactly one end in $X$. A
\emph{path decomposition} of $G$ is a sequence of sets (\emph{bags})
$W_1,\dots,W_s\subseteq V(G)$ such that every edge has both ends in some bag and every
vertex lies in a nonempty set of consecutive bags; its width is $\max_i|W_i|-1$, and
the \emph{pathwidth} $\pw(G)$ is the minimum width of a path decomposition of $G$. The
\emph{cutwidth} $\operatorname{cw}(G)$ is the minimum, over all orderings
$v_1,\dots,v_n$ of $V(G)$, of $\max_ie(\{v_1,\dots,v_i\})$, and the \emph{bisection
width} is the minimum of $e(X)$ over all $X\subseteq V(G)$ with
$|X|=\lfloor n/2\rfloor$.
Put
\[
  c_*=\frac{3}{2\pi}\arcsin\frac13=\frac{3}{2\pi}\arccos\frac{2\sqrt2}{3}\approx0.16226 ;
\]
the two forms agree because an angle with cosine $\frac{2\sqrt2}3$ has sine
$\sqrt{1-\frac89}=\frac13$. Numerically $c_*$ is below $\frac16$ by about $0.0044$,
and this can be checked exactly. The inequality $c_*<\frac16$ means
$\arcsin\frac13<\frac\pi9$, that is, $\frac13<s$ for $s=\sin\frac\pi9$. By the
triple-angle formula, $3s-4s^3=\sin\frac\pi3=\frac{\sqrt3}2$. The function $h(x)=3x-4x^3$
is increasing on $[0,\frac12]$, which contains $\frac13$ and $s<\sin\frac\pi6$, and
$h(\frac13)=\frac{23}{27}<\frac{\sqrt3}2$ because $46^2=2116<2187=3\cdot27^2$. Hence
$\frac13<s$, and $c_*<\frac16$.

\begin{theorem}[cut sweep]\label{thm:cut-sweep}
  Every graph $G$ with $\Delta(G)\le3$ satisfies
  $\pw(G)\le\operatorname{cw}(G)\le c_*n+o(n)$.
\end{theorem}

By considering a set $X$ formed by the first $\lfloor n/2\rfloor$ vertices of
an ordering, we get that the bisection width is at most
$c_*n+o(n)$ as well; for all three parameters the previous bound was
$\frac n6+o(n)$~\cite[Theorem~1]{MonienPreis2006}\cite{FominHoie2006}.
\Cref{thm:intro-bounded,cor:bisection} lower all three to $\cthree n+o(n)$ with
$\cthree\approx0.1398$.

\subsection{The idea}\label{sec:warmup-idea}

Every vertex $w$ draws a real number $Z_w$ from the standard normal distribution
$N(0,1)$, with mean $0$ and variance $1$, independently of all other vertices, and every vertex
$v$ receives the label $Y_v$, a weighted sum of the numbers $Z_w$ over the ball of
radius $R$ around $v$, with coefficients that decrease with $\dist(v,w)$. We order the
vertices by label. Every prefix of this order is a set $S_t=\{v:Y_v\le t\}$ for some
threshold $t$, so the cutwidth of the order is the largest number of edges that
\emph{cross} a threshold, that is, have one end labelled at most $t$ and the other
above $t$.

The two ends of an edge $uv$ see almost the same ball, so $Y_u$ and $Y_v$ are sums of
almost the same noise. Their \emph{correlation} $\rho=\Corr(Y_u,Y_v)$ measures this:
for such sums it is the overlap of the two coefficient sequences, normalized so that
identical sequences have $\rho=1$ (\cref{sec:warmup-toolkit}). So the ends of an edge
have $\rho$ close to $1$. Two standard normal variables with correlation $\rho$ lie on
opposite sides of a threshold with probability at most $\arccos(\rho)/\pi$, which is
small when $\rho$ is close to $1$ (about $\sqrt{2(1-\rho)}/\pi$). Summing over the at
most $\frac32n$ edges, the expected number of edges crossing any fixed threshold is at
most $\frac{3}{2\pi}\arccos(\rho)\,n$. The bound $\frac n6$ corresponds to
$\arccos\rho=\frac\pi9$, since $\frac3{2\pi}\cdot\frac\pi9=\frac16$. The coefficients of
\cref{sec:warmup-lemmas}, for a radius $R$, give every edge of every graph of maximum
degree $3$ correlation at least $\frac{2\sqrt2}3(1-\frac1R)$, and
$\arccos\frac{2\sqrt2}3=\arcsin\frac13<\frac\pi9$.

A bound on the expected number of crossing edges at each threshold is not yet a bound
for one order. We need a single outcome of the random numbers in which \emph{every}
threshold is crossed by few edges. Since every label depends only on the numbers $Z_w$
in a bounded ball, the count at a fixed threshold is a sum of many terms that are
mostly independent of each other, so it stays close to its expectation with
probability close to $1$. We apply this to finitely many thresholds on a grid, which
control all the others, and conclude that some outcome is good at all of them at once.

One may think of $Z_w$ as a fair coin flip of $\pm1$; for large balls the labels are
then approximately normal, and the normal variables make the probability calculation
exact.

The proof has four steps.
\begin{enumerate}[label=\textup{(\arabic*)}]
  \item An order of the vertices gives a path decomposition whose width is at most the
    largest number of edges leaving a prefix (\cref{lem:warmup-layout}).
  \item Each edge crosses each threshold with probability at most $\arccos(\rho)/\pi$
    (\cref{lem:crossing}), where $\rho\ge\frac{2\sqrt2}3(1-\frac1R)$
    (\cref{lem:warmup-edges}). So the expected number of crossing edges is at most
    $(c_*+O(\frac1R))n$ at every threshold, which is below $\frac n6$ for $R$ large.
  \item Each count is close to its expectation with high probability
    (\cref{lem:warmup-variance} and Chebyshev's inequality), and a grid of thresholds
    reduces all thresholds to finitely many counts.
  \item The probability that some count on the grid is far from its expectation is
    less than $1$, so some outcome of the random numbers gives an order of small
    cutwidth.
\end{enumerate}

\subsection{Facts from probability}\label{sec:warmup-toolkit}

This section lists everything about random variables that the proof uses. All random
variables in the proof are functions of finitely many independent standard normal
variables $(Z_w)_{w\in W}$; in \cref{sec:warmup-proof}, $W=V(G)$. For a random variable
$X$ with expectation $\E X$, the \emph{variance} is $\Var X=\E(X-\E X)^2$. For two
random variables the \emph{covariance} is $\Cov(X,X')=\E[(X-\E X)(X'-\E X')]$. If
both variances are positive, the \emph{correlation} is
$\Corr(X,X')=\Cov(X,X')/\sqrt{\Var X\,\Var X'}$, a number in $[-1,1]$.
\begin{enumerate}[label=\textup{(P\arabic*)}]
  \item\label{P:linearity} \emph{Expectations and variances.} $\E(X+X')=\E X+\E X'$
    for all random variables, independent or not (linearity of expectation). Also
    $\Var(\sum_iX_i)=\sum_{i,j}\Cov(X_i,X_j)$, and
    $|\Cov(X,X')|\le\sqrt{\Var X\,\Var X'}$ (the Cauchy--Schwarz inequality).
    Functions of disjoint sets of the $Z_w$ are independent, and independent
    variables have covariance $0$.
  \item\label{P:sums} \emph{Linear sums.} For real coefficients $(a_w)$, the sum
    $\sum_wa_wZ_w$ is normal with mean $0$ and variance $\sum_wa_w^2$. If this variance
    $s^2$ is positive, the sum has the density
    $\frac1{s\sqrt{2\pi}}e^{-x^2/(2s^2)}$, which is symmetric and decreasing in $|x|$;
    in particular the sum takes any given value with probability $0$. The covariance
    of $\sum_wa_wZ_w$ and $\sum_wb_wZ_w$ is $\sum_wa_wb_w$.
  \item\label{P:orth} \emph{Orthogonal sums are independent.} If $\sum_wa_wb_w=0$, then
    $\sum_wa_wZ_w$ and $\sum_wb_wZ_w$ are independent. This is special to sums of
    independent normal variables. For general random variables, covariance $0$ does not
    imply independence.
  \item\label{P:rotation} \emph{Rotation invariance.} If $G_1,G_2$ are independent and
    standard normal, then the direction of the point $(G_1,G_2)$ of the plane is
    uniformly distributed on the unit circle, because the joint density
    $\frac1{2\pi}e^{-(x^2+y^2)/2}$ depends only on the distance from the origin.
  \item\label{P:chebyshev} \emph{Chebyshev's inequality.}
    $\Pr(|X-\E X|\ge\lambda)\le\Var X/\lambda^2$ for every $\lambda>0$.
  \item\label{P:union} \emph{Union bound.} The probability that at least one of
    several events happens is at most the sum of their probabilities. So if these
    probabilities add up to less than $1$, some outcome of the random numbers avoids
    all of the events (the probabilistic method).
\end{enumerate}
We write $\varphi(x)=\frac1{\sqrt{2\pi}}e^{-x^2/2}$ for the standard normal density and
$\Phi(x)=\int_{-\infty}^x\varphi$ for the standard normal distribution function. Since
$\Phi(-3)<0.0014$, a standard normal variable lies outside $[-3,3]$ with probability
less than $0.003$.

\subsection{Four lemmas}\label{sec:warmup-lemmas}

The first lemma turns an ordering into a path decomposition; it is the easy direction
of Kinnersley's theorem that pathwidth equals vertex separation~\cite{Kinnersley1992}.

\begin{lemma}[orderings]\label{lem:warmup-layout}
  Let $G$ be a graph, let $v_1,\dots,v_n$ be an ordering of $V(G)$, and let
  $T_i=\{v_1,\dots,v_i\}$. Then $\pw(G)\le\max_ie(T_i)$. In particular
  $\pw(G)\le\operatorname{cw}(G)$.
\end{lemma}

\begin{proof}
  Let $W_i=\{v_i\}\cup(N(T_i)\setminus T_i)$ for $1\le i\le n$, and let $f_j$ be the
  index of the first neighbor of $v_j$ among $v_1,\dots,v_{j-1}$, or $f_j=j$ if
  there is none; that is, $f_j=\min(\{j\}\cup\{k<j:v_k\in N(v_j)\})$. For $i<j$, $v_j\in W_i$ if and
  only if $v_j$ has a neighbor in $T_i$, that is, $f_j\le i$; also $v_j\in W_j$, and
  $v_j\notin W_i$ for $i>j$ because then $v_j\in T_i\setminus\{v_i\}$. So $v_j$ lies
  exactly in the consecutive bags $W_{f_j},\dots,W_j$. An edge $v_av_b$ with $a<b$ has
  both ends in $W_a$, because $v_b\notin T_a$ has the neighbor $v_a\in T_a$. So the bags
  form a path decomposition. Its width is $\max_i|N(T_i)\setminus T_i|$, since
  $v_i\in T_i$, and $|N(T_i)\setminus T_i|\le e(T_i)$, since every vertex of
  $N(T_i)\setminus T_i$ is the end of an edge leaving $T_i$. Applied to an ordering of
  minimum cutwidth, this gives $\pw(G)\le\operatorname{cw}(G)$.
\end{proof}

The second lemma is the only fact about normal variables that we need beyond
\cref{sec:warmup-toolkit}. Part (a) is Sheppard's formula~\cite{Sheppard1899}, known in
algorithms as the random-hyperplane lemma of Goemans and
Williamson~\cite{GoemansWilliamson1995}. Part (b) says that the median $0$ is the worst
threshold: the two variables straddle $t$ most often where their density is highest.

\begin{lemma}[two Gaussians]\label{lem:crossing}
  Let $(Z_w)_{w\in W}$ be independent standard normal variables, let
  $U=\sum_wa_wZ_w$ and $V=\sum_wb_wZ_w$ with $\sum_wa_w^2=\sum_wb_w^2=1$, and let
  $\rho=\sum_wa_wb_w$, which is the correlation of $U$ and $V$ by \ref{P:sums}. If
  $\rho>-1$, then
  \begin{enumerate}[label=\textup{(\alph*)}]
    \item\label{lem:crossing:a} $\Pr(UV\le0)=\arccos(\rho)/\pi$;
    \item\label{lem:crossing:b} $\Pr\bigl(\min(U,V)\le t<\max(U,V)\bigr)\le\arccos(\rho)/\pi$
      for every $t\in\R$.
  \end{enumerate}
\end{lemma}

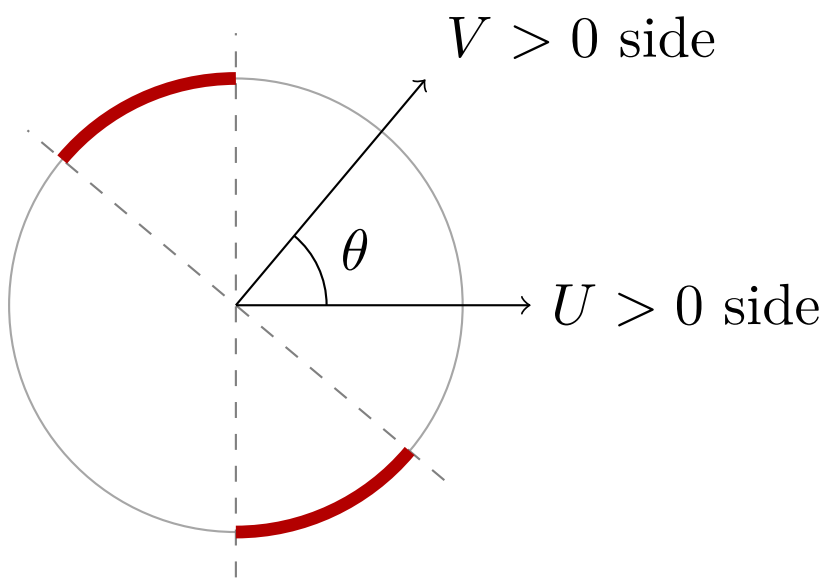
\begin{figure}[ht]
  \centering
  \begin{tikzpicture}[scale=1.5]
    \draw[gray!70] (0,0) circle (1);
    \draw[dashed,gray] (0,-1.2) -- (0,1.2);
    \draw[dashed,gray] (-40:1.2) -- (140:1.2);
    \draw[->] (0,0) -- (1.3,0) node[right] {$U>0$ side};
    \draw[->] (0,0) -- (50:1.3) node[above right] {$V>0$ side};
    \draw[line width=2.4pt,red!70!black] (-90:1) arc[start angle=-90,end angle=-40,radius=1];
    \draw[line width=2.4pt,red!70!black] (90:1) arc[start angle=90,end angle=140,radius=1];
    \draw (0.4,0) arc[start angle=0,end angle=50,radius=0.4];
    \node at (25:0.58) {$\theta$};
  \end{tikzpicture}
  \caption{The proof of \cref{lem:crossing}\ref{lem:crossing:a}. The point
    $(G_1,G_2)$ has a uniformly random direction. $U>0$ when this direction lies in the
    half-circle to the right of the vertical dashed line, and $V>0$ when it lies in the
    same half-circle rotated by $\theta=\arccos\rho$. The signs differ exactly on the
    two thick arcs, each of length $\theta$.}\label{fig:warmup-circle}
\end{figure}

\begin{proof}
  If $\rho=1$, then $\sum_w(a_w-b_w)^2=2-2\rho=0$, so $U=V$, $\Pr(UV\le0)=\Pr(U=0)=0$
  by \ref{P:sums}, and the event of \ref{lem:crossing:b} is empty. Let $\rho<1$ and
  $\theta=\arccos\rho\in(0,\pi)$.

  \ref{lem:crossing:a} We write $V$ in terms of $U$ and a second, independent normal
  variable. Put $a'_w=(b_w-\rho a_w)/\sin\theta$. Then $\sum_wa_wa'_w=0$ and
  $\sum_wa'^2_w=(1-2\rho^2+\rho^2)/\sin^2\theta=1$, so $G_1=U$ and
  $G_2=\sum_wa'_wZ_w$ are standard normal \ref{P:sums} and independent
  \ref{P:orth}, and $V=\cos\theta\,G_1+\sin\theta\,G_2$. Write the point $(G_1,G_2)$
  as $r(\cos\phi,\sin\phi)$ with $r\ge0$. By \ref{P:rotation} the angle $\phi$ is
  uniform on the circle, and $U=r\cos\phi$ and $V=r\cos(\phi-\theta)$. Ignoring the
  outcomes where $U=0$ or $V=0$, which have probability $0$ by \ref{P:sums},
  $U>0$ exactly when $\phi$ lies in the open half-circle centred at angle $0$, and
  $V>0$ exactly when $\phi$ lies in the half-circle centred at angle $\theta$, which is
  the first one rotated by $\theta$. So $UV<0$ exactly when $\phi$ lies in one of the
  two half-circles but not the other. As \cref{fig:warmup-circle} shows,\footnote{Part~\ref{lem:crossing:a}
  also has a proof by calculus alone. For $-1<\rho<1$ let $P(\rho)=\Pr(U>0,\,V>0)$. By
  Plackett's identity~\cite{Plackett1954} for two variables, $P'(\rho)$ is the joint
  density of $(U,V)$ at $(0,0)$, which is $1/(2\pi\sqrt{1-\rho^2})$. Since
  $P(0)=\frac14$, integration gives $P(\rho)=\frac14+\frac{\arcsin\rho}{2\pi}$, and by
  symmetry $\Pr(UV\le0)=1-2P(\rho)=\frac{\arccos\rho}\pi$.} this set
  consists of two arcs of length $\theta$, so
  $\Pr(UV\le0)=2\theta/(2\pi)=\arccos(\rho)/\pi$.

  \ref{lem:crossing:b} We compare the threshold $t$ with the threshold $0$, in four
  steps.

  \emph{Step 1.} Let $S=\frac{U+V}2$ and $D=\frac{U-V}2$. These are sums of the $Z_w$
  with coefficients $\frac{a_w+b_w}2$ and $\frac{a_w-b_w}2$, and
  $\sum_w\frac{(a_w+b_w)(a_w-b_w)}4=\frac14\bigl(\sum_wa_w^2-\sum_wb_w^2\bigr)=0$. So $S$
  and $D$ are independent by \ref{P:orth}. Also $\Var S=\frac{1+\rho}2>0$, so $S$ has a
  density $f$ that is symmetric and decreasing in $|s|$ \ref{P:sums}.

  \emph{Step 2.} The event of \ref{lem:crossing:b} implies $(U-t)(V-t)\le0$, and
  $(U-t)(V-t)=(S-t)^2-D^2$. So its probability is at most $\Pr(|S-t|\le|D|)$.

  \emph{Step 3.} An interval of given length has the largest mass under $f$ when it is
  centred at $0$: for all $c\in\R$ and $w\ge0$,
  $\int_{c-w}^{c+w}f\le\int_{-w}^wf$. Indeed, for $c\ge0$ the difference of the two sides
  is $\int_0^c\bigl(f(u+w)-f(u-w)\bigr)\,du\le0$, as $|u+w|\ge|u-w|$ for $u\ge0$, and the
  case $c<0$ follows by symmetry.

  \emph{Step 4.} Fix a value of $D$. Since $S$ is independent of $D$, the probability
  that $|S-t|\le|D|$ is then the mass of the interval $[t-|D|,t+|D|]$ under $f$, which by
  Step~3 is at most the mass of $[-|D|,|D|]$. Averaging over the values of $D$ gives
  $\Pr(|S-t|\le|D|)\le\Pr(|S|\le|D|)=\Pr(UV\le0)$, since $UV=S^2-D^2$. Now
  \ref{lem:crossing:a} finishes the proof.
\end{proof}

Fix an integer $R\ge1$. For vertices $v,w$ define the \emph{coefficient} $c_v(w)$ and,
for vertices $u,v$, the \emph{overlap} $\sigma(u,v)$ by
\[
  c_v(w)=
  \begin{cases}2^{-\dist(v,w)/2}&\text{if }w\in B_R(v),\\0&\text{otherwise,}
  \end{cases}
  \qquad \sigma(u,v)=\sum_{w\in V(G)}c_u(w)\,c_v(w).
\]
We call $c_v(w)^2=2^{-\dist(v,w)}$ the \emph{weight} of $w$ seen from $v$. So
$\sigma(u,v)$ is a weighted count of the vertices of $B_R(u)\cap B_R(v)$, and
$\sigma(v,v)=\sum_{k=0}^R2^{-k}\,|\{w:\dist(v,w)=k\}|$ is the total weight of the ball
around $v$. In \cref{sec:warmup-proof} the label of $v$ will be
$Y_v=\sigma(v,v)^{-1/2}\sum_wc_v(w)Z_w$. By \ref{P:sums}, the correlation of the labels
of $u$ and $v$ will then be $\sigma(u,v)/\sqrt{\sigma(u,u)\sigma(v,v)}$. The decay
$2^{-k/2}$ of the coefficients balances the spheres: in the $3$-regular tree the sphere
of radius $k\ge1$ around $v$ has $3\cdot2^{k-1}$ vertices, so every sphere carries the
same total weight
$\frac32$, and passing from $v$ to a neighbor loses only part of the outermost of the
$R$ spheres. The third lemma shows that this holds in every graph of maximum degree
$3$, with no assumption on girth, regularity or connectivity. The same coefficients
$(d-1)^{-k/2}$ on spheres are the test functions in Nilli's proof of the Alon--Boppana
bound on the second eigenvalue of $d$-regular graphs~\cite{Nilli1991}, and the
constant $\frac{2\sqrt2}3$ below is the norm $2\sqrt2$ of the adjacency operator of the
$3$-regular tree divided by the degree.

\begin{lemma}[short edges]\label{lem:warmup-edges}
  Let $G$ be a graph with $\Delta(G)\le3$ and let $R\ge1$ be an integer. Then every edge
  $uv$ of $G$ satisfies $\sigma(u,v)\ge r_R\sqrt{\sigma(u,u)\,\sigma(v,v)}$, where
  $r_R=\frac{2\sqrt2}3\bigl(1-\frac1R\bigr)$.
\end{lemma}

\begin{proof}
  The proof has two steps: every shared vertex contributes to $\sigma(u,v)$ nearly its
  full weight from both ends, and the vertices that are not shared carry at most a
  $\frac1R$ fraction of the weight. Write $a_w=c_u(w)$, $b_w=c_v(w)$ and
  $I=B_R(u)\cap B_R(v)$. The coefficients are
  nonnegative and vanish outside the balls, so $\sigma(u,v)=\sum_{w\in I}a_wb_w$. For
  $w\in I$ the distances from $u$ and from $v$ to $w$ differ by at most $1$, so
  $\lambda=a_w/b_w\in[2^{-1/2},2^{1/2}]$; hence $\lambda+\frac1\lambda\le\frac3{\sqrt2}$, that is,
  $a_w^2+b_w^2\le\frac3{\sqrt2}a_wb_w$. Summing over $I$, with
  $A=\sum_{w\in I}a_w^2$ and $B=\sum_{w\in I}b_w^2$ the weights of the shared vertices
  seen from $u$ and from $v$, and using $A+B\ge2\sqrt{AB}$,
  \[
    \sigma(u,v)\ \ge\ \frac{\sqrt2}{3}(A+B)\ \ge\ \frac{2\sqrt2}{3}\sqrt{AB}.
  \]
  It remains to show that the shared vertices carry almost all the weight, that is,
  $A\ge(1-\frac1R)\sigma(u,u)$ and $B\ge(1-\frac1R)\sigma(v,v)$. Let
  $s_j$ be the number of vertices at distance $j$ from $u$. For $j\ge1$, a vertex at
  distance $j$ has a neighbor at distance $j-1$ and hence at most two neighbors at
  distance $j+1$, so $s_{j+1}\le2s_j$, and the sphere weights $s_j2^{-j}$ are
  nonincreasing in $j\ge1$. A vertex $w\in B_R(u)\setminus I$ has $\dist(v,w)>R$ and
  hence $\dist(u,w)\ge\dist(v,w)-1\ge R$, so it lies on the sphere of radius $R$. Since
  $\sigma(u,u)=1+\sum_{j=1}^Rs_j2^{-j}$,
  \[
    \sigma(u,u)-A\ \le\ s_R2^{-R}\ \le\ \frac1R\sum_{j=1}^Rs_j2^{-j}\ \le\ \frac1R\,\sigma(u,u),
  \]
  and likewise $\sigma(v,v)-B\le\frac1R\sigma(v,v)$. Hence
  $\sigma(u,v)\ge r_R\sqrt{\sigma(u,u)\,\sigma(v,v)}$.
\end{proof}

The last lemma is the variance bound for sums of locally dependent terms.

\begin{lemma}[local sums]\label{lem:warmup-variance}
  Let $G$ be a graph with $\Delta(G)\le3$, let $(Z_w)_{w\in V(G)}$ be independent random
  variables, let $\ell\ge0$ be an integer and $\alpha>0$, and let $Q=\sum_{v\in V(G)}I_v$,
  where each $I_v$ takes values in $[0,\alpha]$ and is a function of
  $(Z_w)_{w\in B_\ell(v)}$. Then $\Var Q\le\frac34\alpha^2\,4^{\ell}\,n$.
\end{lemma}

\begin{proof}
  We use \ref{P:linearity} throughout. We have $\Var Q=\sum_{u,v}\Cov(I_u,I_v)$. If $B_\ell(u)\cap B_\ell(v)=\emptyset$,
  then $I_u$ and $I_v$ are functions of disjoint sets of independent variables, hence
  independent, and $\Cov(I_u,I_v)=0$. Otherwise $\dist(u,v)\le2\ell$ and
  $|\Cov(I_u,I_v)|\le\sqrt{\Var I_u\Var I_v}\le\frac{\alpha^2}4$, because a variable $X$
  with values in $[0,\alpha]$ has $\Var X\le\E(X-\frac\alpha2)^2\le\frac{\alpha^2}4$.
  Since $\Delta(G)\le3$, every ball satisfies
  $|B_{2\ell}(u)|\le1+\sum_{j=1}^{2\ell}3\cdot2^{j-1}<3\cdot4^{\ell}$, so at most
  $3\cdot4^\ell n$ pairs $(u,v)$ contribute.
\end{proof}

\subsection{Proof of \texorpdfstring{\cref{thm:cut-sweep}}{the cut-sweep theorem}}\label{sec:warmup-proof}

Let $G$ be a graph with $\Delta(G)\le3$, so that $m\le\frac32n$, and fix integers
$R,K\ge1$. Let $Z=(Z_w)_{w\in V(G)}$ have independent standard normal coordinates and
give every vertex $v$ the \emph{label}
\[
  Y_v=\frac1{\sqrt{\sigma(v,v)}}\sum_{w\in V(G)}c_v(w)\,Z_w ,
\]
a weighted sum of the independent normal numbers in the ball $B_R(v)$. By
\ref{P:sums}, $Y_v$ is standard normal and
$\Corr(Y_u,Y_v)=\sigma(u,v)/\sqrt{\sigma(u,u)\sigma(v,v)}$, and every pair
$(Y_u,Y_v)$ has the form of \cref{lem:crossing} with $a_w=c_u(w)/\sqrt{\sigma(u,u)}$
and $b_w=c_v(w)/\sqrt{\sigma(v,v)}$. For $t\in\R$ let
$S_t=\{v:Y_v\le t\}$ and $e_t=e(S_t)$; an edge $uv$ \emph{crosses} $t$ if exactly one
of its ends lies in $S_t$, that is, if $\min(Y_u,Y_v)\le t<\max(Y_u,Y_v)$.

\emph{Distinct labels.} We need the labels to be distinct, so that sorting by label
gives an order in which every prefix is a set $S_t$. Let $u\ne v$. Among the coefficients of the $Z_w$ in $Y_u$
the largest is the one of $Z_u$, uniquely, and in $Y_v$ it is the one of $Z_v$; so
the two coefficient sequences differ. Hence $Y_u-Y_v$ is a normal variable with
positive variance, so $\Pr(Y_u=Y_v)=0$ by \ref{P:sums}. So all labels are distinct
with probability $1$.

\emph{One threshold.} By \cref{lem:warmup-edges} and
\cref{lem:crossing}\ref{lem:crossing:b}, and since $\arccos$ is decreasing, every edge crosses every threshold with
probability at most $\arccos(r_R)/\pi$. Summing over the $m\le\frac32n$ edges
(linearity of expectation, \ref{P:linearity}) gives
\begin{equation}\label{eq:warmup-mean}
  \E e_t\le\beta_Rn\quad\text{for every }t\in\R,\qquad
  \beta_R=\frac{3}{2\pi}\arccos r_R .
\end{equation}
Since $|\frac{d}{dx}\arccos x|=(1-x^2)^{-1/2}\le3$ for $0\le x\le\frac{2\sqrt2}3$ and
$\frac{2\sqrt2}3-r_R=\frac{2\sqrt2}{3R}$, the mean value theorem gives
$\beta_R\le c_*+\frac3{2\pi}\cdot3\cdot\frac{2\sqrt2}{3R}=c_*+\frac{3\sqrt2}{\pi R}$.

\emph{From one threshold to all thresholds.} The bound \eqref{eq:warmup-mean} holds
for each threshold separately, on average over the random numbers. We need one
outcome in which all thresholds are good at the same time, and there are infinitely
many thresholds. So we first reduce them to finitely many counts on a grid, then show
that each count is close to its mean with high probability, and finally use the union
bound \ref{P:union}.

\emph{A grid of thresholds.} Labels outside $[-3,3]$ are rare
(\cref{sec:warmup-toolkit}), so the grid only needs to cover $[-3,3]$. Let
$\tau_j=-3+j/K$ for $0\le j\le6K$, and for an outcome of the labels let $N_j=|\{v:\tau_j<Y_v\le\tau_{j+1}\}|$,
$C_-=|\{v:Y_v\le-3\}|$ and $C_+=|\{v:Y_v>3\}|$. Let $\tau_j\le\tau<\tau_{j+1}$. An edge
$ab$ with $Y_a\le\tau<Y_b$ either has $Y_a\le\tau_j$, and then it is counted by
$e_{\tau_j}$, or has $\tau_j<Y_a\le\tau<\tau_{j+1}$, and then $a$ is one of the $N_j$
counted vertices, each the end of at most three edges. Hence
$e_\tau\le e_{\tau_j}+3N_j$. In the same way $e_\tau\le3C_-$ for $\tau<-3$, where the
lower end of a crossing edge has label at most $\tau<-3$, and $e_\tau\le3C_+$ for
$\tau\ge3$, where the upper end has label greater than $3$.

\emph{Concentration.} Consider the $12K+2$ counts $e_{\tau_j}$ and $N_j$ for
$0\le j<6K$, $C_-$ and $C_+$. Each label $Y_u$ is a function of $(Z_w)_{w\in B_R(u)}$.
So $e_\tau=\sum_v|\{u\in N(v):Y_v\le\tau<Y_u\}|$, which counts every crossing edge at
its lower end, is a sum over $v$ of $[0,3]$-valued functions of
$(Z_w)_{w\in B_{R+1}(v)}$, and so are $N_j$, $C_-$ and $C_+$. By
\cref{lem:warmup-variance} with $\alpha=3$ and $\ell=R+1$, each count has variance at
most $27\cdot4^Rn$, so by Chebyshev's inequality \ref{P:chebyshev} it deviates from its mean by at least
$n^{7/8}$ with probability at most $27\cdot4^Rn^{-3/4}$. If
\begin{equation}\label{eq:warmup-condition}
  (12K+2)\cdot27\cdot4^R<n^{3/4},
\end{equation}
then the probability that some count deviates from its mean by $n^{7/8}$ or more is
less than $1$. So by \ref{P:union} some outcome of the numbers $Z_w$ has distinct
labels and all $12K+2$ counts within $n^{7/8}$ of their means; fix such an outcome.
Each $Y_v$ is standard normal, and its density is at most $\varphi(0)$, so it lies in
an interval of length $\frac1K$ with probability at most $\varphi(0)/K$. Hence
$\E N_j\le\varphi(0)n/K$, and $\E C_\pm=\Phi(-3)n$. By \eqref{eq:warmup-mean} and the
grid, every $\tau\in[-3,3)$ satisfies
\[
  e_\tau\ \le\ \Bigl(\beta_R+\frac{3\varphi(0)}K\Bigr)n+4n^{7/8},
\]
and every $\tau\notin[-3,3)$ satisfies $e_\tau\le3\Phi(-3)n+3n^{7/8}$, which is
smaller, because $3\Phi(-3)<0.005<c_*\le\beta_R$.

\emph{Conclusion.} Order $V(G)$ by increasing label in the fixed outcome. Every
prefix of this ordering is a set $S_t$, so by \cref{lem:warmup-layout}
\[
  \pw(G)\ \le\ \operatorname{cw}(G)\ \le\ \max_te_t\ \le\
  \Bigl(\beta_R+\frac{3\varphi(0)}K\Bigr)n+4n^{7/8}.
\]
Take $K=R=\lfloor\frac14\log_2n\rfloor$. Then $4^R\le\sqrt n$ and
$12K+2\le3\log_2n+2$, so \eqref{eq:warmup-condition} follows from
$27(3\log_2n+2)<n^{1/4}$, which holds for all $n\ge2^{48}$ (at $n=2^{48}$ it reads
$3942<4096$). For $2^{46}\le n<2^{48}$ we have $K=R=11$, and
\eqref{eq:warmup-condition} reads $134\cdot27\cdot4^{11}<n^{3/4}$, which holds since
$134\cdot27<2^{12}$ and $2^{12}\cdot2^{22}=2^{34}<2^{34.5}\le n^{3/4}$. So
\eqref{eq:warmup-condition} holds for all $n\ge2^{46}$. For these $n$, since
$\frac{3\sqrt2}\pi+3\varphi(0)<2.55$,
\[
  \pw(G)\ \le\ \operatorname{cw}(G)\ \le\ c_*n+\frac{2.55}Rn+4n^{7/8}\ =\ c_*n+o(n).
\]
For $n<2^{46}$ the bound $\operatorname{cw}(G)\le m\le\frac32n$ is absorbed by the
remainder. \qed

\subsection{Larger maximum degree}\label{sec:warmup-degree-p}

The cut sweep extends to every maximum degree. For an integer $p\ge3$ put
\[
  r_{pp}=\frac{2\sqrt{p-1}}{p}\qquad\text{and}\qquad
  \gamma_p=\frac{\arccos r_{pp}}{\pi},
\]
so that $r_{33}=\frac{2\sqrt2}3$, $r_{44}=\frac{\sqrt3}2$, $r_{55}=\frac45$, and
$\gamma_3=\frac23c_*\approx0.10817$, $\gamma_4=\frac16$ and $\gamma_5<0.20484$.

\begin{theorem}[cut sweep, maximum degree $p$]\label{thm:cut-sweep-p}
  For every integer $p\ge3$ there is a function $\eta_p(n)=o(n)$ such that every
  graph $G$ with $\Delta(G)\le p$ satisfies
  \[
    \pw(G)\ \le\ \operatorname{cw}(G)\ \le\ \gamma_p\,m+\eta_p(n).
  \]
  In particular, the bisection width and the cutwidth of every graph $G$ with
  $\Delta(G)\le4$ are at most $\frac m6+o(n)\le\frac n3+o(n)$, and those of every graph
  $G$ with $\Delta(G)\le5$ are at most $\gamma_5m+o(n)\le\frac52\gamma_5n+o(n)<0.5121n+o(n)$.
\end{theorem}

\begin{proof}
  We follow \cref{sec:warmup-proof} and list the changes. Here and below, $O_p(\cdot)$
  hides constants that depend on $p$. Let $c_v(w)=(p-1)^{-\dist(v,w)/2}$ for
  $w\in B_R(v)$, and $c_v(w)=0$ otherwise, and define $Y_v$ from $c_v$ as in
  \cref{sec:warmup-proof}. Then every edge $uv$ satisfies
  $\Corr(Y_u,Y_v)\ge r_{pp}(1-\frac1R)$: the proof of \cref{lem:warmup-edges} goes
  through with $2$ replaced by $p-1$, the ratio $\lambda\in[(p-1)^{-1/2},(p-1)^{1/2}]$,
  $\lambda+\frac1\lambda\le\frac2{r_{pp}}$, and $s_{j+1}\le(p-1)s_j$. (These are the
  padded labels of \cref{def:labels}\ref{def:labels:padded}, and the bound is
  \cref{lem:correlation-padded}.) The labels are distinct with probability $1$, as
  before.

  \emph{One threshold.} By \cref{lem:crossing}\ref{lem:crossing:b}, every edge crosses
  every threshold with probability at most $\arccos(r_{pp}(1-\frac1R))/\pi$, so
  $\E e_t\le\frac m\pi\arccos(r_{pp}(1-\frac1R))$ for every $t$. On $[0,r_{pp}]$ the derivative of
  $\arccos$ has absolute value at most $(1-r_{pp}^2)^{-1/2}=\frac p{p-2}$, and $m\le\frac p2n$,
  so
  \[
    \E e_t\ \le\ \gamma_p\,m+\frac{p^2r_{pp}}{2\pi(p-2)}\cdot\frac nR\qquad\text{for every }t\in\R.
  \]

  \emph{Grid and tails.} With the grid of \cref{sec:warmup-proof}, an edge crossing
  $\tau\in[\tau_j,\tau_{j+1})$ either crosses $\tau_j$ or has its lower end among the
  $N_j$ counted vertices, each the end of at most $p$ edges, so $e_\tau\le e_{\tau_j}+pN_j$.
  For the tails we weight by degree, so that the bound stays proportional to $m$: let
  $C'_-=\sum_{v:\,Y_v\le-3}d(v)$ and $C'_+=\sum_{v:\,Y_v>3}d(v)$. Then
  $e_\tau\le C'_-$ for $\tau<-3$ and $e_\tau\le C'_+$ for $\tau\ge3$, and
  $\E C'_\pm=2m\Phi(-3)$.

  \emph{Concentration.} The counts $e_{\tau_j}$, $N_j$ and $C'_\pm$ are sums over $v$ of
  $[0,p]$-valued functions of $(Z_w)_{w\in B_{R+1}(v)}$. The proof of
  \cref{lem:warmup-variance} applies with $|B_{2\ell}(u)|\le1+\sum_{j=1}^{2\ell}p(p-1)^{j-1}
  \le\frac p{p-2}(p-1)^{2\ell}$, so each count has variance at most
  $\frac{p^3}{4(p-2)}(p-1)^{2R+2}n$. Take $K=R=\lfloor\frac18\log_{p-1}n\rfloor$, so that
  $(p-1)^{2R}\le n^{1/4}$. By Chebyshev's inequality and a union bound over the $12K+2$
  counts, with probability at least $1-O_p(n^{-1/2}\log n)$ all of them lie within
  $n^{7/8}$ of their means, and then every $\tau\in[-3,3)$ satisfies
  \[
    e_\tau\ \le\ \gamma_pm+\Bigl(\frac{p^2r_{pp}}{2\pi(p-2)}+p\,\varphi(0)\Bigr)\frac nR
    +(p+1)n^{7/8},
  \]
  while every $\tau\notin[-3,3)$ satisfies $e_\tau\le2\Phi(-3)m+n^{7/8}$, which is smaller
  because $2\Phi(-3)<0.003<\gamma_3\le\gamma_p$. As in \cref{sec:warmup-proof}, ordering
  the vertices by their labels gives $\pw(G)\le\operatorname{cw}(G)\le\max_te_t$, and the
  error terms are $O_p(n/R)+(p+1)n^{7/8}=o(n)$, since $R\to\infty$; for the finitely many $n$ below the threshold where the
  probability is positive, $\operatorname{cw}(G)\le m\le\frac p2n$ is absorbed into
  $\eta_p$.

  For the bisection width, the first $\lfloor n/2\rfloor$ vertices of the ordering form a
  set $X$ with $e(X)\le\operatorname{cw}(G)$. For $p=4$ we have $r_{44}=\cos\frac\pi6$, so
  $\gamma_4=\frac16$, and $m\le2n$; for $p=5$, $m\le\frac52n$ and $\frac52\gamma_5<0.5121$.
\end{proof}

The proof is a randomized polynomial-time algorithm. The labels need the balls of radius
$R=O(\log n)$, which have polynomial size, and the cut values of an ordering can be
computed in polynomial time and compared with the explicit bound in the display above.
So repeating with fresh Gaussians until the bound holds computes, in expected polynomial
time, an ordering of cutwidth at most $\gamma_pm+o(n)$, for every $p\ge3$. Since the
algorithm checks the bound on the ordering it outputs, its answer is correct whatever
numbers it samples; sampling the Gaussians to finite precision can affect only the
success probability of a round, which we do not analyze. For $p\in\{4,5\}$, \cref{thm:alg-cut-sweep} in
\cref{sec:alg-cutwidth} derandomizes it with the method of conditional expectations and
computes an ordering of cutwidth at most $\gamma_pm+o(n)$, and a bisection with at most
that many edges between its sides, in deterministic time $n^{2+o(1)}$.

For $p=3$, \cref{thm:cut-sweep-p} gives $\gamma_3m\le c_*n$, which \cref{thm:cut-sweep}
states, and the stronger \cref{cor:bisection} gives $\cthree n$. For $p=4$ the pathwidth
bound of \cref{thm:cut-sweep-p}, $\frac n3$, is weaker than $F_4n$ (\cref{thm:deg4}), but
the bounds on bisection width and cutwidth are new: Monien and
Preis~\cite[Theorem~2]{MonienPreis2006} proved that $4$-regular graphs have bisection
width at most $\frac{2n}5+o(n)$, and for maximum degree~$4$ the relations
between cutwidth and pathwidth~\cite[Theorem~3.2]{MakedonSudborough1989}%
\cite[Theorem~2.1]{EllisSudboroughTurner1994} give only
$\operatorname{cw}(G)\le2\pw(G)+O(1)\le2F_4n+o(n)$.

\subsection{Remarks}\label{sec:warmup-remarks}

The argument loses in two places. First, the angle: on cubic graphs of girth greater
than $2R+3$, where every ball $B_{R+1}(v)$ induces a tree, no choice of coefficients
$c_v(w)$ supported on balls of radius $R$ and depending only on $\dist(v,w)$
does better than $\arcsin\frac13$. There the edge correlations of such labels are those
of the $3$-regular tree, at most $\frac{2\sqrt2}3$ because the adjacency operator of
that tree has norm $2\sqrt2$ (\cref{rem:labels-optimal}); so at least $c_*n$ edges
cross the median in expectation, and counting crossing edges cannot give a smaller
constant with such labels. Second, the count: a vertex of $N(S_t)\setminus S_t$ is
counted once for every neighbor in $S_t$. Counting each such vertex once needs the
joint distribution of a label and all its neighbor labels, in which the correlations between two
neighbors are unknown; the star comparison of \cref{sec:stars} handles them and gives
the plain-sweep constant $F_3\approx0.14016$ (\cref{thm:deg4}\ref{thm:deg4:b} with
$n_4=0$), and switching from inner to outer frontiers at threshold $0$ gives $\cthree$
(\cref{thm:cubic}). Finally, for graphs that converge
locally to the $3$-regular tree (for example, cubic graphs of large girth), $c_*$ is
the density of the sign cut (threshold $0$) of the Gaussian wave~\cite{Lyons2017CPC}.

\paragraph{The edge isoperimetric number.} For a $d$-regular graph the
bisection of \cref{thm:cut-sweep-p} gives the edge isoperimetric number
$i(G)=\min\{e(S)/|S|:0<|S|\le\frac n2\}$, where $e(S)$ is the number of edges with
exactly one end in $S$, at most $d\gamma_d+o(1)\le\frac
d2-\frac2\pi\sqrt{d-1}+o(1)$. Alon's upper bound has the form $\frac d2-c\sqrt d$ for
$n>n_0(d)$, with $c=\frac3{16\sqrt2}<0.133$~\cite[Theorem~1.1 and~(11)]{Alon1997};
for $d\ge3$ and $n$ large enough, our bound gives
$\frac2\pi\sqrt{d-1}>0.51\sqrt d$ in place of $c\sqrt d$. For cubic graphs,
\cref{cor:bisection} gives $i(G)\le2\cthree+o(1)<0.2797$, in place of the
bound~$\frac13$ that follows from the theorem of Monien and
Preis~\cite[Theorem~1]{MonienPreis2006}, and
$4$-regular graphs have $i(G)\le\frac23+o(1)$.
{}
\section{Preliminaries}\label{sec:prelim}

This section fixes notation and collects the tools that all later sections share.
Two of them shape the whole of Part~I. The layout lemma turns a labeling of the
vertices into a path decomposition; it is the only way in which path decompositions
are built in Part~I (\cref{sec:prelim-layout}). The reductions on vertices of degree
at most two let the main proofs assume minimum degree at least three, at an additive
loss of $O(\log n)$ in the pathwidth (\cref{sec:prelim-reductions}).

Of the remaining subsections, \cref{sec:prelim-graphs} fixes the notation for graphs,
decompositions and remainders $o(n)$, \cref{sec:prelim-gauss} the Gaussian notation
and comparison tools, and \cref{sec:prelim-concentration} proves a variance bound for
sums of locally dependent variables, which makes frontier sizes concentrate.

\subsection{Graphs, decompositions and remainders}\label{sec:prelim-graphs}

Graphs are finite and simple. We follow Diestel's notation: $V(G)$ and $E(G)$ are the
vertex and edge sets, $uv\in E(G)$ is an edge, $d_G(v)$ or $d(v)$ is the degree,
$N(v)$ and $N[v]=N(v)\cup\{v\}$ are the open and closed neighborhoods,
$N(S)=\bigcup_{v\in S}N(v)$ for $S\subseteq V(G)$, and $\delta(G)$, $\Delta(G)$ are the
minimum and maximum degree. $G[U]$ is the subgraph induced by $U$ and $G-U$ is
obtained by deleting $U$. We write $n=|V(G)|$, $m=|E(G)|$ and $n_d=n_d(G)$ for the
number of vertices of degree $d$. The graph distance $\dist(v,w)\in\N\cup\{\infty\}$
is the number of edges of a shortest $v$--$w$ path, and
$B_k(v)=\{w:\dist(v,w)\le k\}$ is the ball of radius $k$. The sizes of these balls
enter the variance bound of \cref{lem:local-variance} and the analysis of the
algorithms. If $\Delta(G)\le\Delta$ with $\Delta\ge2$, then
\begin{equation}\label{eq:ball-size}
  |B_k(v)|\le 1+\sum_{i=1}^{k}\Delta(\Delta-1)^{i-1}\le\sum_{i=0}^{k}\Delta^{i}\le 2\Delta^{k},
\end{equation}
because a vertex at distance $i\ge1$ from $v$ has at most $\Delta-1$ neighbors at
distance $i+1$, and $\sum_{i=0}^k\Delta^i=(\Delta^{k+1}-1)/(\Delta-1)\le2\Delta^k$
for $\Delta\ge2$. For the maximum degrees $3$, $4$ and $5$, which occur most often
below, evaluating the middle sum gives the sharper bounds
\[
  \begin{aligned}
    |B_k(v)|&\le3\cdot2^k-2&&\text{if }\Delta=3,\\
    |B_k(v)|&\le2\cdot3^k-1&&\text{if }\Delta=4,\\
    |B_k(v)|&\le\tfrac13\bigl(5\cdot4^k-2\bigr)\le2\cdot4^k&&\text{if }\Delta=5.
  \end{aligned}
\]

\paragraph{Decompositions.}
A \emph{tree decomposition} of $G$ is a pair $(T,(W_t)_{t\in V(T)})$ of a tree $T$
and bags $W_t\subseteq V(G)$ such that every edge of $G$ lies in some bag and, for
every vertex $x$, the nodes $t$ with $x\in W_t$ induce a nonempty subtree $T_x$ of
$T$. Its width is $\max_t|W_t|-1$, and the treewidth $\tw(G)$ is the minimum width.
Path decompositions, pathwidth and cutwidth are as in \cref{sec:warmup}; a path
decomposition is a tree decomposition whose tree is a path. The empty graph has $\pw=\tw=-1$. We use the following elementary
facts without further mention: $\tw(G)\le\pw(G)$; both widths are monotone under
taking subgraphs; the pathwidth of a disjoint union is the maximum over its parts
(concatenate the decompositions); and adding a set $X$ of vertices to every bag of a
path decomposition of $G-X$ gives a path decomposition of $G$, so
$\pw(G)\le\pw(G-X)+|X|$. A path decomposition is given in \emph{interval form} if it
is described by the number $s$ of bags and, for each vertex, the indices of the first
and the last bag containing it; the algorithms of \cref{sec:algorithms} output their
decompositions in this form.

\paragraph{Remainders.}
Every remainder $o(n)$ in a statement about a class $\mathcal G$ of graphs means the
following: there is a function $\rho\colon\N\to\R$ with $\rho(n)/n\to0$ such that the
statement holds with $o(n)$ replaced by $\rho(|V(G)|)$ for every $G\in\mathcal G$.
So the remainder depends on $n$ only, and not on the graph. The same convention
applies to $o(m)$. The \emph{running maximum} of a function $\rho\colon\N\to\R$ is
\begin{equation}\label{eq:running-max}
  \bar\rho(x):=\max\bigl(\{0\}\cup\{\rho(k):k\in\N,\ k\le x\}\bigr)\qquad(x\ge0).
\end{equation}
It is nonnegative, nondecreasing and at least $\rho$ on $\N$. If $\rho(n)=o(n)$, then
$\bar\rho(x)=o(x)$: for $\eps>0$ there is $N$ with $\rho(k)\le\eps k$ for $k\ge N$,
so $\bar\rho(x)\le\max\{\bar\rho(N),\eps x\}$ for $x\ge N$, which is at most $\eps x$
for all large $x$. Hence every remainder may be assumed nonnegative and nondecreasing.

\subsection{Layouts from labelings}\label{sec:prelim-layout}

A \emph{labeling} of $G$ is a map $y\colon V(G)\to\R$, $v\mapsto y_v$; it is
\emph{injective} if all labels are distinct. For $t\in\R$ let
$S_t=\{v:y_v\le t\}$ be the sublevel set, and define its outer and inner frontiers
\[
  \outfr S_t=N(S_t)\setminus S_t=\{v:\ y_v>t,\ \exists u\in N(v):\ y_u\le t\},\qquad
  \infr S_t=\{u\in S_t:\ N(u)\not\subseteq S_t\}.
\]
The same notation $\outfr S=N(S)\setminus S$ and
$\infr S=\{u\in S:N(u)\not\subseteq S\}$ is used for arbitrary sets $S\subseteq V(G)$.
Every later bound on pathwidth comes from the following lemma, applied to a random
labeling whose frontiers are small at every threshold. The maxima over $t$ in the
lemma are attained, since $S_t$ takes at most $n+1$ values.

\begin{lemma}[layout]\label{lem:layout}
Let $G$ be a nonempty graph and $y$ an injective labeling of $G$, let
$v_1,\dots,v_n$ be the vertices in increasing order of their labels, and let
$T_i=\{v_1,\dots,v_i\}$. Then the bags $W_i=\{v_i\}\cup\outfr T_i$
\textup{(}$1\le i\le n$\textup{)} form a path decomposition of $G$ of width
$\max_i|\outfr T_i|$. Moreover, $\pw(G)\le\max_{t\in\R}|\outfr S_t|$ and
$\pw(G)\le\max_{t\in\R}|\infr S_t|$.
\end{lemma}

The quantity $\max_i|\infr T_i|$, minimized over all orderings, is the \emph{vertex
separation number} of $G$, and Kinnersley proved that it equals the pathwidth
\cite[Theorem~3.1]{Kinnersley1992}. The bound by inner frontiers is the easy direction
of that theorem, and the bound by outer frontiers is the same statement for the
reversed order. The algorithms of \cref{sec:algorithms} use the explicit bags.

\begin{proof}
The proof of \cref{lem:warmup-layout} shows that the bags $W_i$ form a path
decomposition of width $\max_i|\outfr T_i|$, since $\outfr T_i=N(T_i)\setminus T_i$;
it uses neither the labels nor a degree bound. By injectivity $T_i=S_{y_{v_i}}$, so
$\max_i|\outfr T_i|\le\max_t|\outfr S_t|$; in fact a bound $|\outfr S_t|\le w$ for
all $t$ in any set of thresholds that contains all labels already gives $\pw(G)\le w$.

For the bound by inner frontiers, apply the bound by outer frontiers to the injective
labeling $-y$. Its sublevel sets are the sets
$\{v:y_v\ge-t\}=V(G)\setminus A$, where $A=\{v:y_v<-t\}$ is either empty or equal to
$S_s$ for $s=\max\{y_v:y_v<-t\}$. For every $A\subseteq V(G)$ we have
$\outfr(V(G)\setminus A)=\{a\in A:N(a)\not\subseteq A\}=\infr A$, and $\infr\emptyset=\emptyset$.
Hence the outer frontiers for $-y$ are inner frontiers for $y$, which gives the
claim. \qedhere
\end{proof}

\subsection{Reductions on vertices of degree at most two}\label{sec:prelim-reductions}

The pathwidth bounds of this paper are proved for graphs of minimum degree at least
three and then transferred to all graphs. The transfer uses the classical
series--parallel reductions, which remove vertices of degree at most two. Reduction
rules of this kind are used in the pathwidth bounds of Kneis, M\"olle, Richter and
Rossmanith~\cite{KneisMRR2009} and of Edwards and McDermid~\cite{EdwardsM15}. The
reduction theorem of Edwards and McDermid \cite[Theorem~3.2]{EdwardsM15} interleaves
the reductions with vertex deletions and bounds the effect of all reductions on the
pathwidth once, at the end. Edwards and McDermid take two facts
from~\cite{KneisMRR2009}: that reductions commute with deletions, and that undoing all
reductions increases the pathwidth by at most $O(\log n)$. \Cref{lem:reduction} below
replaces both facts by a single self-contained statement, which needs no uniqueness
of the reduced graph. We first define the reductions and prove two auxiliary facts
about treewidth, then prove \cref{lem:reduction} and derive the transfer from minimum
degree three to all graphs (\cref{cor:min-degree-three}).

\begin{definition}[reduction operations]\label{def:reductions}
The \emph{reduction operations} on a graph $H$ are:
\begin{enumerate}[label=\textup{(R\arabic*)}]
  \item\label{red:R1} delete an isolated vertex;
  \item\label{red:R2} delete a vertex of degree $1$;
  \item\label{red:R3} for a vertex $v$ of degree $2$ whose neighbors $x,y$ are not
  adjacent, delete $v$ and add the edge $xy$ (\emph{suppression});
  \item\label{red:R4} delete a vertex of degree $2$ whose two neighbors are adjacent.
\end{enumerate}
A sequence of reduction operations is \emph{maximal} if no reduction operation
applies to the final graph.
\end{definition}

We use three properties of the reduction operations throughout.
\begin{enumerate}[label=\textup{(RP\arabic*)}]
  \item\label{red:deg} Each operation removes one vertex, produces a simple graph, and
  does not increase the degree of any remaining vertex: \ref{red:R1}, \ref{red:R2} and
  \ref{red:R4} only delete a vertex, and in \ref{red:R3} the vertices $x$ and $y$
  each lose the neighbor $v$ and gain one new neighbor.
  \item\label{red:local} Whether an operation applies at a vertex $v$, and its effect,
  depend only on $v$, its neighbors and the adjacency among them. So operations inside
  one component of a graph are operations on the whole graph.
  \item\label{red:max} By~\ref{red:deg}, every sequence of operations on $H$ has at
  most $|V(H)|$ steps, so maximal sequences exist. The final graph of a maximal
  sequence is empty or has minimum degree at least $3$, since a vertex of degree $0$,
  $1$ or $2$ admits \ref{red:R1}, \ref{red:R2}, or one of \ref{red:R3} and
  \ref{red:R4}.
\end{enumerate}

The proof of \cref{lem:reduction} needs two standard facts about graphs of bounded
treewidth~\cite{Bodlaender1998}.

\begin{lemma}\label{lem:tw-glue}
Let $G=G_1\cup G_2$ where $V(G_1)\cap V(G_2)=K$ and $K$ is either a single vertex or
the two ends of an edge that belongs to both $G_1$ and $G_2$. Then
$\tw(G)\le\max\{\tw(G_1),\tw(G_2)\}$.
\end{lemma}

\begin{proof}
Join tree decompositions of $G_1$ and $G_2$ of minimum width by an edge between a bag
of each that contains $K$. \qedhere
\end{proof}

\begin{lemma}\label{lem:tw-log}
Let $k\ge0$ and let $H$ be a graph with $N\ge1$ vertices and $\tw(H)\le k$. Then
$\pw(H)\le(k+1)\lfloor\log_2N\rfloor$.
\end{lemma}

\begin{proof}
Some bag $X$ of a tree decomposition of width at most $k$ leaves only components with
at most $N/2$ vertices (the standard balanced-separator argument). Decompose these
components by induction, concatenate the decompositions and add $X$ to every bag.
\qedhere
\end{proof}

The next lemma is the bookkeeping device of the paper. Suppose that an arbitrary
interleaving of reduction operations and vertex deletions is applied to a graph. Then
the pathwidth of the original graph exceeds that of the final graph by at most the
number of deleted vertices plus $O(\log n)$, however many reductions were made. The lemma is used for
the passage from minimum degree three to all graphs (\cref{cor:min-degree-three}) and
in the deletion process of \cref{sec:bootstrap}. Its part~\ref{it:reduction:structure} describes how $F-D$
attaches to the final graph; \cref{sec:alg-undo} uses it to compute the decomposition.

\begin{restatable}[undoing reductions]{lemma}{lemreduction}\label{lem:reduction}
Let $F$ be a nonempty graph. Apply to $F$ any finite sequence of steps, each of which
is a reduction operation or the deletion of a vertex of the current graph. Let $D$ be
the set of vertices removed by deletions and $R$ the final graph. Then:
\begin{enumerate}[label=\textup{(\alph*)}]
  \item\label{it:reduction:pw} $\pw(F)\le|D|+\pw(R)+3\log_2|V(F)|+1$;
  \item\label{it:reduction:structure} every edge of $F-D$ with both ends in $V(R)$ is
  an edge of $R$, and for every component $C$ of $(F-D)-V(R)$, the set $N_{F-D}(C)$
  is empty, a single vertex, or the two ends of an edge of $R$, and $\tw(F[C])\le2$.
\end{enumerate}
\end{restatable}

The proof is in \cref{app:reduction}.

In the simplest application no vertex is deleted and the sequence of reduction
operations is maximal, so the final graph is empty or has minimum degree at least
three by~\ref{red:max}. For coefficients that are nondecreasing in the
degree, this transfers every bound for minimum degree three to all graphs.

\begin{corollary}\label{cor:min-degree-three}
Let $k\in\{3,4,\dots\}\cup\{\infty\}$, let $0=a_0=a_1=a_2\le a_3\le a_4\le\cdots$
\textup{(}up to $a_k$\textup{)}, and let $\rho\colon\N\to\R_{\ge0}$ be nondecreasing.
Suppose that every graph $J$ with $\delta(J)\ge3$ and $\Delta(J)\le k$ satisfies
$\pw(J)\le\sum_da_dn_d(J)+\rho(|V(J)|)$. Then every nonempty graph $G$ with
$\Delta(G)\le k$ satisfies
\[
  \pw(G)\le\sum_da_dn_d(G)+\rho(n)+3\log_2n+1 .
\]
\end{corollary}

\begin{proof}
Apply a maximal sequence of reduction operations to $G$ and let $R$ be the final
graph. Degrees do not increase, so $\Delta(R)\le k$, and either $R$ is empty or
$\delta(R)\ge3$. Since the $a_d$ are nonnegative and nondecreasing in $d$, and every
vertex of $R$ has degree in $R$ at most its degree in $G$,
\[
  \sum_da_dn_d(R)=\sum_{v\in V(R)}a_{d_R(v)}\le\sum_{v\in V(R)}a_{d_G(v)}\le\sum_{v\in V(G)}a_{d_G(v)}=\sum_da_dn_d(G).
\]
If $R$ is nonempty, the hypothesis and the monotonicity of $\rho$ give
$\pw(R)\le\sum_da_dn_d(G)+\rho(n)$; if $R$ is empty, $\pw(R)=-1$ satisfies the same
bound. \Cref{lem:reduction} with $D=\emptyset$ completes the proof. \qedhere
\end{proof}

\subsection{Gaussian notation and comparison tools}\label{sec:prelim-gauss}

The star bounds of \cref{sec:stars} and the switch lemma of \cref{sec:cubic} compare
\emph{orthant probabilities}, probabilities of the form $\Pr(Y_i>t_i\ \forall i)$, or
in the switch lemma combinations of them, of Gaussian vectors $Y$ with different correlation matrices. This subsection fixes
the Gaussian notation and states the three classical tools that these comparisons
use: Plackett's identity, Slepian's inequality and Sheppard's formula.

We write $\varphi$ and $\Phi$ for the standard normal density and distribution
function, $N(\mu,\sigma^2)$ for the normal distribution, and
\[
  \varphi_2(x,y;\rho)=\frac{1}{2\pi\sqrt{1-\rho^2}}\exp\Bigl(-\frac{x^2-2\rho xy+y^2}{2(1-\rho^2)}\Bigr)
\]
for the standard bivariate normal density with correlation $\rho\in(-1,1)$. Note
that $\varphi_2(x,y;\rho)\le(2\pi\sqrt{1-\rho^2})^{-1}$.

A random vector is \emph{centered Gaussian} if every linear combination of its
coordinates is a centered normal variable (possibly degenerate); its distribution
$N(0,\Sigma)$ is determined by its covariance matrix $\Sigma$, which may be singular.
A \emph{correlation matrix} is a positive semidefinite matrix with unit diagonal. Every $k\times k$ correlation matrix $C$ is the Gram matrix
$C_{ij}=\langle v_i,v_j\rangle$ of unit vectors $v_1,\dots,v_k\in\R^k$ (for instance the
columns of $C^{1/2}$), and then $U_i=\langle v_i,G\rangle$ with $G\sim N(0,I_k)$ defines a
centered Gaussian vector with correlation matrix $C$.

For $r\in[-1,1]$ we write $s_r=\sqrt{1-r^2}$. In Part~I the letter $r$ always
denotes a correlation level. If $(Y_1,\dots,Y_k)$ is centered Gaussian with
$\Var Y_i>0$ for all $i$, then $\Pr(Y_i=t_i)=0$ for every $i$ and every $t_i\in\R$, so strict and weak inequalities
between coordinates and constants may be exchanged in all probabilities below.

The comparison tool used throughout is Slepian's inequality: raising the
correlations of a Gaussian vector, with its variances fixed, raises its orthant
probabilities. We state it with arbitrary thresholds and singular covariance matrices
allowed, because both cases occur (the ideal stars of \cref{sec:stars} have singular
covariance). It is derived from Plackett's identity,
which also bounds how fast an orthant probability can change with the correlations.

\begin{lemma}[Plackett's identity {\cite{Plackett1954}}]\label{lem:plackett}
Let $k\ge2$ and $t\in\R^k$, and for a positive definite $k\times k$ correlation matrix
$\Sigma$ let $P(\Sigma)=\Pr(Y_i>t_i\ \forall i)$ with $Y\sim N(0,\Sigma)$. Then $P$ is
continuously differentiable in the off-diagonal entries $\Sigma_{ij}=\Sigma_{ji}$
\textup{(}$i<j$\textup{)} on the open set of positive definite correlation matrices,
and
\[
  \frac{\partial P}{\partial\Sigma_{ij}}
  =\varphi_2(t_i,t_j;\Sigma_{ij})\cdot\Pr\bigl(Y_l>t_l\ \forall l\notin\{i,j\}\bigm|Y_i=t_i,\ Y_j=t_j\bigr),
\]
where the conditional probability refers to the nondegenerate Gaussian conditional
distribution and equals $1$ if $k=2$. In particular
$0\le\partial P/\partial\Sigma_{ij}\le(2\pi\sqrt{1-\Sigma_{ij}^2})^{-1}$.
\end{lemma}

The identity is classical \cite{Plackett1954}; for a proof see
\cite[Section~2.2]{AdlerTaylor2007} and the survey \cite{LiShao2001}. It gives
Slepian's inequality, and it is used in the lemma on two exceedances in the proof of
the switch lemma (\cref{lem:cubic-exceed}).

For $k=3$, $\{i,j,l\}=\{1,2,3\}$ and equal thresholds $t_i=t_j=t_l=h$, the
conditional distribution in \cref{lem:plackett} is explicit: the distribution of $Y_l$ given
$Y_i=Y_j=h$ is normal with
\begin{equation}\label{eq:plackett-k3}
  \text{mean}\ \ h\,\frac{\Sigma_{il}+\Sigma_{jl}}{1+\Sigma_{ij}}\qquad\text{and variance}\ \
  \frac{\det\Sigma}{1-\Sigma_{ij}^2}>0 .
\end{equation}
Indeed, the Gaussian conditioning formula gives the conditional mean
$(\Sigma_{il},\Sigma_{jl})\bigl(\begin{smallmatrix}1&\Sigma_{ij}\\\Sigma_{ij}&1\end{smallmatrix}\bigr)^{-1}
\bigl(\begin{smallmatrix}h\\h\end{smallmatrix}\bigr)$, and $(1,1)^{\mathsf T}$ is an
eigenvector of the $2\times2$ matrix with eigenvalue $1+\Sigma_{ij}$; the conditional
variance is the Schur complement of that matrix in $\Sigma$, which is
$\det\Sigma/(1-\Sigma_{ij}^2)$ and positive because $\Sigma$ is positive definite.
\Cref{sec:cubic} uses this formula in the preparation of \cref{lem:cubic-exceed}.

\begin{lemma}[Slepian's inequality {\cite{Slepian1962}}]\label{lem:slepian}
Let $Y=(Y_1,\dots,Y_k)$ and $Y'=(Y'_1,\dots,Y'_k)$ be centered Gaussian vectors with
$\Var Y_i=\Var Y'_i>0$ for all $i$ and $\Cov(Y_i,Y_j)\ge\Cov(Y'_i,Y'_j)$ for all
$i\ne j$. Then, for every $t\in\R^k$,
\[
  \Pr(Y_i>t_i\ \forall i)\ge\Pr(Y'_i>t_i\ \forall i)
  \qquad\text{and}\qquad
  \Pr(Y_i\le t_i\ \forall i)\ge\Pr(Y'_i\le t_i\ \forall i).
\]
\end{lemma}

\begin{proof}
For positive definite covariance matrices this is Slepian's inequality~\cite{Slepian1962}
\cite[Section~2.2]{AdlerTaylor2007}: after scaling to unit variances, integrate
\cref{lem:plackett} along the segment from the correlation matrix of $Y'$ to that of $Y$.
For singular covariance matrices, apply it to $\sqrt{1-\eta}\,Y+\sqrt\eta\,\xi$ and
$\sqrt{1-\eta}\,Y'+\sqrt\eta\,\xi'$ with independent standard Gaussian vectors $\xi,\xi'$,
and let $\eta\to0$; the probabilities converge by dominated convergence, since
$\Pr(Y_i=t_i)=0$. The second inequality is the first one for $-Y$, $-Y'$ and $-t$.
\qedhere
\end{proof}

Sheppard's formula evaluates orthant probabilities at threshold zero. We use it for two
and three coordinates.

\begin{lemma}[Sheppard's formula {\cite{Sheppard1899}}]\label{lem:sheppard}
Let $(X_1,X_2,X_3)$ be centered Gaussian with unit variances and put
$\vartheta_{ij}=\arccos\Corr(X_i,X_j)\in[0,\pi]$. Then
\[
  \Pr(\operatorname{sgn}X_1\ne\operatorname{sgn}X_2)=\frac{\vartheta_{12}}{\pi},\qquad
  \Pr(X_1>0,\ X_2>0)=\frac12-\frac{\vartheta_{12}}{2\pi}=\frac14+\frac{\arcsin\Corr(X_1,X_2)}{2\pi},
\]
and
\[
  \Pr(X_1>0,\ X_2>0,\ X_3>0)=\frac12-\frac{\vartheta_{12}+\vartheta_{13}+\vartheta_{23}}{4\pi}.
\]
\end{lemma}

\begin{proof}
Each $X_i$ is nonzero with probability $1$. The first formula is
\cref{lem:crossing}\ref{lem:crossing:a}, whose proof applies to any centered Gaussian
pair with unit variances (write it as a pair of unit-norm sums of independent standard
normals; the case $\Corr(X_1,X_2)=-1$ is trivial). Since $X$ and $-X$
have the same distribution, the two coordinates are both positive and both negative with equal
probabilities, which gives the second; the last expression follows from
$\arcsin x=\frac\pi2-\arccos x$.

For three coordinates, three nonzero signs have either $0$ or $2$ disagreeing pairs.
Hence $\E\#\{\text{disagreeing pairs}\}=2\Pr(\text{not all signs equal})$, so by the
first formula $\Pr(\text{all signs equal})=1-(\vartheta_{12}+\vartheta_{13}+\vartheta_{23})/(2\pi)$,
and symmetry under $X\mapsto-X$ halves this. \qedhere
\end{proof}

\subsection{Local dependence and concentration}\label{sec:prelim-concentration}

The labels of \cref{sec:labels} are functions of independent variables attached to
the vertices, and each label depends only on the variables in a ball of bounded
radius. Counts such as frontier sizes are then sums of locally dependent indicators,
whose variance is linear in $n$. The plain sweep (\cref{sec:plain-sweep}), the cubic
sweep (\cref{sec:cubic}) and the algorithms (\cref{sec:algorithms}) combine the
following bound with Chebyshev's inequality.

\begin{lemma}[local variance]\label{lem:local-variance}
Let $G$ be a graph with $n$ vertices, let $(Z_w)_{w\in V(G)}$ be independent random
variables, and for every vertex $v$ let $A_v\subseteq V(G)$ and let $I_v$ be a random
variable with values in $[0,a]$ that is a measurable function of $(Z_w)_{w\in A_v}$.
Put $W=\sum_vI_v$. Then
\[
  \Var W\le\frac{a^2}4\cdot\#\bigl\{(u,v)\in V(G)^2:\ A_u\cap A_v\ne\emptyset\bigr\}.
\]
If moreover $\Delta(G)\le\Delta$ for some $\Delta\ge2$ and $A_v\subseteq B_\rho(v)$ for
every $v$, then $\Var W\le\frac{a^2}{4}\,n\max_v|B_{2\rho}(v)|\le\frac{a^2}2\,n\Delta^{2\rho}$
and $\Pr(|W-\E W|\ge\lambda)\le a^2n\Delta^{2\rho}/(2\lambda^2)$ for every $\lambda>0$.
\end{lemma}

\begin{proof}
The first bound is proved as \cref{lem:warmup-variance}: $\Cov(I_u,I_v)=0$ if
$A_u\cap A_v=\emptyset$, and $|\Cov(I_u,I_v)|\le a^2/4$ otherwise. If
$A_v\subseteq B_\rho(v)$ for all $v$, then $A_u\cap A_v\ne\emptyset$ implies
$v\in B_{2\rho}(u)$, so there are at most $n\max_u|B_{2\rho}(u)|\le2n\Delta^{2\rho}$
such pairs by \eqref{eq:ball-size}. The last claim is Chebyshev's inequality. \qedhere
\end{proof}
{}
\section{Gaussian labels and edge correlations}\label{sec:labels}

All sweeps of Part~I order the vertices by random Gaussian labels and bound the
sizes of the frontiers of the resulting order from above with the layout lemma (\cref{lem:layout}). This
section constructs the labels and proves the two properties of them that the rest of
Part~I uses.

First, the labels of a vertex and of its neighbors form a star. For every vertex $v$,
the vector $(Y_v;Y_u)_{u\in N(v)}$ is a centered Gaussian vector with unit variances
(\cref{lem:locality}), and its incident correlations $\Corr(Y_v,Y_u)$ are bounded
below by explicit constants that depend only on the degrees
(\cref{lem:correlation-padded,lem:correlation}). When these constants are positive
and $v$ is not isolated, the vector is therefore a $d(v)$-star in the sense of
\cref{def:star}. The star bounds of \cref{sec:stars} use nothing else about the
labels. They allow arbitrary correlations between two neighbors of $v$, so short
cycles and triangles through $v$ need no separate treatment.

Second, the labels are local and distinct. Each label depends only on the
independent variables in a ball of bounded radius, which makes frontier sizes
concentrate (\cref{lem:local-variance}). The labels are distinct with probability $1$, as
\cref{lem:layout} requires. Both facts are part of \cref{lem:locality}.

\Cref{sec:labels-def} defines the labels and states the three lemmas, and
\cref{sec:labels-locality,sec:labels-ratio,sec:labels-padded,sec:labels-mixed} prove
them. The technical overview (\cref{sec:overview-labels}) explains why the chosen
decay rate is the right one.

\begin{remark}[optimality for radial labels]\label{rem:labels-optimal}
On the infinite $p$-regular tree, the local limit of $p$-regular graphs of large
girth, the adjacency operator has norm $2\sqrt{p-1}$ (Kesten~\cite{Kesten1959};
\cite[Lemma~1.24]{Woess2000}). Hence no Gaussian field $X_v=\sum_wf(\dist(v,w))Z_w$
with radial weights has a neighbor correlation above $\frac{2\sqrt{p-1}}p=r_{pp}$, since
the $p$ neighbor covariances of $v$ sum to $\langle c_v,Ac_v\rangle$, where $c_v$ is the
coefficient vector of $X_v$. The value $r_{pp}$ is attained in the limit by the
Gaussian wave~\cite{CsokaGHV2015,HarangiVirag2015,BackhauszSzegedyVirag2015}, and the
padded labels are its finite-radius approximations: their coefficients
$(p-1)^{-k/2}$ on spheres are the test functions of Nilli's proof of the
Alon--Boppana bound~\cite{Nilli1991}. So \cref{lem:correlation-padded} is
asymptotically sharp for radial labels. For mixed degrees the limiting edge
correlation can lie slightly above $r_{ij}$. On the $(3,4)$-biregular tree, every
generation of the two sides of an edge $uv$ with $d(u)=3$ and $d(v)=4$ carries the same
squared mass, and a ball contains $N_R+O(1)$ generations of each side, with
$N_R=4R/\ln6$; summing over generations gives $\sigma_u^2=\frac32N_R+O(1)$,
$\sigma_v^2=\frac43N_R+O(1)$ and $\Cov(X_u,X_v)=(2^{-1/2}+3^{-1/2})N_R+O(1)$, so the
correlation tends to $(2^{-1/2}+3^{-1/2})/\sqrt2=\frac12+\frac1{\sqrt6}\approx0.9082$,
above $r_{34}\approx0.9074$.
\end{remark}

\subsection{The labels}\label{sec:labels-def}

A label is a normalized weighted sum of independent standard normal variables, one
for each vertex, with weights that decay exponentially in a weighted distance. An
\emph{arc} is an ordered pair $(x,y)$ with $xy\in E(G)$, written $x\to y$.

\begin{definition}[Gaussian labels]\label{def:labels}
Let $G$ be a graph, let $(Z_w)_{w\in V(G)}$ be independent $N(0,1)$ variables, let
every arc $x\to y$ have a length $\ell(x\to y)>0$, and let $\hat R>0$. The
\emph{directed distance} $D(v,w)$ is the minimum of $\sum_{i<k}\ell(x_i\to x_{i+1})$
over all walks $v=x_0,x_1,\dots,x_k=w$, with $D(v,v)=0$ and $D(v,w)=\infty$ if there
is no such walk. Put $B(v)=\{w:D(v,w)\le\hat R\}$ and
\[
  X_v=\sum_{w\in B(v)}e^{-D(v,w)}Z_w,\qquad
  \sigma_v^2=\sum_{w\in B(v)}e^{-2D(v,w)},\qquad
  Y_v=X_v/\sigma_v .
\]
\begin{enumerate}[label=\textup{(\alph*)}]
  \item\label{def:labels:padded} \emph{Padded labels of degree $p$}, for integers
  $p\ge3$ and $R\ge1$ and graphs with $\Delta(G)\le p$: every arc has length
  $\frac12\ln(p-1)$, and $\hat R=\frac R2\ln(p-1)$.
  \item\label{def:labels:mixed} \emph{Mixed labels}, for a real $R>0$ and graphs with
  $\delta(G)\ge3$: $\ell(x\to y)=\frac12\ln\bigl(d(x)-1\bigr)$, and $\hat R=R$.
\end{enumerate}
\end{definition}

By independence, $\sigma_v^2=\Var X_v$. For padded labels,
$D(v,w)=\frac12\ln(p-1)\dist(v,w)$ and $B(v)=B_R(v)$, so
\[
  X_v=\sum_{w\in B_R(v)}(p-1)^{-\dist(v,w)/2}Z_w ;
\]
for $p=3$ we call them the \emph{geometric labels}, with coefficients
$2^{-\dist(v,w)/2}$. Padded labels are used with $p=3$ for graphs of maximum
degree~$3$ (\cref{thm:cubic}) and with $p=4$ for \cref{thm:deg4}\ref{thm:deg4:a}. In
mixed labels the length of an arc depends on its tail only; they are used for graphs
with $\delta(G)\ge3$ and $\Delta(G)\le5$
(\cref{thm:deg4}\ref{thm:deg4:b},\ref{thm:deg4:c} and \cref{thm:deg5}).

A walk of minimum length is a path, because all lengths are positive; since $G$ is
finite, the minimum in the definition of $D$ is attained. The directed distance
satisfies the triangle inequality $D(u,w)\le D(u,x)+D(x,w)$, and in particular
\begin{equation}\label{eq:D-edge}
  D(u,w)\le\ell(u\to v)+D(v,w)\qquad\text{for every edge }uv\text{ and every }w .
\end{equation}
In the mixed case $D$ is not symmetric. Its arc lengths lie between
$\ell_{\min}=\frac12\ln2$ and $\frac12\ln(\Delta(G)-1)$. The decay rate is chosen so
that in both regimes the arcs leaving a vertex, other than one, carry total squared
weight at most one:
\begin{equation}\label{eq:critical}
  \sum_{y\in N(x)\setminus\{y_0\}}e^{-2\ell(x\to y)}\le1
  \qquad\text{for every vertex }x\text{ and every }y_0\in N(x).
\end{equation}
Indeed the sum has $d(x)-1$ terms, each equal to $1/(p-1)$ in case
\ref{def:labels:padded}, where $d(x)\le p$, and to $1/(d(x)-1)$ in case
\ref{def:labels:mixed}. In the mixed case, and in the padded case at vertices of
degree $p$, \eqref{eq:critical} holds with equality. We call this \emph{critical
decay}; \cref{sec:overview-labels} explains its role.

\begin{remark}[path products]\label{rem:P-form}
The algorithms of \cref{sec:algorithms} describe the same labels by vertex weights
instead of arc lengths. Put $q(x)=p-1$ in case \ref{def:labels:padded} and
$q(x)=d(x)-1$ in case \ref{def:labels:mixed}, and let $P(v,w)$ be the minimum of
$\prod_{i<k}q(x_i)$ over all $v$--$w$ paths $v=x_0,\dots,x_k=w$, with $P(v,v)=1$.
Since every arc $x\to y$ has length $\frac12\ln q(x)$, the length of a walk is
$\frac12\ln$ of its product, so
\[
  e^{-D(v,w)}=P(v,w)^{-1/2}\qquad\text{and}\qquad X_v=\sum_{w\in B(v)}P(v,w)^{-1/2}Z_w,
\]
with $B(v)=\{w:P(v,w)\le e^{2\hat R}\}$: the hop ball $B_R(v)$ in case
\ref{def:labels:padded}, and $\{w:P(v,w)\le2^h\}$ in case \ref{def:labels:mixed}
when $R=\frac h2\ln2$. The two descriptions define identical labels.
\end{remark}

The correlation bounds are expressed through the following constants. For integers
$i,j\ge3$ put
\begin{equation}\label{eq:rij}
  r_{ij}=\sech\Bigl(\tfrac14\ln\bigl((i-1)(j-1)\bigr)\Bigr)
  =\frac{2\,g_{ij}}{1+g_{ij}^{2}},\qquad g_{ij}=\bigl((i-1)(j-1)\bigr)^{1/4}.
\end{equation}
They are symmetric and decrease in each index. On the diagonal,
$r_{pp}=\sech(\frac12\ln(p-1))=2\sqrt{p-1}/p$; so $r_{33}=r_*=\frac{2\sqrt2}3$,
$r_{44}=\frac{\sqrt3}2$ and $r_{55}=\frac45$. \Cref{tab:rij} lists the values for
$3\le i\le j\le5$.

\begin{table}[tb]
\centering
\caption{The edge correlation bounds $r_{ij}$ of \eqref{eq:rij} for end degrees
$3\le i\le j\le5$.}\label{tab:rij}
\begin{tabular}{@{}lccc@{}}
\toprule
$(i,j)$ & $(i-1)(j-1)$ & exact value & decimal \\
\midrule
$(3,3)$ & $4$  & $\frac{2\sqrt2}{3}$ & $0.942809$ \\
$(3,4)$ & $6$  & $\frac{2\cdot6^{1/4}}{1+\sqrt6}$ & $0.907430$ \\
$(3,5)$ & $8$  & $\frac{2\cdot8^{1/4}}{1+2\sqrt2}$ & $0.878582$ \\
$(4,4)$ & $9$  & $\frac{\sqrt3}{2}$ & $0.866025$ \\
$(4,5)$ & $12$ & $\frac{2\cdot12^{1/4}}{1+2\sqrt3}$ & $0.833856$ \\
$(5,5)$ & $16$ & $\frac45$ & $0.800000$ \\
\bottomrule
\end{tabular}
\end{table}

The main statements of this section are the following two lemmas. The first covers
the padded labels, and in particular the geometric labels of the cubic case.

\begin{lemma}[edge correlations, padded labels]\label{lem:correlation-padded}
Let $p\ge3$ and $R\ge1$ be integers, let $G$ be a graph with $\Delta(G)\le p$, and let
$(Y_v)$ be the padded labels of degree $p$
\textup{(}\cref{def:labels}\ref{def:labels:padded}\textup{)}. Then every edge $uv$
satisfies $\Corr(Y_u,Y_v)\ge r_{pp}(1-\frac1R)$.
\end{lemma}

The second covers the mixed labels, where the bound depends on the degrees of the two
ends of the edge. For $\Delta_0\ge3$ put $\ell_{\max}=\frac12\ln(\Delta_0-1)$, so that
all arc lengths of a graph with $\delta(G)\ge3$ and $\Delta(G)\le\Delta_0$ lie in
$[\ell_{\min},\ell_{\max}]$, where $\ell_{\min}=\frac12\ln2$ as above. For
$R>2\ell_{\max}$ put
\begin{equation}\label{eq:eps-R}
  \eps_R=\frac{2\ell_{\max}^2}{\ell_{\min}\,(R-2\ell_{\max})}.
\end{equation}

\begin{lemma}[edge correlations, mixed labels]\label{lem:correlation}
Let $\Delta_0\ge3$ and $R>2\ell_{\max}$, let $G$ be a graph with $\delta(G)\ge3$ and
$\Delta(G)\le\Delta_0$, and let $(Y_v)$ be the mixed labels
\textup{(}\cref{def:labels}\ref{def:labels:mixed}\textup{)}. Then every edge $uv$
satisfies $\Corr(Y_u,Y_v)\ge r_{d(u)d(v)}(1-\eps_R)$.
\end{lemma}

For $\Delta_0=5$ we have $\ell_{\max}=\ln2$ and $2\ell_{\max}^2/\ell_{\min}=4\ln2$, so
$\eps_R=\frac{4\ln2}{R-2\ln2}$; with $R=\frac h2\ln2$ for an integer $h>4$, also
$R-2\ell_{\max}=\frac{h-4}2\ln2$ and $\eps_R=\frac8{h-4}$. Since $\eps_R\to0$ as
$R\to\infty$, every edge correlation is asymptotically at least the value $r_{ij}$
determined by the degrees of its ends.

\begin{lemma}[distinctness and locality]\label{lem:locality}
Let $(Y_v)$ be labels as in \cref{def:labels}, in either regime. Put $\rho=R$ for
padded labels and $\rho=\lfloor2R/\ln2\rfloor$ for mixed labels. Then:
\begin{enumerate}[label=\textup{(\roman*)}]
  \item\label{lem:locality:gauss} $(Y_v)_{v\in V(G)}$ is a centered Gaussian vector,
  $\Var Y_v=1$ and $\sigma_v\ge1$ for every $v$;
  \item\label{lem:locality:distinct} the labels are pairwise distinct with probability $1$;
  \item\label{lem:locality:local} $B(v)\subseteq B_\rho(v)$, so $Y_v$ is a function of
  $(Z_w)_{w\in B_\rho(v)}$;
  \item\label{lem:locality:frontier} for every $t\in\R$, the events
  $\{v\in\outfr S_t\}$ and $\{v\in\infr S_t\}$ for the labeling $Y$ are determined by
  $(Z_w)_{w\in B_{\rho+1}(v)}$.
\end{enumerate}
\end{lemma}

So every count of the form $\sum_v\ind\{v\in\outfr S_t\}$ satisfies the hypothesis of
\cref{lem:local-variance} with $A_v=B_{\rho+1}(v)$. The radius $\rho$ is fixed for
fixed $R$; the sweeps let $R$ grow slowly with $n$, so that $\eps_R\to0$ while the
dependence radius stays small enough for concentration.

\subsection{Distinctness and locality}\label{sec:labels-locality}

\begin{proof}[Proof of \cref{lem:locality}]
\ref{lem:locality:gauss} Each $X_v$ is a finite linear combination of the independent
standard normal variables $Z_w$, so $(X_v)$ and $(Y_v)$ are centered Gaussian vectors.
Since $v\in B(v)$ with coefficient $e^{-D(v,v)}=1$, $\sigma_v^2\ge1$, and
$\Var Y_v=1$ by construction.

\ref{lem:locality:distinct} Let $u\ne v$. The coefficient of $Z_u$ in $Y_u-Y_v$ is
$1/\sigma_u-\ind\{u\in B(v)\}e^{-D(v,u)}/\sigma_v$, and the coefficient of $Z_v$ is
$\ind\{v\in B(u)\}e^{-D(u,v)}/\sigma_u-1/\sigma_v$. If both vanish, then $u\in B(v)$
and $\sigma_v=\sigma_ue^{-D(v,u)}$, and $v\in B(u)$ and $\sigma_v=\sigma_ue^{D(u,v)}$.
Hence $D(u,v)+D(v,u)=0$, which is impossible because $u\ne v$ and all arc lengths are
positive. So $Y_u-Y_v$ is a normal variable with positive variance, and
$\Pr(Y_u=Y_v)=0$. A union bound over the finitely many pairs gives the claim.

\ref{lem:locality:local} For padded labels $B(v)=B_R(v)$. For mixed labels, every arc
has length at least $\ell_{\min}=\frac12\ln2$ because $\delta(G)\ge3$, so a walk of
length at most $R$ has at most $R/\ell_{\min}$ arcs, and $D(v,w)\le R$ implies
$\dist(v,w)\le\lfloor R/\ell_{\min}\rfloor=\lfloor2R/\ln2\rfloor$.

\ref{lem:locality:frontier} By definition, $v\in\outfr S_t$ if and only if $Y_v>t$ and
$Y_u\le t$ for some $u\in N(v)$, and $v\in\infr S_t$ if and only if $Y_v\le t$ and
$Y_u>t$ for some $u\in N(v)$. Both are determined by $(Y_x)_{x\in N[v]}$, and by
\ref{lem:locality:local} each such $Y_x$ is a function of $Z$ restricted to
$B_\rho(x)\subseteq B_{\rho+1}(v)$. \qedhere
\end{proof}

\subsection{A ratio inequality}\label{sec:labels-ratio}

Both correlation lemmas reduce to the following inequality for the shared part of two
labels: the weights $a_w$ and $b_w$ that the two labels give to the same variable $Z_w$.

\begin{lemma}\label{lem:sech-ratio}
Let $\alpha,\beta>0$, let $I$ be a finite set, and let
$a_w,b_w>0$ \textup{(}$w\in I$\textup{)} satisfy $b_w/\alpha\le a_w\le\beta b_w$. Then,
with $A=\sum_{w\in I}a_w^2$ and $B=\sum_{w\in I}b_w^2$,
\[
  \sum_{w\in I}a_wb_w\ \ge\ \sech\Bigl(\tfrac12\ln(\alpha\beta)\Bigr)\sqrt{AB}.
\]
\end{lemma}

\begin{proof}
Since $1/\alpha\le a_w/b_w\le\beta$, the inequality of Diaz and
Metcalf~\cite[Theorem~1]{DiazMetcalf1963} gives the first inequality below (it is
the sum of the inequalities $(a_w-b_w/\alpha)(\beta b_w-a_w)\ge0$), and the second is
the inequality of arithmetic and geometric means:
\[
  \Bigl(\beta+\frac1\alpha\Bigr)\sum_{w\in I}a_wb_w\ \ge\ A+\frac\beta\alpha B\ \ge\ 2\sqrt{\frac\beta\alpha}\,\sqrt{AB}.
\]
Finally
$2\sqrt{\beta/\alpha}\big/(\beta+1/\alpha)=2\sqrt{\alpha\beta}/(\alpha\beta+1)
=2/\bigl(\sqrt{\alpha\beta}+1/\sqrt{\alpha\beta}\bigr)=\sech(\frac12\ln(\alpha\beta))$. \qedhere
\end{proof}

For an edge $uv$ we apply the lemma to $a_w=e^{-D(u,w)}$ and $b_w=e^{-D(v,w)}$ on the
common support $I=B(u)\cap B(v)$. By \eqref{eq:D-edge} and its counterpart with $u$ and
$v$ exchanged, $a_w\ge e^{-\ell(u\to v)}b_w$ and $b_w\ge e^{-\ell(v\to u)}a_w$; so the
hypothesis holds with $\alpha=e^{\ell(u\to v)}$ and $\beta=e^{\ell(v\to u)}$, and
$\frac12\ln(\alpha\beta)=\frac12L_{uv}$ with $L_{uv}=\ell(u\to v)+\ell(v\to u)$.
Because coefficients outside $I$ are paired with zero coefficients,
\begin{equation}\label{eq:cov-common}
  \Cov(X_u,X_v)=\sum_{w\in I}a_wb_w\ \ge\ \sech\bigl(\tfrac12L_{uv}\bigr)\sqrt{AB},
  \qquad A=\sum_{w\in I}a_w^2,\quad B=\sum_{w\in I}b_w^2 .
\end{equation}
For padded labels $L_{uv}=\ln(p-1)$ and $\sech(\frac12L_{uv})=r_{pp}$. For mixed labels
$L_{uv}=\frac12\ln((i-1)(j-1))$ with $i=d(u)$, $j=d(v)$, and
$\sech(\frac12L_{uv})=r_{ij}$. It remains to show that $A$ and $B$ are almost all of
$\sigma_u^2$ and $\sigma_v^2$: the outer shells $B(u)\setminus I$ and $B(v)\setminus I$
carry little squared mass.

\subsection{Padded labels: proof of \texorpdfstring{\cref{lem:correlation-padded}}{the padded correlation lemma}}\label{sec:labels-padded}

For $p=3$ these are the geometric labels, and the lemma is \cref{lem:warmup-edges};
the proof of that lemma applies to every $p$ with the base $2$ replaced by $p-1$.

\begin{proof}[Proof of \cref{lem:correlation-padded}]
Write $q=p-1\ge2$ and let $uv$ be an edge. As in the proof of \cref{lem:warmup-edges},
a vertex at distance $j\ge1$ from $u$ has at most $q$ neighbors at distance $j+1$, so
the sphere masses $s_jq^{-j}$ are nonincreasing in $j\ge1$, and the outermost one
carries at most a fraction $\frac1R$ of $\sigma_u^2=1+\sum_{j=1}^Rs_jq^{-j}$. Every
vertex of $B_R(u)\setminus B_R(v)$ lies on the sphere of radius $R$ around $u$. Hence
$A\ge(1-\frac1R)\sigma_u^2$ in \eqref{eq:cov-common}, and likewise
$B\ge(1-\frac1R)\sigma_v^2$. Since $\sech(\frac12\ln q)=2\sqrt q/(q+1)=r_{pp}$,
\eqref{eq:cov-common} gives $\Corr(Y_u,Y_v)\ge r_{pp}(1-\frac1R)$. \qedhere
\end{proof}

No step uses regularity, girth, connectivity or a lower bound on the degrees: vertices
of degree less than $p$ only make the sphere bound $s_{j+1}\le qs_j$ slack, and
vertices in different components have disjoint supports.

\subsection{Mixed labels: proof of \texorpdfstring{\cref{lem:correlation}}{the mixed correlation lemma}}\label{sec:labels-mixed}

For mixed labels the balls $B(v)$ are defined by the directed distance, whose spheres
are not layers of the graph. The sphere argument is therefore replaced by a cut
argument in a shortest-path tree rooted at one end $u$ of the edge $uv$. Step~1
locates the outer shell in a layer of width $\ell_{\max}$ below the boundary of the
ball. Steps~2 and~3 show, using the critical decay \eqref{eq:critical}, that the
squared mass of the cut at directed distance $t$ from $u$ is nonincreasing in $t$.
Step~4 integrates the cut mass over $t$ and compares the shell with the whole ball.

\begin{proof}[Proof of \cref{lem:correlation}]
Let $uv$ be an edge with $d(u)=i$ and $d(v)=j$. All arc lengths lie in
$[\ell_{\min},\ell_{\max}]$, because $3\le d(x)\le\Delta_0$ for every $x$. By
\eqref{eq:cov-common}, it suffices to show
\begin{equation}\label{eq:mixed-shell-goal}
  \sum_{w\in B(u)\setminus B(v)}e^{-2D(u,w)}\le\eps_R\,\sigma_u^2
\end{equation}
and the same statement with $u$ and $v$ exchanged. Indeed, then
$A\ge(1-\eps_R)\sigma_u^2$ and $B\ge(1-\eps_R)\sigma_v^2$, and \eqref{eq:cov-common}
gives $\Corr(Y_u,Y_v)\ge r_{ij}(1-\eps_R)$ if $\eps_R\le1$; if $\eps_R>1$ the claim
holds because $\Cov(X_u,X_v)\ge0$, all coefficients being nonnegative. We prove
\eqref{eq:mixed-shell-goal}; the other statement is symmetric, using
$\ell(u\to v)\le\ell_{\max}$. Throughout, write $D(w)=D(u,w)$.

\emph{Step 1: the shell is thin.} If $w\in B(u)\setminus B(v)$, then $D(v,w)>R$, and
$D(v,w)\le\ell(v\to u)+D(w)$ by \eqref{eq:D-edge}; hence
\begin{equation}\label{eq:shell-location}
  R-\ell_{\max}<D(w)\le R .
\end{equation}

\emph{Step 2: a shortest-path tree.} Let $V_u$ be the set of vertices in the component
of $u$. For $w\in V_u\setminus\{u\}$ choose a shortest $u$--$w$ path and let $p(w)$ be
the vertex before $w$ on it. Its initial segment is a walk from $u$ to $p(w)$, so
$D(p(w))\le D(w)-\ell(p(w)\to w)$, and the triangle inequality gives the reverse; hence
\[
  D(w)=D(p(w))+\ell\bigl(p(w)\to w\bigr),\qquad D(p(w))<D(w).
\]
Following $p$ from any $w$ strictly decreases $D$ and therefore reaches $u$. So the
pairs $(p(w),w)$ form a tree $\mathcal T$ on $V_u$ rooted at $u$, in which $p(w)$ is
the parent of $w$. Write $\operatorname{ch}(x)=\{w:p(w)=x\}$ for the children of $x$. For
$x\ne u$, the children of $x$ are neighbors of $x$ with larger $D$, so
$\operatorname{ch}(x)\subseteq N(x)\setminus\{p(x)\}$, and by \eqref{eq:critical}
\begin{equation}\label{eq:children}
  \sum_{c\in\operatorname{ch}(x)}e^{-2D(c)}=e^{-2D(x)}\sum_{c\in\operatorname{ch}(x)}e^{-2\ell(x\to c)}\le e^{-2D(x)}\qquad(x\ne u).
\end{equation}
Call a set $\mathcal A$ of vertices an \emph{antichain} if no element of $\mathcal A$
is a proper descendant of another in $\mathcal T$. We claim: for every $x\ne u$ and
every antichain $\mathcal A$ of descendants of $x$ (including $x$ itself),
\begin{equation}\label{eq:antichain}
  \sum_{w\in\mathcal A}e^{-2D(w)}\le e^{-2D(x)}.
\end{equation}
We induct on the height of the subtree of $x$. If $x\in\mathcal A$, then
$\mathcal A=\{x\}$. Otherwise $\mathcal A$ is the disjoint union, over
$c\in\operatorname{ch}(x)$, of antichains of descendants of $c$, and by induction and
\eqref{eq:children} its mass is at most $\sum_{c}e^{-2D(c)}\le e^{-2D(x)}$.

\emph{Step 3: moving cuts.} For $t>0$ let
\[
  \mathcal A_t=\bigl\{w\in V_u\setminus\{u\}:\ D(p(w))<t\le D(w)\bigr\},\qquad
  m(t)=\sum_{w\in\mathcal A_t}e^{-2D(w)} .
\]
We show that $m$ is nonincreasing on $(0,\infty)$. Let $0<t<t'$ and $w\in\mathcal A_{t'}$.
Along the path $u=x_0,x_1,\dots,x_k=w$ in $\mathcal T$, $D$ increases strictly from
$0<t$ to $D(w)\ge t'>t$, so there is exactly one index $s\ge1$ with
$D(x_{s-1})<t\le D(x_s)$; that is, $w$ has exactly one ancestor-or-self in
$\mathcal A_t$. For $x\in\mathcal A_t$, the set of $w\in\mathcal A_{t'}$ whose
ancestor-or-self in $\mathcal A_t$ is $x$ is an antichain of descendants of $x$:
if $w'\in\mathcal A_{t'}$ were a proper descendant of $w\in\mathcal A_{t'}$, then
$D(p(w'))\ge D(w)\ge t'$, contradicting $w'\in\mathcal A_{t'}$. By \eqref{eq:antichain},
\[
  m(t')=\sum_{x\in\mathcal A_t}\ \sum_{\substack{w\in\mathcal A_{t'}\\ w\text{ below }x}}e^{-2D(w)}\le\sum_{x\in\mathcal A_t}e^{-2D(x)}=m(t).
\]

\emph{Step 4: integrating the cut.} Each $w\in V_u\setminus\{u\}$ belongs to
$\mathcal A_t$ exactly for $t$ in the interval $J_w=(D(p(w)),D(w)]$, whose length
$\ell(p(w)\to w)$ lies in $[\ell_{\min},\ell_{\max}]$. Since the sums are finite, for
$0\le a<b$
\begin{equation}\label{eq:cut-integral}
  \int_a^bm(t)\,dt=\sum_{w\in V_u\setminus\{u\}}e^{-2D(w)}\,\bigl|J_w\cap(a,b]\bigr| .
\end{equation}
Put $T=R-2\ell_{\max}>0$.
\begin{itemize}
  \item \emph{Upper bound on $m(T)$.} A vertex $w$ contributes to
  \eqref{eq:cut-integral} with $(a,b]=(0,T]$ only if $D(p(w))<T$, and then
  $D(w)<T+\ell_{\max}\le R$, so $w\in B(u)\setminus\{u\}$; its contribution is at most
  $\ell_{\max}e^{-2D(w)}$. Since $m$ is nonincreasing,
  \[
    T\,m(T)\le\int_0^Tm(t)\,dt\le\ell_{\max}\sum_{w\in B(u)\setminus\{u\}}e^{-2D(w)}\le\ell_{\max}\,\sigma_u^2 .
  \]
  \item \emph{The shell as an integral.} Every $w$ in the shell
  $B(u)\setminus B(v)$ satisfies \eqref{eq:shell-location}, so $w\ne u$ and
  $D(p(w))\ge D(w)-\ell_{\max}>R-2\ell_{\max}=T$. Hence $J_w\subseteq(T,R]$, and $w$
  contributes $e^{-2D(w)}|J_w|\ge\ell_{\min}e^{-2D(w)}$ to \eqref{eq:cut-integral} with
  $(a,b]=(T,R]$. All other contributions are nonnegative, and $m(t)\le m(T)$ for
  $t>T$, so
  \[
    \ell_{\min}\sum_{w\in B(u)\setminus B(v)}e^{-2D(w)}\le\int_T^Rm(t)\,dt\le2\ell_{\max}\,m(T).
  \]
\end{itemize}
Combining the two bounds,
\[
  \sum_{w\in B(u)\setminus B(v)}e^{-2D(w)}\le\frac{2\ell_{\max}}{\ell_{\min}}\cdot\frac{\ell_{\max}\sigma_u^2}{R-2\ell_{\max}}=\eps_R\,\sigma_u^2
\]
by the definition \eqref{eq:eps-R} of $\eps_R$, which is \eqref{eq:mixed-shell-goal}.
\qedhere
\end{proof}

The proof uses the degrees only through the critical decay \eqref{eq:critical} and the
bounds $\ell_{\min}\le\ell\le\ell_{\max}$ on the arc lengths. It therefore applies to
any lengths with these two properties.
{}
\section{Star bounds}\label{sec:stars}

A vertex $v$ lies in the outer frontier $\outfr S_\tau$ of the sweep exactly when
$Y_v>\tau$ and some neighbor $u$ of $v$ has $Y_u\le\tau$. This section bounds the
probability of this event from above in terms of lower bounds on the correlations between $Y_v$
and the labels $Y_u$ of the neighbors alone, which is all that the labels of
\cref{sec:labels} provide. These bounds for a single vertex are the input of the
plain sweep (\cref{sec:plain-sweep}), of the cubic sweep (\cref{sec:cubic}) and of the
certificate (\cref{sec:certificate}). The technical overview explains the route
(\cref{sec:overview-stars}).

\Cref{sec:stars-center} defines stars and bounds the frontier probability by the
center bound (\cref{lem:center}), an integral over the label of the center whose
integrand involves the infimum $P_{\min}$ of an orthant probability over all
correlation matrices. \Cref{sec:stars-simplex} shows that for equal thresholds and
three to five coordinates this infimum is attained by the regular simplex vector (the
simplex theorem, \cref{lem:simplex}), and \cref{sec:stars-ideal} defines the ideal
stars and encloses $F_3$ and $F_4$ (\cref{def:ideal-star,prop:F3F4}). The bounds on
$P_{\min}$ for unequal thresholds (\cref{lem:truncation,lem:arcs}), the simplex
recursion (\cref{lem:simplex-recursion}) and the proof of \cref{prop:F3F4} are in
\cref{app:star-tools,app:F3F4}.

The notation and conventions of \cref{sec:prelim-gauss} are in force; in particular
$s_r=\sqrt{1-r^2}$. All coordinates below are standard normal, so strict and weak
inequalities between a coordinate and a constant may be exchanged.

\subsection{Stars and the reduction to the center}\label{sec:stars-center}

\begin{definition}[stars]\label{def:star}
Let $d\ge1$ and $r=(r_1,\dots,r_d)\in(0,1)^d$. A \emph{$d$-star with lower bounds
$r$} is a centered Gaussian vector $(Y_0,Y_1,\dots,Y_d)$ with $\Var Y_k=1$ for
$0\le k\le d$ and $\Corr(Y_0,Y_i)\ge r_i$ for $1\le i\le d$. The coordinate $Y_0$
is the \emph{center}. For a threshold $\tau\in\R$, the \emph{frontier event} of the
star is $\{Y_0>\tau,\ \min_{1\le i\le d}Y_i\le\tau\}$.
\end{definition}

No assumption is made on the correlations $\Corr(Y_i,Y_j)$ between two
non-central coordinates; they may be negative, and the covariance matrix may be
singular. In the sweep, $Y_0$ is the label of a vertex and $Y_1,\dots,Y_d$ are the
labels of its neighbors; the frontier event is the event that the vertex lies in
$\outfr S_\tau$.

Since $\{Y_0>\tau\}$ is the disjoint union of the frontier event and the event
$\{Y_k>\tau\ \text{for}\ 0\le k\le d\}$,
\begin{equation}\label{eq:stars-complement}
  \Pr\bigl(Y_0>\tau,\ \min_iY_i\le\tau\bigr)
  =\Phi(-\tau)-\Pr\bigl(Y_k>\tau\ \text{for}\ 0\le k\le d\bigr).
\end{equation}
So an upper bound on the frontier probability is the same as a lower bound on the
probability that the whole star lies above $\tau$. We now fix notation for the
quantities that the lower bound will involve. For lower bounds $r\in(0,1)^d$, a
threshold $\tau$ and $z\in\R$, put
\[
  s_i=\sqrt{1-r_i^2},\qquad a_i(z)=a_i(z;\tau,r)=\frac{\tau-r_iz}{s_i}\qquad(1\le i\le d).
\]
Each $a_i(z)$ is strictly decreasing in $z$, because $r_i>0$. For
$a\in\R^d$ let
\[
  P_{\min}(a)=P^{(d)}_{\min}(a):=\inf\Bigl\{\Pr\bigl(U_i>a_i\ \text{for}\ 1\le i\le d\bigr)\Bigr\},
\]
where the infimum is over all centered Gaussian vectors $U=(U_1,\dots,U_d)$ with
unit variances (equivalently, over all $d\times d$ correlation matrices, singular
ones included). Finally, the \emph{center bound} is
\begin{equation}\label{eq:center-bound}
  C_d(\tau;r):=\Phi(-\tau)-\int_\tau^\infty\varphi(z)\,P_{\min}\bigl(a(z;\tau,r)\bigr)\,dz .
\end{equation}
By \cref{lem:stars-continuity}(a) below, $P_{\min}$ is continuous, so the integrand
is continuous in $z$ and the integral is well defined.

\begin{lemma}[reduction to the center]\label{lem:center}
Let $(Y_0,\dots,Y_d)$ be a $d$-star with lower bounds $r\in(0,1)^d$, and let
$\tau\in\R$. There is a centered Gaussian vector $U=(U_1,\dots,U_d)$ with unit
variances such that
\begin{equation}\label{eq:center}
  \Pr\bigl(Y_k>\tau\ \text{for}\ 0\le k\le d\bigr)\ \ge\
  \int_\tau^\infty\varphi(z)\,\Pr\bigl(U_i>a_i(z;\tau,r)\ \text{for}\ 1\le i\le d\bigr)\,dz .
\end{equation}
Consequently
\[
  \Pr\bigl(Y_0>\tau,\ \min_{1\le i\le d}Y_i\le\tau\bigr)\le C_d(\tau;r).
\]
\end{lemma}

The proof has three steps. A coupling replaces each neighbor label by one whose
correlation with the center equals its lower bound; this lowers the correlations
between neighbors as well. By Slepian's inequality, the probability that the whole
star lies above $\tau$ can then only decrease. Conditioning on the center finally
gives~\eqref{eq:center}. A separate lower bound for each neighbor is needed for
graphs with several degrees, where the edges at one vertex have different
correlation bounds.

\begin{proof}
Put $Z=Y_0$ and $\rho_i=\Corr(Y_0,Y_i)$, so $0<r_i\le\rho_i\le1$.

\emph{Coupling.} Take independent standard normal variables $\xi_1,\dots,\xi_d$,
independent of $Y$, and put
\[
  Y'_0=Z,\qquad
  Y'_i=Y_i-(\rho_i-r_i)\,Z+\sqrt{\rho_i^2-r_i^2}\;\xi_i\qquad(1\le i\le d).
\]
The square root is real because $\rho_i\ge r_i>0$. The vector $Y'$ is centered
Gaussian. A direct computation gives
\begin{align*}
  \Var Y'_i&=1-2(\rho_i-r_i)\rho_i+(\rho_i-r_i)^2+(\rho_i^2-r_i^2)=1,\\
  \Cov(Y'_0,Y'_i)&=\rho_i-(\rho_i-r_i)=r_i\le\rho_i=\Cov(Y_0,Y_i),\\
  \Cov(Y'_i,Y'_j)&=\Cov(Y_i,Y_j)-\rho_i\rho_j+r_ir_j\le\Cov(Y_i,Y_j)\qquad(i\ne j),
\end{align*}
the last inequality because $0<r_i\le\rho_i$ and $0<r_j\le\rho_j$. The correlations
$\Cov(Y_i,Y_j)$ between neighbors, which may be negative, enter only through the
common term.

\emph{Comparison.} The vectors $Y$ and $Y'$ have the same variances, and every
off-diagonal covariance of $Y$ is at least the corresponding one of $Y'$. By
Slepian's inequality (\cref{lem:slepian}; singular covariance matrices are
allowed),
\[
  \Pr(Y_k>\tau\ \forall k)\ \ge\ \Pr(Y'_k>\tau\ \forall k).
\]

\emph{Conditioning on the center.} Put $U_i=(Y'_i-r_iZ)/s_i$. Since
$Y'_i-r_iZ=Y_i-\rho_iZ+\sqrt{\rho_i^2-r_i^2}\,\xi_i$ has variance
$(1-\rho_i^2)+(\rho_i^2-r_i^2)=s_i^2$ and is uncorrelated with $Z$, the vector
$U=(U_1,\dots,U_d)$ is centered Gaussian with unit variances, $(Z,U)$ is jointly
Gaussian, and $U$ is independent of $Z$. As $Y'_i=r_iZ+s_iU_i$, Fubini's theorem gives
\[
  \Pr(Y'_k>\tau\ \forall k)=\int_\tau^\infty\varphi(z)\,
  \Pr\bigl(r_iz+s_iU_i>\tau\ \forall i\bigr)\,dz
  =\int_\tau^\infty\varphi(z)\,\Pr\bigl(U_i>a_i(z;\tau,r)\ \forall i\bigr)\,dz .
\]
This proves~\eqref{eq:center}. The integrand is at least
$\varphi(z)P_{\min}(a(z;\tau,r))$ by the definition of $P_{\min}$, and the second
claim follows from~\eqref{eq:stars-complement}.
\end{proof}

The center bound depends continuously on the lower bounds, with an explicit
modulus. The labels of \cref{sec:labels} give the lower bounds only up to a factor
$1-\eps_R$ (\cref{lem:correlation}), and the following estimate is how this factor
is removed in the limit $R\to\infty$.

\begin{lemma}[continuity of the center bound]\label{lem:stars-continuity}
Let $d\ge1$ and write $P_{\min}=P^{(d)}_{\min}$. Then:
\begin{enumerate}[label=\textup{(\alph*)}]
  \item $P_{\min}$ is nonincreasing in each coordinate, and
  $|P_{\min}(a)-P_{\min}(a')|\le\varphi(0)\sum_{i=1}^d|a_i-a'_i|$ for all
  $a,a'\in\R^d$.
  \item Let $\bar r\in(0,1)$, let $r,r'\in(0,\bar r]^d$ with $|r_i-r'_i|\le\eps$ for
  all $i$, and let $\tau\in\R$. Then
  \[
    |C_d(\tau;r)-C_d(\tau;r')|\le d\,\varphi(0)\,(1-\bar r^2)^{-3/2}
    \Bigl(|\tau|+\sqrt{2/\pi}\Bigr)\,\eps .
  \]
  In particular, if $d\le5$, $\bar r=r_{33}=\frac{2\sqrt2}{3}$ and $|\tau|\le3$,
  then $|C_d(\tau;r)-C_d(\tau;r')|\le205\,\eps$.
\end{enumerate}
\end{lemma}

\begin{proof}
(a) Fix a centered Gaussian vector $U$ with unit variances. Raising a threshold
shrinks the event $\{U_i>a_i\ \forall i\}$, so its probability is nonincreasing in
each $a_i$; and the two events for $a$ and $a'$ differ only on
$\bigcup_i\{U_i\ \text{lies between}\ a_i\ \text{and}\ a'_i\}$, which has
probability at most $\sum_i|\Phi(a_i)-\Phi(a'_i)|\le\varphi(0)\sum_i|a_i-a'_i|$.
An infimum of functions that are all nonincreasing in each coordinate and all
satisfy the same Lipschitz bound has the same two properties.

(b) For $\rho\in(0,\bar r]$ and fixed $\tau,z$, the map
$\rho\mapsto(\tau-\rho z)/\sqrt{1-\rho^2}$ has derivative
$(\rho\tau-z)/(1-\rho^2)^{3/2}$, whose absolute value is at most
$(1-\bar r^2)^{-3/2}(|\tau|+|z|)$. By the mean value theorem,
$|a_i(z;\tau,r)-a_i(z;\tau,r')|\le(1-\bar r^2)^{-3/2}(|\tau|+|z|)\,\eps$ for each $i$.
By part~(a),
\begin{align*}
  |C_d(\tau;r)-C_d(\tau;r')|
  &\le\int_\tau^\infty\varphi(z)\,d\,\varphi(0)\,(1-\bar r^2)^{-3/2}(|\tau|+|z|)\,\eps\,dz\\
  &\le d\,\varphi(0)\,(1-\bar r^2)^{-3/2}\bigl(|\tau|+\E|N|\bigr)\eps,
\end{align*}
where $N$ is standard normal and $\E|N|=\sqrt{2/\pi}$. For $\bar r=2\sqrt2/3$ we
have $1-\bar r^2=\frac19$ and $(1-\bar r^2)^{-3/2}=27$. With $d\le5$ and
$|\tau|\le3$ the constant is at most
$5\cdot0.39895\cdot27\cdot(3+0.79789)<204.6$.
\end{proof}

\subsection{The simplex theorem}\label{sec:stars-simplex}

The infimum $P_{\min}$ in the center bound is over all correlation matrices. For
equal thresholds it is attained by the most ``spread out'' matrix, that of the
regular simplex. This subsection defines the regular simplex vector and states this
claim for three to five coordinates (\cref{lem:simplex}), as cases of a theorem of
Mulgund~\cite{Mulgund2026}.

For $d\ge2$, let $w_1,\dots,w_d\in\R^{d-1}$ be unit vectors with
$\langle w_i,w_j\rangle=-\frac1{d-1}$ for $i\ne j$ (the vertices of a regular
simplex inscribed in the unit sphere $S^{d-2}$), so that $\sum_iw_i=0$. Let $G$ be
a standard Gaussian vector in $\R^{d-1}$. The \emph{regular simplex vector} is
$U^*=(U^*_1,\dots,U^*_d)$ with $U^*_i=\langle w_i,G\rangle$. It is centered, with
unit variances and all pairwise correlations $-\frac1{d-1}$, and
$\sum_iU^*_i=\langle\sum_iw_i,G\rangle=0$. Its \emph{orthant function} and the
reflection of the latter are
\[
  S_d(a):=\Pr\bigl(U^*_i>a\ \text{for}\ 1\le i\le d\bigr),\qquad
  T_d(h):=S_d(-h)=\Pr\bigl(\max_iU^*_i<h\bigr)\qquad(a,h\in\R),
\]
the last identity because $-U^*$ has the same distribution as $U^*$. Since the coordinates
of $U^*$ sum to zero, they cannot all exceed a number $a\ge0$; hence
\begin{equation}\label{eq:S-zero}
  S_d(a)=0\quad\text{for}\ a\ge0,\qquad T_d(h)=0\quad\text{for}\ h\le0 .
\end{equation}

The bound of \cref{lem:arcs} for three unequal thresholds starts by lowering all
correlations until the correlation matrix is singular.

\begin{lemma}[rank reduction]\label{lem:rank-reduction}
Let $d\ge2$ and let $C$ be a nonsingular $d\times d$ correlation matrix. There is a
singular $d\times d$ correlation matrix $C'$ such that, for $U\sim N(0,C)$ and
$U'\sim N(0,C')$, $\Pr(U_i>a_i\ \forall i)\ge\Pr(U'_i>a_i\ \forall i)$ for every
$a\in\R^d$.
\end{lemma}

\begin{proof}
Let $J$ be the all-ones matrix and $C_t=C-t(J-I)$ for $t\ge0$; it has unit diagonal
and its off-diagonal entries are those of $C$ minus $t$. The set
$\{t\ge0:C_t\succeq0\}$ is closed and convex, contains a neighborhood of $0$ in
$[0,\infty)$ (as $C\succ0$), and is bounded, because
$\mathbf 1^{\mathsf T}C_t\mathbf 1=\mathbf 1^{\mathsf T}C\mathbf 1-d(d-1)t$. So it is
an interval $[0,t^*]$ with $t^*>0$. The matrix $C':=C_{t^*}$ is positive
semidefinite, and it is singular by the maximality of $t^*$: the least eigenvalue of
$C_t$ is continuous in $t$, nonnegative at $t^*$ and negative for $t>t^*$. The
off-diagonal entries of $C'$ are at most those of $C$, so Slepian's inequality
(\cref{lem:slepian}) gives the claim.
\end{proof}

\begin{lemma}[simplex theorem for $d\le5$]\label{lem:simplex}
Let $d\in\{3,4,5\}$ and let $U=(U_1,\dots,U_d)$ be a centered Gaussian vector with
unit variances. Then $\Pr(U_i>a\ \text{for}\ 1\le i\le d)\ge S_d(a)$ for every
$a\in\R$; consequently $P^{(d)}_{\min}(a,\dots,a)=S_d(a)$.
\end{lemma}

\begin{proof}
The second claim follows from the first, since $U^*$ itself is admissible in the
infimum.

Mulgund \cite[Theorem~1.1]{Mulgund2026} proves: for every $n\ge2$,
every $n\times n$ correlation matrix $\Gamma$ (positive semidefinite with unit
diagonal, singular matrices included) and every $t\in\R$,
\[
  \Pr_{X\sim N(0,\Sigma_n)}\bigl(\max_iX_i\le t\bigr)\le\Pr_{X\sim N(0,\Gamma)}\bigl(\max_iX_i\le t\bigr),
\]
where $\Sigma_n$ has unit diagonal and off-diagonal entries $-\frac1{n-1}$. Apply
this with $n=d$, $X=-U$ (whose correlation matrix is that of $U$) and $t=-a$. Since
$-U^*$ has the distribution $N(0,\Sigma_d)$, this gives $\Pr(U_i\ge a\ \forall i)\ge
\Pr(U^*_i\ge a\ \forall i)$, and since every coordinate has a continuous distribution, the
weak inequalities may be replaced by strict ones.
\end{proof}

\begin{remark}[status of the simplex theorem]\label{rem:stars-mulgund}
For $d=3$ the inequality also has an elementary proof: condition on one coordinate,
apply the Fr\'echet bound $\Pr(A\cap B)\ge\Pr A+\Pr B-1$ to the other two, note that
the bound is attained by a planar configuration with the same conditional marginals,
and compare planar configurations by the convexity of the probability in the arcs
between the directions. For $d=4$ the inequality also
follows from the four-cap theorem of L.~Fejes T\'oth~\cite{FejesToth1950}
(\cite[Theorem~5.6.2]{Boroczky2004}) by slicing the Gaussian measure along spheres
centered at the origin, after lowering the correlations until the correlation matrix
is singular (\cref{lem:rank-reduction}); see
\cite[\S\S2--3]{BalitskiyKarasevTsigler2017}. For $d=5$ no other
proof is known to us. The paper \cite{Mulgund2026} is a preprint
(version~2), whose author discloses substantial AI assistance. The three-variable case
enters the proof of \cref{thm:cubic} through \cref{prop:star-cubic}, the four-variable
case only through vertices of degree at least~$4$, and the five-variable case only
through the triples $\tripP,\tripR,\tripS$ of \cref{thm:deg5}, and through $\tripS$
also \cref{thm:main}, \cref{cor:edges} and the results derived from them.
\end{remark}

\subsection{Ideal stars and the constants \texorpdfstring{$F_3$}{F3} and \texorpdfstring{$F_4$}{F4}}\label{sec:stars-ideal}

When all lower bounds of a star are equal, the center bound is the frontier
probability of an explicit star, and this star is the worst case at every
threshold.

\begin{definition}[ideal stars]\label{def:ideal-star}
Let $d\in\{3,4,5\}$ and $r\in(0,1)$. Let $Z$ be standard normal and $U^*$ the
regular simplex vector in $d$ variables, independent of $Z$. The \emph{ideal
$d$-star at correlation $r$} is
\[
  Y^*_0=Z,\qquad Y^*_i=rZ+s_rU^*_i\qquad(1\le i\le d).
\]
Its \emph{ideal star function} is
\[
  f_{d,r}(\tau):=\Pr\bigl(Y^*_0>\tau,\ \min_iY^*_i\le\tau\bigr)\qquad(\tau\in\R).
\]
Put
\[
  F_3:=\sup_{\tau\in\R}f_{3,r_{33}}(\tau),\qquad F_4:=\sup_{\tau\in\R}f_{4,r_{44}}(\tau),
  \qquad r_{33}=\tfrac{2\sqrt2}3,\quad r_{44}=\tfrac{\sqrt3}2 .
\]
\end{definition}

The constants $F_3$ and $F_4$ are those of \cref{thm:deg4,thm:deg5}. The ideal star
is a $d$-star with
all lower bounds equal to $r$: its incident correlations are $r$ and its other
correlations are $r^2-s_r^2/(d-1)=(dr^2-1)/(d-1)$. Since $U^*$ is independent of
$Z$ and $Y^*_i>\tau$ means $U^*_i>(\tau-rZ)/s_r$, Fubini's theorem
and~\eqref{eq:stars-complement} give
\begin{equation}\label{eq:ideal-star-integral}
  f_{d,r}(\tau)=\Phi(-\tau)-\int_\tau^\infty\varphi(z)\,S_d\Bigl(\frac{\tau-rz}{s_r}\Bigr)dz .
\end{equation}
Comparing this with~\eqref{eq:center-bound} and using
$P_{\min}(a,\dots,a)=S_d(a)$ (\cref{lem:simplex}),
\begin{equation}\label{eq:ideal-is-center}
  C_d\bigl(\tau;(r,\dots,r)\bigr)=f_{d,r}(\tau)\qquad(d\in\{3,4,5\},\ r\in(0,1),\ \tau\in\R).
\end{equation}
In the cubic case, $r_{33}=r_*=\frac{2\sqrt2}3$ is the edge correlation bound
$r_*(1-\frac1R)$ of the geometric labels in the limit $R\to\infty$
(\cref{lem:correlation-padded}), and $f_{3,r_*}(0)$ is the constant $\cthree$ of
\cref{thm:cubic} (\cref{lem:kappa}).

\begin{corollary}[ideal star bound]\label{cor:stars-ideal}
Let $d\in\{3,4,5\}$, $r\in(0,1)$ and $0\le k\le d$, and let $(Y_0,Y_1,\dots,Y_k)$ be a
centered Gaussian vector with unit variances and $\Corr(Y_0,Y_i)\ge r$ for
$1\le i\le k$. Then for every $\tau\in\R$,
\[
  \Pr\bigl(Y_0>\tau,\ Y_i\le\tau\ \text{for some}\ 1\le i\le k\bigr)\le f_{d,r}(\tau).
\]
\end{corollary}

\begin{proof}
If $k<d$, add coordinates $Y_j=rY_0+s_r\eta_j$ for $k<j\le d$, with fresh
independent standard normal $\eta_j$. The extended vector is a $d$-star with lower
bounds $(r,\dots,r)$, and the event can only grow. Now apply
\cref{lem:center} and~\eqref{eq:ideal-is-center}.
\end{proof}

In particular $f_{d,r}\le f_{d,r'}$ pointwise whenever $0<r'\le r<1$: apply
\cref{cor:stars-ideal} with lower bound $r'$ to the ideal $d$-star at correlation $r$.

The bound $f_{4,r}$ is used with $r=r_{44}(1-1/R)$ for the padded degree-$4$ labels
(\cref{lem:correlation-padded}); by \eqref{eq:ideal-is-center} and
\cref{lem:stars-continuity}(b), it exceeds $f_{4,r_{44}}$ by at most $178/R$ at every
threshold in $[-3,3]$.

\begin{restatable}[enclosures of $F_3$ and $F_4$]{proposition}{propFthreeFfour}\label{prop:F3F4}
The suprema $F_3$ and $F_4$ of \cref{def:ideal-star} satisfy
\begin{gather*}
  0.1401582523314521636939405140172\ \le\ F_3\ \le\ 0.1401582524,\\
  0.2457719746988\ \le\ F_4\ \le\ 0.24577197475 .
\end{gather*}
\end{restatable}

The proof is computer-assisted, in ball arithmetic with the Arb
library~\cite{Johansson2017}, from one-dimensional integral formulas for the ideal
star functions; it is in \cref{app:F3F4}.

\begin{remark}
The numbers in \cref{prop:F3F4} are used as follows: the lower bounds of $F_3$ and
$F_4$ in tail estimates and in the exact check of the certificate
(\cref{sec:certificate}), and the upper bound of $F_3$ in the certificate for the
triples $\tripR$ and $\tripS$, whose coefficient at degree~$3$ exceeds $F_3$.
Numerically, $F_3$ is attained at $\tau\approx-0.06814$ and $F_4$ at
$\tau\approx-0.16367$; both maximizers are negative, so in the plain sweep the
binding thresholds lie slightly below the median.
\end{remark}
{}
\section{The plain sweep: maximum degree 4 and 5}\label{sec:plain-sweep}

This section bounds the pathwidth of graphs of maximum degree $4$ and $5$ by a
weighted count of their vertices of degree $3$, $4$ and $5$, up to a term $o(n)$. The
two theorems below are its results. In them, $n_d$ is the number of vertices of
degree~$d$, and $F_3$ and $F_4$ are the maxima of the ideal star functions
$f_{3,r_{33}}$ and $f_{4,r_{44}}$ (\cref{def:ideal-star}).

\begin{theorem}[maximum degree $4$]\label{thm:deg4}
Every graph $G$ with $\Delta(G)\le 4$ satisfies
$\pw(G)\le a_3\,n_3+a_4\,n_4+o(n)$ for each of the pairs
\begin{enumerate}[label=\textup{(\alph*)},align=left,leftmargin=2.6em,labelwidth=1.6em,labelsep=1em,itemsep=2pt]
  \item\label{thm:deg4:a} $(a_3,a_4)=(F_4,\,F_4)$, and in particular $\pw(G)\le F_4\,n+o(n)$;
  \item\label{thm:deg4:b} $(a_3,a_4)=(F_3,\,0.2575)$;
  \item\label{thm:deg4:c} $(a_3,a_4)=\bigl(\tfrac{5}{32},\,\tfrac14\bigr)$.
\end{enumerate}
\end{theorem}

\begin{theorem}[maximum degree $5$]\label{thm:deg5}
Every graph $G$ with $\Delta(G)\le 5$ satisfies
$\pw(G)\le a_3\,n_3+a_4\,n_4+a_5\,n_5+o(n)$ for each of the triples
\begin{enumerate}[label={},align=left,leftmargin=2.6em,labelwidth=1.6em,labelsep=1em,itemsep=2pt]
  \item[$\tripP$] $(a_3,a_4,a_5)=(F_3,\,0.2575,\,0.3667)$;
  \item[$\tripR$] $(a_3,a_4,a_5)=(0.1455,\,0.2530,\,0.3600)$;
  \item[$\tripS$] $(a_3,a_4,a_5)=\bigl(\tfrac{5}{32},\,\tfrac14,\,\tfrac{11}{32}\bigr)$.
\end{enumerate}
\end{theorem}

Parts \ref{thm:deg4:b} and \ref{thm:deg4:c} of \cref{thm:deg4} are the cases
$\tripP$ and $\tripS$ of \cref{thm:deg5} for graphs without vertices of degree~$5$;
we state them separately because later sections use them in this form. With the
bootstrap of \cref{sec:bootstrap}, the two theorems give the sequences $\seqF$,
$\tripP$, $\tripR$ and $\tripS$ of \cref{thm:pw-seq}.

The method is the \emph{plain sweep}: draw the Gaussian labels of \cref{sec:labels},
order the vertices by label, and take the layout of \cref{lem:layout}, whose width is
the largest outer frontier $|\outfr S_\tau|$ over all thresholds $\tau$. It therefore
suffices to find one realization of the labels in which every outer frontier has at
most $\sum_da_dn_d+o(n)$ vertices. The proof has three parts.
\begin{enumerate}[label=(\arabic*)]
  \item \emph{Expected frontier at a fixed threshold.} A vertex lies in
  $\outfr S_\tau$ exactly when the frontier event of its star occurs. So
  $\E|\outfr S_\tau|$ is a sum of star probabilities, each bounded by the center
  bound of \cref{lem:center}. The center bound depends on the degree of the vertex
  and on the degrees of its neighbors, which we record as its \emph{type}. A vertex
  of degree~$3$ next to vertices of higher degree has weaker correlation bounds than
  one in a cubic region, and its center bound exceeds $a_3$. The excess is moved
  along the edges to the neighbors of higher degree (\cref{lem:charging}). Checking
  that after these transfers the bound at every vertex is at most the coefficient of
  its degree, at every threshold, is a finite but large computation: the certificate
  of \cref{prop:certificate}.
  \item \emph{Removing the correlation loss.} The labels give the correlation lower
  bounds only up to a factor $1-\eps_R$, where $R$ is the parameter of the labels.
  The explicit continuity of the center bound (\cref{lem:stars-continuity}) turns
  this factor into an additive error that vanishes as $R\to\infty$, uniformly in the
  threshold.
  \item \emph{Concentration.} Frontier sizes at finitely many thresholds are sums of
  local indicators, so they concentrate (\cref{lem:local-counts}). A grid argument
  (\cref{lem:grid}) passes from these thresholds to all thresholds.
\end{enumerate}
The proof of \cref{thm:deg4}\ref{thm:deg4:a} has the same three parts, with the
padded labels of degree~$4$ instead of the mixed labels used for the other bounds
(\cref{def:labels}). There all vertices are
treated alike: the frontier probability of every vertex is at most the ideal star
function $f_{4,r}$ (\cref{cor:stars-ideal}), so no charging and no certificate are
needed.

\paragraph{Setting.} In \cref{sec:types,sec:sweep-certificate,sec:sweep-means}, $H$
is a graph with $\delta(H)\ge3$ and $\Delta(H)\le5$, and $Y=(Y_v)_{v\in V(H)}$ are
the mixed labels of \cref{def:labels}\ref{def:labels:mixed} with parameter $R$. We
use two facts from \cref{sec:labels}.
\begin{enumerate}[label=(L\arabic*)]
  \item\label{it:sweep-L1} For every edge $uv$ of $H$ with $d(u)=i$ and $d(v)=j$,
  $\Corr(Y_u,Y_v)\ge r_{ij}(1-\eps_R)$, where $r_{ij}=\sech(\frac14\ln((i-1)(j-1)))$
  and $\eps_R=\frac{4\ln2}{R-2\ln2}$ for $R>2\ln2$ (\cref{lem:correlation} with
  $\Delta_0=5$); in particular $\eps_R\to0$ as $R\to\infty$.
  \item\label{it:sweep-L2} The labels are pairwise distinct with probability $1$, and each
  $Y_v$ is a function of $(Z_w)_{w\in B_{\varrho_R}(v)}$, where
  $\varrho_R:=\lfloor2R/\ln2\rfloor$ is the locality radius $\rho$ of
  \cref{lem:locality} for mixed labels.
\end{enumerate}
Since $\sech$ is decreasing on $[0,\infty)$, the largest of the
numbers $r_{ij}$ with $3\le i,j\le5$ is $r_{33}=\frac{2\sqrt2}3$.

\subsection{Vertex types and the charging scheme}\label{sec:types}

The center bound of a vertex depends only on its degree and on the degrees of its
neighbors. We record this dependence as a type, define the redistribution of the
excess, and isolate the inequalities that a coefficient vector must satisfy.

\begin{definition}[types, reference bounds and admissible coefficients]\label{def:admissible}
Let $D$ be $\{3,4\}$ or $\{3,4,5\}$.
\begin{enumerate}[label=\textup{(\roman*)}]
  \item A \emph{$D$-type} is a tuple $T=(d;k_3,k_4,k_5)$ of nonnegative integers with
  $d\in D$, $k_3+k_4+k_5=d$, and $k_j=0$ for $j\notin D$. In a graph whose degrees
  all lie in $D$, the \emph{type} of a vertex $v$ is $T(v)=(d(v);k_3,k_4,k_5)$, where
  $k_j$ is the number of neighbors of $v$ of degree $j$.
  \item The \emph{reference bounds} of $T$ are the vector $r_T\in(0,1)^d$ consisting of
  $k_3$ copies of $r_{d3}$, $k_4$ copies of $r_{d4}$ and $k_5$ copies of $r_{d5}$,
  and the \emph{reference star bound} of $T$ is $C_T(\tau):=C_d(\tau;r_T)$, the center
  bound~\eqref{eq:center-bound}.
  \item For a \emph{transfer vector} $t=(t_{34},t_{35},t_{45})\in\R^3$, the
  \emph{transfer term} of $T$ is
  \[
    \lambda_T(t)=\begin{cases}
      -k_4t_{34}-k_5t_{35}, & d=3,\\
      k_3t_{34}-k_5t_{45}, & d=4,\\
      k_3t_{35}+k_4t_{45}, & d=5.
    \end{cases}
  \]
  \item Coefficients $(a_d)_{d\in D}$ are \emph{admissible for $D$} if for every
  $\tau\in[-3,3]$ there is a transfer vector $t$ with
  \[
    C_T(\tau)+\lambda_T(t)\le a_d\qquad\text{for every $D$-type $T=(d;k_3,k_4,k_5)$.}
  \]
\end{enumerate}
\end{definition}

Thus a vertex of degree $i$ sends the amount $t_{ij}$ along each edge to a neighbor of
degree $j>i$. Admissibility says that, at every threshold in $[-3,3]$, some choice of
these transfers makes the reference star bound plus the transfer term of every type
at most the coefficient of its degree. There are $10$ types of degree $3$, $15$ of degree $4$
and $21$ of degree $5$, so $46$ types for $D=\{3,4,5\}$ and $9$ for $D=\{3,4\}$.
The \emph{pure cubic} type $T_0=(3;3,0,0)$ has $C_{T_0}=f_{3,r_{33}}\le F_3$
by~\eqref{eq:ideal-is-center} and \cref{def:ideal-star}, and $\lambda_{T_0}=0$.

\begin{lemma}[charging]\label{lem:charging}
Let $D$ be as in \cref{def:admissible}, let $H$ be a graph all of whose degrees lie
in $D$, and let $t\in\R^3$. Then $\sum_{v\in V(H)}\lambda_{T(v)}(t)=0$. Consequently,
if real numbers $x_v$ ($v\in V(H)$) and $e\ge0$ satisfy
$x_v+\lambda_{T(v)}(t)\le a_{d(v)}+e$ for every $v$, then
\[
  \sum_{v\in V(H)}x_v\ \le\ \sum_{d\in D}a_dn_d(H)+e\,|V(H)| .
\]
\end{lemma}

\begin{proof}
Write $\lambda_{T(v)}(t)$ as a sum over the edges at $v$: by the definition, an edge
$vu$ with $d(v)=i$ and $d(u)=j$ contributes $-t_{ij}$ if $i<j$, $+t_{ji}$ if $i>j$,
and $0$ if $i=j$. So each edge joining degrees $i<j$ contributes $-t_{ij}$ at its
end of degree $i$ and $+t_{ij}$ at its end of degree $j$, and each edge joining
equal degrees contributes nothing. Summing over all vertices therefore gives $0$.
Summing the inequalities $x_v\le a_{d(v)}+e-\lambda_{T(v)}(t)$ over $v$ gives the
second claim.
\end{proof}

\subsection{The certified coefficients}\label{sec:sweep-certificate}

Admissibility is a statement about all thresholds in $[-3,3]$ and about $46$ types at
once, each with a reference star bound that is an integral of an infimum over
correlation matrices. We verify it for three coefficient vectors by a
computer-assisted proof in ball arithmetic and exact rational arithmetic.

\begin{proposition}[certified admissibility]\label{prop:certificate}
Each of the triples
\[
  \tripP=(F_3,\,0.2575,\,0.3667),\qquad
  \tripR=(0.1455,\,0.2530,\,0.3600),\qquad
  \tripS=\bigl(\tfrac5{32},\,\tfrac14,\,\tfrac{11}{32}\bigr)
\]
is admissible for $D=\{3,4,5\}$.
\end{proposition}

If $(a_3,a_4,a_5)$ is admissible for $\{3,4,5\}$, then $(a_3,a_4)$ is admissible for
$\{3,4\}$: every $\{3,4\}$-type is a $\{3,4,5\}$-type with the same reference star
bound, and its transfer term, $-k_4t_{34}$ for $d=3$ and $k_3t_{34}$ for $d=4$, is the
same function of the same transfer vector. In particular $(F_3,0.2575)$ and
$(\frac5{32},\frac14)$ are admissible for $\{3,4\}$.

\emph{What is certified.} For each of the three triples, the accompanying code
partitions $[-3,3]$ into finitely many cells $[t_0,t_1]$. For each cell it stores one
transfer vector $t$ with $|t_{ij}|\le1$ and, for each type $T$, a rational number
$I^{\rm lo}_T(t_1)$. It then verifies, in exact rational arithmetic, that
\begin{equation}\label{eq:cell-inequality}
  \widehat\Phi(-t_0)-I^{\rm lo}_T(t_1)+\lambda_T(t)<a_d\qquad\text{for every type $T\ne T_0$,}
\end{equation}
where $\widehat\Phi(-t_0)$ is a rational upper bound for $\Phi(-t_0)$. The numbers
$I^{\rm lo}_T(t_1)$ are lower bounds for a Riemann sum of the integral
in~\eqref{eq:center-bound} at $\tau=t_1$, in which $P_{\min}$ is replaced by a lower
bound: the bound of \cref{lem:truncation}, or, for the seven degree-$3$ types with
unequal reference bounds, the larger of that bound and the arc bound of
\cref{lem:arcs}. \Cref{sec:certificate} proves that~\eqref{eq:cell-inequality}
implies $C_T(\tau)+\lambda_T(t)\le a_d$ for every $\tau\in[t_0,t_1]$
(\cref{lem:cert-cells,lem:cert-discretize}). The pure cubic type $T_0$ is not
checked: $C_{T_0}\le F_3\le a_3$ for all three triples, and $\lambda_{T_0}=0$.

\emph{How.} Every value of $\varphi$, $\Phi$, $S_d$ and of the arc bound that enters
$I^{\rm lo}$ is a lower or upper bound obtained from an enclosure in Arb ball
arithmetic \cite{Johansson2017}, rounded outward. The arc bound is evaluated through a
closed form in Owen's $T$ function \cite{Owen1956} (\cref{lem:cert-owen}). A separate
program rechecks~\eqref{eq:cell-inequality} for all cells in exact rational
arithmetic. Floating point is used only to make choices, such as the transfer
vectors; no floating-point value enters~\eqref{eq:cell-inequality}. For $\tripP$,
whose coefficient at degree~$3$ is $F_3$ itself, the recheck uses the lower bound of
$F_3$ in \cref{prop:F3F4} in place of $a_3$. For $\tripR$ and $\tripS$, the upper
bound of $F_3$ there gives $F_3\le a_3$. \Cref{sec:cert-arithmetic,sec:cert-results}
give the details, the numbers of cells and the smallest slacks.

\begin{proof}[Proof of \cref{prop:certificate}]
Fix one of the triples and $\tau\in[-3,3]$; let $[t_0,t_1]$ be a cell of the
certificate containing $\tau$ (the cells tile $[-3,3]$, which the code checks), and
$t$ its stored transfer vector. For $T=T_0$, $C_{T_0}(\tau)+\lambda_{T_0}(t)\le F_3\le
a_3$. For every other type $T$ of degree $d$, \cref{lem:cert-cells,lem:cert-discretize}
and the construction of $I^{\rm lo}_T$ described in \cref{sec:certificate} give
$C_T(\tau)\le\Phi(-t_0)-I^{\rm lo}_T(t_1)$, and~\eqref{eq:cell-inequality} gives
$C_T(\tau)+\lambda_T(t)<a_d$.
\end{proof}

\begin{remark}[how close to the method's limit]\label{rem:sweep-optimum}
The certified coefficients are slightly larger than what the star bounds allow
pointwise, because a cell bound loses a term of first order in the cell width. For
orientation, floating-point quadrature at sampled thresholds (not certified) gives
the following values. For $D=\{3,4\}$ with $a_3=F_3$, the smallest possible $a_4$ is
about $0.25659$; it is attained near $\tau\approx-0.089$ by a degree-$4$ vertex all of
whose neighbors have degree~$3$. For $D=\{3,4,5\}$ with $a_3=F_3$, the binding
configuration is a degree-$5$ vertex with five neighbors of degree~$3$, near
$\tau\approx-0.105$, and the smallest $a_5$ is about $0.36621$. The triple $\tripS$
lies about $5\cdot10^{-4}$ inside the corresponding floating-point optimum.

The method does not appear able to bring the coefficient of $n_4$ down to $F_4$. On
$(3,4)$-biregular graphs of large girth, every edge joins a vertex of degree~$3$ to
one of degree~$4$. A numerical evaluation of the frontier density of the sweep on such
graphs, with $a_3=F_3$, requires a coefficient of about $0.2533$ at degree~$4$, above
$F_4\approx0.2458$.
\end{remark}

\subsection{Expected frontier sizes}\label{sec:sweep-means}

We now combine the star bounds, the certificate and the correlation estimate of
\cref{sec:labels}. The explicit continuity estimate of \cref{lem:stars-continuity},
which is uniform in the threshold, turns the factor $1-\eps_R$ in~\ref{it:sweep-L1}
into an additive error of $205\,\eps_R$ per vertex; no slack of the certificate is
needed for this step.

\begin{lemma}[expected frontiers in the mixed case]\label{lem:sweep-means}
Let $(a_3,a_4,a_5)$ be one of $\tripP,\tripR,\tripS$, let $H$ be a graph all of whose
degrees lie in $\{3,4,5\}$, put $\beta(H)=a_3n_3(H)+a_4n_4(H)+a_5n_5(H)$, and let $R$
be such that $\eps_R<1$. Then for every $\tau\in[-3,3]$,
\[
  \E\,|\outfr S_\tau|\ \le\ \beta(H)+205\,\eps_R\,|V(H)| .
\]
\end{lemma}

\begin{proof}
Let $v$ be a vertex of type $T=(d;k_3,k_4,k_5)$, with neighbors $u_1,\dots,u_d$. The
vector $(Y_v,Y_{u_1},\dots,Y_{u_d})$ is centered Gaussian with unit variances, and
by~\ref{it:sweep-L1} it is a $d$-star with lower bounds $r'_T:=(1-\eps_R)r_T$, after
ordering the neighbors as in $r_T$. The event $\{v\in\outfr S_\tau\}$ is its frontier
event. By \cref{lem:center},
\[
  \Pr(v\in\outfr S_\tau)\le C_d(\tau;r'_T).
\]
Every entry of $r_T$ and of $r'_T$ lies in $(0,r_{33}]$, and the entries differ by
$\eps_R(r_T)_i\le\eps_R$. By \cref{lem:stars-continuity}(b), since $|\tau|\le3$ and
$d\le5$,
\[
  \Pr(v\in\outfr S_\tau)\le C_T(\tau)+205\,\eps_R .
\]
By \cref{prop:certificate} there is a transfer vector $t$ with
$C_T(\tau)+\lambda_T(t)\le a_d$ for every $\{3,4,5\}$-type $T$. Hence
$\Pr(v\in\outfr S_\tau)+\lambda_{T(v)}(t)\le a_{d(v)}+205\,\eps_R$ for every vertex
$v$, and \cref{lem:charging} with $x_v=\Pr(v\in\outfr S_\tau)$ and $e=205\,\eps_R$
gives the claim, since $\E|\outfr S_\tau|=\sum_v\Pr(v\in\outfr S_\tau)$.
\end{proof}

\subsection{Concentration}\label{sec:sweep-concentration}

Three lemmas turn bounds on expectations at finitely many thresholds into a bound for
one realization at all thresholds. The first (\cref{lem:grid}) is deterministic: it
bounds a frontier at any threshold by the frontier at a neighboring grid point plus
the number of labels between the two, and outside the grid by the number of labels
beyond its ends. The plain sweep uses only its statements about
outer frontiers; the median-switched sweep of \cref{sec:cubic} also uses those about
inner frontiers.

\begin{lemma}[grid]\label{lem:grid}
Let $y$ be any labeling of a graph $G$, let $\tau_0<\tau_1<\dots<\tau_K$, and for
$0\le j<K$ let $N_j=|\{v:\tau_j<y_v\le\tau_{j+1}\}|$.
\begin{enumerate}[label=\textup{(\roman*)}]
  \item If $\tau_j\le\tau<\tau_{j+1}$, then $|\outfr S_\tau|\le|\outfr S_{\tau_{j+1}}|+N_j$
  and $|\infr S_\tau|\le|\infr S_{\tau_j}|+N_j$.
  \item If $\tau<\tau_0$, then $|\outfr S_\tau|\le\Delta(G)\cdot|\{v:y_v\le\tau_0\}|$
  and $|\infr S_\tau|\le|\{v:y_v\le\tau_0\}|$.
  \item If $\tau\ge\tau_K$, then $|\outfr S_\tau|\le|\{v:y_v>\tau_K\}|$.
\end{enumerate}
\end{lemma}

\begin{proof}
Let $v\in\outfr S_\tau$, so $y_v>\tau$ and some neighbor $u$ (a \emph{witness}) has
$y_u\le\tau$.
(i) If $y_v>\tau_{j+1}$, then $v\in\outfr S_{\tau_{j+1}}$ with the same witness,
because $y_u\le\tau<\tau_{j+1}$. Otherwise $\tau_j\le\tau<y_v\le\tau_{j+1}$, and $v$ is
counted by $N_j$. For the inner frontier, let $u\in\infr S_\tau$, with a neighbor $x$
such that $y_u\le\tau<y_x$. If $y_u\le\tau_j$, then $u\in S_{\tau_j}$ and
$y_x>\tau\ge\tau_j$, so $u\in\infr S_{\tau_j}$. Otherwise
$\tau_j<y_u\le\tau<\tau_{j+1}$, and $u$ is counted by $N_j$.
(ii) Map $v$ to a witness $u$. Then $y_u\le\tau<\tau_0$, and each $u$ is the image of
at most $d(u)\le\Delta(G)$ vertices. For the inner frontier,
$\infr S_\tau\subseteq S_\tau\subseteq S_{\tau_0}$.
(iii) $y_v>\tau\ge\tau_K$.
\end{proof}

The second lemma is the concentration step shared by all sweeps of this paper. It
combines the variance bound of \cref{lem:local-variance}, Chebyshev's inequality and
a union bound over the finitely many counts. The counts of a sweep are frontier sizes,
window counts $|\{v:\tau_j<Y_v\le\tau_{j+1}\}|$ and tail counts. Each is a sum over
the vertices $v$ of indicators that depend only on the labels on $N[v]$, hence on
$(Z_w)_{w\in B_{\rho+1}(v)}$, where $\rho$ is the locality radius of
\cref{lem:locality}.

\begin{lemma}[concentration of local counts]\label{lem:local-counts}
Let $G$ be a graph with $N$ vertices and $\Delta(G)\le\Delta$, where $\Delta\ge2$,
let $(Y_v)$ be labels of $G$ as in \cref{def:labels} with underlying variables
$(Z_w)_{w\in V(G)}$, let $\varrho\ge0$ be an integer, and let $W_1,\dots,W_\mu$ be
sums over $v\in V(G)$ of $[0,1]$-valued measurable functions of
$(Z_w)_{w\in B_\varrho(v)}$. If $\lambda>0$ and $\mu N\Delta^{2\varrho}<2\lambda^2$,
then some realization of $(Z_w)$ has pairwise distinct labels and
$|W_i-\E W_i|<\lambda$ for all $i$.
\end{lemma}

\begin{proof}
By \cref{lem:local-variance} with $a=1$ and $A_v=B_\varrho(v)$, each $W_i$ satisfies
$\Pr(|W_i-\E W_i|\ge\lambda)\le N\Delta^{2\varrho}/(2\lambda^2)$. By the union bound,
the probability that some $W_i$ deviates from its mean by at least $\lambda$ is at
most $\mu N\Delta^{2\varrho}/(2\lambda^2)<1$. By
\cref{lem:locality}\ref{lem:locality:distinct} the labels are pairwise distinct with probability $1$. So the event that the labels are pairwise distinct and every $W_i$ lies
within $\lambda$ of its mean has positive probability, and in particular it is
nonempty.
\end{proof}

The third lemma is the plain sweep itself: bounds on the expected outer frontiers at
the thresholds in $[-3,3]$ give a path decomposition. In it, $\beta$ is the target
width, $e$ is the error per vertex in the expected frontiers, and the grid of
thresholds has $K$ points per unit length. The hypothesis $\Delta\,\Phi(-3)\,N\le\beta$
covers the thresholds outside $[-3,3]$, and the last hypothesis is the condition of
\cref{lem:local-counts} for the $12K+2$ counts used in the proof, with
$\lambda=N^{7/8}$.

\begin{lemma}[plain sweep from expected frontiers]\label{lem:plain-sweep}
Let $G$ be a graph with $N$ vertices and $\Delta(G)\le\Delta$ for some $\Delta\ge2$,
let $(Y_v)$ be labels of $G$ as in \cref{def:labels} with locality radius $\rho$
\textup{(}\cref{lem:locality}\textup{)}, let $K\ge1$ be an integer, and let
$\beta,e\ge0$. Suppose that $\E|\outfr S_\tau|\le\beta+eN$ for every
$\tau\in[-3,3]$, that $\Delta\,\Phi(-3)\,N\le\beta$, and that
$(12K+2)\,\Delta^{2\rho+2}<2N^{3/4}$. Then
\[
  \pw(G)\ \le\ \beta+\Bigl(e+\frac{\varphi(0)}K\Bigr)N+\Delta\,N^{7/8}.
\]
\end{lemma}

\begin{proof}
We fix a realization in which the counts at a grid of thresholds are all close to
their means, and then bound the outer frontier at every threshold by \cref{lem:grid}.
Let $\tau_j=-3+j/K$ for $0\le j\le6K$, and consider the $12K+2$ counts
\begin{gather*}
  A_j=|\outfr S_{\tau_j}|\ \ (1\le j\le6K),\qquad
  N_j=|\{v:\tau_j<Y_v\le\tau_{j+1}\}|\ \ (0\le j<6K),\\
  C_-=|\{v:Y_v\le-3\}|,\qquad C_+=|\{v:Y_v>3\}| .
\end{gather*}
Each is a sum over $v$ of indicators that depend only on the labels on $N[v]$, hence
on $(Z_w)_{w\in B_{\rho+1}(v)}$
(\cref{lem:locality}\ref{lem:locality:local},\ref{lem:locality:frontier}). The
last hypothesis is the condition of \cref{lem:local-counts} with $\varrho=\rho+1$,
$\mu=12K+2$ and $\lambda=N^{7/8}$, so some realization has pairwise distinct labels
and every count within $N^{7/8}$ of its mean. Fix it. Since each $Y_v$ is standard
normal, $\E N_j\le N\varphi(0)/K$ and $\E C_\pm=N\Phi(-3)$, and $\E A_j\le\beta+eN$
by hypothesis. Let $\tau\in\R$.
\begin{itemize}
  \item If $\tau\in[\tau_j,\tau_{j+1})$ for some $j<6K$, then by \cref{lem:grid}(i),
  $|\outfr S_\tau|\le A_{j+1}+N_j\le\beta+(e+\varphi(0)/K)N+2N^{7/8}$.
  \item If $\tau<-3$, then by \cref{lem:grid}(ii),
  $|\outfr S_\tau|\le\Delta\,C_-\le\Delta\,\Phi(-3)N+\Delta N^{7/8}\le\beta+\Delta N^{7/8}$.
  \item If $\tau\ge3$, then by \cref{lem:grid}(iii),
  $|\outfr S_\tau|\le C_+\le\Phi(-3)N+N^{7/8}\le\beta+N^{7/8}$.
\end{itemize}
Since $\Delta\ge2$, every outer frontier has at most
$\beta+(e+\varphi(0)/K)N+\Delta N^{7/8}$ vertices, and \cref{lem:layout} for the
injective labeling of the fixed realization gives the claim ($N\ge1$ by the last
hypothesis).
\end{proof}

\subsection{Proofs of \texorpdfstring{\cref{thm:deg4,thm:deg5}}{the degree-4 and degree-5 bounds}}\label{sec:proof-deg45}

We first prove the mixed bounds for graphs of minimum degree at least $3$, then
remove this assumption with the reductions of \cref{def:reductions}, and finally
prove \cref{thm:deg4}\ref{thm:deg4:a}.

\begin{proposition}[mixed bounds for minimum degree $3$]\label{prop:sweep-mindeg3}
Let $(a_3,a_4,a_5)$ be one of $\tripP,\tripR,\tripS$. There is a function
$\rho^*(N)=o(N)$ such that every graph $H$ on $N$ vertices all of whose degrees lie in
$\{3,4,5\}$ satisfies
\[
  \pw(H)\ \le\ a_3n_3(H)+a_4n_4(H)+a_5n_5(H)+\rho^*(N).
\]
\end{proposition}

\begin{proof}
We let the label parameter $R$ grow slowly with $N$ and apply \cref{lem:plain-sweep}
to the expected frontiers of \cref{lem:sweep-means}.
Write $\beta(H)=a_3n_3(H)+a_4n_4(H)+a_5n_5(H)$. Fix $R_0$ with $\eps_R<1$ for
$R\ge R_0$, and for $N\ge1$ let
\[
  R(N):=\max\Bigl\{R\in\{R_0,\dots,N\}:\ (12R+2)\,25^{\varrho_R+1}\le N^{1/2}\Bigr\},
\]
with $R(N):=0$ if the set is empty. For each fixed $R\ge R_0$ the condition holds for
all large $N$, so $R(N)\to\infty$ as $N\to\infty$. Put
\[
  \rho^*(N):=\begin{cases}
    \bigl(205\,\eps_{R(N)}+\varphi(0)/R(N)\bigr)N+5N^{7/8}, & R(N)\ge R_0,\\
    N, & \text{otherwise.}
  \end{cases}
\]
Since $\eps_R\to0$ and $R(N)\to\infty$, $\rho^*(N)=o(N)$. If $R(N)=0$ the claim holds
because $\pw(H)\le N-1$. So let $R:=R(N)\ge R_0$ and use the labels with parameter
$R$. By \cref{lem:sweep-means}, $\E|\outfr S_\tau|\le\beta(H)+205\,\eps_RN$ for every
$\tau\in[-3,3]$. Every vertex of $H$ has degree at least $3$, and
$a_5\ge a_4\ge a_3\ge F_3>0.14$ for all three triples (\cref{prop:F3F4}), so
$\beta(H)\ge0.14N>5\Phi(-3)N$, as $5\Phi(-3)<0.0068$. Finally
$(12R+2)\,25^{\varrho_R+1}\le N^{1/2}<2N^{3/4}$. Hence \cref{lem:plain-sweep} applies
with $\Delta=5$, the locality radius $\varrho_R$ of~\ref{it:sweep-L2}, $K=R$,
$\beta=\beta(H)$ and $e=205\,\eps_R$, and gives
$\pw(H)\le\beta(H)+(205\,\eps_R+\varphi(0)/R)N+5N^{7/8}=\beta(H)+\rho^*(N)$.
\end{proof}

\begin{proof}[Proof of \cref{thm:deg5} and of \cref{thm:deg4}\ref{thm:deg4:b},\ref{thm:deg4:c}]
Let $(a_3,a_4,a_5)$ be one of $\tripP,\tripR,\tripS$, and put $a_0=a_1=a_2=0$; then
$0=a_0\le a_1\le\dots\le a_5$ in all three cases. Let $\bar\rho$ be the running
maximum~\eqref{eq:running-max} of the function $\rho^*$ of \cref{prop:sweep-mindeg3},
with $\rho^*(0):=0$; it is nonnegative, nondecreasing and $o(n)$. By
\cref{prop:sweep-mindeg3},
every graph $J$ with $\delta(J)\ge3$ and $\Delta(J)\le5$ satisfies
$\pw(J)\le\sum_da_dn_d(J)+\bar\rho(|V(J)|)$. Then \cref{cor:min-degree-three} with
$k=5$ gives, for every nonempty graph $G$ with $\Delta(G)\le5$,
\[
  \pw(G)\le a_3n_3+a_4n_4+a_5n_5+\bar\rho(n)+3\log_2n+1,
\]
which is \cref{thm:deg5}; the remainder depends only on $n$. (That corollary reduces
$G$ by a maximal sequence of the operations of \cref{def:reductions} and applies
\cref{lem:reduction} without deletions.)

\Cref{thm:deg4}\ref{thm:deg4:b} and \ref{thm:deg4:c} are the cases $\tripP$ and
$\tripS$ of \cref{thm:deg5} for graphs with $\Delta(G)\le4$, for which $n_5=0$. In
such graphs no vertex type involves degree~$5$, so only the certified inequalities of
the $\{3,4\}$-types and the transfer $t_{34}$ are used; in particular the case $d=5$
of \cref{lem:simplex} is not needed for \cref{thm:deg4}.
\end{proof}

For \cref{thm:deg4}\ref{thm:deg4:a} we use the padded labels of degree $4$
(\cref{def:labels}\ref{def:labels:padded} with $p=4$) with an integer radius $R\ge2$.
Every arc has length $\frac12\ln3$, so that
$X_v=\sum_{w:\,\dist(v,w)\le R}3^{-\dist(v,w)/2}Z_w$ and $Y_v=X_v/\sqrt{\Var X_v}$. By
\cref{lem:correlation-padded}, every edge $uv$ of a graph with $\Delta\le4$ has
$\Corr(Y_u,Y_v)\ge r_R:=r_{44}(1-1/R)$, and by \cref{lem:locality} the locality radius
is $\rho=R$. The bound $F_4n+o(n)$ needs no reductions: vertices of every degree from
$0$ to $4$ are handled by padding the star (\cref{cor:stars-ideal}).

\begin{proposition}[fixed radius]\label{prop:sweep-deg4a}
Let $R\ge2$ be an integer and let $G$ be a graph with $\Delta(G)\le4$ and $n\ge3^{8R}$
vertices. Then
\[
  \pw(G)\le\Bigl(F_4+\frac{179}R\Bigr)n+4n^{7/8}.
\]
\end{proposition}

\begin{proof}
We apply \cref{lem:plain-sweep} with the ideal star bound. Use the padded labels with
radius $R$. For a vertex $v$ of degree $k\le4$, the vector
$(Y_v;Y_u,u\in N(v))$ has unit variances and incident correlations at least $r_R$.
By \cref{cor:stars-ideal} with $d=4$, \eqref{eq:ideal-is-center} and
\cref{lem:stars-continuity}(b) with $\bar r=r_{33}>r_{44}$ and
$\eps=r_{44}-r_R=r_{44}/R$, for every $\tau\in[-3,3]$,
\[
  \Pr(v\in\outfr S_\tau)\le f_{4,r_R}(\tau)\le f_{4,r_{44}}(\tau)+\frac{205\,r_{44}}R\le F_4+\frac{178}R ,
\]
so $\E|\outfr S_\tau|\le(F_4+178/R)n$ for every $\tau\in[-3,3]$. We check the remaining
hypotheses of \cref{lem:plain-sweep} with $\Delta=4$, $\rho=R$, $K=R$, $\beta=F_4n$
and $e=178/R$. First, $4\Phi(-3)n<0.0055\,n<F_4n$ by \cref{prop:F3F4}. Second,
$8(12R+2)<45^R$ for every $R\ge2$ (at $R=2$ this reads $208<2025$, and the left side
less than doubles when $R$ grows by one), so
$(12R+2)\,16^{R+1}<2\cdot45^R\cdot16^R\le2\cdot729^R=2\cdot3^{6R}\le2n^{3/4}$, since
$n\ge3^{8R}$. Hence \cref{lem:plain-sweep} gives
$\pw(G)\le F_4n+(178+\varphi(0))n/R+4n^{7/8}\le(F_4+179/R)n+4n^{7/8}$, using
$\varphi(0)<0.399$.
\end{proof}

\begin{proof}[Proof of \cref{thm:deg4}\ref{thm:deg4:a}]
For $n\ge3^{16}$ put $R(n)=\lfloor\frac18\log_3n\rfloor\ge2$, so that
$n\ge3^{8R(n)}$, and $\rho(n)=\frac{179}{R(n)}n+4n^{7/8}$; for $n<3^{16}$ put
$\rho(n)=n$. Since $R(n)\to\infty$, $\rho(n)=o(n)$, and so is its running
maximum~$\bar\rho$~\eqref{eq:running-max}. By
\cref{prop:sweep-deg4a} with $R=R(n)$ for $n\ge3^{16}$ and trivially otherwise, $\pw(G)\le F_4n+\rho(n)$
for every graph $G$ with $\Delta(G)\le4$, which is the bound $F_4n+o(n)$. For the bound
with coefficients $(F_4,F_4)$, let $J$ be a graph with $\delta(J)\ge3$ and
$\Delta(J)\le4$; then $|V(J)|=n_3(J)+n_4(J)$, so
$\pw(J)\le F_4\bigl(n_3(J)+n_4(J)\bigr)+\bar\rho(|V(J)|)$. \Cref{cor:min-degree-three}
with $k=4$ and $(a_0,\dots,a_4)=(0,0,0,F_4,F_4)$ gives
$\pw(G)\le F_4n_3+F_4n_4+\bar\rho(n)+3\log_2n+1$ for every nonempty graph $G$ with
$\Delta(G)\le4$.
\end{proof}

{}
\section{The median-switched sweep: maximum degree 3 and 4}\label{sec:cubic}

This section proves the following two theorems, where
$\cthree=\frac{3}{4\pi}\arccos\frac56\approx0.1398$. With the bootstrap of
\cref{sec:bootstrap} they give the sequences $\seqC$ and $\seqM$ of \cref{thm:pw-seq}.

\begin{theorem}[maximum degree $3$]\label{thm:cubic}
Every graph $G$ with $\Delta(G)\le 3$ satisfies $\pw(G)\le \cthree\, n+o(n)$.
\end{theorem}

\begin{restatable}[maximum degree $4$]{theorem}{thmswitchedfour}\label{thm:switched4}
Every graph $G$ with $\Delta(G)\le 4$ satisfies
$\pw(G)\le \cthree\, n_3+0.2627\,n_4+o(n)$.
\end{restatable}

\Cref{thm:switched4} contains \cref{thm:cubic}, but its proof needs more: a switch
lemma for stars with unequal correlation bounds (\cref{lem:switch-mixed}), which we
prove with a computer-assisted certificate (\cref{sec:switch-cert-intro}), and a
computer-assisted charging at nonzero thresholds (\cref{prop:certificate-switched}).
The proof of \cref{thm:cubic} uses no computation.

Compared with the plain sweep of \cref{sec:plain-sweep}, only the layout changes
(\cref{sec:overview-switch}): vertices still enter in increasing order of their
geometric labels, but the layout keeps \emph{inner} frontiers while the threshold is
negative and \emph{outer} frontiers while it is positive, and passes from one to the
other at threshold zero by a short sequence of bags, the \emph{connector}.

After the notation of \cref{sec:cubic-setting}, the proof of \cref{thm:cubic} has four
steps. \Cref{sec:cubic-layout} constructs the layout (\cref{lem:median-layout}), with
an ordering lemma for the connector (\cref{lem:connector}). \Cref{sec:cubic-stars}
bounds the frontier probabilities away from threshold zero (\cref{prop:star-cubic}).
\Cref{sec:cubic-switch} bounds the expected values of $L_0$ and $L'_0$, which control
the connector bags at threshold zero (the switch lemma, \cref{lem:switch}). \Cref{sec:cubic-proof} adds the
concentration argument and completes the proof. The proof does not use Slepian's
inequality: at nonnegative thresholds a pointwise coupling suffices, and negative
thresholds are reduced to nonnegative ones by the symmetry $Y\mapsto-Y$.

The proof of \cref{thm:switched4}, in \cref{sec:switched4}, uses the same layout
with a larger set at threshold zero, together with the mixed labels and the charging
of \cref{sec:plain-sweep}. Finally,
\cref{sec:algorithms} turns the proof of \cref{thm:cubic} into an algorithm, and
\cref{app:alg-switched} that of \cref{thm:switched4}.

\subsection{Setting}\label{sec:cubic-setting}

Throughout, $G$ is a finite simple graph with $\Delta(G)\le3$; it may be disconnected
and may have vertices of degree $0$, $1$ or $2$. Recall $r_*=r_{33}=\frac{2\sqrt2}3$ and
$s_r=\sqrt{1-r^2}$, and put
\begin{equation}\label{eq:cubic-kappa}
  b(r)=\frac{3r^2-1}2,\qquad
  \kappa(r)=\frac3{4\pi}\arccos b(r)\qquad(0<r<1).
\end{equation}
For $r\in(0,1)$ we have $b(r)\in(-\frac12,1)$, so $\kappa$ is well defined and
continuous on $(0,1)$. Since $b(r_*)=\frac{3\cdot\frac89-1}2=\frac56$, we have
$\kappa(r_*)=\cthree$. By \cref{lem:kappa} below, $\kappa(r)$ is the value at
threshold zero of the ideal star function $f_{3,r}$.

\paragraph{Labels.} Fix an integer $R\ge2$ and let $(Y_v)_{v\in V(G)}$ be the
geometric labels at radius $R$, that is, the padded labels of degree $3$
(\cref{def:labels}\ref{def:labels:padded} with $p=3$): for i.i.d.\
standard normal $(Z_w)_{w\in V(G)}$, $X_v=\sum_{w:\,\dist(v,w)\le R}2^{-\dist(v,w)/2}Z_w$
and $Y_v=X_v/\sqrt{\Var X_v}$. Put $r_R=r_*(1-\frac1R)\in(0,r_*)$. We use three
properties.
\begin{enumerate}[label=\textup{(L\arabic*)}]
  \item\label{it:cubic-L1} The vector $(Y_v)_v$ is centered Gaussian with unit
  variances, and every edge $uv$ satisfies $\Corr(Y_u,Y_v)\ge r_R$
  (\cref{lem:correlation-padded}, cubic case).
  \item With probability $1$, the labels are pairwise distinct, also for
  vertices in different components; see
  \cref{lem:locality}\ref{lem:locality:distinct}.
  \item\label{it:cubic-L3} $Y_v$ is a function of $(Z_w)_{w\in B_R(v)}$
  (\cref{lem:locality}\ref{lem:locality:local}, with $\rho=R$); hence functions of $Z$
  restricted to disjoint vertex sets are independent.
\end{enumerate}
(We take $R\ge2$ only so that $r_R>0$.)
Since $-Y$ is again centered Gaussian with the same covariance, $-Y$ has the same
distribution as $Y$; we call this \emph{sign symmetry}. Every label has a continuous distribution, so
for each fixed $t$ the events $\{Y_v=t\}$ are null.

\paragraph{Threshold zero.} For a labeling $y\colon V(G)\to\R$, the sublevel sets
$S_t=\{v:y_v\le t\}$ and their frontiers $\outfr S_t$, $\infr S_t$ are as in
\cref{sec:prelim-layout}. Put $U=\infr S_0$ and $W=\outfr S_0$, and let $D$ be the
set of vertices of degree $3$ all of whose neighbors lie on the other side of zero,
that is, vertices $v$ with $d(v)=3$ and either $y_v\le0<\min_{x\in N(v)}y_x$ or
$y_v>0\ge\max_{x\in N(v)}y_x$. A vertex of $D$ with $y_v\le0$ has a neighbor above zero
and lies in $U$; one with $y_v>0$ has a neighbor at or below zero and lies in $W$.
Hence $D\subseteq U\cup W$. Put $U'=U\setminus D$, $W'=W\setminus D$ and
\begin{equation}\label{eq:cubic-L0}
  L_0=|U|+|D\cap W|=|U'|+|D|,\qquad L_0'=|W|+|D\cap U|=|W'|+|D|.
\end{equation}
The two expressions agree because $U$ and $W$ are disjoint and
$D=(D\cap U)\cup(D\cap W)$. At threshold zero, the width of the median-switched layout
of \cref{sec:cubic-layout} is controlled by $\max(L_0,L_0')$
(\cref{lem:median-layout}).

\subsection{The median-switched layout}\label{sec:cubic-layout}

This subsection proves the deterministic part of \cref{thm:cubic}: a path
decomposition whose width is controlled by inner frontiers below zero, outer frontiers
above zero, and $\max(L_0,L_0')$ at zero (\cref{lem:median-layout}). The
connector at zero needs an ordering lemma, which we state abstractly. Think of the
elements of $Q$ below as the vertices of $W'$, to be added one at a time, and of the
items as the vertices of $U'$, each of which can be dropped once its at most two
neighbors in $W'$ are present.

\begin{lemma}[connector order]\label{lem:connector}
Let $Q$ be a finite set with $k$ elements and let $P$ be a finite set of items, each
item $p$ having a nonempty set $e(p)\subseteq Q$ of at most two endpoints. For an
ordering $q_1,\dots,q_k$ of $Q$ and $0\le j\le k$, let $c(j)$ be the number of items
$p$ with $e(p)\subseteq\{q_1,\dots,q_j\}$. Then some ordering satisfies
\begin{equation}\label{eq:cubic-connector}
  j-c(j)\le\max(0,\,k-|P|)+1\qquad\text{for all }0\le j\le k.
\end{equation}
\end{lemma}

\begin{proof}
We use induction on $k$. If some nonempty proper subset of $Q$ is closed (defined
below), the instance splits into two smaller ones, which we order one after the
other, the one with smaller excess first (Case~1). Otherwise each new element can be
chosen to complete an item whose other endpoint is already present (Case~2).

If $k=0$, then $P=\emptyset$ because endpoint sets are
nonempty, and \eqref{eq:cubic-connector} reads $0\le1$. Let $k\ge1$. Call a set
$Y\subseteq Q$ \emph{closed} if every item $p$ satisfies $e(p)\subseteq Y$ or
$e(p)\subseteq Q\setminus Y$. For a closed set $A$ let $P_A=\{p:e(p)\subseteq A\}$
and define its \emph{excess} $\operatorname{ex}(A)=|A|-|P_A|$; thus
$\operatorname{ex}(Q)=k-|P|$.

\emph{Case 1: some nonempty proper subset $Y$ of $Q$ is closed.} Then $Q\setminus Y$
is closed as well, and since endpoint sets are nonempty, every item lies in exactly one
of $P_Y$ and $P_{Q\setminus Y}$. Let $A$ be the one of $Y,Q\setminus Y$ with the
smaller excess (either one if they are equal) and $B$ the other, so that
\[
  \operatorname{ex}(Q)=\operatorname{ex}(A)+\operatorname{ex}(B)\ge2\operatorname{ex}(A).
\]
The pairs $(A,P_A)$ and $(B,P_B)$ satisfy the hypotheses of the lemma with fewer than
$k$ elements, so by induction they have orderings $a_1,\dots,a_{|A|}$ and
$b_1,\dots,b_{|B|}$ with counting functions $c_A$, $c_B$ satisfying
\eqref{eq:cubic-connector} with $\operatorname{ex}(A)$, respectively
$\operatorname{ex}(B)$, in place of $k-|P|$. Order $Q$ as
$a_1,\dots,a_{|A|},b_1,\dots,b_{|B|}$.

For $j\le|A|$, an item covered by $\{a_1,\dots,a_j\}$ has an endpoint in $A$ and hence
lies in $P_A$, so $c(j)=c_A(j)$ and
$j-c(j)\le\max(0,\operatorname{ex}(A))+1$. If $\operatorname{ex}(A)\le0$ this is at
most $1$; otherwise $\operatorname{ex}(Q)\ge2\operatorname{ex}(A)>\operatorname{ex}(A)$.
In both cases $j-c(j)\le\max(0,\operatorname{ex}(Q))+1$.

For $j=|A|+j'$ with $1\le j'\le|B|$, all items of $P_A$ are covered, and an item of
$P_B$ is covered exactly when $e(p)\subseteq\{b_1,\dots,b_{j'}\}$. Hence
$c(j)=|P_A|+c_B(j')$ and
\[
  j-c(j)=\operatorname{ex}(A)+\bigl(j'-c_B(j')\bigr)
  \le\operatorname{ex}(A)+\max(0,\operatorname{ex}(B))+1 .
\]
If $\operatorname{ex}(B)\ge0$, the right side is
$\operatorname{ex}(Q)+1\le\max(0,\operatorname{ex}(Q))+1$. If
$\operatorname{ex}(B)<0$, then $\operatorname{ex}(A)\le\operatorname{ex}(B)<0$ and the
right side is less than $1$. This proves \eqref{eq:cubic-connector} in Case~1.

\emph{Case 2: no nonempty proper subset of $Q$ is closed.} Choose $q_1\in Q$
arbitrarily. Suppose $q_1,\dots,q_j$ have been chosen with $1\le j<k$, and let
$X=\{q_1,\dots,q_j\}$. Since $X$ is nonempty and proper, it is not closed, so some item
$p$ has endpoints both in $X$ and in $Q\setminus X$. As $|e(p)|\le2$, we have
$e(p)=\{x,y\}$ with $x\in X$ and $y\notin X$; put $q_{j+1}=y$. Then $p$ is covered by
$\{q_1,\dots,q_{j+1}\}$ but not by $X$, so $c(j+1)\ge c(j)+1$. Since $c(1)\ge0$, we
get $c(j)\ge j-1$ for every $1\le j\le k$, and $c(0)=0$. Hence $j-c(j)\le1$ for all
$j$, which implies \eqref{eq:cubic-connector}.
\end{proof}

The layout itself has three phases. The inner phase is the vertex separation layout
of Kinnersley~\cite{Kinnersley1992}: each bag consists of the new vertex and the inner
frontier of the preceding prefix. The outer phase is the layout of \cref{lem:layout}.
The connector between them is new.

\begin{lemma}[median-switched layout]\label{lem:median-layout}
Let $G$ be a graph with $\Delta(G)\le3$, let $y\colon V(G)\to\R$ be injective, and let
$w\ge0$ be a real number such that
\begin{enumerate}[label=\textup{(\roman*)}]
  \item $|\infr S_t|\le w$ for every $t<0$;
  \item $|\outfr S_t|\le w$ for every $t>0$; and
  \item $\max(L_0,L_0')+1\le w$, with $L_0,L_0'$ as in \eqref{eq:cubic-L0}.
\end{enumerate}
Then $\pw(G)\le w$.
\end{lemma}

The proof builds the bags from any ordering of $W'$ that satisfies
\eqref{eq:cubic-connector}, not only from that of \cref{lem:connector}, because the
algorithm of \cref{sec:alg-cubic} uses another one.

\paragraph{The bags.} Let $G$ be a graph with $\Delta(G)\le3$ and let
$y\colon V(G)\to\R$ be injective. Let $v_1,\dots,v_n$ be the vertices in increasing
order of $y$, let $\ell=|S_0|$, and write $T_i=\{v_1,\dots,v_i\}$ for $0\le i\le n$.
Let $P$ be the set of vertices $u\in U'$ for which $e(u):=N(u)\cap W'$ is nonempty;
these are the items, and $e(u)$ is the set of endpoints of $u$. Let $k=|W'|$, and let
$q_1,\dots,q_k$ be an ordering of $W'$. The \emph{inner bags} are
$A_i=\{v_i\}\cup\infr T_{i-1}$ for $1\le i\le\ell$, the \emph{outer bags} are
$A_i=\{v_i\}\cup\outfr T_i$ for $\ell<i\le n$, and the \emph{connector bags} are
\[
  C_{-1}=U\cup(D\cap W),\qquad
  C_j=D\cup\bigl\{u\in P:e(u)\not\subseteq\{q_1,\dots,q_{j-1}\}\bigr\}\cup\{q_1,\dots,q_j\}
  \quad(0\le j\le k).
\]
In words, the connector starts from $C_{-1}=U'\cup D$ and drops the vertices of $U'$
without neighbors in $W'$. It then adds the vertices of $W'$ one at a time and drops
each item one step after its last endpoint has been added. The vertices of $D$ stay
in every connector bag.

\begin{lemma}[the bags of the layout]\label{lem:median-layout-bags}
Let $G$, $y$, $P$ and $q_1,\dots,q_k$ be as above. Then $|e(u)|\le2$ for every
$u\in P$. If moreover $w$
satisfies \textup{(i)}--\textup{(iii)} of \cref{lem:median-layout} and the ordering
$q_1,\dots,q_k$ of $W'$ satisfies \eqref{eq:cubic-connector} for $Q=W'$ and the items
$P$, then $A_1,\dots,A_\ell,C_{-1},C_0,\dots,C_k,A_{\ell+1},\dots,A_n$ is a path
decomposition of $G$ of width at most $w$.
\end{lemma}

\begin{proof}
A vertex $u\in U'$ is not in $D$, so either $d(u)\le2$ or $u$ has a neighbor at or
below zero; in both cases $u$ has at most two neighbors above zero, and
$|e(u)|\le2$.

For the second claim we check the size bound and the three axioms of a path
decomposition (every vertex in a bag, every edge in a bag, the bags containing a
vertex consecutive). Note that $T_i=S_t$ for every $t$ with
$y_{v_i}\le t<y_{v_{i+1}}$ (with $y_{v_0}=-\infty$ and $y_{v_{n+1}}=+\infty$), and
$v_i\in S_0$ if and only if $i\le\ell$.

\emph{Inner bags.} For $1\le i\le\ell$ we have $y_{v_i}\le0$, so $T_{i-1}=S_t$ for
some $t<y_{v_i}\le0$, and $|A_i|\le w+1$ by (i).

\emph{Connector bags.} A vertex of $D$ lies in all connector bags, a vertex $q_j$ of
$W'$ in $C_j,\dots,C_k$, an item $u\in P$ in $C_{-1},\dots,C_J$, where $J\ge1$ is the
largest position of an element of $e(u)$ (so $u$ is dropped one step after its last
endpoint has been added), every other vertex of $U$ in $C_{-1}$ only, and no other
vertex in any connector bag. So the connector bags containing a vertex of $U$ form a
prefix of $C_{-1},C_0,\dots,C_k$, and those containing a vertex of $W$ form a suffix.

We have $|C_{-1}|=L_0$, and $|C_0|=|D|+|P|\le|D|+|U'|=L_0$. For $1\le j\le k$, the
number of items still present in $C_j$ is $|P|-c(j-1)$, where $c$ is the counting
function of \cref{lem:connector} for the ordering $q_1,\dots,q_k$, so by
\eqref{eq:cubic-connector} at $j-1$,
\[
  |C_j|=|D|+|P|-c(j-1)+j\le|D|+|P|+\max(0,k-|P|)+2=|D|+\max(|P|,|W'|)+2 .
\]
Since $|P|\le|U'|$, this is at most $\max(L_0,L_0')+2\le w+1$ by (iii).

\emph{Outer bags.} For $\ell<i\le n$ we have $y_{v_i}>0$, so $T_i=S_t$ for some
$t\ge y_{v_i}>0$, and $|A_i|\le w+1$ by (ii).

Every vertex $v_i$ lies in $A_i$. If $\ell=0$ or $\ell=n$, then $U=W=D=\emptyset$ and
the connector bags are empty; the arguments below remain valid.

\emph{Intervals for lower vertices.} Let $u=v_a$ with $a\le\ell$, and let $b$ be the
largest index of a neighbor of $u$ ($b=0$ if $u$ is isolated). For $i\le\ell$ we have
$u\in A_i$ if and only if $i=a$ or $a\le i-1<b$, so the inner bags containing $u$ are
$A_a,\dots,A_{\min(\max(a,b),\ell)}$. No outer bag contains $u$, because $u\in T_i$
for every $i>\ell$. If $u\notin U$, then $b\le\ell$ and $u$ is in no connector bag,
so its bags are consecutive. If $u\in U$, then $b>\ell$, so its bags are
$A_a,\dots,A_\ell$ followed by a prefix of the connector bags.

\emph{Intervals for upper vertices.} Let $x=v_c$ with $c>\ell$, and let $a$ be the
smallest index of a neighbor of $x$ ($a=\infty$ if $x$ is isolated). No inner bag
contains $x$. For $i>\ell$ we have $x\in A_i$ if and only if $i=c$, or $i<c$ and
$x$ has a neighbor in $T_i$, that is, $a\le i<c$; so the outer bags containing $x$
are $A_{\max(a,\ell+1)},\dots,A_c$ if $a<c$, and $A_c$ alone otherwise. If $x\notin W$,
then $a>\ell$ and $x$ lies in no connector bag, so its bags are consecutive. If
$x\in W$, then $a\le\ell$, so its bags are a suffix of the connector bags followed by
$A_{\ell+1},\dots,A_c$.

\emph{Edges.} Let $v_av_b$ be an edge with $a<b$. If $b\le\ell$, then
$v_a\in\infr T_{b-1}$, because $v_a\in T_{b-1}$ and its neighbor $v_b$ is not, so
both ends lie in $A_b$. If $a>\ell$, then $v_b\in\outfr T_a$, because $v_b\notin T_a$
and its neighbor $v_a$ is in $T_a$, so both ends lie in $A_a$. Otherwise $u=v_a\in U$
and $x=v_b\in W$. If $x\in D\cap W$, both ends lie in $C_{-1}$. If $u\in D\cap U$,
both ends lie in $C_k$, which contains $D$ and all of $W'$. Otherwise $u\in U'$ and
$x\in W'$, so $u\in P$ and $x=q_j\in e(u)$ for some $j$; then
$e(u)\not\subseteq\{q_1,\dots,q_{j-1}\}$, so $u\in C_j$, and $x\in C_j$.

Hence the sequence of bags is a path decomposition with all bags of size at most
$w+1$, that is, of width at most $w$.
\end{proof}

\begin{proof}[Proof of \cref{lem:median-layout}]
By \cref{lem:median-layout-bags}, every item $u\in P$ has one or two endpoints
$e(u)\subseteq W'$, so \cref{lem:connector} applies to $Q=W'$ and the items $P$ and
gives an ordering of $W'$ that satisfies \eqref{eq:cubic-connector}. For this
ordering, \cref{lem:median-layout-bags} gives a path decomposition of $G$ of width at
most $w$, so $\pw(G)\le w$.
\end{proof}

\subsection{The star comparison at nonnegative thresholds}\label{sec:cubic-stars}

This subsection bounds the probability that a vertex lies in an outer frontier at a
threshold $t\ge0$ by $\kappa(r)$ (\cref{prop:star-cubic}). By sign symmetry the same
bound applies to inner frontiers at thresholds $t\le0$. The proof has three
ingredients. A pointwise coupling lowers the incident correlations to $r$; the
simplex theorem for three variables (\cref{lem:simplex}) handles the neighbor residuals; and two facts
about the ideal star function $f_{3,r}$ give the value $\kappa(r)$: its value at zero
(\cref{lem:kappa}) and its maximality at zero among nonnegative thresholds
(\cref{lem:ideal-monotone}).

\paragraph{The ideal $3$-star.} We use the ideal $3$-star of \cref{def:ideal-star}.
Let $U^*=(U^*_1,U^*_2,U^*_3)$ be the regular simplex vector in three variables
(\cref{sec:stars}): centered, with unit variances and all pairwise correlations
$-\frac12$, so that $S_3(a)=\Pr(U^*_i>a\ \forall i)$. It is realized as the
projections of a standard planar Gaussian vector onto three unit vectors at mutual
angles $\frac{2\pi}3$; in particular $U^*_1+U^*_2+U^*_3=0$. For $r\in(0,1)$ the ideal
$3$-star at correlation $r$ is
\[
  Y^*_0\sim N(0,1)\ \text{independent of }U^*,\qquad Y^*_i=rY^*_0+s_rU^*_i\quad(i=1,2,3).
\]
Its incident correlations are $r$, its neighbor-pair correlations are
$r^2-\frac12s_r^2=b(r)$, and
\begin{equation}\label{eq:cubic-sum}
  Y^*_1+Y^*_2+Y^*_3=3rY^*_0 .
\end{equation}
The difference of any two coordinates of the ideal star has positive variance, as
$r<1$ and $s_r>0$, and each coordinate has a continuous distribution. So ties among the
coordinates, or with a fixed threshold, have probability zero, and strict and
non-strict inequalities may be exchanged in all events below. In
particular the ideal star function $f_{3,r}$ of \cref{def:ideal-star}, given by
\eqref{eq:ideal-star-integral} with $d=3$, satisfies
$f_{3,r}(t)=\Pr(Y^*_0\ge t,\ \min_iY^*_i\le t)$.

Both lemmas on $f_{3,r}$ below use Sheppard's formula (\cref{lem:sheppard}).

\begin{lemma}\label{lem:kappa}
For every $r\in(0,1)$, $f_{3,r}(0)=\kappa(r)$.
\end{lemma}

\begin{proof}
If $Y^*_i>0$ for all $i$, then $3rY^*_0>0$ by \eqref{eq:cubic-sum}, so $Y^*_0>0$.
Hence the event $\{\min_iY^*_i>0\}$ is contained in $\{Y^*_0\ge0\}$, and
\[
  f_{3,r}(0)=\Pr(Y^*_0\ge0)-\Pr(Y^*_0\ge0,\ \min_iY^*_i>0)
  =\frac12-\Pr(Y^*_1,Y^*_2,Y^*_3>0).
\]
The three neighbor coordinates have pairwise angles $\arccos b(r)$, so
\cref{lem:sheppard} gives
$f_{3,r}(0)=\frac12-\bigl(\frac12-\frac{3\arccos b(r)}{4\pi}\bigr)=\kappa(r)$.
\end{proof}

The next lemma shows that, among nonnegative thresholds, the ideal star function is
largest at the median $0$. Its proof needs no derivative and no numerical evaluation,
and holds for every $r\in(0,1)$.

\begin{lemma}\label{lem:ideal-monotone}
For every $r\in(0,1)$ and every $t\ge0$, $f_{3,r}(t)\le f_{3,r}(0)$.
\end{lemma}

\begin{proof}
We write $f_{3,r}(0)-f_{3,r}(t)$ as a difference of two probabilities and bound the
event that the smallest neighbor coordinate lies in $(0,t]$ by splitting according to
which coordinate attains the minimum.

Let $M=\min_iY^*_i$. If $M>t\ge0$, then $3rY^*_0>3t$ by \eqref{eq:cubic-sum}, so
$Y^*_0>t/r\ge t$. Hence $\{M>t\}\subseteq\{Y^*_0\ge t\}$, and since
$f_{3,r}(t)=\Pr(Y^*_0\ge t,\ M\le t)$, we get $f_{3,r}(t)=\Pr(Y^*_0\ge t)-\Pr(M>t)$
for all $t\ge0$ and
\[
  f_{3,r}(0)-f_{3,r}(t)=\Pr(0\le Y^*_0<t)-\Pr(0<M\le t).
\]
It remains to show $\Pr(0<M\le t)\le\Pr(0<Y^*_0\le t)$; the two probabilities
$\Pr(0\le Y^*_0<t)$ and $\Pr(0<Y^*_0\le t)$ are equal because $Y^*_0$ has a continuous
distribution.

If $0<M\le t$, then some coordinate attains the minimum, so the event is contained in
the union of the three events
$A_k=\{0<Y^*_k\le t,\ Y^*_k\le Y^*_j\text{ for all }j\in\{1,2,3\}\}$. Fix $k$ and write
$b=b(r)\in(-\frac12,1)$. For $j\ne k$ put $R_j=Y^*_j-bY^*_k$. Then
$\Cov(R_j,Y^*_k)=b-b=0$, so the pair $(R_j)_{j\ne k}$ is independent of $Y^*_k$. On
$A_k$ we have $R_j\ge Y^*_k-bY^*_k=(1-b)Y^*_k>0$ for both $j\ne k$. Hence
\[
  \Pr(A_k)\le\Pr(0<Y^*_k\le t)\cdot\Pr(R_j>0\text{ for both }j\ne k).
\]
The two residuals have variance $1-b^2>0$ and covariance $b-2b^2+b^2=b-b^2$, so their
correlation is $\rho=\frac{b-b^2}{1-b^2}=\frac b{1+b}$. The function
$b\mapsto\frac b{1+b}=1-\frac1{1+b}$ is increasing on $(-1,\infty)$ and equals
$\frac12$ at $b=1$, so $\rho<\frac12$ and $\arccos\rho>\frac\pi3$. By
\cref{lem:sheppard}, applied to the residuals divided by their common standard
deviation (which does not change their signs),
$\Pr(R_j>0\text{ for both }j\ne k)<\frac12-\frac16=\frac13$.
Since $Y^*_k$ and $Y^*_0$ are both standard normal,
$\Pr(A_k)\le\frac13\Pr(0<Y^*_0\le t)$, and summing over $k$ gives the claim.
\end{proof}

We now compare an arbitrary star with the ideal one. The case $d=3$ of the simplex
theorem (\cref{lem:simplex}) states that every centered Gaussian vector
$(\eta_1,\eta_2,\eta_3)$ with unit variances, possibly with a singular correlation
matrix, satisfies $\Pr(\eta_i>\theta\ \forall i)\ge S_3(\theta)$ for every
$\theta\in\R$; equivalently, $\Pr(\min_i\eta_i\le\theta)\le\Pr(\min_iU^*_i\le\theta)$.
The proof below uses it for a vector with a repeated coordinate.

\begin{proposition}[star bound at nonnegative thresholds]\label{prop:star-cubic}
Let $r\in(0,1)$, let $0\le d\le3$, and let $(Y_0,Y_1,\dots,Y_d)$ be a centered
Gaussian vector with unit variances and $\Corr(Y_0,Y_i)\ge r$ for $1\le i\le d$.
Then for every $t\ge0$,
\[
  \Pr\bigl(Y_0\ge t,\ \exists i\in\{1,\dots,d\}:\ Y_i\le t\bigr)\le f_{3,r}(t)\le\kappa(r).
\]
\end{proposition}

\begin{proof}
The second inequality is \cref{lem:ideal-monotone,lem:kappa}. For the first, we
lower the incident correlations to $r$ by a pointwise coupling, condition on $Y_0$,
and apply the simplex theorem to the residuals. Put $a_i=\Corr(Y_0,Y_i)\ge r$. If $a_i=1$ then $Y_i=Y_0$ with probability $1$, and $\{Y_0\ge t\ge Y_i\}$ is contained in the
null event $\{Y_0=t\}$; we discard such indices. If no index remains (in particular
if $d=0$), the probability is $0\le f_{3,r}(t)$. Otherwise, for each remaining $i$ put
$\eta_i=(Y_i-a_iY_0)/s_{a_i}$. The variables $\eta_i$ are standard normal, jointly
Gaussian with $Y_0$, and uncorrelated with $Y_0$, so the vector of the $\eta_i$ is
independent of $Y_0$. If fewer than three indices remain, repeat one of the $\eta_i$
until there are three coordinates $(\eta_1,\eta_2,\eta_3)$; repetition does not change
the event $\{\exists i:\eta_i\le c\}$ for any $c$.

On the event $\{Y_0\ge t\}$ consider $g(a)=(t-aY_0)/\sqrt{1-a^2}$ for $a\in[r,1)$. Its
derivative is
\[
  g'(a)=\frac{-Y_0(1-a^2)+a(t-aY_0)}{(1-a^2)^{3/2}}=\frac{at-Y_0}{(1-a^2)^{3/2}}\le0,
\]
because $at\le t\le Y_0$ (here $t\ge0$ is used). Hence $g(a_i)\le g(r)$, and
\[
  Y_i\le t\iff\eta_i\le g(a_i)\ \Longrightarrow\ \eta_i\le g(r)=\frac{t-rY_0}{s_r}.
\]
By the independence of $\eta$ and $Y_0$ (Fubini) and \cref{lem:simplex} with $d=3$,
\begin{align*}
  \Pr(Y_0\ge t,\ \exists i:Y_i\le t)
  &\le\int_t^\infty\varphi(z)\Pr\Bigl(\min_i\eta_i\le\frac{t-rz}{s_r}\Bigr)dz\\
  &\le\int_t^\infty\varphi(z)\Bigl(1-S_3\Bigl(\frac{t-rz}{s_r}\Bigr)\Bigr)dz
  =f_{3,r}(t),
\end{align*}
the last equality by \eqref{eq:ideal-star-integral}.
\end{proof}

For $t<0$ the monotonicity of $g$ fails on part of the event $\{Y_0\ge t\}$. This is
why the plain sweep needs the general star comparison of \cref{sec:stars}. The
median-switched layout uses outer frontiers only at $t>0$ and inner frontiers only at
$t<0$, so it needs only the pointwise coupling above.

\subsection{Level zero: the switch lemma}\label{sec:cubic-switch}

This subsection proves the switch lemma: if all incident correlations are at least
$r$, each vertex contributes, in expectation, at most $\kappa(r)$ to the quantity
$L_0$ of~\eqref{eq:cubic-L0}, which controls the connector bags at threshold zero.
Applied to the negated labels, the lemma gives the same bound for $L_0'$. This is the
same bound $\kappa(r)$ as at nonzero thresholds (\cref{prop:star-cubic}). For a vertex
$v$ of degree $d$ with neighbors
$x_1,\dots,x_d$, the events $v\in U$ and $v\in D\cap W$ are
$\{Y_v\le0<\max_iY_{x_i}\}$ and $\{d=3,\ Y_v>0\ge\max_iY_{x_i}\}$, so the following
lemma bounds the contribution of $v$ to $\E L_0$.

\begin{lemma}[switch lemma]\label{lem:switch}
Let $r\in(0,1)$, let $0\le d\le3$, and let $(Y_0,Y_1,\dots,Y_d)$ be a centered
Gaussian vector with unit variances and $\Corr(Y_0,Y_i)\ge r$ for $1\le i\le d$. Then
\[
  \Pr\bigl(Y_0\le0,\ \exists i:\ Y_i>0\bigr)
  +\ind_{d=3}\Pr\bigl(Y_0>0,\ Y_i\le0\text{ for all }i\bigr)\le\kappa(r).
\]
\end{lemma}

\paragraph{Idea of the proof.} By sign symmetry the second term equals
$\Pr(Y_0\le0<\min_iY_i)$, so the left side is
$\E\bigl[\ind\{Y_0\le0\}\lceil N/2\rceil\bigr]$, where $N$ is the number of neighbors
with positive label. After conditioning on $Y_0=z<0$ and lowering the incident
correlations to $r$ as in \cref{prop:star-cubic}, it remains to show that, among all
correlation matrices of three standard normal residuals, the probability that at
least two of them exceed a level $h>0$ is smallest for the equilateral matrix of
$U^*$ (\cref{lem:cubic-exceed}). We prove this by a deformation. The smallest pairwise
angles of the residual configuration are raised simultaneously (``water-filling'')
until the angles sum to $2\pi$. Plackett's identity shows that the probability does
not increase along the way. At the end the configuration is planar, and an exact
formula and convexity finish the argument. The simplex theorem cannot be used here:
the event ``at least two exceed $h$'' is not an orthant event, and the comonotone and
countermonotone extremes pull in opposite directions.

We first record the tools: which angle triples are realizable, and Plackett's
identity in the form needed here; the planar formula is part of the proof of
\cref{lem:cubic-exceed}.

\begin{lemma}[angles of three unit vectors]\label{lem:cubic-gram}
Let $\alpha,\beta,\gamma\in[0,\pi]$ and let $C$ be the symmetric $3\times3$ matrix
with unit diagonal and off-diagonal entries $C_{23}=\cos\alpha$, $C_{13}=\cos\beta$,
$C_{12}=\cos\gamma$. Then
\begin{equation}\label{eq:cubic-det}
  \det C=\bigl(\cos\alpha-\cos(\beta+\gamma)\bigr)\bigl(\cos(\beta-\gamma)-\cos\alpha\bigr).
\end{equation}
Moreover, $C$ is positive semidefinite if and only if $\alpha\le\beta+\gamma$,
$\beta\le\alpha+\gamma$, $\gamma\le\alpha+\beta$ and $\alpha+\beta+\gamma\le2\pi$, and
positive definite if and only if these four inequalities are strict.
\end{lemma}

\begin{proof}
The identity is a direct expansion. That the four inequalities suffice follows from
the signs of the two factors in~\eqref{eq:cubic-det}, and that they are necessary
from the triangle inequality for angles between unit vectors.

Expanding, $\det C=1-\cos^2\alpha-\cos^2\beta-\cos^2\gamma+2\cos\alpha\cos\beta\cos\gamma$.
For the right side of \eqref{eq:cubic-det}, use
$\cos(\beta+\gamma)+\cos(\beta-\gamma)=2\cos\beta\cos\gamma$ and
$\cos(\beta+\gamma)\cos(\beta-\gamma)=\cos^2\beta\cos^2\gamma-\sin^2\beta\sin^2\gamma
=\cos^2\beta+\cos^2\gamma-1$; then the product equals
$\cos\alpha\cdot2\cos\beta\cos\gamma-\cos^2\alpha-(\cos^2\beta+\cos^2\gamma-1)$, which
is the expansion.

Assume the four inequalities and put
$m=\min(\beta+\gamma,2\pi-\beta-\gamma)\in[0,\pi]$, so that
$\cos(\beta+\gamma)=\cos m$ and $|\beta-\gamma|\le\alpha\le m$. Since $\cos$ is
strictly decreasing on $[0,\pi]$, the first factor $\cos\alpha-\cos m$ of
\eqref{eq:cubic-det} is nonnegative, and positive exactly when $\alpha<m$, that is,
when $\alpha<\beta+\gamma$ and $\alpha+\beta+\gamma<2\pi$; the second factor
$\cos|\beta-\gamma|-\cos\alpha$ is nonnegative, and positive exactly when
$|\beta-\gamma|<\alpha$, that is, when $\beta<\alpha+\gamma$ and $\gamma<\alpha+\beta$.
So $\det C\ge0$, with equality exactly when one of the four inequalities is an
equality. The $2\times2$ principal minors are $\sin^2\ge0$ and the diagonal entries
are $1$, so all principal minors are nonnegative and $C$ is positive semidefinite.

Conversely, let $C$ be positive semidefinite. Then $C$ is the Gram matrix of unit
vectors $v_1,v_2,v_3$ (for instance the columns of $C^{1/2}$), and
$\alpha,\beta,\gamma$ are the angles between them. The angle between unit vectors is
a metric on the sphere, which gives the three triangle inequalities, and applying the
triangle inequality to $v_1$, $-v_3$, $v_2$ gives
$\gamma\le(\pi-\beta)+(\pi-\alpha)$, the fourth inequality.

Finally, a positive semidefinite matrix is positive definite exactly when its
determinant is positive. By the two preceding paragraphs, $C$ is therefore positive
definite exactly when the four inequalities hold and $\det C>0$, that is, when they
hold strictly.
\end{proof}

The second tool is Plackett's identity (\cref{lem:plackett}), with
$\varphi_2(\cdot,\cdot;c)$ the density of a pair of standard normals with correlation
$c\in(-1,1)$. For positive definite $3\times3$ correlation matrices $C$, a level
$h\in\R$ and $\eta\sim N(0,C)$, put $P_{ij}(C)=\Pr_C(\eta_i>h,\eta_j>h)$ and
$P_{123}(C)=\Pr_C(\eta_1,\eta_2,\eta_3>h)$, as functions of the three off-diagonal
entries. Let $\{i,j,k\}=\{1,2,3\}$. The pair probabilities $P_{ik}$ and $P_{jk}$ depend
only on the distributions of $(\eta_i,\eta_k)$ and $(\eta_j,\eta_k)$, hence not on $C_{ij}$.
\Cref{lem:plackett}, applied to the two coordinates $(\eta_i,\eta_j)$ and to all three
coordinates, gives
\begin{equation}\label{eq:cubic-plackett}
  \frac{\partial P_{ij}}{\partial C_{ij}}=\varphi_2(h,h;C_{ij}),\qquad
  \frac{\partial P_{123}}{\partial C_{ij}}
  =\varphi_2(h,h;C_{ij})\Pr_C\bigl(\eta_k>h\bigm|\eta_i=\eta_j=h\bigr),
\end{equation}
and both derivatives are continuous in $C$. By~\eqref{eq:plackett-k3}, the
conditional distribution of $\eta_k$ given $\eta_i=\eta_j=h$ is normal with mean
$h\,\frac{C_{ik}+C_{jk}}{1+C_{ij}}$ and positive variance.

For a correlation matrix $C$ (positive semidefinite, unit diagonal) and $h>0$ let
\[
  \Pi_C(h)=\Pr_C\bigl(\text{at least two of }\eta_1,\eta_2,\eta_3\text{ exceed }h\bigr),
  \qquad\eta\sim N(0,C),
\]
and let $C_{\mathrm{eq}}$ be the correlation matrix of $U^*$ (all off-diagonal
entries $-\frac12$). The map $C\mapsto\Pi_C(h)$ is continuous on the set of all
correlation matrices: if $C_n\to C$, then $N(0,C_n)\to N(0,C)$ weakly (by
characteristic functions), and the boundary of the event is contained in the union
of the hyperplanes $\{x_i=h\}$, which are null under $N(0,C)$ because every coordinate
is standard normal.

\begin{lemma}[two exceedances]\label{lem:cubic-exceed}
For every $3\times3$ correlation matrix $C$ and every $h>0$,
$\Pi_C(h)\ge\Pi_{C_{\mathrm{eq}}}(h)$.
\end{lemma}

\begin{proof}
We reduce to positive definite $C$, deform $C$ by raising its smallest angles until
the angles sum to $2\pi$, show that $\Pi$ does not increase along this path, and
compare the planar endpoint with $C_{\mathrm{eq}}$.

\emph{Reduction to positive definite $C$.} For $\xi\in(0,1]$, the matrix
$(1-\xi)C+\xi I$ is a positive definite correlation matrix, and it tends to $C$ as
$\xi\downarrow0$. By the continuity of $\Pi$, it suffices to prove the claim for
positive definite $C$, which we now assume.

\emph{The water-filling path.} Write $C_{ij}=\cos\theta_{ij}$ with
$\theta_{ij}\in[0,\pi]$. By \cref{lem:cubic-gram} and positive definiteness, the three
triangle inequalities and the inequality $\theta_{12}+\theta_{13}+\theta_{23}\le2\pi$
hold strictly. For $s\ge0$ let
$\theta_{ij}(s)=\max(\theta_{ij},s)$ and $\Sigma(s)=\sum_{i<j}\theta_{ij}(s)$. The
function $\Sigma$ is continuous and nondecreasing, $\Sigma(0)<2\pi$ and
$\Sigma(\frac{2\pi}3)\ge2\pi$; let $s^*\in(0,\frac{2\pi}3]$ be the smallest $s$ with
$\Sigma(s)=2\pi$, and let $C(s)$ be the matrix with off-diagonal entries
$\cos\theta_{ij}(s)$. On $[0,s^*]$ all angles lie in $[0,\pi]$.

For $0\le s<s^*$, the matrix $C(s)$ is positive definite. Indeed $\Sigma(s)<2\pi$.
For a triangle inequality $\theta_{ij}(s)<\theta_{ik}(s)+\theta_{jk}(s)$: if
$\theta_{ij}(s)=\theta_{ij}$, it follows from
$\theta_{ij}<\theta_{ik}+\theta_{jk}\le\theta_{ik}(s)+\theta_{jk}(s)$; otherwise
$\theta_{ij}(s)=s>0$, and the right side is at least $2s>s$. \Cref{lem:cubic-gram}
gives positive definiteness. Thus $C(s)$ moves within the positive definite matrices
for $s<s^*$, and reaches a matrix $C(s^*)$ with angle sum exactly $2\pi$, which is
positive semidefinite by \cref{lem:cubic-gram} (the non-strict inequalities persist
by continuity).

\emph{Monotonicity along the path.} The breakpoints $\{\theta_{12},\theta_{13},
\theta_{23}\}$ divide $[0,s^*]$ into finitely many intervals. On the interior of each,
the set of \emph{moving} pairs, those with $\theta_{ij}<s$, is fixed; the moving
pairs have $C_{ij}(s)=\cos s$, and the other entries are constant. By inclusion and
exclusion,
\[
  \Pi_C(h)=P_{12}(C)+P_{13}(C)+P_{23}(C)-2P_{123}(C),
\]
because the indicator of ``at least two'' equals the number of exceeding pairs minus
twice the indicator of ``all three'' (check $N=2$: $1-0$; $N=3$: $3-2$). By
\eqref{eq:cubic-plackett}, for $s<s^*$ and each pair $\{i,j\}$ with third index $k$,
\[
  \frac{\partial\Pi_{C(s)}(h)}{\partial C_{ij}}
  =\varphi_2(h,h;C_{ij}(s))\Bigl(1-2\Pr_{C(s)}\bigl(\eta_k>h\bigm|\eta_i=\eta_j=h\bigr)\Bigr).
\]
For a moving pair, $\theta_{ij}(s)=s\le\theta_{ik}(s),\theta_{jk}(s)$, so
$C_{ik}(s),C_{jk}(s)\le C_{ij}(s)$, and $C_{ij}(s)=\cos s\ge-\frac12$. Since $h>0$, the
conditional mean satisfies
\[
  h\,\frac{C_{ik}(s)+C_{jk}(s)}{1+C_{ij}(s)}\le h\,\frac{2C_{ij}(s)}{1+C_{ij}(s)}\le h,
\]
the last step because $2c\le1+c$ for $c\le1$ and $1+c>0$. A normal distribution with mean at most
$h$ gives probability at most $\frac12$ to $(h,\infty)$, so the partial derivative is
nonnegative for every moving pair. By the chain rule, on the interior of each interval
\[
  \frac{d}{ds}\Pi_{C(s)}(h)=\sum_{\text{moving }\{i,j\}}
  \frac{\partial\Pi_{C(s)}(h)}{\partial C_{ij}}\cdot(-\sin s)\le0 ,
\]
since $\sin s\ge0$ for $s\in[0,\pi]$. The function $s\mapsto\Pi_{C(s)}(h)$ is
continuous on $[0,s^*]$ (by the continuity of $\Pi$), so by the mean value theorem on
each interval it is nonincreasing on $[0,s^*]$, and
$\Pi_C(h)\ge\Pi_{C(s^*)}(h)$.

\emph{The planar endpoint.} Let $v_1,v_2,v_3$ be unit vectors with Gram matrix
$C(s^*)$, and let $\theta_{ij}=\theta_{ij}(s^*)$ from now on. By \eqref{eq:cubic-det}
with angle sum $2\pi$, $\cos(\beta+\gamma)=\cos(2\pi-\alpha)=\cos\alpha$, so
$\det C(s^*)=0$ and the vectors lie in a two-dimensional subspace; identify it with
$\R^2$. If $\Gamma$ is a standard Gaussian vector in $\R^2$, then
$(\langle v_i,\Gamma\rangle)_i\sim N(0,C(s^*))$. Order the directions of the $v_i$
around the circle and let $g_1,g_2,g_3\ge0$ be the three circular gaps between
consecutive directions, $g_1+g_2+g_3=2\pi$, labeled so that $g_k$ is the gap between
the directions of $v_i$ and $v_j$ that does not contain the direction of $v_k$. The
angle between $v_i$ and $v_j$ is $\theta_{ij}=\min(g_k,2\pi-g_k)\le g_k$; since the
angles sum to $2\pi=\sum_kg_k$, equality holds throughout, so $g_k=\theta_{ij}\le\pi$
for every $k$.

Write $\Gamma=\varrho(\cos\omega,\sin\omega)$ with $\varrho\ge0$ and $\omega$ uniform on
the circle, independent of $\varrho$. Given $\varrho$, the projection on $v_i$ exceeds
$h$ if and only if $\varrho>h$ and $\omega$ lies within angular distance
$\lambda=\arccos(h/\varrho)\in(0,\frac\pi2)$ of the direction of $v_i$. An arc of
half-width $\lambda$ around each direction gives three arcs of length
$2\lambda<\pi$. No point lies in all three arcs: otherwise the three directions would
lie in an arc of length $2\lambda<\pi$, and the complementary gap would exceed $\pi$.
Two arcs around directions separated by the gap $g_k$ overlap in an arc of length
$(2\lambda-g_k)_+$ on that side, and not at all on the other side, whose length is
$2\pi-g_k\ge\pi>2\lambda$. Therefore
\[
  \Pi_{C(s^*)}(h)=\E\Bigl[\ind\{\varrho>h\}\,\frac1{2\pi}
  \sum_{k=1}^3\bigl(2\lambda(\varrho)-g_k\bigr)_+\Bigr].
\]
The function $g\mapsto(2\lambda-g)_+$ is convex, so by Jensen's inequality
$\sum_k(2\lambda-g_k)_+\ge3(2\lambda-\frac{2\pi}3)_+$, which is the value of the sum for
the equal gaps $\frac{2\pi}3$ of the configuration of $U^*$. The same formula applies to
$C_{\mathrm{eq}}$, so $\Pi_{C(s^*)}(h)\ge\Pi_{C_{\mathrm{eq}}}(h)$. Together with the
monotonicity, $\Pi_C(h)\ge\Pi_{C_{\mathrm{eq}}}(h)$.
\end{proof}

\begin{proof}[Proof of \cref{lem:switch}]
\emph{Reduction to $d=3$.} If $d<3$, the second term is absent. Add coordinates
$Y_j=rY_0+s_r\eta_j$ for $d<j\le3$, with fresh independent standard normal $\eta_j$.
The extended vector satisfies the hypotheses with $d=3$, its first event contains the
original one, and its second term is nonnegative; so it suffices to treat $d=3$.

\emph{The functional.} Let $d=3$ and $N=\#\{i:Y_i>0\}$. By sign symmetry,
$\Pr(Y_0>0,\ Y_i\le0\ \forall i)=\Pr(Y_0<0,\ Y_i\ge0\ \forall i)
=\Pr(Y_0\le0,\ N=3)$, the last step because the events differ by null events. Since
$\lceil N/2\rceil=\ind\{N\ge1\}+\ind\{N=3\}$ for $N\in\{0,1,2,3\}$, the left side of
the lemma equals
\begin{equation}\label{eq:switch-Psi}
  \Psi=\E\bigl[\ind\{Y_0\le0\}\lceil N/2\rceil\bigr].
\end{equation}
We shall also use $\lceil N/2\rceil=N-\ind\{N\ge2\}$, valid for the same values.

\emph{Coupling.} Put $a_i=\Corr(Y_0,Y_i)\ge r$ and $s_i=\sqrt{1-a_i^2}$. For $s_i>0$ let
$\eta_i=(Y_i-a_iY_0)/s_i$, and for $s_i=0$ let $\eta_i$ be an auxiliary standard
normal variable independent of everything else. Then $\eta=(\eta_1,\eta_2,\eta_3)$ is a
centered Gaussian vector with unit variances, independent of $Y_0$; let $C$ be its
correlation matrix. On $\{Y_0<0\}$ (the event $Y_0=0$ is null): if $s_i=0$, then
$Y_i=Y_0<0$ with probability $1$; if $s_i>0$, then $Y_i>0$ if and only if
$\eta_i>|Y_0|a_i/s_i$. The function $a\mapsto a/\sqrt{1-a^2}$ is increasing on $(0,1)$,
so $|Y_0|a_i/s_i\ge|Y_0|r/s_r$. Hence, with probability $1$ on $\{Y_0<0\}$,
\begin{equation}\label{eq:switch-coupling}
  N\le N_{h}:=\#\{i:\eta_i>h\},\qquad h=h(Y_0)=\frac{r|Y_0|}{s_r}>0 .
\end{equation}
Since $n\mapsto\lceil n/2\rceil$ is nondecreasing and $\lceil n/2\rceil=n-\ind\{n\ge2\}$,
independence and Fubini give
\begin{equation}\label{eq:switch-integral}
  \Psi\le\int_{-\infty}^0\varphi(z)\,\E\lceil N_{h(z)}/2\rceil\,dz
  =\int_{-\infty}^0\varphi(z)\Bigl(3\Phi\bigl(-h(z)\bigr)-\Pi_C\bigl(h(z)\bigr)\Bigr)dz,
  \qquad h(z)=\frac{r|z|}{s_r}.
\end{equation}

\emph{Comparison and the ideal value.} By \cref{lem:cubic-exceed},
$\Pi_C(h)\ge\Pi_{C_{\mathrm{eq}}}(h)$ for every $h>0$, so
\begin{equation}\label{eq:switch-compare}
  \Psi\le\int_{-\infty}^0\varphi(z)\Bigl(3\Phi\bigl(-h(z)\bigr)-\Pi_{C_{\mathrm{eq}}}\bigl(h(z)\bigr)\Bigr)dz .
\end{equation}
For the ideal star $(Y^*_0,\dots,Y^*_3)$ the coupling is an identity: its residuals
are $U^*$, its incident correlations are $r$, and $Y^*_i>0$ if and only if
$U^*_i>h(Y^*_0)$ on $\{Y^*_0<0\}$. So the right side equals $\Psi$ for the ideal star,
namely
\[
  \Pr(Y^*_0\le0<\max_iY^*_i)+\Pr(Y^*_0\le0<\min_iY^*_i).
\]
The second probability is $0$: by \eqref{eq:cubic-sum}, $\min_iY^*_i>0$ forces
$Y^*_0>0$. By sign symmetry of the ideal star and the absence of ties, the first
probability equals $\Pr(Y^*_0\ge0,\ \min_iY^*_i\le0)=f_{3,r}(0)$, which is $\kappa(r)$
by \cref{lem:kappa}. Hence $\Psi\le\kappa(r)$.
\end{proof}

\subsection{Concentration and the proof of \texorpdfstring{\cref{thm:cubic}}{the cubic bound}}
\label{sec:cubic-proof}

We can now bound, at each fixed threshold, the expected sizes of the inner frontiers
below zero and of the outer frontiers above zero, and the expected values of $L_0$
and $L_0'$ at zero. To control all thresholds at once we use a finite
grid of thresholds, the concentration of the finitely many grid counts
(\cref{lem:local-counts}), and the deterministic interpolation of \cref{lem:grid}
between grid points, both for inner and for outer frontiers.

\begin{proof}[Proof of \cref{thm:cubic}]
We prove the following uniform statement: for every $\eps>0$ there is $N(\eps)$ such
that every graph $G$ with $\Delta(G)\le3$ and $n\ge N(\eps)$ vertices satisfies
$\pw(G)\le(\cthree+\eps)n$. This implies \cref{thm:cubic}: let $\rho(n)$ be the
maximum of $\max(0,\pw(G)-\cthree n)$ over all graphs $G$ with $\Delta(G)\le3$ on $n$
vertices (a finite set up to isomorphism); then $\rho(n)\le\eps n$ for $n\ge N(\eps)$,
so $\rho(n)=o(n)$, and $\pw(G)\le\cthree n+\rho(n)$ for every such $G$.

For fixed $\eps$, we choose the radius $R$ and a grid of thresholds, bound the means
of finitely many counts, fix a realization in which every count is close to its mean,
and check the three conditions of \cref{lem:median-layout}.

Fix $\eps>0$. Since $\kappa$ is continuous on $(0,1)$ and $r_R\to r_*$ as $R\to\infty$,
choose an integer $R\ge2$ with $\kappa(r_R)\le\cthree+\frac\eps3$. Choose an integer
$K_0\ge1$ with $\varphi(0)/K_0\le\frac\eps3$, and use the grid
$\tau_j=-2+j/K_0$ for $0\le j\le K:=4K_0$. It contains $\tau_{2K_0}=0$ and
$\tau_K=2$. Both $R$ and $K_0$ depend only on $\eps$. Let $G$ have $\Delta(G)\le3$
and $n$ vertices, and let $Y$ be the labels at radius $R$.

\emph{Counts.} Consider the following $8K_0+4$ random variables, each a sum over
$v\in V(G)$ of indicator variables:
\begin{itemize}
  \item $|\infr S_{\tau_j}|$ for $0\le j<2K_0$ (thresholds $\tau_j<0$);
  \item $|\outfr S_{\tau_j}|$ for $2K_0<j\le4K_0$ (thresholds $\tau_j>0$);
  \item $L_0$ and $L_0'$ from \eqref{eq:cubic-L0}, with indicators
  $\ind\{v\in U\}+\ind\{v\in D\cap W\}$ and $\ind\{v\in W\}+\ind\{v\in D\cap U\}$
  (each at most $1$, as the two events are disjoint);
  \item the window counts $M_j:=\#\{v:\tau_j<Y_v\le\tau_{j+1}\}$ for $0\le j<4K_0$
  (the counts $N_j$ of \cref{lem:grid}); and
  \item $\#\{v:Y_v\le-2\}$ and $\#\{v:Y_v>2\}$.
\end{itemize}
The indicator for $v$ in each count is determined by the degree of $v$ and the labels
on $N[v]$, hence by \ref{it:cubic-L3} by $(Z_w)_{w\in B_{R+1}(v)}$.

\emph{Means.} Fix $v$ with neighbors $x_1,\dots,x_d$, $d\le3$. By \ref{it:cubic-L1},
$(Y_v,Y_{x_1},\dots,Y_{x_d})$ and its negative are centered Gaussian vectors with unit
variances and incident correlations at least $r_R$. The events $v\in\outfr S_\tau$
for $\tau>0$ and $v\in\infr S_\tau$ for $\tau<0$ are contained in the events of
\cref{prop:star-cubic} for this vector at $\tau$ and for its negative at $-\tau$. Up
to the null events $\{Y_x=0\}$ with $x\in N[v]$, the two events in the indicator of
$L_0$ are the two events of \cref{lem:switch} for this vector, and those of $L_0'$
are the same events for its negative. So each of these indicators has mean at most
$\kappa(r_R)$. Since moreover
$\Pr(\tau_j<Y_v\le\tau_{j+1})\le\varphi(0)/K_0$ and
$\Pr(Y_v\le-2)=\Pr(Y_v>2)=\Phi(-2)<0.023$, summing over $v$ shows that every inner or
outer grid count and both $L_0$ and $L_0'$ have mean at most $\kappa(r_R)n$, every
$M_j$ has mean at most $\varphi(0)n/K_0$, and both tail counts have mean at most
$0.023n$.

\emph{Concentration.} Choose $N(\eps)$ so large that, for $n\ge N(\eps)$,
\[
  (8K_0+4)\,9^{R+1}<2n^{1/2},\qquad 2n^{3/4}\le\tfrac\eps3n,\qquad
  n^{3/4}+1\le\tfrac\eps3n,\qquad 0.023n+n^{3/4}\le\cthree n;
\]
the last condition is possible because $0.023<\cthree$. All of these conditions depend
only on $\eps$. Let $n\ge N(\eps)$. The first condition is the hypothesis of
\cref{lem:local-counts} for the $8K_0+4$ counts, with $\Delta=3$, $\varrho=R+1$ and
$\lambda=n^{3/4}$, so some realization of $Z$ has pairwise distinct labels and every
count within $n^{3/4}$ of its mean. Fix it and let $y=Y$ for this realization.

\emph{Bounds at every threshold.} Let $w=(\cthree+\eps)n$. For $t\in[-2,0)$, let $j$
be such that $t\in[\tau_j,\tau_{j+1})$; then $j<2K_0$, so $\tau_j<0$, and
\cref{lem:grid}\textup{(i)} gives
\[
  |\infr S_t|\le|\infr S_{\tau_j}|+M_j\le\kappa(r_R)n+\frac{\varphi(0)}{K_0}n+2n^{3/4}
  \le\bigl(\cthree+\tfrac{2\eps}3\bigr)n+2n^{3/4}\le w .
\]
For $t\in(0,2)$, let $j$ be such that $t\in[\tau_j,\tau_{j+1})$; then $j\ge2K_0$, so
$\tau_{j+1}>0$, and the same computation with the outer count at $\tau_{j+1}$ gives
$|\outfr S_t|\le w$. For $t<-2$ and $t\ge2$, \cref{lem:grid}\textup{(ii)},\textup{(iii)}
give (inner, respectively outer) frontiers of
size at most $0.023n+n^{3/4}\le\cthree n\le w$. Finally
$\max(L_0,L_0')+1\le\kappa(r_R)n+n^{3/4}+1\le(\cthree+\frac{2\eps}3)n\le w$.

\emph{Conclusion.} The labeling $y$ is injective, and conditions (i)--(iii) of
\cref{lem:median-layout} hold with $w=(\cthree+\eps)n$. Hence
$\pw(G)\le(\cthree+\eps)n$ for every graph $G$ with $\Delta(G)\le3$ and $n\ge N(\eps)$.
\end{proof}

Every path decomposition is a tree decomposition, so the same bound holds for
treewidth. The proof is existential; \cref{sec:algorithms} turns it into an
algorithm.\label{pg:cubic-proof-end}

{}
\section{Pathwidth sequences and the bootstrap}\label{sec:sequences}\label{sec:bootstrap}

The two previous sections bound the pathwidth of graphs of maximum degree at most
$k\in\{3,4,5\}$ by $\sum_{d=3}^ka_dn_d+o(n)$, for several coefficient vectors
$(a_3,\dots,a_k)$, none of which is best for all graphs
(\cref{thm:deg4,thm:deg5,thm:cubic,thm:switched4}). This section states all
of them in one format, the pathwidth sequence, and extends each of them to all graphs
by the bootstrap lemma (\cref{lem:bootstrap}). In this format, a vertex of degree at
most $2$ adds nothing to the bound, a vertex of degree $d\ge3$ adds its coefficient
$a_d$, and a vertex whose degree is beyond the given coefficients adds $1$, which is
what putting it into every bag adds to the width. Applied to the triple
$\tripS=(\frac5{32},\frac14,\frac{11}{32})$ of \cref{thm:deg5}, the bootstrap lemma
gives the bound for all graphs (\cref{thm:main}) and the bound in terms of edges
(\cref{cor:edges}).

The sequences below use the constants
\[
  \cthree=\frac{3}{4\pi}\arccos\frac56\approx 0.1398221,\qquad
  F_3\approx 0.1401583,\qquad
  F_4\approx 0.2457720,
\]
where $F_3$ and $F_4$ are the maxima of the ideal $3$-star and $4$-star functions
with edge correlations $\frac{2\sqrt2}{3}$ and $\frac{\sqrt3}{2}$, defined in
\cref{sec:stars}; certified enclosures are given in \cref{prop:F3F4}.

\begin{definition}[pathwidth sequence]\label{def:pw-seq}
A sequence $a=(a_d)_{d\ge3}$ of reals is a \emph{pathwidth sequence} if there is a
function $\rho(n)=o(n)$ such that every graph $G$ satisfies
$\pw(G)\le\sum_{d\ge3}a_d\,n_d+\rho(n)$. A finite tuple $(a_3,\dots,a_k)$ stands for
the sequence that continues it with $a_d=1$ for all $d>k$.
\end{definition}

So a finite tuple $(a_3,\dots,a_k)$ is a pathwidth sequence if and only if there is a
function $\rho(n)=o(n)$ with $\pw(G)\le\sum_{d=3}^ka_dn_d+\sum_{d>k}n_d+\rho(n)$ for
every graph $G$.

Two consequences of the definition are used below. A sequence that is entrywise
at least a pathwidth sequence is again one. For a graph $G$ of maximum degree at most
$k$, a pathwidth sequence $(a_3,\dots,a_k)$ gives the bound
$\pw(G)\le\sum_{d=3}^ka_dn_d+o(n)$.

\begin{restatable}[pathwidth sequences]{theorem}{thmpwseq}\label{thm:pw-seq}
Each of the following six tuples is a pathwidth sequence:
\[
  \seqC=(\cthree),\qquad
  \seqM=(\cthree,\,0.2627),\qquad
  \seqF=(F_4,\,F_4),
\]
\[
  \tripS=\bigl(\tfrac{5}{32},\,\tfrac14,\,\tfrac{11}{32}\bigr),\qquad
  \tripP=(F_3,\,0.2575,\,0.3667),\qquad
  \tripR=(0.1455,\,0.2530,\,0.3600).
\]
\end{restatable}

For graphs of maximum degree at most $k$, the six bounds are those of
\cref{thm:cubic} ($\seqC$), \cref{thm:switched4} ($\seqM$),
\cref{thm:deg4}\ref{thm:deg4:a} ($\seqF$) and \cref{thm:deg5} ($\tripP$, $\tripR$,
$\tripS$), stated in \cref{sec:cubic,sec:plain-sweep}. In particular, every graph of
maximum degree $3$ satisfies $\pw(G)\le\cthree n_3+o(n)$, and every graph of maximum
degree $4$ satisfies $\pw(G)\le\cthree n_3+0.2627\,n_4+o(n)$ and
$\pw(G)\le F_4(n_3+n_4)+o(n)$.

The sequence $\seqM$ is entrywise below $\seqC$, so $\seqM$ implies $\seqC$. We list
$\seqC$ nevertheless: its proof uses no computation, while that of $\seqM$ rests on
computer-assisted certificates
(\cref{sec:switched4}). Apart from this pair, none of the six sequences is entrywise
below another, and each of $\seqM,\seqF,\tripP,\tripR,\tripS$ gives the smallest
bound of the six for some graphs (\cref{rem:best-row}).

\subsection{The bootstrap lemma}\label{sec:bootstrap-lemma}

Each of the six sequences has coefficients up to some degree $k\le5$ only, so every
vertex of degree above $k$ adds $1$ to its bound. The bootstrap lemma replaces these
coefficients $1$ by the entries of the following recurrence, which are at most $1$,
and smaller than $1$ in every row of \cref{tab:sequences} except $\seqF$.

Let $k\ge3$ and let $0\le b_3\le\dots\le b_k$ be reals, and put $b_0=b_1=b_2:=0$. For
every $d\ge3$ for which $b_d$ is defined, the \emph{increment} $h_d$ and the
\emph{running minimum} $H_d$ are
\[
  h_d:=b_d-b_{d-1},\qquad H_d:=\min_{3\le i\le d}h_i .
\]
Thus $h_3=b_3$, the sequence $H$ is nonincreasing, and $H_d\le h_d$ and
$H_d\le h_3=b_3$ for every $d\ge3$. The \emph{bootstrap} of $(b_3,\dots,b_k)$ is the
infinite sequence $(b_d)_{d\ge3}$ obtained by defining $b_d$ for $d=k+1,k+2,\dots$ in
turn by
\begin{equation}\label{eq:bootstrap-rec}
  b_d:=\max\Bigl\{\frac{1+(d-1)\,b_{d-1}-H_{d-1}}{d},\;1-d\,H_{d-1}\Bigr\}
  \qquad(d>k).
\end{equation}
This is well defined, since $H_{d-1}$ depends only on $b_3,\dots,b_{d-1}$. For a
graph $G$ we write
\[
  \psi_b(G):=\sum_{v\in V(G)}b_{d_G(v)}=\sum_{d\ge3}b_d\,n_d(G);
\]
vertices of degree at most $2$ contribute nothing. We drop the subscript $b$ when the
sequence is fixed.

\begin{lemma}[bootstrap]\label{lem:bootstrap}\label{lem:bootstrap-seq}
Let $k\ge3$ and let $0\le b_3\le\dots\le b_k$ be reals with $b_k+H_k\le1$. Suppose
that there is a function $\rho(N)=o(N)$ such that every graph $J$ with
$\delta(J)\ge3$ and $\Delta(J)\le k$ satisfies
$\pw(J)\le\sum_{d=3}^kb_d\,n_d(J)+\rho(|V(J)|)$. Then the bootstrap of
$(b_3,\dots,b_k)$ is a pathwidth sequence. It is nondecreasing, and all its entries
are at most $1$.
\end{lemma}

The graphs $J$ in \cref{lem:bootstrap} have no vertex of degree above $k$, so the
bound on them holds whenever $(b_3,\dots,b_k)$ is a pathwidth sequence. Conversely,
under the hypotheses of \cref{lem:bootstrap} the tuple $(b_3,\dots,b_k)$ is entrywise
at least its bootstrap, which agrees with it up to degree $k$ and has all entries at
most $1$; hence the tuple is a pathwidth sequence. This is how \cref{thm:pw-seq}
follows from \cref{thm:cubic,thm:switched4,thm:deg4,thm:deg5}
(\cref{sec:proof-pw-seq}). The bootstrap is used with $k=3$ for the seed $\seqC$,
with $k=4$ for $\seqM$ and $\seqF$, and with $k=5$ for $\tripP$, $\tripR$ and
$\tripS$.

\paragraph{Reading the recurrence.}
By \cref{lem:extension}\ref{it:ext:identity} below, $b_d$ is determined by
\begin{equation}\label{eq:bootstrap-identity}
  1-b_d=(d-1)\,H_d+H_{d-1}\qquad(d>k).
\end{equation}
Heuristically (\cref{sec:overview-bootstrap}), this identity says that deleting a
vertex of degree $d$ that has a neighbor of smaller degree lowers the potential
$\psi$ by at least $1$, while putting the vertex back adds at most $1$ to the width,
since $\pw(G)\le\pw(G-v)+1$. Indeed, the deletion lowers the potential by $b_d$ for
the vertex itself, by at least $H_{d-1}$ for that neighbor and by at least $H_d$ for
each of the other $d-1$ neighbors, whose degrees drop by one; by
\eqref{eq:bootstrap-identity}, these amounts add up to $1$. The two terms of the
maximum in \eqref{eq:bootstrap-rec} are the two cases of
\cref{lem:extension}\ref{it:ext:case}. Regular graphs, in which no neighbor has
smaller degree, are handled by the negative term of the potential $\psi^-$ in
\cref{sec:bootstrap-proof}.

\begin{remark}[the recurrence of Edwards and McDermid]\label{rem:bootstrap-em}
Take $k=3$ and $b_3=\alpha$ with $\frac16\le\alpha\le\frac3{10}$. Then
$H_3=\alpha$ and $b_3+H_3=2\alpha\le1$. By \eqref{eq:bootstrap-rec},
$b_4=\max\{\frac{1+2\alpha}4,\,1-4\alpha\}=\frac{1+2\alpha}4$, because
$1+2\alpha\ge4-16\alpha$ is equivalent to $\alpha\ge\frac16$. So
$b_4=g_\alpha(4)$ and $h_4=\frac{1-2\alpha}4\le\alpha$, whence $H_4=h_4$.
Moreover $H_3\le3H_4$ is equivalent to $4\alpha\le3-6\alpha$, that is,
$\alpha\le\frac3{10}$. Hence \cref{lem:concave} below applies with $j=4$, and its
last part gives $d\,b_d=(d-2)\,b_{d-1}+b_{d-2}+1$ for all $d\ge5$. So $b=g_\alpha$:
\cref{lem:bootstrap} with $k=3$ contains the recurrence of Edwards and McDermid on
exactly their range of $\alpha$. For $\alpha<\frac16$ the second term wins and
$b_4=1-4\alpha$; for $\alpha=\cthree$ this is about $0.4407$, worse than
$g_{1/6}(4)=\frac13$. This is why a smaller cubic coefficient alone does not improve
the coefficients of higher degrees, and why \cref{thm:main} needs the base of
maximum degree $5$ from \cref{thm:deg5}.
\end{remark}

\paragraph{Prior work, and what is new.}
The induction is that of Fomin, Gaspers, Saurabh and
Stepanov~\cite[Lemma~1]{FominGaspersSaurabhStepanov2009}, seeded with the cubic bound
$\frac n6+o(n)$ of Fomin and H{\o}ie~\cite{FominHoie2006} and extended by
Gaspers~\cite[Table~6.1]{GaspersBook2010}, in the form of the reduction theorem of
Edwards and McDermid~\cite[Theorem~3.2]{EdwardsM15}, whose cost function
$g_\alpha$ satisfies $d\,g_\alpha(d)=(d-2)\,g_\alpha(d-1)+g_\alpha(d-2)+1$ and needs
$\frac16\le\alpha\le\frac3{10}$; the proofs of \cref{lem:deletion-step,prop:deletion}
follow theirs step by step. Three things are new. The seeds $b_3\le\dots\le b_k$ are
arbitrary, with the increment of Edwards and McDermid replaced by the running minimum
$H_d$, so that the cubic coefficient $\cthree<\frac16$, which their theorem cannot use,
and the seeds of \cref{thm:deg5} are admissible (\cref{rem:bootstrap-em}). The base is
a per-degree bound for all graphs of minimum degree at least $3$ and maximum degree at
most $k$, not a bound for cubic graphs. And \cref{lem:reduction} replaces the two facts
on reductions taken from Kneis, M{\"o}lle, Richter and Rossmanith~\cite{KneisMRR2009}.

\paragraph{The bootstraps of the six sequences.}
All six sequences of \cref{thm:pw-seq} satisfy the hypotheses of
\cref{lem:bootstrap}. Their bootstraps, rounded up, form \cref{tab:sequences}; the
bootstrap of $\seqC$ is entrywise at least that of $\seqM$ and is omitted.

\begin{restatable}[the bootstraps]{corollary}{cortable}\label{cor:table}
Each row of \cref{tab:sequences} is a pathwidth sequence.
\end{restatable}

\begin{table}[tb]
\centering
\small
\setlength{\tabcolsep}{4.2pt}
\renewcommand{\arraystretch}{1.3}
\caption{The pathwidth sequences of \cref{cor:table}. The row named $X$ is the
bootstrap of the sequence $X$ of \cref{thm:pw-seq}, rounded up to four decimals; in
the last column, the bootstrap is bounded by a closed form. The bootstrap of $\seqC$,
$(0.1399,\,0.4408,\,0.5247,\,\dots)$, is entrywise at least row $\seqM$ and is
omitted.
\colorbox{loud}{\strut\textbf{Highlighted}}: the smallest entry of its column.
\textcolor{muted}{Grey}: produced by the bootstrap~\eqref{eq:bootstrap-rec}. Black
entries are those of \cref{thm:pw-seq}, with $\cthree$, $F_3$ and $F_4$ rounded up.}
\label{tab:sequences}
\begin{tabular}{@{}l*{9}{c}@{}}
\toprule
& $a_3$ & $a_4$ & $a_5$ & $a_6$ & $a_7$ & $a_8$ & $a_9$ & $a_{10}$ & $a_d$, $d\ge11$\\
\midrule
$\seqM$ & \lo{0.1399} & $0.2627$ & \bs{0.3857} & \bs{0.4676} & \bs{0.5319}
  & \bs{0.5824} & \bs{0.6232} & \bs{0.6568} & \bs{1-\frac{3.77}{d+1}}\\
$\seqF$ & $0.2458$ & \lo{0.2458} & \bs{1} & \bs{1} & \bs{1} & \bs{1} & \bs{1} & \bs{1}
  & \bs{1}\\
$\tripP$ & $0.1402$ & $0.2575$ & $0.3667$ & \bs{0.4541} & \bs{0.5196} & \bs{0.5715}
  & \bs{0.6133} & \bs{0.6478} & \bs{1-\frac{3.87}{d+1}}\\
$\tripR$ & $0.1455$ & $0.2530$ & $0.3600$ & \bs{0.4489} & \bs{0.5149} & \bs{0.5673}
  & \bs{0.6096} & \bs{0.6444} & \bs{1-\frac{3.91}{d+1}}\\
$\tripS$ & $\frac{5}{32}$ & $\frac14$ & \lo{\frac{11}{32}} & \lobs{0.4375}
  & \lobs{0.5045} & \lobs{0.5581} & \lobs{0.6012} & \lobs{0.6368}
  & \lobs{1-\frac{4}{d+1}}\\
\bottomrule
\end{tabular}
\end{table}

Among the rows of \cref{tab:sequences}, the bootstrap of $\tripS$ has the smallest
coefficient at every degree $d\ge5$. Rounded up to closed forms, it gives the bound
for all graphs of \cref{thm:main}, and through it the bound in terms of edges of
\cref{cor:edges} (\cref{sec:proof-main,sec:proof-edges}). Which row of
\cref{tab:sequences} is best for a given graph depends only on the proportions of its
degrees (\cref{rem:best-row}, \cref{fig:best-row}).

The decompositions behind the bounds of \cref{thm:pw-seq,cor:table} can be computed
in polynomial time (\cref{thm:algo}). For the bootstrap of a sequence $X$ of
\cref{thm:pw-seq} with entries up to degree $k$, the algorithm runs the deletion
process of the bootstrap lemma and decomposes the remaining components, of maximum
degree at most $k$, by a sweep: the plain sweep for $\seqF$, $\tripP$, $\tripR$ and
$\tripS$, and the median-switched one for $\seqC$ and $\seqM$
(\cref{sec:alg-general,app:alg-switched}).

\paragraph{The proof of the bootstrap lemma.}
The proof involves no probability. A combinatorial induction deletes vertices of
large degree one at a time; putting a deleted vertex back into every bag adds at most
one to the width. The number of deletions is at most the potential $\psi(G)$ of the
input graph $G$: each deletion, together with the reductions that follow it, lowers a
modified potential by at least one. Vertices of degree at most $2$ are removed by the
reduction operations of \cref{def:reductions}, and \cref{lem:reduction} bounds the
effect of all reductions on the pathwidth at once, by an additive term $O(\log n)$.
\Cref{sec:bootstrap-sequence} proves the properties of the bootstrap that the
induction needs, and \cref{sec:bootstrap-proof} runs the deletion process and proves
\cref{lem:bootstrap}. \Cref{sec:proof-main,sec:proof-edges} derive
\cref{thm:main,cor:edges}. \Cref{app:sequences} derives \cref{thm:pw-seq,cor:table}
from the lemma and compares the coefficients with the previous ones, and
\cref{sec:alg-general} makes the deletion process algorithmic.
{}

\subsection{The extended sequence}\label{sec:bootstrap-sequence}

We first show that the recurrence keeps the sequence nondecreasing and bounded by $1$,
and that it satisfies \eqref{eq:bootstrap-identity}; this is all that the deletion
process needs. We also record which term of the maximum is attained, because
\cref{lem:concave} and the proof of \cref{prop:large-degrees} use it. For $d>k$ put
\[
  \theta_d:=\frac{1-b_{d-1}-H_{d-1}}{d},
\]
the increment $h_d$ that the first term of the maximum in \eqref{eq:bootstrap-rec}
would give.

\begin{lemma}[the extended sequence]\label{lem:extension}
Let $k\ge3$, let $0\le b_3\le\dots\le b_k$ be reals with $b_k+H_k\le1$, and extend
them by~\eqref{eq:bootstrap-rec}. Then $h_d\ge0$ and $H_d\ge0$ for every $d\ge3$,
and for every $d>k$:
\begin{enumerate}[label=\textup{(\roman*)}]
  \item\label{it:ext:mono} $b_{d-1}\le b_d\le1$;
  \item\label{it:ext:identity} $1-b_d=(d-1)\,H_d+H_{d-1}$;
  \item\label{it:ext:budget} $b_d+H_d\le1$;
  \item\label{it:ext:case} if $\theta_d\le H_{d-1}$, then $b_d$ is the first term of
  the maximum in \eqref{eq:bootstrap-rec} and $h_d=H_d=\theta_d$; otherwise $b_d$ is
  the second term and $H_d=H_{d-1}<h_d$.
\end{enumerate}
\end{lemma}

\begin{proof}
We first treat the seeds, and then argue by induction on $d>k$, comparing both terms
of the maximum in \eqref{eq:bootstrap-rec} with $b_{d-1}$.

For $3\le d\le k$ we have $h_d=b_d-b_{d-1}\ge0$, because the seeds are nondecreasing
and $b_2=0$; hence also $H_d\ge0$.

Now let $d>k$. The induction hypothesis is that $H_{d-1}\ge0$ and
$b_{d-1}+H_{d-1}\le1$. For $d=k+1$ this holds by the first part and by the
assumption $b_k+H_k\le1$. Put $\mu:=H_{d-1}\ge0$ and $\theta:=\theta_d$, so that
$\theta=(1-b_{d-1}-\mu)/d\ge0$ by the induction hypothesis. Let $p$ and $q$ be the
first and second terms of the maximum in \eqref{eq:bootstrap-rec}. Then
\[
  p-b_{d-1}=\frac{1-b_{d-1}-\mu}{d}=\theta,
\]
and since $1-b_{d-1}=d\theta+\mu$,
\[
  q-b_{d-1}=1-b_{d-1}-d\mu=d\theta-(d-1)\mu=\theta+(d-1)(\theta-\mu).
\]
\emph{Case $\theta\le\mu$.} Then $q\le p$, so $b_d=p$ and $h_d=\theta$. Hence
$H_d=\min\{\mu,\theta\}=\theta$, and
$1-b_d=(1-b_{d-1})-\theta=(d-1)\theta+\mu=(d-1)H_d+H_{d-1}$.

\emph{Case $\theta>\mu$.} Then $q>p$, so $b_d=q$ and
$h_d=d\theta-(d-1)\mu>d\mu-(d-1)\mu=\mu$. Hence $H_d=\mu<h_d$, and
$1-b_d=d\mu=(d-1)H_d+H_{d-1}$.

This proves \ref{it:ext:identity} and \ref{it:ext:case}. In both cases $h_d\ge0$
(as $\theta\ge0$ and $\mu\ge0$), so $b_d\ge b_{d-1}$ and $H_d\ge0$. By
\ref{it:ext:identity} and $d\ge2$,
$1-b_d=(d-1)H_d+H_{d-1}\ge(d-1)H_d\ge H_d\ge0$, which gives $b_d\le1$ and
\ref{it:ext:budget}. In particular the induction hypothesis holds for $d+1$.
\end{proof}

Once two consecutive running minima are within a factor $3$ of each other, the first
term of the maximum is attained at every later degree, and the running minima follow
a linear recurrence. If, in addition, the increment $h_j$ at the later degree $j$ equals $H_j$, then from there on the extended sequence satisfies the recurrence of Edwards
and McDermid. This is how the coefficients of \cref{thm:main} are computed.

\begin{lemma}[the concave regime]\label{lem:concave}
Let $b$ be as in \cref{lem:extension}, and suppose that $H_j\le H_{j-1}\le3H_j$ for
some $j>k$. Then for every $d>j$,
\[
  d\,H_d=(d-3)\,H_{d-1}+H_{d-2},\qquad h_d=H_d,\qquad H_d\le H_{d-1}\le3H_d .
\]
If moreover $h_j=H_j$, then $d\,b_d=(d-2)\,b_{d-1}+b_{d-2}+1$ for every $d>j$.
\end{lemma}

\begin{proof}
We prove the three statements by induction on $d>j$, together with the hypothesis
$H_{d-1}\le H_{d-2}\le3H_{d-1}$, which holds for $d=j+1$ by assumption. Since
$d-1\ge j>k$, \cref{lem:extension}\ref{it:ext:identity} at $d-1$ gives
$1-b_{d-1}=(d-2)H_{d-1}+H_{d-2}$, and therefore
\[
  \theta_d=\frac{1-b_{d-1}-H_{d-1}}{d}=\frac{(d-3)\,H_{d-1}+H_{d-2}}{d}.
\]
The inequality $\theta_d\le H_{d-1}$ is equivalent to $H_{d-2}\le3H_{d-1}$, which
holds. By \cref{lem:extension}\ref{it:ext:case}, $h_d=H_d=\theta_d$; this is the
claimed recurrence for $H_d$, and $H_d\le H_{d-1}$ because $H$ is nonincreasing.
Finally, $H_{d-1}\le3H_d=3\theta_d$ is equivalent to
$d\,H_{d-1}\le3(d-3)H_{d-1}+3H_{d-2}$, that is, to
$(9-2d)\,H_{d-1}\le3H_{d-2}$. This holds because $d\ge j+1\ge k+2\ge5$ makes the
left side nonpositive, while $H_{d-2}\ge0$ by \cref{lem:extension}. So the hypothesis
holds for $d+1$.

For the last statement, let $d>j$. Then $h_d=H_d$ by the first part, and
$h_{d-1}=H_{d-1}$, by assumption if $d-1=j$ and by the first part if $d-1>j$.
Substituting into \cref{lem:extension}\ref{it:ext:identity} gives
$1-b_d=(d-1)(b_d-b_{d-1})+(b_{d-1}-b_{d-2})$, which rearranges to
$d\,b_d=(d-2)\,b_{d-1}+b_{d-2}+1$.
\end{proof}

\subsection{The deletion process and the proof of the bootstrap lemma}
\label{sec:bootstrap-proof}

Throughout this subsection, $b$ is as in \cref{lem:extension}, and
$\rho\colon\N\to[0,\infty)$ is a nondecreasing function such that every graph $J$
with $\delta(J)\ge3$ and $\Delta(J)\le k$ satisfies
\begin{equation}\label{eq:bootstrap-base}
  \pw(J)\le\sum_{d=3}^{k}b_d\,n_d(J)+\rho\bigl(|V(J)|\bigr).
\end{equation}
By \cref{lem:extension}, $b$ is nondecreasing, $0\le b_d\le1$, and
\eqref{eq:bootstrap-identity} holds for all $d>k$. We show that every graph $G$ with
$n\ge1$ vertices satisfies
\begin{equation}\label{eq:bootstrap-explicit}
  \pw(G)\le\sum_{d\ge3}b_d\,n_d+b_3+\rho(n)+6\log_2n+2,
\end{equation}
and deduce \cref{lem:bootstrap}.

The proof of \eqref{eq:bootstrap-explicit} combines three bounds. The deletion
process (\cref{prop:deletion}) deletes vertices of large degree and applies
reductions until every component is basic (\cref{def:basic}); the number of deleted
vertices is at most the decrease of a modified potential $\psi^-$. \Cref{lem:basic}
bounds the pathwidth of basic graphs, and \cref{lem:reduction} bounds the effect of
the reductions. The three bounds are added up after \cref{lem:basic}.

\paragraph{Facts about reductions.}
We use the properties \ref{red:deg}--\ref{red:max} of the reduction operations,
stated after \cref{def:reductions}. Since $b$ is nondecreasing and nonnegative,
\ref{red:deg} implies that an operation does not increase $\psi$.
The effect of the reductions on the pathwidth is bounded by \cref{lem:reduction},
applied to sequences of reduction operations and vertex deletions; recall that the
empty graph has pathwidth $-1$ (\cref{sec:prelim-graphs}).

\paragraph{The potential.}
For a graph $G$ with $\delta(G)\ge3$ and components $G_1,\dots,G_t$ put
\[
  \psi^-(G):=\psi(G)-\sum_{i=1}^{t}H_{\delta(G_i)},
\]
and $\psi^-(G):=0$ for the empty graph. Each component $G_i$ contains a vertex of
degree $\delta(G_i)\ge3$, whose weight is $b_{\delta(G_i)}\ge h_{\delta(G_i)}\ge
H_{\delta(G_i)}$, because $b_{\delta(G_i)-1}\ge0$. Hence $\psi^-(G)\ge0$, and also
$\psi^-(G_i)\ge0$ for each component, and $\psi^-$ is additive over components.
The negative term is the one of Edwards and McDermid \cite{EdwardsM15}. It is
needed when the deleted vertex lies in a regular component, where no neighbor has
smaller degree (Case~1 in the proof of \cref{lem:deletion-step}).

\paragraph{Basic graphs.}
The process stops at components of the following kind.

\begin{definition}[basic graphs]\label{def:basic}
A connected graph $J$ with $\delta(J)\ge3$ is \emph{basic} if $\Delta(J)\le k$, or if
there is a vertex $v\in V(J)$ such that some maximal sequence of reduction operations
applied to $J-v$ ends with the empty graph.
\end{definition}

The deletion process of \cref{prop:deletion} below repeats one step: delete a vertex of maximum degree, then reduce.
The next lemma shows that this step lowers $\psi^-$ by at least one. The algorithm of
\cref{sec:alg-general} uses it directly, without testing whether a component is basic.

\begin{lemma}[one deletion]\label{lem:deletion-step}
Let $G$ be a connected graph with $\delta(G)\ge3$ and $\Delta(G)\ge k+1$. Let $w$ be a
vertex of degree $\Delta(G)$ that has a neighbor of smaller degree, if such a vertex
exists, and any vertex of degree $\Delta(G)$ otherwise. Let $G^*$ be the final graph
of a maximal sequence of reduction operations applied to $G-w$. If $G^*$ is
nonempty, then $1+\psi^-(G^*)\le\psi^-(G)$.
\end{lemma}

\begin{proof}
Write $\Delta:=\Delta(G)$, so $\Delta\ge k+1\ge4$. Let $v_1,\dots,v_\Delta$ be the
neighbors of $w$, ordered so that $d(v_1)\le d(v_2)\le\dots\le d(v_\Delta)$; here and
below, $d(\cdot)$ denotes the degree in $G$. Let $G':=G-w$. As $G^*$ is nonempty,
$\delta(G^*)\ge3$ by \ref{red:max}. As $G$ is connected, the claim reads
\begin{equation}\label{eq:deletion-goal}
  1+\psi^-(G^*)\le\psi^-(G)=\psi(G)-H_{\delta(G)}.
\end{equation}
We bound $1+\psi^-(G^*)$ from above by $\psi(G)$ minus four terms $J_1,\dots,J_4$,
and show by cases that $J_1+\dots+J_4\ge H_{\delta(G)}$.

\emph{Rewriting the left side.} Every neighbor $v_i$ of $w$ loses one degree, from
$d(v_i)\ge3$ to $d(v_i)-1$, and its weight drops by $h_{d(v_i)}$. So
\[
  \psi(G')=\psi(G)-b_\Delta-\sum_{i=1}^{\Delta}h_{d(v_i)}.
\]
Since $H\ge0$ and some component of $G^*$ has minimum degree $\delta(G^*)$, we have
$\psi^-(G^*)\le\psi(G^*)-H_{\delta(G^*)}$. Put
\[
  J_4:=\psi(G')-\psi(G^*)+H_{\delta(G^*)},
\]
so that $\psi^-(G^*)\le\psi(G')-J_4$. Since $\Delta>k$, identity
\eqref{eq:bootstrap-identity} gives $1-b_\Delta=(\Delta-1)H_\Delta+H_{\Delta-1}$.
Combining,
\[
  1+\psi^-(G^*)\le\psi(G)+(\Delta-1)H_\Delta+H_{\Delta-1}
  -\sum_{i=1}^{\Delta}h_{d(v_i)}-J_4
  =\psi(G)-(J_1+J_2+J_3+J_4),
\]
where
\[
  J_1:=h_{d(v_1)}-H_{\Delta-1},\qquad
  J_2:=h_{d(v_2)}-H_\Delta,\qquad
  J_3:=\sum_{i=3}^{\Delta}\bigl(h_{d(v_i)}-H_\Delta\bigr).
\]
Hence \eqref{eq:deletion-goal} follows once we show
\begin{equation}\label{eq:deletion-J}
  J_1+J_2+J_3+J_4\ge H_{\delta(G)} .
\end{equation}

\emph{Signs.} For $3\le j\le\Delta$ we have $h_j\ge H_j\ge H_\Delta$, because $H$
is nonincreasing. As $3\le d(v_i)\le\Delta$, this gives $J_2\ge0$ and $J_3\ge0$.
Similarly $J_1\ge0$ whenever $d(v_1)\le\Delta-1$, since then
$h_{d(v_1)}\ge H_{d(v_1)}\ge H_{\Delta-1}$. Finally $J_4\ge0$: by
\ref{red:deg}, $\psi(G^*)\le\psi(G')$, and $H_{\delta(G^*)}\ge0$.

\emph{Claim.} If some $u\in V(G')$ has $d_{G'}(u)=j\ge3$, then $J_4\ge H_j$.

Indeed, suppose first that $u$ is a vertex of $G^*$. By \ref{red:deg},
$d_{G^*}(u)\le j$, so $3\le\delta(G^*)\le j$ and $H_{\delta(G^*)}\ge H_j$; together
with $\psi(G')-\psi(G^*)\ge0$ this gives $J_4\ge H_j$. Otherwise $u$ was removed by
a reduction. Every vertex of $G^*$ has weight in $G^*$ at most its weight in $G'$ by
\ref{red:deg}, and every removed vertex has nonnegative weight, so
$\psi(G')-\psi(G^*)\ge b_j\ge h_j\ge H_j$; as $H_{\delta(G^*)}\ge0$, again
$J_4\ge H_j$.

\emph{Case 1: $G$ is $\Delta$-regular.} Then $\delta(G)=\Delta$ and
$J_1=h_\Delta-H_{\Delta-1}$. In $G'$ the vertex $v_1$ has degree $\Delta-1\ge3$, so
$J_4\ge H_{\Delta-1}$ by the claim. Hence
$J_1+J_4\ge h_\Delta\ge H_\Delta=H_{\delta(G)}$, and \eqref{eq:deletion-J} follows
from $J_2,J_3\ge0$.

\emph{Case 2: $G$ is not regular.} Since $G$ is connected, some edge joins a vertex
of degree $\Delta$ to a vertex of smaller degree (on a path from a vertex of degree
$\Delta$ to a vertex of smaller degree, take the first such edge). So $w$ was chosen
with a neighbor of smaller degree, $d(v_1)\le\Delta-1$, and $J_1\ge0$. All four
terms are now nonnegative.

\emph{Case 2a: some $u\in V(G')$ has $3\le d_{G'}(u)\le\delta(G)$.} By the claim,
$J_1+J_2+J_3+J_4\ge J_4\ge H_{d_{G'}(u)}\ge H_{\delta(G)}$.

\emph{Case 2b: no such vertex exists.} Since $\delta(G)<\Delta=d(w)$, there is a
vertex $u\ne w$ of degree $\delta(G)$, and $d_{G'}(u)\in\{\delta(G)-1,\delta(G)\}$.
If $\delta(G)\ge4$, then $3\le d_{G'}(u)\le\delta(G)$, which is Case~2a. Hence
$\delta(G)=3$. Every vertex $u$ of degree $3$ is adjacent to $w$, since otherwise
$d_{G'}(u)=3$, which is Case~2a. In particular $d(v_1)=3$, and
$H_{\delta(G)}=H_3=h_3$.
\begin{itemize}
  \item \emph{Case 2b-i: $d(v_\Delta)\ge4$.} Then
  $3\le d_{G'}(v_\Delta)=d(v_\Delta)-1\le\Delta-1$, so by the claim
  $J_4\ge H_{d_{G'}(v_\Delta)}\ge H_{\Delta-1}$. With $J_1=h_3-H_{\Delta-1}$ this
  gives $J_1+J_4\ge h_3=H_{\delta(G)}$.
  \item \emph{Case 2b-ii: $d(v_\Delta)=3$.} Then all neighbors of $w$ have degree $2$
  in $G'$. Let $u$ be any vertex of the nonempty graph $G^*$. By \ref{red:deg},
  $d_{G'}(u)\ge d_{G^*}(u)\ge3$, so $u$ is not a neighbor of $w$; hence
  $d_{G'}(u)=d(u)\in[3,\Delta]$, and the claim gives $J_4\ge H_{d(u)}\ge H_\Delta$.
  With $J_2=h_3-H_\Delta$ (note $d(v_2)=3$) this gives
  $J_2+J_4\ge h_3=H_{\delta(G)}$.
\end{itemize}
In every case \eqref{eq:deletion-J} holds, which proves \eqref{eq:deletion-goal}.
\end{proof}

The following proposition is the heart of the proof. It deletes vertices until every
component is basic, and bounds the number of deleted vertices from above by the
decrease of $\psi^-$.

\begin{proposition}[deletion process]\label{prop:deletion}
Every graph $G$ with $\delta(G)\ge3$ admits a finite sequence of vertex deletions and
reduction operations that transforms $G$ into a graph $R$ all of whose components
are basic, such that the set $X$ of deleted vertices satisfies
\[
  |X|+\psi^-(R)\le\psi^-(G).
\]
\end{proposition}

\begin{proof}
We argue by induction on $|V(G)|$. A disconnected graph is handled component by
component. For a connected graph that is not basic, we delete one vertex $w$ of
maximum degree, reduce, and apply the induction hypothesis to the result; by
\cref{lem:deletion-step}, this deletion, with the reductions after it, lowers
$\psi^-$ by at least one.

If every component of $G$ is basic, take the empty sequence.

\emph{$G$ disconnected.} Each component $G_i$ has minimum degree at least $3$ and
fewer vertices than $G$, so by induction it admits a sequence with deleted set $X_i$
and final graph $R_i$ satisfying $|X_i|+\psi^-(R_i)\le\psi^-(G_i)$. By
\ref{red:local}, performing these sequences one after the other is a sequence of
deletions and reductions on $G$. Its final graph is the disjoint union $R$ of the
$R_i$, whose components are all basic, and its deleted set is the disjoint union $X$
of the $X_i$. Summing, and using additivity of $\psi^-$, gives
$|X|+\psi^-(R)\le\psi^-(G)$.

\emph{$G$ connected and not basic.} Then $\Delta:=\Delta(G)\ge k+1\ge4$, since
otherwise $G$ would be basic. Choose a vertex $w$ of degree $\Delta$ that has a
neighbor of smaller degree, if such a vertex exists, and any vertex of degree
$\Delta$ otherwise, and let $G^*$ be the final graph of a maximal sequence of
reduction operations applied to $G-w$. Since $G$ is not basic (take $v=w$ in
\cref{def:basic}), $G^*$ is not empty, so $\delta(G^*)\ge3$ by \ref{red:max}. As $|V(G^*)|<|V(G)|$, the induction hypothesis gives a sequence
from $G^*$ to a graph $R$ with basic components and deleted set $X^*$, such that
$|X^*|+\psi^-(R)\le\psi^-(G^*)$. The concatenated sequence (delete $w$, reduce
maximally, then follow the sequence for $G^*$) transforms $G$ into $R$ and has
deleted set $X=X^*\cup\{w\}$. By \cref{lem:deletion-step},
$|X|+\psi^-(R)\le1+\psi^-(G^*)\le\psi^-(G)$.
\end{proof}

The pathwidth of a basic graph is bounded either by the hypothesis
\eqref{eq:bootstrap-base} or by \cref{lem:reduction}.

\begin{lemma}[basic graphs]\label{lem:basic}
Let $J$ be a basic graph with $N$ vertices. Then $\psi(J)\le\psi^-(J)+b_3$. If
$\Delta(J)\le k$, then $\pw(J)\le\psi^-(J)+b_3+\rho(N)$. Otherwise
$\pw(J)\le3\log_2N+1$.
\end{lemma}

\begin{proof}
As $J$ is connected, $\psi(J)=\psi^-(J)+H_{\delta(J)}\le\psi^-(J)+b_3$, because
$H_d\le h_3=b_3$ for all $d\ge3$. If $\Delta(J)\le k$, then $\delta(J)\ge3$ and
\eqref{eq:bootstrap-base} give $\pw(J)\le\psi(J)+\rho(N)$, which with the first claim
gives the bound.

Otherwise there is a vertex $v$ such that a maximal sequence of reductions applied to
$J-v$ ends with the empty graph. Apply \cref{lem:reduction} to $F=J$ and the
sequence that deletes $v$ and then performs these reductions: here $D=\{v\}$ and the
final graph is empty, so $\pw(J)\le1+(-1)+3\log_2N+1$.
\end{proof}

\begin{proof}[Proof of \eqref{eq:bootstrap-explicit}]
Let $G$ be a graph with $n\ge1$ vertices. Apply a maximal sequence of reduction
operations to $G$, and let $R_0$ be its final graph. If $R_0$ is not empty, then
$\delta(R_0)\ge3$ by \ref{red:max}; continue with the sequence of
\cref{prop:deletion} for $R_0$. Together this is one sequence of deletions and
reductions from $G$ to a graph $R$, with deleted set $X$; if $R_0$ is empty, then
$R=R_0$ and $X=\emptyset$. We bound the three ingredients.
\begin{itemize}
  \item \emph{The potential.} By \ref{red:deg}, $\psi(R_0)\le\psi(G)$, and
  $\psi^-(R_0)\le\psi(R_0)$ because $H\ge0$. With \cref{prop:deletion},
  $|X|+\psi^-(R)\le\psi^-(R_0)\le\psi(G)$. (If $R_0$ is empty, both sides of the
  first inequality are $0$.)
  \item \emph{The final graph.} If $R$ is empty, then $\pw(R)=-1$. Otherwise every
  component $C$ of $R$ is basic and has at most $n$ vertices. By \cref{lem:basic},
  monotonicity of $\rho$ and $\log_2$, and nonnegativity of $\psi^-(C)$, $b_3$ and
  $\rho$,
  \[
    \pw(C)\le\psi^-(C)+b_3+\rho(n)+3\log_2n+1 .
  \]
  The pathwidth of a graph is the maximum over its components
  (\cref{sec:prelim-graphs}), and $\psi^-(C)\le\psi^-(R)$ by additivity and
  nonnegativity. Hence $\pw(R)\le\psi^-(R)+b_3+\rho(n)+3\log_2n+1$ in all cases.
  \item \emph{The reductions.} By \cref{lem:reduction} applied to the whole sequence,
  with $D=X$, $\pw(G)\le|X|+\pw(R)+3\log_2n+1$.
\end{itemize}
Adding up,
$\pw(G)\le|X|+\psi^-(R)+b_3+\rho(n)+6\log_2n+2\le\psi(G)+b_3+\rho(n)+6\log_2n+2$,
and $\psi(G)=\sum_{d\ge3}b_dn_d$ because $b_0=b_1=b_2=0$.
\end{proof}

\begin{proof}[Proof of \cref{lem:bootstrap}]
Let $k$ and $b_3,\dots,b_k$ be as in \cref{lem:bootstrap}, and let $b$ be their
bootstrap; it is as in \cref{lem:extension}. Let $\rho_0$ be the function $\rho$ of
\cref{lem:bootstrap}, and let $\rho:=\bar\rho_0$ be its running
maximum~\eqref{eq:running-max}, which is nonnegative, nondecreasing, at least
$\rho_0$, and again $o(N)$. Then \eqref{eq:bootstrap-base} holds, so
\eqref{eq:bootstrap-explicit} holds for every graph with $n\ge1$ vertices, and its
remainder $b_3+\rho(n)+6\log_2n+2$ is $o(n)$; the empty graph has pathwidth $-1$.
So $b$ is a pathwidth sequence. By \cref{lem:extension}\ref{it:ext:mono}, $b$ is
nondecreasing and $b_d\le1$ for $d>k$; for $d\le k$, $b_d\le b_k\le1-H_k\le1$,
because $H_k\ge0$ by \cref{lem:extension}.
\end{proof}

\subsection{Proof of \texorpdfstring{\cref*{thm:main}}{the bound for all graphs}}
\label{sec:proof-main}

We apply the explicit bound~\eqref{eq:bootstrap-explicit} with $k=5$ to the triple
$\tripS$ of \cref{thm:deg5}.
Throughout, $\sigma=(\sigma_d)_{d\ge0}$ denotes the bootstrap of $\tripS$: it has
$\sigma_0=\sigma_1=\sigma_2=0$, $(\sigma_3,\sigma_4,\sigma_5)=\tripS$, and it is
extended by \eqref{eq:bootstrap-rec} with $k=5$. We write $h_d$ and $H_d$ for its
increments and their running minima. The seeds are nondecreasing, with increments
$h_3=\frac5{32}$ and $h_4=h_5=\frac3{32}$; so $H_5=\frac3{32}$ and
$\sigma_5+H_5=\frac{14}{32}\le1$, and $\sigma$ is as in \cref{lem:extension}.

We compute $\sigma$ exactly up to degree $8$, and bound it from above by
$1-\frac{111}{28(d+1)}$ from degree $7$ on and by $1-\frac{4}{d+1}$ from degree $11$
on (\cref{prop:large-degrees}). These two bounds use the following lemma on the
recurrence $d\,u_d=(d-2)\,u_{d-1}+u_{d-2}$. In the concave regime of
\cref{lem:concave}, the deficits $u_d=1-b_d$ satisfy this recurrence. We use the
lemma again in the proofs of \cref{lem:below-published} and of \cref{cor:table}.
Its first part says that the recurrence has a conserved quantity $\lambda$, which
reduces it to the first-order rule $d\,u_d=\lambda-u_{d-1}$; the second part turns
this rule into a two-sided bound on the weighted deficits $w_d=(d+1)\,u_d$.

\begin{lemma}[weighted deficits]\label{lem:weights}
Let $d_0\ge3$, and let $(u_d)_{d\ge d_0-2}$ be reals with
$d\,u_d=(d-2)\,u_{d-1}+u_{d-2}$ for all $d\ge d_0$. Put $w_d:=(d+1)\,u_d$ and
$\lambda:=(d_0-1)\,u_{d_0-1}+u_{d_0-2}$. Then:
\begin{enumerate}[label=\textup{(\roman*)}]
\item\label{it:weights:inv} $d\,u_d+u_{d-1}=\lambda$ for all $d\ge d_0-1$;
\item\label{it:weights:bound} if $\lambda>0$ and
  $\frac{d_1+1}{d_1+2}\,\lambda<w_{d_1}<\lambda$ for some $d_1\ge d_0-1$, then
  \[
    \Bigl(1-\frac1{d^2}\Bigr)\lambda<w_d<\lambda\qquad\text{for every }d>d_1 .
  \]
\end{enumerate}
\end{lemma}

\begin{proof}
\ref{it:weights:inv} For $d\ge d_0$,
$(d\,u_d+u_{d-1})-\bigl((d-1)\,u_{d-1}+u_{d-2}\bigr)=d\,u_d-(d-2)\,u_{d-1}-u_{d-2}=0$.

\ref{it:weights:bound} For $d\ge d_0$, the rule $d\,u_d=\lambda-u_{d-1}$ of
\ref{it:weights:inv} and $u_{d-1}=w_{d-1}/d$ give
\[
  w_d=\frac{d+1}{d}\Bigl(\lambda-\frac{w_{d-1}}{d}\Bigr),\qquad\text{so}\qquad
  w_d-\lambda=\frac{d\,\lambda-(d+1)\,w_{d-1}}{d^2} .
\]
We argue by induction on $d>d_1$ under the hypothesis
$\frac{d}{d+1}\lambda<w_{d-1}<\lambda$, which is the assumption for $d=d_1+1$; note
that $d\ge d_0$ throughout. The lower bound on $w_{d-1}$ makes the numerator
negative, so $w_d<\lambda$. The upper bound gives
$w_d-\lambda>\frac{d\lambda-(d+1)\lambda}{d^2}=-\frac{\lambda}{d^2}$. Finally
$1-\frac1{d^2}\ge\frac{d+1}{d+2}$ because $d^2\ge d+2$ for $d\ge2$; hence
$\frac{d+1}{d+2}\lambda<w_d<\lambda$, which is the hypothesis for $d+1$.
\end{proof}

The next statement computes $\sigma$, records where the threshold $d\ge11$ of
\cref{thm:main} comes from and that it is sharp, and shows $\sigma_d\le a_d$ for every
$d\ge3$.

\begin{proposition}[the bootstrap of $\tripS$]\label{prop:large-degrees}
The bootstrap $\sigma$ of $\tripS$ satisfies $\sigma_d=\frac{3d-4}{32}$ for
$3\le d\le6$,
\[
  1-\frac{129}{32\,(d+1)}<\sigma_d<1-\frac{129}{32\,(d+1)}+\frac{129}{32\,d^2(d+1)}
  \qquad(d\ge5),
\]
and, for $d\ge3$, $\sigma_d\le1-\frac{4}{d+1}$ if and only if $d\ge11$. Moreover,
$\sigma_d\le a_d$ for every $d\ge3$, where $a_d$ are the coefficients of
\cref{thm:main}.
\end{proposition}

\begin{proof}
Steps~1 and~2 compute $\sigma$ up to degree $6$ and show that from degree $7$ on it
satisfies the recurrence of Edwards and McDermid. Steps~3 and~4 apply
\cref{lem:weights} to the deficits $u_d=1-\sigma_d$ and give the two closed-form
bounds. The last paragraph collects the remaining claims.

\emph{Step 1: degrees up to $6$.} At $d=6$,
$\theta_6=(1-\sigma_5-H_5)/6=\frac{18/32}{6}=\frac3{32}=H_5$. By
\cref{lem:extension}\ref{it:ext:case}, $\sigma_6$ is the first term of the maximum
and $h_6=H_6=\frac3{32}$; so $\sigma_6=\frac{11}{32}+\frac3{32}=\frac7{16}$. (The
second term $1-6H_5=\frac7{16}$ is equal.) Hence
\[
  \sigma_d=\frac{3d-4}{32}\qquad(3\le d\le6).
\]

\emph{Step 2: the recurrence from degree $7$.} We have $j=6>k=5$ and
$H_6=H_5=\frac3{32}$, so $H_6\le H_5\le3H_6$; also $h_6=H_6$. By \cref{lem:concave},
\[
  d\,\sigma_d=(d-2)\,\sigma_{d-1}+\sigma_{d-2}+1\qquad(d\ge7).
\]
For example, $\sigma_7=\frac{5\cdot\frac{14}{32}+\frac{11}{32}+1}{7}=\frac{113}{224}$
and $\sigma_8=\frac{6\cdot\frac{113}{224}+\frac{98}{224}+1}{8}=\frac{125}{224}$.

\emph{Step 3: the bound $1-\frac{111}{28(d+1)}$.} Put $u_d:=1-\sigma_d$ and
$w_d:=(d+1)\,u_d$. Since $d=(d-2)+1+1$, the recurrence of Step~2 is equivalent to
$d\,u_d=(d-2)\,u_{d-1}+u_{d-2}$ for $d\ge7$; it also holds at $d=6$, because
$6\sigma_6=\frac{21}8=4\sigma_5+\sigma_4+1$. So \cref{lem:weights}\ref{it:weights:inv}
applies with $d_0=6$ and $\lambda=5u_5+u_4=\frac{105}{32}+\frac{24}{32}$:
\[
  d\,u_d+u_{d-1}=\tfrac{129}{32}\qquad(d\ge5).
\]
Moreover $w_5=w_6=\frac{63}{16}=\frac{126}{32}$ and
$\frac78\cdot\frac{129}{32}=\frac{903}{256}<\frac{126}{32}<\frac{129}{32}$, so
\cref{lem:weights}\ref{it:weights:bound} applies with $d_1=6$:
\begin{equation}\label{eq:main-weights}
  \Bigl(1-\frac1{d^2}\Bigr)\frac{129}{32}<w_d<\frac{129}{32}\qquad(d\ge7).
\end{equation}
Now $w_7=8\cdot\frac{111}{224}=\frac{111}{28}$, and for $d\ge8$,
$w_d>\frac{63}{64}\cdot\frac{129}{32}=\frac{8127}{2048}>\frac{111}{28}$. Hence
$w_d\ge\frac{111}{28}$ for every $d\ge7$, that is,
\[
  \sigma_d\le1-\frac{111}{28\,(d+1)}\qquad(d\ge7),
\]
with equality at $d=7$.

\emph{Step 4: the bound $1-\frac{4}{d+1}$.} The inequality $\sigma_d\le1-\frac4{d+1}$
is $w_d\ge4$. For $d\ge12$, \eqref{eq:main-weights} gives
$w_d>\bigl(1-\frac1{144}\bigr)\frac{129}{32}=\frac{18447}{4608}>4$, as
$18447>4\cdot4608=18432$. For $d=11$, the rule $u_d=(\frac{129}{32}-u_{d-1})/d$ gives
$u_7=\frac{111}{224}$, $u_8=\frac{99}{224}$, $u_9=\frac{67}{168}$,
$u_{10}=\frac{2441}{6720}$ and $u_{11}=\frac{24649}{73920}$, so
$w_{11}=\frac{24649}{6160}>4$ because $4\cdot6160=24640$. Hence
\[
  \sigma_d\le1-\frac{4}{d+1}\qquad(d\ge11).
\]

\emph{The remaining claims.} The two-sided bound is \eqref{eq:main-weights} for
$d\ge7$, and for $d\in\{5,6\}$ it reads
$\bigl(1-\frac1{d^2}\bigr)\frac{129}{32}\le\frac{35}{36}\cdot\frac{129}{32}
=\frac{4515}{1152}<\frac{126}{32}=w_d<\frac{129}{32}$. Step~4 gives
$\sigma_d\le1-\frac4{d+1}$ for $d\ge11$. Conversely,
$w_3=4\cdot\frac{27}{32}=\frac{27}8$, $w_4=5\cdot\frac34=\frac{15}4$, and the values of
Step~4 give
\[
  (w_5,\dots,w_{10})=\Bigl(\tfrac{63}{16},\tfrac{63}{16},\tfrac{111}{28},
  \tfrac{891}{224},\tfrac{335}{84},\tfrac{26851}{6720}\Bigr),
\]
all below $4$ (for the last, $4\cdot6720=26880>26851$), so $\sigma_d>1-\frac4{d+1}$
for $3\le d\le10$. Finally, Steps~1, 3 and~4 give $\sigma_d\le a_d$ for every $d\ge3$.
\end{proof}

The statement of \cref{thm:main} rounds the coefficients $\sigma_d$ for $d\ge7$ up to
closed forms. The proof above gives $\sigma_7=\frac{113}{224}$,
$\sigma_8=\frac{125}{224}$, $\sigma_9=\frac{101}{168}$ and
$\sigma_{10}=\frac{4279}{6720}$, which \cref{tab:bootstrap-comparison} compares with
$a_d$.

\begin{proof}[Proof of \cref{thm:main}]
By \cref{thm:deg5} with the triple $\tripS$, there is a function $\rho_0$ with
$\rho_0(N)=o(N)$ such that every graph $J$ with $\Delta(J)\le5$, in particular every
$J$ with $\delta(J)\ge3$ and $\Delta(J)\le5$, satisfies
$\pw(J)\le\sigma_3n_3(J)+\sigma_4n_4(J)+\sigma_5n_5(J)+\rho_0(|V(J)|)$. Its running
maximum $\rho:=\bar\rho_0$ of~\eqref{eq:running-max} is nonnegative, nondecreasing, at
least $\rho_0$ and again $o(N)$, so \eqref{eq:bootstrap-base} holds with $\rho$. As
$\sigma$ is as in \cref{lem:extension}, \eqref{eq:bootstrap-explicit} gives, for every
graph $G$ with $n\ge1$ vertices,
\begin{equation}\label{eq:main-sigma}
  \pw(G)\le\sum_{d\ge3}\sigma_d\,n_d+\rho_4(n),\qquad
  \rho_4(n):=\tfrac5{32}+\rho(n)+6\log_2n+2 .
\end{equation}
The function $\rho_4$ is nondecreasing and $\rho_4(n)=o(n)$. By
\cref{prop:large-degrees}, $\sigma_d\le a_d$ for every $d\ge3$, so
$\pw(G)\le\sum_{d\ge3}a_dn_d+\rho_4(n)$ for every graph $G$.
\end{proof}

\subsection{Edges and average degree}\label{sec:proof-edges}

This subsection derives the bound in terms of edges (\cref{cor:edges}) and a bound in
terms of the average degree (\cref{cor:avg-degree}) from \cref{thm:main}, and
compares the latter with the bound of Edwards and McDermid.
Throughout, $a_d$ ($d\ge3$) are the coefficients of \cref{thm:main},
and $a_0=a_1=a_2:=0$. The edge bound follows from \cref{thm:main} because no
coefficient $a_d$ exceeds $\frac7{48}\cdot\frac d2$; the maximum ratio $2a_d/d$ is
attained at $d=6$.

\begin{proof}[Proof of \cref{cor:edges}]
We first show that
\begin{equation}\label{eq:edges-ratio}
  a_d\le\frac7{48}\cdot\frac d2\qquad\text{for every }d\ge3,
\end{equation}
with equality only at $d=6$, and that $2a_d/d\le\frac{113}{784}$ for $d\ne6$. For
$3\le d\le6$, $a_d=\frac{3d-4}{32}\le\frac{7d}{96}$ is equivalent to $9d-12\le7d$,
that is, to $d\le6$, and $2a_d/d=\frac3{16}-\frac1{4d}\le\frac{11}{80}<\frac{113}{784}$
for $d\le5$. For $d\ge7$ we have $a_d\le1-\frac{111}{28(d+1)}$: with equality for
$7\le d\le10$; for $d\ge11$ because $4>\frac{111}{28}$. Moreover
\[
  \frac{113}{784}\cdot\frac d2-\Bigl(1-\frac{111}{28\,(d+1)}\Bigr)
  =\frac{(d-7)(113d-664)}{1568\,(d+1)}\ge0\qquad(d\ge7),
\]
with equality at $d=7$, and $\frac{113}{784}<\frac7{48}$. This proves both claims.

Now let $G$ be a graph with $m$ edges. Summing \eqref{eq:edges-ratio} over the
vertices of $G$ of degree at least $3$, and using that vertices of degree at most $2$
contribute nonnegative amounts to the right side,
\[
  \sum_{d\ge3}a_d\,n_d\le\frac7{96}\sum_{v\in V(G)}d_G(v)=\frac7{48}\,m .
\]
So \cref{thm:main} gives $\pw(G)\le\frac7{48}m+o(n)$.
\end{proof}

\begin{remark}[remainder in $m$]\label{rem:edges-m}
The remainder of \cref{cor:edges} can also be taken as a function of $m$:
$\pw(G)\le\frac7{48}m+\rho(m+1)$ for every graph $G$ with $m$ edges, where $\rho$
is the remainder of \cref{thm:main}, taken nondecreasing by~\eqref{eq:running-max}.
Indeed, the computation above applied to a component $H$ of $G$ gives
$\pw(H)\le\frac7{48}|E(H)|+\rho(|V(H)|)\le\frac7{48}m+\rho(m+1)$, since
$|V(H)|\le|E(H)|+1\le m+1$, and $\pw(G)$ is the maximum of $\pw(H)$ over the
components. Taking the maximum over components is essential: a sum of remainders over
many small components would not be $o(m)$. The same step shows that every bound of
the form $\sum_d b_dn_d+o(n)$ with $b_d\le\beta d/2$ for all $d$ yields
$\pw(G)\le\beta m+o(m)$. The algorithms of \cref{thm:algo-precise} compute path
decompositions of width at most $\frac7{48}m+\rho(2m)$, a remainder in $m$ as well.
\end{remark}

The same coefficients give a bound in terms of the average degree $\bar d=2m/n$. Let
$A\colon[0,\infty)\to\R$ be given by $A(x):=\frac7{96}x$ for $0\le x\le6$ and, on
$[d,d+1]$ for each integer $d\ge6$, by linear interpolation between $A(d)=a_d$ and
$A(d+1)=a_{d+1}$; the two definitions agree at $6$, since $a_6=\frac7{16}$.

\begin{corollary}[average degree]\label{cor:avg-degree}
Every graph $G$ with $n\ge1$ vertices and average degree $\bar d$ satisfies
$\sum_{d\ge3}a_dn_d\le A(\bar d)\,n$, and hence $\pw(G)\le A(\bar d)\,n+o(n)$, where
$A(\bar d)\le1-\frac{111}{28(\bar d+1)}$ if $\bar d\ge7$ and
$A(\bar d)\le1-\frac{4}{\bar d+1}$ if $\bar d\ge11$. If $\delta(G)\ge2$, then
$\pw(G)\le\frac{3m-2n}{16}+o(n)$.
\end{corollary}

\begin{proof}
We first show that $A$ is concave and lies above every coefficient $a_d$, and apply
Jensen's inequality. The closed-form bounds on $A(\bar d)$ then follow by comparing
$A$ with concave functions, and the bound for $\delta(G)\ge2$ follows from
$a_d\le\frac{3d-4}{32}$.

The function $A$ is continuous and piecewise linear. Its slopes are $\frac7{96}$ on
$[0,6]$, $a_7-a_6=\frac{15}{224}$ on $[6,7]$,
$a_{d+1}-a_d=\frac{111}{28(d+1)(d+2)}$ on $[d,d+1]$ for $7\le d\le9$,
$a_{11}-a_{10}=\frac{25}{924}$ on $[10,11]$, and
$a_{d+1}-a_d=\frac{4}{(d+1)(d+2)}$ on $[d,d+1]$ for $d\ge11$. These slopes are
nonincreasing, because $\frac7{96}>\frac{15}{224}>\frac{111}{28\cdot72}$, the third
expression decreases in $d$ to $\frac{111}{3080}>\frac{25}{924}$ at $d=9$,
$\frac{25}{924}>\frac4{156}$, and the last expression decreases in $d$. Hence $A$ is
concave. Moreover $A(d)\ge a_d$ for every integer $d\ge0$: for $d\le2$ because
$a_d=0$, for $3\le d\le5$ by \eqref{eq:edges-ratio}, and for $d\ge6$ with equality.
Since the weights $n_d/n$ sum to $1$, Jensen's inequality gives
\[
  \sum_{d\ge3}a_dn_d=\sum_{d\ge0}a_dn_d\le\sum_{d\ge0}A(d)\,n_d
  \le n\,A\Bigl(\sum_{d\ge0}d\,\frac{n_d}{n}\Bigr)=n\,A(\bar d),
\]
and \cref{thm:main} gives $\pw(G)\le A(\bar d)\,n+o(n)$. For $\bar d\ge7$, let
$q(x):=1-\frac{111}{28(x+1)}$; it is concave on $[0,\infty)$, and $a_d\le q(d)$ for
$d\ge7$ (with equality for $d\le10$). On each interval $[d,d+1]$ with $d\ge7$, $A$ is
the chord through $(d,a_d)$ and $(d+1,a_{d+1})$, which lies below the chord of the
concave function $q$ and hence below $q$. So $A(\bar d)\le q(\bar d)$. The same
argument with the concave function $1-\frac4{x+1}$, which equals $a_d$ for every
$d\ge11$, gives $A(\bar d)\le1-\frac4{\bar d+1}$ for $\bar d\ge11$.

For the last bound, note that $a_d\le\frac{3d-4}{32}$ for every $d\ge3$: with
equality for $d\le6$, and for $d\ge7$ because $A$ is concave with slope
$\frac{15}{224}<\frac3{32}$ on $[6,7]$, so that
$a_d=A(d)\le a_6+\frac{15}{224}(d-6)\le\frac{3d-4}{32}$. So every graph satisfies
$\pw(G)\le\sum_{d\ge3}\frac{3d-4}{32}\,n_d+o(n)$. Now let $\delta(G)\ge2$, and
write $n_{\ge3}$ for the number of vertices of degree at least $3$. As
$3d-4=2+3(d-2)$, $\sum_{d\ge2}n_d=n$ and $\sum_{d\ge2}d\,n_d=2m$,
\[
  \sum_{d\ge3}a_dn_d\le\sum_{d\ge3}\frac{3d-4}{32}\,n_d
  =\frac{2n_{\ge3}+3\sum_{d\ge2}(d-2)\,n_d}{32}=\frac{n_{\ge3}+3(m-n)}{16}
  \le\frac{3m-2n}{16},
\]
and \cref{thm:main} completes the proof.
\end{proof}

For
$\bar d\le6$ its main term is that of \cref{cor:edges}, since
$A(\bar d)\,n=\frac7{96}\bar d\,n=\frac7{48}m$.
For graphs with $\delta(G)\ge2$ and $\bar d<6$, the bound $\frac{3m-2n}{16}$ is
smaller. The condition $\delta(G)\ge2$ cannot be dropped from it, since adding
isolated vertices makes $3m-2n$ negative without changing the pathwidth.

Edwards and McDermid prove $\pw(G)\le\bigl(g_{1/6}(\bar d)+o(1)\bigr)n$ for
connected graphs and for graphs with $\bar d\ge5$, with $g_{1/6}$ interpolated
linearly \cite[Theorem~5.3]{EdwardsM15}. For $\bar d\ge6$, \cref{cor:avg-degree}
is smaller, since both $A$ and $g_{1/6}$ interpolate linearly between consecutive
integers there and $a_d<g_{1/6}(d)$ for every $d\ge3$ by the following lemma.

\begin{lemma}[comparison with the previous coefficients]\label{lem:below-published}
Let $g_{1/6}(2)=0$, $g_{1/6}(3)=\frac16$ and
$d\,g_{1/6}(d)=(d-2)\,g_{1/6}(d-1)+g_{1/6}(d-2)+1$ for $d\ge4$, the per-degree
coefficients of Edwards and McDermid~\cite[\S3.1.1 and Corollary~5.2]{EdwardsM15};
for $3\le d\le6$ they are $\frac16,\frac13,\frac{13}{30},\frac{23}{45}$, those of Fomin,
Gaspers, Saurabh and Stepanov~\cite[Lemma~1]{FominGaspersSaurabhStepanov2009}. Then
the coefficients $a_d$ of \cref{thm:main} satisfy $a_d<g_{1/6}(d)$ and
$(d+1)\bigl(1-g_{1/6}(d)\bigr)<\frac72$ for every $d\ge3$.
\end{lemma}

\begin{proof}
Let $u_d:=1-g_{1/6}(d)$ and $\omega_d:=(d+1)u_d$. For $d\le6$ the inequality
$a_d<g_{1/6}(d)$ follows from the exact values: $\frac5{32}<\frac16$,
$\frac14<\frac13$, $\frac{11}{32}<\frac{13}{30}$ and $\frac7{16}<\frac{23}{45}$. The
numbers $u_d$ satisfy $d\,u_d=(d-2)\,u_{d-1}+u_{d-2}$ for $d\ge5$, with
$u_3=\frac56$ and $u_4=\frac23$. By \cref{lem:weights}\ref{it:weights:inv} with
$d_0=5$, $d\,u_d+u_{d-1}=4u_4+u_3=\frac72$ for all $d\ge4$, and since
$\frac56\cdot\frac72=\frac{35}{12}<\omega_4=\frac{10}3<\frac72$,
\cref{lem:weights}\ref{it:weights:bound} with $d_1=4$ gives
\[
  \Bigl(1-\frac1{d^2}\Bigr)\frac72<\omega_d<\frac72\qquad(d\ge5),
\]
which is the asymptotic form of Edwards and McDermid with an explicit error term.
Hence $\omega_d<\frac72<\frac{111}{28}\le(d+1)(1-a_d)$ for $d\ge7$, where
$(d+1)(1-a_d)$ is $\frac{111}{28}$ for $7\le d\le10$ and $4$ for $d\ge11$. So
$1-g_{1/6}(d)<1-a_d$, that is, $a_d<g_{1/6}(d)$, for all $d\ge7$. Finally,
$\omega_3=\omega_4=\frac{10}3<\frac72$, so $\omega_d<\frac72$ for every $d\ge3$.
\end{proof}

{}
\section{Computing the decompositions}\label{sec:algorithms}

This section turns the width bounds of
\cref{thm:cubic,thm:deg4,thm:deg5,thm:main,cor:edges}, and with
\cref{app:alg-switched} also those of \cref{thm:switched4,thm:pw-seq,cor:table}, into
algorithms that compute the decompositions. The proofs of these bounds are probabilistic. The frontiers of a random
Gaussian labeling have the right expected size at every threshold and concentrate
around it, and \cref{lem:layout}, or the median-switched layout of
\cref{lem:median-layout}, turns a good labeling into a path decomposition. We state
\cref{thm:algo-precise}, the precise form of \cref{thm:algo}, which gives a Las
Vegas and a deterministic algorithm for each of these bounds, and outline its proof,
which is in \cref{app:algorithms}.
\Cref{part:apps} uses this
theorem, mostly its deterministic part, whenever an algorithm needs a decomposition
of the width given by \cref{part:width}.

\subsection{Statement and outline}\label{sec:alg-statement}

\paragraph{Model.} The input graph $G$ has vertices $1,\dots,n$ and is given by
adjacency lists. Running times count bit operations on integers and rationals in
binary; in the bootstrap and general cases below we also count operations on
integers of $O(\log n)$ bits, which serve as vertex names and indices. The Las Vegas algorithms
have access to independent uniform random bits. A path decomposition
$(W_1,\dots,W_s)$ is output in interval form (\cref{sec:prelim-graphs}), that is, as
the number $s$ and, for each vertex $v$, the indices $\mathrm{first}(v)\le
\mathrm{last}(v)$ of the first and the last bag containing $v$. This takes
$O(n\log n)$ bits when $s=O(n)$, as for all our decompositions. The bags themselves
can have total size of order $n^2$, so the running times below refer to the interval
form.
The terms $o(1)$ in the exponents of running times are functions of $n$ alone.

\paragraph{Cases.} Each case of the following theorem consists of a class $\mathcal C$
of graphs, given by a condition on $G$, and a bound $\beta(G)$ for $G\in\mathcal C$:
\begin{itemize}
  \item \emph{cubic case}: $\Delta(G)\le3$ and $\beta(G)=\cthree n$;
  \item \emph{padded case}: $\Delta(G)\le4$ and $\beta(G)=F_4n$;
  \item \emph{switched case}: $\Delta(G)\le4$ and $\beta(G)=\cthree n_3+0.2627\,n_4$;
  \item \emph{mixed cases}: $\Delta(G)\le5$ and $\beta(G)=a_3n_3+a_4n_4+a_5n_5$ with
  $(a_3,a_4,a_5)\in\{\tripP,\tripR,\tripS\}$;
  \item \emph{bootstrap cases}: all graphs, and $\beta(G)=\sum_{d\ge3}b_dn_d$, where
  $(b_d)_{d\ge3}$ is the bootstrap of one of the six sequences
  $\seqC,\seqM,\seqF,\tripP,\tripR,\tripS$ of \cref{thm:pw-seq};
  \item \emph{general case}: all graphs, and $\beta(G)=\sum_{d\ge3}a_dn_d$ with the
  coefficients $a_d$ of \cref{thm:main}.
\end{itemize}

\begin{restatable}[precise form of \cref{thm:algo}]{theorem}{thmalgoprecise}\label{thm:algo-precise}
In each case listed above there are a nondecreasing function $\rho\colon\N\to[0,\infty)$ with
$\rho(n)=o(n)$ and two algorithms that, given $G\in\mathcal C$, output a path
decomposition of $G$ in interval form, of width at most $\beta(G)+\rho(n)$, and in the
general case also at most $\frac7{48}m+\rho(2m)$:
\begin{enumerate}[label=\textup{(\alph*)}]
  \item\label{it:algo:LV} a Las Vegas algorithm, in expected $n^{1+o(1)}$ bit
  operations;
  \item\label{it:algo:D} a deterministic algorithm, in $n^{2+o(1)}$ bit operations.
\end{enumerate}
In the bootstrap and general cases, each algorithm uses in addition $O(n(n+m))$
operations on integers of $O(\log n)$ bits.
\end{restatable}

Here $\rho$ is the remainder of the width bound. The bounds $\beta$ are those of
\cref{thm:cubic}, \cref{thm:deg4}\ref{thm:deg4:a}, \cref{thm:switched4},
\cref{thm:deg5}, \cref{lem:bootstrap} and \cref{thm:main} without their remainders, and
the second bound of the general case is that of \cref{cor:edges}. A graph with
$\Delta(G)\le4$ has $n_5=0$, so the mixed cases with $\tripP$ and $\tripS$ also give
the bounds of \cref{thm:deg4}\ref{thm:deg4:b} and~\ref{thm:deg4:c}. The bootstrap
cases give the rows of \cref{cor:table} and the sequences of \cref{thm:pw-seq} on all
graphs (\cref{rem:alg-other-rows}); the general case is the bootstrap case of $\tripS$
in the rounded form of \cref{thm:main}. \Cref{thm:algo} follows from
\cref{thm:algo-precise}.

\paragraph{Outline.} The proof of each case follows the same pattern. The algorithm
computes discrete labels $\tilde Y_v$, sorts the vertices by $(\tilde Y_v,v)$, and
builds the layout of the case from the sorted order. The width of that layout is
bounded by a finite family of \emph{counts}, each the number of vertices satisfying a
condition on the labels in their closed neighborhood (\cref{lem:alg-grid} and
\cref{sec:alg-cubic}). Four ingredients, shared by all cases, bound the counts.
\begin{enumerate}[label=\textup{(\arabic*)}]
  \item \emph{Exact local labels} (\cref{sec:alg-model,sec:alg-law}). We round the
  coefficients of \cref{def:labels} to rationals, and we replace the Gaussian
  variables by a finite rational distribution coupled with the standard normal distribution
  (\cref{lem:alg-coupling}).
  \item \emph{Perturbation} (\cref{sec:alg-perturbation}). In this coupling every
  per-vertex probability used in \cref{part:width} changes by $o(1)$, uniformly in the
  threshold (\cref{lem:alg-perturb,cor:alg-transfer}).
  \item \emph{Uniform ideal means} (\cref{sec:alg-means}). The grid of thresholds
  refines as $n$ grows, so the bounds on expected frontier sizes are needed uniformly
  in the threshold (\cref{lem:alg-means}). In the cubic, padded and mixed cases the
  bounds of \cref{part:width} already have this form, with an error term $O(1/h)$; in
  the switched case, \cref{lem:alg-switched-means} makes the error term at threshold
  zero explicit, as $O(\eps_h^{1/3})$.
  \item \emph{Derandomization} (\cref{sec:alg-derand}). The counts are close to their
  means with high probability. The method of conditional expectations (Erd\H{o}s
  and Selfridge~\cite{ErdosSelfridge1973}; Raghavan~\cite{Raghavan1988}), applied to
  the sum of the squared deviations of the counts, fixes the random variables one at a
  time so that every count is close to its mean (\cref{lem:alg-derand}). The locality
  radius grows only like $\frac18\log_2\log_2n$, so every conditional expectation is a
  sum of local terms, each computable exactly in $n^{o(1)}$ bit operations.
\end{enumerate}
Derandomizing a local Gaussian rounding in this way is standard in outline; the work
lies in the exact finite distribution and in the uniformity of the means.

\paragraph{Roadmap.} The proof is in \cref{app:algorithms}.
\Cref{sec:alg-model,sec:alg-law,sec:alg-perturbation,sec:alg-means,sec:alg-derand}
develop the four ingredients in this order, and \cref{sec:alg-undo} makes the
construction of \cref{lem:reduction} algorithmic. \Cref{sec:alg-plain,sec:alg-cubic,sec:alg-general}
assemble the algorithms for the padded and mixed cases, the cubic case, and the
bootstrap and general cases, \cref{app:alg-switched} treats the switched case and the
base layouts of the bootstrap cases $\seqC$ and $\seqM$, and \cref{sec:alg-time}
collects the running times. The rest of this section computes orderings of small
cutwidth for maximum degree~$3$, $4$ and $5$ and discusses the limitations.

\subsection{Linear orderings of small cutwidth}\label{sec:alg-cutwidth}

This subsection computes, for graphs of maximum degree~$3$, linear orderings of
cutwidth at most $\cthree n+o(n)$, and bisections with at most that many edges between
the two sides (\cref{cor:alg-cutwidth}). \Cref{cor:bisection} shows that such
orderings and bisections exist, by bounding the cutwidth through a chain of
known relations. Two links of that chain have constructive proofs, which turn an
ordering of vertex separation $w$, such as the one read off a path decomposition of
\cref{thm:algo-precise}, into an ordering of cutwidth at most $w+2$
(\cref{lem:alg-cutwidth}, proved in \cref{app:alg-cutwidth}).
For maximum degree $4$ and $5$, \cref{thm:alg-cut-sweep} derandomizes the cut sweep of
\cref{thm:cut-sweep-p} directly, with the machinery of
\cref{sec:alg-model,sec:alg-law,sec:alg-perturbation,sec:alg-derand}. The two routes
differ because they start from different bounds. For maximum degree $3$ the cut sweep
gives only $\gamma_3m\le c_*n$, while the route through vertex separation gives
$\cthree n<c_*n$. For maximum degree $4$ the same chain gives only
$\operatorname{cw}\le2\pw+O(1)$~\cite[Theorem~3.2]{MakedonSudborough1989}%
\cite[Theorem~2.1]{EllisSudboroughTurner1994}, that is, $2F_4n+o(n)>\frac n3$.

\begin{restatable}[computing orderings and bisections]{corollary}{coralgcutwidth}\label{cor:alg-cutwidth}
For graphs $G$ with $\Delta(G)\le3$, an ordering of $V(G)$ of cutwidth at most
$\cthree n+o(n)$, and a bisection of $G$ with at most $\cthree n+o(n)$ edges between
its two sides, can be computed in deterministic time $n^{2+o(1)}$.
\end{restatable}

The proof is in \cref{app:alg-cutwidth}.

\begin{restatable}[deterministic cut sweep]{theorem}{thmalgcutsweep}\label{thm:alg-cut-sweep}
Let $p\in\{4,5\}$. For graphs $G$ with $\Delta(G)\le p$ and $m$ edges, an ordering of
$V(G)$ of cutwidth at most $\gamma_pm+o(n)$, and a bisection of $G$ with at most that
many edges between its two sides, can be computed in deterministic time $n^{2+o(1)}$.
For $p=4$ the bound is $\frac m6+o(n)\le\frac n3+o(n)$.
\end{restatable}
The proof is also in \cref{app:alg-cutwidth}.

\subsection{Limitations}\label{sec:alg-limits}

The guarantees of \cref{thm:algo-precise} are asymptotic: our analysis assumes
$n\ge2^{2^{40}}$, and some error terms are below $1$ only when $\log_2\log_2n$ exceeds
$10^4$ or even $10^5$, so the bounds say nothing for practical~$n$. The terms $o(1)$ in
the exponents also hide enumeration factors $e^{O(t^{1/4}\log t)}$ with $t=O(\log n)$,
and practical efficiency is not claimed.
{}

\clearpage
\part{Algorithmic consequences}\label{part:apps}

Part~II derives faster exact algorithms from the width bounds of Part~I. Before the
proofs, which begin in \cref{sec:apps-dp}, this opening gathers the new running-time
bounds and says which of them the two recent preprints that claim to refute the
ETH~\cite{OpenAI2026ETH,Claude2026ETH} (\cref{sec:intro-apps}) would supersede.

\paragraph{The tables.}
\Cref{tab:apps} lists the running-time bounds of this part in
exponential space that, as far as we know, the preprints do not affect, and
\cref{tab:apps-affected} lists those in exponential space that the preprints would
supersede. Both tables follow the conventions stated after
\cref{thm:apps-summary}. ``Previous'' is the fastest bound that we are aware of.

\paragraph{Bounds in polynomial space.}
Neither table lists bounds in polynomial space; we state them here. Separate,
Measure and Conquer with the separator constant $\cthree$ (\cref{thm:smc}) solves Max
$(r,2)$-CSP in polynomial space in time $r^{0.17587m+o(m)}$, in time
$r^{0.16984m+o(m)}$ if $\Delta\le4$ and in time $r^{0.17935n+o(n)}$ on cubic
instances. The previous bounds are $r^{9m/50}$, $r^{7m/40}$ and $r^{n/5}$ up to
subexponential factors~\cite[Theorems~3.1--3.3]{GaspersSorkin2017}. As far as we know,
the preprints do not affect these three bounds for Max $(r,2)$-CSP. The same method
counts the dominating sets of each size of a graph with $\Delta\le3$ in polynomial
space and time $O(1.2178^n)$, against $O(1.2458^n)$~\cite[Theorem~5.1]{GaspersSorkin2017}.

By \cref{thm:is}, proved in \cref{app:is}, the independent sets of a graph can be
counted in polynomial space in time $O(1.1304^n)$ if $\Delta\le3$, $O(1.2037^n)$ if
$\Delta\le4$ and $O(1.2341^n)$ in general, and its chromatic number can be computed in
polynomial space in time $O(2.2341^n)$. The previous bounds, of Gaspers and
Lee~\cite[Lemmas~5 and~6, Theorem~1]{GaspersLee2023}, are $O(1.1389^n)$,
$O(1.2070^n)$, $O(1.2356^n)$ and $O(2.2356^n)$. The previous analysis of the bound
$O(1.1389^n)$ has open points (\cref{app:is}); our bound $O(1.1304^n)$ includes the necessary corrections.

\paragraph{The reach of the preprints.}
The preprints reach other problems through the subexponential reduction families of
Impagliazzo, Paturi and Zane~\cite{ImpagliazzoPZ2001}, which make every problem in
size-constrained SNP solvable in subexponential time if \#3-SAT is, and through
direct counting algorithms~\cite[Corollary~7]{Claude2026ETH},
\cite[Corollary~1.3]{OpenAI2026ETH}. If correct, they supersede seventeen of
our bounds: the twelve of \cref{tab:apps-affected} and the five in polynomial space
above. For the chromatic number this uses inclusion--exclusion~\cite{BjorklundHK2009};
for Max 2-SAT, Max $k$-Vertex Cover, Upper Domination, Edge Dominating Set and Odd
Cycle Transversal on graphs of bounded degree, an encoding in size-constrained SNP or
in $k$-SAT with $O(n)$ variables; for Independent Dominating Set, a result of Ganian,
de Haan, Kanj and Szeider~\cite[Corollary~6]{GanianHKS2016}. As far as we know, the
problems of \cref{tab:apps} lie outside this reach: they have real weights, a global
connectivity or acyclicity constraint (whether Feedback Vertex Set is SERF-reducible
to 3-SAT is open~\cite{GanianHKS2016}), or running times $n^{ck}$ in a pattern
size~$k$, and we know of no SERF reduction from any of them to 3-SAT or \#3-SAT.

\begin{table}[tb]
  \centering
  \caption{The running-time bounds of this part in exponential space that would be
    superseded if the preprints~\cite{OpenAI2026ETH,Claude2026ETH} that claim to refute
    ETH are correct. Letters refer to the notes; the letters a--h have the same meaning
  as in \cref{tab:apps}, and the notes i and j are local to this table.}
  \label{tab:apps-affected}
  \small
  \setlength{\tabcolsep}{2pt}
  \begin{tabular}{@{}lll>{\raggedright\arraybackslash}p{2.2cm}@{}}
    \toprule
    Problem & Previous & New & Where\\
    \midrule
    \multicolumn{4}{@{}l}{\emph{Max 2-SAT, counting dominating sets and independent sets}}\\
    Max 2-SAT & $2^{m/6.321}$ \cite{GaspersSorkin2012} & $2^{m/6.857}$ & \cref{thm:max2csp}\\
    \#Dominating Set, all graphs & $1.5002^n$ \cite{NederlofRD2014}\textsuperscript{a} & $1.4815^n$ & \cref{thm:ds}\\
    \#Dominating Set (every size), $\Delta\le3$ & $3^{n/6}<1.2010^n$ \cite{FominHoie2006} & $3^{\cthree n}<1.1661^n$ & \cref{thm:bdd}\\
    \#Independent Sets, $\Delta\le3$ & $2^{n/6}<1.1225^n$ \cite{FominHoie2006} & $2^{\cthree n}<1.1018^n$ & \cref{thm:bdd}\\
    \#Independent Sets, all graphs & $1.2330^n$ \cite{GaspersLee2023} & $1.2287^n$ & \cref{thm:is}\\
    \midrule
    \multicolumn{4}{@{}l}{\emph{Local problems on graphs of bounded degree (deterministic)}}\\
    Max $k$-Vertex Cover, $\Delta\le3$ & $1.123^n$ \cite{DellaCroceP2012} & $2^{\cthree n}<1.1018^n$ & \cref{thm:bdd}\\
    Min.\ Independent Dominating Set, $\Delta\le3$ & $1.3351^n$ \cite{BourgeoisCEP2013}\textsuperscript{i} & $3^{\cthree n}<1.1661^n$ & \cref{thm:bdd}\\
    Min.\ Independent Dominating Set, $\Delta\le4$ & $1.3351^n$ \cite{BourgeoisCEP2013}\textsuperscript{i} & $3^{F_4n}<1.3100^n$ & \cref{thm:bdd}\\
    Upper Domination, $\Delta\le3$ & $1.3481^n$ \cite{BazganBCFJKLLMP2018} & $6^{\cthree n}<1.2848^n$ & \cref{thm:bdd}\\
    Edge Dominating Set\textsuperscript{j}, $\Delta\le3$ & $3^{n/6}<1.2010^n$ \cite{FominHoie2006} & $3^{\cthree n}<1.1661^n$ & \cref{thm:bdd}\\
    Odd Cycle Transversal, $\Delta\le4$ & $3^{n/3}<1.4423^n$ \cite{FominGaspersSaurabhStepanov2009} & $2^{n/3}<1.2600^n$ & \cref{thm:bdd}\\
    Odd Cycle Transversal, $\Delta\le5$ & $3^{13n/30}<1.6098^n$ \cite{FominGaspersSaurabhStepanov2009} & $2^{0.5121n}<1.4262^n$ & \cref{thm:bdd}\\
    \bottomrule
  \end{tabular}

  \smallskip
  \begin{minipage}{\textwidth}\footnotesize
    \textsuperscript{a}\,A gap in the analysis of this algorithm allowed us only to
    reconstruct a running time of $O(1.5004^n)$ (\cref{sec:nrd-gap}).
    \textsuperscript{i}\,A bound for all graphs.
    \textsuperscript{j}\,Also Minimum Maximal Matching.
  \end{minipage}
\end{table}

\paragraph{Conventions for Part II.}
We write $\Ostar(f)$ for $f\cdot\poly(N)$, where $N$ is the length of the input. All
remainders $o(n)$, $o(m)$ and $o(k)$ are functions of the displayed parameter alone.
The results of Part~II are worst-case asymptotic statements, as are the records they improve.
Numerical bases are rounded \emph{up} in the last printed digit, so that
$c^{n+o(n)}=O(b^n)$ holds for the printed $b$. Running times of the form
$r^{\alpha n+o(n)}$ are stated with the exact exponent; their numerical values come
from the accompanying code.
By the \emph{previous record} for a problem we mean the fastest algorithm we know of.
{}

\section{Dynamic programming over the decompositions}\label{sec:apps-dp}

This section obtains exact algorithms by running dynamic programs over the path
decompositions of \cref{thm:algo}. The running time of such a program is exponential
in the width of the decomposition, so a smaller width bound gives a faster algorithm
directly.

\subsection{Problems, dynamic programs and the decompositions used}
\label{sec:max2csp-def}

\paragraph{Max $(r,2)$-CSP.}
An instance of Max $(r,2)$-CSP consists of a simple graph $G$, the \emph{constraint
graph}, whose vertices are the \emph{variables}; a score function $s_v\colon[r]\to\R$
for every variable $v$; and a score function $s_{uv}\colon[r]^2\to\R$ for every edge
$uv\in E(G)$. The task is to find an assignment $\varphi\colon V(G)\to[r]$ maximizing
\[
  \sum_{v\in V(G)}s_v(\varphi(v))+\sum_{uv\in E(G)}s_{uv}(\varphi(u),\varphi(v)).
\]
Following Scott and Sorkin we write $r$ for the domain size.
We write $n$
for the number of variables and $m$ for the number of constraints. Several constraints
on the same pair of variables are added into one, so the constraint graph is simple and
$|E(G)|\le m$. Max 2-SAT ($r=2$, with $m$ the number of $2$-clauses; unit
clauses become vertex scores) and Max $k$-Colorable Subgraph ($r=k$) are special
cases.

\paragraph{Semiring, ring and polynomial versions.}
Let $(\mathbb S,\oplus,\otimes)$ be a commutative semiring. The \emph{semiring
version} of the problem has score functions with values in $\mathbb S$ and asks for
\begin{equation}\label{eq:apps-partition}
  Z=\bigoplus_{\varphi\colon V(G)\to[r]}\ \bigotimes_{v\in V(G)}s_v(\varphi(v))\otimes
  \bigotimes_{uv\in E(G)}s_{uv}(\varphi(u),\varphi(v)).
\end{equation}
More generally, each variable $v$ may have its own finite domain $D_v$ in place of
$[r]$; the score functions are then defined on $D_v$ and on $D_u\times D_v$, and
\eqref{eq:apps-partition} runs over the assignments with $\varphi(v)\in D_v$ for every
$v$. Max $(r,2)$-CSP is the case of the semiring $(\R\cup\{-\infty\},\max,+)$; counting
the solutions of a binary constraint satisfaction problem is the case
$(\N,+,\times)$ with scores in $\{0,1\}$. Scott and Sorkin \cite{ScottSorkin2009}
call the version over a commutative ring the \emph{ring CSP} (RCSP), and the version
over a polynomial ring the \emph{polynomial CSP} (PCSP). These cover, among others,
counting cuts of every size and the Ising partition function \cite{ScottSorkin2009}. When we say that such a version is solved in time $T$, we mean
$T$ semiring operations; for the applications just named, each ring element has a
representation of polynomial size and each operation takes polynomial time.

The following dynamic programs are standard; we state them for path
decompositions in the form needed below. Part~(a) is the dynamic program behind
\cite[Lemmas~2 and~4]{BerendsohnKM2021} and \cite[\S9]{ScottSorkin2009}; the
$3^{w}$ dynamic program for dominating sets is due, for tree decompositions, to van
Rooij, Bodlaender and Rossmanith \cite{vanRooijBR2009}, and part (c) is standard and folklore.
In this section, vertex weights are real numbers.

\begin{lemma}[dynamic programs over a path decomposition]\label{lem:apps-dp}
  Let $G$ be a graph with $n$ vertices, given with a path decomposition of width $w$ in
  interval form.
  \begin{enumerate}[label=\textup{(\alph*)}]
    \item\label{it:apps-dp-csp} For a semiring version with constraint graph $G$ and
      domains of size at most $r$, the value $Z$ of~\eqref{eq:apps-partition} can be
      computed with $O(n\,(w+2)\,r^{w+1})$ semiring operations, keeping $O(r^{w+1})$
      semiring elements at a time; for Max $(r,2)$-CSP with constraint graph $G$, an
      optimal assignment can be found in time and space $r^{w+1}\poly(n)$.
    \item\label{it:apps-dp-ds} One can count the dominating sets of each size, find a
      dominating set of minimum weight and count the dominating sets of minimum weight,
      in time and space $3^{w+1}\poly(n)$.
    \item\label{it:apps-dp-is} One can count the independent sets, find a maximum
      weight independent set and count the maximum weight independent sets in time and
      space $2^{w+1}\poly(n)$.
  \end{enumerate}
\end{lemma}

Each part keeps a table indexed by the partial solutions on the vertices of the
current bag along a nice sequence of introduce and forget steps.%

\paragraph{The decompositions.}
Most results of this section run one of the dynamic programs of \cref{lem:apps-dp},
the others algorithms from the literature, over a path decomposition. The decompositions are in
interval form and are computed in polynomial time by the deterministic algorithm of
\cref{thm:algo-precise}\ref{it:algo:D}; when we cite \cref{thm:algo} below, we mean
this algorithm. For a graph $G$ the width is at most
$W(G)+\rho(n)$, with $\rho(n)=o(n)$, where $W$ is one of the following bounds
(\cref{thm:cubic,thm:deg4,thm:deg5,thm:main} and \cref{cor:edges}):
\begin{itemize}
  \item $\cthree n$ if $\Delta(G)\le3$, with $\cthree=\frac3{4\pi}\arccos\frac56
    <0.1398222$;
  \item $F_4n$, $F_3n_3+0.2575\,n_4$ or $\frac5{32}n_3+\frac14n_4$ if $\Delta(G)\le4$,
    with $F_4<0.24577198$ (\cref{prop:F3F4}); we write
    $\beta_4(G):=\min\{F_4n,\ F_3n_3+0.2575\,n_4,\ \frac5{32}n_3+\frac14n_4\}$;
  \item $a_3n_3+a_4n_4+a_5n_5$ for each of the triples $\tripP,\tripR,\tripS$ if
    $\Delta(G)\le5$;
  \item $\beta(G):=\sum_{d\ge3}a_dn_d$ for every $G$, with the coefficients $a_d$ of
    \cref{thm:main};
  \item $\frac7{48}m$ for every $G$.
\end{itemize}
Running \cref{thm:algo} once for each of several bounds and keeping the narrowest
decomposition realizes their minimum, for instance $\beta_4(G)$. Since the width is at
most $W(G)$ plus the remainder, a dynamic program with $c^{w+1}\poly(n)$ steps runs in time
$c^{W(G)+o(n)}\poly(n)$ (with $o(m)$ for the bound $\frac7{48}m$), for every constant
$c$ and also for $c=r$ growing with the input.

Two parts of \cref{thm:max2csp} evaluate a bound on the final graph $R$ of a maximal sequence of
reduction operations (\cref{def:reductions}) applied to $G$; then $R$ is empty or has
minimum degree at least~$3$ by~\ref{red:max}. Reduction operations increase no degree
by~\ref{red:deg}, so they do not increase the number of edges either. \Cref{lem:undo},
with no vertex deletions, turns a path decomposition of $R$ of width $w$ in interval
form into one of $G$ of width at most $w+3\log_2n+1$, in polynomial time.

\subsection{Pathwidth in terms of the vertices of degree three}\label{sec:apps-dp-n3}

\Cref{thm:cubic} bounds the pathwidth of a graph of maximum degree~$3$ in terms of the
number of all its vertices. Some algorithms, such as the algorithm for counting
dominating sets of \cref{sec:apps-ds}, need a bound in the number of vertices of
degree~$3$ alone; the polynomial-space analyses of \cref{sec:apps-smc} use the
separator version of such a bound in \cref{lem:sep}. The next lemma converts \cref{thm:cubic} into such a bound: the
vertices of degree at most~$2$ add only a term of order $\log n_3$ to the width.

\begin{restatable}[pathwidth by the number of degree-$3$ vertices]{lemma}{lemnthree}\label{lem:n3}
  There is a nondecreasing function $\rho\colon\N\to\R$ with $\rho(k)=o(k)$ such that
  every graph $H$ with $\Delta(H)\le3$ satisfies
  \[
    \pw(H)\le\cthree\,n_3(H)+\rho\bigl(n_3(H)\bigr).
  \]
  A path decomposition of this width can be computed in polynomial time.
\end{restatable}

The proof, in \cref{app:n3}, removes the trees outside the $2$-core, suppresses the
vertices of degree~$2$ of the core, and applies \cref{thm:cubic} to the resulting cubic
multigraph; putting back the suppressed paths costs $2$ in the width, and the trees
$O(\log n_3)$.

\subsection{Max \texorpdfstring{$(r,2)$}{(r,2)}-CSP and its relatives}\label{sec:apps-max2csp}

The first theorem is the most direct consequence of the width bounds: Max
$(r,2)$-CSP, and every semiring version of it, is solved by the dynamic program of
\cref{lem:apps-dp}\ref{it:apps-dp-csp} over the narrowest available decomposition of
the constraint graph. In terms of the number $m$ of constraints, this lowers the
exponent of the previous record $r^{(13/75)m+o(m)}$~\cite[Cor.~21]{ScottSorkin2007}
to $\frac7{48}m+o(m)$. Part~\ref{it:max2csp-beta} of the theorem is the most general statement; the
other parts are stated in the parameters of the previous records.
Part~\ref{it:max2csp-avg} uses the average degree $\bar d:=2|E(G)|/n$ of the
constraint graph $G$ and the concave function $A\colon[0,\infty)\to\R$ defined
before \cref{cor:avg-degree}: $A(x)=\frac7{96}x$ for $0\le x\le6$, and $A$ interpolates
the coefficients $a_d$ linearly between consecutive integers $d\ge6$.

\begin{theorem}[Max $(r,2)$-CSP]\label{thm:max2csp}
  Let $G$ be the constraint graph of an instance of Max $(r,2)$-CSP, or of a semiring
  version, with $n$ variables and $m$ constraints. The instance can be solved in time
  and space
  \begin{enumerate}[label=\textup{(\arabic*)}]
    \item\label{it:max2csp-beta} $\Ostar\bigl(r^{\beta(R)+o(n)}\bigr)$ for the final
      graph $R$ of some maximal sequence of reduction operations applied to $G$, and hence
      $\Ostar\bigl(r^{\beta(G)+o(n)}\bigr)$;
    \item\label{it:max2csp-edges} $\Ostar\bigl(r^{(7/48)m+o(m)}\bigr)$;
    \item\label{it:max2csp-deg34} $\Ostar\bigl(r^{\cthree n_3+o(n)}\bigr)$ if
      $\Delta(G)\le3$, and $\Ostar\bigl(r^{\beta_4(G)+o(n)}\bigr)$ and
      $\Ostar\bigl(r^{3F_4m/(2+16F_4)+o(m)}\bigr)$ if $\Delta(G)\le4$;
    \item\label{it:max2csp-avg} $\Ostar\bigl(r^{n\,A(\bar d)+o(n)}\bigr)$, and hence
      \[
        \Ostar\Bigl(r^{(1-\frac{111}{28(\bar d+1)})n+o(n)}\Bigr)\ \text{if }\bar d\ge7,
        \qquad
        \Ostar\Bigl(r^{(1-\frac{4}{\bar d+1})n+o(n)}\Bigr)\ \text{if }\bar d\ge11.
      \]
  \end{enumerate}
\end{theorem}

Numerically, $\frac{48}7=6.857\ldots$, so part~\ref{it:max2csp-edges} reads
$r^{m/6.857+o(m)}$, and $\frac{3F_4}{2+16F_4}<0.12429$. For cubic constraint graphs
($m=\frac32n$), part~\ref{it:max2csp-deg34} reads $r^{(2\cthree/3)m+o(m)}$ with
$\frac{2\cthree}3<0.09322$.

\begin{proof}
  In each part we compute a path decomposition of the constraint graph and run
  \cref{lem:apps-dp}\ref{it:apps-dp-csp} with $D_v=[r]$; its running time and space
  are $r^{w+1}\poly(n)$ for a decomposition of width $w$, so it suffices to bound $w$.
  Isolated vertices of $G$ are handled separately in time $O(rn)$ (each contributes a
  factor $\bigoplus_as_v(a)$), so we may assume that $G$ has none; then $n\le2|E(G)|\le
  2m$.

  \ref{it:max2csp-beta} Apply reduction operations to $G$ until none applies, and let
  $R$ be the final graph. By \cref{thm:main} and \cref{thm:algo} applied to $R$, we
  compute a decomposition of $R$ of width at most $\beta(R)+\rho(|V(R)|)$, and by
  \cref{lem:undo} with no deletions one of $G$ of width at most
  $\beta(R)+\rho(|V(R)|)+3\log_2n+1=\beta(R)+o(n)$, since $\rho$ is nondecreasing
  (\cref{thm:algo-precise}). Reduction operations only remove vertices and do not
  increase degrees by~\ref{red:deg}, and $0\le a_d\le a_{d+1}$ for all $d\ge0$ (with
  $a_0=a_1=a_2=0$); hence $\beta(R)\le\beta(G)$.

  \ref{it:max2csp-edges} By \cref{thm:algo-precise} (general case), $w\le\frac7{48}|E(G)|+
  \rho(2|E(G)|)\le\frac7{48}m+o(m)$.

  \ref{it:max2csp-deg34} If $\Delta(G)\le3$, \cref{lem:n3} gives $w\le\cthree
  n_3(G)+o(n)$. If $\Delta(G)\le4$, \cref{thm:deg4} and \cref{thm:algo} give
  $w\le\beta_4(G)+o(n)$, by running the algorithm for each of the three forms and
  keeping the narrowest decomposition. For the bound in $m$, let $R$ be the final graph
  of a maximal sequence of reduction operations applied to $G$; it is empty or has
  $\delta(R)\ge3$, and $\Delta(R)\le4$. If $R$ is empty, $w=O(\log n)=o(m)$ by
  \cref{lem:undo} with no deletions. Otherwise put $n':=|V(R)|$, $m':=|E(R)|$ and
  $x:=n_4(R)/n'\in[0,1]$. Then $n_3(R)=(1-x)n'$ and $2m'=3n_3(R)+4n_4(R)=(3+x)n'$, so
  \[
    \frac{F_4n'}{m'}=\frac{2F_4}{3+x},\qquad
    \frac{\frac5{32}n_3(R)+\frac14n_4(R)}{m'}=\frac{5+3x}{16(3+x)} .
  \]
  The first ratio decreases in $x$, and the second increases, since
  $\frac{d}{dx}\frac{5+3x}{3+x}=\frac4{(3+x)^2}>0$. Hence the minimum of the two is at
  most their common value at the crossing point $x^*=\frac{32F_4-5}3\in(0,1)$, which is
  $\frac{2F_4}{3+x^*}=\frac{3F_4}{2+16F_4}$. This expression increases in $F_4$, and
  $F_4<0.24577198$ gives $\frac{3F_4}{2+16F_4}<0.12429$ (at the crossing,
  $x^*\approx0.9549$). So \cref{thm:deg4}\ref{thm:deg4:a},\ref{thm:deg4:c} and
  \cref{thm:algo} applied to $R$ give a decomposition of $R$ of width at most
  $\frac{3F_4}{2+16F_4}m'+\rho(n')$, where $n'\le\frac23m'$ and $m'\le|E(G)|\le m$,
  because reduction operations do not increase the number of edges. By
  \cref{lem:undo} with no deletions, $w\le\frac{3F_4}{2+16F_4}m+o(m)+3\log_2(2m)+1$, which is
  $\frac{3F_4}{2+16F_4}m+o(m)$.

  \ref{it:max2csp-avg} Part~\ref{it:max2csp-beta} gives the running time
  $\Ostar(r^{\beta(G)+o(n)})$ for the given instance. By \cref{cor:avg-degree}, applied
  to the constraint graph $G$ with all its $n$ vertices (isolated ones included),
  $\beta(G)=\sum_{d\ge3}a_dn_d\le n\,A(\bar d)$, where
  $A(\bar d)\le1-\frac{111}{28(\bar d+1)}$ if $\bar d\ge7$ and
  $A(\bar d)\le1-\frac4{\bar d+1}$ if $\bar d\ge11$.
\end{proof}

This improves the previous records in sparse instances.
Other methods solve dense instances faster. For
integer scores of polynomial size, Williams's algorithm \cite{Williams2005} runs
in time $\Ostar(r^{\omega n/3})$, where $\omega$ is the matrix multiplication
exponent (domain $r$: \cite[Thm.~18]{ScottSorkin2009}).
With $\omega\le\frac94$~\cite{OpenAI2026omega}, which Pass
proves with integer arithmetic~\cite[Theorem~1.1]{Pass2026}, as Williams's algorithm
needs, this algorithm is faster than ours when $\bar d>15$.
For very large $\bar d$, an
algorithm of Golovnev and Kutzkov has a smaller exponent
\cite[p.~942]{EdwardsM15}: for Boolean domains it runs in polynomial space and time
$\Ostar\bigl(2^{(1-\alpha\ln\bar d/\bar d)n}\bigr)$ for every constant
$\alpha\in(0,1)$ and every sufficiently large $\bar d=o(n)$
\cite[Thm.~4]{GolovnevK14}. Their threshold on $\bar d$ is not explicit, so we do not
determine that crossover. Hence part~\ref{it:max2csp-avg} is a record only for
moderate $\bar d$, and for integer scores of polynomial size only for
$\bar d\le15$.

\subsection{Domination, independent sets and vertex covers in graphs of bounded degree}
\label{sec:apps-bdd}

This subsection runs dynamic programs for domination, independent sets and five
further local problems over the decompositions of \cref{lem:n3} (for $\Delta\le3$) and
\cref{thm:deg4} (for $\Delta\le4$), and an algorithm for Odd Cycle Transversal over
the orderings of \cref{thm:cut-sweep-p}. We use the dynamic programs of
\cref{lem:apps-dp}\ref{it:apps-dp-ds},\ref{it:apps-dp-is} and five previous algorithms.
Most previous bounds come from the same dynamic programs over decompositions of width
$\frac n6$ for $\Delta\le3$~\cite{FominHoie2006} and $\frac n3$ for
$\Delta\le4$~\cite{FominGaspersSaurabhStepanov2009}, which the new widths
$\cthree n_3+o(n)\le\cthree n+o(n)$ and $\beta_4(G)+o(n)\le F_4n+o(n)$ replace. For
instance, the dominating sets of each size of
a graph with $\Delta\le3$ are counted in time $O(1.1661^n)$ instead of
$3^{n/6}<1.2010^n$. %

\begin{theorem}[bounded-degree graph problems]\label{thm:bdd}
  Let $G$ be a graph with $n$ vertices.
  \begin{enumerate}[label=\textup{(\alph*)}]
    \item\label{it:bdd-ds} If $\Delta(G)\le3$, then in time and space
      $\Ostar(3^{\cthree n_3+o(n)})=O(1.1661^n)$ one can count the dominating sets of $G$
      of each size, find a minimum weight dominating set and count the dominating sets of
      minimum weight.
    \item\label{it:bdd-mwis} A maximum weight independent set of $G$ can be found, and the
      maximum weight independent sets counted, in time and space $\Ostar(2^{\cthree n_3+o(n)})=O(1.1018^n)$ if $\Delta(G)\le3$, and
      $\Ostar(2^{\beta_4(G)+o(n)})=O(1.1858^n)$ if $\Delta(G)\le4$.
    \item\label{it:bdd-local} If $\Delta(G)\le3$, then, in time and space
      $\Ostar(2^{\cthree n_3+o(n)})=O(1.1018^n)$, one can find $k$ vertices that cover the
      most edges (Max $k$-Vertex Cover); in time and space
      $\Ostar(3^{\cthree n_3+o(n)})=O(1.1661^n)$, a minimum independent dominating set; and
      in time and space $\Ostar(6^{\cthree n_3+o(n)})=O(1.2848^n)$, a minimal dominating
      set of maximum size (Upper Domination). If $\Delta(G)\le4$, a minimum independent
      dominating set can be found in time and space $\Ostar(3^{\beta_4(G)+o(n)})=O(1.3100^n)$.
    \item\label{it:bdd-eds} If $\Delta(G)\le3$, a minimum edge dominating set and a
      minimum maximal matching of $G$ can be found in time and space
      $\Ostar(3^{\cthree n_3+o(n)})=O(1.1661^n)$.
    \item\label{it:bdd-oct} A minimum odd cycle transversal of $G$ can be found by a
      deterministic algorithm in time $\Ostar(2^{m/6+o(n)})=O(1.2600^n)$ if $\Delta(G)\le4$ and in time
      $\Ostar(2^{\gamma_5m+o(n)})=O(1.4262^n)$ if $\Delta(G)\le5$, where $m=|E(G)|$ and
      $\gamma_5<0.20484$ is the constant of \cref{thm:cut-sweep-p}.
  \end{enumerate}
\end{theorem}

\begin{proof}
  If $\Delta(G)\le3$, \cref{lem:n3} gives a path decomposition of width at most
  $\cthree n_3+o(n)$ in polynomial time; if $\Delta(G)\le4$, \cref{thm:deg4} and
  \cref{thm:algo} give one of width at most $\beta_4(G)+o(n)$. Run
  \cref{lem:apps-dp}\ref{it:apps-dp-ds} for~\ref{it:bdd-ds} and
  \cref{lem:apps-dp}\ref{it:apps-dp-is} for~\ref{it:bdd-mwis}. %
  For~\ref{it:bdd-local}
  we run, on these decompositions, the dynamic program for Max $k$-Vertex Cover in time
  $\Ostar(2^{\pw})$ and exponential space that Della Croce and Paschos combine with the
  bound of Fomin and H{\o}ie~\cite[Corollary~2]{DellaCroceP2012}, the dynamic program for
  Independent Dominating Set in time $O(n^33^{\tw})$ of van
  Rooij~\cite[Proposition~11.11]{vanRooij2011}, and that for Upper Domination in time
  $O(6^{\pw}\pw)$ of Dublois, Lampis and Paschos~\cite[Theorem~10]{DubloisLP2022}.
  For~\ref{it:bdd-eds} we run the dynamic program of van Rooij and Bodlaender, which finds
  a minimum weight maximal matching in time $\Ostar(3^{\pw})$ on a path decomposition of
  width $\pw$~\cite[Lemma~3]{vanRooijB2012}. A minimum maximal matching is an edge
  dominating set, and a minimum edge dominating set can be turned in polynomial time into
  a maximal matching of the same size~\cite{YannakakisG1980}.
  For~\ref{it:bdd-oct}, \cref{thm:alg-cut-sweep} computes, in deterministic polynomial
  time, an ordering of cutwidth at most $\frac m6+o(n)\le\frac n3+o(n)$ or
  $\gamma_5m+o(n)\le\frac52\gamma_5n+o(n)$, and Bojikian, Chekan, Hegerfeld and Kratsch
  find a minimum odd cycle transversal deterministically in time
  $2^{\operatorname{ctw}}n^{O(1)}$ for a given ordering of cutwidth
  $\operatorname{ctw}$~\cite[Theorem~36]{BojikianCHK2022arXiv}, with
  $2^{1/3}<1.25993$ and $2^{5\gamma_5/2}<1.42611$, where $\gamma_5=\frac1\pi\arccos\frac45$. The
  numerical bases follow from $n_3\le n$, $F_4n$ being one of the three forms,
  $3^{\cthree}<1.16604$, $2^{\cthree}<1.10177$, $6^{\cthree}<1.28471$, $2^{F_4}<1.18573$ and
  $3^{F_4}<1.30998$.
\end{proof}

\subsection{Counting subgraphs}\label{sec:apps-subgraphs}

Curticapean, Dell and Marx reduce counting the subgraphs of a host graph that are
isomorphic to a pattern to counting homomorphisms from the graphs of the \emph{spasm}
of the pattern (defined below), and the treewidth of these graphs governs the running
time. For a pattern with $k$ edges, their bound $k^{O(k)}n^{0.174k+o(k)}$ comes from
the Scott--Sorkin bound $\frac{13}{75}k+o(k)$ on the treewidth of graphs with $k$
edges; \cref{rem:edges-m} lowers the exponent to $\frac7{48}k+o(k)$
(\cref{thm:subgraphs}).

For a graph $H$, the spasm of $H$ is the set of graphs $H/\pi$ obtained from $H$ by
choosing a partition $\pi$ of $V(H)$ into independent sets and identifying the vertices
of each part, keeping one copy of parallel edges \cite{CurticapeanDM2017}. Since every
part is independent, $H/\pi$ is simple without loops, and every edge of $H/\pi$ is the
image of an edge of $H$, so $|E(H/\pi)|\le|E(H)|$.

\begin{theorem}[counting subgraphs]\label{thm:subgraphs}
  There is a function $\rho(k)=o(k)$ such that, given a pattern graph $H$ with $k$ edges
  and a host graph $G$ with $n$ vertices, the number of subgraphs of $G$ isomorphic to
  $H$ can be computed in time $k^{O(k)}\,n^{\frac7{48}k+\rho(k)}$.
\end{theorem}

In particular this holds for counting paths, cycles and matchings with $k$ edges.
Numerically, $\frac7{48}<0.14584$.

\begin{proof}
  By \cite[Thm.~1.1]{CurticapeanDM2017},
  the number of subgraphs of $G$ isomorphic to $H$ can be computed in time $k^{O(k)}
  n^{t+1}$, where $t$ is the largest treewidth of a graph in the spasm of $H$. Every
  graph $F$ in the spasm is simple and has at most $k$ edges. By \cref{rem:edges-m},
  $\tw(F)\le\pw(F)\le\frac7{48}|E(F)|+\rho(|E(F)|+1)$, where $\rho$ is the
  nondecreasing remainder of \cref{rem:edges-m}; so
  $t\le\frac7{48}k+\rho(k+1)$, and the theorem holds with the function
  $k\mapsto\rho(k+1)+1$ in place of $\rho$.
  (The algorithm needs tree decompositions of the graphs of the spasm; they can be
  computed by brute force in time depending on $k$ only, or by \cref{thm:algo}.)
\end{proof}

\subsection{Bucketing: permutation patterns and \texorpdfstring{$k$}{k}-opt}
\label{sec:apps-bucketing}

This subsection lowers two exponents of the form $\frac k4$ to $\frac{253}{1215}k+o(k)$
(note that $\frac{253}{1215}<0.2083$): for
Permutation Pattern Matching with a pattern of length~$k$, from the bound $n^{k/4}$ of
Berendsohn, Kozma and Marx~\cite{BerendsohnKM2021}, and for $k$-opt Optimization for
TSP, from the bound $n^{(1/4+\eps_k)k}$, with $\eps_k\to0$, of Cygan, Kowalik and
Socała~\cite{CyganKS2019}.
On graphs of bounded degree, \cref{rem:edges-m} also lowers the exponent $(\frac{23}{135}
+\eps_k)k$ of $k$-opt Optimization~\cite{BonnetIJK2019} to $(\frac7{48}+\eps_k)k$
(\cref{thm:perm}). We first define the two problems.

In \emph{Permutation Pattern Matching} the input is a pattern $\pi\colon[k]\to[k]$ and
a text $\tau\colon[n]\to[n]$, both permutations, with $k\le n$. An \emph{occurrence} of
$\pi$ in $\tau$ is a strictly increasing map $f\colon[k]\to[n]$ such that
$\tau(f(i))<\tau(f(j))$ if and only if $\pi(i)<\pi(j)$; the decision version asks
whether an occurrence exists, and the counting version for their number. In
\emph{$k$-opt Optimization} for TSP the input is an edge-weighted graph with a
Hamiltonian cycle, and the task is to find a $k$-opt move (removing $k$ edges of the
cycle and adding $k$ edges so as to obtain a Hamiltonian cycle) of maximum gain
\cite{CyganKS2019,BonnetIJK2019}.

Cygan, Kowalik and Socała \cite{CyganKS2019} introduced a \emph{bucketing} argument for
$k$-opt, and Berendsohn, Kozma and Marx \cite{BerendsohnKM2021} transferred it to
Permutation Pattern Matching. It trades the number of guesses against the pathwidth of
a graph made of a cycle cover (or a path) and some edges of the path $1,2,\dots,k$:
coarser buckets mean fewer guesses but larger domains. Both previous bounds $n^{k/4}$
come from a coefficient $\frac13$ for graphs of maximum degree~$4$. A single smaller
coefficient for maximum degree~$4$, such as $F_4$, improves the bound obtained without
bucketing to $n^{F_4k}$, but it does not combine with bucketing to anything better:
inside the trade-off, the pathwidth bound has to grow with the number of path edges.
The per-degree bounds of \cref{thm:deg4}\ref{thm:deg4:b},\ref{thm:deg4:c} and
\cref{thm:deg5} have exactly this property. We use the triple $\tripR$ of
\cref{thm:deg5}, with $n_5=0$; within our argument, adding the bounds of
\cref{thm:deg4,thm:switched4} does not lower the exponent further (see
\cref{app:bucketing}). We
first bound the pathwidth of the graphs that arise (\cref{lem:cyclecover}), then solve
the trade-off (\cref{lem:tradeoff}), and finally apply both lemmas to the two problems
(\cref{thm:perm}).

For an integer
$k\ge1$ and a real $s\in[0,k]$ put
\[
  p(s):=\min\bigl\{0.291\,s,\ 0.038\,k+0.215\,s\bigr\};
\]
$k$ is fixed wherever $p$ is used. We write $P_k:=\{\{i,i+1\}:i\in[k-1]\}$ for the edge
set of the path $1,2,\dots,k$.

\begin{restatable}[cycle cover plus path edges]{lemma}{lemcyclecover}\label{lem:cyclecover}
  There is a function $\rho(k)=o(k)$ such that the following holds. Let $C$ be a
  multigraph on $[k]$, possibly with loops, of maximum degree at most $2$, and let
  $A\subseteq P_k$ with $|A|=s$. Then the simple graph $H$ underlying $C\cup A$
  satisfies $\pw(H)\le p(s)+\rho(k)$, and a path decomposition of $H$ of this width can
  be computed in polynomial time.
\end{restatable}

The bucketing exponent is a maximum over $s$ of a weighted sum of the number $k-s$ of
guessed buckets and the width bound $p(s)$; the weight $\alpha$ is the exponent of the
bucket width. The next lemma chooses $\alpha$.

\begin{restatable}[the trade-off]{lemma}{lemtradeoff}\label{lem:tradeoff}
  Let $\alpha=\frac{200}{243}$. For every $0\le s\le k$,
  \[
    (1-\alpha)(k-s)+\alpha\,p(s)\le\tfrac{253}{1215}k .
  \]
  Moreover, no $\alpha\in[0,1]$ gives a smaller bound than $\frac{253}{1215}k$ on
  $\max_{0\le s\le k}[(1-\alpha)(k-s)+\alpha p(s)]$.
\end{restatable}

The proofs of the two lemmas, and a check that the other bounds of Part~I do not
lower the exponent within this argument, are in \cref{app:bucketing}.

The two lemmas give the following bounds for the problems defined at the start of this
subsection.

\begin{restatable}[permutation patterns and $k$-opt]{theorem}{thmperm}\label{thm:perm}
  There are a function $\rho(k)=o(k)$ and numbers $\eps_k\to0$ such that:
  \begin{enumerate}[label=\textup{(\alph*)}]
    \item\label{it:perm-ppm} the decision and counting versions of Permutation Pattern
      Matching, for a pattern of length $k$ and a text of length $n$, can be solved in
      time $2^{O(k)}\,n^{\frac{253}{1215}k+\rho(k)}$;
    \item\label{it:perm-kopt} for every fixed $k$, $k$-opt Optimization can be solved in
      time $O(n^{(253/1215+\eps_k)k})$ on $n$-vertex edge-weighted graphs, and in time
      $O(n^{(7/48+\eps_k)k})$ on those of bounded maximum degree.
  \end{enumerate}
\end{restatable}

The proof, in \cref{app:bucketing}, follows Berendsohn, Kozma and Marx for~\ref{it:perm-ppm}
and Cygan, Kowalik and Socała for~\ref{it:perm-kopt}, with \cref{lem:cyclecover,lem:tradeoff}
in place of their pathwidth bounds, and with \cref{rem:edges-m} for bounded degree.

{}
\section{Separate, Measure and Conquer in polynomial space}\label{sec:apps-smc}

The dynamic programs of \cref{sec:apps-dp} keep a table for every bag of a path
decomposition and therefore use exponential space. In polynomial space the width
bounds enter in a different way. A path decomposition of small width contains a small
balanced separator, and a branching algorithm can prefer to branch on the vertices of
that separator: once the separator is exhausted, the instance falls apart into two
independent halves. Gaspers and Sorkin~\cite{GaspersSorkin2017} built a
measure-and-conquer analysis around this idea, which they call \emph{Separate, Measure
and Conquer}, and Gaspers and Lee~\cite{GaspersLee2023} used it to count independent
sets. In both analyses the separator constant---the coefficient $\frac16$ of Fomin and
H{\o}ie~\cite{FominHoie2006}---enters a mathematical program in one or two places. We
replace it by $\cthree$ and solve the programs again. This gives the following
results, all in polynomial space:
\begin{itemize}
  \item Max $(r,2)$-CSP in time $r^{0.17935n}$ on cubic instances, $r^{0.16984m}$ on
    instances of maximum degree~$4$ and $r^{0.17587m}$ on all instances, up to
    $o(\cdot)$ terms in the exponent, and counting dominating sets of subcubic graphs
    in time $O(1.2178^n)$ (\cref{thm:smc}), against the bounds $r^{n/5}$, $r^{7m/40}$,
    $r^{9m/50}$ and $O(1.2458^n)$ of Gaspers and Sorkin;
  \item counting independent sets in time $O(1.1304^n)$ for maximum degree~$3$, in time
    $O(1.2037^n)$ for maximum degree~$4$ and $O(1.2341^n)$ in general, and computing the
    chromatic number in time $O(2.2341^n)$ (\cref{thm:is}),
    against the bounds $O(1.1389^n)$, $O(1.2070^n)$, $O(1.2356^n)$ and $O(2.2356^n)$ of
    Gaspers and Lee.
\end{itemize}
\Cref{app:is} explains the technique, shows where the separator constant enters, and
proves both theorems.
{}
\section{Counting dominating sets}\label{sec:apps-ds}

This section shows how the smaller pathwidth bounds speed up an algorithm that counts
dominating sets by branching first and then switching to dynamic programming: with
the new bounds, the switch can come one step earlier. We start from the
exponential-space algorithm of Nederlof, van Rooij and van
Dijk~\cite{NederlofRD2014} for counting dominating sets in general graphs.
It counts set covers
by branching, and switches to dynamic programming over a path decomposition of the
incidence graph once the instance is sparse. Its stated running time is
$O(1.5002^n)$. We prove the following theorem.

\begin{restatable}[counting dominating sets]{theorem}{thmds}\label{thm:ds}
  The numbers of dominating sets of each size of an $n$-vertex graph can be computed
  deterministically in time and space $O(1.4815^n)$. If the vertices carry real weights
  with $n^{O(1)}$ distinct subset sums, the same holds for the numbers of dominating
  sets of each size and each weight.
\end{restatable}

Our algorithm is the one of Nederlof, van Rooij and van Dijk with two changes. First,
it switches to dynamic programming one step earlier, when all non-exceptional sets
have size at most~$4$ instead of~$3$. Second, it uses a longer list of exceptional
configurations.

While rerunning the original analysis we found a gap in it: one family of branching
cases is not covered by the previous recurrences, and with the previous weights
some of these cases violate the claimed branching factor. The gap is in the proof, not in the
algorithm: re-optimizing the weights with all cases included shows that the original
algorithm runs in time $O(1.5004^n)$ (\cref{prop:ds-repaired}). \Cref{sec:nrd-gap}
describes the gap; our analysis uses the corrected set of recurrences.

\subsection{The algorithm}\label{sec:nrd-algorithm}

\paragraph{Set cover formulation.}
A \emph{set cover instance} consists of a finite universe $\mathcal U$ of
\emph{elements}, a finite family $\mathcal S$ of subsets of $\mathcal U$ called
\emph{sets}, and a \emph{multiplicity} $\operatorname{mult}(s)\ge1$ for every
$s\in\mathcal S$; a set $s$ stands for $\operatorname{mult}(s)$ interchangeable copies.
A \emph{cover of size $\ell$} is a choice of $\ell$ copies, of possibly several sets,
whose union is $\mathcal U$. The task is to compute the numbers $c_\ell$ of covers of
size $\ell$ for all $\ell$. The \emph{incidence graph} $I$ of the instance is the
bipartite graph on $\mathcal S\cup\mathcal U$ in which $s\in\mathcal S$ and
$e\in\mathcal U$ are adjacent when $e\in s$. We write $d(s)=d_I(s)=|s|$ for the
\emph{size} of a set and $d(e)=d_I(e)$ for the \emph{frequency} of an element (the
number of sets containing it), and $N_I(e)$ for the sets containing~$e$. In the thesis
and in the journal version, sets and elements are the red and blue vertices of a
red-blue graph; this is the same object.

A graph $G$ gives the instance with $\mathcal U=V(G)$ and one set $N[v]$ of
multiplicity one for each $v\in V(G)$ (true twins $u,v$, with $N[u]=N[v]$, give two
identical sets, which the first reduction rule below merges). A vertex set $D$ is
dominating if and only if the copies $\{N[v]:v\in D\}$ form a cover, so $c_\ell$ is the
number of dominating sets of size $\ell$.

\paragraph{Exceptional configurations.}
The algorithm leaves certain elements of frequency~$5$ to the dynamic program. The
\emph{type} of an element $e$ of frequency~$5$ all of whose sets have size at most~$5$
is the tuple $(r_1,\dots,r_5)$, where $r_i$ is the number of sets of size $i$
containing~$e$; thus $r_1+\dots+r_5=5$. (Here and below, $r_i$ counts neighbors of a
  vertex of $I$ by degree; it is unrelated to the domain size $r$ of
\cref{sec:apps-dp}.) Let $\mathcal X$ be a list of types and $\sigma\in\{3,4\}$. An
element is \emph{exceptional} if it has frequency~$5$ and its type is in $\mathcal X$.
A set is \emph{exceptional} if its size lies in $\{\sigma+1,\dots,5\}$ and it contains
an exceptional element. We always assume
\begin{equation}\label{eq:ds-list}
  r_i=0\ \text{ for } i>\sigma+1 \text{ and } r_1\le1 \quad\text{for every type } (r_1,\dots,r_5)\in\mathcal X,
\end{equation}
so that every set that contains an exceptional element has size at most $\sigma+1$;
in particular, every exceptional set has size exactly $\sigma+1$.

The previous algorithm has $\sigma=3$ and the list $\mathcal X_{14}$ of fourteen
types of \cref{tab:ds-exceptional}~\cite[Overview~8.1, p.~150]{vanRooij2011},
\cite[Overview~1, p.~727]{NederlofRD2014}; with it, exceptional sets have size~$4$.
Our algorithm has $\sigma=4$ and the list $\mathcal X_{30}$, which adds sixteen types
with $r_1=1$. Both lists satisfy~\eqref{eq:ds-list}: the types of $\mathcal X_{14}$
have $r_5=0$, and those of $\mathcal X_{30}$ have no set of size above~$5$.

\begin{table}[tb]
  \centering
  \caption{Types $(r_1,\dots,r_5)$ of exceptional elements of frequency~$5$. The
    previous list $\mathcal X_{14}$~\cite[Overview~1]{NederlofRD2014} and the sixteen
  types added in $\mathcal X_{30}$.}\label{tab:ds-exceptional}
  \smallskip
  \begin{tabular}{@{}ll@{}}
    \toprule
    $\mathcal X_{14}$ &
    \begin{tabular}[t]{@{}lllll@{}}
      (1,4,0,0,0) & (1,3,1,0,0) & (1,2,2,0,0) & (1,1,3,0,0) & (1,0,4,0,0)\\
      (1,3,0,1,0) & (1,2,1,1,0) & (0,5,0,0,0) & (0,4,1,0,0) & (0,3,2,0,0)\\
      (0,2,3,0,0) & (0,1,4,0,0) & (0,4,0,1,0) & (0,3,1,1,0) &
    \end{tabular}\\
    \midrule
    added in $\mathcal X_{30}$ &
    \begin{tabular}[t]{@{}llll@{}}
      (1,0,0,4,0) & (1,0,1,3,0) & (1,0,2,1,1) & (1,0,2,2,0)\\
      (1,0,3,0,1) & (1,0,3,1,0) & (1,1,0,2,1) & (1,1,0,3,0)\\
      (1,1,1,1,1) & (1,1,1,2,0) & (1,1,2,0,1) & (1,1,2,1,0)\\
      (1,2,0,1,1) & (1,2,0,2,0) & (1,2,1,0,1) & (1,3,0,0,1)
    \end{tabular}\\
    \bottomrule
  \end{tabular}
\end{table}

\paragraph{The algorithm $\mathcal A_{\sigma,\mathcal X}$.}
On an instance $(\mathcal S,\mathcal U,\operatorname{mult})$ the algorithm applies the
first applicable rule of the following list and recurses. We write $C(\cdot)$ for the
vector $(c_\ell)_\ell$ returned by a recursive call, and
\begin{equation}\label{eq:ds-take}
  \operatorname{take}_s(C)_\ell=\sum_{i=1}^{\operatorname{mult}(s)}\binom{\operatorname{mult}(s)}{i}\,c_{\ell-i}
\end{equation}
for the correction that accounts for choosing at least one of the copies of a set $s$
(with $c_j=0$ for $j<0$)~\cite[Proposition~26]{NederlofRD2014}.
\begin{enumerate}[label=(R\arabic*)]
  \item\label{ds:R1} \emph{Identical sets.} If two sets $s_1\ne s_2$ have the same
    elements, delete $s_2$ and set
    $\operatorname{mult}(s_1):=\operatorname{mult}(s_1)+\operatorname{mult}(s_2)$.
  \item\label{ds:R2} \emph{Forced set.} If an element has frequency~$1$, let $s$ be
    its set; return $\operatorname{take}_s(C(\mathcal S\setminus\{s\},\mathcal
    U\setminus s))$, where every remaining set loses its elements in $s$.
  \item\label{ds:R3} \emph{Subsumed element.} If two elements $e_1\ne e_2$ satisfy
    $N_I(e_1)\subseteq N_I(e_2)$ (every set containing $e_1$ contains $e_2$), delete
    $e_2$ from $\mathcal U$ and from all sets.
\end{enumerate}
If no reduction rule applies, let $s$ be a non-exceptional set of maximum size and $e$
a non-exceptional element of maximum frequency (with $d(s)=0$ or $d(e)=0$ if there is
none).
\begin{enumerate}[label=(\Alph*)]
  \item\label{ds:S} \emph{Switch.} If $d(s)\le\sigma$ and $d(e)\le4$, compute a path
    decomposition of $I$ and count the covers by dynamic programming over it.
  \item\label{ds:B1} \emph{Branch on a set.} Otherwise, if $d(s)>d(e)$ or $d(e)\le4$,
    return $\operatorname{take}_s(C(\mathcal S\setminus\{s\},\mathcal U\setminus s))+
    C(\mathcal S\setminus\{s\},\mathcal U)$ (take $s$, or discard it).
  \item\label{ds:B2} \emph{Branch on an element.} Otherwise return
    $C(\mathcal S,\mathcal U\setminus\{e\})-C(\mathcal S\setminus N_I(e),\mathcal
    U\setminus\{e\})$ (make $e$ optional, minus forbid $e$).
\end{enumerate}
This is \cite[Algorithm~8.2, p.~148]{vanRooij2011} and
\cite[Algorithm~4, p.~725]{NederlofRD2014} when $\sigma=3$ and
$\mathcal X=\mathcal X_{14}$; there, the switch is line~11 and line~12,
respectively. The order of the reduction rules plays no role in the
analysis. 

\emph{How the switch works.} The algorithm monitors two quantities: the largest size
of a non-exceptional set and the largest frequency of a non-exceptional element. While
either is large, branching on that set or element lowers the measure~\eqref{eq:ds-measure}
defined below by a large amount in both branches. Once both are small, each further
branching would lower the measure only a little, but the incidence graph is then
sparse (maximum degree~$5$, with the vertices of degree~$5$ confined to the
exceptional configurations), and its pathwidth is small. The threshold $\sigma$
decides whether sets of size $\sigma+1$ are branched on or handed to the dynamic
program. The previous threshold $\sigma=3$ branches on sets of size~$4$; we take
$\sigma=4$ and hand them to the dynamic program. The exceptional configurations make
the same choice for elements of frequency~$5$: branching on an element whose sets are
mostly very small lowers the measure only a little, and the dynamic program handles
such an element at little extra running time.

\begin{lemma}[correctness]\label{lem:ds-correct}
  For every $\sigma\in\{3,4\}$ and every list $\mathcal X$, the algorithm
  $\mathcal A_{\sigma,\mathcal X}$ terminates and returns $(c_\ell)_\ell$, the numbers of
  covers of each size.
\end{lemma}

\begin{proof}
  Every rule produces instances with fewer vertices in the incidence graph, so the
  recursion terminates. Correctness follows by induction on $|V(I)|$; we check each rule
  in turn. \ref{ds:R1}: a choice of
  copies of $s_1$ and $s_2$ is a choice of copies of the merged set, with the same
  union. \ref{ds:R2}: every cover contains a copy of $s$; covers containing $i\ge1$
  copies of $s$ correspond to a choice of $i$ of the $\operatorname{mult}(s)$ copies
  together with a cover of $\mathcal U\setminus s$ by the other sets, which
  gives~\eqref{eq:ds-take}. \ref{ds:R3}: every cover contains a copy of a set
  containing $e_1$, which contains $e_2$; so a choice of copies covers $\mathcal U$ if
  and only if it covers $\mathcal U\setminus\{e_2\}$. \ref{ds:B1}: covers either contain
  a copy of $s$ (counted as in~\ref{ds:R2}) or none. \ref{ds:B2}: a choice of copies
  that covers $\mathcal U\setminus\{e\}$ either covers $e$ or uses no set containing
  $e$, and the latter choices are exactly the covers of $\mathcal U\setminus\{e\}$ by
  $\mathcal S\setminus N_I(e)$; so the number of covers of $\mathcal U$ of each size is
  the difference. \ref{ds:S}: the dynamic program over a path decomposition of width
  $p$ counts the covers of each size, with multiplicities, for every instance, in time
  $O^*(2^{p})$~\cite[Proposition~8.10]{vanRooij2011},
  \cite[Proposition~27]{NederlofRD2014}; vertices of degree~$0$ of $I$ (empty sets and
  elements of frequency~$0$) are handled by a factor $(1+x)^{\operatorname{mult}(s)}$ in
  the size-generating function, or by returning~$0$.
\end{proof}

\paragraph{The measure.}
The analysis measures an instance by
\begin{equation}\label{eq:ds-measure}
  k=\sum_{e\in\mathcal U}v\bigl(d(e)\bigr)+\sum_{s\in\mathcal S}w\bigl(d(s)\bigr),
\end{equation}
where merged sets count once. The weights satisfy the following conditions, which are
those of the thesis~\cite[p.~151]{vanRooij2011}; write $\Delta v(i)=v(i)-v(i-1)$ and
$\Delta w(i)=w(i)-w(i-1)$.
\begin{enumerate}[label=(W\arabic*)]
  \item\label{ds:W1} $v(0)=v(1)=w(0)=0$; $v(i)=v(7)$ and $w(i)=w(7)$ for $i\ge7$;
    $v(2)>0$ and $w(1)>0$.
  \item\label{ds:W2} $\Delta v(i)\ge0$ for $i\ge2$ and $\Delta w(i)\ge0$ for $i\ge1$.
  \item\label{ds:W3} $\Delta v(i)\ge\Delta v(i+1)$ for $i\ge2$ and
    $\Delta w(i)\ge\Delta w(i+1)$ for $i\ge1$.
  \item\label{ds:W4} $2\,\Delta v(5)\le v(2)$.
\end{enumerate}
By~\ref{ds:W1}--\ref{ds:W2} the weights are nonnegative and nondecreasing, so none
of~\ref{ds:R1}--\ref{ds:R3} increases the measure: each deletes vertices of $I$ and
lowers sizes or frequencies of others. A graph with $n$ vertices gives an instance of
measure at most $(v(7)+w(7))\,n$.

\subsection{Outline of the analysis}\label{sec:ds-outline}

The running-time analysis follows the measure-and-conquer scheme of the
thesis~\cite[\S8.5]{vanRooij2011} in three steps, which \cref{app:ds} carries out for
both thresholds at once. First, branching recurrences bound the number of instances of
each measure on which the dynamic program is called~\cite[Lemma~8.11]{vanRooij2011};
the branching vectors of degree above~$9$ reduce to those of degree at most~$9$.
Second, a linear program in the numbers of sets and elements of each degree bounds the
pathwidth of these instances~\cite[Lemma~8.12]{vanRooij2011}; its objectives are the
pathwidth bounds of Part~I, applied to the incidence graph. Third, the two bounds
combine into the running time $\alpha^{(v(7)+w(7))n}n^{O(1)}$ whenever $2^L<\alpha$
for the value $L$ of the linear program~\cite[Theorem~8.13]{vanRooij2011}.

For $\mathcal A_{4,\mathcal X_{30}}$ we take $\alpha=\frac{13}{10}$ and rational
weights with $v(7)+w(7)=1.4980028$ (\cref{tab:ds-weights}). All $30{,}972$ branching
vectors of degree at most~$9$ hold at $\alpha$, the linear program has value
$L\approx0.37851$ with $2^L<\alpha$, and $\alpha^{1.4980028}<1.4815$. Each of these
steps is checked in exact rational or interval arithmetic. The decisive input is the
coefficient $\frac14$ of \cref{thm:main} at degree~$4$, in place of the $\frac13$ of
Fomin, Gaspers, Saurabh and Stepanov~\cite[Lemma~1]{FominGaspersSaurabhStepanov2009}:
the worst instances of the linear program are close to incidence graphs of cubic
graphs. For the weighted counts, the multiplicity of a set becomes a generating
polynomial in size and weight, and the measure and running time are those of the
unweighted case (\cref{sec:ds-weighted}).
{}
\section{Connectivity problems}\label{sec:apps-connectivity}

Many problems with a global connectivity or acyclicity constraint can be solved in time
single exponential in the width of a given decomposition, and several of them are
NP-hard on graphs of maximum degree $3$ or $4$. This section runs such algorithms on
the decompositions of \cref{thm:algo} and on the orderings of
\cref{cor:alg-cutwidth} and \cref{thm:cut-sweep-p}. This gives new running-time bounds
for these problems on graphs of maximum degree $3$ to $6$, on graphs with a given number
of edges, and for three parameterized problems (\cref{thm:connectivity} and the
paragraph on Almost Forest Deletion at the end of this section).

The algorithms that we use are the following. The Cut\&Count technique of Cygan,
Nederlof, Pilipczuk, Pilipczuk, van Rooij and Wojtaszczyk gives Monte Carlo algorithms
in time $c^{\tw}n^{O(1)}$ for a given tree decomposition of width
$\tw$~\cite[Theorem~1.2]{CyganNPPRW2011arXiv}. The base is $c=3$ for Feedback Vertex
Set, Connected Vertex Cover and Steiner Tree, $c=4$ for Connected Dominating Set and
eight further problems on undirected graphs, and $c=6$ for the directed versions of
Min Cycle Cover, Longest Path and Longest Cycle and for Exact $k$-Leaf Outbranching.
Cygan, Kratsch and Nederlof decide Hamiltonicity by a Monte Carlo algorithm in time
$(2+\sqrt2)^{\pw}n^{O(1)}$ for a given path decomposition of width
$\pw$~\cite[Theorem~4.3]{CyganKN2012arXiv}, and Longest Path in the same
time~\cite[Theorem~4.6, stated with a proof sketch]{CyganKN2012arXiv}. The rank-based
approach of Bodlaender, Cygan, Kratsch and Nederlof gives deterministic algorithms, for
instance in time $(2+2^{\omega/2})^{\pw}n^{O(1)}$ for the traveling salesperson problem
with nonnegative integer edge weights, where $\omega$ is the matrix multiplication
exponent~\cite[Theorem~3.18]{BodlaenderCKN2015}. Bojikian, Chekan, Hegerfeld and
Kratsch~\cite{BojikianCHK2022arXiv} and Bojikian, Chekan and
Kratsch~\cite{BojikianCK2025arXiv} give algorithms of the same kind for a given linear
ordering of cutwidth $\operatorname{ctw}$, with smaller bases.

Combined with the pathwidth bound $\frac n6$ of Fomin and H{\o}ie~\cite{FominHoie2006}
and the per-degree bounds of Fomin, Gaspers, Saurabh and
Stepanov~\cite{FominGaspersSaurabhStepanov2009} ($\frac n3$ for maximum degree~$4$ and
$\frac{13n}{30}$ for maximum degree~$5$), these algorithms gave the fastest algorithms
known to us for several of these problems on graphs of maximum degree $3$, $4$
and~$5$. Cygan et al.\ list the Cut\&Count
bounds~\cite[Corollary~1.9]{CyganNPPRW2011arXiv}. Cygan, Kratsch and
Nederlof~\cite[Corollary~4.5]{CyganKN2012arXiv} state theirs for maximum degree~$3$,
and Bodlaender et al.~\cite[Corollary~3.19]{BodlaenderCKN2015} theirs for cubic
graphs. Kowalik and Majewski tabulate those for Hamiltonicity and the traveling
salesperson problem~\cite[Table~1]{KowalikM2020arXiv}. The pathwidth bound of Fomin,
Gaspers, Saurabh and Stepanov in terms of the numbers of vertices of each degree also
bounds the width of the instances that remain after branching in parameterized
algorithms for Co-Path Packing~\cite{LiuXiao2026} and $k$-Co-Path
Set~\cite{SullivanvdP2016}. With the bounds of Part~I we obtain the following.

\begin{restatable}[connectivity problems]{theorem}{thmconnectivity}\label{thm:connectivity}
Polynomial factors are omitted, all algorithms use exponential space, and the Monte
Carlo algorithms have one-sided error at most $\frac12$.
\begin{enumerate}[label=\textup{(\alph*)}]
  \item\label{it:conn-3tw} Feedback Vertex Set and Connected Vertex Cover can be
  solved by Monte Carlo algorithms in time $2^{m/6+o(n)}\le2^{n/3+o(n)}=O(1.2600^n)$ on
  graphs with $m$ edges and maximum degree $4$, and in time
  $2^{\gamma_5m+o(n)}=O(1.4262^n)$ on graphs of maximum degree $5$, where
  $\gamma_5<0.20484$ is the constant of \cref{thm:cut-sweep-p}. Steiner Tree with unit
  weights can be solved by Monte Carlo algorithms in time
  $3^{\cthree n+o(n)}=O(1.1661^n)$ and $3^{F_4n+o(n)}=O(1.3100^n)$ on graphs of maximum
  degree $3$ and $4$. Feedback Vertex Set can be solved on graphs of maximum degree $6$
  by a Monte Carlo algorithm in time $3^{7n/16+o(n)}=O(1.6172^n)$. Feedback Vertex Set
  can be solved on graphs of maximum degree $4$ deterministically in time
  $5^{F_4n+o(n)}=O(1.4853^n)$.
  \item\label{it:conn-cds} Connected Dominating Set, Connected Feedback Vertex Set,
  Connected Odd Cycle Transversal, Exact $k$-Leaf Spanning Tree (in particular Minimum
  and Maximum Leaf Spanning Tree), Maximum Full Degree Spanning Tree, Graph Metric
  Traveling Salesperson, and the undirected versions of Min Cycle Cover and Longest
  Cycle can be solved by Monte Carlo algorithms in time
  $4^{\cthree n+o(n)}=O(1.2139^n)$, $4^{F_4n+o(n)}=O(1.4060^n)$ and
  $4^{11n/32+o(n)}=O(1.6105^n)$ on graphs of maximum degree $3$, $4$ and $5$.
  Minimum Leaf Spanning Tree (equivalently, Maximum Internal Spanning Tree) can be
  solved on graphs of maximum degree $6$ by a Monte Carlo algorithm in time
  $4^{7n/16+o(n)}=O(1.8341^n)$, and Connected Dominating Set on graphs of maximum
  degree $3$ by a Monte Carlo algorithm in time $3^{\cthree n+o(n)}=O(1.1661^n)$.
  Maximum Leaf Spanning Tree has the bounds of Connected Dominating Set.
  \item\label{it:conn-ham} Hamiltonian Cycle can be solved by Monte Carlo algorithms
  in time $(1+\sqrt2)^{\cthree n+o(n)}=O(1.1312^n)$ on graphs of maximum degree $3$,
  in time $(1+\sqrt2)^{m/6+o(n)}\le(1+\sqrt2)^{n/3+o(n)}=O(1.3416^n)$ on graphs with
  $m$ edges and maximum degree $4$, and in time
  $(2+\sqrt2)^{11n/32+o(n)}=O(1.5252^n)$ on graphs of maximum degree $5$.
  Undirected Longest Path can be solved by Monte Carlo algorithms in time
  $(2+\sqrt2)^{\cthree n+o(n)}=O(1.1874^n)$, $(2+\sqrt2)^{F_4n+o(n)}=O(1.3523^n)$ and
  $(2+\sqrt2)^{11n/32+o(n)}=O(1.5252^n)$ on graphs of maximum degree $3$, $4$ and $5$.
  \item\label{it:conn-tsp} With $\omega\le\frac94$~\cite{OpenAI2026omega,AmanovBZ2026},
  the traveling salesperson problem with nonnegative integer edge weights can be solved
  deterministically in time $(1+2^{\omega/2})^{\cthree n+o(n)}=O(1.1757^n)$ on graphs
  of maximum degree $3$, and in time $(2+2^{\omega/2})^{F_4n+o(n)}=O(1.4214^n)$,
  $(2+2^{\omega/2})^{11n/32+o(n)}=O(1.6352^n)$ and
  $(2+2^{\omega/2})^{7n/16+o(n)}=O(1.8699^n)$ on graphs of maximum degree $4$, $5$
  and $6$.
  \item\label{it:conn-directed} The directed versions of Min Cycle Cover, Longest Path
  and Longest Cycle, and Exact $k$-Leaf Outbranching, can be solved by Monte Carlo
  algorithms in time $6^{\cthree n+o(n)}=O(1.2848^n)$ and $6^{F_4n+o(n)}=O(1.5533^n)$
  on directed graphs whose underlying undirected graph has maximum degree $3$ and $4$.
  \item\label{it:conn-param} Co-Path Packing (delete at most $k$ vertices so that every
  component of the remaining graph is a path) can be solved by a Monte Carlo algorithm
  in time $\Ostar(2.8192^k)$, and $k$-Co-Path Set (delete at most $k$ edges so that
  every component of the remaining graph is a path) by a Monte Carlo algorithm in time
  $\Ostar(1.5423^k)$.
  \item\label{it:conn-edges} On graphs with $m$ edges, Hamiltonian Cycle can be solved
  by a Monte Carlo algorithm in time $(2+\sqrt2)^{7m/48+o(m)}=O(1.1962^m)$, and the
  traveling salesperson problem with nonnegative integer edge weights deterministically in
  time $(2+2^{\omega/2})^{7m/48+o(m)}=O(1.2320^m)$, with $\omega$ as
  in~\ref{it:conn-tsp}. For average degree at most $d$, that is $m\le\frac d2n$, these
  are $O(1.0937^{dn})$ and $O(1.1100^{dn})$.
\end{enumerate}
\end{restatable}

The proof, in \cref{app:connectivity}, runs the algorithms listed above on the path
decompositions of \cref{thm:algo-precise} and on the orderings of
\cref{cor:alg-cutwidth,thm:alg-cut-sweep}. The path-decomposition bounds for maximum degree $5$ and $6$
and those of~\ref{it:conn-param} and~\ref{it:conn-edges} use \cref{thm:main}, and
hence \cref{thm:deg5}. The cutwidth bounds for maximum degree $4$ and $5$ use the
orderings of \cref{thm:cut-sweep-p}, which \cref{thm:alg-cut-sweep} computes in
deterministic polynomial time.

\paragraph{Almost Forest Deletion.} Gowda, Lonkar, Panolan, Patel and Saurabh ask
for at most $k$ vertices whose deletion leaves a graph that becomes a forest after the
deletion of at most $\ell$ further edges. One of their Monte Carlo algorithms runs in
time $\Ostar(3^k\cdot1.78^\ell)$ and exponential
space~\cite[Theorem~1.2(4)]{GowdaLPPS2024}. It applies the
Cut\&Count algorithm to a tree decomposition of width
$k+\frac3{5.769}\ell+O(\log\ell)$, obtained from the bound of Kneis et al.\ on a graph
with at most $3\ell$ edges~\cite[Lemma~3.34]{GowdaLPPS2024}. With
\cref{rem:edges-m} in place of that bound, the width is $k+\frac7{16}\ell+o(\ell)$, and
since $3^{7/16}<1.6172$ the running time becomes $\Ostar(3^k\cdot1.62^\ell)$.

{}
\section{Conclusion and open problems}\label{sec:conclusion}

For twenty years, the bound $\frac n6+o(n)$ of Fomin and H{\o}ie~\cite{FominHoie2006}
was the best bound on the pathwidth of $n$-vertex graphs of maximum degree~$3$, and
many exact algorithms stated their running times through it. We replaced the
bisection behind it by a threshold sweep through correlated Gaussian labels. Comparing
each star with an ideal star, and switching from inner to outer frontiers at the
median, brought the bound down to $\cthree n+o(n)$ with
$\cthree=\frac3{4\pi}\arccos\frac56\approx0.1398$. A charging scheme certified by computer and a bootstrap to large degrees
extended the method to all graphs, with the coefficient $\frac14$ in place of
$\frac13$ at degree~$4$ and the bound $\frac7{48}m$ per edge in place of
$\frac{13}{75}m$. The decompositions can be computed in polynomial time. This solves
the open problem of van Rooij~\cite[Open Problem~14.2]{vanRooij2011}, which asks for
tree decompositions of cubic graphs of width $c_tn+o(n)$, for some constant
$c_t<\frac16$, computable in polynomial time: we obtain $c_t=\cthree$, even for path
decompositions. With the new decompositions, many exact algorithms whose running times
depend on these widths have smaller running-time bounds (\cref{part:apps}). We close
with the questions that the Gaussian sweep leaves open.

\paragraph{The constant for maximum degree $3$.} What is the smallest constant $c$ such
that every graph of maximum degree~$3$ has pathwidth at most $cn+o(n)$? It lies in
$[0.1032,\cthree]$, and the lower end comes from random cubic
graphs~\cite{LichevMitsche2023}. Since $\cthree$ is exactly what the median-switched
sweep gives on cubic graphs of large girth, an improvement must change the rule at
threshold zero, for instance by letting a vertex above the threshold carry the
crossing edges of all lower neighbors whose only upper neighbor it is. On the
\emph{ideal wave}, the centered Gaussian vector on the rooted tree $T_2$ (root $o$
with three children $u_1,u_2,u_3$, each with two children) with covariance
$2^{-d/2}(1+\frac d3)$ at distance $d$, which is the limit of our labels on cubic graphs
of large girth, this rule saves $3p_{\mathrm{ch}}$ per vertex at threshold zero, where
$p_{\mathrm{ch}}=0.00252477819777\ldots$ (certified in interval arithmetic) is the
probability that $u_1$ and $u_2$ both have $o$ as their only neighbor above
zero.\footnote{This is the ideal-tree value of the ``isolated helper'' event of
  Lichev and Mitsche~\cite{LichevMitsche2023}; the number $0.002818666$ printed there is
  not this value, and its authors have confirmed that it was not accurate (L.~Lichev and
D.~Mitsche, personal communication, 2026).}
\begin{conjecture}[ideal-wave bound]\label{conj:ideal-wave}
  Every cubic graph $G$ satisfies
  $\pw(G)\le(\cthree-3p_{\mathrm{ch}})n+o(n)$, where
  $\cthree-3p_{\mathrm{ch}}=0.13224778941\ldots$.
\end{conjecture}
Proving it, or any improvement of $\cthree$, seems to require a quantitative
stability version of the star comparison.

\paragraph{Maximum degree $4$.} For maximum degree~$3$, the median switch lowers the
coefficient of the plain sweep, the maximum $F_3\approx0.1402$ of the ideal $3$-star
function, to the value $\cthree$ of that function at threshold zero. For maximum degree~$4$ we prove $F_4n+o(n)$, where
$F_4\approx0.2458$ is the maximum of the ideal $4$-star function; its value at
threshold zero is $0.2423$. Is there a rule at threshold zero whose width bound on
$4$-regular graphs approaches $0.2423n$? Our median switch, analyzed star by star,
cannot approach $0.2423$: the switch quantity $\Psi$ of a vertex with four neighbors
of degree~$4$ can equal $\frac14>F_4$ (see the discussion after
\cref{lem:switch-mixed}). It does give the bound $\cthree n_3+0.2627\,n_4$
(\cref{thm:switched4}), which is smaller than the bounds of \cref{thm:deg4}
when $n_4<\frac{F_3-\cthree}{0.0052}\,n_3\approx0.0646\,n_3$. This form of the method
cannot lower $0.2627$ much below $0.2618$ (\cref{rem:switched4-coefficient}).

\paragraph{Degree $6$ and the bound per edge.} Can the coefficient $a_6=\frac7{16}$ of
\cref{thm:main} be lowered? For degrees $6$ and more, all our coefficients come from
the bootstrap. The constant $\frac7{48}$ of \cref{cor:edges} is $2a_6/6$, and every
other ratio $2a_d/d$ is at most $\frac{113}{784}$, attained at $d=7$ (proof of
\cref{cor:edges}); so a better coefficient at degree~$6$ alone would lower the bound
per edge.

\paragraph{Treewidth.} For any of the classes considered here, is the worst-case
coefficient of treewidth smaller than that of pathwidth? The Gaussian sweep produces
path decompositions, and it has given us no gain specific to treewidth at leading
order.

{}
\section*{Acknowledgments}
This research was supported by the Australian Government through the Australian Research Council's
Discovery Projects funding scheme (project DP210103849). This research includes computations using the
computational cluster Katana supported by Research Technology Services at UNSW Sydney~\cite{Katana}. We thank the UNSW AI Institute for supporting
the co-algorithmist framework, a framework for human--AI research collaboration in algorithms and theoretical
computer science that is under development.
We thank Yuexi He for interesting discussions around the treewidth and pathwidth of structured (but not inherently sparse) graphs occurring in the context of a parallel research project.
{}
\section*{AI methodology statement}
\addcontentsline{toc}{section}{AI methodology statement}

This research was carried out with assistance from large language model
systems: Codex (OpenAI) in the early stages, and Claude (Anthropic) thereafter, in the
versions ChatGPT 6 Astra, Sol 5.6, Claude Opus~5.5, Sonnet~5.5, Haiku~5.5 and Fable~5.1.
GitHub Copilot assisted in project maintenance. These tools materially affected every
section of the paper.
\begin{itemize}
  \item They proposed and developed ideas in discussion with the author, searched the literature, carried out mathematical
    derivations and proofs, wrote the code for the experiments, the computer certificates and the Lean development, drafted, reviewed, and revised
    the text.
  \item OpenAI Codex was also used for developing the
    research-workflow tooling of the project.
  \item GitHub Copilot was used for code completion, reviews of pull requests and commit messages. Pull
    requests were also reviewed by Claude Haiku 5.5.
\end{itemize}
The author directed the research, contributed ideas, checked the
proofs and the Lean statements, edited the manuscript and extended the literature search. The author verified the correctness and originality of all content, including
references.

{}

\clearpage
\appendix
\crefalias{section}{appendix}
\crefalias{subsection}{subappendix}
\crefalias{subsubsection}{subsubappendix}
\section{Deferred proofs}\label{app:deferred}

This appendix collects the proofs deferred from the main text. The first five
subsections restate the deferred statements, with their numbers in the main text, and
prove them: \cref{lem:reduction} in \cref{app:reduction}; \cref{prop:F3F4} and the
tools of the certificate behind \cref{prop:certificate} in \cref{sec:certificate};
\cref{thm:switched4} in \cref{sec:switch-cert}; \cref{thm:pw-seq,cor:table} in
\cref{app:sequences}; and \cref{thm:algo-precise,cor:alg-cutwidth,thm:alg-cut-sweep}
in \cref{app:algorithms}. The next, \cref{app:alg-switched}, restates nothing: it
completes the proof of \cref{thm:algo-precise} in the cases that use the
median-switched layout. The last, \cref{app:part2}, proves
\cref{lem:n3,lem:cyclecover,lem:tradeoff,thm:perm,thm:connectivity} of Part~II; the
analyses behind \cref{thm:smc,thm:is,thm:ds} have appendices of their own
(\cref{app:is,app:ds}).

\addtocontents{toc}{\protect\setcounter{tocdepth}{2}}
\subsection{Proof of the reduction lemma}\label{app:reduction}

This subsection proves \cref{lem:reduction}, with the reduction operations of
\cref{def:reductions}, their properties \ref{red:deg}--\ref{red:max} and the two
facts \cref{lem:tw-glue,lem:tw-log} of \cref{sec:prelim-reductions}.

\lemreduction*

\begin{proof}
The proof follows the sequence step by step. The vertices removed so far by
reductions induce components in $F$ minus the vertices deleted so far. We maintain
as an invariant that each of these components has treewidth at most $2$ and attaches
to the current graph at no vertex, at one vertex, or at the two ends of an edge. At
the end, a path decomposition of the final graph $R$ is extended to one of $F-D$ by
inserting path decompositions of these components, each of width at most
$3\log_2|V(F)|$ by \cref{lem:tw-log}, next to bags that contain their attachments.
Adding $D$ to every bag then gives the claim.

\emph{The invariants.} Write $n=|V(F)|$. Let $H_t$ be the current graph after $t$
steps, $D_t$ the set of vertices deleted in the first $t$ steps, and $F_t:=F-D_t$. Then
$V(H_t)\subseteq V(F_t)$, but $H_t$ need not be a subgraph of $F_t$, since
\ref{red:R3} adds edges. For a component $C$ of $F_t-V(H_t)$ put $S_C:=N_{F_t}(C)$;
these are vertices of $H_t$. We maintain two invariants:
\begin{enumerate}[label=\textup{(I\arabic*)}]
  \item\label{inv:I1} every edge of $F_t$ with both ends in $V(H_t)$ is an edge of $H_t$;
  \item\label{inv:I2} for every component $C$ of $F_t-V(H_t)$: $|S_C|\le2$; if
  $|S_C|=2$ then $S_C$ spans an edge of $H_t$; and the graph $Q^+_C$ obtained from
  $Q_C:=F_t[C\cup S_C]$ by adding the edge on $S_C$ when $|S_C|=2$ has treewidth at most $2$.
\end{enumerate}
Both hold for $t=0$, when $H_0=F_0=F$ and there are no components.

\emph{A deletion of a vertex $w\in V(H_t)$} removes $w$ from both graphs. Edges of
$F_{t+1}$ inside $V(H_{t+1})$ are edges of $F_t$ inside $V(H_t)$ avoiding $w$, hence
edges of $H_t-w=H_{t+1}$; so \ref{inv:I1} holds. The graph $F_t-V(H_t)$ does not
change, each $S_C$ loses at most the vertex $w$, an edge on $S_C$ that avoids $w$
survives in $H_{t+1}$, and each new $Q^+_C$ is a subgraph of the old one. So
\ref{inv:I2} holds.

\emph{A reduction operation at a vertex $v$.} Let $N:=N_{H_t}(v)$, so $|N|\le2$.
Here $F_{t+1}=F_t$ and $V(H_{t+1})=V(H_t)\setminus\{v\}$. Edges of $F_t$ inside
$V(H_{t+1})$ are edges of $H_t$ not incident with $v$, and these survive in
$H_{t+1}$; so \ref{inv:I1} holds. The components $C$ of $F_t-V(H_t)$ with
$v\notin S_C$ remain components, with the same $S_C$, the same $Q^+_C$, and an edge
on $S_C$ that avoids $v$ and therefore survives. The components $C$ with $v\in S_C$
merge with $v$ into one new component $C'$. We check \ref{inv:I2} for $C'$.
\begin{itemize}
  \item $S_{C'}\subseteq N$. A neighbor of $v$ in $F_t$ outside $C'$ lies in
  $V(H_t)$, so by \ref{inv:I1} it lies in $N$. A neighbor of a merged $C$ outside
  $C'$ lies in $S_C\setminus\{v\}$; if this set is nonempty, $S_C=\{v,u\}$ spans an
  edge of $H_t$ by \ref{inv:I2}, so $u\in N$.
  \item If $|S_{C'}|=2$, then $N=S_{C'}=\{x,y\}$, so $v$ had degree $2$ and the
  operation was \ref{red:R3} or \ref{red:R4}. In both cases $xy\in E(H_{t+1})$.
  \item Let $K$ be the graph on $\{v\}\cup N$ with the edges from $v$ to $N$, and
  with the edge $xy$ if $N=\{x,y\}$; it has at most three vertices, so $\tw(K)\le2$.
  Every edge of $Q_{C'}=F_t[C'\cup S_{C'}]$ lies in $K$ or in some $Q_C$ with $C$
  merged: edges at $v$ go to $N$ or into some merged $C$; edges at a vertex of a merged
  $C$ stay in $C\cup S_C$; and an edge inside $S_{C'}$ exists only if
  $S_{C'}=N=\{x,y\}$, in which case it is the edge $xy$ of $K$. The added edge on
  $S_{C'}$ (if $|S_{C'}|=2$) is also an edge of $K$. Hence $Q^+_{C'}$ is a subgraph of
  the union of $K$ and the graphs $Q^+_C$ for the merged $C$. Each $Q^+_C$ meets $K$
  exactly in $S_C$ (the set $C$ is disjoint from $V(H_t)\supseteq\{v\}\cup N$, and
  $S_C\subseteq\{v\}\cup N$ by the first item), and different $Q^+_C$ meet only inside
  $K$. Each $S_C$ is the vertex $v$ or an edge $vu$ with $u\in N$, which is an edge of
  both $K$ and $Q^+_C$. Gluing the graphs $Q^+_C$ to $K$ one at a time,
  \cref{lem:tw-glue} gives $\tw(Q^+_{C'})\le2$.
\end{itemize}

\emph{The final decomposition.} At the end, $H_T=R$ and $F_T=F-D$. Every component
$C$ of $(F-D)-V(R)$ induces a graph $F[C]\subseteq Q^+_C$ of treewidth at most $2$ on at most $n$ vertices, so
$\pw(F[C])\le3\log_2n$ by \cref{lem:tw-log}. Take a path decomposition
$W_1,\dots,W_s$ of $R$ of width $\pw(R)$ (no bags if $R$ is empty). For each $C$ with
$S_C\ne\emptyset$ choose $i$ with $S_C\subseteq W_i$, which is possible because $S_C$
is a vertex or spans an edge of $R$ by \ref{inv:I2}. Immediately after $W_i$ insert
the bags $W_i\cup P$ for the bags $P$ of a path decomposition of $F[C]$ of minimum
width, the components assigned to the same $i$ one after another. Append the path
decompositions of the components with $S_C=\emptyset$ at the end. The result is a path
decomposition of $F-D$:
\begin{itemize}
  \item every edge inside $V(R)$ is an edge of $R$ by \ref{inv:I1} and lies in some
  $W_i$; every other edge has an end in some $C$ and its other end in $C\cup S_C$, so
  it lies in an inserted bag $W_i\cup P$ or in a bag of $F[C]$;
  \item a vertex of $C$ occurs only in the consecutive bags inserted for $C$;
  \item a vertex $x\in V(R)$ occurs in a consecutive run of the bags $W_i$, and every
  bag inserted after $W_i$ contains $W_i$; so $x$ occurs in the bags inserted after
  $W_i$ exactly when $x\in W_i$, and its bags remain consecutive.
\end{itemize}
Every bag has at most $(\pw(R)+1)+(3\log_2n+1)$ vertices, so
$\pw(F-D)\le\pw(R)+3\log_2n+1$; this also holds when $R$ is empty, since then
$\pw(R)=-1$. Adding $D$ to every bag gives
$\pw(F)\le|D|+\pw(R)+3\log_2n+1$, which implies~\ref{it:reduction:pw}.

Part~\ref{it:reduction:structure} is \ref{inv:I1} and \ref{inv:I2} for $t=T$, since
$H_T=R$, $F_T=F-D$, $S_C=N_{F-D}(C)$ and $F[C]\subseteq Q^+_C$. \qedhere
\end{proof}
{}
\subsection{The star-bound certificate}\label{sec:certificate}

This subsection describes the computer-assisted proof of \cref{prop:certificate}:
the triples $\tripP,\tripR,\tripS$ are admissible for $D=\{3,4,5\}$
(\cref{def:admissible}). Admissibility asks that for \emph{every} threshold
$\tau\in[-3,3]$ there is a transfer vector $t$ such that, for every type $T$ of
degree $d$, the reference star bound $C_T(\tau)$ plus the transfer term
$\lambda_T(t)$ is at most the coefficient $a_d$. The certificate replaces the
continuum of thresholds by finitely many cells. It bounds each reference star bound
from above through a lower Riemann sum of explicit lower bounds for the orthant
infimum $P_{\min}$ of \cref{sec:stars}. The code also certifies the pair
$(F_3,0.2575)$ for $D=\{3,4\}$ directly, although its admissibility follows from that
of $\tripP$ by restriction to $D=\{3,4\}$.

\Cref{app:star-tools,app:F3F4} collect the tools of the star bounds that the certificate
uses and prove the enclosures of $F_3$ and $F_4$. \Cref{sec:cert-method} proves that
the replacement by cells and sums is valid.
\Cref{sec:cert-arithmetic} describes the rigorous arithmetic,
\cref{sec:cert-results} the search for cells and the results, \cref{sec:cert-repro}
how the runs are reproduced, and \cref{sec:cert-limits} what is trusted.

\subsubsection{Tools from the star bounds}\label{app:star-tools}

This subsection collects the lower bounds on $P_{\min}$ for unequal thresholds and the
recursion for the simplex orthant that the certificate and \cref{app:F3F4} use; the
notation is that of \cref{sec:stars}.

\paragraph{Unequal thresholds.}

When the lower bounds $r_i$ of a star differ, the thresholds
$a_i(z)=(\tau-r_iz)/s_i$ of the center bound differ, and the simplex theorem
(\cref{lem:simplex}) does not apply directly. The certificate of
\cref{sec:certificate} uses the following two lower bounds on the infimum
$P_{\min}$. The first holds for every $d\in\{3,4,5\}$ and is exact for equal
thresholds; the second is specific to $d=3$ and is much sharper for unequal
thresholds. For $d\in\{3,4,5\}$ and $a\in\R^d$ put
\[
  \Lambda_d(a):=\max\Bigl\{0,\ 1-\sum_i\Phi(a_i),\ \max_{1\le j\le d}\Bigl(S_d(a_j)-\sum_{i:\,a_i>a_j}\bigl(\Phi(a_i)-\Phi(a_j)\bigr)\Bigr)\Bigr\}.
\]

\begin{lemma}[simplex truncation]\label{lem:truncation}
Let $d\in\{3,4,5\}$ and $a\in\R^d$. Then $P^{(d)}_{\min}(a)\ge\Lambda_d(a)$.
\end{lemma}

\begin{proof}
Let $U$ be a centered Gaussian vector in $\R^d$ with unit variances. The probability
$\Pr(U_i>a_i\ \forall i)$ is nonnegative, and it is at least $1-\sum_i\Phi(a_i)$ by
the union bound $\Pr(\exists i:\ U_i\le a_i)\le\sum_i\Phi(a_i)$. Let $\lambda\in\R$.
If $U_i>\lambda$ for all $i$ but $U_k\le a_k$ for some $k$, then $\lambda<U_k\le a_k$,
so in particular $a_k>\lambda$. Hence
\[
  \{U_i>\lambda\ \forall i\}\ \subseteq\ \{U_i>a_i\ \forall i\}\ \cup
  \bigcup_{k:\,a_k>\lambda}\{\lambda<U_k\le a_k\},
\]
and $\Pr(\lambda<U_k\le a_k)=\Phi(a_k)-\Phi(\lambda)$. With
$\Pr(U_i>\lambda\ \forall i)\ge S_d(\lambda)$ from \cref{lem:simplex}, this gives
\[
  \Pr(U_i>a_i\ \forall i)\ \ge\ S_d(\lambda)-\sum_{i:\,a_i>\lambda}\bigl(\Phi(a_i)-\Phi(\lambda)\bigr).
\]
Taking $\lambda=a_j$ for each $j$ and then the infimum over $U$ gives
$P^{(d)}_{\min}(a)\ge\Lambda_d(a)$.
\end{proof}

For equal thresholds the bound is exact: every sum over $\{i:a_i>a_j\}$ is empty,
and $S_d(a)\ge\max\{0,1-d\,\Phi(a)\}$ by the union bound applied to $U^*$, so
$\Lambda_d(a,\dots,a)=S_d(a)=P^{(d)}_{\min}(a,\dots,a)$ by \cref{lem:simplex}.

For three variables, the rank reduction of \cref{lem:rank-reduction}, which allows
unequal thresholds, leads to a planar picture that can be counted exactly. The
following bound is used for the stars of a degree-$3$ vertex whose neighbors do not
all have the same degree.

\begin{lemma}[rank reduction and arc count, $d=3$]\label{lem:arcs}
Let $a\in\R^3$. For $\rho>0$ put
$\gamma_i(\rho)=\arccos\clip(-a_i/\rho,-1,1)\in[0,\pi]$, where
$\clip(x,-1,1)=\max\{-1,\min\{1,x\}\}$. Then for all $0\le\rho_a\le\rho_b$,
\[
  P^{(3)}_{\min}(a)\ \ge\ \int_{\rho_a}^{\rho_b}\Bigl(1-\sum_{i=1}^3\frac{\gamma_i(\rho)}{\pi}\Bigr)\rho\,e^{-\rho^2/2}\,d\rho .
\]
\end{lemma}

\begin{proof}
Let $U$ be a centered Gaussian vector in $\R^3$ with unit variances and correlation
matrix $C$.

\emph{Rank reduction.} By \cref{lem:rank-reduction} it suffices to bound
$\Pr(U_i>a_i\ \forall i)$ from below when $C$ is singular, that is,
$\operatorname{rank}C\le2$. Then $U$ has the distribution of $(\langle v_i,G\rangle)_i$ for
unit vectors $v_1,v_2,v_3\in\R^2$ and a standard Gaussian vector $G$ in $\R^2$.

\emph{Arc count.} Write $G=\rho(\cos\vartheta,\sin\vartheta)$, so $\rho$ has density
$\rho e^{-\rho^2/2}$ on $(0,\infty)$ and $\vartheta$ is uniform and independent of
$\rho$. Given $\rho>0$, the event $\{U_i\le a_i\}$ says
$\langle-v_i,(\cos\vartheta,\sin\vartheta)\rangle\ge-a_i/\rho$. If $-a_i/\rho>1$ this
set of directions is empty, if $-a_i/\rho<-1$ it is the whole circle, and otherwise
it is a closed arc of half-angle $\arccos(-a_i/\rho)$ around the direction of
$-v_i$. In all three cases its normalized length is $\gamma_i(\rho)/\pi$ (up to a
null set). By the union bound, the conditional probability of
$\{U_i>a_i\ \forall i\}$ given $\rho$ is at least
$\max\{0,1-\sum_i\gamma_i(\rho)/\pi\}$. Integrating against the density of $\rho$,
and dropping the nonnegative contribution of $\rho\notin[\rho_a,\rho_b]$,
\[
  \Pr(U_i>a_i\ \forall i)\ \ge\ \int_{\rho_a}^{\rho_b}\Bigl(1-\sum_i\frac{\gamma_i(\rho)}{\pi}\Bigr)\rho e^{-\rho^2/2}\,d\rho .
\]
The right side does not depend on $U$, so it also bounds the infimum $P_{\min}(a)$.
\end{proof}

The interval $[\rho_a,\rho_b]$ is a free parameter; any choice gives a valid bound.
The certificate takes $\rho_a=\max\{0,a_1,a_2,a_3\}$, below which some
$\gamma_i=\pi$ and the integrand is negative, and $\rho_b$ close to the first zero
of the integrand beyond $\rho_a$. The integral has a closed form in terms of Owen's
$T$ function (\cref{lem:cert-owen}).

\paragraph{Evaluating the simplex orthant.}

The functions $S_d$ are evaluated through the reflection $T_d(h)=S_d(-h)$, the
distribution function of the maximum of the regular simplex vector. It satisfies a
one-dimensional recursion in $d$, which is the basis of all numerical evaluations
of $S_3,S_4,S_5$ in this paper.

\begin{lemma}[simplex recursion]\label{lem:simplex-recursion}
For $h\ge0$, $T_2(h)=2\Phi(h)-1$, and for $d\ge3$,
\[
  T_d(h)=\int_0^h d\,\varphi(x)\,T_{d-1}\Bigl(\sqrt{\tfrac d{d-2}}\;x\Bigr)\,dx ;
\]
for $h\le0$, $T_d(h)=0$. In particular each $T_d$ is continuous and nondecreasing,
and $S_d(a)=T_d(-a)$ is continuous and nonincreasing.
\end{lemma}

\begin{proof}
Let $X=U^*$ be the regular simplex vector in $d$ variables. For $d=2$ we have
$X_2=-X_1$ (the correlation is $-1$), so $\max_iX_i=|X_1|$ and
$T_2(h)=\Pr(|X_1|<h)=2\Phi(h)-1$ for $h\ge0$. The case $h\le0$ is~\eqref{eq:S-zero}.

Let $d\ge3$ and $h\ge0$, and put $\rho=-\frac1{d-1}$. For $j\ne k$,
$\Var(X_j-X_k)=2-2\rho>0$, so the maximum is attained at a unique index almost
surely. By symmetry,
\[
  T_d(h)=\sum_{i=1}^d\Pr\bigl(X_i<h,\ X_j\le X_i\ \forall j\ne i\bigr)
  =d\,\Pr\bigl(X_1<h,\ X_j\le X_1\ \forall j\ge2\bigr).
\]
Write $X_j=\rho X_1+R_j$ for $j\ge2$. The residual vector $(R_2,\dots,R_d)$ is
centered Gaussian and independent of $X_1$, with
$\Var R_j=1-\rho^2$ and $\Cov(R_j,R_k)=\rho-\rho^2$ for $j\ne k$; its correlations
are $(\rho-\rho^2)/(1-\rho^2)=\rho/(1+\rho)=-\frac1{d-2}$. So
$\tilde R=R/\sqrt{1-\rho^2}$ is the regular simplex vector in $d-1$ variables. The
condition $X_j\le X_1$ reads $R_j\le(1-\rho)X_1$, that is,
$\tilde R_j\le X_1\sqrt{(1-\rho)/(1+\rho)}=X_1\sqrt{d/(d-2)}$. By the independence of
$X_1$ and $\tilde R$ and Fubini's theorem,
\[
  T_d(h)=d\int_{-\infty}^h\varphi(x)\,\Pr\Bigl(\max_j\tilde R_j\le\sqrt{\tfrac d{d-2}}\,x\Bigr)dx
  =\int_{-\infty}^hd\,\varphi(x)\,T_{d-1}\Bigl(\sqrt{\tfrac d{d-2}}\,x\Bigr)dx,
\]
using that the maximum of $\tilde R$ has a continuous distribution. The integrand vanishes
for $x\le0$ by~\eqref{eq:S-zero}, which gives the formula. Continuity and
monotonicity follow by induction on $d$.
\end{proof}

\subsubsection{The constants \texorpdfstring{$F_3$}{F3} and \texorpdfstring{$F_4$}{F4}}\label{app:F3F4}

The constants $F_3$ and $F_4$ are evaluated with explicit one-dimensional integral
formulas, which we now record. For $F_3$ we use the geometry of the planar regular
simplex; for $F_4$ we use the recursion of \cref{lem:simplex-recursion}.

\begin{lemma}[formula for the ideal $3$-star]\label{lem:stars-f3-formula}
Let $r\in(0,1)$, $s=s_r$ and $\tau\in\R$. Put $z_\ell=\max\{\tau,\tau/r\}$ and, for
$\vartheta\in[0,\pi/3]$, $\sigma=s\cos\vartheta$ and $q=\sigma^2+r^2$. Then
$P_{3,r}(\tau):=\Pr(Y^*_k>\tau\ \text{for}\ 0\le k\le3)$ equals
\[
  P_{3,r}(\tau)=\Phi(-z_\ell)-\frac3\pi\int_0^{\pi/3}\frac{\sigma}{\sqrt q}\,
  e^{-\tau^2/(2q)}\,\Phi\Bigl(-\frac{\sqrt q}{\sigma}\Bigl(z_\ell-\frac{r\tau}q\Bigr)\Bigr)d\vartheta,
\]
and $f_{3,r}(\tau)=\Phi(-\tau)-P_{3,r}(\tau)$. Moreover $P_{3,r}$ is nonincreasing.
\end{lemma}

\begin{proof}
Given $Z=z$, all $Y^*_i$ exceed $\tau$ iff $\langle-w_i,G\rangle<h:=(rz-\tau)/s$
for $i=1,2,3$. Since $\sum_i\langle-w_i,G\rangle=0$, this is impossible if $h\le0$.
For $h>0$ it says that $G$ lies in the open equilateral triangle $T_h$ with inradius
$h$ and outer unit normals $-w_1,-w_2,-w_3$. In polar coordinates, the ray in a
direction at angle $\vartheta\in[-\pi/3,\pi/3]$ from the nearest outer normal leaves
$T_h$ at distance $h/\cos\vartheta$, and $\Pr(|G|<x)=1-e^{-x^2/2}$; by the six-fold
symmetry,
\[
  \mu(h):=\Pr(G\in T_h)=\frac3\pi\int_0^{\pi/3}\Bigl(1-e^{-h^2/(2\cos^2\vartheta)}\Bigr)d\vartheta .
\]
The conditions $z>\tau$ (for $Y^*_0$) and $h>0$ together say $z>z_\ell$. By Tonelli's
theorem,
\[
  P_{3,r}(\tau)=\int_{z_\ell}^\infty\varphi(z)\,\mu\Bigl(\frac{rz-\tau}s\Bigr)dz
  =\Phi(-z_\ell)-\frac3\pi\int_0^{\pi/3}\int_{z_\ell}^\infty\varphi(z)\,
  e^{-(rz-\tau)^2/(2\sigma^2)}\,dz\,d\vartheta .
\]
Completing the square, $z^2+(rz-\tau)^2/\sigma^2=(q/\sigma^2)(z-r\tau/q)^2+\tau^2/q$,
because $q-r^2=\sigma^2$. Hence the inner integral equals
$(\sigma/\sqrt q)\,e^{-\tau^2/(2q)}\,\Phi(-(\sqrt q/\sigma)(z_\ell-r\tau/q))$, which
is the formula. The identity for $f_{3,r}$ is~\eqref{eq:stars-complement}, and
$P_{3,r}$ is nonincreasing because the event shrinks as $\tau$ grows.
\end{proof}

\begin{lemma}[formula for the ideal $4$-star]\label{lem:stars-f4-formula}
Let $r=r_{44}=\frac{\sqrt3}2$ and $s=s_r=\frac12$. Put $g(x)=4\varphi(x)\,T_3(\sqrt2\,x)$
for $x\ge0$ and $x_0(\tau)=\max\{0,(r-1)\tau/s\}$. Then
$P_4(\tau):=\Pr(Y^*_k>\tau\ \text{for}\ 0\le k\le4)$ satisfies
\[
  P_4(\tau)=\Phi(-\tau)\,T_4\bigl(x_0(\tau)\bigr)+\int_{x_0(\tau)}^\infty g(x)\,
  \Phi\Bigl(-\frac{\tau+sx}r\Bigr)dx,
\]
$f_{4,r}=\Phi(-\tau)-P_4(\tau)$ is continuously differentiable, and
\[
  f_{4,r}'(\tau)=-\varphi(\tau)+\varphi(\tau)\,T_4\bigl(x_0(\tau)\bigr)
  +\frac1r\int_{x_0(\tau)}^\infty g(x)\,\varphi\Bigl(\frac{\tau+sx}r\Bigr)dx .
\]
\end{lemma}

\begin{proof}
By \cref{lem:simplex-recursion}, $T_4(h)=\int_0^hg(x)\,dx$ for $h\ge0$ and $T_4=0$
on $(-\infty,0]$; recall $T_4(h)=S_4(-h)$. Conditioning on $Z$ as in
\eqref{eq:ideal-star-integral},
\[
  P_4(\tau)=\int_\tau^\infty\varphi(z)\,T_4\Bigl(\frac{rz-\tau}s\Bigr)dz
  =\int_0^\infty g(x)\int\varphi(z)\,\ind\Bigl\{z>\tau,\ z>\frac{\tau+sx}r\Bigr\}\,dz\,dx
\]
by Tonelli's theorem, since $T_4((rz-\tau)/s)=\int_0^\infty g(x)\ind\{x<(rz-\tau)/s\}dx$.
The inner integral is $\Phi(-\max\{\tau,(\tau+sx)/r\})$, and $(\tau+sx)/r\ge\tau$
exactly when $x\ge(r-1)\tau/s$. Splitting the outer integral at $x_0(\tau)$ gives
the formula. For the derivative, differentiate under the integral sign (the
integrand and its $\tau$-derivative are dominated by $g$ and $g\,\varphi(0)/r$). When
$x_0(\tau)>0$, the two terms coming from the moving limit $x_0$ cancel, because
$(\tau+sx_0)/r=\tau$ there; when $x_0=0$ they are absent, and $T_4(0)=0$. This gives
$P_4'(\tau)=-\varphi(\tau)T_4(x_0)-\frac1r\int_{x_0}^\infty g(x)\varphi((\tau+sx)/r)\,dx$,
which is continuous in $\tau$, and $f'_{4,r}=-\varphi-P_4'$.
\end{proof}

\propFthreeFfour*

The proof is computer-assisted. All four bounds are obtained in ball arithmetic
with the Arb library \cite{Johansson2017} (through its Python interface
\texttt{python-flint}~0.9.0), which encloses every real number computed in an
interval with rigorously controlled rounding, and integrates analytic functions
with rigorous error bounds. Nothing else is trusted beyond the formulas of
\cref{lem:stars-f3-formula,lem:stars-f4-formula} and the correctness of Arb.

\begin{proof}
Each lower bound comes from an enclosure of the ideal star function at one
threshold. Each upper bound combines a tail estimate outside a bounded interval of
thresholds with a bound on each cell of a finite cover of that interval.

\emph{Lower bounds.} A supremum is at least any value. Evaluating $f_{3,r_{33}}$ at
a point $\hat\tau\approx-0.0681371311$ with the formula of
\cref{lem:stars-f3-formula} (for $\tau\le0$ it simplifies to
$f_{3,r}(\tau)=\frac3\pi\int_0^{\pi/3}\frac{\sigma}{\sqrt q}e^{-\tau^2/(2q)}
\Phi(-\tau\frac{q-r}{\sigma\sqrt q})\,d\vartheta$), in $200$-bit Arb arithmetic,
gives an enclosure of radius below $10^{-59}$ around
\[
  0.14015825233145216369394051401725\ldots.
\]
Evaluating $f_{4,r_{44}}$ at $-0.16367$
with the formula of \cref{lem:stars-f4-formula} (with $T_3,T_4$ computed by nested
Arb integrals from \cref{lem:simplex-recursion}) gives
$0.2457719746989\pm7.9\cdot10^{-14}$.

\emph{Upper bound for $F_3$.} Write $f=f_{3,r_{33}}$ and $P=P_{3,r_{33}}$. For
$\tau\le-2$, $f(\tau)\le\Pr(\min_iY^*_i\le\tau)\le3\Phi(-2)<0.069$; for $\tau\ge2$,
$f(\tau)\le\Phi(-2)<0.023$. On an interval $[\alpha,\beta]\subseteq[-2,2]$,
\[
  f(\tau)=\Phi(-\tau)-P(\tau)\le\Phi(-\alpha)-P(\beta)\qquad(\tau\in[\alpha,\beta]),
\]
because $\Phi(-\cdot)$ and $P$ are nonincreasing (\cref{lem:stars-f3-formula}). The
computation covers $[-2,0]$ and $[0,2]$ by $830{,}610$ intervals, obtained by
bisection, on each of which the Arb enclosure of $\Phi(-\alpha)-P(\beta)$ lies below
$0.1401582524$. This proves $F_3\le0.1401582524$ without any information on the
location of the maximizer.

\emph{Upper bound for $F_4$.} Write $f=f_{4,r_{44}}$. For $\tau\ge0.7$,
$f(\tau)\le\Phi(-\tau)\le\Phi(-0.7)<0.24197$; for $\tau\le-2.5$,
$f(\tau)\le\Pr(\min_iY^*_i\le\tau)\le4\Phi(-2.5)<0.0249$. The interval
$[-2.5,0.7]$ is split at $0$ and covered by $97$ cells, obtained by bisection from
cells of width $\frac1{16}$ (the smallest has width $2^{-17}$). On a cell
$[t_0,t_1]$ with midpoint $m$, the mean value theorem gives
$f(\tau)\le f(m)+\max\{D^+(t_1-m),D^-(t_0-m)\}$, where $[D^-,D^+]$ is an Arb
enclosure of $f'$ on the cell, computed from \cref{lem:stars-f4-formula}. The
integrals over $x$ are cut at $x=10$; the omitted part is at most
$\int_{10}^\infty g\le4\Phi(-10)$ (respectively $4\varphi(0)\Phi(-10)/r$ for $f'$),
and is added to the enclosure as an interval. Every cell bound lies below
$0.24577197475$; the largest, $0.2457719747237\ldots$, occurs on the cell of width
$2^{-16}$ around $-0.16364$.
\end{proof}

\subsubsection{Cells and discretization}\label{sec:cert-method}

Two lemmas reduce the check to finitely many computations: one replaces a cell of
thresholds by its two endpoints, and the other replaces the integral in the
reference star bound by a finite sum. Fix a type $T$ of degree $d$ with reference
bounds $r=r_T\in(0,1)^d$ (\cref{def:admissible}), and write $a(z;t)=a(z;t,r)$ for the
vector of thresholds $a_i(z;t)=(t-r_iz)/s_i$, $s_i=\sqrt{1-r_i^2}$, of
\cref{lem:center} at threshold $t$. Put
\[
  J_T(t):=\int_t^\infty\varphi(z)\,P_{\min}\bigl(a(z;t)\bigr)\,dz,
  \qquad\text{so that}\qquad C_T(t)=\Phi(-t)-J_T(t).
\]
The first lemma shows that a bound for a whole cell of thresholds needs only one
evaluation of $J_T$, at the right end of the cell.

\begin{lemma}[cells]\label{lem:cert-cells}
For $t_0\le\tau\le t_1$, $C_T(\tau)\le\Phi(-t_0)-J_T(t_1)$.
\end{lemma}

\begin{proof}
First, $\Phi(-\tau)\le\Phi(-t_0)$. Second, $J_T$ is nonincreasing: each threshold
$a_i(z;t)=(t-r_iz)/s_i$ is increasing in $t$, $P_{\min}$ is nonincreasing in each
threshold (\cref{lem:stars-continuity}(a)), the integrand is nonnegative, and the
domain of integration shrinks as $t$ grows. Hence $J_T(\tau)\ge J_T(t_1)$.
\end{proof}

The bound of \cref{lem:cert-cells} can exceed $C_T(\tau)$ by a term of first order in
the cell width, roughly $(d+2)\varphi(t_0)(t_1-t_0)$. This is why the certified
coefficients are slightly larger than the floating-point optima of the same star
bounds (\cref{rem:sweep-optimum}).

\begin{lemma}[discretization]\label{lem:cert-discretize}
For every $t\in\R$ and all points $t=z_0<z_1<\dots<z_N$,
\[
  J_T(t)\ \ge\ \sum_{k=0}^{N-1}P_{\min}\bigl(a(z_k;t)\bigr)\bigl(\Phi(z_{k+1})-\Phi(z_k)\bigr).
\]
\end{lemma}

\begin{proof}
Since $r_i>0$, each $a_i(z;t)$ is decreasing in $z$, so $z\mapsto P_{\min}(a(z;t))$
is nondecreasing (\cref{lem:stars-continuity}(a)). For every $z\in[z_k,z_{k+1}]$,
the integrand of $J_T(t)$ is therefore at least $\varphi(z)P_{\min}(a(z_k;t))$.
Integrate over each interval $[z_k,z_{k+1}]$ and drop the nonnegative contribution of
$z>z_N$.
\end{proof}

In the sum of \cref{lem:cert-discretize}, each value $P_{\min}(a(z_k;t))$ is
replaced by a computable lower bound $L_T(a(z_k;t))$:
\begin{itemize}
  \item if all reference bounds of $T$ are equal, that is, all neighbors have the same
  degree ($9$ types), the thresholds are equal and
  $L_T(a,\dots,a)=\max\{0,\ 1-d\,\Phi(a),\ S_d(a)\}=S_d(a)$, which equals $P_{\min}$
  (\cref{lem:simplex});
  \item if $d\in\{4,5\}$ and the reference bounds are not all equal, $L_T=\Lambda_d$,
  the simplex truncation of \cref{lem:truncation};
  \item if $d=3$ and the reference bounds are not all equal (seven types),
  $L_T=\max\{\Lambda_3,A\}$, where $A$ is the arc bound of \cref{lem:arcs} with radii
  $\rho_a=\max\{0,a_1,a_2,a_3\}$ and $\rho_b\in(\rho_a,12]$. The radius $\rho_b$ is
  located by floating-point bisection near the first zero of the integrand beyond
  $\rho_a$; any $\rho_b$ gives a valid bound. To reduce the running time, $A$ is
  evaluated only at the points where a floating-point estimate of it exceeds
  $\Lambda_3$ by more than $10^{-10}$; at the other points $L_T=\Lambda_3$. Either way
  $L_T\le P_{\min}$, by \cref{lem:truncation,lem:arcs}; the estimate only decides
  which valid bound is computed.
\end{itemize}
For the arc bound, the integral over $\rho$ is evaluated in closed form with Owen's
function~\cite{Owen1956}
\[
  T_{\mathrm{Owen}}(h,x)=\frac1{2\pi}\int_0^x\frac{e^{-h^2(1+y^2)/2}}{1+y^2}\,dy
  \qquad(h,x\in\R).
\]

\begin{lemma}[closed form of the arc integral]\label{lem:cert-owen}
Let $|c|\le\rho_1\le\rho_2$. If $c\ne0$, then
\[
  \int_{\rho_1}^{\rho_2}\arccos\frac c\rho\;\rho e^{-\rho^2/2}\,d\rho
  =\Bigl[-e^{-\rho^2/2}\arccos\frac c\rho\Bigr]_{\rho_1}^{\rho_2}
  +\operatorname{sign}(c)\,2\pi\bigl(T_{\mathrm{Owen}}(|c|,y_2)-T_{\mathrm{Owen}}(|c|,y_1)\bigr),
\]
where $y_j=\sqrt{\rho_j^2-c^2}/|c|$; if $c=0$, the integral equals $\frac\pi2(e^{-\rho_1^2/2}-e^{-\rho_2^2/2})$.
Consequently, for $a\in\R^3$, $c_i=-a_i$ and $\rho_a\ge\max\{0,a_1,a_2,a_3\}$, the
bound of \cref{lem:arcs} equals
\[
  e^{-\rho_a^2/2}-e^{-\rho_b^2/2}-\frac1\pi\sum_{i=1}^3\int_{\max\{\rho_a,|c_i|\}}^{\rho_b}
  \arccos\frac{c_i}\rho\;\rho e^{-\rho^2/2}\,d\rho,
\]
where an integral with lower limit above $\rho_b$ is read as $0$.
\end{lemma}

\begin{proof}
For $c=0$ the integrand is $\frac\pi2\rho e^{-\rho^2/2}$. For $c\ne0$, integrate by
parts with $v=-e^{-\rho^2/2}$; since
$\frac d{d\rho}\arccos\frac c\rho=\frac c{\rho\sqrt{\rho^2-c^2}}$ for $\rho>|c|$
(integrable at $\rho=|c|$), the remaining integral is
$\int_{\rho_1}^{\rho_2}e^{-\rho^2/2}\frac{c}{\rho\sqrt{\rho^2-c^2}}\,d\rho$. The
substitution $\rho^2=c^2(1+y^2)$, $y\ge0$, gives $\rho\,d\rho=c^2y\,dy$ and
$\sqrt{\rho^2-c^2}=|c|y$, so the integrand becomes
$\operatorname{sign}(c)\,e^{-c^2(1+y^2)/2}(1+y^2)^{-1}dy$, whose integral from $y_1$
to $y_2$ is $\operatorname{sign}(c)\,2\pi(T_{\mathrm{Owen}}(|c|,y_2)-T_{\mathrm{Owen}}(|c|,y_1))$.

For the second claim, $\int_{\rho_a}^{\rho_b}\rho e^{-\rho^2/2}d\rho=e^{-\rho_a^2/2}-
e^{-\rho_b^2/2}$, and $\gamma_i(\rho)=\arccos\clip(c_i/\rho,-1,1)$. If $c_i>0$ and
$\rho<c_i$, then $\gamma_i(\rho)=\arccos1=0$, so the integral over $\gamma_i$ starts at
$\max\{\rho_a,c_i\}$. If $c_i\le0$, then $|c_i|=\max\{0,a_i\}\le\rho_a$, so on
$[\rho_a,\rho_b]$ the clip is inactive.
\end{proof}

\subsubsection{Rigorous arithmetic}\label{sec:cert-arithmetic}

For each type $T$ and each right end $t_1$ of a cell, the accompanying code computes a
rational number $I^{\rm lo}_T(t_1)$ such that
\begin{equation}\label{eq:cert-Ilo}
  I^{\rm lo}_T(t_1)\ \le\ \sum_{k=0}^{N-1}P_{\min}\bigl(a(z_k;t_1)\bigr)\bigl(\Phi(z_{k+1})-\Phi(z_k)\bigr)
  \ \le\ J_T(t_1),
\end{equation}
the second inequality being \cref{lem:cert-discretize}. The first inequality rests on
the following facts about the computation.
\begin{enumerate}[label=(\alph*)]
  \item \emph{Grids.} All cell endpoints are multiples of $2^{-17}$ in $[-3,3]$. The
  points of \cref{lem:cert-discretize} are $z_k=t_1+k\delta_z$, $0\le k\le N$, with
  $N\delta_z=10$ and $\delta_z=2^{-14}$ ($\delta_z=2^{-13}$ for the seven types that
  use the arc bound). So every $z_k$ is a multiple of $2^{-17}$ in $[-3,13]$.
  \item \emph{Masses.} Enclosures of $\Phi$ and $\varphi$ at all multiples of $2^{-17}$
  in $[-8,13]$ are computed once in Arb \cite{Johansson2017} ($80$-bit precision) and
  rounded outward to double precision. The mass $\Phi(z_{k+1})-\Phi(z_k)$ is bounded
  below by the lower enclosure at $z_{k+1}$ minus the upper enclosure at $z_k$,
  rounded down and clamped at $0$.
  \item \emph{Values of $\Phi$ elsewhere.} At a point $a$ off the grid, with $x$ the
  grid point below $a$ and $\delta=a-x\in[0,2^{-17})$, Taylor's theorem and
  $|\varphi'|\le\varphi(1)<0.242$ give
  $|\Phi(a)-\Phi(x)-\varphi(x)\delta|\le0.121\,\delta^2$. Upper and lower bounds for
  $\Phi(a)$ are formed from the grid enclosures accordingly, with outward rounding.
  Below $-8$ and above $13$ the trivial monotone bounds are used. These bounds were
  also compared with Arb at $800$ random points.
  \item \emph{Thresholds.} The reference bounds $r_{ij}$ and the numbers
  $s_{ij}=\sqrt{1-r_{ij}^2}$ are double-precision numbers within $4\cdot10^{-16}$ of
  their Arb values (checked at start-up).
  Each computed threshold $(t_1-r_iz_k)/s_i$ is increased by $3\cdot10^{-13}$, which
  exceeds its rounding error for $|z|\le13$ and $s_i\ge\frac13$. So the code evaluates
  its lower bound at thresholds $a'\ge a(z_k;t_1)$ coordinatewise, and
  $P_{\min}(a')\le P_{\min}(a(z_k;t_1))$ by monotonicity.
  \item \emph{Simplex tables.} The simplex functions enter through the functions $T_d$
  of \cref{lem:simplex-recursion}, with $S_d(a)=T_d(-a)$. Lower tables
  $T_d^{\rm lo}\le T_d$ for $d=2,3,4,5$ are built on the grid $x_k=k\,2^{-19}$,
  $0\le x_k\le12$. For $d=2$, $T_2^{\rm lo}$ is the lower Arb enclosure of
  $2\Phi(x_k)-1$. For $d\ge3$, \cref{lem:simplex-recursion} writes $T_d$ as an
  integral of $d\,\varphi(x)T_{d-1}(\sqrt{d/(d-2)}\,x)$. On $[x_j,x_{j+1}]$ this
  integrand is at least $d\,\varphi(x_{j+1})\,T^{\rm lo}_{d-1}(g_j)$, where $g_j$ is
  the largest grid point not exceeding $\sqrt{d/(d-2)}\,x_j$, because $\varphi$ is
  decreasing on $[0,\infty)$ and $T_{d-1}$ is nondecreasing. These terms, with
  $\varphi(x_{j+1})$ replaced by the
  lower end of its Arb enclosure and the product rounded down, are floored to
  multiples of $2^{-61}$ and summed exactly in integers. A lookup of $S_d(a)=T_d(-a)$
  rounds $-a$ down to the grid (and uses $0$ for $-a\le0$ and $T^{\rm lo}_d(12)$ for
  $-a>12$), which is valid because $T_d$ is nondecreasing.
  \item \emph{Lower bounds for $P_{\min}$.} The functions $\Lambda_d$ and $S_d$ are
  evaluated from the tables and the bounds of~(c), each term rounded in the direction
  that lowers the result. The arc bound is evaluated from the closed form of
  \cref{lem:cert-owen} as an Arb ball at $64$-bit precision; its inputs $a_i$,
  $\rho_a$ and $\rho_b$ are double-precision numbers and hence exact. The
  exponentials, arccosines and square roots are Arb functions. The Owen term
  $\int_{y_1}^{y_2}e^{-c^2(1+y^2)/2}(1+y^2)^{-1}\,dy=2\pi\bigl(T_{\mathrm{Owen}}(|c|,y_2)-T_{\mathrm{Owen}}(|c|,y_1)\bigr)$
  is enclosed by Arb's rigorous numerical integration of this integrand, which is
  analytic off the poles $\pm i$, between the exact midpoints of the balls for $y_1$
  and $y_2$. The result is then widened by the radii of these balls, which is valid
  because the integrand is at most $1$. For $|c|<2^{-30}$ the enclosure
  $[\arctan y_2-\arctan y_1-\frac{c^2}2(y_2-y_1),\,\arctan y_2-\arctan y_1]$, which
  follows from $1-u\le e^{-u}\le1$, is used instead. The lower end of the resulting
  ball, rounded down, is the value of the arc bound. In the run reported below, the
  arc bound was enclosed at about $1.2\cdot10^8$ points, all with radius below
  $10^{-13}$.
  \item \emph{The sum.} Each product of a lower bound and a mass is rounded down,
  floored to a multiple of $2^{-62}$, and the products are summed exactly in
  integers.
\end{enumerate}

\subsubsection{Charging on cells and results}\label{sec:cert-results}

The main run treats four \emph{problems}, one for each coefficient vector: the
pair $(F_3,0.2575)$ for $D=\{3,4\}$ and the triples $\tripP$, $\tripR$, $\tripS$ for
$D=\{3,4,5\}$. Each problem is solved by a search for cells with transfer vectors,
and its output is checked by a second program.

\emph{Cell search.} For each problem the interval $[-3,3]$ starts as $192$ cells of
width $\frac1{32}$. For a cell $[t_0,t_1]$ the code forms, for every type $T$ other
than the pure cubic type $T_0$, the number
$\widehat C_T=\widehat\Phi(-t_0)-I^{\rm lo}_T(t_1)$, where $\widehat\Phi(-t_0)$ is the
upper Arb enclosure of $\Phi(-t_0)$ in double precision. By
\cref{lem:cert-cells} and~\eqref{eq:cert-Ilo}, $C_T(\tau)\le\widehat C_T$ for all
$\tau\in[t_0,t_1]$. The type $T_0$ needs no check, since $C_{T_0}\le F_3\le a_3$ and
$\lambda_{T_0}=0$. The code then solves, in floating point, the linear program that
minimizes the largest violation of the inequalities
$\widehat C_T+\lambda_T(t)\le a_d$ over transfer vectors $t\in[-1,1]^3$; we used the
HiGHS solver through SciPy \cite{Virtanen2020}. The solver's output is used only as a
candidate: the cell is accepted if, in exact rational arithmetic, every inequality
holds for this $t$ with slack at least $10^{-12}$. Otherwise the cell is bisected.

\emph{Exact recheck.} A second, short program reads only the stored cells,
transfers and numbers $I^{\rm lo}_T(t_1)$. It checks in exact rational arithmetic
that the cells tile $[-3,3]$, that $|t_{ij}|\le1$, and that every inequality
$\widehat\Phi(-t_0)-I^{\rm lo}_T(t_1)+\lambda_T(t)<a_d$ holds. Here
$\widehat\Phi(-t_0)$ is a fresh Arb upper bound for $\Phi(-t_0)$, computed at
$100$-bit precision and rounded up to a multiple of $10^{-15}$. For the pair and for $\tripP$, whose coefficient at
degree~$3$ is $F_3$ itself, the degree-$3$ inequalities are checked with
$a_3=0.1401582523314521636939405140172$, a lower bound for $F_3$
(\cref{prop:F3F4}), so that they prove $\widehat C_T+\lambda_T(t)\le F_3$. For
$\tripR$ and $\tripS$, the omission of $T_0$ uses $F_3\le0.1401582524<a_3$.

\emph{Results.} All four problems are certified: in each, the accepted cells tile
$[-3,3]$, no cell failed, and the recheck passes. The smallest slack is the minimum
over all cells and all checked types of
$a_d-\widehat C_T-\lambda_T(t)$, as computed by the recheck. The column ``types'' gives the number of types
checked (all $D$-types except $T_0$), and ``min.\ width'' the smallest cell width.
\begin{center}
\begin{tabular}{llcccc}
\toprule
Problem & coefficients & types & cells & min.\ width & min.\ slack\\
\midrule
pair & $(F_3,\,0.2575)$ & $8$ & $890$ & $2^{-13}$ & $2.76\cdot10^{-7}$\\
$\tripP$ & $(F_3,\,0.2575,\,0.3667)$ & $45$ & $977$ & $2^{-13}$ & $1.04\cdot10^{-7}$\\
$\tripR$ & $(0.1455,\,0.2530,\,0.3600)$ & $45$ & $375$ & $2^{-10}$ & $7.09\cdot10^{-6}$\\
$\tripS$ & $(\frac5{32},\,\frac14,\,\frac{11}{32})$ & $45$ & $1012$ & $2^{-14}$ & $1.24\cdot10^{-7}$\\
\bottomrule
\end{tabular}
\end{center}
The slack is a property of the cells after bisection stopped, not the distance of a
coefficient vector from the boundary of the admissible set. Smaller coefficient
vectors were also tried. For example, $(0.1444,0.2514,0.3583)$ in place of $\tripR$
could not be certified: cells still failed at width $2^{-17}$, so this vector is at or
beyond the reach of this certificate.

A fifth problem, used for \cref{thm:switched4}, is run separately by the same
program. It is the pair $(\cthree,0.2627)$ for $D=\{3,4\}$ on $[0,\frac{35}{32}]$
only, with $\cthree$ replaced by its rational lower bound $0.139822124008$. The type
$T_0$ is omitted, because at nonnegative thresholds \cref{prop:star-cubic} bounds it
instead. The search starts from $35$ cells of width $\frac1{32}$ and certifies the
problem with $62$ cells, the smallest of width $2^{-8}$, and smallest slack
$1.06\cdot10^{-4}$ (\cref{prop:certificate-switched}). The exact recheck covers this
problem as well.

\subsubsection{Reproducibility}\label{sec:cert-repro}

A single driver script in the accompanying code reproduces every computation used
in the proofs of \cref{thm:deg4,thm:deg5}, and hence of \cref{thm:main} and
\cref{cor:edges}, as well as the charging certificate of \cref{thm:switched4}. The
switch-lemma certificates of \cref{sec:switch-cert-intro} have their own programs and
are not part of this script. The driver has two modes for these computations:
\begin{itemize}
  \item \emph{quick} (about one minute): the exact recheck of the stored cells for
  all four problems, and exact rational checks of the coefficient arithmetic used
  elsewhere in the paper;
  \item \emph{full} (about $75$ minutes on $16$ cores): in addition, the Arb
  computations of the lower and upper bounds of $F_3$ and of the enclosure of $F_4$
  (\cref{prop:F3F4}), and the certificate computation itself, including the fifth
  problem of \cref{prop:certificate-switched} (about $4$ minutes). The certificate
  computation rewrites the stored cell files, and a second exact recheck follows.
\end{itemize}
The runs reported here used CPython~3.14.7, \texttt{python-flint}~0.9.0,
NumPy~2.5.3, SciPy~1.18.1 and mpmath~1.4.1 on a $16$-core Apple M3 Max computer. The
certificate computation took $3661$\,s with $14$ worker processes, the enclosure of
$F_4$ $475$\,s, the global upper bound of $F_3$ $189$\,s, and each exact recheck about
one second. Repeated runs reproduced the stored certificate data (the cells,
transfers and $I^{\rm lo}$ values of the four problems, and the cells of the $F_4$
enclosure) bit for bit, as verified by SHA-256 checksums.

\subsubsection{What is trusted}\label{sec:cert-limits}

The certificate relies on the following software, data, outputs and mathematical
inputs.
\begin{itemize}
  \item \emph{Arb and exact arithmetic.} The computation rests on the correctness of
  Arb/FLINT, including its rigorous numerical integration, and of Python's exact
  rational arithmetic, and on the derivations of \cref{sec:stars} and of this
  subsection. Floating point is used only for choices. The linear program for the
  transfers is solved in floating point, but its output enters only through the exact
  check. The radius $\rho_b$ and the points at which the arc bound is evaluated are
  chosen in floating point, and every choice gives a valid lower bound. No
  floating-point value of a special function enters an inequality.
  \item \emph{Stored data.} The exact recheck uses the stored numbers
  $I^{\rm lo}_T(t_1)$, which only the main certificate program computes.
  \item \emph{Reporting.} The Arb programs for $F_3$ and $F_4$ report their
  enclosures in their output; the driver checks only exit codes, so their outputs
  must be read.
  \item \emph{Mathematical inputs outside the computation.} The certificate uses the
  simplex theorem (\cref{lem:simplex}) through $S_d$ and $\Lambda_d$. For $d=4$ and
  $d=5$ this rests on Mulgund's theorem~\cite{Mulgund2026}, whose status
  \cref{rem:stars-mulgund} describes.
\end{itemize}
{}
\subsection{Maximum degree 4: proof of \texorpdfstring{\cref{thm:switched4}}{Theorem 8.2} and its certificates}\label{sec:switch-cert}

This subsection proves \cref{thm:switched4}, which we restate below
(\cref{sec:switched4}). The charging at nonzero thresholds that the proof uses
(\cref{prop:certificate-switched}) is computed by the programs of
\cref{sec:certificate}. The remaining parts,
\cref{sec:switch-cert-intro,sec:switch-cert-formulas,sec:switch-cert-boxes,sec:switch-cert-results},
describe the computer-assisted part of the proof of \cref{lem:switch-mixed}.

\thmswitchedfour*

\subsubsection{Proof of \texorpdfstring{\cref{thm:switched4}}{Theorem 8.2}}\label{sec:switched4}

The proof of \cref{thm:cubic} has three ingredients: the layout of
\cref{lem:median-layout}, the bound $\kappa(r)$ at nonzero thresholds
(\cref{prop:star-cubic}), and the switch lemma at threshold zero (\cref{lem:switch}).
For maximum degree~$4$ we keep the layout, with a larger set at threshold zero, and
use the mixed labels of \cref{def:labels}\ref{def:labels:mixed}, whose edge
correlations depend on the degrees of the two ends. A vertex of degree $3$ with a
neighbor of degree $4$ then has weaker correlation bounds than one in a cubic region,
and the bound on its expected contribution to the frontiers exceeds $\cthree$. As in
\cref{sec:plain-sweep}, the excess is sent to the neighbors of degree~$4$, and the
coefficient $0.2627$ must absorb it.
\begin{itemize}
  \item At threshold zero, the expected contribution of a vertex is a switch-lemma
  quantity for its star, whose correlation bounds may be unequal. For a vertex of
  degree~$3$ whose neighbors all have the same degree, the bounds are equal and
  \cref{lem:switch} applies. For the other types, two of degree~$3$ and five of
  degree~$4$, we prove explicit bounds with a computer-assisted certificate
  (\cref{lem:switch-mixed}). Each vertex of degree~$3$ sends the same amount
  $s\approx0.01308$ to each neighbor of degree~$4$.
  \item At nonzero thresholds, the frontier probabilities are bounded by the star
  bounds of \cref{sec:plain-sweep}, except for vertices of degree~$3$ all of whose
  neighbors have degree~$3$, which are bounded by $\cthree$ through
  \cref{prop:star-cubic}. The charging is certified as in \cref{prop:certificate},
  now only on $[0,\frac{35}{32}]$, with $\cthree$ in place of $F_3$
  (\cref{prop:certificate-switched}).
\end{itemize}
Throughout, $c_3=\frac3{4\pi}\arccos\frac56$, the numbers $r_{ij}$ are those
of~\eqref{eq:rij}, and $s$ is the excess of the bound of
\cref{lem:switch-mixed}\ref{it:switch-mixed:a} over $\cthree$:
\[
  s:=0.1529003-\cthree\approx0.0130782 .
\]

\paragraph{The layout.} For a graph $G$ with $\Delta(G)\le4$ and a labeling $y$, let
$U=\infr S_0$ and $W=\outfr S_0$ as in \cref{sec:cubic-setting}, and let $\hat D$ be
the set of vertices with at least three neighbors on the other side of zero, that is,
vertices $v$ with $y_v\le0$ and at least three neighbors above zero, or $y_v>0$ and at
least three neighbors at or below zero. For $\Delta(G)\le3$ this is the set $D$ of
\cref{sec:cubic-setting}. A vertex of $\hat D$ with $y_v\le0$ lies in $U$, and one with
$y_v>0$ lies in $W$, so $\hat D\subseteq U\cup W$. Put
\[
  \hat L_0=|U|+|\hat D\cap W|,\qquad \hat L_0'=|W|+|\hat D\cap U| .
\]

\begin{lemma}[median-switched layout, maximum degree $4$]\label{lem:median-layout4}
\Cref{lem:median-layout} holds for graphs $G$ with $\Delta(G)\le4$, with $\hat L_0$
and $\hat L_0'$ in place of $L_0$ and $L_0'$.
\end{lemma}

\begin{proof}
The proofs of \cref{lem:median-layout-bags,lem:median-layout} use the degree bound at
one point only: in the first claim of \cref{lem:median-layout-bags}, that every vertex
$u\in U\setminus D$ has at most two neighbors above zero, so that \cref{lem:connector}
applies to the items $u\in P$ with $e(u)=N(u)\cap W'$. For $u\in U\setminus\hat D$
this holds by the definition of $\hat D$. The rest of the two proofs uses only that
$\hat D\subseteq U\cup W$, the identities
$\hat L_0=|U\setminus\hat D|+|\hat D|$ and $\hat L_0'=|W\setminus\hat D|+|\hat D|$
(which follow as in~\eqref{eq:cubic-L0}), and the placement of $\hat D$ in every
connector bag $C_0,\dots,C_k$ and of $\hat D\cap W$ in $C_{-1}$. In particular, a
crossing edge $uw$ with $y_u\le0<y_w$ lies in $C_{-1}$ if $w\in\hat D$, in $C_k$ if
$u\in\hat D$, and otherwise in the bag $C_j$ with $w=q_j\in e(u)$; none of these steps
uses the degree of $u$ or $w$.
\end{proof}

\paragraph{Threshold zero.} For a centered Gaussian vector $(Y_0,Y_1,\dots,Y_d)$ with
unit variances let $N=\#\{i\ge1:Y_i>0\}$ and
\[
  \Psi(Y_0;Y_1,\dots,Y_d)=\E\bigl[\ind\{Y_0\le0\}\,\lceil N/2\rceil\bigr].
\]
For $d\le4$ we have $\lceil N/2\rceil=\ind\{N\ge1\}+\ind\{N\ge3\}$. If $Y$ is a labeling
with the sign symmetry of \cref{sec:cubic-setting} and continuous marginals, then for a
vertex $v$ with neighbors $x_1,\dots,x_d$, $d\le4$,
\begin{equation}\label{eq:switch4-psi}
\begin{aligned}
  \Pr(v\in U)+\Pr(v\in\hat D\cap W)
  &=\Pr(Y_v\le0<\max_iY_{x_i})+\Pr\bigl(Y_v>0,\ \#\{i:Y_{x_i}\le0\}\ge3\bigr)\\
  &=\Psi(Y_v;Y_{x_1},\dots,Y_{x_d}),
\end{aligned}
\end{equation}
because the second probability equals $\Pr(Y_v<0,\ N\ge3)$ by sign symmetry. For
$d\le3$ the left side of \cref{lem:switch} is the same quantity, so \cref{lem:switch}
states $\Psi\le\kappa(r)$ when all incident correlations are at least $r$.

\begin{lemma}[switch lemma, unequal bounds]\label{lem:switch-mixed}
Let $(Y_0,Y_1,\dots,Y_d)$ be a centered Gaussian vector with unit variances and
$\Corr(Y_0,Y_i)\ge r_i$ for $1\le i\le d$.
\begin{enumerate}[label=\textup{(\alph*)}]
  \item\label{it:switch-mixed:a} If $d=3$ and $(r_1,r_2,r_3)$ is a permutation of
  $(r_{33},r_{33},r_{34})$, then $\Psi\le0.1529003$.
  \item\label{it:switch-mixed:b} If $d=3$ and $(r_1,r_2,r_3)$ is a permutation of
  $(r_{33},r_{34},r_{34})$, then $\Psi\le0.165616$.
  \item\label{it:switch-mixed:c} If $d=4$, $0\le j\le4$, and $j$ of the bounds $r_i$
  equal $r_{34}$ and the other $4-j$ equal $r_{44}$, then $\Psi\le0.2627-j\,s$.
\end{enumerate}
\end{lemma}

The numerical maxima of $\Psi$ in the five patterns of \ref{it:switch-mixed:c} are
$0.250000$, $0.245466$, $0.235687$, $0.221375$ and $0.208185$ for $j=0,\dots,4$, and
those in \ref{it:switch-mixed:a} and \ref{it:switch-mixed:b} are $0.1528953$ and
$0.1656115$. So the bound in \ref{it:switch-mixed:a} is within $5\cdot10^{-6}$ of the
numerical maximum, and \ref{it:switch-mixed:c} is tight to within $10^{-3}$ at $j=2$.
Unlike in \cref{lem:switch}, the maximizers are not the regular simplex in general;
for $j=0$, for instance, three equal residuals and a fourth opposite give
$\Psi=\frac14$.

\begin{proof}
\emph{Reduction to equality in the incident correlations.} Write $Z=Y_0$ and
$a_i=\Corr(Z,Y_i)\ge r_i$. For $a_i<1$ put $W_i=(Y_i-a_iZ)/s_{a_i}$, which is standard
normal and independent of $Z$; for $a_i=1$ let $W_i$ be an auxiliary standard normal
variable independent of everything else. Then $W=(W_1,\dots,W_d)$ is a centered
Gaussian vector with unit variances, independent of $Z$. On $\{Z<0\}$ and for
$a_i<1$, $Y_i>0$ holds exactly when $W_i>a_i|Z|/s_{a_i}$; since $a\mapsto a/s_a$ is
increasing, this implies $Y_i':=r_iZ+s_{r_i}W_i>0$. For $a_i=1$, $Y_i=Z<0$ is never
positive on $\{Z<0\}$. Hence on $\{Z<0\}$ the number $N$ for $Y$ is at most the number
$N'$ for $Y'=(Z,Y'_1,\dots,Y'_d)$, and since $\lceil\cdot/2\rceil$ is nondecreasing and
$\{Z=0\}$ is null, $\Psi(Y)\le\Psi(Y')$. The vector $Y'$ has incident correlations
exactly $r_i$, and its residuals have the correlation matrix $C$ of $W$. So it
suffices to bound, for each pattern, the function
\begin{equation}\label{eq:switch4-G}
  G_r(C)=\Psi\bigl(Z;\,r_1Z+s_{r_1}W_1,\dots,r_dZ+s_{r_d}W_d\bigr),\qquad
  W\sim N(0,C)\ \text{independent of}\ Z\sim N(0,1),
\end{equation}
over the compact convex set $E_d$ of $d\times d$ correlation matrices $C$.

\emph{The bounds on $E_d$.} These are proved by computer-assisted certificates in
Arb ball arithmetic, described in
\cref{sec:switch-cert-intro,sec:switch-cert-formulas,sec:switch-cert-boxes,sec:switch-cert-results}.
Each certificate covers a parametrization of $E_d$ by boxes. On each box, either
$G_r$ is bounded through exact formulas for its value and gradient (Plackett's
identity), or the box is shown to miss $E_d$. Every bound is a strict inequality in
Arb.
\end{proof}

The labels satisfy the bounds of \cref{lem:switch,lem:switch-mixed} only up to a
factor $1-\eps_R$. For $d\le4$ and $r\in(0,1)^d$ let $M_d(r)$ be the supremum of
$\Psi$ over all centered Gaussian vectors $(Y_0,\dots,Y_d)$ with unit variances and
$\Corr(Y_0,Y_i)\ge r_i$ for all $i$. The next lemma shows that lowering the bounds by
a factor $1-\gamma$ raises $M_d$ only slightly.

\begin{lemma}[lowering the bounds]\label{lem:switch-cont}
For $d\le4$, $r\in(0,1)^d$ and $0<\gamma<1$,
$M_d\bigl((1-\gamma)r\bigr)\le M_d(r)+3d\,\gamma^{1/3}$.
\end{lemma}

\begin{proof}
Let $(Y_0,\dots,Y_d)$ be centered Gaussian with unit variances and
$a_i=\Corr(Y_0,Y_i)\ge(1-\gamma)r_i$. Write $Y_i=\langle u_i,X\rangle$ with unit
vectors $u_0,\dots,u_d\in\R^{d+1}$ and $X$ standard Gaussian. If $a_i\ge r_i$ put
$u'_i=u_i$. Otherwise $0<a_i<r_i<1$, the vector $w_i=u_i-a_iu_0$ has
$|w_i|=s_{a_i}>0$, and $u'_i=r_iu_0+s_{r_i}w_i/|w_i|$ is a unit vector with
$\langle u_0,u'_i\rangle=r_i$ and
\[
  |u_i-u'_i|^2=2-2\bigl(a_ir_i+s_{a_i}s_{r_i}\bigr)\le2-2\bigl(a_ir_i+s_{r_i}^2\bigr)
  =2r_i(r_i-a_i)\le2\gamma .
\]
Put $Y'_i=\langle u'_i,X\rangle$; the vector $(Y_0,Y'_1,\dots,Y'_d)$ is one of the
vectors in the supremum that defines $M_d(r)$. Let $t=\gamma^{1/3}$. The events $\{Y_i>0\}$ and $\{Y'_i>0\}$ differ only if
$|Y_i|\le t$ or $|Y_i-Y'_i|>t$. Since $Y_i$ is standard normal and $Y_i-Y'_i$ is
centered normal with variance at most $2\gamma$, this has probability at most
$2t\varphi(0)+2\gamma/t^2<3\gamma^{1/3}$. Changing the signs of $k$ neighbors changes
$\lceil N/2\rceil$ by at most $k$, so
$\Psi(Y_0;Y_1,\dots,Y_d)\le\Psi(Y_0;Y'_1,\dots,Y'_d)+3d\gamma^{1/3}\le
M_d(r)+3d\gamma^{1/3}$.
\end{proof}

\paragraph{Setting.} For the rest of this subsection $H$ is a graph with
$\delta(H)\ge3$ and $\Delta(H)\le4$, with $N$ vertices, and
$\beta(H):=\cthree n_3(H)+0.2627\,n_4(H)$. The labels $Y$ are the mixed labels of
\cref{def:labels}\ref{def:labels:mixed} with parameter $R$. Properties
\ref{it:sweep-L1} and \ref{it:sweep-L2} of \cref{sec:plain-sweep} hold, since
$\Delta(H)\le5$. We use the $\{3,4\}$-types, reference bounds $r_T$ and transfer
terms $\lambda_T$ of \cref{def:admissible}, with the transfer vector written
$t=(t_{34},t_{35},t_{45})$; only $t_{34}$ matters here. By
\cref{lem:locality}\ref{lem:locality:gauss}, $-Y$ has the same distribution as $Y$, so the sign
symmetry of \cref{sec:cubic-setting} holds.

\begin{lemma}[threshold zero, maximum degree $4$]\label{lem:switch4-zero}
For every $\eta>0$ there is $R_0$ such that for every $R\ge R_0$ and every graph $H$
with $\delta(H)\ge3$ and $\Delta(H)\le4$ on $N$ vertices,
\[
  \E\hat L_0\le\beta(H)+\eta N\qquad\text{and}\qquad\E\hat L_0'\le\beta(H)+\eta N .
\]
\end{lemma}

\begin{proof}
We bound the expected contribution of a vertex to $\hat L_0$ for each type, sum these
bounds with \cref{lem:charging}, and obtain $\hat L_0'$ by sign symmetry. Let $v$
have type $T$, with $k$ neighbors of degree $4$ if $d(v)=3$ and $j$ neighbors of
degree $3$ if $d(v)=4$. By~\eqref{eq:switch4-psi},~\ref{it:sweep-L1} and the
definition of $M_d$, $\Pr(v\in U)+\Pr(v\in\hat D\cap W)\le M_{d(v)}((1-\eps_R)r_T)$.
By \cref{lem:switch} (applied with $r=r_{33}$ when $k=0$, and with $r=r_{34}$ when
$k=3$) and by \cref{lem:switch-mixed}, $M_{d(v)}(r_T)\le\beta_T$,
where
\begin{center}
\small
\renewcommand{\arraystretch}{1.15}
\begin{tabular}{@{}lccccc@{}}
\toprule
type & $d=3$, $k=0$ & $k=1$ & $k=2$ & $k=3$ & $d=4$, $j$\\
\midrule
$\beta_T$ & $\kappa(r_{33})=\cthree$ & $\cthree+s$ & $0.165616$
  & $\kappa(r_{34})<0.1778364$ & $0.2627-j\,s$\\
$\beta_T+\lambda_T(s,0,0)$ & $\cthree$ & $\cthree$ & $<0.1394597$ & $<0.1386019$
  & $0.2627$\\
\bottomrule
\end{tabular}
\end{center}
with $\lambda_T(s,0,0)=-k\,s$ for $d=3$ and $j\,s$ for $d=4$. The second row is at most
$a_{d(v)}$ for $(a_3,a_4)=(\cthree,0.2627)$, because $\cthree>0.1398221$. There are
nine types, and by \cref{lem:switch-cont} and $\eps_R\to0$ there is $R_0$ such that
$M_{d}((1-\eps_R)r_T)\le\beta_T+12\,\eps_R^{1/3}\le\beta_T+\eta$ for every type $T$ and every $R\ge R_0$. Hence
$\Pr(v\in U)+\Pr(v\in\hat D\cap W)+\lambda_{T(v)}(s,0,0)\le a_{d(v)}+\eta$ for every
vertex $v$, and \cref{lem:charging} with $D=\{3,4\}$ and $e=\eta$ gives the bound on
$\E\hat L_0=\sum_v\bigl(\Pr(v\in U)+\Pr(v\in\hat D\cap W)\bigr)$. The events
$\{v\in W\}$ and $\{v\in\hat D\cap U\}$ are, up to null events, the events
$\{v\in U\}$ and $\{v\in\hat D\cap W\}$ for the labeling $-Y$, which has the same distribution;
so the same bound holds for $\E\hat L_0'$.
\end{proof}

\paragraph{Nonzero thresholds.} Away from zero the frontier events are those of
\cref{sec:plain-sweep}. Only thresholds $\tau\ge0$ are needed, since negative ones
follow by sign symmetry. At these thresholds the pure cubic type $T_0=(3;3,0,0)$ is
bounded by $\cthree$ through \cref{prop:star-cubic} instead of by $F_3$, and above
$\frac{35}{32}$ no charging is needed at all.

\begin{proposition}[certified charging at nonnegative thresholds]\label{prop:certificate-switched}
Let $(a_3,a_4)=(\cthree,0.2627)$. For every $\tau\in[0,\frac{35}{32}]$ there is a
transfer vector $t$ with $C_T(\tau)+\lambda_T(t)\le a_d$ for every $\{3,4\}$-type
$T=(d;k_3,k_4,0)$ other than $T_0$.
\end{proposition}

\begin{proof}
This is a further problem of the certificate of \cref{prop:certificate}, computed and
rechecked by the same programs (\cref{sec:certificate}), with $[0,\frac{35}{32}]$ in
place of $[-3,3]$ and the eight $\{3,4\}$-types other than $T_0$;
\cref{sec:cert-results} reports it as the fifth problem. The degree-$3$
inequalities are checked against the rational number $0.139822124008<\cthree$, a lower
bound for $\cthree$. The $62$ cells tile $[0,\frac{35}{32}]$, their smallest width is
$2^{-8}$, and the smallest slack in~\eqref{eq:cell-inequality} is
$1.06\cdot10^{-4}$. Let $\tau$ lie in a cell $[t_0,t_1]$ with stored transfer vector
$t$, and let $T\ne T_0$ be a type of degree $d$. Then
\cref{lem:cert-cells,lem:cert-discretize} and the construction of $I^{\rm lo}_T$ give
$C_T(\tau)\le\Phi(-t_0)-I^{\rm lo}_T(t_1)$, and~\eqref{eq:cell-inequality} gives
$C_T(\tau)+\lambda_T(t)<a_d$.
\end{proof}

The factor $1-\eps_R$ in the correlation bounds adds an error term at nonzero
thresholds. For $\eps_R<1$ put
$e_R:=205\,\eps_R+\kappa\bigl(r_{33}(1-\eps_R)\bigr)-\cthree$, where the constant
$205$ comes from \cref{lem:stars-continuity}(b). Since $\kappa$ is continuous and
decreasing on $(0,1)$ (as $b(r)$ in \eqref{eq:cubic-kappa} is increasing), $e_R\ge0$
and $e_R\to0$ as $R\to\infty$.

\begin{lemma}[expected frontiers, maximum degree $4$]\label{lem:switch4-means}
Let $\eps_R<1$. Then $\E|\outfr S_\tau|\le\beta(H)+e_RN$ for every $\tau>0$, and
$\E|\infr S_\tau|\le\beta(H)+e_RN$ for every $\tau<0$.
\end{lemma}

\begin{proof}
First let $0<\tau\le\frac{35}{32}$, and let $t$ be the transfer vector of
\cref{prop:certificate-switched}. Let $v$ have type $T\ne T_0$. By~\ref{it:sweep-L1},
its star is a $d(v)$-star with lower bounds $(1-\eps_R)r_T$, so \cref{lem:center} and
\cref{lem:stars-continuity}(b) (as $\tau<3$, $d(v)\le4\le5$, and the entries of $r_T$
lie in $(0,r_{33}]$) give $\Pr(v\in\outfr S_\tau)\le C_T(\tau)+205\,\eps_R$, and hence
$\Pr(v\in\outfr S_\tau)+\lambda_T(t)\le a_{d(v)}+e_R$. A vertex of type $T_0$ has
three neighbors of degree $3$, so by~\ref{it:sweep-L1} the vector
$(Y_v,Y_{x_1},Y_{x_2},Y_{x_3})$ satisfies the hypothesis of \cref{prop:star-cubic}
with $r=r_{33}(1-\eps_R)$, and $\Pr(v\in\outfr S_\tau)\le\kappa(r_{33}(1-\eps_R))
\le\cthree+e_R$, with $\lambda_{T_0}=0$. \Cref{lem:charging} with $e=e_R$ gives the
claim. If $\tau>\frac{35}{32}$, then $\Pr(v\in\outfr S_\tau)\le\Pr(Y_v>\tau)\le
\Phi(-\frac{35}{32})<0.13704<\cthree$ for every $v$, so $\E|\outfr S_\tau|\le\cthree
N\le\beta(H)$. Finally, for $\tau<0$ the event $\{v\in\infr S_\tau\}$ is, up to null
events, the event $\{v\in\outfr S_{-\tau}\}$ for the labeling $-Y$, which has the
same distribution.
\end{proof}

\begin{proof}[Proof of \cref{thm:switched4}]
The proof has two parts. The first proves the bound with remainder $\eta N$ for
graphs of minimum degree at least~$3$, by a concentration argument as for
\cref{thm:cubic}; the second extends it to all graphs of maximum degree~$4$ with
\cref{cor:min-degree-three}.

\emph{Minimum degree $3$.} We first show: for every $\eta>0$ there is $N(\eta)$ such
that every graph $H$ with $\delta(H)\ge3$, $\Delta(H)\le4$ and $N\ge N(\eta)$ vertices
satisfies $\pw(H)\le\beta(H)+\eta N$. Fix $\eta>0$. We begin by fixing the parameters and
the counts to be controlled. Choose $R$ with $\eps_R<1$,
$e_R\le\frac\eta4$, and $R\ge R_0$ for $\frac\eta4$ in
\cref{lem:switch4-zero}; choose an integer $K_0\ge1$ with
$\varphi(0)/K_0\le\frac\eta4$; and let $\tau_j=-2+j/K_0$ for $0\le j\le4K_0$. Consider
the $8K_0+4$ counts $|\infr S_{\tau_j}|$ for $0\le j<2K_0$, $|\outfr S_{\tau_j}|$ for
$2K_0<j\le4K_0$, $\hat L_0$ and $\hat L_0'$, the window counts
$M_j=\#\{v:\tau_j<Y_v\le\tau_{j+1}\}$ for $0\le j<4K_0$, and the tail counts
$\#\{v:Y_v\le-2\}$ and $\#\{v:Y_v>2\}$. Each is a sum over $v$ of $\{0,1\}$-valued
terms (for $\hat L_0$ and $\hat L_0'$ because $U\cap W=\emptyset$) that depend only on
the labels on $N[v]$ and the degrees, hence, by~\ref{it:sweep-L2}, only on
$(Z_w)_{w\in B_{\varrho_R+1}(v)}$. Choose $N(\eta)$ such that for $N\ge N(\eta)$
\[
  (8K_0+4)\,16^{\varrho_R+1}<2N^{1/2},\qquad
  2N^{3/4}\le\tfrac\eta4N,\qquad 0.023N+N^{3/4}\le\cthree N .
\]
Next, by \cref{lem:local-counts} with $\Delta=4$, $\varrho=\varrho_R+1$ and
$\lambda=N^{3/4}$, some realization has pairwise distinct labels and every count
within $N^{3/4}$ of its mean; fix it. By
\cref{lem:switch4-means,lem:switch4-zero}, every inner count at $\tau_j<0$ and every
outer count at $\tau_j>0$ is at most $\beta(H)+\frac\eta4N+N^{3/4}$, both $\hat L_0$
and $\hat L_0'$ are at most $\beta(H)+\frac\eta4N+N^{3/4}$, every window count is at
most $\frac\eta4N+N^{3/4}$, and the two tail counts are at most $0.023N+N^{3/4}$.

Finally, we bound every frontier of this realization. Let
$t\in[-2,2)\setminus\{0\}$, and let $j$ be such that $t\in[\tau_j,\tau_{j+1})$.
If $t<0$, then $j<2K_0$, and \cref{lem:grid}\textup{(i)} bounds $|\infr S_t|$ by the
inner count at $\tau_j$ plus $M_j$. If $t>0$, then $\tau_{j+1}>0$, and
\cref{lem:grid}\textup{(i)} bounds $|\outfr S_t|$ by the outer count at $\tau_{j+1}$ plus $M_j$. In both cases the bound
is at most $\beta(H)+\frac\eta2N+2N^{3/4}\le\beta(H)+\eta N$. For $t<-2$ and $t\ge2$,
\cref{lem:grid}\textup{(ii)},\textup{(iii)} bound the inner, respectively outer,
frontier by a tail count, which is at most $0.023N+N^{3/4}\le\cthree N\le\beta(H)$
because every vertex of $H$ has degree $3$ or $4$. Also
$\max(\hat L_0,\hat L_0')+1\le\beta(H)+\frac\eta4N+N^{3/4}+1\le\beta(H)+\eta N$.
\Cref{lem:median-layout4} with $w=\beta(H)+\eta N$ gives $\pw(H)\le\beta(H)+\eta N$.

\emph{All graphs.} Let $\rho_0(N)$ be the largest value of $\max\{0,\pw(H)-\beta(H)\}$
over the graphs $H$ on $N$ vertices with $\delta(H)\ge3$ and $\Delta(H)\le4$ (zero if
there are none). By the first part, $\rho_0(N)\le\eta N$ for $N\ge N(\eta)$, so
$\rho_0(N)=o(N)$, and so is its running maximum $\bar\rho$~\eqref{eq:running-max},
which is nondecreasing.
\Cref{cor:min-degree-three} with $k=4$ and $(a_0,\dots,a_4)=(0,0,0,\cthree,0.2627)$
gives $\pw(G)\le\cthree n_3+0.2627\,n_4+\bar\rho(n)+3\log_2n+1$ for every nonempty
graph $G$ with $\Delta(G)\le4$.
\end{proof}

\begin{remark}[the coefficient $0.2627$]\label{rem:switched4-coefficient}
The coefficient $0.2627$ is a target, not the limit of the method. The numerical
maxima after \cref{lem:switch-mixed} show that, in this form, the bounds of
\cref{lem:switch-mixed}\ref{it:switch-mixed:c} cannot be lowered below about
$0.2618-j\,s$, the binding pattern being $j=2$. The charging at nonzero thresholds
has slack about $10^{-4}$ (\cref{prop:certificate-switched}). The
target lies just below $(1+6\cthree)/7\approx0.2627047$, the value at which the two
branches of the bootstrap recurrence~\eqref{eq:bootstrap-rec} meet at $d=5$. Compared with
\cref{thm:deg4}, \cref{thm:switched4} gives the smaller bound exactly when
$n_4<\frac{F_3-\cthree}{0.0052}\,n_3\approx0.0646\,n_3$ (against \ref{thm:deg4:b};
\cref{rem:best-row}). \Cref{app:alg-switched} turns the proof into an algorithm
(\cref{thm:algo-precise}, switched case).
\end{remark}

\subsubsection{The certificates of the switch lemma}\label{sec:switch-cert-intro}

The rest of this subsection describes the computer-assisted part of the proof of
\cref{lem:switch-mixed}. After the reduction to equal incident correlations in that
proof, it remains to bound the function $G_r$ of~\eqref{eq:switch4-G} on the set
$E_d$ of $d\times d$ correlation matrices. This is done for $d=3$ and the two
patterns of \cref{lem:switch-mixed}\ref{it:switch-mixed:a},\ref{it:switch-mixed:b},
and for $d=4$ and the five patterns $j=0,\dots,4$ of
\cref{lem:switch-mixed}\ref{it:switch-mixed:c}. Write $V$ for the bound to be proved
and $s_i=s_{r_i}=\sqrt{1-r_i^2}$. The two cases use different parametrizations and
formulas (\cref{sec:switch-cert-formulas}) but the same kind of certificate
(\cref{sec:switch-cert-boxes}). \Cref{sec:switch-cert-results} gives the results and
lists what is trusted.

\subsubsection{Parametrization and formulas}\label{sec:switch-cert-formulas}

Let $W\sim N(0,C)$ be independent of $Z\sim N(0,1)$, let $Y_i=r_iZ+s_iW_i$, and let
$\Sigma$ be the correlation matrix of $(Y_1,\dots,Y_d)$, so that
$\Sigma_{ij}=r_ir_j+s_is_jC_{ij}$ for $i\ne j$.

\emph{Degree $3$: angles.} Write $C_{ij}=\cos\theta_{ij}$ with
$\theta=(\theta_{12},\theta_{13},\theta_{23})\in[0,\pi]^3$. By
\cref{lem:cubic-gram}, $C$ is a correlation matrix exactly when $\theta$ lies in the
convex polytope $\Theta$ cut out by the three triangle inequalities and
$\theta_{12}+\theta_{13}+\theta_{23}\le2\pi$, and $C$ is positive definite exactly in
the interior of $\Theta$. Let $\Sigma^{(0)}$ be the correlation matrix of
$(Z,Y_1,Y_2,Y_3)$. It is positive definite whenever $C$ is, since its Schur
complement in $Z$ is $\mathrm{diag}(s_i)\,C\,\mathrm{diag}(s_i)$. Since
$\lceil N/2\rceil=\ind\{N\ge1\}+\ind\{N=3\}$ for $N\le3$,
\[
  G_r=\Pr(Z\le0,\ N\ge1)+\Pr(Z\le0,\ N=3)
  =\tfrac12-P_4\bigl(\Sigma^{(0)}\bigr)+P_4\bigl(\Sigma^{(0)\prime}\bigr),
\]
where $P_4$ is the probability of the positive orthant and $\Sigma^{(0)\prime}$ is
$\Sigma^{(0)}$ with the signs of the row and column of $Z$ reversed. Along the path
$R_t=(1-t)I+t\Sigma^{(0)}$, \cref{lem:plackett} gives the value as an integral,
\[
  P_4\bigl(\Sigma^{(0)}\bigr)=\frac1{16}+\int_0^1\sum_{i<j}
  \frac{\Sigma^{(0)}_{ij}}{2\pi\sqrt{1-t^2(\Sigma^{(0)}_{ij})^2}}
  \Bigl(\frac14+\frac{\arcsin\pi_{kl}(t)}{2\pi}\Bigr)dt,
\]
where $\{k,l\}$ is the complementary pair of indices and $\pi_{kl}(t)$ is the partial
correlation of $k$ and $l$ given $i$ and $j$ under $R_t$. Here the conditional
orthant probability of two coordinates is given by \cref{lem:sheppard}. In the
interior of $\Theta$, \cref{lem:plackett} and the chain rule
$\partial\Sigma_{ij}/\partial\theta_{ij}=-s_is_j\sin\theta_{ij}$ give
\[
  \frac{\partial G_r}{\partial\theta_{ij}}
  =-s_is_j\sin\theta_{ij}\,\varphi_2(0,0;\Sigma_{ij})\,\bigl(p'_{ij}-p_{ij}\bigr),
\]
where $p_{ij}$ and $p'_{ij}$ are the conditional probabilities, given $Y_i=Y_j=0$, that
the two remaining coordinates of the vectors with correlation matrices $\Sigma^{(0)}$
and $\Sigma^{(0)\prime}$ are positive; both are closed forms by \cref{lem:sheppard}. When
$r_i=r_j$, the factor $s_is_j\sin\theta_{ij}\,\varphi_2(0,0;\Sigma_{ij})$ is rewritten as
$s_i\sqrt{1+\cos\theta_{ij}}/\bigl(2\pi\sqrt{1+\Sigma_{ij}}\bigr)$ (using
$1-\Sigma_{ij}=s_i^2(1-\cos\theta_{ij})$), which stays bounded as
$\theta_{ij}\to0$.

\emph{Degree $4$: the elliptope.} Use the six entries
$c=(c_{12},c_{13},c_{14},c_{23},c_{24},c_{34})$ of $C$ as coordinates. The parameter set
is the elliptope $E_4\subset[-1,1]^6$, which is compact and convex, with the positive
definite matrices as its interior. For every $v\in\R^4$, $v^{\mathsf T}Cv\ge0$ is a
linear inequality in $c$ valid on $E_4$; with $v$ a vector of binary fractions it has
rational coefficients, and these \emph{cuts} are used to describe $E_4$ from outside.
Since $G_r$ is invariant under the permutations of the neighbors that preserve the
bounds, we may also impose linear symmetry constraints: for $j\in\{0,4\}$,
$c_{12}$ is the largest entry and $c_{13}\ge c_{14},c_{23},c_{24}$; for
$j\in\{1,3\}$, the three neighbors with equal bounds are $2,3,4$ and
$c_{12}\ge c_{13}\ge c_{14}$; for $j=2$, the neighbors with bound $r_{34}$ are $1,2$
and $c_{13}\ge c_{14},c_{23},c_{24}$. The next lemma reduces the value of $\Psi$, and
hence of $G_r$, to a sign moment.

\begin{lemma}[sign moment]\label{lem:switch-sign}
For a centered Gaussian vector $(Z,Y_1,\dots,Y_4)$ with unit variances and
$\Corr(Z,Y_i)=a_i$,
\[
  \Psi(Z;Y_1,\dots,Y_4)=\sum_{i=1}^4\frac{\arccos a_i}{4\pi}+\frac18-\frac K8,\qquad
  K=\E\bigl[\operatorname{sgn}Y_1\operatorname{sgn}Y_2\operatorname{sgn}Y_3\operatorname{sgn}Y_4\bigr].
\]
\end{lemma}

\begin{proof}
With probability $1$, all coordinates are nonzero. Write
$\lceil N/2\rceil=\frac N2+\frac12\ind\{N\text{ odd}\}$. By \cref{lem:sheppard},
$\E[\ind\{Z\le0\}N]=\sum_i\Pr(Z\le0<Y_i)=\sum_i\frac{\arccos a_i}{2\pi}$. Next,
$\ind\{N\text{ odd}\}=\frac12(1-\Pi)$ with $\Pi=\prod_i\operatorname{sgn}Y_i$, because
$\Pi=(-1)^{4-N}=(-1)^N$; and
$\ind\{Z\le0\}=\frac12(1-\operatorname{sgn}Z)$. The means of $\operatorname{sgn}Z$
and of $\operatorname{sgn}Z\cdot\Pi$ vanish, as both are odd functions of a centered
Gaussian vector. Hence
$\E[\ind\{Z\le0\}\ind\{N\text{ odd}\}]=\frac14(1-K)$, and the identity follows.
\end{proof}

After the reduction to equal incident correlations, $a_i=r_i$, so the first term in
\cref{lem:switch-sign} is a constant, and $K$ is a function of $\Sigma$. Along
$R_t=(1-t)I+t\Sigma$, \cref{lem:plackett} gives
\[
  K(\Sigma)=\int_0^1\sum_{p<q}\frac4{\pi^2}\,
  \frac{\Sigma_{pq}\arcsin\pi_{kl}(t)}{\sqrt{1-t^2\Sigma_{pq}^2}}\,dt,
\]
since $\partial K/\partial\Sigma_{pq}$ is $4\varphi_2(0,0;\Sigma_{pq})$ times the
conditional mean of $\operatorname{sgn}Y_k\operatorname{sgn}Y_l$ given $Y_p=Y_q=0$,
which is $\frac2\pi\arcsin$ of the conditional correlation. In the interior of
$E_4$, therefore,
\[
  \frac{\partial G_r}{\partial c_{pq}}=-\frac{s_ps_q}{2\pi^2}\,
  \frac{\arcsin\pi_{kl\cdot pq}}{\sqrt{1-\Sigma_{pq}^2}},
\]
where $\pi_{kl\cdot pq}$ is the partial correlation in $\Sigma$. If $r_p=r_q$, then
$1-\Sigma_{pq}=s_p^2(1-c_{pq})$, and the derivative is at most
$K_{pq}/\sqrt{1-c_{pq}}$ in absolute value, with
$K_{pq}=s_p/\bigl(4\pi\sqrt{1+\Sigma_{pq}^{\min}}\bigr)$ for the smallest value
$\Sigma^{\min}_{pq}$ of $\Sigma_{pq}$ on the region considered. Integrated along a
segment from $x'$ to $x$, this coordinate contributes at most
$2K_{pq}|\sqrt{1-x_{pq}}-\sqrt{1-x'_{pq}}|$. This square-root bound handles the
corners where two residuals with equal bounds coincide. Near such corners the
gradient has no limit, and the following bound is used instead.

\begin{lemma}[pair bound]\label{lem:switch-pair}
In the setting of \cref{lem:switch-sign}, for every pair $\{p,q\}$ with complementary
pair $\{k,l\}$,
\[
  \Psi\le\sum_{i=1}^4\frac{\arccos a_i}{4\pi}+\frac18
  -\frac{\arcsin\Sigma_{kl}}{4\pi}+\frac{\arccos\Sigma_{pq}}{4\pi}.
\]
\end{lemma}

\begin{proof}
With probability $1$, $\operatorname{sgn}Y_p\operatorname{sgn}Y_q=1-2\cdot\ind\{Y_pY_q<0\}$, so
\[
  K=\E[\operatorname{sgn}Y_k\operatorname{sgn}Y_l]
  -2\,\E[\operatorname{sgn}Y_k\operatorname{sgn}Y_l;\ Y_pY_q<0].
\]
By
\cref{lem:sheppard} the first term is $\frac2\pi\arcsin\Sigma_{kl}$, and the second
has absolute value at most $2\Pr(Y_pY_q<0)=\frac2\pi\arccos\Sigma_{pq}$. Substitute
into \cref{lem:switch-sign}.
\end{proof}

On a box, the right side is largest at the lower ends of $c_{kl}$ and $c_{pq}$, so
the pair bound needs only one evaluation per box.

\subsubsection{Box certificates}\label{sec:switch-cert-boxes}

In this subsection the \emph{domain} is $\Theta$ for $d=3$, and for $d=4$ the set of
points of $E_4$ that satisfy the symmetry constraints.

The root box ($[0,\pi]^3$ for $d=3$, $[-1,1]^6$ for $d=4$) is covered by finitely many
closed boxes forming a guillotine partition, found by a floating-point
branch-and-bound search. A separate program checks that the stored boxes tile the
root box exactly. It rebuilds the partition by finding, at each step, a coordinate
value that separates the boxes, so the certificate does not rely on the bookkeeping of
the search.

Some points of the domain need no boxes, because $G_r$ is continuous.
Continuity holds because the distribution of the vector in~\eqref{eq:switch4-G}
depends weakly continuously on $C$, and the function
$y\mapsto\ind\{y_0\le0\}\lceil N(y)/2\rceil$ is bounded and continuous outside the
hyperplanes $\{y_i=0\}$, a null set for every distribution in the family, so the continuous
mapping theorem applies. For $d=3$, the floating-point root box misses the points of
$\Theta$ with a coordinate in $(\mathrm{fl}(\pi),\pi]$, where $\mathrm{fl}(\pi)<\pi$ is
the floating-point value of $\pi$. These points are limits of points of $\Theta$
inside the root box, since $\Theta$ is convex with nonempty interior, so they need no
boxes. For $d=4$, the half-open boxes $\prod_m[\mathrm{lo}_m,\mathrm{hi}_m)$
partition $[-1,1)^6$, which contains the interior of the domain. Since the domain is
convex with nonempty interior and $G_r$ is continuous, it suffices to bound $G_r$ on
the interior.

Every box $B$ is then verified in Arb ball arithmetic, regardless of how the search
classified it, in one of the following ways; a box that fails is subdivided and the
parts are verified again.
\begin{itemize}
  \item \emph{Empty.} The box misses the domain. For $d=3$ one constraint of $\Theta$
  is violated on all of $B$, checked exactly. For $d=4$ a Farkas certificate, a
  nonnegative combination of the box bounds, the symmetry constraints and the stored
  cuts, is checked in exact rational arithmetic; if the stored cuts do not suffice,
  cuts at eigenvectors of non-positive-definite points are added. For $d=4$ a box can
  also be discarded because an exact weak-duality bound shows $x_m\ge\mathrm{hi}_m$ on
  the domain for some coordinate $m$; then the domain meets the half-open box in no
  point. This occurs where symmetry and the box bounds force an equality.
  \item \emph{Pair bound} ($d=4$). \Cref{lem:switch-pair} for some pair, evaluated at
  the lower box ends, is $<V$.
  \item \emph{Bounded.} Fix a point $x_1$ in the interior of the domain, in or near
  $B$ (for $d=4$, positive definiteness is checked by leading principal minors in Arb,
  and $x_1$ satisfies the symmetry constraints exactly). Let $H$ be the smallest box
  containing $B$ and $x_1$. Let $[g_{c,m}-g_{r,m},\,g_{c,m}+g_{r,m}]$ enclose the
  partial derivative of $G_r$ in the coordinate $m$ over the intersection of $H$ with
  the interior of the domain. Let $M$ be an upper bound for $\max g_c\cdot x$ over the points $x$ of
  $B$ that satisfy the linear constraints; it is certified by a dual feasible solution
  of the linear program, evaluated exactly or in ball arithmetic. For $x$ in $B$ and
  in the domain, the half-open segment $[x_1,x)$ lies in the interior of the domain
  (by convexity), where $G_r$ is differentiable, so integrating the gradient along it
  gives
  \[
    G_r(x)\le G_r(x_1)+\bigl(M-g_c\cdot x_1\bigr)+\sum_mg_{r,m}\max_{y\in B}|y_m-x_{1,m}|,
  \]
  plus, for $d=4$, the term $\sum2K_{pq}\max_{y\in B}|\sqrt{1-y_{pq}}-\sqrt{1-x_{1,pq}}|$
  for the coordinates treated by the square-root bound instead of the gradient
  enclosure. The box is accepted if the right side is $<V$ in Arb.
  \item \emph{Pushed to a face} ($d=3$). Near the face
  $F=\{\theta_{12}+\theta_{13}+\theta_{23}=2\pi\}$ the conditional variances
  degenerate and the gradient formula fails. Moving along $(1,1,1)$ preserves the
  triangle inequalities, so every $x\in B\cap\Theta$ is joined to a point $x^*\in F$
  by a segment in $\Theta$. The verifier certifies, by affine-arithmetic gradient
  enclosures on sub-boxes, that $\sum_m\partial G_r/\partial\theta_m>0$ on the prism
  swept by $B$ towards $F$, so $G_r(x)\le G_r(x^*)$. It then bounds $G_r$ on the face
  patch by covering it with parallelogram cells, each handled like a bounded box with
  a linear term maximized at a vertex.
\end{itemize}

\emph{Enclosures.} Values $G_r(x_1)$ are enclosed by integrating the Plackett
representations above with Arb's rigorous integration (\texttt{acb.integral}, whose
\texttt{analytic} flags certify analyticity of the integrand) on $[0,1-10^{-8}]$. The
rest of the path is bounded by the elementary integral
$\int dt/\sqrt{1-t^2\rho^2}$, an arcsine difference, since every conditional term is
bounded ($\frac14+\frac{\arcsin}{2\pi}\in[0,\frac12]$, $|\arcsin|\le\frac\pi2$). For
$d=4$ the integrand is evaluated without matrix inverses: along $R_t$ each
conditional covariance given $(p,q)$ is $A_{ab}(t)/(1-t^2\Sigma_{pq}^2)$ for a cubic
polynomial $A_{ab}$, and the substitution $t=1-e^{-u}$ moves poles near $t=1$ away
from the path. Gradients are enclosed in ball arithmetic and, for $d=4$, also in
affine arithmetic over Arb (with rigorous remainders for products, reciprocals and
square roots), on up to $16$ sub-boxes; the results are intersected. An undetermined
partial correlation is replaced by the whole interval $[-1,1]$ (equivalently, a
conditional probability by $[0,1]$), and if every enclosure fails for $d=4$, the
analytic bound $s_ps_q/(4\pi\sqrt{1-\max|\Sigma_{pq}|^2})$ on the derivative is used.

\subsubsection{Results and what is trusted}\label{sec:switch-cert-results}

All seven bounds are certified. Every comparison is a strict inequality in Arb, and
every box list passes the coverage check.
\begin{center}
\small
\setlength{\tabcolsep}{4pt}
\begin{tabular}{@{}llrrr@{}}
\toprule
pattern & $V$ & boxes & largest certified bound & numerical maximum\\
\midrule
$(r_{33},r_{33},r_{34})$ & $0.1529003$ & $8\,498$ & $0.1529002993$ & $0.1528953$\\
$(r_{33},r_{34},r_{34})$ & $0.165616$ & $21\,082$ & $0.1656159972$ & $0.1656115$\\
$d=4$, $j=0$ & $0.2627$ & $368\,779$ & $0.2626999832$ & $0.250000$\\
$d=4$, $j=1$ & $0.2627-s$ & $2\,366\,198$ & $0.2496218225$ & $0.245466$\\
$d=4$, $j=2$ & $0.2627-2s$ & $5\,450\,439$ & $0.2365436480$ & $0.235687$\\
$d=4$, $j=3$ & $0.2627-3s$ & $5\,415\,446$ & $0.2234654716$ & $0.221375$\\
$d=4$, $j=4$ & $0.2627-4s$ & $2\,738\,755$ & $0.2103872958$ & $0.208185$\\
\bottomrule
\end{tabular}
\end{center}
The column ``boxes'' counts the boxes of the search; the verification subdivides some
of them further. The largest certified bounds lie just below $V$ because the search
accepts a box as soon as its bound is below $V$; they are not estimates of the
maximum. For $d=3$, $27$ boxes of the first pattern touch or approach the face $F$,
where the numerical maximum $0.1528953$ lies $5\cdot10^{-6}$ below $V$; they were
closed by the push to the face, with $331$ face cells. For $d=4$, the first
verification pass left $14$ boxes open, and recursive subdivision with the same
certificates closed them. Twelve of these had been labeled empty by the search
although they contain feasible points, because its cutting-plane loop had stopped
early. The verifier for $d=4$ compared its bounds with a value of $V$ computed in
double precision, with $\cthree$ rounded to $0.139822124008$. This value differs
from the exact $V$ by less than
$2\cdot10^{-14}$, far below the smallest margin, $2.7\cdot10^{-11}$ at $j=2$.

\emph{What is trusted.} As in \cref{sec:cert-limits}, the certificates rest on the
correctness of Arb/FLINT (\texttt{python-flint}~0.9.0), including its rigorous
integration, of exact rational arithmetic, and of the derivations above. The HiGHS
linear-program solver is used only to find dual and Farkas certificates, which are
rechecked exactly or in ball arithmetic. Each verifier has a single implementation. An earlier
verification of the case $d=3$ had a defect: an undetermined partial correlation
(a not-a-number enclosure) was turned into an exact value, which collapsed the gradient
enclosures near the face $F$. It was found by an audit and corrected, and the results
above are those of the corrected verifier, which treats an undetermined enclosure as
the whole interval. The box lists for $d=3$ and per-part summaries for $d=4$ are part
of the accompanying code. The full box lists for $d=4$, about $11$\,GB, are not; the
search programs regenerate them.
{}
\subsection{Pathwidth sequences: proofs, and comparison with previous bounds}
\label{app:sequences}

This subsection proves the two results on pathwidth sequences stated in
\cref{sec:sequences}: \cref{thm:pw-seq}, which lists six pathwidth sequences, and
\cref{cor:table}, which states that every row of \cref{tab:sequences} is a pathwidth
sequence; the rows are the bootstraps of five of the six sequences, rounded up, with the
last column bounded by closed forms. \Cref{sec:bootstrap-remarks}
then compares the coefficients of \cref{thm:main} with the previous ones.

We use the notation of \cref{sec:sequences}: pathwidth sequences are as
in \cref{def:pw-seq}, the bootstrap is the extension~\eqref{eq:bootstrap-rec}, $h_d$ and
$H_d$ are the increments of a bootstrap and their running minima
(\cref{sec:bootstrap-lemma}), and $\sigma$ is the bootstrap of $\tripS$
(\cref{sec:proof-main}). The constants $\cthree$, $F_3$ and $F_4$ are those of
\cref{sec:sequences}.

\subsubsection{Pathwidth sequences: proofs of
\texorpdfstring{\cref*{thm:pw-seq,cor:table}}{the results on
pathwidth sequences}}
\label{sec:proof-pw-seq}

Both statements follow from \cref{lem:bootstrap}, applied with the six sequences of
\cref{thm:pw-seq} as seeds. The first proof checks the hypotheses of the lemma; the
second computes the bootstraps.

\thmpwseq*

\begin{proof}[Proof of \cref{thm:pw-seq}]
For each of the six sequences $X=(X_3,\dots,X_k)$ we check the hypotheses of
\cref{lem:bootstrap} with $(b_3,\dots,b_k)=X$.

First, all six are nondecreasing with $X_3\ge0$.

Next, $X_k+H_k\le1$ in all six cases. For $\seqC$ ($k=3$), $X_3+H_3=2\cthree$. For
$\seqM$ ($k=4$), $H_4=h_4=0.2627-\cthree<h_3$, so $X_4+H_4=0.2627+(0.2627-\cthree)$.
For $\seqF$ ($k=4$), $H_4=h_4=0$, so $X_4+H_4=F_4$. For $\tripP$, $\tripR$ and
$\tripS$ ($k=5$), $X_5+H_5$ equals $0.3667+0.1092$, $0.3600+0.1070$ and
$\frac{11}{32}+\frac{3}{32}$, respectively; the increments of $\tripP$ and $\tripR$
are listed in \cref{rem:bootstrap-refinements}(a), and those of $\tripS$ are
$\frac5{32},\frac3{32},\frac3{32}$. All these values are at most $1$.

Finally, we check the bound on the graphs $J$ in \cref{lem:bootstrap}. Let $J$ be a
graph with $\delta(J)\ge3$ and $\Delta(J)\le k$. If $X=\seqC$, then $J$ is cubic,
$|V(J)|=n_3(J)$, and \cref{thm:cubic} gives $\pw(J)\le\cthree n_3(J)+o(|V(J)|)$. If
$X=\seqM$, \cref{thm:switched4} gives
$\pw(J)\le\cthree n_3(J)+0.2627\,n_4(J)+o(|V(J)|)$. If $X=\seqF$, then
$|V(J)|=n_3(J)+n_4(J)$, and \cref{thm:deg4}\ref{thm:deg4:a} gives
$\pw(J)\le F_4(n_3(J)+n_4(J))+o(|V(J)|)$. If $X\in\{\tripP,\tripR,\tripS\}$,
\cref{thm:deg5} gives $\pw(J)\le\sum_{d=3}^5X_dn_d(J)+o(|V(J)|)$.

So \cref{lem:bootstrap} shows that the bootstrap $b$ of $X$ is a pathwidth sequence
with all entries at most $1$. Since $b_d=X_d$ for $d\le k$, the sequence $X$, which
continues with $X_d=1$ for $d>k$ (\cref{def:pw-seq}), is entrywise at least $b$, hence
a pathwidth sequence.
\end{proof}

\cortable*

\begin{proof}[Proof of \cref{cor:table}]
By \cref{lem:bootstrap}, whose hypotheses were checked in the previous proof, the
bootstraps of the six sequences are pathwidth sequences. A sequence that is entrywise
at least a pathwidth sequence is one, so it suffices to show that every entry of
\cref{tab:sequences} is at least the corresponding entry of the bootstrap. For the
black entries this holds by $\cthree<0.1399$, $F_3<0.1402$ and $F_4<0.2458$
(\cref{prop:F3F4}). The grey entries for $4\le d\le10$ are the exact values of
\eqref{eq:bootstrap-rec} rounded up to four decimals. The accompanying script
\texttt{bootstrap\_table.py} computes them in exact rational arithmetic and checks the
branch of the maximum at every step; it also computes the bootstrap of $\seqC$, which
the table omits. We record what the computation consists of, row by row, and then
bound the entries of the last column.

\emph{Row $\seqF$.} Here $H_4=0$, so \eqref{eq:bootstrap-rec} gives
$b_5=\max\{\frac{1+3F_4}{5},1\}=1$, then $H_5=0$ again, and $b_d=1$ for all $d\ge5$.

\emph{Row $\seqC$.} Here $k=3$ and $H_3=\cthree$. Since $\frac18<\cthree<\frac17$,
the recurrence takes the second term of the maximum at $d=4$ and the first term at
$d=5,6$:
\[
  b_4=1-4\cthree,\qquad b_5=1-\tfrac{17}{5}\cthree,\qquad b_6=1-\tfrac{44}{15}\cthree,
\]
with $H_4=\cthree$ (as $h_4=1-5\cthree>\cthree$), $H_5=h_5=\frac35\cthree$ and
$H_6=h_6=\frac{7}{15}\cthree$. So $H_6\le H_5\le3H_6$ and $h_6=H_6$, and
\cref{lem:concave} with $j=6$ gives $d\,b_d=(d-2)\,b_{d-1}+b_{d-2}+1$ for $d\ge7$.
Hence $u_d:=1-b_d=\gamma_d\cthree$ for $d\ge4$, with rationals $\gamma_d$ that satisfy
$d\gamma_d=(d-2)\gamma_{d-1}+\gamma_{d-2}$ for $d\ge7$, and $b_d$ is rounded up using a
rigorous lower bound for $\cthree$.

\emph{Row $\seqM$.} Here $k=4$ and $H_4=0.2627-\cthree$. Since
$0.2627<\frac{1+6\cthree}7$, the recurrence takes the second term of the maximum at
$d=5$, and then the first term:
\[
  b_5=1-5\,(0.2627-\cthree),\qquad
  b_6=\tfrac{13}{3}\cthree-\tfrac{4151}{30000},
\]
with $H_5=H_4$ (as $h_5=5\cthree-0.5762>H_4$) and
$H_6=h_6=\frac{5254}{30000}-\frac23\cthree$. So $H_6\le H_5\le3H_6$, and
\cref{lem:concave} with $j=6$ gives $d\,b_d=(d-2)\,b_{d-1}+b_{d-2}+1$ for $d\ge7$.
Every entry is of the form $\alpha_d+\beta_d\cthree$ with rationals $\alpha_d,\beta_d$.
The script decides each branch of the maximum and each running minimum at both ends
of a rigorous enclosure of $\cthree$, and rounds up with the same enclosure.

\emph{Rows $\tripP$, $\tripR$ and $\tripS$.} For $\tripP$ and $\tripR$ the computation
of \cref{rem:bootstrap-refinements}(a) applies: the first term of the maximum is
attained from $d=6$ on, and $d\,b_d=(d-2)\,b_{d-1}+b_{d-2}+1$ for $d\ge7$. For $\tripP$
the entries for $d\ge6$ do not depend on the exact value of $F_3$, since $H_5=h_5$
whenever $0.1092\le F_3\le0.1483$. Row $\tripS$ is the bootstrap $\sigma$ of
\cref{sec:proof-main}.

\emph{The last column.} For $X\in\{\seqC,\seqM,\tripP,\tripR,\tripS\}$ put
$u_d:=1-b_d$ and $w_d:=(d+1)u_d$. By the above, $d\,u_d=(d-2)\,u_{d-1}+u_{d-2}$ for
$d\ge7$, so \cref{lem:weights} applies with $d_0=7$ and
\[
  \lambda=6u_6+u_5=21\cthree,\qquad\tfrac{81437}{10000}-31\cthree,\qquad
  \tfrac{3909}{1000},\qquad\tfrac{3947}{1000},\qquad\tfrac{129}{32}
\]
for $\seqC$, $\seqM$, $\tripP$, $\tripR$ and $\tripS$ respectively; for $\seqM$,
$\lambda\approx3.80921$. The hypothesis $\frac{11}{12}\lambda<w_{10}<\lambda$ of
\cref{lem:weights}\ref{it:weights:bound} with $d_1=10$ holds in all five cases
(checked by the script). Hence
$w_d>(1-\frac1{d^2})\lambda\ge\frac{120}{121}\lambda$ for every $d\ge11$, that is,
$b_d<1-\frac{120\lambda/121}{d+1}$. Since
$\frac{120}{121}\cdot21\cthree>2.91$, $\frac{120}{121}\cdot3.8092>3.77$,
$\frac{120}{121}\cdot3.909>3.87$ and $\frac{120}{121}\cdot3.947>3.91$, the entries of
the last column for $\seqM$, $\tripP$ and $\tripR$, and the closed form
$1-\frac{2.91}{d+1}$ for $\seqC$, bound the bootstrap from above. For
$\tripS$, $b_d=\sigma_d\le1-\frac4{d+1}$ for $d\ge11$ by \cref{prop:large-degrees},
and for $\seqF$ the entry is $1$.
\end{proof}

We also justify the highlighting in \cref{tab:sequences}: the highlighted entries are
the column minima. At $d=3$ and $d=4$ this can be read off the table. For $d\ge5$, row
$\tripS$ is below rows $\tripP$ and $\tripR$ (\cref{rem:bootstrap-refinements}(a)). It
is also below row $\seqM$: for $5\le d\le10$ by the table, and for $d\ge11$ because the
bootstrap of $\seqM$ has $w_d<\lambda<3.81$ by \cref{lem:weights}, while that of
$\tripS$ has $w_d\ge4$. The omitted bootstrap of $\seqC$ is entrywise at least row
$\seqM$: the script checks this exactly for $d\le60$, and for $d>60$ its weights
satisfy $w_d<21\cthree<2.94$, while those of $\seqM$ satisfy
$w_d>\frac{120}{121}\cdot3.8092$. In the last column, the closed forms all have the
shape $1-\frac{c}{d+1}$, so the minimum there is the one with the largest $c$.

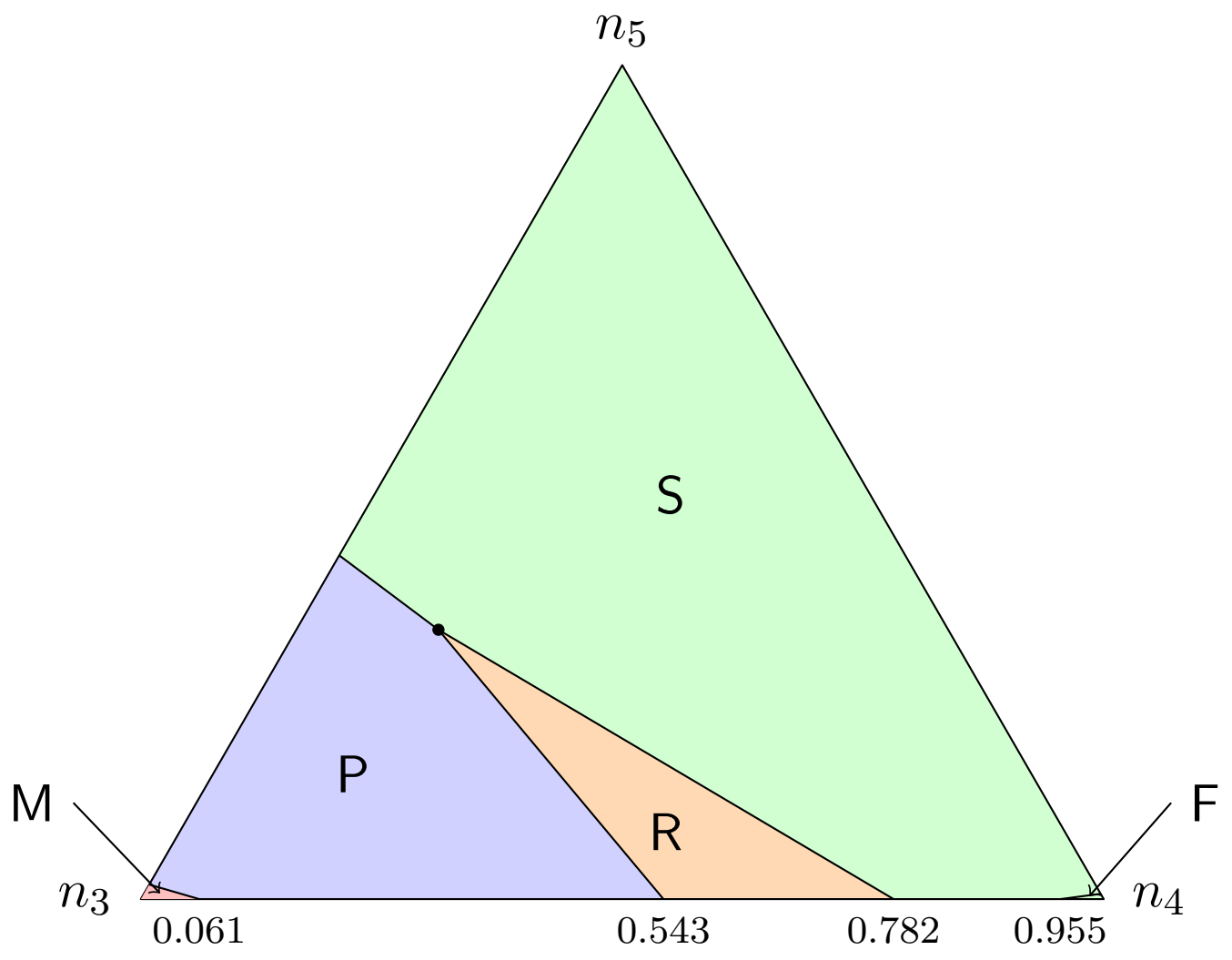
\begin{figure}[t]
\centering
\begin{tikzpicture}[scale=7,line join=round]
  \fill[blue!18]   (0,0) -- (0.5428,0) -- (0.3091,0.2797) -- (0.2061,0.3569) -- cycle;
  \fill[orange!30] (0.3091,0.2797) -- (0.5428,0) -- (0.7818,0) -- cycle;
  \fill[green!18]  (0.2061,0.3569) -- (0.3091,0.2797) -- (0.7818,0) -- (1,0)
                   -- (0.5,0.8660) -- cycle;
  \draw (0,0) -- (1,0) -- (0.5,0.8660) -- cycle;
  \draw[thin] (0.5428,0) -- (0.3091,0.2797) -- (0.7818,0);
  \draw[thin] (0.3091,0.2797) -- (0.2061,0.3569);
  \draw[thin] (0.9549,0) -- (0.9968,0.0055);
  \fill[red!25] (0,0) -- (0.0607,0) -- (0.0087,0.0151) -- cycle;
  \draw[thin] (0.0607,0) -- (0.0087,0.0151);
  \fill (0.3091,0.2797) circle (0.18pt);
  \node at (0.22,0.13) {$\tripP$};
  \node at (0.545,0.07) {$\tripR$};
  \node at (0.55,0.42) {$\tripS$};
  \draw[->,thin] (-0.07,0.10) node[left] {$\seqM$} -- (0.02,0.006);
  \draw[->,thin] (1.07,0.10) node[right] {$\seqF$} -- (0.985,0.003);
  \node[left] at (-0.01,0) {$n_3$};
  \node[right] at (1.01,0) {$n_4$};
  \node[above] at (0.5,0.8660) {$n_5$};
  \foreach \x/\l in {0.0607/0.061,0.5428/0.543,0.7818/0.782,0.9549/0.955}
    \node[below,font=\scriptsize] at (\x,0) {\l};
\end{tikzpicture}
\caption{The row of \cref{tab:sequences} with the smallest bound for graphs of maximum
degree at most $5$, as a function of $(n_3,n_4,n_5)/(n_3+n_4+n_5)$
(\cref{rem:best-row}; computed in floating point). A corner stands for the graphs all
of whose vertices of degree at least $3$ have that degree; the labels on the bottom
edge give the proportion of degree-$4$ vertices among those of degree $3$ or $4$. The
region of $\seqF$ is too thin to be visible.}
\label{fig:best-row}
\end{figure}

\begin{remark}[which row gives the smallest bound]\label{rem:best-row}
The rows of \cref{tab:sequences} are pairwise incomparable, and the best of them for a
given graph depends only on the proportions of its degrees. Two rows $x,y$ compare as
$\sum_dx_dn_d\le\sum_dy_dn_d$ if and only if $\sum_d(y_d-x_d)n_d\ge0$. The numbers
below are computed in floating point from the exact sequences and rounded; they are
illustrations, not claims used elsewhere. With the abbreviation
$\delta^X_d:=b^X_d-b^{\tripS}_d$, where $b^X$ is the bootstrap of $X$,
\begin{align*}
  \tripP\le\tripS &\iff 0.01609\,n_3\ge0.0075\,n_4+0.02295\,n_5
     +\textstyle\sum_{d\ge6}\delta^{\tripP}_d\,n_d,\\
  \tripR\le\tripS &\iff 0.01075\,n_3\ge0.003\,n_4+0.01625\,n_5
     +\textstyle\sum_{d\ge6}\delta^{\tripR}_d\,n_d,\\
  \tripP\le\tripR &\iff 0.00534\,n_3\ge0.0045\,n_4+0.0067\,n_5
     +\textstyle\sum_{d\ge6}(\delta^{\tripP}_d-\delta^{\tripR}_d)\,n_d .
\end{align*}

\emph{Maximum degree $3$.} Row $\seqM$ is best; it coincides with $\seqC$ there.

\emph{Maximum degree $4$.} Each row is best on an interval of the ratio $n_4/n_3$:
\begin{center}
\small
\renewcommand{\arraystretch}{1.15}
\begin{tabular}{@{}l>{$}l<{$}>{$}l<{$}@{}}
\toprule
best row & n_4/n_3 & \text{bound per vertex of degree $3$ or $4$}\\
\midrule
$\seqM$ & [0,\ 0.0646] & 0.1398\text{ to }0.1473\\
$\tripP$ & [0.0646,\ 1.187] & 0.1473\text{ to }0.2038\\
$\tripR$ & [1.187,\ 3.583] & 0.2038\text{ to }0.2295\\
$\tripS$ & [3.583,\ 21.17] & 0.2295\text{ to }0.2458\\
$\seqF$ & [21.17,\ \infty] & 0.2458\\
\bottomrule
\end{tabular}
\end{center}
So $\seqM$ is best only if fewer than about $6.1\%$ of the vertices of degree $3$
or $4$ have degree $4$, and $\seqF$ only if fewer than about $4.5\%$ of the vertices of degree at least $3$ have
degree $3$. The bound per vertex of degree $3$ or $4$ never exceeds $F_4$.

\emph{Maximum degree $5$.} \Cref{fig:best-row} shows the best row as a function of
the proportions of $n_3,n_4,n_5$. Rows $\tripP$, $\tripR$ and $\tripS$ meet at
$(n_3,n_4,n_5)\approx(0.529,0.148,0.323)$. On the edge $n_4=0$ of the triangle, row
$\seqM$ is best for $n_5/n_3\le0.0178$, row $\tripP$ for
$0.0178\le n_5/n_3\le0.70$, and row $\tripS$ beyond.

\emph{Larger degrees.} Without the bootstrap, every seed would have coefficient $1$ at
these degrees. With it, the differences between the rows are small: $\delta^{\tripP}_d$ is
$0.0166$, $0.0151$, $0.0134$, $0.0121$, $0.0111$ and $\delta^{\tripR}_d$ is $0.0114$,
$0.0105$, $0.0093$, $0.0084$, $0.0076$ for $d=6,\dots,10$, and both behave like
$(\frac{129}{32}-\lambda)/(d+1)$, where the conserved quantity
$\lambda=6(1-b^X_6)+(1-b^X_5)$ of \cref{lem:weights} is $3.909$ for $X=\tripP$ and
$3.947$ for $X=\tripR$. So row $\tripP$ remains best for graphs with many vertices of
degree $3$ and a fair number of vertices of large degree. For example, if every vertex
has degree $3$ or $10$, row $\tripP$ is best for $0.0374\,n_3\le n_{10}\le1.46\,n_3$;
row $\seqM$ is best below this range and row $\tripS$ above it.
\end{remark}

\subsubsection{Comparison with previous bounds, and refinements}
\label{sec:bootstrap-remarks}

This subsection compares the coefficients $a_d$ of \cref{thm:main} with the previous
ones (\cref{rem:bootstrap-comparison}) and records refinements that follow from the
same argument (\cref{rem:bootstrap-refinements}).

\begin{remark}[comparison with previous coefficients]
\label{rem:bootstrap-comparison}
The previous per-degree coefficients are those of Fomin, Gaspers, Saurabh and
Stepanov \cite[Lemma~1]{FominGaspersSaurabhStepanov2009} for $d\le6$, namely
$\frac16,\frac13,\frac{13}{30},\frac{23}{45}$ at degrees $3$ to $6$, and the
sequence $g_{1/6}(d)$ of Edwards and McDermid \cite[Corollary~5.2]{EdwardsM15}
(recalled in \cref{lem:below-published}), which agrees with them for $d\le6$ and
continues with
$g_{1/6}(7)=\frac{359}{630}$, $g_{1/6}(8)=\frac{1553}{2520}$,
$g_{1/6}(9)=\frac{14827}{22680}$, $g_{1/6}(10)=\frac{155273}{226800}$.
Edwards and McDermid show $g_{1/6}(d)=1-\frac{7/2}{d+1}+O(d^{-3})$
\cite[Section~3.1.2]{EdwardsM15}. The previous bound in terms of edges is
$\frac{13}{75}m+o(m)$: for treewidth by Scott and Sorkin
\cite[Corollary~21]{ScottSorkin2007}, and for pathwidth by Kneis, M{\"o}lle,
Richter and Rossmanith~\cite{KneisMRR2009}; it also follows from the bound of Edwards
and McDermid, since $\max_d2g_{1/6}(d)/d=\frac{13}{75}$, attained at $d=5$.
\Cref{tab:bootstrap-comparison} compares these values with the bootstrap $\sigma$ of
$\tripS$ and the coefficients $a_d$ of \cref{thm:main}.

\begin{table}[tb]
\centering
\small
\renewcommand{\arraystretch}{1.2}
\caption{Per-degree coefficients, rounded to five decimals: the previous
$g_{1/6}(d)$ of \cite{EdwardsM15}, which for $d\le6$ are those of
\cite{FominGaspersSaurabhStepanov2009} (where every degree $d\ge7$ has coefficient
$1$), the
bootstrap $\sigma_d$ of $\tripS$, and the coefficients $a_d$ of \cref{thm:main}. The
last row gives, for each sequence $x$, the coefficient $\max_d2x_d/d$ of $m$ (their
ratio in the last column).}
\label{tab:bootstrap-comparison}
\begin{tabular}{@{}lcccc@{}}
\toprule
$d$ & $g_{1/6}(d)$ & $\sigma_d$ & $a_d$ & $a_d/g_{1/6}(d)$\\
\midrule
$3$ & $0.16667$ & $0.15625$ & $0.15625$ & $0.938$\\
$4$ & $0.33333$ & $0.25000$ & $0.25000$ & $0.750$\\
$5$ & $0.43333$ & $0.34375$ & $0.34375$ & $0.793$\\
$6$ & $0.51111$ & $0.43750$ & $0.43750$ & $0.856$\\
$7$ & $0.56984$ & $0.50446$ & $0.50446$ & $0.885$\\
$8$ & $0.61627$ & $0.55804$ & $0.55952$ & $0.908$\\
$9$ & $0.65375$ & $0.60119$ & $0.60357$ & $0.923$\\
$10$ & $0.68463$ & $0.63676$ & $0.63961$ & $0.934$\\
$d\ge11$ & $1-\frac{7/2}{d+1}+O(d^{-3})$ & $1-\frac{129/32}{d+1}+O(d^{-3})$
  & $1-\frac{4}{d+1}$ & $\to1$\\
\midrule
per edge & $\frac{13}{75}\approx0.17333$ & $\frac{7}{48}\approx0.14583$
  & $\frac{7}{48}$ & $0.841$\\
\bottomrule
\end{tabular}
\end{table}

By \cref{lem:below-published}, the coefficients of \cref{thm:main} are below
$g_{1/6}(d)$ at every degree $d\ge3$, and the previous sequence never reaches the
coefficient $1-\frac4{d+1}$ that \cref{thm:main} has from degree $11$ on.

The relative improvement over $g_{1/6}$ (one minus the last column of
\cref{tab:bootstrap-comparison}) is largest at degree $4$ and shrinks for large
degrees. The deficit
$1-a_d$ is $\frac{111/28}{d+1}$ for $7\le d\le10$ and $\frac{4}{d+1}$ for $d\ge11$,
instead of the deficit $\frac{7/2}{d+1}+O(d^{-3})$ of $g_{1/6}$. The bootstrap
$\sigma$ itself has deficit $\frac{w_d}{d+1}$ with
$\frac{111}{28}\le w_d<\frac{129}{32}$ for $d\ge7$ and $w_d\to\frac{129}{32}$ by
\cref{prop:large-degrees}, so the ratio of the deficit of $\sigma$ to that of
$g_{1/6}$ tends to $\frac{129/32}{7/2}=\frac{129}{112}\approx1.152$. At degree $3$
the coefficient $\frac5{32}$ of \cref{thm:main} is above the cubic constant
$\cthree\approx0.13982$ of \cref{thm:cubic}; graphs in which vertices of degree $3$
dominate get smaller bounds from the refinements of \cref{rem:bootstrap-refinements}.
\end{remark}

\begin{remark}[refinements]\label{rem:bootstrap-refinements}
The following bounds follow from \cref{thm:deg5,lem:bootstrap,lem:concave}, each with
a remainder $o(n)$. Part~(a) is the
computation behind rows $\tripP$ and $\tripR$ of \cref{cor:table}.

\emph{(a) The bootstraps of $\tripP$ and $\tripR$.} Apply \cref{lem:bootstrap} with
$k=5$ to the triples $\tripP=(F_3,0.2575,0.3667)$ and $\tripR=(0.1455,0.2530,0.3600)$
of \cref{thm:deg5}. Both triples are nondecreasing. Their increments are
$(h_3,h_4,h_5)\approx(0.14016,\,0.11734,\,0.10920)$ for $\tripP$ (using
$F_3<0.1401583$) and $(0.1455,\,0.1075,\,0.1070)$ for $\tripR$, so $H_5=h_5$ equals
$0.1092$ and $0.1070$, and $b_5+H_5\le1$. At $d=6$, $\theta_6=(1-b_5-H_5)/6$
equals $0.08735$ for $\tripP$ and $0.08883\ldots$ for $\tripR$; in both cases
$\theta_6\le H_5\le3\theta_6$. By \cref{lem:extension}\ref{it:ext:case},
$h_6=H_6=\theta_6$, so $b_6=b_5+\theta_6$, and \cref{lem:concave} with $j=6$ gives
$d\,b_d=(d-2)\,b_{d-1}+b_{d-2}+1$ for $d\ge7$. Rounded up, the resulting sequences
$\sigma^{\tripP}$ and $\sigma^{\tripR}$ are rows $\tripP$ and $\tripR$ of
\cref{tab:sequences}; in particular $\sigma^{\tripP}_6=0.45405$ and
$\sigma^{\tripR}_6=0.44883\ldots$.

Every graph satisfies
$\pw(G)\le\sum_{d\ge3}\sigma^X_dn_d+o(n)$ for each $X\in\{\tripP,\tripR,\tripS\}$,
where $\sigma^{\tripS}:=\sigma$, hence
$\pw(G)\le\min_X\sum_{d\ge3}\sigma^X_dn_d+o(n)$. The sequences from $\tripP$ and
$\tripR$ improve on $\sigma$ only at degree $3$: for $d\ge4$,
$\sigma_d<\sigma^{\tripR}_d$ and $\sigma_d<\sigma^{\tripP}_d$. (For $4\le d\le6$ this
follows from $\sigma_4=\frac14<0.2530$, $\sigma_5=\frac{11}{32}<0.3600$ and
$\sigma_6=\frac7{16}<0.4488$. For $d\ge7$ the differences satisfy the homogeneous
recurrence $d\,e_d=(d-2)\,e_{d-1}+e_{d-2}$, whose coefficients are positive, and they
are positive at $d=5,6$.) They do not improve \cref{cor:edges} either, since
$\sigma^{\tripP}_6$ and $\sigma^{\tripR}_6$ exceed $\frac7{16}$, so
$\max_d2\sigma^X_d/d\ge\sigma^X_6/3>\frac7{48}$.

\emph{(b) Average degree.} The increments of $\sigma^{\tripP}$, $\sigma^{\tripR}$ and
$\sigma$ are nonincreasing from $d=3$ on: for $d\le6$ by the increments computed in
(a) and by $\sigma_d=\frac{3d-4}{32}$ (\cref{prop:large-degrees}), and for $d\ge7$
because $h_d=H_d\le H_{d-1}=h_{d-1}$ by \cref{lem:concave}. So for each $X$ the linear
interpolation $\bar\sigma^X\colon[2,\infty)\to\R$ of the points $(d,\sigma^X_d)$,
$d\ge2$, where $\sigma^X_2=0$, is concave. Recall the average degree $\bar d=2m/n$
(\cref{sec:proof-edges}). Let $n_{\ge3}$ be the number of vertices of degree at least
$3$ and, if $n_{\ge3}\ge1$, let $\hat d:=\frac1{n_{\ge3}}\sum_{d\ge3}d\,n_d$ be their
average degree. Jensen's inequality for $\bar\sigma^X$ bounds
$\sum_{d\ge3}\sigma^X_dn_d$ from above in two ways: by $\bar\sigma^X(\bar d)\,n$, with
the weights $n_d/n$ for $d\ge2$, if $\delta(G)\ge2$; and by
$\bar\sigma^X(\hat d)\,n_{\ge3}$, with the weights $n_d/n_{\ge3}$ for $d\ge3$. With
(a) this gives
\[
  \pw(G)\le\min_X\bar\sigma^X(\bar d)\,n+o(n)\quad\text{if }\delta(G)\ge2,
  \qquad
  \pw(G)\le\min_X\bar\sigma^X(\hat d)\,n_{\ge3}+o(n)\quad\text{if }n_{\ge3}\ge1.
\]
The second bound needs no condition on the vertices of degree at most $2$. The
condition $\delta(G)\ge2$ in the first cannot be dropped, since adding isolated
vertices can lower $\bar d$ to $2$, where the bound is $o(n)$, without changing the
pathwidth.

The minimum in the first bound is attained by $\tripP$ for $2\le\bar d\le3.5428$, by
$\tripR$ for $3.5428\le\bar d\le3.7818$, and by $\tripS$ from $3.7818$ on; the
breakpoints are numerical. Indeed, on $[2,3]$ all three interpolations pass through
$(2,0)$, and $\sigma^{\tripP}_3$ is the smallest value at $3$; on $[3,4]$ the
crossings are those of the three chords; and from $4$ on, $\sigma_d$ is the smallest
value at every integer, by (a). For example, the minimum is $F_3<0.1402$ at
$\bar d=3$ and about $0.1988$ at $\bar d=3.5$, compared with $\frac7{96}\bar d$, that
is, $0.21875$ and $0.2552\ldots$, from \cref{cor:edges}.
\end{remark}
{}
\subsection{Proof of the algorithmic theorem}\label{app:algorithms}

This subsection proves \cref{thm:algo-precise}, restated below, following the outline
and the roadmap of \cref{sec:alg-statement}, and then
\cref{cor:alg-cutwidth,thm:alg-cut-sweep}.

\thmalgoprecise*

Besides the Las Vegas and deterministic algorithms of the theorem, the case proofs also analyze \emph{the
randomized algorithm}, which takes one sample of the labels within a budget of random
draws and outputs its layout without checking the counts. The Las Vegas algorithm
draws samples in the same way, without the budget, and checks the counts of each
sample until one passes. The analysis also gives the following statement, which
\cref{thm:algo-precise} does not need.

\begin{remark}[a randomized algorithm]\label{rem:alg-randomized}
In each case of \cref{thm:algo-precise}, the randomized algorithm outputs, in
$n(\log n)^{O(1)}$ bit operations (in the bootstrap and general cases, with the
$O(n(n+m))$ further operations of the theorem), a path decomposition of width at most
$\beta(G)+\rho(n)$, and in the general case also at most $\frac7{48}m+\rho(2m)$,
except with probability $\pi(n)$, or $\pi(n_{\ge1})$ in the bootstrap
and general cases, where $\pi(k)\to0$ as $k\to\infty$ and $n_{\ge1}$ is the number of
non-isolated vertices of $G$. This follows from the case proofs below and
\cref{sec:alg-time}.
\end{remark}

\subsubsection{Model, parameters and exact local labels}\label{sec:alg-model}

This subsection fixes the parameters as functions of $n$, recalls the labels of
\cref{def:labels} in the multiplicative form used by the algorithms, and shows how
their coefficients are computed. The algorithms use these coefficients, rounded to
rationals, together with the finite distribution of \cref{sec:alg-law} in place of the
Gaussian variables.

\paragraph{Parameters.} For an integer $n\ge3$ put
\[
  t:=\lceil\log_2(n+2)\rceil,\qquad
  h:=\max\bigl\{1,\lfloor\tfrac18\log_2t\rfloor\bigr\},\qquad
  K:=h,\qquad
  \gamma:=t^{-2},\qquad
  M:=2t^3+1,
\]
and
\[
  \xi:=12\,t^{-7/4},\qquad p_t:=12\,t^{-11/4},\qquad
  \omega_d:=2(d+1)\,\varphi(0)\,\xi+p_t\quad(d\ge0).
\]
Here $h$ is the locality radius of the labels, $1/K$ is the step of the grid of
thresholds, $\gamma$ is the step of the finite distribution and $M$ its number of points, and
$\xi$, $p_t$, $\omega_d$ are the error terms of
\cref{lem:alg-perturb,cor:alg-transfer}. The parameters are always
computed from the order of the \emph{input} graph (in the bootstrap and general
cases from $n_{\ge1}$), also when they are used for a subgraph.

\paragraph{The threshold $n_2$.} Put $n_2:=2^{2^{40}}$. For $n\ge n_2$ we have
$t\ge2^{40}$, so $h=\lfloor\frac18\log_2t\rfloor\ge5$ and
\begin{equation}\label{eq:alg-4h}
  4^h=2^{2h}\le2^{\frac14\log_2t}=t^{1/4}.
\end{equation}
Moreover the following inequalities hold for all $n\ge n_2$:
\begin{enumerate}[label=\textup{(N\arabic*)}]
  \item\label{it:alg-N1} $\xi<1/K$;
  \item\label{it:alg-N2} $\xi\varphi(0)+p_t\le5\cdot10^{-5}$;
  \item\label{it:alg-N3} $8\mu\,t^{1/2}n^{-1/2}\le\frac12$ for $\mu=12K+3$, and hence
  for every $\mu\le12K+3$.
\end{enumerate}
Indeed, $K=h\le\log_2t\le t$, so $\xi K\le12t^{-3/4}<1$, and
$\xi\varphi(0)+p_t\le24\,t^{-7/4}\le24\cdot2^{-70}$. For \ref{it:alg-N3}, put
$L=\log_2n\ge2^{40}$; then $t\le L+1\le2L$ and $\mu\le12t+3\le13t$, so
$8\mu t^{1/2}\le104\,t^{3/2}\le104\,(2L)^{3/2}\le2^{L/2-1}=\frac12n^{1/2}$.

For $n<n_2$ we shall put $\rho(n):=n$ and $\pi(n):=0$; then every width claim is trivial,
because every path decomposition of a graph with $n$ vertices has width at most
$n-1$. All probabilistic and quantitative estimates below therefore assume
$n\ge n_2$.

\paragraph{Weights and path products.} Each case fixes an integer weight
$q(x)\in\{2,3,4\}$ for every vertex $x$ of the graph to be labeled:
\begin{itemize}
  \item cubic case: $q\equiv2$; padded case: $q\equiv3$; padded case of degree $5$,
  used only for the cut sweep of \cref{thm:alg-cut-sweep}: $q\equiv4$;
  \item mixed cases, applied to graphs with $\delta\ge3$ and $\Delta\le5$, and switched
  case, applied to graphs with $\delta\ge3$ and $\Delta\le4$: $q(x)=d(x)-1$.
\end{itemize}
For vertices $v,w$ let $P(v,w)$ be the minimum of the products
$\prod_{i=0}^{k-1}q(x_i)$ over all $v$--$w$ paths $v=x_0,x_1,\dots,x_k=w$, with
$P(v,v)=1$. The \emph{support} $B(v)$
of $v$ is the ball $B_h(v)$ in the cubic and padded cases (of degree $4$ and $5$), and
$\{w:P(v,w)\le2^h\}$ in the mixed and switched cases. Since $q\ge2$, a path with $k$ edges has
product at least $2^k$; hence every $w\in B(v)$ satisfies $\dist(v,w)\le h$, so
$B(v)\subseteq B_h(v)$ in all cases.

\paragraph{Ideal labels.} Let $(Z_w)_w$ be independent standard normal variables and
put
\[
  X_v:=\sum_{w\in B(v)}P(v,w)^{-1/2}Z_w,\qquad
  \sigma_v^2:=\Var X_v=\sum_{w\in B(v)}\frac1{P(v,w)}\ \ (\ge1),\qquad
  Y_v:=X_v/\sigma_v .
\]
These are the labels of \cref{def:labels} with truncation radius $h$.
\begin{itemize}
  \item In the cubic case $P(v,w)=2^{\dist(v,w)}$, so the coefficients are the
  geometric coefficients $2^{-\dist(v,w)/2}$, truncated at distance $h$.
  \item In the padded case $P(v,w)=3^{\dist(v,w)}$, so the coefficients are
  $3^{-\dist(v,w)/2}$, truncated at distance $h$; in the padded case of degree $5$,
  $P(v,w)=4^{\dist(v,w)}$.
  \item In the mixed and switched cases, the arc length
  $\ell(x\to y)=\frac12\ln(d(x)-1)=\frac12\ln q(x)$ gives the directed distance $D(v,w)=\frac12\ln P(v,w)$, so
  $e^{-D(v,w)}=P(v,w)^{-1/2}$, and $P(v,w)\le2^h$ is the truncation
  $D(v,w)\le R$ with $R:=\frac h2\ln2$.
\end{itemize}
This is the path-product form of \cref{rem:P-form}. Thus the integer $h$ plays the
role of the radius parameter $R$ of \cref{sec:labels}. The cubic and padded cases use
the padded labels of \cref{def:labels}\ref{def:labels:padded} (with $p=3$ and $p=4$, and
$p=5$ in the padded case of degree $5$)
with $R=h$, and the mixed cases use the mixed labels of
\cref{def:labels}\ref{def:labels:mixed} with $R=\frac h2\ln2$. In this subsection we
write $h$ throughout.

The next lemma restates the edge correlation bounds of \cref{sec:labels} for these
labels. For $h\ge5$ put
$\eps_h:=1/h$ in the cubic and padded cases and $\eps_h:=8/(h-4)$ in the mixed and
switched cases, and for an edge $uv$ put $r_{uv}:=r_*=r_{33}$ in the cubic case,
$r_{uv}:=r_{44}$ in the padded case, $r_{uv}:=r_{55}$ in the padded case of degree $5$,
and $r_{uv}:=r_{d(u)d(v)}$ in the mixed and switched cases.

\begin{lemma}[edge correlations]\label{lem:alg-correlation}
Let $h\ge5$. Every edge $uv$ satisfies $\Corr(Y_u,Y_v)\ge r_{uv}(1-\eps_h)$.
\end{lemma}

\begin{proof}
In the cubic and padded cases this is \cref{lem:correlation-padded} with radius $h$
(with $p=3$, $4$ and $5$).
In the mixed and switched cases it is \cref{lem:correlation} with $\Delta_0=5$ and
radius $R=\frac h2\ln2$: the arc lengths lie between $\ell_{\min}=\frac12\ln2$ and
$\ell_{\max}=\frac12\ln4=\ln2$, and its error term~\eqref{eq:eps-R} is
\[
  \eps_R=\frac{2\ell_{\max}^2}{\ell_{\min}(R-2\ell_{\max})}
  =\frac{2(\ln2)^2}{\frac12\ln2\cdot(\frac h2-2)\ln2}=\frac8{h-4},
\]
which is positive and finite for $h>4$. \qedhere
\end{proof}

\paragraph{Ball sizes.} For a graph with $\Delta\le5$, the first bound of
\eqref{eq:ball-size} gives $|B_k(v)|\le1+5\sum_{i=0}^{k-1}4^i\le2\cdot4^k$ for every
$k\ge0$. With \eqref{eq:alg-4h}, for $n\ge n_2$,
\begin{equation}\label{eq:alg-balls}
  |B(v)|\le2t^{1/4},\qquad|B_{h+1}(v)|\le8t^{1/4},\qquad|B_{2h+2}(v)|\le32\,t^{1/2}.
\end{equation}

\paragraph{Computing the coefficients.} For each $v$, a breadth-first search to depth
$h$ finds $B_h(v)$. In the cubic and padded cases $P(v,w)=q^{\dist(v,w)}$. In the
mixed and switched cases we run Dijkstra's algorithm from $v$ inside $B_h(v)$, with path products
of integers in place of path sums (all factors are at least $1$, so Dijkstra's
algorithm applies). This computes $P(v,w)$ correctly for every $w$ with
$P(v,w)\le2^h$: a minimizing path has product at most $2^h$, hence at most $h$
edges, so all its vertices lie in $B_h(v)$. For the other vertices of $B_h(v)$ the
computed value is at least the true value and so also exceeds $2^h$.

The algorithms use rational roundings of the coefficients and of the standard
deviations. Put
\[
  \tilde c_{vw}:=\frac{\lfloor t^3P(v,w)^{-1/2}\rfloor}{t^3}\quad(w\in B(v)),
  \qquad
  \tilde\sigma_v:=\frac{\lfloor t^3\sigma_v\rfloor}{t^3}.
\]
Both are computed with integer square roots, using
$\lfloor\sqrt{\lfloor x\rfloor}\rfloor=\lfloor\sqrt x\rfloor$ for $x\ge0$:
$\lfloor t^3P^{-1/2}\rfloor=\lfloor\sqrt{\lfloor t^6/P\rfloor}\rfloor$ and
$\lfloor t^3\sigma_v\rfloor=\lfloor\sqrt{\lfloor t^6\sigma_v^2\rfloor}\rfloor$, where
$\sigma_v^2=\sum_{w\in B(v)}1/P(v,w)$ is an exact rational. Then
\begin{equation}\label{eq:alg-coeff}
  0\le\tilde c_{vw}\le1,\qquad|\tilde c_{vw}-P(v,w)^{-1/2}|\le t^{-3},\qquad
  \tilde\sigma_v\ge1,\qquad|\tilde\sigma_v-\sigma_v|\le t^{-3},
\end{equation}
because $P\ge1$ and $\sigma_v\ge1$. The numbers $P(v,w)\le4^h\le t^{1/4}$ and the
numerators of $\tilde c_{vw},\tilde\sigma_v$ have $O(\log t)$ bits. For $n\ge n_2$
the work per vertex is $t^{O(1)}$ bit operations by \eqref{eq:alg-balls}; for
$n<n_2$ it is bounded by a constant. In total, computing the coefficients takes
$n(\log n)^{O(1)}$ bit operations.

\subsubsection{A finite distribution for the Gaussian variables}\label{sec:alg-law}

The algorithms replace each $Z_w$ by a random variable with finitely many rational
values and rational probabilities, so that labels can be sampled and compared, and
conditional expectations computed, in exact arithmetic. This subsection defines that
distribution, couples it with the standard normal distribution, and defines the discrete labels that the
algorithms sort.

\paragraph{The distribution.} The finite distribution puts on each point of a grid of step $\gamma$ in
$[-t,t]$ a weight proportional to a rational approximation of the standard normal
density. For $|j|\le t^3$ put $x_j:=j\gamma$, so that $|x_j|\le t$; there are
$M=2t^3+1$ such points. Let $g_j:=1/T_j$, where $T_j=\sum_{k<4t^2}(x_j^2/2)^k/k!$ is the Taylor
polynomial of $e^{x_j^2/2}$ with $4t^2$ terms. Since $y:=x_j^2/2\le t^2/2$ and
$k!\ge(k/e)^k$, the omitted terms satisfy
\[
  0\le e^y-T_j\le\sum_{k\ge4t^2}\Bigl(\frac{ey}{k}\Bigr)^k
  \le\sum_{k\ge4t^2}\Bigl(\frac e8\Bigr)^k\le2\cdot0.34^{4t^2},
\]
so $T_j=e^y(1-\delta_j)$ with $0\le\delta_j\le2\cdot0.34^{4t^2}\le\frac12$, and
$|g_j-e^{-y}|=e^{-y}\delta_j/(1-\delta_j)\le4\cdot0.34^{4t^2}\le t^{-8}$ for
$t\ge2$. Put
\[
  A_j:=\lfloor t^8g_j\rfloor,\qquad A:=\sum_{|i|\le t^3}A_i,
\]
and let $\zeta$ be the random variable with $\Pr(\zeta=x_j)=A_j/A$. Since $g_0=1$,
we have $A_0=t^8>0$. Every $A_j$ is an integer in $[0,t^8]$ and $A\le3t^{11}$, so
the probabilities have $O(\log t)$ bits. Computing $T_j$ exactly takes $t^{O(1)}$
bit operations (it is a sum of $4t^2$ rationals whose numerators and denominators
have $O(t^2\log t)$ bits), so computing the whole distribution takes $t^{O(1)}=(\log n)^{O(1)}$
bit operations.

\begin{lemma}[coupling]\label{lem:alg-coupling}
Let $t\ge5$. There is a coupling of $\zeta$ with a standard normal variable $Z$ such
that $\Pr(|Z-\zeta|>\gamma)\le t^{-3}$.
\end{lemma}

\begin{proof}
We round $Z$ to the grid, show that the distribution of the rounding is within total variation
distance $t^{-3}/10$ of the distribution of $\zeta$, and then glue a maximal coupling of the
two distributions to $Z$.

Let $\hat Z$ be the rounding of $Z$ to the grid: $\hat Z=x_j$ if
$Z\in[x_j-\frac\gamma2,x_j+\frac\gamma2)$ for some $|j|\le t^3$, and $\hat Z=\star$
(an extra point) otherwise. Write $\nu$ for the distribution of $\hat Z$, so
$\nu(x_j)=p_j:=\Pr(Z\in[x_j-\frac\gamma2,x_j+\frac\gamma2))$, and
$p_{\mathrm{out}}:=\nu(\star)\le\Pr(|Z|\ge t)\le e^{-t^2/2}$.

\emph{Midpoint rule.} Put $\bar q_j:=\gamma\varphi(x_j)$. Since
$\varphi''(z)=(z^2-1)\varphi(z)$ satisfies $\sup|\varphi''|=\varphi(0)<0.4$, the midpoint
rule gives $|p_j-\bar q_j|\le\frac{\gamma^3}{24}\cdot0.4=\frac{t^{-6}}{60}$. Summed over
the $M\le3t^3$ points this is at most $t^{-3}/20$.

\emph{Rational weights.} Put $\tilde a_j:=\gamma A_j/(t^8\sqrt{2\pi})$. Since
$|A_j/t^8-g_j|\le t^{-8}$ and $|g_j-e^{-x_j^2/2}|\le t^{-8}$, we have
$|A_j/t^8-e^{-x_j^2/2}|\le2t^{-8}$, so
$|\tilde a_j-\bar q_j|\le\gamma\cdot2t^{-8}/\sqrt{2\pi}\le t^{-10}$; summed over $j$
this is at most $3t^{-7}$.

\emph{Total variation.} Hence
\[
  \sum_j|\tilde a_j-p_j|+p_{\mathrm{out}}\le e_t:=\frac{t^{-3}}{20}+3t^{-7}+e^{-t^2/2}
  \le\frac{t^{-3}}{10}\qquad(t\ge5),
\]
because $3t^{-7}\le t^{-3}/40$ and $e^{-t^2/2}\le t^{-3}/40$ for $t\ge5$. The distribution of
$\zeta$ is $(\tilde a_j/\sum_i\tilde a_i)_j$, since $\tilde a_j$ is proportional to
$A_j$. Normalizing changes the $\ell_1$ distance to $\nu$ by at most
$|1-\sum_i\tilde a_i|=|\sum_i(\tilde a_i-p_i)-p_{\mathrm{out}}|\le e_t$. So the
$\ell_1$ distance between the distribution of $\zeta$ and $\nu$ (on the points $x_j$ and
$\star$) is at most $2e_t$, and the total variation distance is at most
$e_t\le t^{-3}/10$.

\emph{Coupling.} Take a maximal coupling $(\hat Z',\zeta)$ of $\nu$ and the distribution of
$\zeta$, so $\Pr(\hat Z'\ne\zeta)\le t^{-3}/10$, and glue it to $(Z,\hat Z)$ along
$\hat Z=\hat Z'$ (draw $Z$ from its conditional distribution given $\hat Z=\hat Z'$). On the
event $\zeta=\hat Z$, the common value is a grid point $x_j$, because $\zeta$ never
equals $\star$, and then $|Z-\zeta|\le\gamma/2$. Hence
$\Pr(|Z-\zeta|>\gamma)\le\Pr(\zeta\ne\hat Z)\le t^{-3}$. \qedhere
\end{proof}

\paragraph{Discrete labels.} Let $(\tilde Z_w)_w$ be independent copies of $\zeta$
and
\[
  \tilde X_v:=\sum_{w\in B(v)}\tilde c_{vw}\tilde Z_w,\qquad
  \tilde Y_v:=\tilde X_v/\tilde\sigma_v .
\]
Each $\tilde Y_v$ is a rational of the form $a/(b\,t^2)$ with integers $a,b$ of
$O(\log t)$ bits, so two labels, or a label and a rational threshold with $O(\log t)$
bits, are compared in $(\log t)^{O(1)}$ bit operations. The algorithms order the
vertices by $(\tilde Y_v,v)$ lexicographically; ties between labels are broken by the
vertex index. For a rational $\tau$ write
\[
  \tilde S_\tau:=\{v:\ \tilde Y_v<\tau\},
\]
a prefix of this order. Note the strict inequality.

\paragraph{Sampling.} To sample $\zeta$, draw $\lceil\log_2A\rceil$ random bits,
forming an integer $U$; if $U\ge A$, draw again; otherwise return the $x_j$ with
$\sum_{i<j}A_i\le U<\sum_{i\le j}A_i$, found by binary search in the prefix sums. A
draw succeeds with probability $A/2^{\lceil\log_2A\rceil}>\frac12$, so a sample needs
fewer than two draws in expectation, each of $O(\log t)$ bits and $O(\log M)$
comparisons. To keep the running time of the randomized algorithms bounded in the
worst case, they draw at most $2N\lceil\log_2n\rceil$ times in total for $N$ samples;
if this budget is exhausted they output the trivial decomposition with one bag
$V(G)$. By Markov's inequality the budget is exhausted with probability less than
$1/\lceil\log_2n\rceil$. The Las Vegas algorithms have no budget.

\subsubsection{Perturbation}\label{sec:alg-perturbation}

We couple the discrete labels with the ideal labels so that, in the closed
neighborhood of any fixed vertex, all labels move by at most $\xi$ except with
probability at most $p_t$ (\cref{lem:alg-perturb}). As a consequence, the per-vertex
probabilities that bound the counts transfer from the ideal to the discrete labels,
up to a shift of the threshold by $\xi$ and an additive error that tends to zero as
$n\to\infty$ (\cref{cor:alg-transfer}). So the analysis of the algorithms can use
bounds on the ideal labels, which \cref{sec:alg-means} derives from
\cref{part:width}.

\begin{lemma}[perturbation]\label{lem:alg-perturb}
Let $n\ge n_2$, and let $H$ be a graph with $\Delta(H)\le5$ whose labels are built
with the parameters of $n$. There is a coupling of $(\tilde Y_v)_v$ with the ideal
labels $(Y_v)_v$ such that every vertex $v$ satisfies
\[
  \Pr\bigl(|\tilde Y_u-Y_u|>\xi\ \text{for some}\ u\in N[v]\bigr)\le p_t .
\]
\end{lemma}

\begin{proof}
Couple each pair $(\tilde Z_w,Z_w)$ by \cref{lem:alg-coupling} ($t\ge2^{40}\ge5$),
independently over $w$. The labels in $N[v]$ use at most
$|N[v]|\cdot2t^{1/4}\le12t^{1/4}$ variables by \eqref{eq:alg-balls}, so with
probability at least $1-12t^{1/4}t^{-3}=1-p_t$ all of them satisfy
$|\tilde Z_w-Z_w|\le\gamma=t^{-2}$. On this event, for $u\in N[v]$, using
\eqref{eq:alg-coeff}, $|\tilde Z_w|\le t$ and $|B(u)|\le2t^{1/4}$,
\[
\begin{split}
  |\tilde X_u-X_u|&\le\sum_{w\in B(u)}\Bigl(|\tilde c_{uw}-P(u,w)^{-1/2}|\,|\tilde Z_w|
  +P(u,w)^{-1/2}|\tilde Z_w-Z_w|\Bigr)\\
  &\le2t^{1/4}(t^{-2}+t^{-2})=4t^{-7/4}.
\end{split}
\]
Also $|\tilde X_u|\le2t^{1/4}\cdot t$, and
$|1/\tilde\sigma_u-1/\sigma_u|=|\sigma_u-\tilde\sigma_u|/(\sigma_u\tilde\sigma_u)\le t^{-3}$
since $\sigma_u,\tilde\sigma_u\ge1$. Hence
\[
  |\tilde Y_u-Y_u|\le\frac{|\tilde X_u-X_u|}{\sigma_u}
  +|\tilde X_u|\,\Bigl|\frac1{\tilde\sigma_u}-\frac1{\sigma_u}\Bigr|
  \le4t^{-7/4}+2t^{5/4}t^{-3}=6t^{-7/4}\le\xi.\qedhere
\]
\end{proof}

The coupling transfers per-vertex probabilities from ideal to discrete labels. Recall
that $S_\tau=\{v:Y_v\le\tau\}$ for the ideal labels, and $\tilde S_\tau=\{v:\tilde
Y_v<\tau\}$ for the discrete ones.

\begin{corollary}[transfer]\label{cor:alg-transfer}
In the setting of \cref{lem:alg-perturb}, let $v$ be a vertex of degree $d$ and
$\tau\in\R$. Then:
\begin{enumerate}[label=\textup{(\alph*)}]
  \item\label{it:transfer:out}
  $\Pr(v\in\outfr\tilde S_\tau)\le\Pr(v\in\outfr S_{\tau-\xi})+\omega_d$;
  \item\label{it:transfer:in}
  $\Pr(v\in\infr\tilde S_\tau)\le\Pr(v\in\infr S_{\tau+\xi})+\omega_d$;
  \item\label{it:transfer:sign} for every family $\mathcal A$ of subsets of $N[v]$,
  $\Pr(\{u\in N[v]:\tilde Y_u<0\}\in\mathcal A)\le
  \Pr(\{u\in N[v]:Y_u<0\}\in\mathcal A)+\omega_d$;
  \item\label{it:transfer:window} for $\tau\le\tau'$:
  $\Pr(\tau\le\tilde Y_v<\tau')\le(\tau'-\tau+2\xi)\varphi(0)+p_t$, and
  $\Pr(\tilde Y_v<\tau)\le\Phi(\tau)+\xi\varphi(0)+p_t$,
  $\Pr(\tilde Y_v\ge\tau)\le\Phi(-\tau)+\xi\varphi(0)+p_t$.
\end{enumerate}
\end{corollary}

\begin{proof}
Let $\mathcal E$ be the event that $|\tilde Y_u-Y_u|\le\xi$ for all $u\in N[v]$, and
$\mathcal W$ the event that no $Y_u$ with $u\in N[v]$ lies in $[\tau-\xi,\tau+\xi]$.
Each $Y_u$ is standard normal, so $\Pr(\mathcal W^c)\le(d+1)\cdot2\xi\varphi(0)$, and
$\Pr(\mathcal E^c)\le p_t$ by \cref{lem:alg-perturb}. Hence
$\Pr(\mathcal E^c\cup\mathcal W^c)\le\omega_d$, and it suffices to prove the
inclusions (a)--(c) on $\mathcal E\cap\mathcal W$.

(a) Let $v\in\outfr\tilde S_\tau$: $\tilde Y_v\ge\tau$ and some neighbor $u$ has
$\tilde Y_u<\tau$. Then $Y_v\ge\tau-\xi$, so $Y_v>\tau+\xi>\tau-\xi$ on $\mathcal
W$; and $Y_u<\tau+\xi$, so $Y_u<\tau-\xi$ on $\mathcal W$. Thus $v\in\outfr
S_{\tau-\xi}$.

(b) Let $v\in\infr\tilde S_\tau$: $\tilde Y_v<\tau$ and some neighbor $x$ has
$\tilde Y_x\ge\tau$. Then $Y_v<\tau+\xi$, and $Y_x\ge\tau-\xi$, so $Y_x>\tau+\xi$ on
$\mathcal W$. Thus $v\in\infr S_{\tau+\xi}$.

(c) Take $\tau=0$. On $\mathcal E\cap\mathcal W$, every $u\in N[v]$ has
$|Y_u|>\xi\ge|\tilde Y_u-Y_u|$, so $\tilde Y_u<0$ if and only if $Y_u<0$. Hence the
two sets in~(c) coincide on $\mathcal E\cap\mathcal W$.

(d) uses only the event $|\tilde Y_v-Y_v|\le\xi$, which fails with probability at
most $p_t$: on it, $\tau\le\tilde Y_v<\tau'$ implies $\tau-\xi\le Y_v<\tau'+\xi$, an
event of probability at most $(\tau'-\tau+2\xi)\varphi(0)$; $\tilde Y_v<\tau$ implies
$Y_v<\tau+\xi$; and $\tilde Y_v\ge\tau$ implies $Y_v\ge\tau-\xi$. \qedhere
\end{proof}

\subsubsection{Uniform ideal means}\label{sec:alg-means}

This subsection bounds the expected frontier sizes of the ideal labels uniformly in
the threshold (\cref{lem:alg-means}). The counts of the layouts are taken at the
thresholds of a grid whose step $1/K$ tends to zero with $n$, and
\cref{cor:alg-transfer} shifts these thresholds by $\xi$. We therefore need the
bounds of \cref{part:width} on expected frontier sizes for all thresholds in an
interval at once, with an error that tends to zero as $h\to\infty$. In the cubic,
padded and mixed cases the bounds of \cref{part:width} have this form. In the cubic
and padded cases the per-vertex bounds hold for every threshold. In the mixed cases \cref{lem:sweep-means} holds for every
threshold in $[-3,3]$, because the continuity of the center bound
(\cref{lem:stars-continuity}) turns the factor $1-\eps_R$ in the correlation bounds
into an additive error $205\,\eps_R$ per vertex. The lemma below restates these
bounds for the ideal labels of radius $h$.

In the lemma, $r_h:=r_*(1-1/h)$, where $r_*=r_{33}$ is the edge correlation bound of
the cubic case, and $\eps_{\mathrm{mix}}(h):=205\,\eps_h=1640/(h-4)$, where
$\eps_h=8/(h-4)$ is the error term of \cref{lem:alg-correlation} in the mixed cases.

\begin{lemma}[uniform ideal means]\label{lem:alg-means}
Let $h\ge5$ and let $(Y_v)$ be the ideal labels of radius $h$ of a graph $H$ with $N$
vertices.
\begin{enumerate}[label=\textup{(\alph*)}]
  \item\label{it:means:cubic} \emph{Cubic case}, $\Delta(H)\le3$. Every vertex $v$
  satisfies $\Pr(v\in\outfr S_\tau)\le\kappa(r_h)$ for every $\tau\ge0$,
  $\Pr(v\in\infr S_\tau)\le\kappa(r_h)$ for every $\tau\le0$, and
  \begin{align*}
    \Pr\bigl(Y_v<0,\ \exists u\in N(v):Y_u\ge0\bigr)
    +\ind_{d(v)=3}\Pr\bigl(Y_v\ge0,\ Y_u<0\ \forall u\in N(v)\bigr)&\le\kappa(r_h),\\
    \Pr\bigl(Y_v\ge0,\ \exists u\in N(v):Y_u<0\bigr)
    +\ind_{d(v)=3}\Pr\bigl(Y_v<0,\ Y_u\ge0\ \forall u\in N(v)\bigr)&\le\kappa(r_h).
  \end{align*}
  \item\label{it:means:padded} \emph{Padded case}, $\Delta(H)\le4$. For every
  $\tau\in\R$, $\E|\outfr S_\tau|\le(F_4+178/h)\,N$.
  \item\label{it:means:mixed} \emph{Mixed cases}, $\delta(H)\ge3$ and $\Delta(H)\le5$.
  For every $\tau\in[-3,3]$, $\E|\outfr S_\tau|\le\beta(H)+\eps_{\mathrm{mix}}(h)\,N$.
\end{enumerate}
\end{lemma}

\begin{proof}
\ref{it:means:cubic} By \cref{lem:alg-correlation}, the labels of $v$ and its
$d(v)\le3$ neighbors form a centered Gaussian vector with unit variances and
incident correlations at least $r_h$. For $\tau\ge0$,
$\{v\in\outfr S_\tau\}\subseteq\{Y_v\ge\tau,\ \exists u\in N(v):Y_u\le\tau\}$, whose
probability is at most $\kappa(r_h)$ by \cref{prop:star-cubic}. For $\tau\le0$,
apply \cref{prop:star-cubic} to the labels $-Y$, which have the same incident
correlations, at the threshold $-\tau\ge0$: the event $v\in\infr S_\tau$ is
contained in $\{-Y_v\ge-\tau,\ \exists x\in N(v):-Y_x\le-\tau\}$. The first of the two
displayed inequalities is \cref{lem:switch}, because its events differ from those of
\cref{lem:switch} only by events on which some label equals $0$, which have
probability zero. The second is \cref{lem:switch} applied to $-Y$, in the same way.

\ref{it:means:padded} By \cref{lem:alg-correlation}, the labels of $v$ and its
$d(v)\le4$ neighbors form a vector as in \cref{cor:stars-ideal} with $d=4$ and
$r=r_{44}(1-1/h)$. So $\Pr(v\in\outfr S_\tau)\le f_{4,r_{44}(1-1/h)}(\tau)$. For $|\tau|\le3$ this is
at most $f_{4,r_{44}}(\tau)+205\,r_{44}/h\le F_4+178/h$ by \eqref{eq:ideal-is-center},
\cref{lem:stars-continuity}(b) with $\bar r=r_{33}$ and \cref{def:ideal-star}; for
$|\tau|>3$ it is at most $4\Phi(-3)<F_4$ (\cref{prop:F3F4}). Sum over $v$.

\ref{it:means:mixed} If $h\le12$, then $\eps_{\mathrm{mix}}(h)\ge205$ and the claim is
trivial, since $\E|\outfr S_\tau|\le N$. Let $h\ge13$, so that $\eps_h<1$. In the
mixed cases the ideal labels are the mixed labels of
\cref{def:labels}\ref{def:labels:mixed} with $R=\frac h2\ln2$, and $\eps_R=\eps_h$ for
this $R$ (proof of \cref{lem:alg-correlation}). The triple of the case is one of
$\tripP,\tripR,\tripS$, and all degrees of $H$ lie in $\{3,4,5\}$. So
\cref{lem:sweep-means} gives
$\E|\outfr S_\tau|\le\beta(H)+205\,\eps_hN=\beta(H)+\eps_{\mathrm{mix}}(h)\,N$ for
every $\tau\in[-3,3]$.
\end{proof}

\subsubsection{Counts and derandomization}\label{sec:alg-derand}

This subsection shows that the counts are close to their means with high probability,
and that the method of conditional expectations finds discrete labels for which they
are close to their means for certain. The randomized and Las Vegas algorithms use
the first statement, and the deterministic algorithm uses the second.

\paragraph{Counts.} Let $H$ be a graph with $N$ vertices and $\Delta(H)\le5$, labeled
with the parameters of some $n\ge\max\{N,n_2\}$. A \emph{count} is a random variable
$X_c=\sum_{v\in V(H)}I_{c,v}$, where each $I_{c,v}\in\{0,1\}$ is a function of
$(\tilde Z_w)_{w\in B_{h+1}(v)}$ that can be evaluated from these values in
$t^{O(1)}$ bit operations. For example, $\ind\{v\in\outfr\tilde S_\tau\}$ is such an
indicator for every rational $\tau$ with $O(\log t)$ bits, because it depends on the
labels in $N[v]$, and $\tilde Y_u$ depends on $\tilde Z_w$ for $w\in B(u)\subseteq
B_h(u)\subseteq B_{h+1}(v)$.

\paragraph{Greedy fixing.} Let $X_1,\dots,X_\mu$ be counts, $m_c:=\E X_c$, and
\[
  \Psi:=\sum_{c=1}^\mu(X_c-m_c)^2 .
\]
\emph{Greedy fixing} is the following deterministic procedure: fix $\tilde Z_w$ for
$w\in V(H)$ one at a time, each time choosing a value $x_j$ with $A_j>0$ that minimizes
the conditional expectation of $\Psi$ given the values fixed so far.

\begin{lemma}[concentration and derandomization]\label{lem:alg-derand}
In this setting the following hold.
\begin{enumerate}[label=\textup{(\alph*)}]
  \item\label{it:derand:var} $\E\Psi=\sum_c\Var X_c\le8\mu Nt^{1/2}$.
  \item\label{it:derand:rand} For every $\lambda>0$, the probability that
  $X_c>m_c+\lambda$ for some $c$ is at most $8\mu Nt^{1/2}/\lambda^2$.
  \item\label{it:derand:det} Greedy fixing ends with values of all $\tilde Z_w$, each
  of positive probability, for which $\Psi\le8\mu Nt^{1/2}$, and hence
  $|X_c-m_c|\le\sqrt{8\mu N}\,t^{1/4}$ for every $c$.
  \item\label{it:derand:cost} The means $m_c$ can be computed exactly with
  $\mu N\,n^{o(1)}$ bit operations, and greedy fixing uses $\mu N^2n^{o(1)}$ bit
  operations.
\end{enumerate}
\end{lemma}

\begin{proof}
\ref{it:derand:var} Apply \cref{lem:local-variance} with $a=1$ and
$A_v=B_{h+1}(v)$: two indicators $I_{c,u}$ and $I_{c,v}$ can be dependent only if
$B_{h+1}(u)\cap B_{h+1}(v)\ne\emptyset$, that is, if $v\in B_{2h+2}(u)$. By
\eqref{eq:alg-balls} there are at most $32t^{1/2}N$ such ordered pairs, so
$\Var X_c\le8Nt^{1/2}$, and $\E\Psi$ is the sum of these variances.

\ref{it:derand:rand} Chebyshev's inequality and a union bound over the $\mu$ counts.

\ref{it:derand:det} Let $\mathcal F$ be the values fixed before the step that fixes
$\tilde Z_w$. Then
\[
  \E[\Psi\mid\mathcal F]=\sum_j\frac{A_j}{A}\,\E[\Psi\mid\mathcal F,\ \tilde Z_w=x_j],
\]
an average over the values of positive probability, so the minimizing value chosen by
greedy fixing does not increase the conditional expectation. Initially it is $\E\Psi\le8\mu Nt^{1/2}$ by
\ref{it:derand:var}. At the end all variables are fixed and $\Psi$ is a number at most
$8\mu Nt^{1/2}$; each of its terms $(X_c-m_c)^2$ is at most $\Psi$.

\ref{it:derand:cost} Given fixed values $\mathcal F$,
\[
  \E\bigl[(X_c-m_c)^2\bigm|\mathcal F\bigr]
  =\Bigl(\sum_v\E[I_{c,v}\mid\mathcal F]-m_c\Bigr)^2
  +\sum_{u}\ \sum_{v\in B_{2h+2}(u)}\Cov(I_{c,u},I_{c,v}\mid\mathcal F),
\]
because the unfixed variables are independent, so the conditional covariance vanishes
when $B_{h+1}(u)\cap B_{h+1}(v)=\emptyset$. Each term involves at most
$|B_{h+1}(u)\cup B_{h+1}(v)|\le16t^{1/4}$ variables by \eqref{eq:alg-balls}, and is
computed exactly by enumerating the at most $M^{16t^{1/4}}$ assignments of the
unfixed ones among them, with probabilities that are products of the rationals
$A_j/A$. Since $M\le3t^3$ and $t\le2\log_2n$,
\[
  M^{16t^{1/4}}=\exp\bigl(O(t^{1/4}\log t)\bigr)
  =\exp\bigl(O((\log n)^{1/4}\log\log n)\bigr)=n^{o(1)}.
\]
All numbers involved are rationals whose denominators are powers of $A$ with
exponent $O(t^{1/4})$, and whose numerators have $n^{o(1)}$ bits; each indicator is
evaluated in $t^{O(1)}$ bit operations. So each term takes $n^{o(1)}$ bit operations,
and one evaluation of $\E[\Psi\mid\mathcal F]$, with its $O(\mu Nt^{1/2})$ terms,
takes $\mu N\,n^{o(1)}$. The means $m_c$ are computed in the same way with nothing
fixed. Greedy fixing makes $N$ steps with at most $M$ candidates each, which gives
$\mu N^2n^{o(1)}$ bit operations. \qedhere
\end{proof}

In all our layouts $\mu=O(K)=O(\log\log n)$, so the bounds in
\ref{it:derand:cost} become $N\,n^{o(1)}$ and $N^2n^{o(1)}$ bit operations. Below we
use only upper bounds of the form $X_c\le m_c+\lambda$, in the following form.

\paragraph{The three algorithms.} Let $n\ge n_2$, $N\le n$ and $\mu\le12K+3$, and put
$\theta_{\mathrm D}(n):=\sqrt{8\mu n}\,t^{1/4}$ and $\theta_{\mathrm R}(n):=n^{3/4}$.
By \cref{lem:alg-derand}\ref{it:derand:det}, greedy fixing ends with every count at
most its mean plus $\theta_{\mathrm D}(n)$. By \cref{lem:alg-derand}\ref{it:derand:rand}
with $\lambda=n^{3/4}$, one sample has every count at most its mean plus
$\theta_{\mathrm R}(n)$ except with probability at most $8\mu t^{1/2}n^{-1/2}$, which
is at most $\frac12$ by~\ref{it:alg-N3}. So the randomized algorithm, with its sampling
budget, fails with probability at most
$\pi(n):=8(12K+3)t^{1/2}n^{-1/2}+1/\lceil\log_2n\rceil$ (as $\mu\le12K+3$), and the
Las Vegas algorithm,
which resamples until every count is at most its mean plus $n^{3/4}$, needs at most two
samples in expectation. In the padded, mixed and cubic cases below, it therefore
suffices to bound the width when every count is at most its mean plus
$\theta\in\{\theta_{\mathrm D}(n),\theta_{\mathrm R}(n)\}$.

\subsubsection{Undoing reductions}\label{sec:alg-undo}

In the switched, mixed, bootstrap and general cases the algorithm first applies the
reduction operations of \cref{def:reductions}, possibly interleaved with vertex
deletions, and decomposes the final graph. The proof of \cref{lem:reduction} constructs from this a
decomposition of the original graph. This subsection shows that the construction can
be carried out in near-linear time (\cref{lem:undo}). If the final graph is empty,
its path decomposition has no bags and width $-1$.

\begin{lemma}[undoing reductions]\label{lem:undo}
Let $F$, $D$ and $R$ be as in \cref{lem:reduction}, where $F$ has $n$ vertices and $m$
edges. Given $F$, $D$, $V(R)$ and a path decomposition of $R$ of width $w$ with $s$
bags, a path decomposition of $F$ of width at most $|D|+w+3\log_2n+1$ with at most
$s+n$ bags can be computed with $O(m+s+n\log n)$ operations on integers of
$O(\log n)$ bits, where both decompositions are in interval form.
\end{lemma}

The algorithm does not need the sequence of operations itself; the sequence only
guarantees the structure of \cref{lem:reduction}\ref{it:reduction:structure}. We first record two
standard facts, in the form in which the algorithm uses them. Reduction operations
take a graph of treewidth at most~$2$ to the empty graph and yield a tree
decomposition of width at most~$2$ (\cref{cl:alg-elim}). Every tree decomposition of
a graph with $N'$ vertices has a bag whose removal leaves components of at most
$N'/2$ vertices, and such a bag can be found in time linear in the size of the decomposition
(\cref{cl:alg-balanced}).

\paragraph{Elimination bags.} Suppose that reduction operations are applied to a
graph $Q$ until none applies, and that the final graph is empty. Let $u_1,u_2,\dots$ be the
vertices of $Q$ in the order of their removal, let $\Gamma_i$ be the set of neighbors
of $u_i$ in the current graph when $u_i$ is removed, and put
$X_i:=\{u_i\}\cup\Gamma_i$. The parent of node $i$ is the node $j$ of smallest index
with $u_j\in\Gamma_i$ (then $j>i$), and node $i$ is a root if $\Gamma_i=\emptyset$. We
call the sets $X_i$, with this forest, the \emph{elimination bags} of the sequence.

\begin{claim}\label{cl:alg-elim}
Let $Q$ be a graph with $\tw(Q)\le2$. Applying reduction operations to $Q$ in any
order until none applies ends with the empty graph, and the elimination bags, with
the roots of their forest joined into a tree, form a tree decomposition of $Q$ of
width at most $2$.
\end{claim}

\begin{proof}
First, the process ends with the empty graph. Each operation produces a minor of the
current graph, since~\ref{red:R3} contracts the
edge $vx$ and the others delete a vertex; so every current graph has treewidth at most
$2$.
A nonempty graph of treewidth at most $k$ has a vertex of degree at most $k$: in a
tree decomposition of width at most $k$ in which no bag is contained in a neighboring
bag (contract tree edges violating this), either there is one bag, or a leaf bag
contains a vertex outside its neighbor's bag, and all neighbors of that vertex lie in
the leaf bag. By~\ref{red:max}, every vertex of degree at most $2$ admits an
operation, so the process continues until the graph is empty.

Next, we show that the elimination bags form a tree decomposition. In every
operation, $\Gamma_i$ is a clique of the graph after the operation: it has at most
one vertex, or it is $\{x,y\}$ and $xy$ is present in~\ref{red:R4} and added
in~\ref{red:R3}. Every edge of $Q$ joins $u_i$ to $\Gamma_i$, where $u_i$ is its end removed
first, because edges disappear only with their ends; so it lies in $X_i$. Let
$u_a\in \Gamma_i$ with $a>i$, and let $j$ be the parent of $i$. If $j\ne a$, then $j<a$,
and $u_a,u_j\in \Gamma_i$ are adjacent after step $i$; this edge survives until step $j$,
so $u_a\in \Gamma_j$. Repeating, the nodes containing $u_a$ include a path from $i$ to $a$
in the forest, so the nodes containing any vertex form a connected subtree. Joining
the roots by arbitrary edges keeps this property. \qedhere
\end{proof}

\begin{claim}\label{cl:alg-balanced}
Let $Q$ be a graph with $N'\ge1$ vertices and let $(\mathcal T,(X_\nu)_\nu)$ be a tree
decomposition of $Q$. One can find, in time linear in $|V(\mathcal T)|+\sum_\nu|X_\nu|$, a
node $\nu$ such that every component of $Q-X_\nu$ has at most $N'/2$ vertices.
\end{claim}

\begin{proof}
Root $\mathcal T$, and assign each vertex $x$ to the node $\mathrm{top}(x)$ closest to
the root among the nodes whose bags contain $x$. For a node $\nu$ let $\omega(\nu)$ be the
number of vertices assigned to the subtree $\mathcal T_\nu$ rooted at $\nu$. Both are
computed by one traversal. For a child $\nu'$ of $\nu$, the vertices assigned to
$\mathcal T_{\nu'}$ are exactly the vertices that occur in $\mathcal T_{\nu'}$ but not in $X_\nu$:
the nodes containing a vertex form a subtree, so a vertex occurring both in
$\mathcal T_{\nu'}$ and outside it occurs in $X_\nu$. Similarly, the vertices outside $X_\nu$
not assigned to $\mathcal T_\nu$ occur only outside $\mathcal T_\nu$. A component of
$Q-X_\nu$ is connected and avoids $X_\nu$, so all its vertices occur in the same
component of $\mathcal T-\nu$; hence it has at most $\omega(\nu')$ vertices for some child
$\nu'$, or at most $N'-\omega(\nu)$ vertices.

To find $\nu$, start at the root, and while the current node
has a child $\nu'$ with $\omega(\nu')>N'/2$, move to that child. At the final node $\nu$, all
children have $\omega(\nu')\le N'/2$, and $N'-\omega(\nu)<N'/2$ (if $\nu$ is not the root,
we moved to $\nu$ because $\omega(\nu)>N'/2$; if $\nu$ is the root, $\omega(\nu)=N'$). \qedhere
\end{proof}

\begin{proof}[Proof of \cref{lem:undo}]
For each component $C$ of $(F-D)-V(R)$ put $S_C:=N_{F-D}(C)\subseteq V(R)$. By
\cref{lem:reduction}\ref{it:reduction:structure}, $S_C$ is empty, a single vertex, or the two ends of an
edge of $R$, and $\tw(F[C])\le2$. The algorithm decomposes each $F[C]$ separately,
with width $O(\log n)$, inserts this decomposition, with that bag added to each of its bags, after a bag of
the decomposition of $R$ that contains $S_C$ (at the end if $S_C=\emptyset$), and adds
$D$ to every bag.
In detail, it has four steps.
\begin{enumerate}[label=\textup{(\arabic*)}]
  \item \emph{Components.} Compute the components $C$ of $(F-D)-V(R)$ and their sets
  $S_C$ by breadth-first search, in $O(n+m)$ time.
  \item \emph{Decomposing a component.} Fix $C$, with $N_C=|C|$. Apply reduction
  operations to $F[C]$ in any order, keeping the vertices of degree at most $2$ in a
  queue and testing adjacency for \ref{red:R3} and \ref{red:R4} with a balanced search
  tree of edges. By \cref{cl:alg-elim} this ends with the empty graph, and its
  elimination bags form a tree decomposition of $F[C]$ of width at most $2$ with $N_C$
  nodes. A graph of treewidth at most $2$ has fewer than $2N_C$ edges, and every
  operation adds at most one edge, so this step takes $O(N_C\log n)$ operations.
  \item \emph{Balanced recursion.} Given a connected graph $Q=F[C']$, $C'\subseteq C$,
  with a tree decomposition of width at most $2$, find by \cref{cl:alg-balanced} a bag
  $X$ such that every component of $Q-X$ has at most $|C'|/2$ vertices. For each
  component $C''$ of $Q-X$, restrict the tree decomposition to $C''$: intersect every
  bag with $C''$ and keep the nodes whose bags meet $C''$. These nodes form a subtree,
  because $Q[C'']$ is connected and the nodes containing the two ends of an edge share
  a node, so this is a tree decomposition of $Q[C'']$ of width at most $2$. Recurse on
  each component, concatenate the resulting path decompositions, and add $X$ to every
  bag (if $Q-X$ is empty, output the single bag $X$). By induction on $|C'|$, the
  result has width at most $3\lfloor\log_2|C'|\rfloor$: if $Q-X$ is empty, the
  width is $|X|-1$, which is $0$ if $|C'|=1$ and at most $2$ otherwise; if not, each
  $C''$ has $\lfloor\log_2|C''|\rfloor\le\lfloor\log_2|C'|\rfloor-1$ because
  $|C''|\le|C'|/2$, and adding the at most three vertices of $X$ to bags of width at
  most $3(\lfloor\log_2|C'|\rfloor-1)$ gives the bound. It has at most $|C'|$ bags,
  since the bags $X$ are nonempty. At one level of the recursion the components are disjoint, and a bag of
  at most three vertices meets at most three of them, so the work per level is linear
  in $N_C$; there are at most $\lfloor\log_2N_C\rfloor+1$ levels. In interval form,
  the vertices of $X$ receive the interval of the whole block produced by the call.
  This step takes $O(N_C\log N_C)$ operations.
  \item \emph{Insertion.} Let $W_1,\dots,W_s$ be the decomposition of $R$. For each $C$
  with $S_C\ne\emptyset$ choose the index $i_C:=\max_{x\in S_C}\mathrm{first}(x)$;
  then $S_C\subseteq W_{i_C}$, because $S_C$ is a vertex, or the two ends of an edge
  of $R$, whose intervals intersect. Immediately after $W_{i_C}$ insert the bags
  $W_{i_C}\cup P$ for the bags $P$ of the decomposition of $F[C]$ from step 3, the
  components with the same index one after another. Append the decompositions of the
  components with $S_C=\emptyset$ at the end, and finally add $D$ to every bag.
\end{enumerate}
\emph{Correctness.} The result is a path decomposition of $F$. Every vertex lies in a
bag and $D$ in all bags, which covers the edges at $D$. An edge of $F-D$ inside
$V(R)$ is an edge of $R$ by \cref{lem:reduction}\ref{it:reduction:structure}, so it lies in some $W_i$.
Any other edge of $F-D$ joins a vertex of some $C$ to a vertex of $C\cup S_C$, so it
lies in a bag $W_{i_C}\cup P$ or $P$. A vertex of $C$ occurs only in bags added for
$C$, and there in consecutive bags, since step 3 gives a path decomposition of
$F[C]$. A vertex $x\in V(R)$ occurs in a run of consecutive bags $W_i$ and in the
bags inserted after them, since a bag inserted after $W_i$ contains $x$ exactly when
$x\in W_i$. As step 3 gives width at most $3\log_2n$, the width is at most
$|D|+w+3\log_2n+1$, with at most $s+\sum_CN_C\le s+n$ bags. (If no bag results, which
happens only when $F-D$ is empty, output the single bag $D$.)

\emph{Interval form and running time.} Let $\ell_i$ be the total number of bags
inserted after $W_i$, and
$\mathrm{new}(i):=i+\sum_{i'<i}\ell_{i'}$ the new index of $W_i$. A vertex $x$ of $R$
with interval $[\mathrm{first}(x),\mathrm{last}(x)]$ gets
$[\mathrm{new}(\mathrm{first}(x)),\mathrm{new}(\mathrm{last}(x))+\ell_{\mathrm{last}(x)}]$,
because the bags inserted after $W_i$ contain $x$ exactly when $x\in W_i$. A vertex of
a component $C$ gets its interval within the block of $C$, shifted by the position of
the block, and every vertex of $D$ gets the interval of all bags. Computing the
$\ell_i$, the prefix sums and the intervals takes $O(n+s)$ time. Summing the running
times of the four steps, using $\sum_CN_C\le n$, gives $O(m+s+n\log n)$. \qedhere
\end{proof}

\subsubsection{The plain sweep: padded and mixed cases}\label{sec:alg-plain}

This subsection proves \cref{thm:algo-precise} in the padded and mixed cases. The
layout is the one of \cref{lem:layout} for the order of the discrete labels, and its
width is controlled by frontier sizes on the grid
\[
  \tau_j:=-3+\frac jK\qquad(0\le j\le6K),
\]
together with the numbers of labels in the windows between grid points and in the two
tails. \Cref{lem:alg-grid} bounds the width by these counts, and
\cref{prop:alg-sweep} bounds the width when every count is at most its mean plus some
$\theta\ge0$, using the bounds on the means from
\cref{sec:alg-perturbation,sec:alg-means}. The algorithms ensure this condition, with
high probability or for certain, by \cref{lem:alg-derand}; the proof of
\cref{thm:algo-precise} in these cases ends the subsection.

\paragraph{The layout.} Let $H$ be a graph with $N$ vertices and $\Delta(H)\le5$,
with $\delta(H)\ge3$ in the mixed cases, labeled with the parameters of some
$n\ge N$. Let $v_1,\dots,v_N$ be the vertices in increasing order of
$(\tilde Y_v,v)$ and $T_i:=\{v_1,\dots,v_i\}$. The \emph{plain-sweep layout}
consists of the bags $W_i:=\{v_i\}\cup\outfr T_i$, $1\le i\le N$. By
\cref{lem:layout}, applied to any injective labeling that induces this order, it is a
path decomposition of $H$ of width $\max_i|\outfr T_i|$. By the definition of the
bags, $v_j\in W_i$ if and only if $i=j$, or $i<j$ and $v_j$ has a neighbor in $T_i$.
So $v_j$ occupies the bags $W_{f_j},\dots,W_j$, where $f_j$ is the smallest index
$k$ with $k=j$ or $v_k\in N(v_j)$, and the interval form is computed in $O(N)$ time
from the order.

\paragraph{Counts.} The plain sweep uses the $\mu=12K+3$ counts
\begin{align*}
  X^{\mathrm{fr}}_j&:=|\outfr\tilde S_{\tau_j}|=\bigl|\{v:\ \tilde Y_v\ge\tau_j,\ \exists u\in N(v):\tilde Y_u<\tau_j\}\bigr|
  &&(0\le j\le6K),\\
  X^{\mathrm{win}}_j&:=\bigl|\{v:\ \tau_j\le\tilde Y_v<\tau_{j+1}\}\bigr|&&(0\le j<6K),\\
  X^{\mathrm{lo}}&:=\bigl|\{v:\ \tilde Y_v<-3\}\bigr|,\qquad
  X^{\mathrm{hi}}:=\bigl|\{v:\ \tilde Y_v\ge3\}\bigr| .
\end{align*}
Each is a count in the sense of \cref{sec:alg-derand}.

The following analogue of \cref{lem:grid} works directly with the discrete labels,
whose sublevel sets $\tilde S_\tau$ are defined by a strict inequality.

\begin{lemma}[grid]\label{lem:alg-grid}
Every $i$ satisfies
$|\outfr T_i|\le\max\bigl\{5X^{\mathrm{lo}},\ X^{\mathrm{hi}},\
\max_{0\le j<6K}(X^{\mathrm{fr}}_{j+1}+X^{\mathrm{win}}_j)\bigr\}$.
\end{lemma}

\begin{proof}
Let $y:=\tilde Y_{v_i}$. Every vertex of $T_i$ has discrete label at most $y$, and
every vertex outside $T_i$ has discrete label at least $y$. Let $u\in\outfr T_i$, with
a neighbor $z\in T_i$; so $\tilde Y_u\ge y\ge\tilde Y_z$.
\begin{itemize}
  \item If $y<-3$, then $z$ is counted by $X^{\mathrm{lo}}$. Each such $z$ is a
  neighbor of at most $5$ vertices $u$, so $|\outfr T_i|\le5X^{\mathrm{lo}}$.
  \item If $y\ge3$, then $u$ is counted by $X^{\mathrm{hi}}$.
  \item Otherwise $\tau_j\le y<\tau_{j+1}$ for some $0\le j<6K$. If $\tilde
  Y_u\ge\tau_{j+1}$, then $u\in\outfr\tilde S_{\tau_{j+1}}$, because $\tilde Y_z\le
  y<\tau_{j+1}$; so $u$ is counted by $X^{\mathrm{fr}}_{j+1}$. Otherwise
  $\tau_j\le y\le\tilde Y_u<\tau_{j+1}$, and $u$ is counted by $X^{\mathrm{win}}_j$.
  \qedhere
\end{itemize}
\end{proof}

Let $\eps(h)$ be the error term of \cref{lem:alg-means}: $\eps(h):=178/h$ in the
padded case and $\eps(h):=\eps_{\mathrm{mix}}(h)$ in the mixed cases. Put
\[
  \rho_0(n):=\eps(h)+14\,\xi\varphi(0)+2p_t+\frac{\varphi(0)}K .
\]

\begin{proposition}[plain sweep]\label{prop:alg-sweep}
Let $n\ge n_2$ and $\theta\ge0$. If every count of the plain sweep is at most its mean
plus $\theta$, then the plain-sweep layout of $H$ has width at most
$\beta(H)+\rho_0(n)N+5\theta$.
\end{proposition}

\begin{proof}
\emph{Frontier counts.} Let $1\le j\le6K$. By \cref{cor:alg-transfer}\ref{it:transfer:out},
$\E X^{\mathrm{fr}}_j\le\E|\outfr S_{\tau_j-\xi}|+\omega_5N$. Since
$\tau_j\in[-3+1/K,3]$ and $\xi<1/K$ by \ref{it:alg-N1}, the threshold $\tau_j-\xi$
lies in $[-3,3]$, and \cref{lem:alg-means} gives
$\E X^{\mathrm{fr}}_j\le\beta(H)+(\eps(h)+12\xi\varphi(0)+p_t)N$ (in the padded case
$\beta(H)=F_4N$).

\emph{Windows.} By \cref{cor:alg-transfer}\ref{it:transfer:window},
$\E X^{\mathrm{win}}_j\le(\varphi(0)/K+2\xi\varphi(0)+p_t)N$.

Hence $X^{\mathrm{fr}}_{j+1}+X^{\mathrm{win}}_j\le\beta(H)+\rho_0(n)N+2\theta$ for
every $0\le j<6K$.

\emph{Tails.} By \cref{cor:alg-transfer}\ref{it:transfer:window} and \ref{it:alg-N2},
$\E X^{\mathrm{lo}}$ and $\E X^{\mathrm{hi}}$ are at most
$(\Phi(-3)+\xi\varphi(0)+p_t)N\le0.0014\,N$. So $5X^{\mathrm{lo}}$ and
$X^{\mathrm{hi}}$ are at most $0.007N+5\theta$. In every case $\beta(H)\ge0.14N$: in
the padded case $F_4>0.14$, and in the mixed cases every vertex has degree at least
$3$ and every coefficient is at least $0.14$. So $0.007N+5\theta\le\beta(H)+5\theta$.

By \cref{lem:alg-grid}, every $|\outfr T_i|$, and hence the width, is at most
$\beta(H)+\rho_0(n)N+5\theta$. \qedhere
\end{proof}

\paragraph{The algorithms.} Let $G$ be a graph of the padded or a mixed case, with
$n\ge3$ vertices (graphs with $n\le2$ are decomposed by one bag). The randomized, Las
Vegas and deterministic algorithms consist of the following four steps and differ
only in step 3.
\begin{enumerate}[label=\textup{(\arabic*)}]
  \item In the padded case put $H:=G$. In the mixed cases apply reduction operations
  to $G$ until none applies and let $H$ be the final graph; it is empty or has
  $\delta(H)\ge3$ by~\ref{red:max}, and $\Delta(H)\le\Delta(G)$ by~\ref{red:deg}.
  If $H$ is empty, go to step 4.
  \item Compute the parameters from $n$, the coefficients $\tilde c_{vw}$ and
  $\tilde\sigma_v$ of $H$, and the finite distribution.
  \item \emph{Randomized:} sample all $\tilde Z_w$, compute the labels, sort, and
  build the plain-sweep layout. \emph{Las Vegas:} also compute the means $m_c$ of the
  counts; if $n\ge n_2$ and some count exceeds $m_c+n^{3/4}$, sample again.
  \emph{Deterministic:} fix the $\tilde Z_w$ by greedy fixing
  (\cref{lem:alg-derand}\ref{it:derand:det}), then build the layout.
  \item In the mixed cases apply \cref{lem:undo} to $F=G$, the performed reductions
  ($D=\emptyset$), and the layout of $H$ (no bags if $H$ is empty).
\end{enumerate}

\begin{proof}[Proof of \cref{thm:algo-precise} in the padded and mixed cases]
Let $n\ge n_2$ (for $n<n_2$ the claims are trivial, as explained in
\cref{sec:alg-model}). Since operations remove vertices and do not increase degrees
by~\ref{red:deg}, and the coefficients are nonnegative and nondecreasing in the degree with
$a_0=a_1=a_2=0$, we have $\beta(H)\le\beta(G)$. Let
\[
  \theta_{\mathrm D}(n):=\sqrt{8(12K+3)n}\,t^{1/4},\qquad
  \theta_{\mathrm R}(n):=n^{3/4},
\]
\[
  \rho(n):=\rho_0(n)\,n+5\max\{\theta_{\mathrm D}(n),\theta_{\mathrm R}(n)\}+3\log_2n+1
\]
for $n\ge n_2$, and $\rho(n):=n$ for $n<n_2$; we replace $\rho$ by its running
maximum~\eqref{eq:running-max}. Then $\rho(n)=o(n)$: we have $\rho_0(n)\to0$, because
$h\to\infty$, $\eps(h)\to0$, $\xi,p_t\to0$ and $K=h$; and
$\theta_{\mathrm D}(n)=O(\sqrt{n\log\log n}\,(\log n)^{1/4})$.

When every count is at most its mean plus $\theta\in\{\theta_{\mathrm
D}(n),\theta_{\mathrm R}(n)\}$, with $\mu=12K+3$ in the paragraph ``The three
algorithms'' of \cref{sec:alg-derand}, \cref{prop:alg-sweep} bounds the width of the
layout of $H$ by $\beta(G)+\rho_0(n)n+5\theta$, and \cref{lem:undo} adds at most
$3\log_2n+1$; so the width is at most $\beta(G)+\rho(n)$. That paragraph gives the
three algorithms, with failure probability $\pi(n)\to0$ for the randomized one.

The running times are derived in \cref{sec:alg-time}. \qedhere
\end{proof}

\subsubsection{The median-switched layout: cubic case}\label{sec:alg-cubic}

This subsection proves \cref{thm:algo-precise} in the cubic case. The layout is the
median-switched layout of \cref{lem:median-layout}: inner frontiers below threshold
zero, a connector at zero, and outer frontiers above zero. Two points need care.
First, the layout must be built for the discrete order, which is a lexicographic order
with ties broken by vertex index rather than an order of distinct real labels; an
auxiliary injective labeling $y$ handles this. Second, the connector order of
\cref{lem:connector} must be computed efficiently (\cref{lem:connector-linear}). The
counts and their means then bound the width as in \cref{sec:alg-plain}.

Let $G$ be a graph with $\Delta(G)\le3$ and $n\ge3$ vertices; vertices of degree $0$,
$1$ and $2$ are allowed and no reductions are applied. The three algorithms are those
of \cref{sec:alg-plain} with $H:=G$ and without steps 1 and 4, and with the layout
below in place of the plain-sweep layout and the counts below in place of its
counts. The labels use the weight $q\equiv2$, and the counts use the grid
\[
  \tau_j:=-2+\frac jK\qquad(0\le j\le4K),\qquad\text{so that }\tau_{2K}=0 .
\]

\paragraph{An injective labeling for the discrete order.} Let $\Sigma$ be the finite
set $\{\tilde Y_v:v\in V(G)\}\cup\{0\}\cup\{\tau_j:0\le j\le4K\}$, let $\vartheta>0$ be
so small that $\vartheta n$ is less than every positive difference of two elements of
$\Sigma$, and put
\[
  y_v:=\tilde Y_v+\vartheta v .
\]
The algorithms never compute $y$; it is used only in the analysis.
\begin{itemize}
  \item \emph{$y$ is injective and induces the order of $(\tilde Y_v,v)$.} If
  $\tilde Y_u<\tilde Y_v$, then $\tilde Y_v-\tilde Y_u>\vartheta n>\vartheta(u-v)$, so
  $y_u<y_v$; if $\tilde Y_u=\tilde Y_v$ and $u<v$, then $y_u<y_v$.
  \item \emph{Sublevel sets at the grid.} For every $\tau\in\Sigma$,
  \[
    S_\tau=\{v:\ y_v\le\tau\}=\tilde S_\tau=\{v:\ \tilde Y_v<\tau\}.
  \]
  If $\tilde Y_v<\tau$, then $\tau-\tilde Y_v>\vartheta n\ge\vartheta v$, so
  $y_v<\tau$; if $\tilde Y_v\ge\tau$, then $y_v>\tilde Y_v\ge\tau$, since $v\ge1$.
\end{itemize}
Here and below, $S_\tau$, $\infr S_\tau$, $\outfr S_\tau$, $U$, $W$, $D$, $U'$, $W'$,
$L_0$ and $L_0'$ are as in \cref{sec:cubic-setting}, for the labeling $y$. We write
$\tau$ for the threshold called $t$ there, because $t$ is a parameter of this
section. By the second item,
\begin{align*}
  U&=\infr S_0=\{u:\ \tilde Y_u<0,\ \exists x\in N(u):\tilde Y_x\ge0\},\\
  W&=\outfr S_0=\{x:\ \tilde Y_x\ge0,\ \exists u\in N(x):\tilde Y_u<0\},
\end{align*}
and $D$ is the set of vertices of degree $3$ all of whose neighbors lie on the other
side of $0$, where the two sides are $\{\tilde Y<0\}$ and $\{\tilde Y\ge0\}$.

\paragraph{The connector in linear time.} The connector of \cref{lem:median-layout}
needs an order of $W'$ satisfying \eqref{eq:cubic-connector}. The proof of
\cref{lem:connector} is a recursion on closed parts; the following order is computed
directly. As in \cref{lem:connector}, let $Q$ be a finite set and $P$ a finite set of
items, each item $p$ having a nonempty set $e(p)\subseteq Q$ of at most two
endpoints. Let $Q_1,\dots,Q_r$ be the vertex sets of the connected components of the
hypergraph on $Q$ whose edges are the sets $e(p)$, $p\in P$, indexed so that their
\emph{excesses}
$\mathrm{ex}_l:=|Q_l|-|\{p\in P:e(p)\subseteq Q_l\}|$ are nondecreasing in $l$. A
\emph{greedy order} of $Q_l$ starts with any element of $Q_l$ and, as long as some
item has one endpoint among the elements chosen so far and one outside them, adds the
latter. A \emph{component order} of $Q$ is a concatenation of greedy orders of
$Q_1,\dots,Q_r$, in this order.

\begin{lemma}[connector in linear time]\label{lem:connector-linear}
Every component order is an ordering of $Q$ that satisfies
\eqref{eq:cubic-connector}, and a component order can be computed in $O(|Q|+|P|)$
time.
\end{lemma}

\begin{proof}
Every endpoint set is nonempty, connected in the hypergraph, and hence contained in
exactly one $Q_l$; so $\sum_l\mathrm{ex}_l=|Q|-|P|$. Fix $l$ and let $X$ be the set of
chosen elements of $Q_l$ at some point, with $\emptyset\ne X\subsetneq Q_l$. Since
$Q_l$ is connected, some edge $e(p)$ meets both $X$ and $Q_l\setminus X$; as
$|e(p)|\le2$, $p$ has one endpoint in $X$ and one outside. So the greedy order
continues until it exhausts $Q_l$, and each step after the first covers an item that
was not covered before; in particular, a component order is an ordering of $Q$. Let
$c_l(j')$ be the number of items with $e(p)$ contained in the first $j'$ chosen
elements of $Q_l$; then $c_l(j')\ge j'-1$ for $j'\ge1$, so $j'-c_l(j')\le1$ for all
$0\le j'\le|Q_l|$. In particular $\mathrm{ex}_l=|Q_l|-c_l(|Q_l|)\le1$.

Now let $j=\sum_{l<i}|Q_l|+j'$ with $0\le j'\le|Q_i|$. The first $j$ elements cover
all items of $Q_1,\dots,Q_{i-1}$, and the items of $Q_i$ counted by $c_i(j')$; no
other item, since each item lies in one component. Hence
\[
  j-c(j)=\sum_{l<i}\mathrm{ex}_l+\bigl(j'-c_i(j')\bigr)\le\sum_{l<i}\mathrm{ex}_l+1 .
\]
If $\sum_{l<i}\mathrm{ex}_l>0$, then $\mathrm{ex}_l>0$ for some $l<i$, so
$\mathrm{ex}_{l'}\ge\mathrm{ex}_l>0$ for all $l'\ge l$ by the ordering, and
$\sum_{l<i}\mathrm{ex}_l\le\sum_l\mathrm{ex}_l=|Q|-|P|$. Otherwise the sum is at most
$0$. In both cases $j-c(j)\le\max(0,|Q|-|P|)+1$.

The components are found by breadth-first search in $O(|Q|+|P|)$ time; the greedy
orders are breadth-first searches as well. The excesses are integers in $[-|P|,1]$, so
counting sort orders the components in $O(|Q|+|P|)$ time. \qedhere
\end{proof}

\paragraph{The layout.} Let $v_1,\dots,v_n$ be the vertices in increasing order of
$(\tilde Y_v,v)$, which is the increasing order of $y$, and $\ell:=|\tilde S_0|$. The
algorithms build the bags of \cref{lem:median-layout-bags}: the inner bags
$A_1,\dots,A_\ell$, the connector bags $C_{-1},C_0,\dots,C_k$ with $k=|W'|$, and the
outer bags $A_{\ell+1},\dots,A_n$. The connector bags use a component order of $W'$,
computed by \cref{lem:connector-linear} with $Q=W'$ and the items
$P=\{u\in U':N(u)\cap W'\ne\emptyset\}$, where $e(u)=N(u)\cap W'$; each item has at
most two endpoints by \cref{lem:median-layout-bags}. By
\cref{lem:connector-linear,lem:median-layout-bags}, under the hypotheses of
\cref{lem:median-layout} these bags form a path decomposition of width at most $w$.
The sets $U$, $W$, $D$, $U'$, $W'$ and $P$ are computed in $O(n)$ time from the signs
of the labels. By the definitions of the bags, those containing a vertex are
determined by its index, the smallest and largest indices of its neighbors, its
position in the connector order, and the largest position of an element of $e(u)$.
So the interval form is computed in $O(n)$ time after sorting.

\paragraph{Counts.} The cubic case uses the $\mu=8K+4$ counts
\begin{align*}
  X^{\mathrm{in}}_j&:=|\infr\tilde S_{\tau_j}|=\bigl|\{u:\ \tilde Y_u<\tau_j,\ \exists x\in N(u):\tilde Y_x\ge\tau_j\}\bigr|&&(0\le j<2K),\\
  X^{\mathrm{out}}_j&:=|\outfr\tilde S_{\tau_j}|&&(2K<j\le4K),\\
  X^{\mathrm{win}}_j&:=\bigl|\{v:\ \tau_j\le\tilde Y_v<\tau_{j+1}\}\bigr|&&(0\le j<4K),\\
  X^{\mathrm{lo}}&:=\bigl|\{v:\ \tilde Y_v<-2\}\bigr|,\qquad
  X^{\mathrm{hi}}:=\bigl|\{v:\ \tilde Y_v\ge2\}\bigr|,
\end{align*}
and $L_0$, $L_0'$. The last two are counts in the sense of \cref{sec:alg-derand}:
$L_0=\sum_v\ind\{v\in U\cup(D\cap W)\}$, a disjoint union since $U\cap W=\emptyset$,
and each indicator depends only on the set $\{u\in N[v]:\tilde Y_u<0\}$; similarly for
$L_0'=\sum_v\ind\{v\in W\cup(D\cap U)\}$.

\paragraph{Means.} Let $n\ge n_2$ and $r_h=r_*(1-1/h)$, as in
\cref{sec:alg-means}. Then:
\begin{itemize}
  \item for $j<2K$ we have $\tau_j+\xi\le-1/K+\xi<0$ by \ref{it:alg-N1}, so
  \cref{cor:alg-transfer}\ref{it:transfer:in} and \cref{lem:alg-means}\ref{it:means:cubic}
  give $\E X^{\mathrm{in}}_j\le(\kappa(r_h)+\omega_3)n$;
  \item for $j>2K$ we have $\tau_j-\xi\ge1/K-\xi>0$, so
  \cref{cor:alg-transfer}\ref{it:transfer:out} and \cref{lem:alg-means}\ref{it:means:cubic}
  give $\E X^{\mathrm{out}}_j\le(\kappa(r_h)+\omega_3)n$;
  \item \cref{cor:alg-transfer}\ref{it:transfer:sign} and the two switch inequalities
  of \cref{lem:alg-means}\ref{it:means:cubic} give $\E L_0,\E L_0'\le(\kappa(r_h)+\omega_3)n$;
  \item \cref{cor:alg-transfer}\ref{it:transfer:window} gives
  $\E X^{\mathrm{win}}_j\le(\varphi(0)/K+2\xi\varphi(0)+p_t)n$, and with
  \ref{it:alg-N2}, $\E X^{\mathrm{lo}},\E X^{\mathrm{hi}}\le(\Phi(-2)+\xi\varphi(0)+p_t)n<0.023\,n$.
\end{itemize}

\begin{proof}[Proof of \cref{thm:algo-precise} in the cubic case]
Let $n\ge n_2$ and suppose that every count is at most its mean plus $\theta$. We
check the hypotheses of \cref{lem:median-layout} for $y$ with
\[
  w:=\bigl(\kappa(r_h)+\omega_3+\varphi(0)/K+2\xi\varphi(0)+p_t\bigr)n+2\theta+1 .
\]
\emph{Inner frontiers.} Let $\tau<0$. If $\tau<-2$, then
$\infr S_\tau\subseteq S_\tau\subseteq S_{-2}=\tilde S_{-2}$, so
$|\infr S_\tau|\le X^{\mathrm{lo}}\le0.023n+\theta$. Otherwise
$\tau\in[\tau_j,\tau_{j+1})$ for some $0\le j<2K$. Let $u\in\infr S_\tau$, so
$y_u\le\tau<y_x$ for some neighbor $x$. If $y_u\le\tau_j$, then
$u\in\infr S_{\tau_j}$; otherwise $\tau_j<y_u\le\tau_{j+1}$. Since
$S_{\tau_j}=\tilde S_{\tau_j}$ and
$S_{\tau_{j+1}}\setminus S_{\tau_j}=\{v:\tau_j\le\tilde Y_v<\tau_{j+1}\}$, this gives
$|\infr S_\tau|\le X^{\mathrm{in}}_j+X^{\mathrm{win}}_j\le w-1$.

\emph{Outer frontiers.} Let $\tau>0$. If $\tau\ge2$, then $\outfr S_\tau\subseteq
V(G)\setminus S_2$, so $|\outfr S_\tau|\le X^{\mathrm{hi}}\le0.023n+\theta$. Otherwise
$\tau\in[\tau_j,\tau_{j+1})$ for some $2K\le j<4K$. Let $x\in\outfr S_\tau$, so
$y_x>\tau\ge y_u$ for some neighbor $u$. If $y_x>\tau_{j+1}$, then
$x\in\outfr S_{\tau_{j+1}}$, and $j+1>2K$; otherwise $\tau_j\le\tau<y_x\le\tau_{j+1}$.
So $|\outfr S_\tau|\le X^{\mathrm{out}}_{j+1}+X^{\mathrm{win}}_j\le w-1$.

\emph{Threshold zero.} $\max(L_0,L_0')+1\le(\kappa(r_h)+\omega_3)n+\theta+1\le w$.

\emph{Tails.} $0.023n+\theta\le w$, because $\kappa(r_h)>\cthree>0.023$; indeed
$\kappa$ is decreasing, since $b(r)$ is increasing and $\arccos$ is decreasing.

So \cref{lem:median-layout-bags} applies, and the decomposition built by the
algorithms has width at most $w$. Put
\[
  \rho(n):=\bigl(\kappa(r_h)-\cthree+\omega_3+\varphi(0)/K+2\xi\varphi(0)+p_t\bigr)n
  +2\max\{\theta_{\mathrm D}(n),\theta_{\mathrm R}(n)\}+1
\]
for $n\ge n_2$, with $\theta_{\mathrm D}(n)=\sqrt{8(8K+4)n}\,t^{1/4}$ and
$\theta_{\mathrm R}(n)=n^{3/4}$, and $\rho(n):=n$ for $n<n_2$; we replace $\rho$ by
its running maximum~\eqref{eq:running-max}. Since $\kappa$ is continuous and
$r_h\to r_*$, we have $\kappa(r_h)\to\cthree$, and $\rho(n)=o(n)$.

The paragraph ``The three algorithms'' of \cref{sec:alg-derand}, with $\mu=8K+4$,
shows that the algorithms ensure this with $\theta\in\{\theta_{\mathrm
D}(n),\theta_{\mathrm R}(n)\}$, the randomized one except with probability $\pi(n)$;
then the width is at most $\cthree n+\rho(n)$. \qedhere
\end{proof}

\subsubsection{The deletion process: bootstrap and general cases}\label{sec:alg-general}

This subsection proves \cref{thm:algo-precise} in the bootstrap cases and in the
general case, by making the deletion process of \cref{sec:bootstrap} algorithmic. The
bootstrap lemma is proved through the explicit bound~\eqref{eq:bootstrap-explicit},
and the proof of that bound, through \cref{prop:deletion}, is already a procedure: delete a vertex
of maximum degree, reduce, and repeat until every component is basic. The induction
of Edwards and McDermid~\cite{EdwardsM15}, which \cref{prop:deletion} generalizes, is
of the same kind. The algorithm below makes the procedure explicit, with one
simplification: it never tests whether a component is basic, which would require
trying every vertex. Instead, it stops at a component of maximum degree at most $k$
(defined below), which it decomposes by a sweep, and at a component that becomes
empty when the chosen vertex is deleted and reduction operations are applied.

Throughout, $X$ is one of the six sequences of \cref{thm:pw-seq}, the \emph{seed},
with base degree $k$: $k=3$ for $\seqC$, $k=4$ for $\seqM$ and $\seqF$, and $k=5$ for
$\tripP$, $\tripR$ and $\tripS$. Let $(b_d)_{d\ge0}$ be the bootstrap of $X$, with
$b_0=b_1=b_2=0$ and $(b_3,\dots,b_k)=X$; it satisfies the hypotheses of
\cref{lem:extension} (proof of \cref{thm:pw-seq}). For $X=\tripS$ it is denoted
$\sigma$ in \cref{sec:proof-main}; we keep the letter $b$ of
\cref{sec:bootstrap-lemma}, because $\sigma_v$ denotes standard deviations here. The
sum $\psi(G)=\sum_{d\ge3}b_dn_d(G)$, the potential $\psi^-$ and the running minima
$H_d$ of the increments of $b$ are as in
\cref{sec:bootstrap-lemma,sec:bootstrap-proof}. The general case is the bootstrap
case of $\tripS$ with the coefficients rounded up: by \cref{prop:large-degrees},
$b_d\le a_d$ for every $d\ge3$ when $X=\tripS$, where $a_d$ are the coefficients of
\cref{thm:main}.

\paragraph{Base layouts.} Each seed comes with a layout for graphs $J$ with
$\delta(J)\ge3$ and $\Delta(J)\le k$, labeled with the parameters of some
$n\ge\max\{|V(J)|,n_2\}$:
\begin{itemize}
  \item $\seqC$: the median-switched layout with the labels of the cubic case
  ($q\equiv2$), as in \cref{sec:alg-cubic};
  \item $\seqM$: the median-switched layout with the labels of the switched case
  ($q(x)=d(x)-1$), as in \cref{app:alg-switched};
  \item $\seqF$: the plain-sweep layout with the labels of the padded case ($q\equiv3$);
  \item $\tripP$, $\tripR$, $\tripS$: the plain-sweep layout with the labels of the
  mixed cases ($q(x)=d(x)-1$) and the triple $X$.
\end{itemize}
For such a graph $J$, $\psi(J)=\sum_{d=3}^kb_dn_d(J)$, and this is the bound $\beta(J)$
of the case whose labels are used; for $\seqC$ and $\seqF$ because every vertex of $J$
has degree $3$, respectively degree $3$ or $4$. Each base layout has at most $12K+3$
counts in the sense of \cref{sec:alg-derand}, and if every count is at most its mean
plus some $\theta\ge0$, then the layout is a path decomposition of $J$ of width at
most
\begin{equation}\label{eq:alg-base}
  \psi(J)+\rho_X(n)\,|V(J)|+5\theta+1,
\end{equation}
where $\rho_X(n)$ is the function $\rho_0(n)$ of \cref{sec:alg-plain} for the plain
sweep (with the error term $\eps(h)$ of the case) and $\rho_{\mathrm{med}}(n)$ of
\cref{app:alg-switched} for the median-switched layout. This is
\cref{prop:alg-sweep} for the plain sweep and \cref{cor:alg-switched-base} for the
median-switched layout.

\paragraph{The algorithm.} Let $G_0$ be $G$ without its isolated vertices, and
$n_{\ge1}:=|V(G_0)|$. If $n_{\ge1}\le2$, output one bag $V(G_0)$ and one bag for each isolated
vertex. Otherwise compute all parameters from $n_{\ge1}$, and an adjacency matrix of
$G_0$.
\begin{enumerate}[label=\textup{(\arabic*)}]
  \item Apply reduction operations to $G_0$ until none applies, recording the removed
  vertices; let $R_0$ be the final graph.
  \item Keep a list of components, initially those of $R_0$, and a deleted set
  $D:=\emptyset$. Repeat the following \emph{round} while some component $J$ in the
  list is not marked final:
  \begin{itemize}
    \item if $\Delta(J)\le k$, mark $J$ final (a \emph{sweep component});
    \item otherwise choose a vertex $w$ of degree $\Delta(J)$ that has a neighbor of
    smaller degree, if one exists, and any vertex of degree $\Delta(J)$ otherwise.
    Apply reduction operations to $J-w$ until none applies; let $J^*$ be the final
    graph.
    \begin{itemize}
      \item If $J^*$ is empty, mark $J$ final (a \emph{thin component}). The deletion
      of $w$ and these reductions are kept for step 3 only; they are not part of the
      global sequence. The reductions may have changed the shared adjacency matrix,
      so they are recorded in an undo log while they are applied and rolled back in
      time $O(|V(J)|+|E(J)|)$, which restores $J$.
      \item Otherwise \emph{commit}: add $w$ to $D$, record the deletion and the
      reductions as part of the global sequence, and replace $J$ in the list by the
      components of $J^*$.
    \end{itemize}
  \end{itemize}
  \item Let $R$ be the union of the final components. Decompose each sweep component
  $J$ by the base layout of $X$, with the parameters of
  $n_{\ge1}$ (randomized, Las Vegas or deterministic, as below). Decompose each thin
  component $J$ by \cref{lem:undo}, applied to $F=J$, the deleted set $\{w\}$ and the
  empty final graph. Concatenate these decompositions into a decomposition of $R$.
  \item Apply \cref{lem:undo} to $F=G_0$, the global sequence (deleted set $D$, final
  graph $R$) and the decomposition of $R$. Append one bag for each isolated vertex of
  $G$.
\end{enumerate}
By \ref{red:local}, reductions inside a component of the current graph are
reductions of the whole current graph, so the global sequence is a sequence of
reductions and deletions from $G_0$ to $R$, as \cref{lem:undo} requires. Every
component in the list has minimum degree at least $3$ by~\ref{red:max}, since it is a
component of the final graph of a maximal sequence of reductions. Every final
component is basic in the sense of \cref{def:basic}: a sweep component has
$\Delta\le k$, and a thin component has the witness $w$.

\begin{lemma}[the deletion process]\label{lem:alg-deletion}
The algorithm makes at most $n_{\ge1}$ commits and at most $2n_{\ge1}$ rounds, and
$|D|+\psi^-(R)\le\psi(G)$.
\end{lemma}

\begin{proof}
Every commit deletes a vertex. Every other round marks a component final, and final
components are disjoint, so there are at most $n_{\ge1}$ of them.

Consider a commit at a component $J$. Then $J$ is connected, $\delta(J)\ge3$,
$\Delta(J)\ge k+1$, the vertex $w$ was chosen by the rule of
\cref{lem:deletion-step}, and $J^*$ is the final graph of a maximal sequence of
reductions applied to $J-w$, and is nonempty. So \cref{lem:deletion-step} gives
$1+\psi^-(J^*)\le\psi^-(J)$. Hence the quantity $|D|+\sum_J\psi^-(J)$, summed over
the components in the list, does not increase at a commit, and it does not change
when a component is marked final. At the start it is $\psi^-(R_0)$ (by additivity),
and at the end it is $|D|+\psi^-(R)$. Finally
$\psi^-(R_0)\le\psi(R_0)\le\psi(G_0)=\psi(G)$, by \ref{red:deg}, $H\ge0$, and
because isolated vertices have weight $b_0=0$.
\end{proof}

\begin{proof}[Proof of \cref{thm:algo-precise} in the bootstrap and general cases]
Let $n_{\ge1}\ge n_2$; otherwise the claims are trivial, because the output has width at
most $n_{\ge1}-1$ and we put $\rho(N):=N$ and $\pi(N):=0$ for $N<n_2$. Put
\begin{align*}
  \rho_{\mathrm{sw}}(N)&:=\rho_X(N)\,N+5\max\{\theta_{\mathrm D}(N),\theta_{\mathrm R}(N)\}+1,\\
  \rho(N)&:=b_3+\rho_{\mathrm{sw}}(N)+6\log_2N+2\qquad(N\ge n_2),
\end{align*}
with $\theta_{\mathrm D}(N)=\sqrt{8(12K+3)N}\,t^{1/4}$ and
$\theta_{\mathrm R}(N)=N^{3/4}$ (parameters of $N$); we replace $\rho$ by its running
maximum~\eqref{eq:running-max}. Then $\rho(N)=o(N)$, because $\rho_X(N)\to0$.

We first bound the widths of the decompositions of the final components, of $R$ and
of $G_0$, assuming that every count of every sweep component is at most its mean
plus some $\theta\le\max\{\theta_{\mathrm D}(n_{\ge1}),\theta_{\mathrm R}(n_{\ge1})\}$.
Then we show that the three algorithms ensure this assumption, for certain or with
the stated probability.

\emph{Sweep components.} Let $J$ be a sweep component with $N_J$ vertices, and
suppose that all its counts are at most their means plus
$\theta\le\max\{\theta_{\mathrm D}(n_{\ge1}),\theta_{\mathrm R}(n_{\ge1})\}$.
By~\eqref{eq:alg-base}, applied with the parameters of $n_{\ge1}\ge N_J$, its
decomposition has width at most
$\psi(J)+\rho_X(n_{\ge1})N_J+5\theta+1\le\psi(J)+\rho_{\mathrm{sw}}(n_{\ge1})$.
Since $J$ is basic, $\psi(J)\le\psi^-(J)+b_3$ by \cref{lem:basic}.

\emph{Thin components.} By \cref{lem:undo} with deleted set $\{w\}$ and $R$ empty,
the width is at most $1+(-1)+3\log_2n_{\ge1}+1$.

\emph{The final graph.} The width of the concatenation is the maximum over the
components. Since $\psi^-$ is nonnegative and additive, the decomposition of $R$ has
width at most $\psi^-(R)+b_3+\rho_{\mathrm{sw}}(n_{\ge1})+3\log_2n_{\ge1}+1$.

\emph{Undoing.} By \cref{lem:undo} and \cref{lem:alg-deletion}, the decomposition of
$G_0$ has width at most
\[
  |D|+\psi^-(R)+b_3+\rho_{\mathrm{sw}}(n_{\ge1})+6\log_2n_{\ge1}+2
  \le\psi(G)+\rho(n_{\ge1})\le\sum_{d\ge3}b_dn_d(G)+\rho(n),
\]
using $n_{\ge1}\le n$ and the monotonicity of $\rho$. The bags of isolated vertices
do not increase the width. This is the width bound of \cref{thm:algo-precise} in the
bootstrap case of $X$. In the general case, $X=\tripS$ and $b_d\le a_d$ give the first
width bound, $\sum_{d\ge3}a_dn_d(G)+\rho(n)$.

\emph{Edge bound} (general case). Every vertex of $G_0$ has degree at least $1$, so
$n_{\ge1}\le2m$. By \eqref{eq:edges-ratio}, $a_d\le\frac7{48}\cdot\frac d2$ for every
$d\ge3$, and $b_0=b_1=b_2=0$; so
$\psi(G)=\sum_db_dn_d(G)\le\frac7{48}\cdot\frac12\sum_vd(v)=\frac7{48}m$. Hence the
same decomposition also has width at most
$\frac7{48}m+\rho(n_{\ge1})\le\frac7{48}m+\rho(2m)$.

\emph{Probability.} The randomized algorithm samples the variables of different sweep
components independently. We apply \cref{lem:alg-derand}\ref{it:derand:rand} to each
sweep component $J$ with the \emph{global} threshold $\lambda=n_{\ge1}^{3/4}$: the
probability that some count of $J$ exceeds its mean plus $n_{\ge1}^{3/4}$ is at most
$8\mu N_Jt^{1/2}n_{\ge1}^{-3/2}$, with $\mu\le12K+3$. The sweep components are disjoint, so
summing over them gives at most $8(12K+3)t^{1/2}n_{\ge1}^{-1/2}$. With the sampling budget,
which is $2n_{\ge1}\lceil\log_2n_{\ge1}\rceil$ draws for the at most $n_{\ge1}$ samples, the
failure probability is at most $\pi(n_{\ge1})$ with
$\pi(N):=8(12K+3)t^{1/2}N^{-1/2}+1/\lceil\log_2N\rceil$ (parameters of $N$). A union
bound with each component's own threshold $N_J^{3/4}$ would not suffice, because
there may be many components of moderate size.

\emph{Deterministic.} Greedy fixing (\cref{lem:alg-derand}\ref{it:derand:det}) is run
on each sweep component separately; the deviations are at most
$\sqrt{8\mu N_J}\,t^{1/4}\le\theta_{\mathrm D}(n_{\ge1})$.

\emph{Las Vegas.} Steps 1, 2 and 4 are deterministic. In step 3, each sweep component
is sampled, and resampled until all its counts are at most their means plus
$n_{\ge1}^{3/4}$. By \cref{lem:alg-derand}\ref{it:derand:rand} and \ref{it:alg-N3}, a sample
of $J$ is accepted with probability at least $1-8\mu N_Jt^{1/2}n_{\ge1}^{-3/2}\ge\frac12$,
so the expected number of samples per component is at most $2$. \qedhere
\end{proof}

\begin{remark}[the rows and the sequences on all graphs]\label{rem:alg-other-rows}
The bootstrap of a seed $X$ is entrywise at most the row $X$ of \cref{tab:sequences},
which rounds it up (for $X\ne\seqC$; the bootstrap of $\seqC$ is given in the caption),
and at most the sequence $X$ of \cref{thm:pw-seq}, which agrees with it up to degree
$k$ and has coefficient $1$ above $k$ (\cref{lem:bootstrap}). So the bootstrap cases
compute, for every graph, decompositions of width at most the bound of each row of
\cref{cor:table} and of each sequence of \cref{thm:pw-seq}, plus the remainder
$\rho(n)$. No row is
best for all graphs (\cref{rem:best-row}); running the algorithm for each of the six
seeds and keeping the narrowest decomposition realizes the minimum of the six bounds.
For a graph of maximum degree at most $k$, the bootstrap case of $X$ never commits a
deletion, and its bound $\sum_{d=3}^kb_dn_d$ improves on the bound $\beta(G)$ of the
cubic and padded cases when $G$ has vertices of degree at most $2$.
\end{remark}

\subsubsection{Running times}\label{sec:alg-time}

The width claims of \cref{thm:algo-precise}, and the probability claims of \cref{rem:alg-randomized}, are proved in
\cref{sec:alg-plain,sec:alg-cubic,sec:alg-general} and \cref{app:alg-switched}. This subsection bounds the running
times of the steps of the algorithms, which completes the proof. Recall that
$t=O(\log n)$, that $h,K=O(\log\log n)$, and that the number of counts is
$\mu=O(K)$.

\paragraph{Cubic, padded, switched and mixed cases.} Here $m\le\frac52n$.
\begin{itemize}
  \item \emph{Reductions} (switched and mixed cases). Degrees never exceed $5$, so
  each operation,
  including the adjacency test of \ref{red:R3} and \ref{red:R4}, takes $O(1)$ time
  with a queue of vertices of degree at most $2$; in total $O(n)$ operations on
  $O(\log n)$-bit integers, that is $O(n\log n)$ bit operations.
  \item \emph{Coefficients and finite distribution.} $n(\log n)^{O(1)}$ and $(\log n)^{O(1)}$
  bit operations (\cref{sec:alg-model,sec:alg-law}).
  \item \emph{Labels.} Each $\tilde Y_v$ is a sum of at most $2t^{1/4}$ products of
  $O(\log t)$-bit rationals; in total $n\,t^{O(1)}$ bit operations.
  \item \emph{Sampling} (randomized). At most $2n\lceil\log_2n\rceil$ draws of
  $O(\log t)$ bits, each followed by a binary search over $M$ prefix sums; in total
  $n(\log n)^{O(1)}$ bit operations in the worst case.
  \item \emph{Sorting.} $O(n\log n)$ comparisons of labels, each in $(\log t)^{O(1)}$
  bit operations.
  \item \emph{Layout.} $O(n)$ operations for the plain sweep; for the median-switched
  layout, $O(n)$ operations by \cref{lem:connector-linear} and the interval rules.
  \item \emph{Undoing.} $O(n\log n)$ operations by \cref{lem:undo}, since the layout of
  $H$ has $s\le2N+2\le3n$ bags: $N$ in the plain sweep, and $N+|W'|+2$ in the
  median-switched layout.
\end{itemize}
So one sample (the randomized algorithm) uses $n(\log n)^{O(1)}$ bit operations. The Las Vegas
algorithm additionally computes the means in $\mu n\,n^{o(1)}=n^{1+o(1)}$ bit
operations (\cref{lem:alg-derand}\ref{it:derand:cost}), checks the counts of each
sample in $n(\log n)^{O(1)}$ bit operations, and takes at most two samples in
expectation; so its expected running time is $n^{1+o(1)}$ (for $n<n_2$ it accepts the
first sample). The deterministic algorithm replaces sampling by greedy fixing
(\cref{lem:alg-derand}\ref{it:derand:det}), which uses $\mu n^2n^{o(1)}=n^{2+o(1)}$ bit
operations.

\paragraph{Bootstrap and general cases.} Removing isolated vertices takes $O(n)$
operations, and building the adjacency matrix of $G_0$ takes $O(n_{\ge1}^2)$.
\begin{itemize}
  \item \emph{Steps 1 and 2.} With adjacency lists that support deletion in $O(1)$
  time, the adjacency matrix for the tests and insertions of \ref{red:R3} and
  \ref{red:R4}, and a queue of vertices of degree at most $2$, each reduction takes
  $O(1)$ operations. A component's reductions after the deletion of $w$ take
  $O(|V(J)|+|E(J)|)$ operations, and so do the choice of $w$ and the computation of
  the components of $J^*$ by breadth-first search. By \cref{lem:alg-deletion} there
  are at most $2n_{\ge1}$ rounds, so steps 1 and 2 take $O(n_{\ge1}(n_{\ge1}+m))$
  operations.
  \item \emph{Step 3.} The sweep components are disjoint and have maximum degree at
  most $5$, and each is decomposed by the base layout of the seed, so the bounds of
  the previous paragraph, summed over them, give
  $n_{\ge1}(\log n_{\ge1})^{O(1)}$ bit operations (randomized), $n_{\ge1}^{1+o(1)}$ in
  expectation (Las Vegas), and $\sum_JN_J^2n_{\ge1}^{o(1)}\le n_{\ge1}^{2+o(1)}$
  (deterministic). Each thin component takes $O(|E(J)|+|V(J)|\log n_{\ge1})$
  operations by \cref{lem:undo}.
  \item \emph{Step 4.} $O(m+n_{\ge1}\log n_{\ge1})$ operations by \cref{lem:undo}, since the
  decomposition of $R$ has at most $3n_{\ge1}$ bags: a component $J$ contributes at most
  $2|V(J)|+2\le3|V(J)|$ of them, as $|V(J)|\ge4$.
\end{itemize}
In total, the bootstrap and general cases use $O(n(n+m))$ operations on integers of
$O(\log n)$ bits, plus $n(\log n)^{O(1)}$, expected $n^{1+o(1)}$, or $n^{2+o(1)}$ bit operations
for one sample, the Las Vegas and the deterministic algorithm. This completes the proof of \cref{thm:algo-precise}.
\hfill$\square$

\subsubsection{Orderings of small cutwidth}\label{app:alg-cutwidth}

\paragraph{Notation.} We use the inner frontier
$\infr X=\{u\in X:N(u)\not\subseteq X\}$ of \cref{sec:prelim-layout}. Let $e(X)$ be
the number of edges with exactly one end in $X$. For an ordering $(u_1,\dots,u_n)$ of $V(G)$ with
prefixes $U_i=\{u_1,\dots,u_i\}$, the \emph{vertex separation} of the ordering is
$\max_i|\infr U_i|$ and its \emph{cutwidth} is $\max_ie(U_i)$.

\begin{lemma}[cutwidth from vertex separation]\label{lem:alg-cutwidth}
Let $G$ be a graph with $\Delta(G)\le3$ and $n\ge1$ vertices, and let
$(v_1,\dots,v_n)$ be an ordering of $V(G)$ of vertex separation $w$. Then an ordering
of $V(G)$ of cutwidth at most $w+2$ can be computed in $O(n)$ operations on integers
of $O(\log n)$ bits.
\end{lemma}

\begin{proof}
The relations $s(G)\le\operatorname{vs}(G)+2$ (\cite[Lemma~2.2]{EllisSudboroughTurner1994},
half of their Theorem~2.1) and $\operatorname{cw}(G)\le s(G)$ for maximum degree~$3$
(\cite[Theorem~3.2]{MakedonSudborough1989}, which gives their Corollary~3.3), where $s$
is the search number, have constructive proofs; we check that they apply to the given
ordering and run in linear time. Both papers work with connected multigraphs.
Cutwidth and vertex separation of an ordering are maxima over the components (restrict
the ordering to each component, then concatenate), so let $G$ be connected.

\emph{Reduction.} Let $G'$ be obtained from $G$ by suppressing, while possible, a
vertex of degree $2$ with two distinct neighbors (deleting it and joining its
neighbors by an edge). This creates no loops, keeps all degrees, and takes $O(n)$
operations. In $G'$ every vertex of degree $2$ is joined to a single vertex by two
parallel edges, which is the form that the proof
of~\cite[Theorem~3.2]{MakedonSudborough1989} brings the graph into (the reduction there,
followed by subdividing every loop once). Restrict the given ordering to $V(G')$. If
$u\in V(G')$ lies in a prefix and has a $G'$-neighbor outside it, then the path of $G$
that the corresponding edge replaces leaves the prefix, and its last vertex in the
prefix, counted from $u$, has a $G$-neighbor outside it. This vertex is $u$ or an
inner vertex of a path at $u$, so distinct such $u$ give distinct vertices, and the
restricted ordering of $G'$ has vertex separation at most $w$.

\emph{Searching.} The procedure \texttt{search1} in the proof
of~\cite[Lemma~2.2]{EllisSudboroughTurner1994} takes the vertices of $G'$ in this order,
places a searcher on each, clears its edges to earlier vertices with one further
searcher, and then removes the searchers from vertices with no uncleared edge. By the
two invariants stated in that proof it never uses more than $w+2$ searchers and never
recontaminates an edge. This gives a search strategy $S$ for $G'$ with at most $k=w+2$
searchers, without recontamination, with $O(n)$ steps, and in which the step that
clears each edge is known.

\emph{Ordering.} The proof of~\cite[Theorem~3.2]{MakedonSudborough1989} orders the
vertices of $G'$ by the first step of $S$ after which at least half of their incident
edges are cleared, breaking ties by an explicit rule for vertices of degree $2$, and
shows that this ordering has cutwidth at most
$\lfloor\Delta(G')/2\rfloor(k-1)+1=k$. It is computed from the clearing steps by a
bucket sort in $O(n)$ operations.

\emph{Back to $G$.} Undo the suppressions in reverse order: an edge $yz$ with $y\ne z$
that replaced a vertex $x$, with $y$ before $z$, becomes the path $yxz$ with $x$ placed
directly after $y$. The new cut between $y$ and $x$ is crossed by the same edges as
the cut after $y$ was, with $yx$ in place of $yz$; every other cut that $yz$ crossed
is now crossed by $xz$ instead. So the cutwidth does not change, and the resulting
ordering of $V(G)$ has cutwidth at most $w+2$.
\end{proof}

\coralgcutwidth*

\begin{proof}
\Cref{thm:algo-precise}\ref{it:algo:D} computes a path decomposition
$(W_1,\dots,W_r)$ of $G$ of width at most $\cthree n+\rho(n)$, in interval form, in
$n^{2+o(1)}$ bit operations. Order the vertices by $\mathrm{first}(v)$, breaking ties
arbitrarily. Let $L$ be a prefix of this ordering other than $V(G)$, let $u'$ be the
first vertex after $L$, and let $g=\mathrm{first}(u')$, which is the least value of
$\mathrm{first}(y)$ over $y\notin L$. If $x\in L$ has a neighbor $y\notin L$, then
$\mathrm{first}(x)\le g\le\mathrm{first}(y)\le j$ for a bag $W_j$ that contains $x$
and $y$, so $x\in W_g$. Hence $\infr L\subseteq W_g\setminus\{u'\}$; since
$\infr V(G)=\emptyset$, the ordering has vertex separation at most
$\cthree n+\rho(n)$. \Cref{lem:alg-cutwidth} turns it into an ordering of cutwidth at
most $\cthree n+\rho(n)+2$ in $O(n\log n)$ bit operations. Splitting this ordering after its $\lfloor n/2\rfloor$-th vertex gives
a bisection with at most as many edges between its two sides.
\end{proof}

\thmalgcutsweep*

\begin{proof}
We use the padded labels of \cref{sec:alg-model}: the padded case ($q\equiv3$) for
$p=4$ and the padded case of degree $5$ ($q\equiv4$) for $p=5$, with the parameters of
$n$; for $n<n_2$ every ordering has cutwidth at most $m\le\frac p2n$, which the
remainder absorbs. Let $n\ge n_2$, let $v_1,\dots,v_n$ be the vertices in increasing
order of $(\tilde Y_v,v)$, let $T_i=\{v_1,\dots,v_i\}$, and let
$\tau_j=-3+j/K$ for $0\le j\le6K$.

\emph{Counts.} For $0\le j<6K$ and $1\le k\le p$ let
\begin{align*}
  X^{\mathrm{cut}}_{j,k}&:=\bigl|\{v:\ \tilde Y_v<\tau_j,\ |\{u\in N(v):\tilde Y_u\ge\tau_j\}|\ge k\}\bigr|,\\
  X^{\mathrm{win}}_{j,k}&:=\bigl|\{v:\ \tau_j\le\tilde Y_v<\tau_{j+1},\ d(v)\ge k\}\bigr|,
\end{align*}
and let $X^{\mathrm{lo}}_k$ and $X^{\mathrm{hi}}_k$ be the numbers of vertices $v$ with
$d(v)\ge k$ and $\tilde Y_v<-3$, respectively $\tilde Y_v\ge3$. These $\mu=p(12K+2)$
random variables are counts in the sense of \cref{sec:alg-derand}, since each indicator
depends only on the labels in $N[v]$. Summing over $k$ turns the conditions $c\ge k$
into the numbers $c$, so
\[
  C_j:=\sum_kX^{\mathrm{cut}}_{j,k}=e(\tilde S_{\tau_j}),\qquad
  D_j:=\sum_kX^{\mathrm{win}}_{j,k}=\sum_{v:\,\tau_j\le\tilde Y_v<\tau_{j+1}}d(v),
\]
and $D^{\mathrm{lo}}:=\sum_kX^{\mathrm{lo}}_k$ and $D^{\mathrm{hi}}:=\sum_kX^{\mathrm{hi}}_k$
are the degree sums over the two tails. (Each edge leaving $\tilde S_{\tau_j}$ is
counted once, at its end in $\tilde S_{\tau_j}$.) We split the cut and the degree sums
into counts because \cref{lem:alg-derand} needs indicators.

\emph{Grid.} Every prefix satisfies
\begin{equation}\label{eq:alg-cut-grid}
  e(T_i)\le\max\Bigl\{D^{\mathrm{lo}},\ D^{\mathrm{hi}},\ \max_{0\le j<6K}(C_j+D_j)\Bigr\}.
\end{equation}
Indeed, let $y=\tilde Y_{v_i}$ and let $zu$ be an edge with $z\in T_i$ and $u\notin
T_i$, so $\tilde Y_z\le y\le\tilde Y_u$. If $y<-3$, we charge the edge to $z$, which
has $\tilde Y_z<-3$ and receives at most $d(z)$ edges; if $y\ge3$, we charge it to $u$,
which has $\tilde Y_u\ge3$. Otherwise $\tau_j\le y<\tau_{j+1}$ for some $j$. If
$\tilde Y_z<\tau_j$, the edge leaves $\tilde S_{\tau_j}$, because
$\tilde Y_u\ge y\ge\tau_j$, and is counted by $C_j$; otherwise
$\tau_j\le\tilde Y_z<\tau_{j+1}$, and we charge the edge to $z$, counted by $D_j$. This
is the charging of the grid step in \cref{sec:warmup-proof}, for the discrete labels.

\emph{Means.} Fix an edge $uv$ and $\tau\in\R$, and couple the discrete and ideal
labels as in \cref{lem:alg-perturb}. Let $\mathcal E$ be the event that
$|\tilde Y_x-Y_x|\le\xi$ for $x\in\{u,v\}$, which fails with probability at most $p_t$
since $\{u,v\}\subseteq N[u]$, and let $\mathcal W$ be the event that neither $Y_u$ nor
$Y_v$ lies in $[\tau-\xi,\tau+\xi]$, which fails with probability at most
$4\xi\varphi(0)$. If $\tilde Y_u<\tau\le\tilde Y_v$, then on $\mathcal E\cap\mathcal W$
we have $Y_u<\tau+\xi$, hence $Y_u<\tau-\xi$, and $Y_v\ge\tau-\xi$, hence
$Y_v>\tau+\xi$: the edge crosses $\tau$ for the ideal labels. The ideal labels have
correlation at least $r_{pp}(1-\frac1h)$ (\cref{lem:alg-correlation}), so by
\cref{lem:crossing}\ref{lem:crossing:b}
\[
  \Pr(uv\text{ leaves }\tilde S_\tau)\le\tfrac1\pi\arccos\bigl(r_{pp}(1-\tfrac1h)\bigr)
  +4\xi\varphi(0)+p_t=\gamma_p+o(1),
\]
uniformly in $\tau$, because $h\to\infty$ and $\xi,p_t\to0$. Summing over the edges,
$\E C_j\le(\gamma_p+o(1))m$. By \cref{cor:alg-transfer}\ref{it:transfer:window} and
\ref{it:alg-N2}, a vertex lies in $[\tau_j,\tau_{j+1})$ with probability at most
$\varphi(0)/K+o(1)$, and below $-3$ or at least $3$ with probability at most $0.0014$; as
$\sum_vd(v)=2m$, this gives $\E D_j\le(2\varphi(0)/K+o(1))m=o(m)$ and
$\E D^{\mathrm{lo}},\E D^{\mathrm{hi}}\le0.0028\,m<\gamma_pm$.

\emph{Greedy fixing.} By \cref{lem:alg-derand}\ref{it:derand:det}, greedy fixing on
these counts ends with values of the $\tilde Z_w$ for which every count is within
$\theta=\sqrt{8\mu n}\,t^{1/4}=o(n)$ of its mean; so each of $C_j$, $D_j$,
$D^{\mathrm{lo}}$ and $D^{\mathrm{hi}}$ is within $p\theta$ of its mean. By
\eqref{eq:alg-cut-grid}, the sorted order has cutwidth at most $\gamma_pm+o(n)$, and its
first $\lfloor n/2\rfloor$ vertices give the bisection. Since $\mu=O(K)$, greedy
fixing takes $n^{2+o(1)}$ bit operations by \cref{lem:alg-derand}\ref{it:derand:cost},
and the other steps take $n(\log n)^{O(1)}$ (\cref{sec:alg-model,sec:alg-law}).
Finally $\gamma_4=\frac16$ and $m\le2n$ for $p=4$.
\end{proof}
{}
\subsection{Computing the median-switched decompositions}\label{app:alg-switched}

This subsection completes the proof of \cref{thm:algo-precise} where the
median-switched layout of \cref{sec:cubic-layout} is used outside
\cref{sec:alg-cubic}: in the switched case, which makes \cref{thm:switched4}
algorithmic, and in the base layouts of the bootstrap cases $\seqC$ and $\seqM$
(\cref{sec:alg-general}). The proof follows \cref{sec:alg-cubic}, and two points are
new. First, the layout must work for graphs whose labels are built with the
parameters of an integer $n$ larger than their number of vertices, as the sweep
components of the deletion process require. Second, the existence proof of
\cref{thm:switched4} bounds $\E\hat L_0$ and $\E\hat L_0'$ only in the limit of a
large radius, so we make the error term explicit; together with the bounds at
nonzero thresholds, this gives expected frontier sizes that are bounded uniformly in
the threshold (\cref{lem:alg-switched-means}). We use the notation and the
parameters of \cref{sec:algorithms}.

\paragraph{Two settings.} Let $n\ge n_2$, and let $H$ be a graph with $N\le n$
vertices whose names are distinct positive integers. We consider two settings.
\begin{itemize}
  \item \emph{Cubic setting:} $\Delta(H)\le3$, the labels are those of the cubic case
  (weight $q\equiv2$), and $\beta(H):=\cthree N$.
  \item \emph{Switched setting:} $\delta(H)\ge3$ and $\Delta(H)\le4$, the labels are
  those of the switched case (weight $q(x)=d(x)-1$, as in the mixed cases), and
  $\beta(H):=\cthree n_3(H)+0.2627\,n_4(H)$.
\end{itemize}
In both settings the labels are built with the parameters of $n$. The construction of
\cref{sec:alg-model,sec:alg-law} and \cref{lem:alg-correlation,lem:alg-perturb} and
\cref{cor:alg-transfer} apply, since they require only $\Delta(H)\le5$, and
$\delta(H)\ge3$ for the weights $q(x)=d(x)-1$. In the switched setting the edge correlation bounds
are $r_{uv}=r_{d(u)d(v)}$ with $\eps_h=8/(h-4)$. In both settings
$\beta(H)\ge\cthree N$: in the cubic setting by definition, and in the switched
setting because every vertex has degree $3$ or $4$ and both coefficients are at least
$\cthree$.

\paragraph{Layout and counts.} Let $\hat D$ be the set of vertices with at least three
neighbors on the other side of zero, as in \cref{sec:switched4}, and
$\hat L_0=|U|+|\hat D\cap W|$ and $\hat L_0'=|W|+|\hat D\cap U|$; in the cubic setting $\hat D$ is the set $D$ of
\cref{sec:cubic-setting}, and $\hat L_0=L_0$, $\hat L_0'=L_0'$. The
\emph{median-switched layout} of $H$ consists of the bags of
\cref{lem:median-layout-bags}, built as in \cref{sec:alg-cubic} from the order of
$(\tilde Y_v,v)$, with $\hat D$ in place of $D$ and a component order of $W'=W\setminus
\hat D$ (\cref{lem:connector-linear}) for the items $u\in U\setminus\hat D$ with
$e(u)=N(u)\cap W'\ne\emptyset$; each item has at most two endpoints by the definition
of $\hat D$. Its interval form is computed in $O(N)$ time after sorting, as in
\cref{sec:alg-cubic}. The layout uses the grid $\tau_j=-2+j/K$ ($0\le j\le4K$) and
the $\mu=8K+4$ counts of \cref{sec:alg-cubic}: the inner counts $X^{\mathrm{in}}_j$
($j<2K$), the outer counts $X^{\mathrm{out}}_j$ ($j>2K$), the windows
$X^{\mathrm{win}}_j$, the tails $X^{\mathrm{lo}}$, $X^{\mathrm{hi}}$, and $\hat L_0$,
$\hat L_0'$ for the discrete labels. The last two are counts in the sense of
\cref{sec:alg-derand}: $\hat L_0=\sum_v\ind\{v\in U\cup(\hat D\cap W)\}$, a disjoint
union because $U\cap W=\emptyset$, and the indicator depends only on the set
$\{u\in N[v]:\tilde Y_u<0\}$; similarly for $\hat L_0'$.

\paragraph{Error terms.} Recall $r_h=r_*(1-1/h)$ and $\kappa$ from
\eqref{eq:cubic-kappa}. For $h\ge13$ we have $\eps_h=8/(h-4)\le\frac89$; put
\[
  e_h:=205\,\eps_h+\kappa\bigl(r_{33}(1-\eps_h)\bigr)-\cthree,\qquad
  \eps_{\mathrm{sw}}(h):=\max\bigl\{e_h,\ 12\,\eps_h^{1/3}\bigr\}\qquad(h\ge13),
\]
and $\eps_{\mathrm{sw}}(h):=1$ for $5\le h<13$. Let $\eps_{\mathrm{med}}(h)$ be
$\kappa(r_h)-\cthree$ in the cubic setting and $\eps_{\mathrm{sw}}(h)$ in the switched
setting. Since $\kappa$ is continuous and $\eps_h\to0$, both tend to $0$ as
$h\to\infty$. Put
\[
  \rho_{\mathrm{med}}(n):=\eps_{\mathrm{med}}(h)+\omega_4+\frac{\varphi(0)}K
  +2\xi\varphi(0)+p_t .
\]

\subsubsection{Uniform means in the switched setting}

The proof of \cref{thm:switched4} bounds the expected frontiers at every nonzero
threshold with the same error term $e_R$ (\cref{lem:switch4-means}), so it is
already uniform in the threshold. At threshold zero, \cref{lem:switch-cont} makes the
error term of \cref{lem:switch4-zero} explicit.

\begin{lemma}[uniform ideal means, switched setting]\label{lem:alg-switched-means}
Let $h\ge5$, and let $(Y_v)$ be the ideal labels of radius $h$ of a graph $H$ with $N$
vertices in the switched setting, and let $S_\tau$, $U$, $W$, $\hat D$, $\hat L_0$
and $\hat L_0'$ be defined for the labeling $Y$. Then
\begin{align*}
  \E|\outfr S_\tau|&\le\beta(H)+\eps_{\mathrm{sw}}(h)\,N\quad(\tau>0),\qquad
  \E|\infr S_\tau|\le\beta(H)+\eps_{\mathrm{sw}}(h)\,N\quad(\tau<0),\\
  \E\hat L_0&\le\beta(H)+\eps_{\mathrm{sw}}(h)\,N,\qquad
  \E\hat L_0'\le\beta(H)+\eps_{\mathrm{sw}}(h)\,N .
\end{align*}
\end{lemma}

\begin{proof}
For $h<13$ the claims are trivial, because each of the four quantities is at most $N$
($\hat L_0\le|U\cup W|$ since $U\cap W=\emptyset$, and similarly $\hat L_0'$). Let
$h\ge13$, so that $\eps_h<1$. By \cref{lem:alg-correlation}, $(Y_v)$ is a centered
Gaussian vector with unit variances and $\Corr(Y_u,Y_v)\ge r_{d(u)d(v)}(1-\eps_h)$ for
every edge $uv$. This is property~\ref{it:sweep-L1} of \cref{sec:plain-sweep} with
$\eps_h$ in place of $\eps_R$. Moreover $-Y$ has the same distribution as $Y$, because $Y$ is a
linear function of the independent standard normal variables $Z_w$, and every $Y_v$
has a continuous distribution, because $Y_v$ has variance $1$. The proofs of
\cref{lem:switch4-means,lem:switch4-zero} use the labels only through these
properties: \ref{it:sweep-L1}, the sign symmetry and the continuity of the marginals.

\emph{Nonzero thresholds.} \Cref{lem:switch4-means}, with $\eps_h$ in place of
$\eps_R$, gives $\E|\outfr S_\tau|\le\beta(H)+e_hN$ for $\tau>0$ and
$\E|\infr S_\tau|\le\beta(H)+e_hN$ for $\tau<0$, and $e_h\le\eps_{\mathrm{sw}}(h)$.

\emph{Threshold zero.} Let $v$ be a vertex of $\{3,4\}$-type $T$, of degree $d\le4$,
with reference bounds $r_T$ and transfer terms $\lambda_T$ as in \cref{def:admissible}. As in the
proof of \cref{lem:switch4-zero}, \eqref{eq:switch4-psi} and~\ref{it:sweep-L1} give
$\Pr(v\in U)+\Pr(v\in\hat D\cap W)\le M_d\bigl((1-\eps_h)r_T\bigr)$, and
\cref{lem:switch-cont} with $\gamma=\eps_h$ gives
$M_d((1-\eps_h)r_T)\le M_d(r_T)+3d\,\eps_h^{1/3}\le\beta_T+12\,\eps_h^{1/3}$, with
$\beta_T$ as in the table in that proof. By the second row of the table, with the
transfer $s=0.1529003-\cthree$ of \cref{sec:switched4},
\[
  \Pr(v\in U)+\Pr(v\in\hat D\cap W)+\lambda_T(s,0,0)\le
  \begin{cases}\cthree+12\,\eps_h^{1/3}&\text{if }d=3,\\
  0.2627+12\,\eps_h^{1/3}&\text{if }d=4.\end{cases}
\]
All degrees of $H$ lie in $\{3,4\}$, so \cref{lem:charging}, with the coefficients
$\cthree$ and $0.2627$ for the degrees $3$ and $4$ and with $e=12\,\eps_h^{1/3}$, gives
$\E\hat L_0\le\beta(H)+12\,\eps_h^{1/3}N$. Up to null events, the events
$\{v\in W\}$ and $\{v\in\hat D\cap U\}$ are the events $\{v\in U\}$ and
$\{v\in\hat D\cap W\}$ for the labeling $-Y$, which has the same distribution; so the same
bound holds for $\E\hat L_0'$.
\end{proof}

In the cubic setting, \cref{lem:alg-means}\ref{it:means:cubic} gives the analogous
per-vertex bounds, with $\kappa(r_h)$ in place of $\cthree+\eps_{\mathrm{sw}}(h)$.

\subsubsection{The width of the median-switched layout}

\begin{proposition}[median-switched sweep]\label{prop:alg-switched}
Let $n\ge n_2$ and $\theta\ge0$, and let $H$ be a graph with $N\le n$ vertices in the
cubic or the switched setting. If every count of the median-switched layout is at most
its mean plus $\theta$, then this layout is a path decomposition of $H$ of width at
most $\beta(H)+\rho_{\mathrm{med}}(n)\,N+2\theta+1$.
\end{proposition}

\begin{proof}
We first bound the means of the counts, then show that the layout is the path
decomposition of \cref{lem:median-layout-bags} for an injective labeling that induces
the order of $(\tilde Y_v,v)$, and finally check the hypotheses of that lemma.

\emph{Means.} In the cubic setting the means are bounded in \cref{sec:alg-cubic},
with $N$ in place of $n$ in the bounds and $\omega_3\le\omega_4$; the argument does not
use $N=n$. In the switched setting we argue in the same way, with
\cref{lem:alg-switched-means} in place of \cref{lem:alg-means}\ref{it:means:cubic}:
\begin{itemize}
  \item for $j<2K$, \cref{cor:alg-transfer}\ref{it:transfer:in} at the threshold
  $\tau_j$, for which $\tau_j+\xi<0$ by~\ref{it:alg-N1}, gives
  $\E X^{\mathrm{in}}_j\le\E|\infr S_{\tau_j+\xi}|+\omega_4N$, since every degree is at
  most $4$ and $\omega_d$ is increasing in $d$;
  \item for $j>2K$, similarly \cref{cor:alg-transfer}\ref{it:transfer:out} gives
  $\E X^{\mathrm{out}}_j\le\E|\outfr S_{\tau_j-\xi}|+\omega_4N$ with $\tau_j-\xi>0$;
  \item \cref{cor:alg-transfer}\ref{it:transfer:sign}, applied to the indicator of
  $v\in U\cup(\hat D\cap W)$ as a function of the set of vertices of $N[v]$ with
  negative label (for the ideal labels, $Y_u<0$ and $Y_u\le0$ differ by a null
  event), gives $\E\hat L_0\le\E\hat L_0^{\mathrm{id}}+\omega_4N$, where
  $\hat L_0^{\mathrm{id}}$ is $\hat L_0$ for the ideal labels; similarly for
  $\hat L_0'$;
  \item the windows and tails are bounded by \cref{cor:alg-transfer}\ref{it:transfer:window}
  and~\ref{it:alg-N2}: $\E X^{\mathrm{win}}_j\le(\varphi(0)/K+2\xi\varphi(0)+p_t)N$ and
  $\E X^{\mathrm{lo}},\E X^{\mathrm{hi}}<0.023N$.
\end{itemize}
With \cref{lem:alg-switched-means} (switched setting) and
\cref{lem:alg-means}\ref{it:means:cubic} (cubic setting), in both settings
\[
  \E X^{\mathrm{in}}_j,\ \E X^{\mathrm{out}}_j,\ \E\hat L_0,\ \E\hat L_0'
  \le\beta(H)+(\eps_{\mathrm{med}}(h)+\omega_4)\,N .
\]

\emph{An injective labeling.} Let $\nu$ be the largest name of a vertex of $H$, let
$\Sigma$ be the finite set $\{\tilde Y_v:v\in V(H)\}\cup\{0\}\cup\{\tau_j:0\le j\le4K\}$,
let $\vartheta>0$ be so small that $\vartheta\nu$ is less than every positive
difference of two elements of $\Sigma$, and put $y_v:=\tilde Y_v+\vartheta v$. Exactly
as in \cref{sec:alg-cubic}, with $\nu$ in place of $n$, the labeling $y$ is injective,
induces the order of $(\tilde Y_v,v)$, and has sublevel sets $S_\tau=\tilde S_\tau$
for every $\tau\in\Sigma$. In particular $U$, $W$ and $\hat D$ are the same for $y$
as for the discrete labels. By \cref{lem:median-layout-bags} in the cubic setting, and
by the same lemma with $\hat D$ in place of $D$ in the switched setting (which is
established in the proof of \cref{lem:median-layout4}), the layout is a path
decomposition of $H$ of width at most $w$ whenever $w$ satisfies
\textup{(i)}--\textup{(iii)} of \cref{lem:median-layout} for $y$, with $\hat L_0$ and
$\hat L_0'$ in~\textup{(iii)}; the component order satisfies
\eqref{eq:cubic-connector} by \cref{lem:connector-linear}.

\emph{The hypotheses.} Put $w:=\beta(H)+\rho_{\mathrm{med}}(n)N+2\theta+1$, and assume
that every count is at most its mean plus $\theta$. For a threshold $\tau<0$ with
$\tau\ge-2$, let $\tau\in[\tau_j,\tau_{j+1})$ with $j<2K$. The argument of
\cref{sec:alg-cubic} gives $|\infr S_\tau|\le X^{\mathrm{in}}_j+X^{\mathrm{win}}_j$,
which is at most
\[
  \beta(H)+(\eps_{\mathrm{med}}(h)+\omega_4)N+\bigl(\varphi(0)/K+2\xi\varphi(0)+p_t\bigr)N
  +2\theta=w-1 .
\]
For $0<\tau<2$, likewise $|\outfr S_\tau|\le X^{\mathrm{out}}_{j+1}+X^{\mathrm{win}}_j\le
w-1$, where $\tau\in[\tau_j,\tau_{j+1})$ with $2K\le j<4K$. For $\tau<-2$ and
$\tau\ge2$, the frontiers are bounded by the tails, which are at most
$0.023N+\theta\le\cthree N+\theta\le\beta(H)+\theta\le w-1$.
Finally, $\max(\hat L_0,\hat L_0')+1\le\beta(H)+(\eps_{\mathrm{med}}(h)+\omega_4)N
+\theta+1\le w$. So \textup{(i)}--\textup{(iii)} hold, and the width is at most $w$.
\end{proof}

For $H=G$ in the cubic setting, \cref{prop:alg-switched} is the width bound proved in
\cref{sec:alg-cubic}, with $\omega_4$ in place of $\omega_3$.

\subsubsection{The switched case}

\paragraph{The algorithms.} Let $G$ be a graph with $\Delta(G)\le4$ and $n\ge3$
vertices (graphs with $n\le2$ are decomposed by one bag). The randomized, Las Vegas
and deterministic algorithms are those of \cref{sec:alg-plain}, with the following
changes. In step~1, reduction operations are applied to $G$ until none applies; the
final graph $H$ is empty or has $\delta(H)\ge3$ by~\ref{red:max}, and
$\Delta(H)\le4$ by~\ref{red:deg}, so it is in the switched setting. In step~2 the
coefficients of $H$ use the weights $q(x)=d_H(x)-1$. In step~3 the algorithms build the
median-switched layout of $H$ and use its $8K+4$ counts. Step~4 applies
\cref{lem:undo} as in the mixed cases.

\begin{proof}[Proof of \cref{thm:algo-precise} in the switched case]
Let $n\ge n_2$; for $n<n_2$ the claims are trivial (\cref{sec:alg-model}). The
coefficients $(0,0,0,\cthree,0.2627)$ are nonnegative and nondecreasing in the
degree, and reductions remove vertices and increase no degree by~\ref{red:deg}, so
$\beta(H)\le\beta(G)$. Put
\[
  \rho(n):=\rho_{\mathrm{med}}(n)\,n+2\max\{\theta_{\mathrm D}(n),\theta_{\mathrm R}(n)\}
  +3\log_2n+2\qquad(n\ge n_2),
\]
with $\theta_{\mathrm D}(n)=\sqrt{8(8K+4)n}\,t^{1/4}$ and
$\theta_{\mathrm R}(n)=n^{3/4}$, and $\rho(n):=n$ for $n<n_2$; we replace $\rho$ by
its running maximum~\eqref{eq:running-max}. Then $\rho(n)=o(n)$, because
$\rho_{\mathrm{med}}(n)\to0$. The three algorithms ensure that every count is at most
its mean plus $\theta$, exactly as in \cref{sec:alg-plain,sec:alg-cubic}:
deterministically with $\theta=\theta_{\mathrm D}(n)$ by
\cref{lem:alg-derand}\ref{it:derand:det}; in the randomized algorithm with
$\theta=n^{3/4}$, except with probability at most
$\pi(n):=8(8K+4)t^{1/2}n^{-1/2}+1/\lceil\log_2n\rceil$; and in the Las Vegas algorithm
with $\theta=n^{3/4}$, which accepts a sample with probability at least $\frac12$ by
\cref{lem:alg-derand}\ref{it:derand:rand} and~\ref{it:alg-N3}. In each case
\cref{prop:alg-switched}, applied with $N=|V(H)|\le n$, bounds the width of the layout
of $H$ by $\beta(G)+\rho_{\mathrm{med}}(n)n+2\theta+1$, and \cref{lem:undo} adds at
most $3\log_2n+1$. So the width is at most $\beta(G)+\rho(n)$. The running times are
those of the cubic case (\cref{sec:alg-time}): the reductions take $O(n)$
operations, since degrees never exceed $4$, and the coefficients are computed as in
the mixed cases.
\end{proof}

\subsubsection{The sweep components of the bootstrap cases \texorpdfstring{$\seqC$}{C} and \texorpdfstring{$\seqM$}{M}}

In the bootstrap case of a seed $X$ (\cref{sec:alg-general}), the deletion process
stops at sweep components $J$ with $\delta(J)\ge3$ and $\Delta(J)\le k$, where $k=3$
for $X=\seqC$ and $k=4$ for $X=\seqM$, and decomposes them with the parameters of the
number $n_{\ge1}\ge|V(J)|$ of non-isolated vertices of the input graph.

\begin{corollary}[base layouts for $\seqC$ and $\seqM$]\label{cor:alg-switched-base}
Let $X\in\{\seqC,\seqM\}$ with base degree $k$, let $n\ge n_2$ and $\theta\ge0$, and let
$J$ be a graph with $\delta(J)\ge3$, $\Delta(J)\le k$ and $N_J\le n$ vertices, labeled
in the cubic setting if $X=\seqC$ and in the switched setting if $X=\seqM$, with the
parameters of $n$. If every count of the median-switched layout of $J$ is at most its
mean plus $\theta$, then the layout is a path decomposition of $J$ of width at most
$\sum_{d=3}^kb_dn_d(J)+\rho_{\mathrm{med}}(n)N_J+2\theta+1$, where $(b_3,\dots,b_k)=X$.
\end{corollary}

\begin{proof}
For $X=\seqC$ the graph $J$ is $3$-regular, so $\beta(J)=\cthree N_J=\cthree n_3(J)$.
For $X=\seqM$, $\beta(J)=\cthree n_3(J)+0.2627\,n_4(J)$. In both cases
\cref{prop:alg-switched} gives the claim.
\end{proof}
{}
\subsection{Proofs for Part II}\label{app:part2}

This subsection proves %
\cref{lem:n3} (\cref{app:n3}), the statements on bucketing
(\cref{app:bucketing}) and \cref{thm:connectivity} (\cref{app:connectivity}). The
analyses for \cref{sec:apps-smc,sec:apps-ds} are in \cref{app:is,app:ds}.

\subsubsection{Pathwidth in terms of the vertices of degree three}\label{app:n3}

\lemnthree*

\begin{proof}
  Let $\rho_1\colon\N\to\R$ be the larger of the remainder of \cref{thm:cubic} and the
  remainder of the deterministic algorithm of \cref{thm:algo} for that bound; replacing
  it by its running maximum~\eqref{eq:running-max}, we take $\rho_1$ nonnegative and
  nondecreasing. We prove the lemma with
  \[
    \rho(k):=\rho_1(k)+\log_2(2k+2)+6\qquad(k\in\N).
  \]
  Write $n_3:=n_3(H)$. The proof has four steps: we remove the trees outside the $2$-core
  of $H$, suppress the vertices of degree~$2$ of the core and decompose the
  simple graph underlying the resulting cubic multigraph by \cref{thm:cubic}, and then put back first the suppressed
  paths and then the trees.

  \emph{Step 1: the $2$-core.} Repeatedly delete a vertex of degree at most $1$ until
  none is left, and call the resulting graph $H'$ (possibly empty). No degree
  increases, so $n_3(H')\le n_3$. Every component $T$ of $H-V(H')$ is a tree with at
  most one edge to $H'$. Otherwise $T$ together with $H'$ contains a cycle through a
  vertex of $T$, or a path between two vertices of $H'$ whose interior lies in $T$; the
  first vertex of this cycle or path to be deleted had two neighbors at the time of its
  deletion, a contradiction.

  \emph{Step 2: the core.} Every vertex of $H'$ has degree $2$ or $3$. Components of
  $H'$ that are cycles have pathwidth at most $2$. In every other component, every
  maximal path whose interior vertices have degree $2$ joins two vertices of degree $3$
  (possibly the same one). Suppressing all vertices of degree $2$ of these components
  therefore gives a multigraph $M$, possibly with loops and parallel edges, on the
  $n_3(H')$ vertices of degree $3$ of $H'$. Its underlying simple graph $G_M$ has
  maximum degree at most $3$ and $n_3(H')$ vertices. By \cref{thm:cubic} and
  \cref{thm:algo} we can compute a path decomposition of $G_M$ of width at most
  $\cthree n_3(H')+\rho_1(n_3(H'))\le\cthree n_3+\rho_1(n_3)$.

  \emph{Step 3: restoring the subdivisions.} An edge $e$ of $M$ with ends $u,w$
  (possibly $u=w$) stands for a path $u\,x_1\cdots x_q\,w$ of $H'$ with $q\ge0$. If
  $q\ge1$, choose a bag $B$ of the current decomposition that contains $u$ and $w$ (it
  exists, because $uw\in E(G_M)$ or $u=w$). Directly after $B$ insert the bags
  \[
    B\cup\{x_1\},\ B\cup\{x_1,x_2\},\ B\cup\{x_2,x_3\},\ \dots,\ B\cup\{x_{q-1},x_q\}.
  \]
  These cover the edges $ux_1$, $x_ix_{i+1}$ and $x_qw$. Each $x_i$ occurs in
  consecutive bags, and every vertex of $B$ still occurs in an interval of bags, since
  the inserted bags contain $B$. Paths that use the same bag $B$ are inserted one after
  another. The width grows by at most $2$. Finally append, for each cycle component of
  $H'$, a decomposition of width $2$. The result is a decomposition of $H'$ of width at
  most $\cthree n_3+\rho_1(n_3)+2$.

  \emph{Step 4: the trees.} Let $T$ be a component of $H-V(H')$ with $N$ vertices, and
  let $t\le n_3$ be the number of vertices of degree $3$ in $T$. If $N\ge2$, then $T$
  has exactly $t+2$ leaves (count $\sum_{v}(d_T(v)-2)=-2$), and suppressing its
  vertices of degree $2$ gives a tree $T^\circ$ with at most $2t+2$ vertices; if $N=1$,
  put $T^\circ=T$. A tree on $N^\circ\ge1$ vertices has pathwidth at most
  $\log_2N^\circ$: by induction, put a centroid into every bag of a concatenation of
  decompositions of the components of the tree minus the centroid, each of which has at
  most $N^\circ/2$ vertices. Restoring the subdivisions of $T^\circ$ as in Step~3 adds
  at most $2$, so $\pw(T)\le\log_2(2t+2)+2$. If $T$ has an edge to a vertex $v$ of
  $H'$, take a bag $B\ni v$ of the decomposition of $H'$ and insert directly after it
  the bags $B\cup Y$, for the bags $Y$ of a decomposition of $T$ in order; this covers
  the edge between $T$ and $v$, and the width grows by at most $\pw(T)+1$. Trees
  attached at the same bag are inserted one after another, so the growth is the maximum,
  not the sum, over the trees. Trees without an edge to $H'$ are appended as separate
  segments.

  Combining the steps,
  \[
    \pw(H)\le\cthree n_3+\rho_1(n_3)+2+\log_2(2n_3+2)+3\le\cthree n_3+\rho(n_3),
  \]
  and the bound also holds if $H'$ is empty. Every step takes polynomial time; Step~2
  computes the decomposition of $G_M$ by the deterministic algorithm of \cref{thm:algo}.
  The function
  $\rho$ is nondecreasing and $\rho(k)=o(k)$.
\end{proof}

\subsubsection{Bucketing}\label{app:bucketing}

We use $p(s)$ and $P_k$ as defined before \cref{lem:cyclecover}.

\lemcyclecover*

\begin{proof}
  Every vertex lies on at most two edges of $P_k$, so $d_H(v)\le d_C(v)+d_A(v)\le4$.
  Let $n_d$ be the number of vertices of degree $d$ in $H$, which has $k$ vertices; since
  $\Delta(H)\le4$, $n_5=0$. \Cref{thm:deg5} with the triple $\tripR$ and \cref{thm:algo} give a
  path decomposition of $H$ of width at most $0.1455\,n_3+0.2530\,n_4+\rho(k)$, where
  $\rho$ is the remainder of this algorithm. A vertex of degree $d\ge3$ in $H$ has
  $d_A(v)\ge d-2$, so $n_3+2n_4\le\sum_vd_A(v)=2s$; also $n_3+n_4\le k$. Hence
  \begin{align*}
    0.1455\,n_3+0.2530\,n_4&=0.1455\,(n_3+2n_4)-0.038\,n_4\le0.291\,s
    \quad\text{and}\\
    0.1455\,n_3+0.2530\,n_4&=0.038\,(n_3+n_4)+0.1075\,(n_3+2n_4)\le
    0.038\,k+0.215\,s. \qedhere
  \end{align*}
\end{proof}

\lemtradeoff*

\begin{proof}
  Here $1-\alpha=\frac{215}{1215}$ and $\alpha=\frac{1000}{1215}$. Since $\alpha\ge0$ and
  $p(s)\le0.038\,k+0.215\,s$,
  \[
    (1-\alpha)(k-s)+\alpha\,p(s)\le\frac{215(k-s)+38k+215s}{1215}=\frac{253k}{1215}.
  \]
  For the second statement, let $s^*:=\frac{962}{1215}k$. Then $k-s^*=\frac{253}{1215}k$.
  The two terms in the definition of $p$ are equal at $s=k/2$, and the second is the
  smaller one for $s\ge k/2$; since $s^*\ge k/2$,
  $p(s^*)=0.038\,k+0.215\,s^*=\frac{253}{1215}k$. So at $s=s^*$ the expression equals
  $\frac{253}{1215}k$ for every $\alpha\in[0,1]$.
\end{proof}

The three bounds of \cref{thm:deg4} and the bound of \cref{thm:switched4} also give
upper bounds on $\pw(H)$ in \cref{lem:cyclecover}: $F_4(n_3+n_4)$,
$F_3n_3+0.2575\,n_4$, $\frac5{32}n_3+\frac14n_4$ and $\cthree n_3+0.2627\,n_4$, up to
$o(k)$. As long as only the constraints $n_3+2n_4\le2s$ and $n_3+n_4\le k$ of the proof
of \cref{lem:cyclecover} are used, these four bounds do not lower the exponent
$\frac{253}{1215}k$, even together with the bound of $\tripR$. Indeed, let
$s^*=\frac{962}{1215}k$ as in the proof above, and take $n_3=\frac{506}{1215}k$ and
$n_4=\frac{709}{1215}k$, which satisfy both constraints at $s=s^*$ with equality. At
these values $0.1455\,n_3+0.2530\,n_4=\frac{253}{1215}k=k-s^*<0.20824k$, while the four
other bounds are at least $0.2457k$, $0.2086k$, $0.2109k$ and $0.2114k$, by the lower
bounds on $F_4$ and $F_3$ of \cref{prop:F3F4} and by $\cthree>0.1397$. So the largest
value, over all reals $n_3,n_4\ge0$ that satisfy both constraints at $s=s^*$, of the
minimum of the five bounds is $\frac{253}{1215}k=p(s^*)$: it is at least this value at
the point above, and at most the bound of $\tripR$, which is at most $p(s^*)$ by the
proof of \cref{lem:cyclecover}. With this value in place of $p(s^*)$, the expression of
\cref{lem:tradeoff} at $s=s^*$ still equals $\frac{253}{1215}k$ for every
$\alpha\in[0,1]$.

\thmperm*

\begin{proof}
  For~\ref{it:perm-ppm} we follow Berendsohn, Kozma and Marx: we write the problem as a
  binary constraint satisfaction problem, split the text positions into intervals
  (strips), guess which entries of the pattern are leftmost in their strip and the strips
  of these entries, and solve each restricted
  instance by the dynamic program of \cref{lem:apps-dp}; \cref{lem:cyclecover,lem:tradeoff}
  bound the total time. For~\ref{it:perm-kopt} we substitute \cref{lem:cyclecover,lem:tradeoff} into
  the exponent of Cygan, Kowalik and Socała, and \cref{rem:edges-m} into the algorithm of
  Bonnet, Iwata, Jansen and Kowalik.

  \ref{it:perm-ppm} \emph{The constraint satisfaction formulation.} Following
  Berendsohn, Kozma and Marx \cite[\S3]{BerendsohnKM2021}, take variables $x_1,\dots,x_k$
  with domain $[n]$; the value of $x_i$ is the text position $f(i)$. Impose the
  \emph{index constraints} $x_i<x_{i+1}$ for $i\in[k-1]$, and the \emph{value
  constraints} $\tau(x_a)<\tau(x_b)$ for $a=\pi^{-1}(v)$, $b=\pi^{-1}(v+1)$ and
  $v\in[k-1]$. The solutions are exactly the occurrences: the index constraints say that
  $f$ is strictly increasing, and the value constraints say that
  $\tau\circ f\circ\pi^{-1}$ is strictly increasing, which is equivalent to the order
  condition. The constraint graph is the union of the path $P_k$ (index constraints) and
  the path $Q$ on $[k]$ with edges $\{\pi^{-1}(v),\pi^{-1}(v+1)\}$ (value constraints).

  \emph{Bucketing.} Let $\alpha=\frac{200}{243}$, $q:=\lceil n^\alpha\rceil$, and split
  $[n]$ into the $J:=\lceil n/q\rceil\le n^{1-\alpha}+1\le2n^{1-\alpha}$ intervals
  (\emph{strips}) $I_j:=\{(j-1)q+1,\dots,jq\}\cap[n]$, each of size at most $q\le
  2n^\alpha$. (Berendsohn, Kozma and Marx use $n^{1/4}$ strips of width $n^{3/4}$;
  nothing else in their argument depends on the width.) For an occurrence $f$ let
  $\sigma(i)$ be the index of the strip containing $f(i)$; since $f$ is increasing,
  $\sigma$ is nondecreasing. Let
  \[
    X(f):=\{1\}\cup\{i\in\{2,\dots,k\}:\sigma(i-1)<\sigma(i)\},
  \]
  the entries that are leftmost in their strip. For every $i$, $\sigma(i)=\sigma(i^*)$,
  where $i^*$ is the largest element of $X(f)$ with $i^*\le i$, because $\sigma$ does not
  increase between consecutive elements of $X(f)$.

  The algorithm enumerates all \emph{guesses}: sets $X\subseteq[k]$ with $1\in X$ and
  strictly increasing maps $\sigma_X\colon X\to[J]$. For a guess it solves the
  \emph{restricted instance}: the domain of $x_i$ is $I_{\sigma_X(i^*)}$, with $i^*$ the
  largest element of $X$ that is at most $i$; all value constraints are kept; the index
  constraint between $i$ and $i+1$ is kept if $i+1\notin X$ and dropped if $i+1\in X$.
  We claim that the solutions of the restricted instance are exactly the occurrences
  $f$ with $X(f)=X$ and $\sigma|_X=\sigma_X$. Let $f$ be a solution. A dropped
  constraint holds anyway: if $i+1\in X$, then $f(i)$ lies in the strip
  $\sigma_X(i^*)<\sigma_X(i+1)$, so $f(i)<f(i+1)$. Hence $f$ is an occurrence. Its strips
  are $\sigma(i)=\sigma_X(i^*)$, so for $i\in X\setminus\{1\}$ the strip strictly
  increases at $i$, and for $i\notin X$ the entries $i-1$ and $i$ share a strip; thus
  $X(f)=X$ and $\sigma|_X=\sigma_X$. Conversely, an occurrence with $X(f)=X$ and
  $\sigma|_X=\sigma_X$ satisfies all constraints, and $f(i)$ lies in the strip
  $\sigma(i)=\sigma(i^*)=\sigma_X(i^*)$. So every occurrence is a solution of exactly one
  restricted instance. For the decision version we test each guess; for the counting
  version we add the numbers of solutions over all guesses.

  \emph{Running time.} The constraint graph of a restricted instance is the union of $Q$
  and the set $A_X:=\{\{i,i+1\}:i+1\notin X\}$ of $s:=k-|X|$ edges of $P_k$ (the
  $|X|-1$ elements of $X\setminus\{1\}$ each drop one index constraint). $Q$ is a path, so
  \cref{lem:cyclecover} with $C=Q$ gives, in polynomial time, a path decomposition of
  width $w\le p(s)+\rho_1(k)$, where $\rho_1$ is the remainder there. Every domain has
  size at most $2n^\alpha$. We may assume $w\le k-1$ (otherwise use a single bag), so
  \cref{lem:apps-dp}\ref{it:apps-dp-csp} over the counting
  semiring $(\N,+,\times)$, with scores in $\{0,1\}$ given by the constraints, takes
  $(2n^\alpha)^{w+1}\poly(n)$ steps; the restricted instance is built in polynomial time.
  (This is \cite[Lemmas~2 and~4]{BerendsohnKM2021}.) There are at most $J^{|X|}\le
  (2n^{1-\alpha})^{|X|}$ maps $\sigma_X$ for each $X$, and at most $2^k$ sets $X$. Since
  $|X|\le k$ and $w+1\le k$, the total time is at most
  \[
    \sum_{X}(2n^{1-\alpha})^{|X|}(2n^{\alpha})^{p(k-|X|)+\rho_1(k)+1}\poly(n)
    \le2^{3k}\,
    n^{\max_{0\le s\le k}[(1-\alpha)(k-s)+\alpha p(s)]+\rho_1(k)+1+O(1)} ,
  \]
  and \cref{lem:tradeoff} bounds the maximum by $\frac{253}{1215}k$. With $\rho(k):=
  \rho_1(k)+c$, for $\rho_1$ taken nondecreasing by~\eqref{eq:running-max} and a
  suitable constant $c$, this is $2^{O(k)}n^{\frac{253}{1215}k+\rho(k)}$. Since $k\le n$, also
  $2^{O(k)}=n^{O(k/\log_2 n)}\le n^{O(k/\log_2 k)}=n^{o(k)}$, so the bound can be
  written $n^{\frac{253}{1215}k+o(k)}$, as Berendsohn, Kozma and Marx do for $n^{k/4}$.

  \ref{it:perm-kopt} \emph{General graphs.} Cygan, Kowalik and Socała
  \cite[Lemma~10]{CyganKS2019} solve $k$-opt
  Optimization in time $2^{O(k\log k)}n^{c(k)}$, where
  \[
    c(k)=\max_{M}\ \min_{\alpha\in[0,1]}\ \max_{A\subseteq P_k}\Bigl[(1-\alpha)(k-|A|)
    +\alpha\bigl(\tw(I_M\cup A)+1\bigr)\Bigr],
  \]
  the outer maximum ranges over the valid connection patterns $M$, and $I_M$ is a graph
  on $[k]$ associated with $M$. Every component of $I_M$ is a cycle or a single edge
  \cite[Proposition~3]{CyganKS2019}, and, as in the proof of
  \cite[Lemma~8]{CyganKS2019}, doubling every single-edge component gives a multigraph $I_M^*\supseteq I_M$ on $[k]$ in which every
  vertex has degree~$2$. Hence $I_M\cup A$ is a subgraph of the simple graph underlying
  $I_M^*\cup A$, and \cref{lem:cyclecover} with $C=I_M^*$ gives $\tw(I_M\cup A)\le
  p(|A|)+\rho_1(k)$. Taking $\alpha=\frac{200}{243}$ for every $M$, \cref{lem:tradeoff}
  gives $c(k)\le\frac{253}{1215}k+\rho_1(k)+1$. Put $\eps_k:=(\rho_1(k)+1)/k\to0$; for fixed
  $k$ the factor $2^{O(k\log k)}$ is a constant, so the time is $O(n^{(253/1215+\eps_k)k})$.

  \emph{Bounded degree.} Bonnet, Iwata, Jansen and Kowalik \cite[Thm.~3]{BonnetIJK2019}
  solve $k$-opt Optimization on graphs of maximum degree at most a constant by running,
  for each of at most $(2k)!$ patterns, a dynamic program in time $O(n^{\tw(C_M)+1})$ for
  fixed $k$ and maximum degree, where $C_M$ is a simple graph that is a subgraph of an
  even multigraph with $k$ edges. So $C_M$ has at most $k$ edges, and \cref{rem:edges-m}
  gives $\tw(C_M)\le\pw(C_M)\le\frac7{48}k+o(k)$. The time is
  $O(n^{(7/48+\eps'_k)k})$ with $\eps'_k\to0$. The larger of $\eps_k$ and $\eps'_k$
  serves for both bounds.
\end{proof}

\subsubsection{Connectivity problems}\label{app:connectivity}

\thmconnectivity*

\begin{proof}
  For a graph of maximum degree $4$, \cref{thm:algo-precise}\ref{it:algo:D} computes a
  path decomposition of width at most $F_4n+o(n)$ (the bound of \cref{thm:deg4}(a)) in
  polynomial time, and for maximum degree $3$ one of width at most $\cthree n+o(n)$ (the
  bound of \cref{thm:cubic}); by \cref{thm:main} and \cref{thm:algo}, it computes one of
  width at most $\frac{11}{32}n+o(n)$ for maximum degree $5$ and $\frac7{16}n+o(n)$ for
  maximum degree $6$, since $a_d\le a_5$ for $d\le5$ and $a_d\le a_6$ for $d\le6$. A
  path decomposition is a tree decomposition of the same width. For maximum degree $3$,
  \cref{cor:alg-cutwidth} computes an ordering of cutwidth at most $\cthree n+o(n)$ in
  polynomial time. For a graph with $m$ edges and maximum degree $4$ or $5$,
  \cref{thm:alg-cut-sweep} computes, in deterministic polynomial time, an ordering
  of cutwidth at most $\frac m6+o(n)\le\frac n3+o(n)$ or $\gamma_5m+o(n)\le\frac52\gamma_5
  n+o(n)$, respectively. For a graph with $m$ edges, \cref{thm:algo-precise} (general case)
  gives a path decomposition of width at most $\frac7{48}m+o(m)$ in polynomial time. We run
  the following algorithms on these decompositions and orderings.

  \ref{it:conn-3tw} Bojikian, Chekan, Hegerfeld and Kratsch solve Connected Vertex Cover
  and Feedback Vertex Set by Monte Carlo algorithms with one-sided error in time
  $2^{\operatorname{ctw}}n^{O(1)}$ for a given ordering of cutwidth
  $\operatorname{ctw}$~\cite[Theorems~12 and~43]{BojikianCHK2022arXiv}, on all graphs.
  We use them for maximum degree $4$ and $5$; the numerical bases follow from
  $2^{1/3}<1.25993$ and $2^{5\gamma_5/2}<1.42611$, where $\gamma_5=\frac1\pi\arccos\frac45$. Cygan
  et al.~\cite[Theorem~1.2]{CyganNPPRW2011arXiv} solve the three problems in time
  $3^{\tw}n^{O(1)}$, which we use for Steiner Tree and for Feedback Vertex Set at maximum
  degree $6$, and Bodlaender et al.\ solve Feedback Vertex Set
  deterministically in time $5^{\pw}n^{O(1)}$~\cite[Theorem~4.5]{BodlaenderCKN2015}.

  \ref{it:conn-cds} Cygan et al.\ solve these problems in time
  $4^{\tw}n^{O(1)}$~\cite[Theorem~1.2]{CyganNPPRW2011arXiv}. For maximum degree $3$,
  Bojikian et al.\ solve Connected Dominating Set in time
  $3^{\operatorname{ctw}}n^{O(1)}$ for a given ordering of cutwidth
  $\operatorname{ctw}$~\cite[Theorem~23]{BojikianCHK2022arXiv}. Maximum Leaf Spanning
  Tree has the bounds of Connected Dominating Set because a connected graph with $n\ge3$
  vertices has a spanning tree with at least $\ell$ leaves if and only if it has a
  connected dominating set of at most $n-\ell$ vertices. Indeed, the vertices that are
  not leaves of a spanning tree form a connected dominating set, and a spanning tree of
  the subgraph induced by a connected dominating set extends to a spanning tree in which
  every other vertex is a leaf. Finally, a spanning tree has $n-\ell$ internal vertices
  if it has $\ell$ leaves, so Minimum Leaf Spanning Tree and Maximum Internal Spanning
  Tree have the same optimal solutions.

  \ref{it:conn-ham} Bojikian, Chekan and Kratsch decide Hamiltonicity by a Monte Carlo
  algorithm with one-sided error in time $(1+\sqrt2)^{\operatorname{ctw}}n^{O(1)}$ for a
  given ordering of cutwidth $\operatorname{ctw}$, on all
  graphs~\cite[Theorem~47]{BojikianCK2025arXiv}. We use it for maximum degree $3$ and $4$,
  where $(1+\sqrt2)^{1/3}<1.34151$. (The proof of
    \cite[Corollary~4.5]{CyganKN2012arXiv} gives the same base for a path decomposition of a
  graph of maximum degree~$3$.) For maximum degree $5$ we use the algorithm
  of~\cite[Theorem~4.3]{CyganKN2012arXiv}; there the ordering of \cref{thm:cut-sweep-p}
  would give only $(1+\sqrt2)^{5\gamma_5n/2}>1.570^n$. For Longest Path, Cygan, Kratsch and Nederlof
  decide whether a graph has a path of length $k$ in time
  $(2+\sqrt2)^{\pw}n^{O(1)}$ for a given path decomposition of width
  $\pw$~\cite[Theorem~4.6]{CyganKN2012arXiv}, with an algorithm that extends the Monte Carlo
  algorithm of their Theorem~4.3 by a counter; we run it for each $k$.

  \ref{it:conn-tsp} For maximum degree $4$, $5$ and $6$ we use the algorithm
  of~\cite[Theorem~3.18]{BodlaenderCKN2015}. Its exponent $\omega$ enters through
  minimum weight row bases over $\mathrm{GF}(2)$~\cite[Lemma~3.15]{BodlaenderCKN2015},
  which reduce to products of matrices over $\mathrm{GF}(2)$. The bound
  $\omega\le\frac94$ is proved over the complex numbers in~\cite{OpenAI2026omega}, and
  Amanov, van den Berg and Zuiddam prove it over every field~\cite[Theorem~1.1]{AmanovBZ2026},
  in particular over $\mathrm{GF}(2)$. For cubic graphs, Bodlaender et al.\
  obtain their bound $1.2186^nn^{O(1)}$~\cite[Corollary~3.19]{BodlaenderCKN2015}
  by running that algorithm on a path decomposition with fewer states. By an
  observation of Cut\&Count~\cite{CyganNPPRW2011arXiv}, they discard the states in which
  the degree of a vertex in a partial solution is incompatible with the number of its
  edges introduced so far. In a graph of maximum degree $3$, a vertex of a bag that has at most one
  introduced edge then has degree $0$ or $1$, and one with two introduced edges has
  degree $1$ or $2$. So besides degree $1$ each vertex has one possible degree instead of
  two, and the count of states in the proof
  of~\cite[Theorem~3.18]{BodlaenderCKN2015} gives the base $1+2^{\omega/2}$ instead
  of $2+2^{\omega/2}$. With the width $\frac n6+o(n)$ this is
  $(1+2^{\omega/2})^{n/6+o(n)}=O(1.2187^n)$ with $\omega<2.3727$, which they state as
  $1.2186^n$, and $O(1.2128^n)$ with $\omega\le\frac94$. On a path
  decomposition of width $\cthree n+o(n)$, the same algorithm runs in time
  $(1+2^{\omega/2})^{\cthree n+o(n)}$.

  \ref{it:conn-directed} Cygan et al.\ solve these problems in time $6^{\tw}n^{O(1)}$
  for a tree decomposition of width $\tw$ of the underlying undirected
  graph~\cite[Theorem~1.2]{CyganNPPRW2011arXiv}.

  \ref{it:conn-edges} We run the algorithms of~\cite[Theorem~4.3]{CyganKN2012arXiv} and
  \cite[Theorem~3.18]{BodlaenderCKN2015} on a path decomposition of width
  $\frac7{48}m+o(m)$. Numerically $(2+\sqrt2)^{7/48}<1.19612$ and
  $(2+2^{\omega/2})^{7/48}<1.23198$ for $\omega\le\frac94$, and the bounds in $dn$ are their square roots.

  \ref{it:conn-param} \emph{Co-Path Packing.} Liu and Xiao~\cite{LiuXiao2026} branch,
  with branching factor $2.8192$, until the remaining instance $(G,k')$, with $k'\le k$,
  is a \emph{proper} graph~\cite[Section~3.1]{LiuXiao2026}: its maximum degree is at
  most $4$, and if it has a solution of size at most $k'$, then $|V(G)|\le100k'$ and
  $\frac{n_3}6+\frac{n_4}3\le\frac{2k'}3$~\cite[Lemma~2]{LiuXiao2026}, where $n_d$ is
  the number of vertices of degree $d$. On a proper graph, if these two conditions hold,
  they run a Monte Carlo algorithm in time $5^{\pw}n^{O(1)}$~\cite[Theorem~4]{LiuXiao2026}
  on a path decomposition of width $\frac{2k'}3+o(k')$, which the bound of Fomin,
  Gaspers, Saurabh and Stepanov gives; otherwise they reject the instance. If these two
  conditions hold, \cref{thm:main} and \cref{thm:algo} give instead a path decomposition
  of width at most
  \[
    \tfrac5{32}n_3+\tfrac14n_4+o(n)
    =\tfrac{15}{16}\cdot\tfrac{n_3}6+\tfrac34\cdot\tfrac{n_4}3+o(n)
    \le\tfrac{15}{16}\cdot\tfrac{2k'}3+o(k')=\tfrac58k'+o(k'),
  \]
  and the dynamic program takes time $5^{5k'/8+o(k')}<2.7344^{k'}$ for large $k'$. Since
  $2.7344$ is below the branching factor $2.8192$, the whole algorithm runs in time
  $\Ostar(2.8192^k)$.

  \emph{$k$-Co-Path Set.} Sullivan and van der Poel~\cite[\S4.3, Algorithm~2]{SullivanvdP2016}
  first reduce the instance, in polynomial time, to an equivalent instance $(G',k')$ with
  $k'\le k$ and at most $6k'$ vertices (a kernel of Feng, Zhou and Wang that they cite).
  They then branch on the vertices of degree greater than~$10$ and solve each resulting
  instance by a Monte Carlo algorithm in time $4^{\tw}n^{O(1)}$ on a tree decomposition of
  width $\tw$~\cite[Theorem~4]{SullivanvdP2016}, which applies to every graph. With our
  width bound the branching is not needed. A yes-instance $(G',k')$ satisfies
  $\sum_{d\ge3}(d-2)n_d\le2k'$, since at least $d-2$ edges at each vertex of degree $d\ge3$
  are deleted~\cite[Lemma~10]{SullivanvdP2016}; we reject $(G',k')$ if this inequality
  fails. Otherwise, since $a_d\le\frac5{32}(d-2)$ for every $d\ge3$, \cref{thm:main} and
  \cref{thm:algo} give a path decomposition of $G'$ of width at most
  $\frac5{32}\cdot2k'+o(k')=\frac5{16}k'+o(k')$, in place of the bound
  $\frac{k}3+o(n)$ that their Lemma~11 gives for the instances after branching, and their
  algorithm runs on it in time $4^{5k'/16+o(k')}n^{O(1)}\le2^{5k/8+o(k)}n^{O(1)}$. Since
  $2^{5/8}<1.5423$, this is $\Ostar(1.5423^k)$.
\end{proof}
{}
\addtocontents{toc}{\protect\setcounter{tocdepth}{1}}

\section{Formal verification}\label{app:formal}
\label{rem:cubic-lean}\label{rem:deg4-lean}\label{rem:switched4-lean}%
\label{rem:main-lean}\label{rem:correlation-lean}%

Several results of Part~I, among them \cref{thm:cut-sweep}, have formal proofs in the Lean~4
proof assistant (version 4.34.0) with the Mathlib library (commit \texttt{5ed2965}).
In addition, a Lean port of the implementation of the algorithms of \cref{thm:algo},
for the cases described in \cref{sec:alg-implementation}, is proved to output path decompositions that attain the bounds of
\cref{thm:cubic,thm:deg4,thm:deg5,thm:main} and \cref{cor:edges} for suitable
labelings, under the same hypotheses. The development is part of the accompanying
code, and the driver script of \cref{sec:cert-repro} has a third mode, \emph{lean},
that rebuilds the formal proof of \cref{thm:cubic}.
\Cref{tab:formal} lists the formal proofs; the text after the table states the
hypotheses they assume, describes the formal statements and what is not formalized,
and records where the formal proofs follow other routes than the written ones.
\Cref{sec:alg-implementation} describes the implementation, its tests and its Lean
port.
No result of \cref{part:apps} or of \cref{app:part2,app:is,app:ds} is
formalized.

\paragraph{What each result rests on.} The results rest on three kinds of evidence,
and we say for each headline result which kind it needs.
\begin{enumerate}[label=\textup{(\alph*)},align=left,leftmargin=2.6em,labelwidth=1.6em,labelsep=1em,itemsep=2pt]
  \item \Cref{thm:intro-bounded}\ref{thm:intro-bounded:a}, \cref{cor:bisection} and
    the switch lemma behind them (\cref{lem:switch}) use no computation; their
    written proof uses the case $d=3$ of \cref{lem:simplex}, which is a case of
    Mulgund's theorem~\cite{Mulgund2026} and also has an elementary proof
    (\cref{rem:stars-mulgund}). \Cref{thm:cubic}, which is
    \cref{thm:intro-bounded}\ref{thm:intro-bounded:a}, has a complete formal proof in
    Lean, without hypotheses. Its written proof
    takes pages~\pageref{sec:cubic-setting}--\pageref{pg:cubic-proof-end} and lemmas
    from \cref{sec:warmup,sec:prelim,sec:labels,sec:stars,sec:plain-sweep}; the formal proof
    follows it with the differences listed under ``Routes'' below.
  \item \Cref{thm:intro-bounded}\ref{thm:intro-bounded:b} uses two inputs. The first
    is, for vertices of degree~$4$, the extremality of the regular simplex among four
    Gaussian variables, which is a case of Mulgund's theorem~\cite{Mulgund2026}
    and also follows from the four-cap theorem of Fejes
    T\'oth~\cite{FejesToth1950} (\cite[Theorem~5.6.2]{Boroczky2004}). The second is a certified enclosure of
    $F_4$ (\cref{prop:F3F4}).
  \item \Cref{thm:main} and \cref{cor:edges} are derived from the triple
    $(\frac5{32},\frac14,\frac{11}{32})$ for maximum degree~$5$ (\cref{thm:deg5}).
    This triple rests on a computer-assisted charging certificate, verified in ball
    arithmetic and exact rational arithmetic (\cref{prop:certificate}), and on the
    case of five variables of the simplex theorem. That case is Mulgund's
    theorem~\cite{Mulgund2026}, proved in a preprint with a Lean formalization that
    we compiled and checked against the inequality we use (see H1 below).
    \Cref{thm:deg4}\ref{thm:deg4:b},\ref{thm:deg4:c} are restrictions of triples of
    \cref{thm:deg5} and rest on the same certificate.
\end{enumerate}
The bootstrap from degree~$5$ to all degrees (\cref{lem:bootstrap}) is a
combinatorial induction without computation. \Cref{thm:switched4} rests on two
further computer-assisted certificates. \Cref{thm:intro-bounded}\ref{thm:intro-bounded:b},
\cref{thm:main}, \cref{cor:edges} and \cref{thm:deg5,thm:switched4} also have formal
proofs in Lean, conditional on the certified and external inputs named above.

For every entry of \cref{tab:formal}, the final Lean theorem has no hypotheses
beyond those of the written statement and those listed, and Lean's
\texttt{\#print axioms} reports only the three standard axioms \texttt{propext},
\texttt{Classical.choice} and \texttt{Quot.sound}. The development contains no
unproved steps (\texttt{sorry}), no additional axioms and no trusted native
computation. The last column of the table says what each formal statement was
checked against: the written statement it formalizes, and for the implementation
also the code it ports.

Since Mathlib has no notion of pathwidth, the development defines path
decompositions directly: a path decomposition is a nonempty finite sequence of bags
that covers every vertex and every edge, such that the bags containing any vertex
are consecutive. Graphs are Mathlib's simple graphs on the vertex set
$\{0,\dots,n-1\}$, and a bound $\pw(G)\le w$ means that $G$ has a path decomposition
whose bags have at most $w+1$ vertices. Treewidth is not formalized. Unless said
otherwise below, a formal statement is the uniform $\eps$--$N$ form of the written
one: for every real $\eps>0$ there is $N$ such that every graph of the class with
$n\ge N$ vertices satisfies $\pw(G)\le\beta(G)+\eps n$, where $\beta$ is the bound
without its remainder. All theorem names below live in the Lean namespace
\texttt{CubicPathwidth}.

\begin{table}[p]
\centering
\caption{The formal proofs. ``Axioms'' lists what \texttt{\#print axioms} reports
beyond the three standard axioms, which every entry uses. The hypotheses H1, H2,
H3, H6 and H7 are stated, with their status, after the table.}
\label{tab:formal}
\footnotesize
\setlength{\tabcolsep}{4pt}
\renewcommand{\arraystretch}{1.15}
\begin{tabular}{@{}>{\raggedright\arraybackslash}p{3.4cm}>{\raggedright\arraybackslash}p{4.4cm}>{\raggedright\arraybackslash}p{2.9cm}>{\raggedright\arraybackslash}p{1.1cm}>{\raggedright\arraybackslash}p{3.4cm}@{}}
\toprule
Result & Lean theorem & Hypotheses & Axioms & Checked against\\
\midrule
\Cref{thm:cubic}, that is \cref{thm:intro-bounded}\ref{thm:intro-bounded:a}
  & \texttt{C3.c3MainClaim}, proving the proposition \texttt{C3MainClaim}
  & none & none & \cref{thm:cubic}\\
\Cref{thm:cut-sweep}
  & \texttt{CutSweep.cutSweep}, proving the proposition \texttt{CutSweepClaim}
  & none & none & \cref{thm:cut-sweep,lem:warmup-layout}\\
\Cref{lem:correlation-padded} for $p=3$
  & \texttt{Locality.}\allowbreak\texttt{graph\_correlation\_lower}
  & none & none & a lemma inside the development of \cref{thm:cubic}; not
  separately checked\\
\Cref{thm:deg4}\ref{thm:deg4:a}
  & \texttt{T2.target2a}
  & H1 & none & \cref{thm:deg4}\\
\Cref{thm:deg4}\ref{thm:deg4:b}, \ref{thm:deg4:c}
  & \texttt{T2.target2b}, \texttt{T2.target2c}
  & H2 at the pair & none & \cref{thm:deg4}\\
\Cref{thm:deg5}, triples $\tripP$, $\tripR$, $\tripS$
  & \texttt{T3.target3P}, \texttt{T3.target3R}, \texttt{T3.target3S}
  & H3 at the triple & none & \cref{thm:deg5}\\
\Cref{thm:main}; \cref{cor:edges}
  & \texttt{T4.target4}; \texttt{T4.target5}
  & H3 at $\tripS$ & none & \cref{thm:main}; \cref{cor:edges}\\
\Cref{def:pw-seq} (its $\eps$--$N$ form is equivalent to its $\rho(n)=o(n)$ form);
  \cref{lem:bootstrap}: its first claim; the whole lemma when $(a_3,\dots,a_k)$ is a
  pathwidth sequence
  & \texttt{PwSeq.targetRhoForm}; \texttt{PwSeq.targetBootstrapLemma};
  \texttt{PwSeq.targetLemma2}
  & none & none & \cref{def:pw-seq,lem:bootstrap}\\
\Cref{thm:pw-seq} for $\seqC$, $\seqF$, $\tripP$, $\tripR$, $\tripS$
  & \texttt{PwSeq.targetThm1C}, \texttt{targetThm1F}, \texttt{targetThm1P},
  \texttt{targetThm1R}, \texttt{targetThm1S}
  & none for $\seqC$; H1 for $\seqF$; H3 at the triple for $\tripP$, $\tripR$,
  $\tripS$ & none & \cref{thm:pw-seq}\\
\Cref{thm:switched4}; \cref{thm:pw-seq} for $\seqM$
  & \texttt{Sw4.targetThm62}; \texttt{Sw4.targetThm1M}
  & H6 and H7 & none & \cref{thm:switched4}; \cref{thm:pw-seq}\\
Implementation of \cref{sec:alg-implementation}: every output is a path decomposition
  & \texttt{Alg.algoT1Valid}, \texttt{algoT2aValid}, \texttt{algoMixedValid},
  \texttt{algoT4Valid}
  & none & none & the Lean port against the implementation, and the statements
  against \cref{sec:alg-implementation}\\
Implementation: no fallback on the inputs of these cases of \cref{thm:algo-precise}; the width
  bounds of \cref{lem:connector-linear,lem:undo} and of \cref{lem:alg-deletion} for the
  bootstrap of $\tripS$; the bounds of
  Part~I attained for some labeling
  & \texttt{Alg.}\allowbreak\texttt{algoT1RealizesTheorem1}, \dots,
  \texttt{Alg.}\allowbreak\texttt{algoT4RealizesCorollary5}, with $29$ statements in
  all
  & those of the attained theorem (none, H1, H2, H3) & none & the statements
  against \cref{sec:alg-implementation} and the theorems they realize\\
\bottomrule
\end{tabular}
\end{table}

\paragraph{The hypotheses and their status.} The five hypotheses are mathematics
established outside the development, stated as explicit hypotheses of the final
theorems rather than as axioms.
\begin{description}[leftmargin=2.2em,labelindent=0pt,itemsep=2pt]
  \item[H1] (\texttt{SimplexFourClaim}) The case $d=4$ of \cref{lem:simplex},
  written exactly as the theorem \texttt{FSC.cdf\_simplex\_le} of Mulgund's
  formalization~\cite{Mulgund2026} for four variables. Status: proved in that
  formalization, by the author of~\cite{Mulgund2026}. We compiled it at the commit
  its author designates as certified, and checked that its main theorem depends only
  on the three standard axioms and that, for $n\in\{3,4,5\}$ variables and after the
  substitution $X=-U$, its statement is exactly the inequality of \cref{lem:simplex}.
  The distribution in that statement is Mathlib's multivariate Gaussian measure,
  which for a positive semidefinite matrix is the centered Gaussian distribution with
  that covariance, singular matrices included. The reproducibility package does not
  rebuild this formalization. It uses Lean 4.31.0 and another version of Mathlib, so
  our development cannot import it; H1 therefore restates the theorem word for word.
  The written proof uses the same theorem.
  \item[H2] (\texttt{AdmissiblePair}, at $(F_3,0.2575)$ and at
  $(\frac5{32},\frac14)$) Admissibility for $D=\{3,4\}$ as in \cref{def:admissible},
  with the exact center bounds; these pairs are the restrictions of $\tripP$ and
  $\tripS$ noted after \cref{prop:certificate}. Status: certified by the
  computer-assisted proof of \cref{prop:certificate} in ball arithmetic and exact
  rational arithmetic (\cref{sec:certificate}). The computation is not repeated inside
  Lean, because that would require verified enclosures of $\Phi$, of the simplex
  functions $S_d$ and of Owen's $T$ function on about a thousand cells.
  \item[H3] (\texttt{AdmissibleTriple}, at $\tripP$, $\tripR$ and $\tripS$)
  \Cref{prop:certificate} as stated. Status: as for H2. The written method of this
  certificate uses the cases $d=4$ and $d=5$ of \cref{lem:simplex}, which rest on
  Mulgund's preprint (\cref{rem:stars-mulgund}). The simplex theorem for
  five variables enters the formal proofs only through this certificate.
  \item[H6] (\texttt{SwitchMixedClaim}) \Cref{lem:switch-mixed} as stated, for
  centered Gaussian vectors described by their correlation matrices and for the
  bounds in every order, with $\Psi$ written as the expectation
  $\E[\ind\{Y_0\le0\}\lceil N/2\rceil]$, where $N$ counts the neighbors with positive
  label. Status: proved in the paper by a coupling reduction and the box
  certificates of \cref{sec:switch-cert-intro} in ball arithmetic; not formalized.
  \item[H7] (\texttt{SwitchedChargingClaim}) \Cref{prop:certificate-switched} with
  $\cthree$ exact and the center bounds of H2 and H3. Status: a further problem of
  the certificate of \cref{prop:certificate} (\cref{sec:certificate}); not
  formalized.
\end{description}

\paragraph{The formal statements.} \texttt{C3MainClaim} is the uniform $\eps$--$N$
form of \cref{thm:cubic} with $\cthree$ written as the exact real expression
$\frac3{4\pi}\arccos\frac56$: for every real $\eps>0$ there is $N$ such that for
every $n\ge N$, every simple graph on the vertex set $\{0,\dots,n-1\}$ all of whose
degrees are at most $3$ has a path decomposition in which every bag has at most
$(\cthree+\eps)n+1$ vertices. The proposition also contains the inequality
$\cthree<2c_*/(1+6c_*)\approx0.16443$, where
$c_*=\frac3{2\pi}\arccos\frac{2\sqrt2}3$ is the constant of \cref{thm:cut-sweep}; this compares $\cthree$ with a weaker
coefficient proved earlier in the same development and plays no role here. The
statement \texttt{graph\_correlation\_lower} is the inequality
$\Corr(Y_u,Y_v)\ge r_*(1-\frac1R)$ for every graph of maximum degree at most $3$,
every radius $R\ge1$ and every edge $uv$.

In the statements for \cref{thm:deg4} and
\cref{thm:deg5}, $F_4$ and $F_3$ are defined as the suprema over the threshold of
the ideal star functions \eqref{eq:overview-f}, and \cref{thm:deg4}\ref{thm:deg4:a}
is stated in the form $\pw(G)\le F_4n+o(n)$. The formal target for \cref{cor:edges} is the
$\eps$--$M$ form of \cref{rem:edges-m}, which implies \cref{cor:edges}: for every
$\eps>0$ there is $M$ such that every graph with $m\ge M$ edges, on any number of
vertices, satisfies $\pw(G)\le(\frac7{48}+\eps)m$. It implies the $o(n)$ form: \cref{lem:warmup-layout} gives
$\pw(G)\le m$, as at most $m$ edges leave any prefix of an ordering, and
$\pw(G)\le n-1<\frac7{48}m$ if $m>\frac{48}7n$, so
$\pw(G)\le\frac7{48}m+\max\{M,\frac{48}7\eps n\}$ for every graph.

The formal definition of a pathwidth sequence is the finite-tuple phrasing given after
\cref{def:pw-seq}, for tuples $(a_3,\dots,a_k)$ with $k\in\{3,4,\dots\}\cup\{\infty\}$,
with the remainder in uniform $\eps$--$N$ form: for every $\eps>0$ there is $N$ such
that every graph with $n\ge N$ vertices satisfies
$\pw(G)\le\sum_{d=3}^ka_dn_d+\sum_{d>k}n_d+\eps n$. The theorem
\texttt{PwSeq.targetRhoForm} proves this equivalent to the form of \cref{def:pw-seq}
with one remainder function $\rho(n)=o(n)$. The statements for \cref{thm:pw-seq}
bound every graph, with coefficient $1$ for each vertex of degree greater than $k$.
\texttt{CutSweepClaim} states \cref{thm:cut-sweep} in four parts:
$c_*=\frac3{2\pi}\arcsin\frac13$ for $c_*$ written as $\frac3{2\pi}\arccos\frac{2\sqrt2}3$;
$c_*<\frac16$; $\pw(G)\le w$ whenever some ordering of $V(G)$ has at most $w$ edges
leaving each of its prefixes, for every graph (the generality of
\cref{lem:warmup-layout}); and the uniform $\eps$--$N$ form of
$\operatorname{cw}(G)\le c_*n+o(n)$ for maximum degree at most~$3$.

\paragraph{What is not formalized.} The following are not formalized:
\cref{cor:bisection}; \cref{thm:cut-sweep-p}; \cref{cor:table}; \cref{cor:avg-degree}; the enclosures of \cref{prop:F3F4}; the
certificates behind H2, H3, H6 and H7; the explicit remainders of
\cref{lem:reduction} and \eqref{eq:bootstrap-explicit}; the explicit threshold
$n\ge2^{46}$ and the explicit constants of the proof of \cref{thm:cut-sweep}; the
computation of the labels
(\cref{sec:alg-law,sec:alg-perturbation,sec:alg-means,sec:alg-derand}) and the
running times of \cref{thm:algo-precise}; \cref{cor:alg-cutwidth,thm:alg-cut-sweep}; and treewidth.

\paragraph{Routes.} Each formal proof follows the written proof, with the following
differences. The formal proof of \cref{thm:cubic} proves the switch lemma
(\cref{lem:switch}) by a different deformation than \cref{lem:cubic-exceed}: it
rotates one vector of the residual configuration at a time, controls each rotation by a
Plackett-type derivative computation, and reaches a planar configuration after finitely
many rotations by a potential-function argument; this route exists only in the formal
development. It also proves the case $d=3$ of \cref{lem:simplex} directly, by the
elementary argument of \cref{rem:stars-mulgund}, instead of taking it from Mulgund's
theorem. All formal proofs prove only $o(n)$ remainders, so they choose their
parameters from $\eps$, replace explicit rates by continuity or elementary estimates
(for \cref{thm:deg4,thm:deg5}, they also prove the inequalities
$5\Phi(-3)\le\frac1{50}\le F_3\le\frac{103}{400}$ by elementary estimates instead of
the certified values of \cref{prop:F3F4}), and, except for
\cref{thm:cut-sweep}, bound the expected total absolute deviation of all counts
instead of applying Chebyshev's inequality to each count. For vertices of degree at
most~$2$, the formal proofs of \cref{thm:deg4,thm:deg5,thm:switched4,thm:main} do not
follow \cref{lem:reduction}: they keep a smallest set of vertices to which every
component of the rest, of treewidth at most~$2$, attaches at most twice, and lift the
decomposition by a torso construction, with the remainder $12\sqrt n+1$. The formal
bootstrap proves $\sigma_d\le a_d$ through the conserved deficit
$d(1-\sigma_d)+(1-\sigma_{d-1})=\frac{129}{32}$ instead of \cref{lem:weights}; with base
degree $k\ge3$ in place of $5$ it proves \cref{lem:bootstrap}, and with it
\cref{thm:pw-seq} for $\seqC$, $\seqF$, $\tripP$, $\tripR$ and $\tripS$, and for $\seqM$
under H6 and H7 with the base of \cref{thm:switched4}.

\subsection{The implementation}\label{sec:alg-implementation}

The accompanying code implements the randomized algorithms of the cubic, padded, mixed
and general cases of \cref{thm:algo-precise}, with floating-point labels, and its
outputs were valid path decompositions on all test inputs; for the sweeps, the widths
were between $1.19$ and $1.63$ times the bounds without remainder at $n\le2000$. A Lean port of these
algorithms, which agrees with the implementation on $16{,}650$ test calls, is
proved to output path decompositions that, for suitable labelings, attain
\cref{thm:cubic,thm:deg4,thm:deg5,thm:main} and \cref{cor:edges} under the hypotheses
of their formal proofs.
{}
\section{Separate, Measure and Conquer: the analyses}\label{app:is}

This appendix contains the analyses behind \cref{sec:apps-smc}.
\Cref{sec:smc-technique} explains the technique operationally and shows where the
separator constant enters, and \cref{app:smc-computer} gives the conventions for the
computer-assisted steps. \Cref{sec:smc-separators} derives balanced separators from our
decompositions. \Cref{app:smc-max2csp} reruns the linear program of Gaspers and Sorkin
for cubic Max 2-CSP; we show that the optimum of their linear program is exactly
$\frac{2c}{1+4c}$ as a function of the separator constant~$c$.
\Cref{app:is-statement,app:is3,app:is-general,app:is-g1,app:is-g2} prove the bounds
for counting independent sets and coloring.

Throughout this appendix $r\ge2$ denotes the domain size of Max $(r,2)$-CSP, as in
\cref{sec:max2csp-def}. All logarithms without a base are binary.

\subsection{The technique}\label{sec:smc-technique}

This subsection describes Separate, Measure and Conquer in the generality needed
below, following \cite[Sections~1.1 and~4]{GaspersSorkin2017}, and isolates the one
constraint in which the separator constant appears. Section, lemma and equation
numbers of \cite{GaspersSorkin2017} refer to its arXiv
version~\cite{GaspersSorkin2017v2}. The algorithm of \cref{app:smc-max2csp},
and the subcubic part of the algorithm of \cref{app:is-statement}, are instances of this
scheme.

A \emph{separation} of a graph $G$ is a partition $(L,S,R)$ of $V(G)$ such that no
edge joins $L$ and $R$; $S$ is the \emph{separator}, $L$ and $R$ are the left and
right sides.

\paragraph{Branching on the separator.} Consider a branching algorithm that removes a
vertex in every branch, as most algorithms for Max 2-CSP, counting independent sets or
counting dominating sets do. Suppose it is given a separation $(L,S,R)$ of the input
with $|S|\le cn$ and $|L|,|R|\le\frac{1-c}{2}n$, and that it always branches on a
vertex of~$S$. Used on its own, the separator gives a weak bound: if branching on each
separator vertex multiplies the running time by $r$ and nothing else is gained, the
running time satisfies
$t(n)=r^{cn}\cdot 2\,t(\frac{1-c}{2}n)$, whose solution is $r^{\frac{2c}{1+c}n}$,
that is $r^{2n/7}$ for $c=\frac16$. This is slower than the simple $r^{n/4}$
algorithm for cubic Max 2-CSP~\cite{ScottSorkin2007}, which branches on any vertex
and uses that its three neighbors become contractible.

A better bound comes from combining the two effects. Locally, branching on a separator
vertex simplifies the instance as much as branching on any other vertex, and the split
at the end comes in addition. In the cubic Max 2-CSP setting, each branching on a separator vertex removes
it and makes its three neighbors contractible. If these neighbors fall on the two
sides evenly, each side shrinks by $\frac32$ per separator vertex, and the
\emph{optimistic recurrence} of \cite[Section~1.1]{GaspersSorkin2017} is
\[
  t(n)=r^{cn}\cdot 2\,t\Bigl(\frac{1-c}{2}n-\frac32\,cn\Bigr).
\]
With $t(n)=r^{xn}$ this reads $x=c+x\cdot\frac{1-4c}{2}$, so $x=\frac{2c}{1+4c}$: for
$c=\frac16$ this is the $\frac15$ of Gaspers and Sorkin, and for $c=\cthree$ it is
$0.179341$. Edwards~\cite{Edwards2016JCSS}, whose polynomial-space algorithm is also
separator-based, remarks that an improved coefficient $c<\frac16$ for cubic graphs
would give the exponent $\frac{2c}{1+4c}$ by his argument.

\paragraph{The measure.} The optimistic recurrence is not a proof: the branchings need
not split evenly between the sides, the algorithm must sometimes move vertices between
$L$, $S$ and $R$ to keep branching on the separator, and the split takes effect only at
the end, all at once. Gaspers and Sorkin make it rigorous with a measure that
assigns weight only to the separator and to the larger side. Let $\mu_s$ and $\mu_r$
assign nonnegative weights to vertex sets (``separator'' and ``right'' weights,
depending on the current labeled graph), and let the sides be named so that
$\mu_r(R)\ge\mu_r(L)$. The measure is \cite[(26)]{GaspersSorkin2017}
\begin{equation}\label{eq:smc-measure}
  \mu(L,S,R)=\mu_s(S)+\mu_r(R)
  +\max\Bigl\{0,\,B-\frac{\mu_r(R)-\mu_r(L)}{2}\Bigr\}
  +(1+B)\log_{1+\eta}\bigl(\mu_r(R)+\mu_s(S)\bigr),
\end{equation}
where $B$ is a constant larger than the change of the imbalance $\mu_r(R)-\mu_r(L)$
in any single transformation except the separation step below, and $\eta>0$ is a
small constant. The vertices of $L$ carry no weight: when the separator is
exhausted, the remaining running time is at most twice that for the larger side $R$,
and the smaller side is accounted for by the halving of the weight of $R$. The third
term is a penalty for balanced instances. Gaspers and Sorkin combine it with a
condition on the transformations \cite[condition~(30)]{GaspersSorkin2017}: if
$\mu_r(R)-\mu_r(L)>B$, a transformation must decrease $\mu_r(R)$ at least as much as
$\mu_r(L)$. Under this condition every transformation decreases $\mu_r(R)$ plus the
penalty by at least half of its total progress on the two sides. The logarithmic term
offsets the penalty $B$ that may appear each time a new separation is computed, and it
contributes only a polynomial factor to the running time.

\paragraph{Where the separator constant enters.} Every transformation of the algorithm
must satisfy the usual measure-and-conquer condition (\cref{lem:smc-measure} below).
For the local branching and simplification rules this gives constraints that do not
involve the separator. The only transformation that does is the \emph{separation
step}: when $S=\emptyset$, the instance on $L\cup R$ is solved as two independent
instances, and for each of them (say the one on $R$) a new separation $(L',S',R')$ is
computed. The measure must not increase; the logarithmic term offsets the new
penalty provided that (the case $S=\emptyset$ of~\cite[(29)]{GaspersSorkin2017})
\begin{equation}\label{eq:smc-29}
  \mu_r(R)\ge(1+\eta)\bigl(\mu_r(R')+\mu_s(S')\bigr).
\end{equation}

Suppose that every vertex has right weight $w_r$ and separator weight at most $w_s$,
where $w_s\ge\frac{w_r}2$, and that the new separation has $|S'|\le c|R|$ and
$|L'|=|R'|$, so that $|R'|=\frac{|R|-|S'|}2$. Then
$\mu_r(R')+\mu_s(S')\le w_s|S'|+w_r\frac{|R|-|S'|}2$, which is nondecreasing in $|S'|$
because $w_s\ge\frac{w_r}2$, so it is at most $(c\,w_s+\frac{1-c}2w_r)|R|$ (this
monotonicity argument is carried out in \cref{lem:gs-separation}). Hence
\eqref{eq:smc-29} holds as soon as
\begin{equation}\label{eq:smc-ratio}
  c\,w_s+\frac{1-c}{2}\,w_r<w_r,\qquad\text{equivalently}\qquad
  w_s<\frac{1+c}{2c}\,w_r .
\end{equation}
The separations of \cref{sec:smc-separators} have $|S'|\le c|R|$ only up to a
lower-order term; this is absorbed by taking $c$ slightly larger than the separator
constant and $|R|$ large.

This is the mechanism by which the separator constant controls the running time. A
branching on a separator vertex removes that vertex from $S$, so its large weight
$w_s$ contributes to the decrease of the measure; the branching constraints therefore
need $w_s$ large
compared with the right weights. The time bound is governed by $w_r$, the weight of a
vertex in the initial instance $(\emptyset,\emptyset,V)$. Constraint
\eqref{eq:smc-ratio} caps the ratio $w_s/w_r$ by $\frac{1+c}{2c}$, which is $3.5$ for
$c=\frac16$ and $4.076$ for $c=\cthree$. A smaller separator constant relaxes the cap,
and $w_r$ can decrease. Everything else in the analysis is unchanged. Gaspers and
Sorkin note that any improvement of their separator lemma improves their running
times \cite[Section~3.3]{GaspersSorkin2017}; \cref{lem:lp} quantifies this exactly for
cubic Max 2-CSP.

The analyses rest on the following lemma, which bounds the running time of a branching
algorithm by a measure that decreases suitably along every branching, and allows a
hand-over to a subroutine with its own measure. In it, $A$ is an algorithm for a
problem, $\mathcal{B}$ is an algorithm for a class $\mathcal{C}$ of instances, and
$\mu$, $\mu_{\mathcal{B}}$ and $\nu$ are nonnegative measures on instances.

\begin{lemma}[{\cite[Lemma~4.2]{GaspersSorkin2017}, \cite[Lemma~1]{GaspersLee2023}}]
  \label{lem:smc-measure}
  Let $c_0\ge0$ and $\lambda>1$ be constants, and let $\mu_{\mathcal{B}}(I)\le\mu(I)$ for
  every $I\in\mathcal{C}$. Suppose that on every instance $I$,
  either $I\in\mathcal C$ and $A$ solves $I$ by calling $\mathcal{B}$ in time
  $O(\nu(I)^{c_0+1}\lambda^{\mu_{\mathcal{B}}(I)})$, or $A$ reduces $I$ to instances
  $I_1,\dots,I_k$, solves them recursively and combines the solutions, using time
  $O(\nu(I)^{c_0})$ outside the recursive calls, where $\nu(I_i)\le\nu(I)-1$ for every $i$
  and
  \[
    \sum_{i=1}^{k}\lambda^{\mu(I_i)}\le\lambda^{\mu(I)}.
  \]
  Then $A$ solves every instance $I$ in time $O(\nu(I)^{c_0+1}\lambda^{\mu(I)})$.
\end{lemma}

The case without a subroutine $\mathcal B$ is \cite[Lemma~2.3]{GaspersSorkin2017}.

For a two-way branching with $\lambda=2$ whose branches decrease the measure by at
least $a$ and $b$, the condition reads $2^{-a}+2^{-b}\le1$; we call $(a,b)$ a
\emph{branching vector}. For Max $(r,2)$-CSP the branching creates $r$ instances with
the same measure $\mu'$, and with $\lambda=r$ the condition reads $1+\mu'-\mu\le0$.

\subsection{Computer-assisted steps}\label{app:smc-computer}

The analyses of \cref{app:smc-max2csp} and of
\cref{app:is-statement,app:is3,app:is-general,app:is-g1,app:is-g2} end in
finitely many explicit inequalities between explicit rational weights. We list every
inequality and every weight, so that each can be checked by hand or by machine. The accompanying code checks
them as follows. Linear inequalities are checked in exact rational arithmetic
(Python \texttt{fractions}). Every branching vector $(a,b)$ is checked to satisfy
$2^{-a}+2^{-b}\le1$ in interval arithmetic (\texttt{mpmath.iv}, 40 decimal digits),
with the weights entered as exact rationals, except in the finite enumerations of
configurations of \cref{app:is-g2} (\texttt{g2\_enumeration.py}): there the vectors
are computed in exact rational arithmetic and the condition is checked in floating
point, with $40$-digit decimal arithmetic for sums close to~$1$. Floating-point optimization (\texttt{scipy}) is used only to find the
weights and to report optima that the theorems do not use; nothing in a proof depends
on it.

\subsection{Balanced separators from the decompositions}\label{sec:smc-separators}

The previous analyses obtain their separators from the bounds of Fomin and
H{\o}ie and of Fomin, Gaspers, Saurabh and
Stepanov~\cite{FominHoie2006,FominGaspersSaurabhStepanov2009}: Gaspers and Sorkin
use separators of size $\alpha_\Delta|V|+o(|V|)$ with
$\alpha_3,\alpha_4,\alpha_5,\alpha_6=\frac16,\frac13,\frac{13}{30},\frac{23}{45}$
\cite[Lemmas~1.2 and~4.1]{GaspersSorkin2017}, and Gaspers and Lee use $\frac{n_3}{6}$
\cite{GaspersLee2023}. The following lemma provides the same kind of
separators from our decompositions. Part~(a) is used for Max 2-CSP, part~(b) for
counting independent sets.

Let $a^*_3=\cthree$, $a^*_4=F_4$, $a^*_5=\frac{11}{32}$ and $a^*_6=\frac7{16}$. For a
graph $G$ and $\theta>0$, a function $\mu_r$ from the subsets of $V(G)$ to $[0,\infty)$ is
\emph{$\theta$-Lipschitz} if $\mu_r(\emptyset)=0$ and $|\mu_r(X\cup\{v\})-\mu_r(X)|\le\theta$
for all $X\subseteq V(G)$ and $v\in V(G)$. We write $V_3$ for the set of vertices of
degree~$3$ of $G$, so that $n_3=|V_3|$.

\begin{lemma}[balanced separators]\label{lem:sep}
  Let $\theta>0$.
  \begin{enumerate}[label=\textup{(\alph*)}]
    \item Let $3\le\Delta\le6$. There is a function $\rho'_\Delta(x)=o(x)$ such that for
      every graph $G$ with $\Delta(G)\le\Delta$ and every $\theta$-Lipschitz $\mu_r$
      computable in polynomial time, a separation $(L,S,R)$ of $G$ with
      $|S|\le a^*_\Delta|V(G)|+\rho'_\Delta(|V(G)|)$ and $|\mu_r(L)-\mu_r(R)|\le\theta$ can
      be computed deterministically in polynomial time; if $\mu_r(X)=|X|$, it can moreover
      have $|L|=|R|$.
    \item There is a function $\rho_3(x)=o(x)$ such that for every graph $G$ with
      $\Delta(G)\le3$ and $\delta(G)\ge2$ and every $\theta$-Lipschitz $\mu_r$ computable
      in polynomial time with $\mu_r(X)=\mu_r(X\cap V_3)$ for all $X$, a separation
      $(L,S,R)$ of $G$ with
      $S\subseteq V_3$, $|S|\le\cthree n_3+\rho_3(n_3)$ and $|\mu_r(L)-\mu_r(R)|\le\theta$ can
      be computed deterministically in polynomial time.
  \end{enumerate}
\end{lemma}

For comparison, $\cthree<\frac16$, $F_4<0.2458<\frac13$, $\frac{11}{32}<\frac{13}{30}$
and $\frac7{16}<\frac{23}{45}$.
The proof is the argument of \cite[Lemma~4.1]{GaspersSorkin2017}:
a nice path decomposition gives a sequence of separations whose separators are its
bags, and the difference of $\mu_r$ between the two sides changes by at most $\theta$
from one separation to the next. Part~(b) applies this to the cubic multigraph
obtained by suppressing the vertices of degree~$2$.

\begin{proof}
  We first show how a path decomposition yields the separation. Let $H$ be a graph and
  $(W_1,\dots,W_t)$ a path decomposition of $H$ of width $w$ in which consecutive bags
  differ in exactly one vertex (a \emph{nice} path decomposition). Every path
  decomposition, given for instance in the interval form of \cref{thm:algo-precise}, can be
  turned into a nice one of the same width with at most $2|V(H)|$ bags in polynomial
  time: list the bags, and between two consecutive bags first remove the vertices of
  $W_i\setminus W_{i+1}$ one at a time, then add those of $W_{i+1}\setminus W_i$ one at
  a time; start with the empty bag and end by removing the last vertices
  one at a time (empty bags are discarded at the end). For $1\le i\le t$ let
  \[
    L_i=\Bigl(\bigcup_{j<i}W_j\Bigr)\setminus W_i,\qquad R_i=V(H)\setminus(L_i\cup W_i).
  \]
  Then $(L_i,W_i,R_i)$ is a separation: a vertex of $L_i$ occurs only in bags before
  $W_i$ and a vertex of $R_i$ only in bags after $W_i$, so by the interval property no
  bag contains both, and they are not adjacent. Let $f(i)=\mu_r(L_i)-\mu_r(R_i)$. If
  $W_{i+1}=W_i\cup\{v\}$, then $v\in R_i$, $L_{i+1}=L_i$ and $R_{i+1}=R_i\setminus\{v\}$;
  if $W_{i+1}=W_i\setminus\{v\}$, then $L_{i+1}=L_i\cup\{v\}$ and $R_{i+1}=R_i$. In both
  cases $|f(i+1)-f(i)|\le\theta$, and in both cases $|L_i|-|R_i|$ increases by exactly one.
  Moreover $L_1=\emptyset$ and $R_t=\emptyset$, so $f(1)=-\mu_r(R_1)\le0\le\mu_r(L_t)=f(t)$.
  Let $i$ be the largest index with $f(i)\le0$. If $i=t$ then $f(t)=0$; otherwise
  $f(i+1)>0$ and $f(i)\ge f(i+1)-\theta>-\theta$. Either way $|f(i)|\le\theta$. For $\mu_r(X)=|X|$ the
  integer $|L_i|-|R_i|$ increases by exactly one at each step, from at most $0$ to at
  least~$0$, so it vanishes at some~$i$. All of this takes polynomial time, and the
  separator $W_i$ has at most $w+1$ vertices.

  (a) Let $n=|V(G)|$. By \cref{thm:algo-precise}\ref{it:algo:D} a path decomposition of
  $G$ of width at most $W(G)+\rho(n)$, with $\rho(n)=o(n)$, is computed
  deterministically in polynomial time, where $W(G)$ is the respective bound: $\cthree n$ by \cref{thm:cubic} if
  $\Delta\le3$; $F_4n$ by \cref{thm:deg4}\ref{thm:deg4:a} if $\Delta\le4$;
  $\frac5{32}n_3+\frac14n_4+\frac{11}{32}n_5\le\frac{11}{32}n$ by \cref{thm:deg5} with
  the triple $\tripS$ if $\Delta\le5$; and $\sum_{d\le6}a_dn_d\le a_6n=\frac7{16}n$ by
  \cref{thm:main} if $\Delta\le6$. (Vertices of degree at most~$2$ have coefficient~$0$
  in these bounds, and the coefficients are nondecreasing in~$d$.) So $W(G)\le
  a^*_\Delta n$, and the construction above gives a separation with
  $|S|\le a^*_\Delta n+\rho(n)+1$. This is one algorithm for each $\Delta$, and $\rho$
  depends only on $\Delta$ (through the case of \cref{thm:algo-precise} used); take
  $\rho'_\Delta(x)=\rho(x)+1$.

  (b) Suppress the vertices of degree~$2$: every component of $G$
  that is a cycle contains no vertex of $V_3$, and in every other component each vertex
  of degree~$2$ lies on a unique path $u\,x_1\cdots x_q\,w$ whose ends $u,w$ (possibly
  equal) are in $V_3$ and whose inner vertices have degree~$2$. Replacing every such path
  by an edge $uw$ gives a multigraph $M$ (possibly with loops and parallel edges) on
  $V_3$ in which every vertex has degree~$3$. Let $G_M$ be the underlying simple graph,
  without loops; $\Delta(G_M)\le3$. By \cref{thm:cubic} and
  \cref{thm:algo-precise}\ref{it:algo:D} a path decomposition of $G_M$ of width at most
  $\cthree n_3+\rho(n_3)$ is computed deterministically in polynomial time, and
  the construction above, applied with $\mu_r$ restricted to $V_3$, yields a separation
  $(L_M,S,R_M)$ of $G_M$ with $|S|\le\cthree n_3+\rho(n_3)+1$ and
  $|\mu_r(L_M)-\mu_r(R_M)|\le\theta$. Extend it to $G$: put every cycle component and every
  suppressed path whose ends lie in $L_M\cup S$ into $L$, and every other suppressed path
  into $R$; the ends of a suppressed path are adjacent or equal in $M$, so they never lie
  in $L_M$ and $R_M$ respectively. Then $(L,S,R)$ with $L\supseteq L_M$, $R\supseteq R_M$
  is a separation of $G$ with $S\subseteq V_3$, and $\mu_r(L)=\mu_r(L_M)$,
  $\mu_r(R)=\mu_r(R_M)$ because $\mu_r$ depends only on degree-$3$ vertices. Take
  $\rho_3(x)=\rho(x)+1$.
\end{proof}

\subsection{Max 2-CSP}\label{app:smc-max2csp}

This subsection reruns the analysis of Gaspers and
Sorkin~\cite[Section~3]{GaspersSorkin2017} for cubic Max $(r,2)$-CSP with the
separators of \cref{lem:sep}(a), and lifts the result to all instances with the
theorem of Scott and Sorkin. For cubic instances the exponent drops from the
$\frac n5$ of Gaspers and Sorkin to $\frac{2\cthree}{1+4\cthree}n+o(n)<0.179341n+o(n)$
(\cref{thm:smc}). The algorithm is that of Gaspers and Sorkin, rules R1--R9, with the separations of
\cref{lem:sep}(a). Only one constraint of their linear program involves the separator
constant (\cref{lem:gs-separation}), and we solve the program exactly as a function of
that constant~$c$: its optimum is $\frac{2c}{1+4c}$, with explicit weights and a
three-line dual certificate (\cref{lem:lp}). At $c=\frac16$ these are the weights of
Gaspers and Sorkin.

Put
\[
  \gamma=\frac{2\cthree}{1+4\cthree}\qquad\text{and}\qquad\alpha=\frac{2\gamma}{3}.
\]
RCSP and PCSP are the ring and polynomial versions of \cref{sec:max2csp-def}, and
polynomially bounded PCSP is meant in the sense of Scott and
Sorkin~\cite{ScottSorkin2009}.

\begin{theorem}[polynomial-space Max 2-CSP]\label{thm:smc}
  Max $(r,2)$-CSP with $n$ variables and $m$ constraints can be solved in polynomial
  space and time
  \begin{enumerate}[label=\textup{(\alph*)}]
    \item\label{thm:smc:a} $r^{\gamma n+o(n)}=r^{\alpha m+o(m)}$ if the constraint graph
      is cubic;
    \item\label{thm:smc:b} $r^{(\frac18+\frac38\alpha)m+o(m)}$ if the constraint graph has
      maximum degree at most~$4$;
    \item\label{thm:smc:c} $r^{(\frac7{50}+\frac3{10}\alpha)m+o(m)}$ for all instances.
  \end{enumerate}
  The same bounds hold for the number of ring operations of RCSP and for the running
  time of polynomially bounded PCSP. The numbers of dominating sets of each size of an
  $n$-vertex graph of maximum degree at most~$3$ can be computed in polynomial space and
  time $3^{\gamma n+o(n)}$.
\end{theorem}

Here $\gamma<0.179341$, $\alpha<0.119561$, $\frac18+\frac38\alpha<0.169836$,
$\frac7{50}+\frac3{10}\alpha<0.175869$ and $3^{\gamma}<1.21778$, so the last bound is
$O(1.2178^n)$. The polynomially bounded PCSPs include counting cuts by size and the
Ising partition function.

The previous bounds are $r^{n/5}$, $r^{7m/40}$, $r^{9m/50}$ and $3^{n/5}=O(1.2458^n)$
\cite[Theorems~3.1--3.3 and~5.1]{GaspersSorkin2017}; Edwards~\cite{Edwards2016JCSS}
also obtains $r^{9m/50+o(m)}$ in polynomial space. In exponential space,
\cref{thm:max2csp} is faster. In polynomial space, this route does not improve Max
2-SAT, which is solved in time $\Ostar(2^{m/6.321})$~\cite[Theorem~4]{GaspersSorkin2012},
nor counting the dominating sets of general graphs, which takes time
$O(1.5183^n)$~\cite[Theorem~5.2]{GaspersSorkin2017}: a separator constraint of that
analysis is not tight, so the new separators do not lower this bound.

\paragraph{Reductions.} Instances are as in \cref{sec:max2csp-def}. Scott and
Sorkin~\cite{ScottSorkin2007} use four reductions, each computable in polynomial time:
a vertex of degree~$0$ is deleted after maximizing its score (Reduction~0); a vertex
of degree~$1$ is folded into its neighbor (Reduction~1); a vertex of degree~$2$ is
replaced by an edge between its two neighbors, merged with an existing edge if there
is one (Reduction~2); and a vertex $v$ of degree at least~$3$ is \emph{branched on}
(Reduction~3): for each color $C\in[r]$, $v$ is deleted and its score functions are
absorbed into those of its neighbors, which gives $r$ instances with the common
constraint graph $G-v$. An optimal solution of the instance is obtained in polynomial
time from optimal solutions of the subinstances.

\paragraph{The algorithm.} The algorithm maintains a separation $(L,S,R)$ of the
current constraint graph $G$, initially $(\emptyset,\emptyset,V)$. Let $L_3,S_3,R_3$
denote the vertices of degree~$3$ in $L,S,R$, let $S_2$ denote the vertices of
degree~$2$ in $S$, and name the sides so that $|L_3|\le|R_3|$. The measure is
\cite[(6)]{GaspersSorkin2017}
\begin{multline}\label{eq:gs-measure}
  \mu(L,S,R)=w_s|S_3|+w_{s,2}|S_2|+w_r|R_3|+w_b\ind(|R_3|=|L_3|)\\
  +w_c\ind(|R_3|=|L_3|+1)+w_d\log_{3/2}(|R_3|+|S_3|),
\end{multline}
a special case of \eqref{eq:smc-measure} with integral imbalance: $w_b$ and $w_c$
play the role of the penalty, and the logarithm that of the last term. The algorithm
applies the first applicable rule of the following list; when a rule is described, the
earlier rules are assumed not to apply. The constraint numbers refer to
\cref{tab:gs-constraints}.
\begin{enumerate}[label=\textup{R\arabic*.},leftmargin=*]
  \item \emph{Degree $0$.} Apply Reduction~0. The measure does not change.
  \item \emph{Half-edge deletion.} If some vertex has degree~$1$, or a vertex of
    $L\cup R$ has degree~$2$, apply Reduction~1 or~2. Viewing every edge deletion as two
    half-edge deletions, each lowering one degree, the measure does not increase under
    constraints (7)--(11). From now on the graph is cubic except for vertices of
    degree~$2$ in~$S$.
  \item \emph{Separation.} If $S=\emptyset$: the instance is the disjoint union of the
    instances on $L$ and $R$, which are solved independently. For each of them a new
    separation is computed by \cref{lem:sep}(a) with $\mu_r(X)=|X|$. Constraint
    $(12_c)$ below.
  \item \emph{Degree $2$ in $S$.} If $s\in S$ has degree~$2$, apply Reduction~2. If
    its neighbors are $u_1\in L$ and $u_2\in R$, the new edge $u_1u_2$ would join the
    sides, so $u_2$ is moved into~$S$. Constraints (13)--(15).
  \item \emph{No neighbor in $L$.} If $s\in S$ has no neighbor in $L$, move it into
    $R$ (``drag'' it). Constraint (16).
  \item \emph{No neighbor in $R$.} If $s\in S$ has no neighbor in $R$, move it into
    $L$ (swapping the names of the sides if $|R_3|=|L_3|$). Constraints (17), (18).
  \item \emph{One neighbor in each of $L$, $S$, $R$.} Branch on such an $s$.
    Constraint (19).
  \item \emph{Two neighbors in $L$.} Let $s\in S$ have two neighbors in $L$ and one,
    $u$, in $R$. If $|R_3|\le|L_3|+1$, branch on $s$; constraints (20), (21).
    Otherwise move $s$ into $L$ and $u$ into $S$; constraint (22).
  \item \emph{Two neighbors in $R$.} Otherwise every $s\in S$ has two neighbors in $R$
    and one in $L$; branch on such an $s$. Constraints (23), (24), and (20) after
    swapping the sides when $|R_3|=|L_3|$.
\end{enumerate}
Every rule takes polynomial time. The rules that only move vertices between $L$, $S$
and $R$ decrease the secondary measure $\nu=3|S_3|+2|R_3|+|L_3|+2|E(G)|$
\cite[end of Section~3.2]{GaspersSorkin2017}, and so do all rules that delete
vertices; \cref{lem:gs-separation} treats the separation step R3.

\begin{table}[tb]
  \centering
  \caption{The constraints of the cubic Max 2-CSP analysis
    \cite[Section~3.2]{GaspersSorkin2017v2}, each in the form ``expression $\le0$''. In
    $(12_c)$ the previous value $c=\frac16$ is replaced by the separator constant~$c$.
    The last column gives the expression for the weights of \cref{lem:lp}, multiplied by
  $D=1+4c$; a constant term~$1$ becomes~$D$.}
  \label{tab:gs-constraints}
  \small
  \begin{tabular}{@{}lll>{$}l<{$}@{}}
    \toprule
    No. & Rule (case) & Constraint & D\cdot\text{(value at \cref{lem:lp})}\\
    \midrule
    (7) & measure nonnegative & $w_s,w_{s,2},w_r,w_b,w_c,w_d\ge0$ & \text{holds}\\
    (8) & R2, degree $3\to2$ in $L$ & $-w_b+w_c$ & -c\\
    (9) & R2, degree $3\to2$ in $S$ & $-w_s+w_{s,2}$ & -c\\
    (10) & R2, degree $3\to2$ in $R$ & $-w_r+w_c$ & -c\\
    (11) & R2, degree $3\to2$ in $R$ & $-w_r+w_b-w_c$ & -c\\
    $(12_c)$ & R3, separation & $c\,w_s+\frac{1-c}{2}w_r-w_r$ & 0\\
    (13) & R4, balanced & $-w_{s,2}+w_s-w_b+w_c$ & 0\\
    (14) & R4, imbalanced & $-w_{s,2}+w_s-w_r+w_c$ & 0\\
    (15) & R4, imbalanced & $-w_{s,2}+w_s-w_r+w_b-w_c$ & 0\\
    (16) & R5 & $-w_s+w_r$ & c-1\\
    (17) & R6 & $-w_s+w_c$ & -1\\
    (18) & R6 & $-w_s+w_b-w_c$ & -1\\
    (19) & R7, branch & $1-2w_s+w_{s,2}-w_r$ & 0\\
    (20) & R8, branch & $1-w_s-w_r-w_b+w_c$ & 0\\
    (21) & R8, branch & $1-w_s-w_r-w_c$ & 0\\
    (22) & R8, drag & $-w_r+w_b$ & 0\\
    (23) & R9, branch & $1-w_s-2w_r-w_c+w_b$ & 0\\
    (24) & R9, branch & $1-w_s-2w_r+w_c$ & 0\\
    \bottomrule
  \end{tabular}
\end{table}

\Cref{tab:gs-constraints} lists the constraints. They are derived in
\cite[Section~3.2]{GaspersSorkin2017} from the condition of \cref{lem:smc-measure}:
for the rules that create one instance the measure must not increase, and for a
branching on a vertex, which creates $r$ instances of the same measure $\mu'$, the
condition $r\cdot r^{\mu'}\le r^{\mu}$ is $1+\mu'-\mu\le0$. For example, in R7 the
separator loses the degree-$3$ vertex $s$, its neighbor in $S$ drops to degree~$2$, and
one degree-$3$ vertex of $R$ and one of $L$ drop to degree~$2$; the balance is
unchanged, which gives (19). We do not repeat the other derivations; the only one that
involves the separator is that of $(12_c)$. In the following lemma, an instance with
fewer than $N_0$ vertices that arises in the Separation rule R3 is counted as a leaf
of measure~$0$.

\begin{lemma}[the separation constraint]\label{lem:gs-separation}
  Let $0<c<\frac13$, and suppose that R3 uses separations with $|S'|\le c|V|$ and
  $|L'|=|R'|$ on every cubic graph with $|V|\ge N_0$ to which it applies. If the weights
  satisfy (7), $(12_c)$ and (16) of \cref{tab:gs-constraints} and $w_d=w_b+1$, then on
  every instance with $|R|\ge N_0$, R3 satisfies the condition of \cref{lem:smc-measure}
  and decreases the secondary measure~$\nu$.
\end{lemma}

\begin{proof}
  We follow \cite[Section~3.2, Separation]{GaspersSorkin2017}. Let the instance be
  $(L,\emptyset,R)$ with $N=|R|\ge|L|$. Since R1 and R2 do not apply and $S$ is empty,
  every vertex has degree~$3$, so $N=|R_3|$, and $\mu\ge w_rN+w_d\log_{3/2}N$ because the
  penalty terms are nonnegative. The graph $G[R]$ is cubic, since $S=\emptyset$ and no
  edge joins $L$ and~$R$. Let $(L',S',R')$ be the new separation of $G[R]$ and
  $s=|S'|\le cN$. It is balanced ($|L'_3|=|R'_3|$), so its measure is
  \[
    \mu'=w_ss+w_r\frac{N-s}{2}+w_b+w_d\log_{3/2}\frac{N+s}{2}.
  \]
  The derivative of $w_ss+w_r\frac{N-s}{2}$ in $s$ is $w_s-\frac{w_r}2\ge0$ by (16) and
  (7), so this part is at most $(c\,w_s+\frac{1-c}2w_r)N$. Since $c<\frac13$ we have
  $\frac{N+s}{2}\le\frac{1+c}{2}N\le\frac23N$, hence
  $w_d\log_{3/2}\frac{N+s}2\le w_d\log_{3/2}N-w_d$. With $w_d=w_b+1$,
  \[
    \mu'\le\Bigl(c\,w_s+\frac{1-c}{2}w_r\Bigr)N+w_d\log_{3/2}N-1
    \le w_rN+w_d\log_{3/2}N-1\le\mu-1
  \]
  by $(12_c)$. The instance on $L$ has measure $\mu''\le\mu-1$ by the same bound with $|L|\le N$ in
  place of $N$, since the bound is nondecreasing in $N$; if $|L|<N_0$, it is a leaf of
  measure $\mu''=0\le\mu-1$. Since $r\ge2$, $r^{\mu'}+r^{\mu''}\le2r^{\mu-1}\le r^{\mu}$.
  For the secondary measure: if $L\ne\emptyset$, splitting $(L,\emptyset,R)$ into
  $(\emptyset,\emptyset,R)$ and $(\emptyset,\emptyset,L)$ decreases $\nu$ because
  $|L|\le|R|$ and both sides are nonempty; if $L=\emptyset$, as for the initial
  instance, this first step is void. Replacing $(\emptyset,\emptyset,R)$ by
  $(L',S',R')$ changes $\nu$ by
  $3s+2|R'|+|L'|-2N=s-|L'|=s-\frac{N-s}{2}<0$, because $s\le cN<\frac N3$.
\end{proof}

Thus the separator constant enters the linear program only through $(12_c)$, and the
proof once more through $c<\frac13$ (the logarithmic estimate and the secondary
measure). The previous derivation estimates $|R'_3|\le\frac5{12}|R_3|$ and
$|S'_3|\le\frac16|R_3|+o(|R_3|)$ separately; the monotonicity in $s$ used above shows
that the combined form $(12_c)$ is what is needed for every separator fraction
$c<\frac13$.

\begin{lemma}[the linear program]\label{lem:lp}
  For every $c\in(0,1]$ the weights
  \[
    w_s=\frac{1+c}{1+4c},\quad w_{s,2}=\frac{1}{1+4c},\quad
    w_r=w_b=\frac{2c}{1+4c},\quad w_c=\frac{c}{1+4c},\quad w_d=w_b+1
  \]
  satisfy every constraint of \cref{tab:gs-constraints}. Conversely, every solution of
  (12$_c$), (21) and (24) has $w_r\ge\frac{2c}{1+4c}$. Hence the minimum of $w_r$ subject to the constraints of
  \cref{tab:gs-constraints} is exactly $\frac{2c}{1+4c}$.
\end{lemma}

\begin{proof}
  Put $D=1+4c$ and multiply all weights by $D$: the scaled weights are
  $w_s=1+c$, $w_{s,2}=1$, $w_r=w_b=2c$ and $w_c=c$, and the constant~$1$ in (19)--(21), (23)
  and (24) becomes $D$. Substituting gives the last column of
  \cref{tab:gs-constraints}:
  (8)--(11) each equal $-c$; $(12_c)$ equals $c(1+c)+\frac{1-c}{2}\cdot2c-2c=0$;
  (13) equals $-1+(1+c)-2c+c=0$, (14) equals (13) because $w_r=w_b$, and (15) equals
  $-1+(1+c)-2c+2c-c=0$; (16) equals $c-1\le0$; (17) and (18) equal $-1$;
  (19) equals $D-2(1+c)+1-2c=0$; (20) equals $D-(1+c)-2c-2c+c=0$; (21) equals
  $D-(1+c)-2c-c=0$; (22) equals $0$; (23) equals $D-(1+c)-4c-c+2c=0$; and (24) equals
  $D-(1+c)-4c+c=0$. All weights are nonnegative, so (7) holds.

  For the converse, $(12_c)$ says $w_s\le\frac{1+c}{2c}w_r$, and the sum of (21) and (24)
  says $2\le2w_s+3w_r$. Together, $2\le(\frac{1+c}{c}+3)w_r=\frac{1+4c}{c}w_r$.
\end{proof}

At $c=\frac16$ these are the previous weights
$(w_s,w_{s,2},w_r,w_b,w_c)=(0.7,0.6,0.2,0.2,0.1)$ \cite[Section~3.2]{GaspersSorkin2017},
and the tight constraints are those listed there, (12)--(15) and (19)--(24).
Gaspers and Sorkin solved the program numerically. The value $\frac{2c}{1+4c}$ is the
exponent of the optimistic recurrence, and Edwards notes that his argument would give
it for an improved separator coefficient (\cref{sec:smc-technique}); \cref{lem:lp}
adds that it is the exact optimum of the
linear program of Gaspers and Sorkin for every~$c$, with the explicit weights and the
certificate $(12_c)+(21)+(24)$. The accompanying code confirms both (exact check of the
weights for several rational $c$, and numerical optima at $c=\frac16,\cthree,0.1,0.3$).

We now prove \cref{thm:smc}.

\begin{proof}
  We prove~(a) with the weights of \cref{lem:lp} and \cref{lem:smc-measure}, and obtain
  (b) and~(c) from~(a) by the lifting theorem of Scott and Sorkin. The ring versions
  reuse the search tree of~(a), and the count of dominating sets reuses its analysis
  with $r=3$.

  (a) The algorithm is that of Gaspers and Sorkin, rules R1--R9, with the separations
  of \cref{lem:sep}(a) for $\Delta=3$ and $\mu_r(X)=|X|$. Since that lemma gives one
  algorithm, this algorithm has no parameter; $\eps$ and $N_0$ below enter only the
  analysis. Fix
  $\eps\in(0,\frac1{10})$, put $c=\cthree+\eps<\frac13$, and choose $N_0$ such that
  $\rho'_3(x)\le\eps x$ for $x\ge N_0$. For the analysis, an instance with fewer
  than $N_0$ vertices that arises in the Separation rule is a leaf: the algorithm runs on
  it as usual, and its running time is bounded by a constant $C_\eps$. By \cref{lem:sep}(a) the separations computed on
  larger instances satisfy $|S'|\le c|V|$ and $|L'|=|R'|$. By \cref{lem:lp} the weights
  of the lemma for this $c$ satisfy every constraint of \cref{tab:gs-constraints}, and
  $w_d=w_b+1$. By \cref{lem:gs-separation} and the derivations of
  \cite[Section~3.2]{GaspersSorkin2017} for the other rules, every rule satisfies the
  condition of \cref{lem:smc-measure} with $\lambda=r$, and $\nu$ decreases in every
  step. So the number of leaves of the search tree is at most $r^{\mu}$ for the initial
  measure
  \[
    \mu(\emptyset,\emptyset,V)=w_r n+w_d\log_{3/2}n=\frac{2c}{1+4c}\,n+O(\log n),
  \]
  and, since $\nu$ is polynomial in $n$ and each step takes polynomial time, the
  running time is $C_\eps\,r^{\frac{2c}{1+4c}n}\poly(n)$. The function
  $c\mapsto\frac{2c}{1+4c}$ is continuous, so letting $\eps\to0$ gives
  $r^{\gamma n+o(n)}$ for the one algorithm. The recursion depth is polynomial and each
  call uses polynomial space, so the space is polynomial. For a cubic graph $m=\frac32n$,
  so $\gamma n=\alpha m$.

  (b), (c) Scott and Sorkin~\cite[Theorem~22]{ScottSorkin2007} show that an algorithm
  solving cubic instances in time $O^*(r^{\alpha' m})$ with $\frac19\le\alpha'\le\frac15$
  yields one solving instances of maximum degree~$4$ in time
  $O^*(r^{(\frac18+\frac38\alpha')m})$ and arbitrary instances in time
  $O^*(r^{(\frac7{50}+\frac3{10}\alpha')m})$; the new algorithm uses the cubic one as a
  subroutine and is in polynomial space if the
  subroutine is \cite[Section~3.3]{GaspersSorkin2017}. This is how Gaspers and Sorkin
  obtain $\frac7{40}$ and $\frac9{50}$ from $\alpha'=\frac2{15}$. Our cubic algorithm
  runs in time $O^*(r^{(\alpha+\eps)m})$ for every $\eps>0$, and
  $\alpha\approx0.1195606\in[\frac19,\frac15]$, so the lifted algorithm runs in time
  $O^*(r^{(\frac18+\frac38(\alpha+\eps))m})$ and
  $O^*(r^{(\frac7{50}+\frac3{10}(\alpha+\eps))m})$ for every small $\eps>0$; again the
  algorithm does not depend on $\eps$.

  RCSP and PCSP: Gaspers and Sorkin~\cite[Theorems~3.2 and~3.3]{GaspersSorkin2017}
  obtain their bounds by running the same algorithm with the ring versions of
  Reductions~0--3 of Scott and Sorkin~\cite{ScottSorkin2009}, which change the
  constraint graph exactly as the original reductions and cost $O(r^3)$ ring operations
  each. The search tree, and hence the analysis, is that of~(a); the lift to maximum
  degree~$4$ and to all instances is the same.

  Dominating sets: Gaspers and Sorkin~\cite[Section~5.1, Theorem~5.1]{GaspersSorkin2017}
  count the dominating sets of every size of a subcubic graph $G$ by branching, with a
  three-way branching (``in'', ``optional'', ``forbidden''), on the vertices of the
  cubic structure of $G$, the cubic multigraph obtained by contracting the edges at
  vertices of degree at most~$2$. The pivot is chosen exactly as in the Max 2-CSP
  algorithm, the analysis is that of cubic Max $(3,2)$-CSP, and instances whose cubic
  structure has at most a constant number $A$ of vertices are solved in polynomial time
  by dynamic programming over a tree decomposition of width at most~$A$. We use the
  separations of \cref{lem:sep}(a) of the underlying simple graph of the cubic structure
  (a separation of it is one of the multigraph) and take $A\ge N_0$. The analysis of~(a)
  with $r=3$ gives $3^{\gamma n+o(n)}$.
\end{proof}

\subsection{Counting independent sets, and graph coloring}\label{app:is-statement}

Gaspers and Lee~\cite{GaspersLee2023} count the independent sets of a graph in
polynomial space with a branching algorithm whose subcubic part is a Separate, Measure
and Conquer subroutine. It branches on the vertices of a balanced separator, computed
from a path decomposition (\cref{lem:sep}), until the graph falls apart into two halves
that are counted independently; a measure that gives the separator vertices extra
weight (\cref{sec:smc-technique}) makes the measure drop at each branching on a
separator vertex by enough to cover the extra weight of the separator. They extend the
measure to maximum degree~$4$ and to all graphs with Wahlstr\"om's compound
measures~\cite{Wahlstrom2008}, and they compute the chromatic number from these counts
with the inclusion--exclusion method of Bj\"orklund, Husfeldt and
Koivisto~\cite{BjorklundHK2009}. The separator constant $\frac16$ enters their
subcubic analysis in two places, the separator constraint and the function $g(d)$
of~\cite[Lemma~3]{GaspersLee2023}. A better pathwidth bound shortens the separator
phase, and the gain carries over to the larger degrees
and to the chromatic number. With $\cthree$ in place of $\frac16$
(\cref{lem:sep}(b)) and the changes to the algorithm described below, their analysis,
adapted, gives the following bounds.

\begin{theorem}[counting independent sets, and coloring]\label{thm:is}
  Let $G$ be a graph with $n$ vertices. The number of independent sets of $G$ can be
  computed
  \begin{enumerate}[label=\textup{(\alph*)}]
    \item\label{thm:is:3} in polynomial space and time $O(1.1304^n)$ if $\Delta(G)\le3$;
    \item\label{thm:is:4} in polynomial space and time $O(1.2037^n)$ if $\Delta(G)\le4$;
    \item\label{thm:is:poly} in polynomial space and time $O(1.2341^n)$, and the chromatic
      number of $G$ in polynomial space and time $O(2.2341^n)$;
    \item\label{thm:is:exp} in time and space $O(1.2287^n)$.
  \end{enumerate}
\end{theorem}

The previous bounds are $O(1.1389^n)$, $O(1.2070^n)$, $O(1.2356^n)$, $O(2.2356^n)$ and
$O(1.2330^n)$ \cite[Lemmas~5 and~6, Theorems~1 and~2]{GaspersLee2023}. The
proof of \cref{thm:is}, in the rest of this appendix, is computer-assisted. It runs the analysis of
Gaspers and Lee with $\cthree$ in place of $\frac16$, with three changes. First, the
analysis of their subcubic subroutine lacks condition~(30) of Gaspers and
Sorkin~\cite{GaspersSorkin2017} (\cref{sec:smc-technique}), without which a step that
crosses the balance threshold of the measure can decrease the measure by less than its
branching vector counts. Their local analysis also has some further minor gaps. Fixing
condition~(30) increases the running time given by the analysis; for maximum
degree~$3$ we obtain $O(1.1304^n)$. In exchange, the new pathwidth bound makes the special
treatment of so-called spider vertices (the separator vertices to which their subroutine
gives a larger weight) unnecessary: in our subroutine every separator vertex has the same
weight, which gives a simpler algorithm and analysis. Second, above maximum degree~$3$
their branching on lazy $2$-separators need not satisfy the measure condition of
\cref{lem:smc-measure} (\cref{prop:isg-counterexample}); our rule tests that condition
in exact rational arithmetic. Third, we prove two steps that their paper leaves
unproved, the branching vectors of their Lemma~8 for degrees $5$ and~$6$ and a vector
for the lazy rule (\cref{prop:is-general-steps}).

A recent claim of $O^*(1.2321^n)$ for counting independent
sets~\cite{DeItaLuna2026} would beat part~\ref{thm:is:poly} of \cref{thm:is}, but not
part~\ref{thm:is:exp}. Its proof is not valid. For instance, its Lemma~3 asserts
$|E(G)|\le c\,|V(G)|$ for every graph $G$ and vertex $v$, with
$c=\max\{\delta(N[v])/(2|N[v]|),\,\Delta(G-N[v])/2\}$ and $\delta(N[v])$ the sum of the
degrees of the vertices of $N[v]$; for $G=K_{2,3}$ and $v$ on the side of size~$2$,
$c=\frac98$ and $|E(G)|=6>\frac{45}8$. Moreover, the analysis takes the maximum of bounds
obtained in two different measures.

We prove part~\ref{thm:is:3} of \cref{thm:is} in \cref{app:is3}, and
parts~\ref{thm:is:4}--\ref{thm:is:exp} in \cref{app:is-general}, using two steps proved
in \cref{app:is-g1,app:is-g2}. Numbering refers to the journal version of Gaspers and
Lee~\cite{GaspersLee2023}. The computer-assisted steps follow the paragraph
conventions of \cref{app:smc-computer}; the scripts named below are in
\texttt{experiments/algorithmic-consequences} of the accompanying code.

\paragraph{The algorithm of Gaspers and Lee.} The algorithm \#IS
\cite[Section~3]{GaspersLee2023} computes a weighted count of independent sets, the
weights (a \emph{cardinality function}) recording earlier steps. It solves graphs of
maximum degree at most~$2$ in polynomial time and splits a disconnected graph into its
components. Otherwise it picks a vertex $x$ of maximum degree that maximizes the
associated average degree of Wahlstr\"om~\cite{Wahlstrom2008}, and applies, in this
order: the \emph{multiplier reduction}, which folds a part of constant size attached at
a cut vertex~$v$ into the cardinality function of~$v$ and deletes it; a branching for a
\emph{lazy $2$-separator} (defined in \cref{app:is-general}); if $\Delta(G)=3$ and no
vertex of degree~$3$ has all neighbors of degree~$3$ (a \emph{$(3,3,3)$ vertex}), a
Separate, Measure and Conquer subroutine; otherwise it \emph{branches} on~$x$, counting
the independent sets of $G-x$ and of $G-N[x]$. All rules only delete vertices, so
degrees never increase, and for $\Delta(G)\le3$ no rule creates a $(3,3,3)$ vertex
(it would have been one before).

\subsection{Maximum degree 3}\label{app:is3}

\paragraph{Skeletons.} Let $G$ be a graph with $\Delta(G)\le3$ and $\delta(G)\ge2$,
and let $V_3$ be its set of vertices of degree~$3$. A \emph{$2$-path} is a path of
length at least~$2$ whose inner vertices have degree~$2$ and whose ends lie in~$V_3$.
The \emph{skeleton} of $G$ is the multigraph on $V_3$ with one edge for every edge of
$G$ inside $V_3$ and one edge for every $2$-path. Every vertex of the skeleton has
degree~$3$, a loop counted twice; a loop at $v$ makes $v$ a cut vertex. The far end of
a skeleton edge at $v$ is a \emph{skeleton neighbor} of~$v$. For a separation $(L,S,R)$
with $S\subseteq V_3$, the inner vertices of a $2$-path lie all in $L$ or all in $R$,
so no skeleton edge joins $L$ and~$R$. For $v\in S$ let $e_L(v)$, $e_R(v)$ and $e_S(v)$
be the numbers of skeleton edges at $v$ whose far end lies in $L$, $R$ and $S$.

\paragraph{The measure.} Let $r_3,s_3>0$ be weights, $B=42r_3$ and $J=40r_3$. For a
separation $(L,S,R)$ let $a=r_3|R\cap V_3|$, $b=r_3|L\cap V_3|$,
$\Xi(a,b)=\max\{a,b,\frac{a+b}2+B\}$, let $\eta>0$, and let
$\Lambda(x)=(1+B)\log_{1+\eta}\max\{1,x\}$. The measure of a separated instance is
\begin{equation}\label{eq:gl-measure}
  \mu_{\mathrm{sep}}(L,S,R)=s_3|S|+\Xi(a,b)+\Lambda\bigl(\max\{a,b\}+s_3|S|\bigr).
\end{equation}
This is \eqref{eq:smc-measure} with $\mu_r(X)=r_3|X\cap V_3|$ and $\mu_s(X)=s_3|X|$,
written symmetrically in the sides. It is nondecreasing in $a$, $b$ and $|S|$. We name
the sides so that $a\ge b$ and call $t=a-b$ the \emph{imbalance}. Each of the three
linear \emph{pieces} $a$, $b$, $\frac{a+b}2+B$ of $\Xi$ is at most $\Xi$, and one of
them attains it. Hence, if $(a,b)$ changes to $(a',b')$ and $\pi$ is a piece that
attains $\Xi(a',b')$,
\begin{equation}\label{eq:gl-max}
  \Xi(a',b')-\Xi(a,b)\le\pi(a',b')-\pi(a,b).
\end{equation}
This inequality, together with the pull rule~\ref{it:S6} below, replaces condition~(30)
of Gaspers and Sorkin~\cite{GaspersSorkin2017} (\cref{sec:smc-technique}).

\paragraph{The algorithm.} Let $N_0$ be a constant and $K_0=8$. To \emph{count
directly} a graph of maximum degree at most~$3$ with at most $N_0$ vertices of
degree~$3$ means to run the standard dynamic program over a path decomposition of
width $N_0+2$ (deleting the vertices of degree~$3$ leaves paths and cycles), in time
$2^{N_0+3}n^{O(1)}$ and polynomial space. The multiplier reduction is applied only to
parts with at most $K_0$ vertices of degree~$3$; such a part with its cut vertex has
pathwidth at most $K_0+3$, so the reduction takes polynomial time. An instance is a
graph $G$ with $\Delta(G)\le3$ and a cardinality function, and either no separation (a
\emph{fresh} instance) or a separation $(L,S,R)$ with $S\subseteq V_3$ (a
\emph{separated} instance). The input is fresh. The algorithm applies the first rule
that applies; for a fresh instance:
\begin{enumerate}[label=\textup{(F\arabic*)},leftmargin=*]
  \item\label{it:F1} if $\Delta(G)\le2$ or $n_3\le N_0$, count directly;
  \item\label{it:F2} if $G$ is disconnected, solve its components as fresh instances
    and multiply the results;
  \item\label{it:F3} apply the multiplier reduction;
  \item\label{it:F4} if $G$ has a $(3,3,3)$ vertex $x$ (chosen as by Gaspers and Lee):
    if exactly two skeleton edges join $N(x)$ to vertices outside $N[x]$, with far ends
    $y$ and $z$, branch on $y$, and in each branch fold what remains of the part cut off
    by $\{y,z\}$ that contains $x$ into $z$ by the multiplier reduction (or count it
    directly if it has become a component); otherwise branch on~$x$;
  \item\label{it:F5} otherwise compute a separation by \cref{lem:sep}(b) with
    $\mu_r(X)=r_3|X\cap V_3|$, which is $r_3$-Lipschitz, and continue with the separated
    instance.
\end{enumerate}
For a separated instance, with the sides named so that $a\ge b$:
\begin{enumerate}[label=\textup{(S\arabic*)},leftmargin=*]
  \item\label{it:S1} if $\Delta(G)\le2$ or $n_3\le N_0$, count directly;
  \item\label{it:S2} if $S=\emptyset$, solve the components of $G$ as fresh instances
    and multiply the results;
  \item\label{it:S3} if some component of $G$ has at most $N_0$ vertices of degree~$3$,
    count it directly and delete it;
  \item\label{it:S4} apply the multiplier reduction;
  \item\label{it:S5} \emph{(move)} if some $v\in S$ has $e_L(v)=0$, move $v$ into $R$,
    together with the inner vertices of its $2$-paths that lie in~$L$; symmetrically if
    $e_R(v)=0$;
  \item\label{it:S6} \emph{(pull)} if $t>2B-J$ and some $v\in S$ has $e_L(v)>e_R(v)$,
    let $q$ be the far end of the skeleton edge at $v$ that ends in~$R$; move $v$ and the
    inner vertices of that skeleton edge into $L$, and $q$ into~$S$;
  \item\label{it:S7} otherwise branch on any $v\in S$, keeping the separation
    restricted to the remaining vertices.
\end{enumerate}
In \ref{it:F2} and~\ref{it:S2}, components with at most $N_0$ vertices of degree~$3$
are counted directly. Every step that deletes vertices is followed, as part of the same
step, by a \emph{clean-up}: vertices of degree~$1$ are folded into their neighbors by
the multiplier reduction and vertices of degree~$0$ are counted, as long as there are
any; then, in a separated instance, the separator is \emph{normalized}:
\begin{enumerate}[label=\textup{(N)},leftmargin=*]
  \item\label{it:N1} while some $u\in S$ has degree~$2$, let $P$ be the maximal path or
    cycle through $u$ whose inner vertices have degree~$2$, with ends $f_1,f_2\in V_3$ if
    it is a path; if no end lies in $R$, move the inner vertices of $P$ into $L$; else, if
    no end lies in $L$, move them into $R$; else, say $f_1\in L$ and $f_2\in R$, move them
    into $L$ and $f_2$ into~$S$.
\end{enumerate}
Each operation of~\ref{it:N1} gives a separation and lowers the number of vertices of
degree~$2$ in $S$, so afterwards $S\subseteq V_3$. Every rule computes the weighted
count exactly; the separation only steers the choice of vertices.

\begin{lemma}[moves and deletions]\label{lem:is3-moves}
  Let $s_3>r_3$, and let $I'$ be obtained from a separated instance $I$ by \ref{it:S3},
  \ref{it:S4}, \ref{it:S5} or~\ref{it:S6}, including the clean-up. Then
  $\mu_{\mathrm{sep}}(I')\le\mu_{\mathrm{sep}}(I)$, and the argument $\max\{a,b\}+s_3|S|$ of $\Lambda$ does
  not increase.
\end{lemma}

\begin{proof}
  \ref{it:S3}, \ref{it:S4} and the removal of vertices of degree at most~$1$ delete
  vertices and lower degrees, so none of $a$, $b$, $|S|$ increases. An operation
  of~\ref{it:N1} moves vertices of weight~$0$ and either lowers $|S|$ or keeps $|S|$ and
  lowers $a$ by~$r_3$. In~\ref{it:S5} with $e_L(v)=0$, the result is a separation, $|S|$
  drops by one, $a$ grows by $r_3$ and $b$ is unchanged; by \eqref{eq:gl-max} $\Xi$ grows
  by at most $r_3$, and the argument of $\Lambda$ changes by at most $r_3-s_3<0$.
  In~\ref{it:S6}, $e_R(v)\ge1$ and $e_L(v)\ge2$ force $(e_L,e_R,e_S)(v)=(2,1,0)$, so $q$
  is unique and the result is a separation; $|S|$ is unchanged, $a'=a-r_3$, $b'=b+r_3$
  and $a'-b'=t-2r_3>2B-J-2r_3>0$. So the piece $b$ is not active at $(a',b')$, and by
  \eqref{eq:gl-max} $\Xi$ does not increase; the argument of $\Lambda$ drops by $r_3$.
\end{proof}

For $G$ with $\Delta(G)\le3$ and $\delta(G)\ge2$, the \emph{closure} of
$X_0\subseteq V_3$ is the smallest $Z\supseteq X_0$ that contains every vertex of $V_3$
joined to $Z$ by at least two skeleton edges. Let $\partial Z$ be the set of skeleton
edges with exactly one end in $Z$, and $Y$ the set of vertices of $V_3\setminus Z$
joined to $Z$ by exactly one skeleton edge.

\begin{lemma}[the clean-up after a deletion]\label{lem:is3-closure}
  Let $\Delta(G)\le3$, $\delta(G)\ge2$, let $\emptyset\ne X_0\subseteq V_3$ have closure
  $Z$, and let $G'$ be obtained from $G$ by deleting $X_0$ and some neighbors of
  degree~$2$ of vertices of $X_0$, and then removing vertices of degree at most~$1$ as
  long as there are any. Then $G'$ is $G$ minus $Z$ and the inner vertices of the skeleton
  edges with an end in~$Z$; in $G'$ the vertices of $Y$ have degree~$2$ and those of
  $V_3\setminus(Z\cup Y)$ degree~$3$. Moreover $|Y|=|\partial Z|\le|\partial X_0|$,
  $|Z|\le|X_0|+|\partial X_0|$, and $\partial Z\ne\emptyset$ if some component of the
  skeleton that meets $X_0$ has more than $|X_0|+|\partial X_0|$ vertices.
\end{lemma}

\begin{proof}
  Let $W$ be $Z$ with the inner vertices of the skeleton edges at~$Z$. The removals
  leave the largest subgraph of minimum degree at least~$2$, whatever their order. In
  $G-W$ a vertex of $V_3\setminus Z$ has lost one edge per skeleton edge to $Z$, so it
  has degree~$2$ if it is in $Y$ and $3$ otherwise; all deleted vertices lie in~$W$, so
  no vertex outside $W$ is removed. Conversely, $Z$ arises from $X_0$ by adding, one at a
  time, vertices joined to the current set by two or more skeleton edges; once the
  current set is gone, the inner vertices of its skeleton edges are removed one after
  another, and then the next added vertex has degree at most~$1$. So all of $W$
  disappears. Adding a vertex joined to the set by $k\ge2$ skeleton edges changes the
  number of leaving skeleton edges by $(3-k)-k\le-1$, so
  $|Y|=|\partial Z|\le|\partial X_0|-|Z\setminus X_0|$. If $\partial Z=\emptyset$, then
  $Z$ contains every component of the skeleton that meets~$X_0$.
\end{proof}

\begin{lemma}[branchings]\label{lem:is3-credits}
  Let $N_0\ge17$, $s_3+\frac32r_3\ge1$, $2s_3\ge1$ and $s_3\ge\frac{r_3}2$. Let $I$ be a
  separated instance to which none of \ref{it:S1}--\ref{it:S6} applies, and let $I'$ be
  either branch of~\ref{it:S7}, including the clean-up. Then
  $\mu_{\mathrm{sep}}(I')\le\mu_{\mathrm{sep}}(I)-1$, and the argument of $\Lambda$ does not increase.
\end{lemma}

\begin{proof}
  Let $v\in S$ be the branching vertex. As \ref{it:S1}--\ref{it:S4} do not apply,
  $\delta(G)\ge2$, every component of $G$ meeting $V_3$ has more than $N_0\ge17$ vertices
  of degree~$3$, and the skeleton has no loop; as \ref{it:S5} does not apply,
  $e_L(v),e_R(v)\ge1$. Let $X_0=\{v\}$ in the branch $G-v$, and let $X_0$ be $v$ with its
  neighbors of degree~$3$ in the branch $G-N[v]$. With $Z,Y$ as in
  \cref{lem:is3-closure}, $|\partial X_0|\le12$, so $|Y|\le12$, $|Z|\le16$ and
  $\partial Z\ne\emptyset$.

  We have $|Z\setminus\{v\}|+|Y|\ge3$: if $Z=\{v\}$, the three skeleton neighbors of $v$
  are distinct and form $Y$; if $Z=\{v,u\}$, at most two skeleton edges join $v$ and $u$,
  so $|\partial Z|\ge2$; if $|Z|\ge3$, use $|Y|\ge1$. After the clean-up the vertices of
  degree~$2$ in $S$ are those of $Y\cap S$. Each path or cycle $P$ of~\ref{it:N1} through
  $k\ge1$ of them takes these $k$ vertices out of $S$ and puts at most one end $f\in R$
  into~$S$; each such $f$ serves one $P$ only. Give a \emph{credit} to every vertex of
  $(Z\setminus\{v\})\cup Y$: an \emph{$a$-credit} if it lies in $R$, a
  \emph{$b$-credit} if it lies in $L$, an \emph{$s$-credit} if it lies in
  $(Z\cap S)\setminus\{v\}$; on each $P$, $k-1$ of its vertices of $Y\cap S$ get an
  $s$-credit and the last gets an $a$-credit (for $f$) if an end enters $S$ and an
  $s$-credit otherwise. With $\#_a,\#_b,\#_s$ the numbers of credits,
  $\#_a+\#_b+\#_s\ge3$, and, as no vertex of degree~$3$ enters $L$ or~$R$,
  \[
    |S'|\le|S|-1-\#_s,\qquad a'\le a-r_3\#_a,\qquad b'\le b-r_3\#_b .
  \]
  So the argument of $\Lambda$ does not increase, and the imbalance changes by at most
  $r_3(|Z|+2|Y|)\le40r_3=J$.

  If $t\le2B-J$, then $|a'-b'|\le2B$, $\Xi=\frac{a+b}2+B$ before and after, and the
  measure drops by at least $s_3(1+\#_s)+\frac{r_3}2(\#_a+\#_b)\ge s_3+\frac32r_3\ge1$.
  If $t>2B-J$, then, as \ref{it:S6} does not apply, $(e_R,e_L,e_S)(v)$ is $(2,1,0)$ or
  $(1,1,1)$. We claim $\#_s\ge1$ or $\#_a\ge2$. Suppose $\#_s=0$; then $Z\cap S=\{v\}$
  and every vertex of $Y\cap S$ has an $a$-credit, so every skeleton edge from
  $Z_R\cup\{v\}$, where $Z_R=Z\cap R$, to a vertex of $(R\cup S)\setminus Z$ yields its
  own $a$-credit. If $Z_R=\emptyset$, the $e_R(v)$ edges into $R$ give $e_R(v)$ credits,
  and if $e_R(v)=1$ the skeleton neighbor of $v$ in $S$ gives another. If $|Z_R|\ge2$ we
  are done. If $Z_R=\{z\}$, at most two of the three skeleton edges at $z$ go to $v$, and
  the others end in $(R\cup S)\setminus Z$. Now if $\#_s\ge1$, $|S|$ drops by two and the
  measure by $2s_3\ge1$. Otherwise $\#_a\ge2$, $\#_a+\#_b\ge3$ and
  $a'-b'\ge t-J>2B-2J>0$, so $b$ is not active after the step and, by \eqref{eq:gl-max},
  $\Xi$ drops by at least $\min\{2r_3,\frac32r_3\}$; the measure drops by at least
  $s_3+\frac32r_3\ge1$.
\end{proof}

\begin{lemma}[initial measure]\label{lem:is3-initial}
  Let $c\in(\cthree,\frac13)$, $s_3\ge\frac{r_3}2$, $\kappa=c\,s_3+\frac{1-c}2r_3$ and
  $\kappa<r_3$. There are $N'_0$ and $\eta'>0$ such that for every $\eta>0$ and every
  connected $G$ with $\Delta(G)\le3$, $\delta(G)\ge2$ and $n_3\ge N'_0$, the separation of
  \ref{it:F5} satisfies
  $\mu_{\mathrm{sep}}(L,S,R)\le\kappa n_3+2B+\Lambda\bigl(r_3n_3/(1+\eta')\bigr)$, and the
  argument of $\Lambda$ in it is at most $r_3n_3/(1+\eta')$.
\end{lemma}

\begin{proof}
  By \cref{lem:sep}(b), $k=|S|\le\cthree n_3+\rho_3(n_3)\le cn_3$ for large $n_3$,
  $|a-b|\le r_3$ and $a+b=r_3(n_3-k)$. As $\max\{a,b\}=\frac{a+b}2+\frac{|a-b|}2$, both
  $s_3k+\Xi(a,b)-B$ and $\max\{a,b\}+s_3k$ are at most
  $s_3k+\frac{r_3(n_3-k)}2+\frac{r_3}2$, which is nondecreasing in $k$; so the first
  is at most $\kappa n_3+B$ (as $\frac{r_3}2<B$), and the second at most
  $\kappa n_3+\frac{r_3}2\le r_3n_3/(1+\eta')$ for a suitable $\eta'$ and large $n_3$, as
  $\kappa<r_3$. The bound on the measure follows as $\Lambda$ is nondecreasing.
\end{proof}

\paragraph{Measures of all instances.} Graphs with a $(3,3,3)$ vertex are measured
with weights $w'_2,w'_3$ for the vertices of degree $2$ and~$3$, as in
\cite[Lemma~5]{GaspersLee2023}. Let $\ell(G)=w'_2n_2+w'_3n_3$,
$m(G)=\min\{\ell(G),\kappa n_3\}$, and let $\eta'$ be as in \cref{lem:is3-initial}.
The measure $\mu(I)$ of an instance $I$ on $G$ is $0$ if $\Delta(G)\le2$ or $n_3\le
N_0$; otherwise it is $\mu_3(G)=\ell(G)+\Lambda(r_3n_3)$ for a fresh instance with a
$(3,3,3)$ vertex, $\mu_f(G)=m(G)+2B+\Lambda\bigl(r_3n_3/(1+\eta')\bigr)$ for a fresh
instance without one, and $\mu_{\mathrm{sep}}(L,S,R)$ for a separated instance.

\begin{table}[tb]
  \centering
  \caption{The constraints of the subcubic analysis and their values at the weights
    \eqref{eq:is3-weights}, with $c=0.13985$ and $\kappa=c\,s_3+\frac{1-c}2r_3$. Linear
    constraints read ``expression $\ge0$'' and are checked exactly; for the branching
    vectors $(a,b)$ the last column is an upper bound on $2^{-a}+2^{-b}$ in interval
  arithmetic. Decimals are rounded.}
  \label{tab:gl-linear}
  \small
  \begin{tabular}{@{}lll>{$}r<{$}@{}}
    \toprule
    No. & Origin & Expression & \text{value}\\
    \midrule
    (U1) & \cref{lem:is3-credits} & $s_3+\frac32r_3-1$ & 0\\
    (U2) & \cref{lem:is3-moves,lem:is3-credits} & $s_3-r_3$;\ $2s_3-1$ & 0.5516;\ 0.4619\\
    (U3) & \cref{lem:is3-initial}, \eqref{eq:smc-ratio} & $r_3-\kappa$ & 1.108\cdot10^{-7}\\
    (U4) & hand-over \ref{it:F5} & $w'_2$;\ $2w'_3+w'_2-2\kappa$ & 0.0052;\ 4.22\cdot10^{-7}\\
    \midrule
    (B11) & \ref{it:F4}, branch on $x$ & $(4w'_3-3w'_2,\ 8w'_3-4w'_2)$ & 0.9999956\\
    (B12) & \ref{it:F4}, branch on $y$ & $(6w'_3-w'_2,\ 6w'_3-w'_2)$ & 0.9623865\\
    \bottomrule
  \end{tabular}
\end{table}

\paragraph{Weights and constants.} Let $c=0.13985$, so $\cthree<0.1398222<c<\frac13$,
and
\begin{equation}\label{eq:is3-weights}
  r_3=0.179364,\quad s_3=0.730954=1-\tfrac32r_3,\qquad
  w'_2=0.0052182,\quad w'_3=0.176755 ,
\end{equation}
exact decimals. They satisfy the constraints of \cref{tab:gl-linear}, checked by
\texttt{is3\_uniform\_repair.py} (part~1); in particular $w'_2\le w'_3$ and
$\kappa<r_3$. Let $N'_0,\eta'$ be as in \cref{lem:is3-initial}. We fix $\eta>0$ with
$(1+B)\log_{1+\eta}(1+\eta')\ge2B+1$, and $N_0\ge\max\{N'_0,17\}$ with $w'_3N_0\ge1$,
$\kappa N_0\ge1$ and $r_3N_0\ge1+\eta'$. Then every graph with $n_3>N_0$ satisfies
\begin{equation}\label{eq:is3-mu-f}
  \mu_f(G)\le\mu_3(G)-1,
\end{equation}
as $m\le\ell$ and $\Lambda(r_3n_3/(1+\eta'))=\Lambda(r_3n_3)-(1+B)\log_{1+\eta}(1+\eta')$.
For reals $x_1,\dots,x_p\ge1$, induction on
$2^{x+y}-2^x-2^y=(2^x-1)(2^y-1)-1\ge0$ gives
\begin{equation}\label{eq:is3-superadd}
  \textstyle\sum_{j=1}^p2^{x_j}\le2^{x_1+\dots+x_p}.
\end{equation}

\begin{proof}[Proof of \cref{thm:is}\ref{thm:is:3}]
  We apply \cref{lem:smc-measure} with $\lambda=2$, no subroutine, the measure $\mu$ and
  the secondary measure $\nu$ defined below, and check its condition rule by rule.
  \ref{it:F1}, \ref{it:S1} and the direct counts make no recursive call.
  \ref{it:S3}--\ref{it:S6}: \cref{lem:is3-moves}. \ref{it:F3} deletes vertices and lowers
  degrees; as $0\le w'_2\le w'_3$, neither $\ell$ nor $n_3$ grows, no $(3,3,3)$ vertex
  appears, and if the $(3,3,3)$ vertices disappear the measure changes from $\mu_3$ to
  $\mu_f\le\mu_3-1$. \ref{it:S7}: \cref{lem:is3-credits}, with the vector $(1,1)$.

  \ref{it:F2}: let $C_1,\dots,C_p$ be the components with more than $N_0$ vertices of
  degree~$3$. If $G$ has a $(3,3,3)$ vertex, $\mu(C_j)\le\ell(C_j)+\Lambda(r_3n_3(G))$ by
  \eqref{eq:is3-mu-f} and $\sum_j\ell(C_j)\le\ell(G)$. Otherwise
  $\mu(C_j)\le m(C_j)+2B+\Lambda(r_3n_3(G)/(1+\eta'))$ and $\sum_jm(C_j)\le m(G)$. As
  $\ell(C_j)\ge m(C_j)\ge1$, \eqref{eq:is3-superadd} gives $\sum_j2^{\mu(C_j)}\le2^{\mu(G)}$.

  \ref{it:F4}: $G$ is connected with $n_3>N_0$ and $\delta(G)\ge2$, its skeleton has no
  loop, the new instances are fresh with measure at most $\ell+\Lambda(r_3n_3(G))$ by
  \eqref{eq:is3-mu-f}, so it suffices that $\ell$ drops by the components of (B11) or
  (B12). Let $e$ be the number of skeleton edges inside $N(x)$; then $6-2e$ skeleton
  edges leave $N[x]$, and $e=3$ would make $n_3=4$. If $6-2e\ge4$ we branch on $x$. In
  $G-x$, $\ell$ drops by $4w'_3-3w'_2$. In $G-N[x]$ let $Z,Y$ be as in
  \cref{lem:is3-closure} for $X_0=N[x]$; as $n_3>17>|X_0|+|\partial X_0|$, $\ell$ drops by
  at least $|Z|w'_3+|Y|(w'_3-w'_2)$ with $|Z|\ge4$ and $|Y|\ge1$, which is at least
  $8w'_3-4w'_2$ if $|Z|+|Y|\ge8$. If $Z=N[x]$, then $|Y|=6-2e\ge4$. Otherwise $|Z|\ge5$
  and $|Y|=|\partial Z|\equiv3|Z|\pmod2$, so $|Z|+|Y|\ge8$ unless $|Z|=5$, $|Y|=1$; then
  the vertex of $Y$ cuts off a part with five vertices of degree~$3$, and \ref{it:F3}
  would apply as $K_0\ge5$. If $6-2e=2$, let $H$ be $N[x]$ with the inner vertices of
  the skeleton edges at $N(x)$; it has four vertices of degree~$3$, and all edges leaving
  $H$ end in $y$ or~$z$. If $y=z$, \ref{it:F3} would apply at~$y$; so $y\ne z$ and
  $d(y)=d(z)=3$. In both branches $y$ and the four vertices of degree~$3$ of $H$ are
  deleted and $z$ loses an edge or is deleted, so $\ell$ drops by at least
  $5w'_3+(w'_3-w'_2)$, which is (B12).

  \ref{it:F5}: $G$ is connected, $\delta(G)\ge2$, $n_3>N_0\ge N'_0$, and every vertex of
  degree~$3$ has a neighbor of degree~$2$, so $n_3\le2n_2$. If $w'_3\ge\kappa$, then
  $\ell(G)\ge\kappa n_3$; otherwise
  $\ell(G)-\kappa n_3\ge(\frac{w'_2}2+w'_3-\kappa)n_3\ge0$ by (U4). So
  $m(G)=\kappa n_3$, and \cref{lem:is3-initial} gives $\mu(I')\le\mu_f(G)$.

  \ref{it:S2}: let the instance be $(L,\emptyset,R)$ with $a\ge b$, so $\mu\ge
  a+\Lambda(a)$. A component $C$ with $n_3(C)>N_0$ lies in $L$ or in $R$, has no
  $(3,3,3)$ vertex, and $1+\eta'\le r_3n_3(C)\le a$; by $\kappa<r_3$ and the choice
  of~$\eta$,
  \[
    \mu(C)\le\kappa n_3(C)+2B+\Lambda(r_3n_3(C))-(1+B)\log_{1+\eta}(1+\eta')\le
    r_3n_3(C)+\Lambda(a)-1 .
  \]
  As $\sum_{C\subseteq R}r_3n_3(C)\le a$ and $\sum_{C\subseteq L}r_3n_3(C)\le b\le a$,
  \eqref{eq:is3-superadd} gives $\sum_C2^{\mu(C)}\le2^{\Lambda(a)-1}(2^a+2^b)\le2^\mu$.
  If $G$ is connected, the new instance is $G$ itself, with measure at most $\mu-1$.

  \emph{Secondary measure.} For an instance on a graph with $n$ vertices and $m$ edges
  let $p=n+m$, $T=3(n+1)^2$, $M=\lfloor n+2B+\Lambda(n)\rfloor\ge\lfloor\mu\rfloor$ (as
  $s_3+r_3<1$ and $w'_3<1$), $\tau=(2n+2)|S|+|a-b|/r_3<T$ for a separated and $\tau=T$
  for a fresh instance, and
  $\nu=(M+1)(T+1)p+(T+1)\lfloor\mu\rfloor+\tau$. A step that deletes a vertex or an edge
  lowers $p$ and so $\nu$. Of the steps that keep the graph, none raises $\mu$;
  \ref{it:F5} lowers $\tau$ from $T$; \ref{it:S2} on a connected graph lowers
  $\lfloor\mu\rfloor$ and raises $\tau$ by at most $T$; \ref{it:S5} lowers $\tau$ by at
  least $2n+1$, and \ref{it:S6} by~$2$. So $\nu$ drops in every step, and the rules
  terminate.

  \emph{Conclusion.} The input has $\mu\le\mu_3(G)\le w'_3n+\Lambda(n)$. By
  \cref{lem:smc-measure} the running time is $2^{w'_3n}n^{O(1)}$, and
  $2^{0.176755}<1.130339$. Every step takes polynomial time and space, and the recursion
  depth is at most the initial $\nu$, so the space is polynomial.
\end{proof}

\subsection{Maximum degree at least 4}\label{app:is-general}

\paragraph{The algorithm.} On graphs of maximum degree at most~$3$ the algorithm is
that of \cref{app:is3}. Fix constants $K\ge65$, $K_m\ge6K+1$ and $N_1=7K+2$. In a
graph of maximum degree at least~$4$ the algorithm splits a disconnected graph into its
components, solves every component with at most $N_1$ vertices directly in constant
time, and otherwise selects a vertex $x$ of maximum degree: as by Gaspers and Lee if
$\Delta(G)=4$ or the component is regular, and otherwise, if $\Delta(G)\ge5$, a vertex
of maximum degree with a neighbor of smaller degree (one exists in every connected
non-regular graph). It then applies the first of the following rules that applies.
\begin{itemize}
  \item \emph{Multiplier reduction} at a cut vertex $v$ such that some component of
    $G-v$ has at most $K_m$ vertices.
  \item \emph{Hybrid rule}, if $x$ has a \emph{lazy $2$-separator}: distinct vertices
    $y,z$ with $d(y),d(z)\ge3$ such that $G-\{y,z\}$ is disconnected and
    $N[x]\setminus\{y,z\}$ lies in one component of $G-\{y,z\}$ with at most $K$
    vertices ($x$ may belong to the pair). For each $v\in\{y,z,x\}$ in this order it
    computes the two instances of the branching on $v$, and it branches on the first $v$
    whose instances $I_1,I_2$ satisfy $2^{\mu(I_1)}+2^{\mu(I_2)}\le2^{\mu(G)}$ for the
    measures below without their sublinear term $\Lambda$, and on $y$ if there is none.
    The weights are rational, so this is a polynomial-time test in exact arithmetic. As
    $\Lambda(I_j)\le\Lambda(G)$, a branching that passes the test satisfies the same
    inequality for the full measures.
  \item Otherwise branch on~$x$.
\end{itemize}
The \emph{instances} of a branching on $v$ are $G-v$ and $G-N[v]$, each taken after the
reductions that make no recursive call, applied as long as one applies: the removal of
vertices of degree at most~$1$, the multiplier reduction, and the direct solution of
components with at most $N_1$ vertices. Our lazy $2$-separator is a reading of the rule
of Gaspers and Lee~\cite[Algorithm~\#IS]{GaspersLee2023}, which branches on~$y$.
\Cref{prop:isg-counterexample} below shows that under this reading their branching
alone can violate the condition of \cref{lem:smc-measure}.

\paragraph{The measures.} We follow \cite[Lemmas~6 and~8, Theorem~1]{GaspersLee2023}.
All measures have the form $\sum_iw_in_i$ with $w_0=w_1=0$, plus, for maximum degree
at least~$4$, the sublinear term $\Lambda(r_3|V(G)|)$, which no step increases and which
is at least the term $\Lambda(r_3n_3)$ of $\mu_3$ of every graph of maximum degree at
most~$3$ that arises from $G$; we omit it below. We write $\Delta w_i=w_i-w_{i-1}$. From now on a $2$-path is a
path of length at least~$2$ whose inner vertices have degree~$2$ and whose ends have
degree at least~$3$; its \emph{far end} from one end is the other.

\emph{Maximum degree $4$.} The pivots $p_2,\dots,p_5=3,\frac{16}5,\frac72,\frac{15}4$
cut the interval $[2,4]$ of average degrees $d(G)=2|E(G)|/|V(G)|$ into five
\emph{sections}. Section~$1$, $[2,3]$, has the weights
$(w^{(1)}_2,w^{(1)}_3)=(w'_2,w'_3)$ of \eqref{eq:is3-weights} and a weight $w^{(1)}_4$,
and its measure $\mu^{(1)}$ has the additional term $\psi\cdot\mathrm{ind}(G)$, where
$\psi\ge0$ and $\mathrm{ind}(G)=1$ if all degrees of $G$ are $2$ or $4$ and no two
vertices of degree~$4$ are adjacent, and $\mathrm{ind}(G)=0$ otherwise. Section $k\ge2$
has $w^{(k)}_i=w^{(k-1)}_i-(i-p_k)\alpha_k$ ($i=2,3,4$) with $\alpha_k\ge0$, so that
\begin{equation}\label{eq:gl-section-diff}
  \mu^{(k)}_{\mathrm{lin}}(G)-\mu^{(k-1)}_{\mathrm{lin}}(G)=-\alpha_k|V(G)|\,(d(G)-p_k),
\end{equation}
where $\mu^{(k)}_{\mathrm{lin}}=\sum_iw^{(k)}_in_i$ and $\mu^{(k)}=\mu^{(k)}_{\mathrm{lin}}$
for $k\ge2$. A graph $G$ of maximum degree~$4$ has the measure
$\mu_4(G)=\min_k\mu^{(k)}(G)$, and its \emph{minimizing section} is the smallest $k$
with $\mu^{(k)}(G)=\mu_4(G)$. Let $\tau_k$ be the maximum of $\sum_ix_iw^{(k)}_i$ over
the degree distributions $x$ on $\{2,3,4\}$ with average degree in section~$k$. The
degree-$4$ branchings are those of \cite[Table~1]{GaspersLee2023}, listed by the
degrees of the neighbors of $x$ (the \emph{type} of $x$), each with a branching vector
and a \emph{highest average degree}~$h$, the largest $d(G)$ of a graph in which the
selection rule can pick a vertex of this type (Wahlstr\"om's averaging
argument~\cite{Wahlstrom2008}, which we use as Gaspers and Lee do). Its
\emph{group} is the section that contains $h$, the lower of the two sections if $h$ is
a common end of two sections. Two entries of that table are
misprinted (as in Figure~4 of \cite{GaspersLee2016arXiv}): the first component for
type $(2,2,2,3)$ is $4w_4-2w_3+2w_2$, and in cases (a) and (b) of type $(2,2,2,2)$
both components are $5w_4-4w_3+4w_2+\psi$ and $4w_4-2w_3+3w_2-\psi$, respectively
(the signs of $\psi$ are swapped in print). Case~(a) is a vertex of type $(2,2,2,2)$
all of whose $2$-paths end at vertices of degree~$4$; case~(b) has a $2$-path to a
vertex of degree~$3$.

\emph{Degrees $5$ and $6$.} One measure $\sum_iw_in_i$ with weights $w_2,\dots,w_6$,
at least those of section~$5$ on degrees $2,3,4$, increasing and concave. Branching on a
vertex $v$ with $d(v)\in\{5,6\}$ has the vector \cite[Lemma~8]{GaspersLee2023}
\begin{equation}\label{eq:gl-lemma8}
  \Bigl(w_{d(v)}+\sum_{u\in N(v)}w_{d(u)}+\mathrm{out}(v)\,\Delta w_{d(v)},\ \
  w_{d(v)}+\sum_{u\in N(v)}\Delta w_{d(u)}+\deg_2(v)\,\Delta w_{d(v)}\Bigr),
\end{equation}
where $\deg_2(v)$ is the number of neighbors of degree~$2$, $\mathrm{out}(v)=5$ for the
neighbor degrees $(2,2,2,2,2)$ and $(2,2,2,2,2,3)$, $\mathrm{out}(v)=6$ for
$(2,\dots,2)$ with $d(v)=6$, and otherwise $\mathrm{out}(v)=3$ if
$\sum_{u\in N(v)}d(u)-d(v)$ is odd and $4$ if it is even.

\emph{Degree at least $7$.} The measure is $w_6|V(G)|$; a branching deletes one vertex
or at least eight.

\paragraph{Cross-section cases.} A type with highest average degree $h$ can be selected
in every graph with $d(G)\le h$. By \eqref{eq:gl-section-diff}, a graph whose average
degree lies inside section~$k$ has minimizing section~$k$, and a graph with $d(G)=p_k$
has minimizing section at most $k-1$. So in a graph with minimizing section $k$ the
selection rule picks only types whose group is at least $k$; we call these types
\emph{admissible} for section~$k$, and we check the vector of every type under the
weights of its own section and of every earlier section. A graph with
$\mathrm{ind}(G)=1$ has $4n_4\le2n_2$, so $d(G)\le\frac83$, and its minimizing section
is~$1$, since $\mu^{(2)}-\mu^{(1)}\ge\alpha_2|V(G)|(3-d(G))-\psi\ge\frac{\alpha_2}3-\psi>0$
for our weights.

\paragraph{The potential.} The printed vectors account for $\mathrm{ind}$ only in
cases (a) and~(b). Case~(a) is selected only when every vertex of degree~$4$ has type
$(2,2,2,2)$, the type of smallest associated average degree, and no vertex has
degree~$3$ (otherwise \#IS would branch on a case-(b) vertex), so $\mathrm{ind}(G)=1$;
case~(b) occurs only when $\mathrm{ind}(G)=0$. Other section-$1$ branchings may switch
the indicator from $0$ to~$1$. The
indicator is evaluated after the removal of vertices of degree at most~$1$; each
further reduction in the definition of the instances deletes at least two vertices of
degree at least~$2$, so it lowers the linear part of $\mu^{(1)}$ by at least
$2w^{(1)}_2=0.0104364>\psi$, and no $\mu^{(k)}$ increases even if the indicator
switches. Subtracting $\psi$ from both components keeps every section-$1$ vector
other than (a), (b) and $(2,2,2,3)$ valid (sums at most $0.90706$), and every
later-group vector under section-$1$ weights (sums at most $0.89110$). For
$(2,2,2,3)$, \cref{lem:psi-2223} gives the excess
$\varepsilon=(w^{(1)}_3-w^{(1)}_2)-(w^{(1)}_4-w^{(1)}_3)=0.0122914>\psi$ in each branch
where the indicator switches; with $\varepsilon-\psi$ added in the first, the second or
both components, the sums are at most $0.99543$, $0.99544$ and $0.99331$. These checks
are part~2 of \texttt{gaspers\_lee\_general\_uniform.py}.

In the next lemma $w_i=w^{(1)}_i$, the instance $G'$ of a branch is taken after the
removal of vertices of degree at most~$1$, and the \emph{printed component} of a
branch is the corresponding component of the vector of $(2,2,2,3)$ in
\cite[Table~1]{GaspersLee2023}. The \emph{decrease} $\sum_v(w_{d_G(v)}-w_{d_{G'}(v)})$,
with $w_0=w_1=0$ and deleted vertices of degree~$0$, is a sum of nonnegative terms.

\begin{lemma}[the potential in case $(2,2,2,3)$]\label{lem:psi-2223}
  Let $G$ be a connected graph with more than $N_1$ vertices, minimizing section~$1$,
  $\Delta(G)\le4$ and $\delta(G)\ge2$, to which neither the multiplier reduction nor the
  hybrid rule applies, and let $x$ have type $(2,2,2,3)$. If the instance $G'$ of a branch
  on $x$ has indicator~$1$, its decrease exceeds the printed component by at
  least~$\varepsilon$.
\end{lemma}

\begin{proof}
  Let $u$ be the neighbor of degree~$3$ of $x$, $y_1,y_2,y_3$ the others, and $t_i$ the
  far end of the $2$-path that starts with $xy_i$. The printed components charge $w_4$ to
  $x$, $w_2$ to each $y_i$, $w_4-w_3$ to each $t_i$ (the least drop of a vertex of degree
  at least~$3$ that loses an edge, as $w_4-w_3<w_3-w_2$), and $w_3-w_2$ (branch $G-x$)
  or $w_3$ (branch $G-N[x]$) to $u$. In $G'$ no vertex has degree~$3$.

  \emph{Closed parts.} Call $P\supseteq\{x,y_1,y_2,y_3\}$ \emph{closed} if it consists
  otherwise of vertices of degree~$3$ and inner vertices of $2$-paths, and deleting $x$
  and then removing vertices of degree at most~$1$ removes all of~$P$. Below, $P$ is a
  component of $G$ minus at most two vertices, so $|P|>b$ with $b=K$, $K_m$ or $N_1$
  (otherwise the hybrid rule, the multiplier reduction or the rule for small components
  would apply). Then $P$ disappears in both branches, and the excess is at least
  $\sum_{v\in P\setminus\{x,y_1,y_2,y_3,u\}}w_{d(v)}-3(w_4-w_3)$, which is at least
  $387w_2-3(w_4-w_3)=1.5417072>\varepsilon$ if $b\ge K_m\ge391$, and at least
  $2w_3+60w_2-3(w_4-w_3)=0.1888658>\varepsilon$ if $b=K\ge65$, $u\notin P$ and $P$ has two
  vertices of degree~$3$.

  $G$ has an even number of vertices of degree~$3$, so one, $z$, besides $u$; choose it
  with the largest number $k$ of indices $i$ with $t_i=z$. The charges at $z$ are
  $k(w_4-w_3)$, and $z$ is deleted or loses an edge. The printed vectors use the drop of
  one vertex to cover a charge at another only for a far end of degree~$3$ at which two
  $2$-paths end (the drop along its third skeleton edge), which is then $z$ with $k=2$,
  and when $u$ is a far end, treated last. If $k=3$, $x$, $z$ and the three $2$-paths
  form a closed part, a component of $G-u$. If $k=0$, $z$ drops by at least
  $w_3-w_2>\varepsilon$. If $k=1$, $z$ drops by $w_3-w_2=(w_4-w_3)+\varepsilon$ or by
  $w_3>(w_4-w_3)+\varepsilon$. If $k=2$, $z$ is deleted, and the far end $z''$ of its
  third $2$-path loses an edge and has degree~$2$ or is deleted in $G'$. If $z''$ is not
  charged, the excess at $z,z''$ is at least $w_3+(w_3-w_2)-2(w_4-w_3)=0.0298010$. If
  $z''$ is the third far end and $d(z'')=4$, it is at least
  $w_3+(w_4-w_2)-3(w_4-w_3)=0.0298010$. If $z''$ is the third far end, $d(z'')=3$ and
  $f$ is the far end of its third $2$-path, then $x,z,z''$ and the $2$-paths between them
  and to $f$ form a closed part, a component of $G-\{u,f\}$ (of $G-u$ if $f=u$), with two
  vertices of degree~$3$. If $z''=u$: in $G-x$, $u$ is deleted and the far end $q$ of
  its third $2$-path loses an edge; in $G-N[x]$ the drop at $q$ is not charged; arguing
  for $q$ as for $z''$ (if $q$ is the third far end and $d(q)=3$, $x,u,z$ and their
  $2$-paths form a closed part, a component of $G-q$), the excess at $z,u,q$ is at least
  $2w_3-2(w_4-w_3)=0.0350192$ in $G-x$ and $w_3+(w_3-w_2)-2(w_4-w_3)$ in $G-N[x]$.

  If $u=t_i$, the printed vectors use the drop at the third skeleton neighbor $q$ of $u$
  for the charges at $u$. If $z\ne q$, the cases above apply, with $z''\ne q$ for $k=2$
  (otherwise $x,u,z$ and their $2$-paths form a closed part, a component of $G-q$). If
  $z=q$ and $k=2$, then $G$ is a closed part with $b=N_1$. If $z=q$ and $k\le1$, the
  excess at $u$ and $z$ is at least $w_3-(w_4-w_3)$ and $\varepsilon$ in the two
  branches if $k=0$, and $2w_3-2(w_4-w_3)$ and $w_3+(w_3-w_2)-2(w_4-w_3)$ if $k=1$. The
  excesses of the cases $k\le2$ and $u=t_i$ are checked exactly in part~2 of
  \texttt{gaspers\_lee\_general\_uniform.py}.
\end{proof}

\paragraph{Regular neighborhoods.} The vectors \eqref{eq:gl-lemma8} are not needed for
the types $(5,5,5,5,5)$ and $(6,\dots,6)$: a vertex of maximum degree all of whose
neighbors have this degree is selected only in a regular component. A $\Delta$-regular
component with $\Delta\in\{5,6,7\}$ arises at most once for each $\Delta$ on a
root-to-leaf path of the search tree, since degrees never increase and components are
solved separately; so these branchings cost a constant factor in the running time.

\paragraph{Weights.} The polynomial-space algorithms use the weights of
\cref{tab:gl-sections} and, for degrees $5$ and~$6$,
\begin{equation}\label{eq:is-general-weights}
  (w_2,\dots,w_6)=(0.1136655,\ 0.1986263,\ 0.2673669,\ 0.295498,\ 0.3034362).
\end{equation}
The exponential-space algorithm of part~\ref{thm:is:exp} uses one measure for degrees
$5$ to~$7$, with
\begin{equation}\label{eq:is-exp-weights}
  (w_2,\dots,w_7)=(0.1103289,\ 0.1932144,\ 0.2500011,\ 0.2867833,\ 0.296828,\ 0.2970601),
\end{equation}
and $w_7|V(G)|$ for maximum degree at least~$8$. All decimals are exact.

\begin{table}[tb]
  \centering
  \caption{The weights of the five sections at maximum degree~$4$; section~1 also has
  $\psi=0.0061458$. All decimals are exact, and $\tau_k$ is computed exactly.}
  \label{tab:gl-sections}
  \small
  \begin{tabular}{@{}l>{$}l<{$}>{$}l<{$}>{$}l<{$}>{$}l<{$}>{$}l<{$}>{$}l<{$}@{}}
    \toprule
    $k$ & \text{section} & \alpha_k & w^{(k)}_2 & w^{(k)}_3 & w^{(k)}_4 & \tau_k\\
    \midrule
    1 & [2,3] & - & 0.0052182 & 0.176755 & 0.3360004 & 0.176755\\
    2 & [3,\frac{16}5] & 0.0484291 & 0.0536473 & 0.176755 & 0.2875713 & 0.19891826\\
    3 & [\frac{16}5,\frac72] & 0.0142955 & 0.0708019 & 0.1796141 & 0.2761349 & 0.2278745\\
    4 & [\frac72,\frac{15}4] & 0.0122574 & 0.089188 & 0.1857428 & 0.2700062 & 0.24894035\\
    5 & [\frac{15}4,4] & 0.0105598 & 0.10766765 & 0.19366265 & 0.26736625 & 0.26736625\\
    \bottomrule
  \end{tabular}
\end{table}

The analysis needs two steps that the paper of Gaspers and Lee does not prove: the
vectors of their Lemma~8 rest on a bound on the number of edges leaving $N[v]$ (their
Lemma~7) stated for $3$-connected graphs, and the lazy $2$-separator branching has no
analyzed vector. We prove them for our algorithm in \cref{app:is-g1,app:is-g2}.

\begin{proposition}\label{prop:is-general-steps}
  For the algorithm and the constants above, the following hold.
  \begin{enumerate}[label=\textup{(G\arabic*)},leftmargin=*]
    \item\label{it:G1} Let $v$ be a vertex of maximum degree $d(v)\in\{5,6\}$ with the
      weights \eqref{eq:is-general-weights}, or $d(v)\in\{5,6,7\}$ with the weights
      \eqref{eq:is-exp-weights}, on which the algorithm branches other than in the hybrid
      rule. Then \eqref{eq:gl-lemma8} is a branching vector of this branching for
      $\sum_iw_in_i$, the instances being taken after the removal of vertices of degree at
      most~$1$.
    \item\label{it:G2} Every branching of the hybrid rule in a graph of maximum degree
      $\Delta\ge4$ whose component, if $\Delta\in\{5,6\}$, is not $\Delta$-regular,
      satisfies the condition of \cref{lem:smc-measure} for the measures above.
  \end{enumerate}
\end{proposition}

\begin{proof}[Proof of \cref{thm:is}\ref{thm:is:4}--\ref{thm:is:exp}]
  \ref{thm:is:4}, \ref{thm:is:poly}: we apply \cref{lem:smc-measure} with $\lambda=2$ to
  the algorithm above with the measures above. The following are checked by part~1 of
  \texttt{gaspers\_lee\_general\_uniform.py}: every vector of \cite[Table~1]{GaspersLee2023},
  corrected as above, under the weights of its own section and of every earlier section
  (interval arithmetic; largest sums $0.9999997$ and $0.9671$); every vector
  \eqref{eq:gl-lemma8} for $d(v)\in\{5,6\}$ and all neighbor degrees in
  $\{2,\dots,d(v)\}$ except the regular neighborhoods (interval arithmetic); that the
  weights \eqref{eq:is-general-weights} are at least the section-$5$ weights on degrees
  $2,3,4$, increasing and concave, and that $\alpha_k,\psi\ge0$ (exact); and the values
  $\tau_k$, by exact enumeration of the vertices of the polytope of degree distributions
  with average degree in the section. By \cref{lem:isg-one-section} below and the
  paragraphs ``Cross-section cases'' and ``The potential'', the degree-$4$ branchings
  satisfy the condition of \cref{lem:smc-measure}, and the hand-overs between the
  measures are covered by \cref{lem:isg-one-section}; so do those at degrees $5$ and~$6$ by
  \cref{prop:is-general-steps}\ref{it:G1} and \cref{lem:isg-one-section}, the
  reductions, which only delete vertices, and the branchings of the hybrid rule by
  \cref{prop:is-general-steps}\ref{it:G2}, except in regular components, which cost a
  constant factor. For maximum degree at least~$7$, the measure $w_6|V(G)|$ dominates
  $\sum_iw_in_i$, and the vector $(w_6,8w_6)$ is valid because $2^{w_6}>1.23408$ exceeds
  the positive root $1.232055\ldots$ of $\lambda^8=\lambda^7+1$. So graphs of maximum
  degree at most~$4$ are solved in time $2^{\max_k\tau_kn}n^{O(1)}$ and
  $2^{0.26736625}<1.203609$, and all graphs in time $2^{w_6n}n^{O(1)}$ and
  $2^{0.3034362}<1.234081$, in polynomial space.

  For the chromatic number, Bj\"orklund, Husfeldt and Koivisto~\cite{BjorklundHK2009}
  show that $G$ is $k$-colorable if and only if
  $\sum_{X\subseteq V}(-1)^{|V\setminus X|}i(G[X])^k>0$, where $i(H)$ is the number of
  independent sets of $H$. Computing every $i(G[X])$ in polynomial space takes time
  $\sum_XO(1.2341^{|X|})=O(2.2341^n)$, and binary search over $k$ adds a polynomial
  factor.

  \ref{thm:is:exp}: if the current graph has maximum degree at most~$4$, compute a path
  decomposition of width $\frac5{32}n_3+\frac14n_4+\rho(|V|)$ by
  \cref{thm:deg4}\ref{thm:deg4:c} and \cref{thm:algo-precise}\ref{it:algo:D}, with $|V|$
  the number of vertices of the current graph, and count by the standard dynamic program
  in time $2^{\frac5{32}n_3+\frac14n_4+\rho(|V|)}\poly(n)$. Above maximum degree~$4$ use the
  weights \eqref{eq:is-exp-weights}. They are at least $(0,\frac5{32},\frac14)$ on
  degrees $2,3,4$, increasing and concave, and satisfy the vectors
  \eqref{eq:gl-lemma8} for degrees $5$ to~$7$ (largest sum $0.9999990$, interval
  arithmetic; part~4 of \texttt{gaspers\_lee\_general.py}). These vectors are branching
  vectors by \cref{prop:is-general-steps}\ref{it:G1}, and the hybrid rule is covered by
  \cref{prop:is-general-steps}\ref{it:G2}, except in regular components. For maximum
  degree at least~$8$, $(w_7,9w_7)$ is valid since the positive root of
  $\lambda^9=\lambda^8+1$ is below $1.21315<2^{w_7}$. We apply \cref{lem:smc-measure}
  with the dynamic program as the subroutine $\mathcal B$ and the measures
  $\mu+\rho(|V|)$ and $\mu_{\mathcal B}+\rho(|V|)$, where
  $\mu_{\mathcal B}=\frac5{32}n_3+\frac14n_4\le\mu$. No step increases $|V|$ and $\rho$ is
  nonnegative and nondecreasing (\cref{thm:algo-precise}), so every branching condition
  for $\mu$ remains valid. The time is $2^{w_7n+\rho(n)}\poly(n)$, and
  $2^{0.2970601}<1.228639$.
\end{proof}

\subsection{Proof of \texorpdfstring{\ref{it:G1}}{(G1)}}\label{app:is-g1}

Throughout \cref{app:is-g1,app:is-g2}, $\mu(G)=\sum_{u\in V(G)}w_{d(u)}$ for weights
$w_0=w_1=0<w_2\le\dots\le w_D$, which are \emph{concave from~$2$} if
$\Delta w_2\ge\Delta w_3\ge\dots\ge\Delta w_D>0$. For $X\subseteq V(G)$ let $G\ominus X$
be the largest subgraph of $G-X$ of minimum degree at least~$2$, which the removal of
vertices of degree at most~$1$, in any order, leaves; let $W(X)=V(G)\setminus(X\cup
V(G\ominus X))$. For $A,B\subseteq V(G)$, $e(A,B)$ is the number of edges between $A$
and $B$, and $e(A)$ the number inside~$A$.

\begin{lemma}[decrease of the measure]\label{lem:isg-decrease}
  Let $X\subseteq V(G)$, $G'=G\ominus X$ and $W=W(X)$. Then
  \[
    \mu(G)-\mu(G')=\sum_{u\in X\cup W}w_{d(u)}
    +\sum_{s\in V(G')}\bigl(w_{d(s)}-w_{d_{G'}(s)}\bigr),
  \]
  with nonnegative terms, and $2\le d_{G'}(s)\le d(s)-e(\{s\},X\cup W)$ for
  $s\in V(G')$. If $X\subseteq X'$, then $\mu(G\ominus X')\le\mu(G\ominus X)$.
\end{lemma}

\begin{proof}
  The identity sums the weights over $X$, $W$ and $V(G')$; the terms are nonnegative as
  the weights are nondecreasing. $G\ominus X'$ is a subgraph of $G-X$ of minimum degree
  at least~$2$, hence of $G\ominus X$.
\end{proof}

Let $\theta(v)=e(N[v],V(G)\setminus N[v])$. Counting edges at $N(v)$,
\begin{equation}\label{eq:isg-parity}
  \textstyle\sum_{u\in N(v)}d(u)=d(v)+2e(N(v))+\theta(v),
\end{equation}
so in \eqref{eq:gl-lemma8} the parity cases have $\theta(v)\equiv\mathrm{out}(v)\pmod 2$.

\begin{lemma}[the situation of a branching]\label{lem:isg-situation}
  Let the algorithm branch on its selected vertex $v$, other than in the hybrid rule,
  with $5\le d(v)=\Delta(G)\le7$. Then $G$ is connected, $|V(G)|>N_1$, $\delta(G)\ge2$,
  and:
  \begin{enumerate}[label=\textup{(\arabic*)}]
    \item\label{it:isg-sit-cut} if $P\subseteq V(G)\setminus\{v\}$ is a nonempty union of
      components of $G-v$ with $|P|\le K_m$, then $v$ has no neighbor outside $P$; and if
      every edge leaving $N[v]$ ends at one vertex $c$, then $c$ has no neighbor outside
      $N[v]$;
    \item\label{it:isg-sit-pair} there are no distinct $y,z\notin N[v]$ with
      $d(y),d(z)\ge3$ such that every edge leaving $N[v]$ ends in $\{y,z\}$ and
      $G-\{y,z\}$ has a vertex outside $N[v]$.
  \end{enumerate}
\end{lemma}

\begin{proof}
  Small components are solved directly and vertices of degree at most~$1$ are removed.
  In~\ref{it:isg-sit-cut}, a neighbor of $v$ outside $P$, or of $c$ outside $N[v]$, would
  make $v$ or $c$ a cut vertex with a part of at most $K_m$ vertices, and the multiplier
  reduction would apply. In~\ref{it:isg-sit-pair}, $\{y,z\}$ would be a lazy
  $2$-separator for $v$, as $|N[v]|\le8\le K$.
\end{proof}

\begin{lemma}[the branch $G-v$]\label{lem:isg-g1-minus-v}
  Let the weights be concave from~$2$ with $w_2\ge2\Delta w_D$, and let $G,v$ be as in
  \cref{lem:isg-situation} with $d(v)=D$. Then
  $\mu(G)-\mu(G\ominus\{v\})\ge w_D+\sum_{u\in N(v)}\Delta w_{d(u)}+\deg_2(v)\Delta w_D$.
\end{lemma}

\begin{proof}
  Use \cref{lem:isg-decrease} with $X=\{v\}$. Let $U_2$ and $U_3$ be the neighbors of
  degree~$2$ and of degree at least~$3$. Each $u\in U_2$ lies in $W$ and contributes
  $w_2=\Delta w_{d(u)}$; let $t(u)$ be its other neighbor. By
  \cref{lem:isg-situation}\ref{it:isg-sit-cut}, $t(u)\notin U_2$. For $s\ne v$ let
  $m(s)=|\{u\in U_2:t(u)=s\}|$; it suffices that every $s$ contributes at least
  $[s\in U_3]\Delta w_{d(s)}+m(s)\Delta w_D$. If $s$ survives, it loses at least
  $[s\in U_3]+m(s)$ edges, and its term is a sum of that many increments $\Delta w_i$
  with $3\le i\le d(s)$, the largest being $\Delta w_{d(s)}$. If $s\in W$, its term is
  $w_{d(s)}$, and $w_{d(s)}-\Delta w_{d(s)}\ge(d(s)-1)\Delta w_D\ge m(s)\Delta w_D$ for
  $s\in U_3$, while $w_{d(s)}\ge w_2+(d(s)-2)\Delta w_D\ge d(s)\Delta w_D\ge m(s)\Delta w_D$
  otherwise.
\end{proof}

\begin{lemma}[the branch $G-N{[v]}$]\label{lem:isg-g1-minus-nv}
  Let the weights and $G,v,D$ be as in \cref{lem:isg-g1-minus-v}. Then
  $\mu(G)-\mu(G\ominus N[v])\ge w_D+\sum_{u\in N(v)}w_{d(u)}+\mathrm{out}(v)\,\Delta w_D$.
\end{lemma}

\begin{proof}
  Let $X=N[v]$, $W=W(X)$, $\theta=\theta(v)$ and $k_X(t)=e(\{t\},X)$. By
  \cref{lem:isg-decrease} the decrease is $w_D+\sum_uw_{d(u)}+Q$, where, as every
  surviving $s$ loses $e(\{s\},X\cup W)$ edges and each edge leaving $X$ ends in $W$ or
  at a survivor,
  \[
    Q\ge\sum_{t\in W}w_{d(t)}+e(X\cup W,V(G\ominus X))\,\Delta w_D\ge
    \theta\,\Delta w_D+\sum_{t\in W}\bigl(w_{d(t)}-k_X(t)\Delta w_D\bigr),
  \]
  and every summand $w_{d(t)}-k_X(t)\Delta w_D$ is at least $w_2-2\Delta w_D\ge0$, and at
  least $w_2-\Delta w_D$ if $k_X(t)\le1$. We show $Q\ge\mathrm{out}(v)\Delta w_D$. As
  $|V(G)|>N_1\ge D+3$ and $G$ is connected, $\theta\ge1$.

  \emph{$\mathrm{out}(v)\in\{5,6\}$.} By \cref{lem:isg-situation}\ref{it:isg-sit-cut} there
  is no edge inside the set $U_2$ of neighbors of degree~$2$, and no two edges from $U_2$
  to a neighbor of degree~$3$. So $e(N(v))=0$, or $e(N(v))\le1$ for $(2,2,2,2,2,3)$, and
  \eqref{eq:isg-parity} gives $\theta\ge\mathrm{out}(v)$.

  \emph{$\mathrm{out}(v)\in\{3,4\}$.} If $\theta\ge\mathrm{out}(v)$ we are done; by
  parity, otherwise $(\theta,\mathrm{out}(v))$ is $(1,3)$ or $(2,4)$. If a leaving edge
  $ab$ ($a\in N(v)$) has $d(b)=2$, the vertices of degree~$2$ from $b$ away from $a$,
  $b=t_1,\dots,t_p$, end at a vertex $b'$ of degree at least~$3$; if no other leaving edge
  ends on this walk, then $t_1,\dots,t_p\in W$. If $\theta=1$, with leaving edge $ab$,
  then $d(b)=2$ by \cref{lem:isg-situation}\ref{it:isg-sit-cut}, and
  $Q\ge w_2+\Delta w_D\ge3\Delta w_D$ ($b'$ is removed or loses an edge). If $\theta=2$,
  with leaving edges $a_1b_1$, $a_2b_2$:
  if $b_1=b_2=b$, then $d(b)\ge3$ contradicts \cref{lem:isg-situation}\ref{it:isg-sit-cut},
  and $d(b)=2$ makes $X\cup\{b\}$ a component of $D+2<|V(G)|$ vertices; if $b_1\ne b_2$ have
  degree at least~$3$, \cref{lem:isg-situation}\ref{it:isg-sit-pair} is contradicted; if
  $d(b_1)=d(b_2)=2$, then $b_1,b_2\in W$ with $k_X(b_i)=1$ and $Q\ge2\Delta w_D+
  2(w_2-\Delta w_D)\ge4\Delta w_D$; if $d(b_1)=2<d(b_2)$, the walk from $b_1$ ends at
  $b'$ ($b'=b_2$ possible), and $Q\ge pw_2+w_3\ge4\Delta w_D$ if $b_2\in W$,
  $Q\ge w_2+2\Delta w_D$ if $b_2\notin W$ and $b'=b_2$, and
  $Q\ge w_2+\Delta w_D+\min\{w_3,\Delta w_D\}$ otherwise.
\end{proof}

\begin{proof}[Proof of \ref{it:G1}]
  By \cref{lem:isg-g1-minus-v,lem:isg-g1-minus-nv} it suffices that the weights are
  concave from~$2$ with $w_2\ge2\Delta w_D$ for $D=d(v)$. This is checked exactly by
  part~A of \texttt{g1\_check.py}: both weight vectors are increasing and concave from
  $2$, $w_2=0.1136655\ge4\Delta w_5\ge4\Delta w_6$ for \eqref{eq:is-general-weights},
  and $w_2=0.1103289\ge2\Delta w_5=0.0735644$ with $\Delta w_6,\Delta w_7<\Delta w_5$ for
  \eqref{eq:is-exp-weights}.
\end{proof}

\subsection{Proof of \texorpdfstring{\ref{it:G2}}{(G2)}}\label{app:is-g2}

A \emph{lazy configuration} is the current graph $G$ with $\Delta=\Delta(G)\ge4$ when the
hybrid rule applies, with the selected vertex $x$ and its lazy $2$-separator $\{y,z\}$,
whose component $C\supseteq N[x]\setminus\{y,z\}$ of $G-\{y,z\}$ has $|C|\le K$. Let
$C^+$ be the union of the components of $G-\{y,z\}$ with at most $K$ vertices,
$R=V(G)\setminus(C^+\cup\{y,z\})$, $o_y=|N(y)\cap R|$, $o_z=|N(z)\cap R|$,
$H=G[C^+\cup\{y,z\}]$ and $\bar e=|C^+\setminus N[x]|$.

\begin{lemma}[the configuration]\label{lem:isg-config}
  In a lazy configuration, $\delta(G)\ge2$; the neighbors of $C^+$ lie in
  $C^+\cup\{y,z\}$; every component of $G[R]$ has more than $K$ vertices; every component
  of $G[C^+]$ contains a neighbor of $y$ and one of~$z$; and $|C^+|\le\Delta K$. If
  $\Delta\le7$, then $o_y,o_z\ge1$ and $x\notin\{y,z\}$.
\end{lemma}

\begin{proof}
  A component of $G[C^+]$ without a neighbor of $z$ would be a component of $G-y$ with at
  most $K\le K_m$ vertices; so $G[C^+]$ has at most $\min\{d(y),d(z)\}$ components. The
  rest of the first sentence follows from the definitions. Let $\Delta\le7$. $G$ is
  connected with $|V(G)|>N_1=7K+2\ge\Delta K+2$, so $R\ne\emptyset$ and not both $o_y,o_z$
  vanish. If $o_z=0<o_y$, then $C^+\cup\{z\}$ is a union of components of $G-y$ with at
  most $(d(y)-1)K+1\le K_m$ vertices, and the multiplier reduction would apply; similarly
  if $o_y=0<o_z$. If $x=y$, all neighbors of $x$ but $z$ lie in $C$, so $o_y=0$.
\end{proof}

In the branch $G-y$ the vertex $z$ cuts off every component of $G[C^+]$, which the
multiplier reduction deletes; in $G-N[y]$ the components of $G[C^+\setminus N[y]]$ are
deleted by the multiplier reduction at $z$ or, if $yz\in E(G)$, solved directly. So the
instances arise from $I_1=G\ominus(\{y\}\cup C^+)$ and $I_2=G\ominus(N[y]\cup C^+)$ by
further reductions, which by \cref{lem:isg-decrease} do not increase $\mu$; the same
holds for $z$, and for $x$ with $G\ominus\{x\}$ and $G\ominus N[x]$.

\begin{lemma}[credit from a large part]\label{lem:isg-large-credit}
  Let $w_2>0$ and $\Delta w_3\ge\dots\ge\Delta w_D>0$ with $D\ge\Delta(G)$. Let
  $X_0\subseteq V(G)$, $S\subseteq V(G\ominus X_0)$, and let $P$ be a connected set of
  vertices of $G-X_0$, disjoint from $S$, of vertices of degree at least~$2$, with
  $N(P)\subseteq P\cup X_0\cup S$. Let $W_P=P\cap W(X_0)$. Then the terms of the
  vertices of $P$ in \cref{lem:isg-decrease} add up to at least
  $e(P,X_0)\Delta w_D+|W_P|w_2-2[W_P=P]\Delta w_D$, and so to at least
  $e(P,X_0)\Delta w_D$ if $|P|>K$ and $Kw_2\ge2\Delta w_D$.
\end{lemma}

\begin{proof}
  Let $P'=P\setminus W_P$. When $t\in W_P$ is removed it has at
  most one remaining neighbor, its \emph{out-neighbor}, in $W_P$, $P'$ or $S$; let
  $N_{\mathrm{out}},N_S,N_{\mathrm{iso}}$ count the $t$ whose out-neighbor lies in $P'$,
  lies in $S$, or does not exist. Counting the edges at $W_P$ at their later endpoints,
  $e(X_0,W_P)=\sum_{t\in W_P}(d(t)-2)+N_{\mathrm{out}}+N_S+2N_{\mathrm{iso}}$. The vertices
  of $P'$ contribute at least $(N_{\mathrm{out}}+e(X_0,P'))\Delta w_D$, and $t\in W_P$
  contributes $w_{d(t)}\ge w_2+(d(t)-2)\Delta w_D$; so $P$ contributes at least
  $|W_P|w_2+(e(P,X_0)-N_S-2N_{\mathrm{iso}})\Delta w_D$. A component $A$ of $G[W_P]$ has
  $e(A)=|A|-N_{\mathrm{out}}(A)-N_S(A)-N_{\mathrm{iso}}(A)\ge|A|-1$; if $W_P\ne P$, its
  edge to $P'$ is an out-edge, so $N_S(A)=N_{\mathrm{iso}}(A)=0$. If $W_P=P$, then
  $G[W_P]$ is connected and $N_S+2N_{\mathrm{iso}}\le2$.
\end{proof}

\begin{lemma}[one section suffices]\label{lem:isg-one-section}
  Let $I_1,I_2$ be the instances of a branching of $G$. If $\Delta(G)=4$, $G$ has
  minimizing section $k$, and
  $a_j=\mu^{(k)}_{\mathrm{lin}}(G)-\mu^{(k)}_{\mathrm{lin}}(I_j)-[k=1]\psi$ satisfy
  $2^{-a_1}+2^{-a_2}\le1$, then $2^{\mu(I_1)}+2^{\mu(I_2)}\le2^{\mu(G)}$. The same holds
  for $\Delta(G)\ge5$ with $a_j=\sum_iw_in_i(G)-\sum_iw_in_i(I_j)$ for the weights
  \eqref{eq:is-general-weights} or \eqref{eq:is-exp-weights}.
\end{lemma}

\begin{proof}
  It suffices that $\mu(G)-\mu(I_j)\ge a_j$. For $\Delta(G)=4$,
  $\mu_4(G)=\mu^{(k)}(G)\ge\mu^{(k)}_{\mathrm{lin}}(G)$, and $\mu(I_j)\le
  \mu^{(k)}_{\mathrm{lin}}(I_j)+[k=1]\psi$: for $\Delta(I_j)=4$ as $\mu_4\le\mu^{(k)}$, and
  for $\Delta(I_j)\le3$ as $\mu(I_j)\le\mu^{(1)}_{\mathrm{lin}}(I_j)\le
  \mu^{(k)}_{\mathrm{lin}}(I_j)$ by \eqref{eq:gl-section-diff}. For $\Delta(G)\ge5$, an
  instance of maximum degree at most~$4$ has measure at most $\mu^{(5)}_{\mathrm{lin}}$,
  or at most $\frac5{32}n_3+\frac14n_4$, and both are at most $\sum_iw_in_i$.
\end{proof}

\paragraph{The three candidates.} For $X_0\subseteq V(G)$, the \emph{$H$-clean-up}
deletes $X_0\cap V(H)$ from $H$ and then removes vertices of degree at most~$1$, where
$y$ and $z$ count $o_y$ and $o_z$ extra edges to vertices never removed. Let $W'$ be the
removed vertices, $d'(v)$ the final degree of a surviving $v\in V(H)$, and $\ell_R$ the
number of edges between $R$ and $\{y,z\}\cap(X_0\cup W')$. Let
\[
  \beta(X_0)=\sum_{v\in(X_0\cap V(H))\cup W'}w_{d(v)}
  +\sum_{v\in V(H)\setminus(X_0\cup W')}\bigl(w_{d(v)}-w_{d'(v)}\bigr)+\ell_R\,\kappa',
\]
with $\kappa'=\Delta w_D$ if $X_0\subseteq V(H)$ and $\kappa'=\min\{w_2,\Delta w_D\}$ if
$X_0=N[v]\cup C^+$ with $v\in\{y,z\}$. The \emph{candidate vectors} are
$\bigl(\beta(\{v\}\cup C^+),\max\{\beta(\{v\}\cup C^+)+o_vw_2,\beta(N[v]\cup C^+)\}\bigr)$
for $v\in\{y,z\}$ and $(\beta(\{x\}),\beta(N[x]))$, each component reduced by $\psi$ in
section~$1$.

\begin{lemma}[the candidates]\label{lem:isg-bounds}
  In a lazy configuration, let the weights be nondecreasing with $w_2>0$ and
  $\Delta w_3\ge\dots\ge\Delta w_D>0$, $D\ge\Delta$, and $(K+1-\Delta)w_2\ge2\Delta w_D$.
  Then each candidate vector is a lower bound on the decreases of
  $\mu_{\mathrm{lin}}=\sum_iw_in_i$ in the two branches of its branching. Consequently,
  by \cref{lem:isg-one-section}, if one candidate vector $(a,b)$, computed with the
  weights of the minimizing section for $\Delta=4$ and with \eqref{eq:is-general-weights}
  or \eqref{eq:is-exp-weights} for $\Delta\ge5$, has $2^{-a}+2^{-b}\le1$, the hybrid rule
  branches validly.
\end{lemma}

The hypothesis holds for $K\ge65$ and all our weights: $2\Delta w_D/w_2$ is largest
for $\Delta=D=4$ in section~$1$, where it is $61.03<62\le K+1-\Delta$.

\begin{proof}
  By \cref{lem:isg-decrease} the decrease from $G$ to $G\ominus X_0$ is a sum of
  nonnegative terms. \emph{The $H$-side:} by induction on the removal order,
  $W'\subseteq W(X_0)$, since a vertex of $W'$ has at most one neighbor left in $G-X_0$
  when it is removed in $H$ (vertices of $C^+$ have no neighbor in $R$). So every
  $v\in(X_0\cap V(H))\cup W'$ contributes $w_{d(v)}$, and every other $v\in V(H)$ either
  lies in $W(X_0)$ and contributes $w_{d(v)}\ge w_{d(v)}-w_{d'(v)}$, or survives with
  $d_{G\ominus X_0}(v)\le d'(v)$, as it loses its edges to $X_0\cup W'$; so $H$
  contributes at least the first two sums of $\beta(X_0)$. \emph{The $R$-side,
  $X_0\subseteq V(H)$:} with $X'_0=X_0\cup(W(X_0)\cap V(H))$, $G\ominus X'_0=G\ominus X_0$,
  and each component $P$ of $G[R]$ satisfies \cref{lem:isg-large-credit} with $X'_0$ and
  $S=\{y,z\}\setminus X'_0$; so $R$ contributes at least $\ell_R\,\Delta w_D$.
  \emph{The $R$-side, $X_0=N[y]\cup C^+$:} each $r\in N(y)\cap R$ is deleted and
  contributes $w_{d(r)}$; if $z\in X_0\cup W'$, each $s\in(N(z)\cap R)\setminus N(y)$
  contributes $w_2$ if removed and $\Delta w_{d(s)}$ otherwise, and $s\in N(y)\cap
  N(z)\cap R$ has $d(s)\ge3$ (else $\{s\}\subseteq C^+$) and contributes
  $w_{d(s)}\ge2\min\{w_2,\Delta w_D\}$; so $R$ contributes at least $\ell_R\kappa'$.
  \emph{The term $o_yw_2$:} with $X=\{y\}\cup C^+$ and $X'=N[y]\cup C^+$, $G\ominus X'=
  (G\ominus X)\ominus(X'\cap V(G\ominus X))$, so by \cref{lem:isg-decrease} the second
  branch gains at least $w_2$ over the first for each $s\in N(y)\cap R$ that survives in
  $G\ominus X$. Each other $s\in N(y)\cap R$ lies in $W_P$ for a component $P$ of
  $G[R]$, and by \cref{lem:isg-large-credit}, applied in the first branch with
  $X^\circ=X\cup(W(X)\cap V(H))$ in place of $X_0$, $P$ contributes at least
  $e(P,X^\circ)\Delta w_D+|W_P\cap N(y)|w_2$: trivially if $W_P\ne P$, and if $W_P=P$
  because $|P\setminus N(y)|\ge K+1-\Delta$ and $(K+1-\Delta)w_2\ge2\Delta w_D$. The
  case of $z$ is symmetric.
\end{proof}

\begin{lemma}[the branching on $y$ in closed form]\label{lem:isg-closed-form}
  In a lazy configuration, with the weights as in \cref{lem:isg-bounds}, let
  $\zeta=w_{d(z)}-w_{o_z}$ if $o_z\ge2$ and $\zeta=w_{d(z)}+\Delta w_D$ if $o_z=1$, and
  \[
    L=w_{d(y)}+\sum_{c\in C^+}w_{d(c)}+\zeta+o_y\,\Delta w_D.
  \]
  Then the candidate vector of $y$ is at least $(L,L+o_yw_2)$. $L$ does not decrease when
  vertices are added to $C^+$ or degrees grow, with $o_y,o_z$ fixed. If
  $L-[k=1]\psi\ge1$, with $k$ the minimizing section when $\Delta=4$, the branching on
  $y$ satisfies the condition of \cref{lem:smc-measure}.
\end{lemma}

\begin{proof}
  For $X_0=\{y\}\cup C^+$, $W'\subseteq\{z\}$; if $o_z\ge2$, $z$ survives with
  $d'(z)=o_z$ and $\ell_R=o_y$, and if $o_z=1$, $z$ is removed and $\ell_R=o_y+1$. So
  $\beta(X_0)=L$. The monotonicity holds as the weights are nondecreasing and
  $w_2>0$, and the last claim by \cref{lem:isg-bounds}, as $2\cdot2^{-1}=1$.
\end{proof}

\begin{corollary}[maximum degree at least $7$]\label{cor:isg-high-degree}
  The branching on $y$ of a lazy configuration satisfies the condition of
  \cref{lem:smc-measure} if $\Delta=7$ with the weights \eqref{eq:is-exp-weights}, and
  for the measures $w_6|V|$ ($\Delta\ge7$) and $w_7|V|$ ($\Delta\ge8$).
\end{corollary}

\begin{proof}
  For $\Delta=7$ and \eqref{eq:is-exp-weights}, \cref{lem:isg-config} gives $x\in C^+$
  and $|N(x)\cap C^+|\ge5$, so $L>w_3+w_7+5w_2=1.041919$. For the measures $w_6|V|$ and
  $w_7|V|$, both branches delete $\{y\}\cup C^+$, which contains $y$, $x$ and
  $N(x)\setminus\{y,z\}$, at least $\Delta$ vertices, and for $\Delta=7$ the branch
  $G-N[y]$ also deletes the $o_y\ge1$ neighbors of $y$ in $R$. With $\lambda<2^{w_6}$ the
  root of $\lambda^8=\lambda^7+1$, $2^{-7w_6}+2^{-8w_6}<\lambda^{-1}+\lambda^{-8}=1$, and
  $2\cdot2^{-8w_6}<0.38$, $2\cdot2^{-8w_7}<0.39$.
\end{proof}

For $\Delta\ge5$ a type is \emph{admissible} if it contains a degree smaller than
$\Delta$, which holds in non-regular components.

\begin{lemma}[thresholds for the branching on $y$]\label{lem:isg-thresholds}
  The branching on $y$ satisfies the condition of \cref{lem:smc-measure} in every lazy
  configuration with an admissible type of $x$ in which
  \begin{enumerate}[label=\textup{(\arabic*)}]
    \item\label{it:isg-thr-45} $\Delta=4$ and the minimizing section is $4$ or~$5$;
    \item\label{it:isg-thr-23} $\Delta=4$, the minimizing section is $2$ or~$3$, and
      $\bar e\ge2$ or some vertex of $C^+\setminus N[x]$ has degree at least~$3$;
    \item\label{it:isg-thr-1} $\Delta=4$, the minimizing section is~$1$, and some vertex
      of $C^+\setminus\{x\}$ has degree at least~$3$;
    \item\label{it:isg-thr-56} $\bar e\ge1$, and $\Delta\in\{5,6\}$ with
      \eqref{eq:is-general-weights} or $\Delta=6$ with \eqref{eq:is-exp-weights};
    \item\label{it:isg-thr-5exp} $\bar e\ge2$, $\Delta=5$ and \eqref{eq:is-exp-weights}.
  \end{enumerate}
\end{lemma}

\begin{proof}
  By \cref{lem:isg-closed-form} with $D=\Delta$ it suffices that $L-[k=1]\psi\ge1$. By
  \cref{lem:isg-config}, $d(x)=\Delta$, the degrees of $N(x)\setminus\{y,z\}$ are those
  of the type, $3\le d(y),d(z)\le\Delta$, $1\le o_y\le d(y)-1$, $1\le o_z\le d(z)-1$, and
  the $\bar e$ vertices of $C^+\setminus N[x]$ contribute at least $\bar ew_2$, or $w_3$
  for one of degree at least~$3$. As $L$ is monotone, the remaining parameters (type,
  whether $y,z\in N(x)$, $d(y),d(z),o_y,o_z$) take finitely many values, and
  \texttt{g2\_thresholds.py} (parts~A and~B) computes the minima of $L$ exactly. They are
  $1.0800248$ and $1.2168722$ in~\ref{it:isg-thr-45}; at least $1.0236558$ in
  \ref{it:isg-thr-23}; $L-\psi\ge1.0070736$ in~\ref{it:isg-thr-1}; at least $1.0050485$
  in~\ref{it:isg-thr-56}; and $1.1052066$ in~\ref{it:isg-thr-5exp}.
\end{proof}

\begin{theorem}[the hybrid rule on bounded configurations]\label{thm:isg-enumeration}
  In every lazy configuration of the following classes, one of the three candidate
  vectors $(a,b)$ satisfies $2^{-a}+2^{-b}\le1$:
  \begin{enumerate}[label=\textup{(\arabic*)}]
    \item\label{it:isg-class1} $\Delta=4$, any minimizing section $k$, an admissible type
      for section~$k$, and $C^+=(N[x]\setminus\{y,z\})\cup E$ with $|E|\le2$;
    \item\label{it:isg-class2} $\Delta\in\{5,6\}$, an admissible type, and
      $C^+=N[x]\setminus\{y,z\}$;
    \item\label{it:isg-class3} $\Delta=5$, an admissible type, and
      $C^+=(N[x]\setminus\{y,z\})\cup E$ with $|E|\le1$;
  \end{enumerate}
  in classes~\ref{it:isg-class2} and~\ref{it:isg-class3} with both weight vectors
  \eqref{eq:is-general-weights} and \eqref{eq:is-exp-weights}, and in each class
  optionally with one further vertex $p\in C^+$ with $N(p)=\{y,z\}$.
\end{theorem}

\begin{proof}
  The proof is computer-assisted (\texttt{g2\_enumeration.py}, parts~A and~B). For each
  class the program generates every graph $H$ on $C^+\cup\{y,z\}$ with the prescribed
  degrees (all at most $\Delta$, with $d(y),d(z)\in[3,\Delta]$ enumerated independently
  of the type), and every pair $o_y,o_z\ge1$. It skips $H$ if some $c\in C^+$ leaves a
  component of $H-c$ without $y$ and $z$ (a part of $G-c$ with at most $K\le K_m$
  vertices, so the multiplier reduction would apply), or if a component of $G[C^+]$
  misses $y$ or~$z$ (\cref{lem:isg-config}). For the others it computes the three candidate vectors in
  exact rational arithmetic and tests $2^{-a}+2^{-b}\le1$ in floating point with margin
  $10^{-9}$, and with $40$ decimal digits in the doubtful band. About $8.7\cdot10^6$
  configurations are tested in class~\ref{it:isg-class1}, $4.1\cdot10^7$ in
  class~\ref{it:isg-class2} and $2.1\cdot10^7$ in class~\ref{it:isg-class3} (the last
  two for each weight vector); none fails.
\end{proof}

\begin{lemma}[the $2$-path family]\label{lem:isg-two-paths}
  Let a lazy configuration with $\Delta=4$ have all vertices of $C^+\setminus\{x\}$ of
  degree~$2$. Replace every maximal path of vertices of degree~$2$ in $C$ that starts at
  a neighbor $u$ of $x$ by $u$ alone, joined to the end ($y$ or $z$) of the path, and
  every other component of $G[C^+]$, which is a path from a neighbor of $y$ to a
  neighbor of $z$, by one vertex $p$ with $N(p)=\{y,z\}$. The result $H'$, with the
  same $o_y,o_z$, satisfies all conditions that the proof of \cref{thm:isg-enumeration}
  tests for class~\ref{it:isg-class1}, with the same type, $d(y),d(z),o_y,o_z$, and each
  candidate vector of $H$ is componentwise at least that of~$H'$.
\end{lemma}

\begin{proof}
  From $u\in N(x)\setminus\{y,z\}$ the vertices of degree~$2$ lead to $x$, $y$ or $z$; to
  $x$ would give a component of $G-x$ of at most $K_m$ vertices. Two such paths are
  disjoint, and they cover $C$. A component of $G[C^+]$ other than $C$ is not a cycle and
  has neighbors of $y$ and $z$ (\cref{lem:isg-config}), so it is a path from $y$ to $z$;
  there are at most $d(y)-o_y-1\le2$ of them. So $H'$ has the same degrees at $x,y,z$,
  at most two vertices outside $N[x]$, no vertex skipped by the proof of
  \cref{thm:isg-enumeration}, and its part $\{x\}\cup(N(x)\setminus\{y,z\})$ keeps a
  neighbor of $y$ and one of $z$, as the paths keep their ends. A path of degree-$2$ vertices is removed by a clean-up as
  soon as an attachment vertex in $\{x,y,z\}$ or a deleted vertex of the path goes, and
  keeps all degrees~$2$ otherwise; so the $H$- and
  $H'$-clean-ups remove the same vertices among $x,y,z$ and the same paths, with the same
  $\ell_R$, and $\beta$ for $H$ exceeds $\beta$ for $H'$ by the weights $w_2\ge0$ of the
  extra removed vertices.
\end{proof}

\begin{proof}[Proof of \ref{it:G2}]
  By \cref{lem:isg-bounds} it suffices that one candidate vector satisfies
  $2^{-a}+2^{-b}\le1$; for $\Delta\le7$, $x\notin\{y,z\}$ and
  $C^+\supseteq N[x]\setminus\{y,z\}$ (\cref{lem:isg-config}). For $\Delta=4$ let $k$ be
  the minimizing section; the type of $x$ is admissible for $k$ (paragraph
  ``Cross-section cases''). Sections $4,5$:
  \cref{lem:isg-thresholds}\ref{it:isg-thr-45}. Sections $2,3$:
  \cref{lem:isg-thresholds}\ref{it:isg-thr-23}, or class~\ref{it:isg-class1} of
  \cref{thm:isg-enumeration}. Section~$1$: \cref{lem:isg-thresholds}\ref{it:isg-thr-1},
  or \cref{lem:isg-two-paths} and \cref{thm:isg-enumeration}. For $\Delta\in\{5,6\}$ the
  component is not regular, so $x$ has an admissible type:
  \cref{lem:isg-thresholds}\ref{it:isg-thr-56},\ref{it:isg-thr-5exp} or
  classes~\ref{it:isg-class2},~\ref{it:isg-class3}. For $\Delta\ge7$:
  \cref{cor:isg-high-degree}.
\end{proof}

The hybrid rule cannot be replaced by the branching on $y$ alone:

\begin{proposition}[the branching on $y$ alone fails]\label{prop:isg-counterexample}
  For $K=65$ there is a connected graph $G$ with $488$ vertices, $\Delta(G)=4$,
  $\delta(G)=2$, no cut vertex and minimizing section~$2$, all of whose vertices of
  degree~$4$ have type $(2,3,3,4)$, with a vertex $x$ and a lazy $2$-separator
  $\{y,z\}\subseteq N(x)$, $d(y)=3$, for which both instances of the branching on $y$ are
  $G-\{x,u_1,u_2,y,r\}$ and $2^{\mu_4(I_1)}+2^{\mu_4(I_2)}=1.0210061\cdot2^{\mu_4(G)}$.
\end{proposition}

\begin{proof}
  Take $m=80$ and the vertices $a_i,a'_i,b_{i,1},b_{i,2},b_{i,3}$
  ($i\in\mathbb Z_m$), $c_i$ ($i\in\mathbb Z_m\setminus\{0\}$) and
  $x,u_1,u_2,y,z,r,s_1,s_2,s_3$, with the edges $a_ia'_i$,
  $a_ib_{i,1}$, $a_ib_{i,2}$, $a'_ib_{i,2}$, $a'_ib_{i,3}$, $b_{i,1}b_{i,3}$,
  $b_{i,2}b_{i,3}$ for all $i$; $a'_ic_i$, $c_ia_{i+1}$ for $i\ne0$; $b_{i,1}b_{i+1,1}$
  for even $i\ne0$; and $a'_0r$, $ry$, $yx$, $yu_2$, $xu_1$, $xu_2$, $u_1u_2$, $xz$,
  $zs_1$, $s_1a_1$, $zs_2$, $zs_3$, $s_2s_3$, $s_2b_{0,1}$, $s_3b_{1,1}$. In $G-y$, $z$
  cuts off the part $\{x,u_1,u_2\}$, which the multiplier reduction deletes, and $r$
  becomes pendant; in $G-N[y]=G-\{y,x,u_2,r\}$, $u_1$ loses both neighbors. So in both
  branches the vertices $y,x,u_1,u_2,r$ disappear and $z,a'_0$ drop from degree~$4$ to~$3$, so the
  decrease is $2w^{(2)}_2+2w^{(2)}_3+w^{(2)}_4+2\Delta w^{(2)}_4=0.9700085$. The
  remaining properties are checked by part~C of \texttt{g2\_enumeration.py}.
\end{proof}
{}
\section{Counting dominating sets: the analysis}\label{app:ds}

This appendix carries out the running-time analysis outlined in
\cref{sec:ds-outline}, in the notation of \cref{sec:nrd-algorithm}, and proves
\cref{thm:ds}:

\thmds*

\Cref{sec:nrd-analysis} proves the three steps of the analysis for a general
threshold and list of exceptional types, \cref{sec:ds-proof} applies them to
$\mathcal A_{4,\mathcal X_{30}}$, and \cref{sec:ds-weighted} treats the weighted
counts. \Cref{sec:nrd-gap} applies the analysis to the previous algorithm.

\subsection{The three steps of the analysis}\label{sec:nrd-analysis}

The analysis follows the measure-and-conquer scheme of the thesis, in three steps.
First, a set of branching recurrences bounds the number of instances of each measure
on which the dynamic program is called \cite[Lemma~8.11]{vanRooij2011}
(\cref{lem:ds-branching,lem:ds-high,lem:ds-leaves}). Next, a linear program bounds the
pathwidth of these instances \cite[Lemma~8.12]{vanRooij2011}
(\cref{lem:ds-pendant,lem:ds-dp}). Finally, the two bounds are combined
\cite[Theorem~8.13]{vanRooij2011} (\cref{lem:ds-time}). We prove the three steps for
both thresholds at once.

For $a,b>0$ let $\tau(a,b)$ be the unique $x>1$ with $x^{-a}+x^{-b}=1$, the
\emph{branching factor} of $(a,b)$. A vector $(a,b)$ \emph{holds at $\alpha$} if
$\alpha^{-a}+\alpha^{-b}\le1$, that is, if $\tau(a,b)\le\alpha$.

\paragraph{Branching vectors.} For $d\ge5$ and $r=(r_1,\dots,r_d)$ with
$\sum_ir_i=d$, $r_1\le1$, and $(r_1,\dots,r_5)\notin\mathcal X$ when $d=5$, the
\emph{element vector} $E_{d,r}$ is
\begin{equation}\label{eq:ds-element}
  \Bigl(v(d)+\sum_ir_i\,\Delta w(i),\ \ v(d)+\sum_ir_i\,w(i)+\Delta v(d)\sum_i(i-1)r_i\Bigr).
\end{equation}
For $d\ge\sigma+1$, $f(d)=\max\{d-1,4\}$ and $r=(r_2,\dots,r_{f(d)})$ with
$\sum_ir_i=d$, the \emph{set vector} $S_{d,r}$ is
\begin{equation}\label{eq:ds-set}
  \Bigl(w(d)+\sum_ir_i\,\Delta v(i)+r_2\,w(1),\ \ w(d)+\sum_ir_i\,v(i)+\Delta w(d)\sum_i(i-1)r_i\Bigr).
\end{equation}
These are the recurrences of \cite[p.~152]{vanRooij2011} and
\cite[Lemma~29, pp.~728--729]{NederlofRD2014}, with the range of cases that
the algorithm actually produces; \cref{sec:nrd-gap} shows that the previous range
misses some of these cases.

\begin{lemma}[branching recurrences]\label{lem:ds-branching}
Let $\sigma\in\{3,4\}$, let $\mathcal X$ satisfy~\eqref{eq:ds-list}, and let the
weights satisfy~\ref{ds:W1}--\ref{ds:W4}. Whenever $\mathcal A_{\sigma,\mathcal X}$
branches on an instance of measure $k$, there is an element vector $(a,b)=E_{d,r}$ or
a set vector $(a,b)=S_{d,r}$ such that the two child instances, after the reduction
rules have been applied exhaustively, have measures at most $k-a$ and $k-b$.
\end{lemma}

\begin{proof}
Since the reduction rules do not increase the measure, it suffices to bound the
measure right after the branching step, except for the term $r_2w(1)$ in the set
vector, which comes from reductions and is treated separately. When the algorithm
branches, no reduction rule applies, so every element has frequency at least~$2$,
distinct sets have distinct elements, and no two elements satisfy
$N_I(e_1)\subseteq N_I(e_2)$. We treat the two branching rules in turn. For each, we
first bound the sizes and frequencies that can occur, which shows that the vector is
one of those defined above, and then compute the decrease of the measure in the two
branches.

\emph{Branching on an element $e$ (rule~\ref{ds:B2}).} Rule~\ref{ds:B1} did not apply,
so $d:=d(e)\ge5$ and $d(s)\le d$. Every set containing $e$ has size at most $d$:
non-exceptional sets have size at most $d(s)$, and exceptional ones have size
$\sigma+1\le5$. Every element has frequency at most $d$: non-exceptional ones by the
choice of $e$, exceptional ones have frequency $5$. At most one set containing $e$ has
size~$1$, since two such sets would both equal $\{e\}$. If $d=5$, the type of $e$ is not
in $\mathcal X$, since $e$ is not exceptional. Let $r_i$ be the number of sets of size
$i$ containing~$e$.

In the optional branch, $e$ is deleted, which removes $v(d)$, and each set containing
$e$ shrinks by one, which removes $\Delta w(i)$ for a set of size $i$ (a set of size~$1$
becomes empty and loses $w(1)=\Delta w(1)$). This is the first entry
of~\eqref{eq:ds-element}. In the forbidden branch, $e$ and the sets containing it are
deleted, which removes $v(d)+\sum_ir_iw(i)$, and every other element $f$ of these sets
loses frequency. The sets containing $e$ have $\sum_i(i-1)r_i$ incidences with other
elements in total. We claim that an element of frequency $j$ that loses $t\ge1$ of its
sets loses at least $t\,\Delta v(d)$ in weight; summing over the elements gives the
second entry. We have $2\le j\le d$ and $1\le t\le j$. If $j-t\ge1$, then
$v(j)-v(j-t)=\sum_{l=j-t+1}^{j}\Delta v(l)$ with $2\le l\le d$ (recall $v(1)=0$), and
each term is at least $\Delta v(d)$ by~\ref{ds:W3}. If $j-t=0$, then $t=j$ and
$v(j)-v(0)=v(2)+\sum_{l=3}^{j}\Delta v(l)\ge2\Delta v(d)+(j-2)\Delta v(d)=t\,\Delta v(d)$,
using $v(2)\ge2\Delta v(5)\ge2\Delta v(d)$ from~\ref{ds:W4},~\ref{ds:W3} and $d\ge5$.

\emph{Branching on a set $s$ (rule~\ref{ds:B1}).} Let $d=d(s)$. First, $d\ge\sigma+1$:
if $d\le\sigma$, then either $d(e)\le4$, and the switch~\ref{ds:S} applies, or
$d(e)\ge5>d(s)$, and the algorithm branches on $e$. Next, $s$ contains no exceptional
element: if $d\le5$, a set of size in $\{\sigma+1,\dots,5\}$ containing an exceptional
element would be exceptional, and $s$ is not; if $d\ge6$, $s$ cannot contain an
exceptional element, because by~\eqref{eq:ds-list} exceptional elements lie only in
sets of size at most $\sigma+1\le5$. Hence every element of $s$ is non-exceptional and
has frequency at most $d(e)$. Since rule~\ref{ds:B1} applies, $d(e)<d$ or $d(e)\le4$, so
every element of $s$ has frequency in $\{2,\dots,f(d)\}$. Every set has size at most
$d$: non-exceptional sets by the choice of $s$, exceptional sets have size
$\sigma+1\le d$. Let $r_i$ be the number of elements of $s$ of frequency $i$.

In the take branch, $s$ and its elements are deleted, which removes
$w(d)+\sum_ir_iv(i)$, and every other set loses the elements it shares with $s$; these
are $\sum_i(i-1)r_i$ incidences. A set of size $j\le d$ that loses $t\ge1$ elements
loses $w(j)-w(j-t)=\sum_{l=j-t+1}^{j}\Delta w(l)\ge t\,\Delta w(d)$, since $1\le l\le d$
and~\ref{ds:W3}. This is the second entry of~\eqref{eq:ds-set}.

In the discard branch, $s$ is deleted, which removes $w(d)$, and each element of $s$ of
frequency $i$ loses $\Delta v(i)$ (for $i=2$ this is $v(2)$). This gives
$w(d)+\sum_ir_i\Delta v(i)$ before reductions. Let $E_2$ be the elements of $s$ of
frequency~$2$, and for $e\in E_2$ let $s_e\ne s$ be the other set containing~$e$. The
sets $s_e$ are distinct: if $s_e=s_{e'}$ for $e\ne e'$, then $N_I(e)=\{s,s_e\}=N_I(e')$
before branching, and~\ref{ds:R3} would have applied. We show that every $s_e$ is
deleted during the reductions that follow. Then the final measure is at most the
measure before reductions minus $\sum_{e\in E_2}w(d(s_e))\ge r_2w(1)$, because no weight
increases during reductions, each $s_e$ is a distinct vertex of weight at least $w(1)$
before reductions, and it is absent at the end.

During these reductions no element reaches frequency~$0$: elements of $s$ have
frequency at least~$1$ after the discard; \ref{ds:R2} and~\ref{ds:R3} delete elements
but do not lower the frequency of the remaining ones; and~\ref{ds:R1} lowers the
frequency of an element only if it lies in two identical sets, leaving frequency at
least~$1$. Suppose some $s_e$ is never deleted. We may assume that it is never removed
by~\ref{ds:R1} either, by following the merged representative, which has the same
elements. Then $e$ is not deleted by~\ref{ds:R2}, since that would delete its only set
$s_e$. If $e$ survives to the end, it has frequency~$1$ there, and~\ref{ds:R2} would
apply, a contradiction. So $e$ is deleted by~\ref{ds:R3}, because of an element $e_1$
with $\emptyset\ne N_I(e_1)\subseteq N_I(e)=\{s_e\}$; then $e_1\in s_e$ has frequency~$1$,
and we repeat the argument with $e_1$ in place of $e$. As there are finitely many
elements, this is a contradiction. So every $s_e$ is deleted, which completes the
proof of the first entry of~\eqref{eq:ds-set}.
\end{proof}

\begin{lemma}[high degrees]\label{lem:ds-high}
Assume~\ref{ds:W1}--\ref{ds:W3}. For every element or set vector of degree $d\ge10$
there is a vector of the same kind and of degree $d-1$ that is componentwise at most
the given one. Hence, if every vector of degree at most $9$ holds at $\alpha$, then
every vector holds at~$\alpha$.
\end{lemma}

\begin{proof}
By~\ref{ds:W1}, $\Delta v(i)=\Delta w(i)=0$ and $v(i)=v(7)$, $w(i)=w(7)$ for $i\ge8$.

\emph{Element vectors.} Let $d\ge10$. Delete one of the $d$ neighbors from $r$, and
replace every remaining neighbor of size $d$ by one of size $d-1$; the result is a
neighborhood $r'$ of $d-1$ sets of sizes at most $d-1$ with $r'_1\le r_1\le1$, so
$E_{d-1,r'}$ is an element vector ($d-1\ge9>5$). Since $d-1\ge9$, the weights
$v(d)=v(d-1)$, $\Delta v(d)=\Delta v(d-1)=0$, $w(d)=w(d-1)$ and
$\Delta w(d)=\Delta w(d-1)=0$ do not change, and deleting a set of size $i$ lowers the
two entries by $\Delta w(i)\ge0$ and $w(i)\ge0$.

\emph{Set vectors.} Let $d\ge10$, so $f(d)=d-1$ and $f(d-1)=d-2$. Delete one element
from $r$ and replace every remaining element of frequency $d-1$ by one of frequency
$d-2$; this gives a set vector $S_{d-1,r'}$. Since $d-2\ge8$, we have
$v(d-1)=v(d-2)$ and $\Delta v(d-1)=\Delta v(d-2)=0$; also $w(d)=w(d-1)$ and
$\Delta w(d)=\Delta w(d-1)=0$. Deleting an element of frequency $i$ lowers the entries
by $\Delta v(i)$ (plus $w(1)$ if $i=2$) and by $v(i)$, both nonnegative.

Since $\tau$ is decreasing in each argument, a vector componentwise at least a vector
that holds at $\alpha$ also holds at $\alpha$; induction on $d$ gives the last
statement.
\end{proof}

The bound $d\ge10$ cannot be lowered to $d\ge9$ for set vectors: at $d=9$, elements of
frequency $8$ would be replaced by elements of frequency $7$, and $\Delta v(7)>0=\Delta
v(8)$, so the discard entry may increase. Vectors of degree $9$ are therefore checked
directly.

\begin{lemma}[leaves]\label{lem:ds-leaves}
Let $\sigma$, $\mathcal X$ and the weights be as in \cref{lem:ds-branching}, and let
every element and set vector hold at $\alpha>1$. For every instance $J$ of measure $k$
with incidence graph $I$, the recursion tree of $J$ has at most $\alpha^{k-h}$ calls of
the dynamic program on instances of measure $h$, for every real $h\ge0$, and at most
$2(1+|V(I)|)$ times as many nodes as leaves.
\end{lemma}

\begin{proof}
Let $N_h$ be the number of calls of the dynamic program on instances of measure
exactly $h$ in the recursion tree of an instance $J$ of measure $k$. We show
$N_h\le\alpha^{k-h}$ by induction on the size of the recursion tree. If $J$ is itself a
leaf, $N_h\le1$, and $N_h=0$ unless $k=h$. If $J$ is reduced, its child has measure at
most $k$, and $N_h$ is the same. If $J$ branches, its two children, after
reductions, have measures at most $k-a$ and $k-b$ by \cref{lem:ds-branching}, so
$N_h\le\alpha^{k-a-h}+\alpha^{k-b-h}\le\alpha^{k-h}$. For the second statement, every
branching node has two children, so there are fewer branching nodes than leaves. The
reduction steps form chains, each ending at a branching node or a leaf, so there are at
most twice as many chains as leaves, and each chain has length at most $|V(I)|$, since
each step deletes a vertex.
\end{proof}

\paragraph{The instances handed to the dynamic program.} Let $J$ be an instance on
which the switch~\ref{ds:S} applies, and let $I$ be its incidence graph. Then every
non-exceptional set has size at most $\sigma$, every non-exceptional element has
frequency at most~$4$, every exceptional element has frequency $5$ and a type in
$\mathcal X$, and every set of size larger than $\sigma$ is exceptional and has size
$\sigma+1$. In particular $\Delta(I)\le5$. Let $I'$ be obtained from $I$ by deleting
the sets of size~$1$ and the isolated vertices.

\begin{lemma}[pendant vertices]\label{lem:ds-pendant}
With $I$ and $I'$ as above, $\pw(I)\le\pw(I')+1$.
\end{lemma}

\begin{proof}
Isolated vertices can be added as one-vertex bags. A set of size~$1$ is a vertex of
degree~$1$ adjacent to its element $f$. Take a path decomposition of $I'$, and for
each such vertex $s$ choose a bag $B\ni f$; insert directly after $B$ the bag
$B\cup\{s\}$, and insert the bags for vertices attached at the same $B$ one after
another. Every new bag has at most $|B|+1$ vertices, every edge $sf$ is covered, and
every vertex still occurs in consecutive bags. This is the argument
of~\cite[Proposition~28]{NederlofRD2014} and \cite[Remark~8.1]{vanRooij2011}.
\end{proof}

We now bound $\pw(I')$ by \emph{pathwidth forms}. A form is a nondecreasing function
$\varphi:\{1,\dots,5\}\to[0,1]$ such that every graph $H$ without isolated vertices and
with $\Delta(H)\le5$ satisfies
\begin{equation}\label{eq:ds-form}
  \pw(H)\le\sum_{x\in V(H)}\varphi\bigl(d_H(x)\bigr)+\rho_\varphi\bigl(|V(H)|\bigr)
\end{equation}
with a remainder $\rho_\varphi(t)=o(t)$, which we take nondecreasing
(\cref{sec:prelim-graphs}). A form is \emph{computable} if for every such $H$ a path
decomposition of $H$ of width at most the right side of~\eqref{eq:ds-form} can be
computed in polynomial time. We use the following forms, written as
$(\varphi(1),\dots,\varphi(5))$; the decimal values are rational upper bounds for the
constants, as used in the certificates below.
\begin{itemize}
  \item $\varphi_{\tripS}=(0,0,\frac5{32},\frac14,\frac{11}{32})$, by \cref{thm:main}
  (or \cref{thm:deg5} with $\tripS$);
  \item $\varphi_{\tripR}=(0,0,0.1455,0.2530,0.3600)$ and
  $\varphi_{\tripP}=(0,0,0.140159,0.2575,0.3667)$, by \cref{thm:deg5}, using
  $F_3\le0.140159$ (\cref{prop:F3F4});
  \item $\varphi_{a}=(F_4,F_4,F_4,F_4,1)$ with $F_4\le0.24577197475$
  (\cref{prop:F3F4}): by \cref{thm:deg4}\ref{thm:deg4:a} applied to $H-V_5$, where
  $V_5$ is the set of vertices of degree~$5$, and $\pw(H)\le\pw(H-V_5)+|V_5|$
  (\cref{sec:prelim-graphs}); \cref{thm:deg4}\ref{thm:deg4:a} counts every vertex of
  $H-V_5$, whatever its degree;
  \item $\varphi_{b}=(0,0,0.140159,0.2575,1)$: by \cref{thm:deg4}\ref{thm:deg4:b} applied
  to $H-V_5$;
  \item $\varphi_{3}=(0,0,0.139823,1,1)$: by \cref{lem:n3} applied to $H-V_{\ge4}$,
  where $V_{\ge4}$ is the set of vertices of degree at least~$4$, with
  $\cthree\le0.139823$.
\end{itemize}
In the last three items, let $X$ be the deleted set, $V_5$ or $V_{\ge4}$. The degrees
in $H-X$ are at most the degrees in $H$, and the cited bounds have coefficients
nondecreasing in the degree, so~\eqref{eq:ds-form} holds with the degrees of $H$. All
these forms are computable: a decomposition of the stated width is computed
deterministically in polynomial time by \cref{thm:algo-precise}\ref{it:algo:D} or by
\cref{lem:n3}, and in the last three items $X$ is then added to all bags.

Assign to each vertex of $I'$ a \emph{nominal degree}: a set gets its size, a
non-exceptional element its frequency, and an exceptional element of type $c$ gets
$5-[c_1>0]$, its degree after the deletion of its set of size~$1$ (it has at most one).
The nominal degree is at least the degree in $I'$, since an element lies in at most one
set of size~$1$ (two such sets would be identical); the two differ only for elements in
a set of size~$1$. Since each form is nondecreasing and $I'$ has no isolated vertices
and maximum degree at most~$5$, $\pw(I')$ is at most the sum of the form over the
nominal degrees of the vertices of $I'$, plus the remainder.

\paragraph{The linear program.} For a leaf instance of measure $h>0$, divide the
numbers of vertices of each kind by $h$: let $x_i$ ($1\le i\le\sigma+1$) be the scaled
number of sets of size $i$, $y_i$ ($2\le i\le4$) that of non-exceptional elements of
frequency $i$, and $e_c$ ($c\in\mathcal X$) that of exceptional elements of type $c$.
These numbers satisfy
\begin{align}
  &\sum_{i=1}^{\sigma+1}w(i)\,x_i+\sum_{i=2}^{4}v(i)\,y_i+v(5)\sum_{c}e_c=1,
  \label{eq:ds-lp-measure}\\
  &\sum_{i=1}^{\sigma+1}i\,x_i=\sum_{i=2}^{4}i\,y_i+5\sum_ce_c,
  \label{eq:ds-lp-edges}\\
  &\sum_c c_j\,e_c\le j\,x_j\qquad(1\le j\le\sigma+1),
  \label{eq:ds-lp-upper}\\
  &x_{\sigma+1}\le\sum_c c_{\sigma+1}\,e_c,
  \label{eq:ds-lp-lower}
\end{align}
and all variables are nonnegative. Here~\eqref{eq:ds-lp-measure} is the measure
(sets of size~$0$ and elements of frequency~$0$ have weight~$0$, and elements of
frequency~$1$ do not occur), and~\eqref{eq:ds-lp-edges} counts the edges of $I$ from
both sides. Inequality~\eqref{eq:ds-lp-upper} holds because the $\sum_cc_je_c$
incidences between exceptional elements and sets of size $j$ are at most $j$ per such
set, and~\eqref{eq:ds-lp-lower} holds because every set of size $\sigma+1$ is
exceptional and so contains an exceptional element.

For a form $\varphi$ let
$\Phi_\varphi(x,y,e)$ be the sum of $\varphi$ over the nominal degrees, scaled:
\[
  \Phi_\varphi=\sum_{i=2}^{\sigma+1}\varphi(i)\,x_i+\sum_{i=2}^4\varphi(i)\,y_i
  +\sum_c\varphi\bigl(5-[c_1>0]\bigr)\,e_c ,
\]
where sets of size~$1$ contribute nothing. Let $L(\sigma,\mathcal X,\Psi)$ be the
maximum of $\min_{\varphi\in\Psi}\Phi_\varphi$ over the polytope
\eqref{eq:ds-lp-measure}--\eqref{eq:ds-lp-lower}, for a set $\Psi$ of forms. This is the
linear program of \cite[Lemma~8.12]{vanRooij2011} and
\cite[Lemma~30, pp.~729--731]{NederlofRD2014}, with several forms and with the
constraints~\eqref{eq:ds-lp-upper}--\eqref{eq:ds-lp-lower} for a general threshold.

\begin{lemma}[pathwidth at the switch]\label{lem:ds-dp}
Let $\sigma$, $\mathcal X$ and the weights be as in \cref{lem:ds-branching}, let $\Psi$
be a finite set of computable forms, and let $L\ge L(\sigma,\mathcal X,\Psi)$. Then
for every instance of measure $h$ on which $\mathcal A_{\sigma,\mathcal X}$ calls the
dynamic program, a path decomposition of its incidence graph of width at most
$L\,h+o(h)$ can be computed in polynomial time, where the $o(h)$ depends only on the
weights and on $\Psi$.
\end{lemma}

\begin{proof}
The scaled counts of a leaf instance are feasible, so
$\min_\varphi\Phi_\varphi\le L$, that is, $\min_\varphi\sum_x\varphi(\cdot)\le Lh$
before scaling. By the discussion of nominal degrees and~\eqref{eq:ds-form},
$\pw(I')\le Lh+\max_\varphi\rho_\varphi(|V(I')|)$, and $\pw(I)\le\pw(I')+1$ by
\cref{lem:ds-pendant}. Every vertex of $I'$ has weight at least
$\min\{v(2),w(1)\}>0$ in the measure, so $|V(I')|\le h/\min\{v(2),w(1)\}$, and the
remainder is $o(h)$. The decomposition is the narrowest of those computed for the
forms, followed by the insertions of \cref{lem:ds-pendant}.
\end{proof}

\paragraph{Certificates for $L$.} Write the program in terms of the vector
$z=(x,y,e)$: \eqref{eq:ds-lp-measure} as $\mathbf m^{\top}z=1$, \eqref{eq:ds-lp-edges}
as $\mathbf a^{\top}z=0$, \eqref{eq:ds-lp-upper}--\eqref{eq:ds-lp-lower} as $Bz\le0$,
and $\Phi_\varphi(z)=\mathbf f_\varphi^{\top}z$. By weak duality, a bound
$L\ge L(\sigma,\mathcal X,\Psi)$ is certified by $\lambda_\varphi\ge0$
($\varphi\in\Psi$) with $\sum_\varphi\lambda_\varphi=1$, a real $\mu$ and a vector
$\nu\ge0$ such that, componentwise,
\begin{equation}\label{eq:ds-dual}
  \sum_{\varphi\in\Psi}\lambda_\varphi\,\mathbf f_\varphi\le L\,\mathbf m+\mu\,\mathbf a+B^{\top}\nu .
\end{equation}
Indeed, the inner product of~\eqref{eq:ds-dual} with a feasible point $z\ge0$ gives
$\min_\varphi\Phi_\varphi(z)\le\sum_\varphi\lambda_\varphi\Phi_\varphi(z)\le
L\,\mathbf m^{\top}z+\mu\,\mathbf a^{\top}z+\nu^{\top}Bz\le L$, since $\nu\ge0$ and
$Bz\le0$.

\begin{lemma}[running time]\label{lem:ds-time}
Let $\sigma$, $\mathcal X$ and the weights be as in \cref{lem:ds-branching}, let
$\Psi$ be a finite set of computable forms, and let $\alpha>1$ with
$2^{L(\sigma,\mathcal X,\Psi)}<\alpha$. If every branching vector of degree at most $9$
holds at $\alpha$, then $\mathcal A_{\sigma,\mathcal X}$, with the dynamic program run
on the narrowest decomposition of \cref{lem:ds-dp}, counts the dominating sets of each
size of an $n$-vertex graph in time and space
$O\bigl(\alpha^{(v(7)+w(7))n}\,n^{O(1)}\bigr)$.
\end{lemma}

\begin{proof}
Fix a real $L\ge L(\sigma,\mathcal X,\Psi)$ with $2^L<\alpha$. By \cref{lem:ds-high}
every branching vector holds at $\alpha$, so by
\cref{lem:ds-leaves} at most $\alpha^{k-h}$ leaves have measure $h$, where
$k\le(v(7)+w(7))n$ is the initial measure. By \cref{lem:ds-dp} and the dynamic program,
a leaf of measure $h$ takes time $2^{Lh+o(h)}n^{O(1)}$. Since $2^{L}<\alpha$ and the
remainder is $o(h)$, there is a constant $C$ with $2^{Lh+o(h)}\le C\alpha^{h}$ for all
$h\ge0$. The measure is determined by the numbers of sets of each size $1,\dots,7$
(sizes at least $7$ together) and of elements of each frequency $2,\dots,7$; these are
at most $2n$, so at most $(2n+1)^{13}$ values of $h$ occur. Every vertex of the
recursion tree costs polynomial time besides the dynamic program, and there are
polynomially many nodes per leaf by \cref{lem:ds-leaves}. Summing,
\[
  T\le n^{O(1)}\sum_{h}\alpha^{k-h}\,C\,\alpha^{h}\le n^{O(1)}\,\alpha^{(v(7)+w(7))n}.
\]
The space is dominated by the tables of the dynamic program, at most
$2^{Lh+o(h)}n^{O(1)}\le C\alpha^{h}n^{O(1)}$ at a leaf of measure~$h\le k$.
\end{proof}

\subsection{Proof of Theorem~\ref{thm:ds}}\label{sec:ds-proof}

The algorithm of \cref{thm:ds} is $\mathcal A_{4,\mathcal X_{30}}$. Compared with the
previous one, it switches to dynamic programming when every non-exceptional set has
size at most~$4$ (rather than~$3$) and every non-exceptional element has frequency at
most~$4$. Its exceptional elements are those with a type in $\mathcal X_{30}$, and its
exceptional sets are the sets of size~$5$ that contain one. The dynamic program runs
on the narrowest of the path decompositions for the forms
$\Psi=\{\varphi_{\tripS},\varphi_{\tripR},\varphi_{\tripP},\varphi_a,\varphi_b,\varphi_3\}$
listed before \cref{lem:ds-dp}, computed by \cref{thm:algo-precise}\ref{it:algo:D} and
\cref{lem:n3}, instead of the decomposition of Fomin, Gaspers, Saurabh and
Stepanov~\cite[Lemma~1]{FominGaspersSaurabhStepanov2009}. Correctness is
\cref{lem:ds-correct}.

\begin{table}[tb]
\centering
\caption{Weights of the measure~\eqref{eq:ds-measure}: previous
(\cite[p.~151]{vanRooij2011}, \cite[p.~728]{NederlofRD2014}), certified for the
previous algorithm with all cases
(\cref{prop:ds-repaired}), and certified for \cref{thm:ds}. All values are exact
decimals; $v(i)=v(7)$ and $w(i)=w(7)$ for $i\ge7$.}\label{tab:ds-weights}
\smallskip
\small
\setlength{\tabcolsep}{3pt}
\begin{tabular}{@{}lccccccc@{}}
\toprule
$i$ & 1 & 2 & 3 & 4 & 5 & 6 & 7\\
\midrule
previous $v(i)$ & 0 & 0.275603 & 0.551206 & 0.658484 & 0.719009 & 0.745509 & 0.749538\\
previous $w(i)$ & 0.354816 & 0.625964 & 0.785112 & 0.923325 & 0.986711 & 1 & 1\\
\midrule
repaired $v(i)$ & 0 & 0.2351251 & 0.4702502 & 0.5842398 & 0.6447407 & 0.6729014 & 0.6766715\\
repaired $w(i)$ & 0.3126782 & 0.5477033 & 0.6838767 & 0.8200501 & 0.8626895 & 0.8695487 & 0.8695487\\
\midrule
\cref{thm:ds} $v(i)$ & 0 & 0.2175171 & 0.4350342 & 0.5562242 & 0.6220752 & 0.6539177 & 0.6655099\\
\cref{thm:ds} $w(i)$ & 0.3062246 & 0.5338365 & 0.6754745 & 0.7616247 & 0.8178813 & 0.8324929 & 0.8324929\\
\bottomrule
\end{tabular}
\end{table}

\begin{proof}[Proof of \cref{thm:ds} for unweighted counting]
We apply \cref{lem:ds-time} with $\sigma=4$, $\mathcal X=\mathcal X_{30}$,
$\alpha=\frac{13}{10}$, the weights of \cref{tab:ds-weights} and the set $\Psi$ of six
computable forms above. Each step is checked by
the accompanying code, as follows.

\emph{Weights.} The weights are rational and satisfy~\ref{ds:W1}--\ref{ds:W4}, checked
in exact rational arithmetic. Explicitly, the increments
\begin{align*}
  (\Delta v(2),\dots,\Delta v(7))&=(0.2175171, 0.2175171, 0.1211900, 0.0658510, 0.0318425, 0.0115922),\\
  (\Delta w(1),\dots,\Delta w(7))&=(0.3062246, 0.2276119, 0.1416380, 0.0861502, 0.0562566, 0.0146116, 0)
\end{align*}
are nonnegative and nonincreasing, and $2\Delta v(5)=0.131702\le v(2)$. The list $\mathcal X_{30}$ satisfies~\eqref{eq:ds-list}.

\emph{Branching.} By \cref{lem:ds-branching} the branching vectors are the element
vectors $E_{d,r}$ ($d\ge5$) and the set vectors $S_{d,r}$ ($d\ge5$, frequencies at most
$d-1$). There are $30{,}972$ of them of degree at most~$9$. Their entries are exact
rationals; the code evaluates $\alpha^{-a}+\alpha^{-b}$ in interval arithmetic (mpmath,
120-bit precision), and the largest upper end is $0.999999610<1$. By \cref{lem:ds-high},
every branching vector holds at $\alpha$.

\emph{Dynamic programming.} The code computes an exact rational solution
of~\eqref{eq:ds-dual} for the program with $\sigma=4$, $\mathcal X_{30}$ and $\Psi$, which
certifies $L(4,\mathcal X_{30},\Psi)\le L$ with $L\approx0.378511274$, and
$2^{L}<1.2999997<\alpha$, by interval arithmetic. In floating point, the optimum is
attained at $x_4=y_4\approx0.7492$ with small amounts of sets of size $2$ and $3$ and of
exceptional elements of type $(0,1,4,0,0)$; so the worst instances are close to
incidence graphs of cubic graphs. The exact certificate puts weight only on
$\varphi_{\tripS}$ (\cref{thm:main}) and $\varphi_a$
(\cref{thm:deg4}\ref{thm:deg4:a} after deleting the vertices of degree~$5$), with
$\lambda_{\varphi_{\tripS}}=\frac{460610214}{535261133}\approx0.8605$ and
$\lambda_{\varphi_a}=\frac{74650919}{535261133}\approx0.1395$, so it also certifies
the bound for these two forms alone; with $\varphi_{\tripS}$ alone the value is
$0.3794062>\log_2\alpha$ in floating point.

\emph{Combination.} By \cref{lem:ds-time} the running time is
$O(\alpha^{(v(7)+w(7))n}n^{O(1)})$ with $v(7)+w(7)=1.4980028$, and
$\alpha^{1.4980028}<1.481452<1.4815$, which absorbs the polynomial factor. The
algorithm is deterministic, since \cref{thm:algo-precise}\ref{it:algo:D} and
\cref{lem:n3} are.
\end{proof}

Since the exact certificate puts weight only on $\varphi_{\tripS}$ and $\varphi_a$, it
also certifies the bound with $\Psi=\{\varphi_{\tripS},\varphi_a\}$; with this $\Psi$
the proof uses, among the results of this paper, only \cref{thm:main},
\cref{thm:deg4}\ref{thm:deg4:a} and \cref{thm:algo-precise}. The other forms are kept
because adding a form to $\Psi$ never raises $L(\sigma,\mathcal X,\Psi)$, and its
decomposition is computed in polynomial time.

\subsection{Weighted versions}\label{sec:ds-weighted}

Let the vertices of $G$ have real weights, and let the set
$\Sigma=\{\sum_{v\in X}\mathrm{wt}(v):X\subseteq V(G)\}$ of subset sums have size
polynomial in~$n$. We count
dominating sets by size and weight; the least weight with a nonzero count, and that
count, solve Minimum Weight Dominating Set and \#Minimum Weight Dominating Set. The
algorithm and its measure stay the same; only the bookkeeping changes.

The difficulty is the identical-sets rule~\ref{ds:R1}: true twins of different weights
give identical sets of different weights, so after merging a multiplicity no longer
describes the copies. We therefore replace the multiplicity of a set $s$ by the
generating polynomial
\[
  P_s(x,z)=\prod_{c}\bigl(1+x\,z^{\mathrm{wt}(c)}\bigr)-1
\]
of the nonempty selections of its copies $c$ (vertices of $G$), by number ($x$) and
weight ($z$); initially $P_{N[v]}=x\,z^{\mathrm{wt}(v)}$. All copies have the same
elements, so every nonempty selection covers the same elements. The algorithm returns
a polynomial $\sum_{\ell,\omega}c_{\ell,\omega}x^\ell z^\omega$, and the rules become:
\begin{itemize}
  \item \ref{ds:R1}: the merged set gets $(1+P_{s_1})(1+P_{s_2})-1$;
  \item \ref{ds:R2}, the take branch of~\ref{ds:B1}, and the dynamic program when it puts
  $s$ into the solution: multiply by $P_s$ in place of~\eqref{eq:ds-take}; discarding $s$
  multiplies by~$1$;
  \item \ref{ds:R3} and~\ref{ds:B2} involve no weights and are unchanged.
\end{itemize}

\emph{Correctness.} Different sets have disjoint sets of copies, since the rules only
merge or remove sets. Call a set $Y$ of copies a \emph{cover} if
the sets of its copies have union $\mathcal U$; its size is $|Y|$ and its weight is
$\sum_{c\in Y}\mathrm{wt}(c)$. By induction on $|V(I)|$, the algorithm returns the
generating polynomial of the covers by size and weight; the recursion terminates,
since every rule produces instances with fewer vertices of $I$. The nonempty selections of copies of a set $s$ have generating
polynomial $P_s$, and each covers exactly the elements of $s$. \ref{ds:R1}: a selection
of copies of the merged set is a pair of selections of copies of $s_1$ and of $s_2$.
\ref{ds:R2} and the take branch of~\ref{ds:B1}: a cover that contains a copy of $s$ is
a nonempty selection of copies of $s$ together with a cover of $\mathcal U\setminus s$
by the other sets, and in~\ref{ds:R2} every cover contains a copy of $s$. The discard
branch of~\ref{ds:B1}: the covers without a copy of $s$ are the covers by
$\mathcal S\setminus\{s\}$. \ref{ds:R3}: every cover contains a copy of a set
containing $e_1$, which contains $e_2$, so a set of copies covers $\mathcal U$ if and
only if it covers $\mathcal U\setminus\{e_2\}$. \ref{ds:B2}: a set of copies that covers
$\mathcal U\setminus\{e\}$ either covers $e$ or has no copy of a set containing $e$,
and the latter are the covers of $\mathcal U\setminus\{e\}$ by
$\mathcal S\setminus N_I(e)$. \ref{ds:S}: the dynamic program
of~\cite[Proposition~27]{NederlofRD2014} handles each set with its polynomial when the
set is processed, multiplying by $P_s$ when it puts $s$ into the solution; an empty set
$s$ contributes the factor $1+P_s$ (any selection of its copies, in place of
$(1+x)^{\operatorname{mult}(s)}$), and an element of frequency~$0$ gives~$0$. Every
polynomial
that occurs is the generating function of selections of pairwise disjoint sets of
vertices of $G$ (products combine disjoint sets of copies), so its exponents lie in
$\{0,\dots,n\}\times\Sigma$, and every arithmetic operation on these polynomials takes
time polynomial in $n$ and $|\Sigma|$. The measure, the recursion tree
and the decompositions are those of the unweighted case, so the running time is again
$O(1.4815^n)$. This proves the weighted part of \cref{thm:ds}.

Nederlof, van Rooij and van Dijk state only Minimum Weight Dominating
Set~\cite[Corollary~32]{NederlofRD2014}, with a proof that refers to their
polynomial-space algorithm, which has no merged sets; van Rooij's thesis also states
the counting version~\cite[Corollary~8.14]{vanRooij2011}.

The accompanying code checks this bookkeeping (not the running time) against brute force on small random graphs with true twins of different weights.

\subsection{Remarks}\label{sec:ds-remarks}

\begin{remark}[variants]\label{rem:ds-variants}
The following values come from the model of \cref{sec:nrd-analysis} (recurrences and
linear program), adapted to variants of the algorithm. Certified values are marked; the others are floating-point
optima of the weights at $\alpha=1.3$ (the objective $\alpha^{v(7)+w(7)}$ does not
depend on the scale of the weights).
\begin{itemize}
  \item With the previous list $\mathcal X_{14}$ and $\sigma=4$, the same proof gives
  $O(1.4835^n)$ (certified: base $<1.483408$, with $L\approx0.378511246$ and $30{,}988$
  vectors). With the form $\varphi_{\tripS}$ of \cref{thm:main} alone, the bases are
  $1.4837$ ($\mathcal X_{14}$) and $1.4818$ ($\mathcal X_{30}$).
  \item The improvement comes from handing sets of size~$4$ to the dynamic program.
  With the previous $\sigma=3$, the new width bounds do not lower the running-time
  bound (as noted after \cref{prop:ds-repaired}); with the coefficients of Fomin,
  Gaspers, Saurabh and Stepanov~\cite[Lemma~1]{FominGaspersSaurabhStepanov2009}, the
  switch at $\sigma=4$ gives only $1.5997^n$.
  \item Other thresholds are worse: switching when sets have size at most~$3$ and
  elements frequency at most~$5$ gives $1.4930^n$; sets at most~$4$ and elements at
  most~$5$, $1.5117^n$; sets at most~$5$ and elements at most~$4$, $1.5240^n$ (all
  without exceptional elements). With weights that are constant only from degree~$9$
  on, the optimum for $\mathcal A_{4,\mathcal X_{30}}$ is unchanged.
\end{itemize}
The list $\mathcal X_{30}$ was found by a greedy search; we did not try to optimize it
further.
\end{remark}

\begin{remark}[status of the computations]\label{rem:ds-status}
The floating-point optimizer only produces candidate weights; every constant in the
proofs of \cref{prop:ds-repaired} and \cref{thm:ds} is certified as described there.
A separate recheck, with code written apart from the main script, of all branching
vectors up to degree~$12$ ($1{,}747{,}700$ vectors) found the largest value
$0.99999961$ of $\alpha^{-a}+\alpha^{-b}$, consistent with \cref{lem:ds-high}.
\end{remark}

\subsection{A gap in the previous analysis}\label{sec:nrd-gap}

This subsection describes a gap in the previous running-time analysis of the
algorithm of Nederlof, van Rooij and van Dijk for counting dominating
sets~\cite{NederlofRD2014}, which also appears in van Rooij's
thesis~\cite[\S8.5]{vanRooij2011}. In the notation of \cref{sec:apps-ds}, it is the
algorithm $\mathcal A_{3,\mathcal X_{14}}$, with switching threshold $\sigma=3$ and
the previous list $\mathcal X_{14}$ of exceptional types (\cref{sec:nrd-algorithm}).
The previous recurrences do not cover branchings on a set of size~$4$ that contains
an element of frequency~$4$, and with the previous weights some of these cases
violate the claimed branching factor (\cref{prop:ds-gap}). We show that the
algorithm does branch on such sets, and that with all cases included it runs in time
$O(1.5004^n)$ (\cref{prop:ds-repaired}). The analysis of \cref{thm:ds} in
\cref{sec:ds-proof} uses the corrected set of recurrences.

Throughout, the notation is that of \cref{sec:nrd-algorithm,sec:nrd-analysis}: the
measure $k$ of~\eqref{eq:ds-measure} with weights $v$ and $w$ and their increments
$\Delta v$ and $\Delta w$, the set vectors $S_{d,r}$ and element vectors $E_{d,r}$
of~\eqref{eq:ds-set} and~\eqref{eq:ds-element}, the branching factor $\tau(a,b)$, and
the linear program for the pathwidth at the switch.

\paragraph{The previous analysis.}
The thesis \cite[Lemma~8.11, p.~152]{vanRooij2011} and the journal version
\cite[Lemma~29, pp.~728--729]{NederlofRD2014} state that, with the previous weights
of \cref{tab:ds-weights} (\cite[p.~151]{vanRooij2011}, \cite[p.~728]{NederlofRD2014}),
the branching produces at most $1.26089^{k-h}$ subproblems of measure~$h$, where $k$ is
the measure of the input instance. The linear program of
\cite[Lemma~8.12]{vanRooij2011} and \cite[Lemma~30]{NederlofRD2014} has value
$0.325782$, so that the dynamic program on a subproblem of measure $h$ runs in time
$O(1.25334^{h})$. From these, \cite[Theorem~8.13]{vanRooij2011} and
\cite[Theorem~31]{NederlofRD2014} conclude the running time
$O(1.26089^{(0.749538+1)n})=O(1.5002^n)$.

We reproduced the previous analysis exactly, including the value $0.325782$ of the
linear program. With the previous weights, every branching vector of
\cref{lem:ds-branching} for $\sigma=3$ has $\tau\le1.260893$, except four set vectors
$S_{4,r}$ of degree~$4$, where $r_i$ is the number of elements of the set that have
frequency~$i$.

\begin{proposition}[the uncovered case]\label{prop:ds-gap}
For $\sigma=3$ and the previous weights, the set vector $S_{4,r}$ with $r_4=4$ is
\[
  \bigl(w(4)+4\Delta v(4),\ w(4)+4v(4)+12\Delta w(4)\bigr)=(1.352437,\ 5.215817),
\]
and $\tau(1.352437,5.215817)=1.27583>1.26089$. Also $\tau>1.26089$ for the set
vectors $S_{4,r}$ with $(r_3,r_4)$ equal to $(3,1)$, $(2,2)$ or $(1,3)$.
\end{proposition}

\begin{proof}
With the previous weights, $\Delta v(4)=0.658484-0.551206=0.107278$ and
$\Delta w(4)=0.923325-0.785112=0.138213$, so the entries are
$0.923325+0.429112=1.352437$ and $0.923325+2.633936+1.658556=5.215817$. The values of
$\tau$ were computed numerically by the accompanying code; $\tau(1.352437,5.215817)$
lies in $(1.2758,1.2759)$, since $1.2758^{-1.352437}+1.2758^{-5.215817}>1$ and
$1.2759^{-1.352437}+1.2759^{-5.215817}<1$. The largest value of $\tau$ for the other
three vectors is $1.27004$.
\end{proof}

\paragraph{The algorithm branches on these sets.} The switch
of~\cite[Algorithm~8.2]{vanRooij2011} requires $d(s)\le3$, so when $d(s)=4$ and
$d(e)\le4$ the algorithm branches on the set (rule~\ref{ds:B1}), as the journal version
says~\cite[pp.~726--727]{NederlofRD2014}. The case already occurs at the root for every
cubic graph without true twins, where every set has size~$4$, every element has
frequency~$4$ and no reduction rule applies. The coefficient $12\,\Delta w(4)$ of the
printed recurrence is the correct one, since no exceptional type in $\mathcal X_{14}$
has a set of size~$5$.

\paragraph{A repair.}
We do not know whether the algorithm is slower than $O(1.5002^n)$; the gap concerns
the proof. Including all cases of \cref{lem:ds-branching} and re-optimizing the
weights gives the following repair.

\begin{proposition}[the previous algorithm, repaired]\label{prop:ds-repaired}
The algorithm $\mathcal A_{3,\mathcal X_{14}}$ of Nederlof, van Rooij and van Dijk,
with the decompositions of~\cite[Lemma~1]{FominGaspersSaurabhStepanov2009}, counts the
dominating sets of each size of an $n$-vertex graph in time and space $O(1.5004^n)$.
\end{proposition}

\begin{proof}
We apply \cref{lem:ds-time} with $\sigma=3$, $\mathcal X=\mathcal X_{14}$ (which
satisfies~\eqref{eq:ds-list}), $\alpha=\frac{13}{10}$, the repaired weights of
\cref{tab:ds-weights}, and $\Psi=\{\varphi_\eps\}$, and check its hypotheses.

\emph{The form.} Let $\varphi_{\mathrm{FGSS}}=(0,0,\frac16,\frac13,\frac{13}{30})$ be
the coefficients of~\cite[Lemma~1]{FominGaspersSaurabhStepanov2009} for degrees
$1,\dots,5$, and let $\varphi_\eps=\varphi_{\mathrm{FGSS}}+\eps$ on every degree, for
a fixed $\eps>0$ chosen below. For fixed $\eps$, Lemma~1
of~\cite{FominGaspersSaurabhStepanov2009} gives a path decomposition of width at most
$\sum_x\varphi_{\mathrm{FGSS}}(d_H(x))+\eps|V(H)|+O(1)$ in polynomial time, so
$\varphi_\eps$ is a computable form.

\emph{The branching vectors.} The accompanying code checks~\ref{ds:W1}--\ref{ds:W4}
in exact rational arithmetic, and all $31{,}003$ branching vectors of degree at most
$9$ in interval arithmetic (mpmath, 120-bit precision); the largest upper end of
$\alpha^{-a}+\alpha^{-b}$ is $0.999999551$.

\emph{The linear program.} The code also computes an exact rational solution
of~\eqref{eq:ds-dual} for $\{\varphi_{\mathrm{FGSS}}\}$, which certifies
$L(3,\mathcal X_{14},\{\varphi_{\mathrm{FGSS}}\})\le L$ with $L\approx0.371341949$ and
$2^{L}<1.2935555<\alpha$. Every vertex counted in $\Phi_\varphi$ has weight at least
$m_0=\min\{v(2),w(1)\}$ in~\eqref{eq:ds-lp-measure}, so
$\Phi_{\varphi_\eps}\le\Phi_{\varphi_{\mathrm{FGSS}}}+\eps/m_0$ on the polytope and
$L(3,\mathcal X_{14},\Psi)\le L+\eps/m_0$. A fixed $\eps>0$ with
$2^{L+\eps/m_0}<\alpha$ exists because $2^L<1.2936<\alpha$.

\emph{The running time.} Finally $v(7)+w(7)=1.5462202$ and
$\alpha^{1.5462202}<1.500312$.
\end{proof}

The pathwidth bounds of this paper do not improve this running time. With the forms
$\varphi_{\tripS},\dots,\varphi_3$ listed before \cref{lem:ds-dp} in place of
$\varphi_{\mathrm{FGSS}}$, the optimum for $\sigma=3$ is still $1.500312$ in floating
point, because the dynamic program is not the bottleneck for $\sigma=3$.

\paragraph{The source of the gap.} The case range of the previous recurrences appears
to come from the polynomial-space analysis of the same
authors~\cite[Theorem~8.5]{vanRooij2011}, \cite[Theorem~11]{NederlofRD2014}, where set
branchings are needed only for elements of frequency at most $d(s)-1$; re-optimizing
the weights with that range reproduces the previous $1.50016^n$ in floating point.
The journal version has the same switching rule, weights and recurrences, so the gap is
there too. Two printing slips, which also occur in the thesis~\cite{vanRooij2011}, do
not affect the results: constraints~2--4 on the
weights~\cite[p.~728]{NederlofRD2014} are meant for $i\ge1$, $i\ge1$ and $i\ge2$, and
the subsumption rule~\cite[Algorithm~4, line~7]{NederlofRD2014} is printed with closed
neighborhoods, $N[e_1]\subseteq N[e_2]$, where~\ref{ds:R3} uses open ones.
{}
{}

\phantomsection
\addcontentsline{toc}{section}{References}
\bibliographystyle{alphaurl-onelink}
\bibliography{references}

\end{document}